\documentclass{aa}
\usepackage{graphicx}
\usepackage{txfonts}
\usepackage[usenames,dvipsnames]{xcolor}
\usepackage{adjustbox}
\usepackage{hyperref}
\usepackage[final]{pdfpages}
\usepackage{natbib}
\usepackage{indentfirst}
\usepackage{threeparttable, tablefootnote}
\usepackage{makecell}
\usepackage{amsmath}
\usepackage{gensymb}
\usepackage{booktabs}
\usepackage{longtable}
\usepackage{caption}
\usepackage{subcaption}
\usepackage{multirow}
\usepackage{bigdelim}
\usepackage{tabu}
\usepackage{pdflscape}
\usepackage{geometry}
\usepackage{pgffor}
\usepackage{multicol}
\usepackage{lscape}
\usepackage{placeins}
\usepackage[morefloats=500]{morefloats}
\definecolor{Blue}{rgb}{0,0,1}
\hypersetup{colorlinks, citecolor=Blue, linkcolor=Blue, urlcolor=Blue}

\DeclareUnicodeCharacter{2212}{-}

\begin{document}

   \title{Charting Circumstellar Chemistry of Carbon-rich AGB Stars\thanks{Tables \ref{tab:alma_line_detections} and \ref{tab:apex_line_detections} are available in electronic form at the CDS, along with the reprocessed ALMA cubes and line spectra, via anonymous ftp to \url{cdsarc.u-strasbg.fr} (\url{130.79.128.5}) or via \url{http://cdsweb.u-strasbg.fr/cgi-bin/qcat?J/A+A/}.}}

   \subtitle{I. ALMA 3\,mm spectral surveys}

   \author{R. Unnikrishnan
          \inst{1}
          \and
          E. De Beck\inst{1}
          \and
          L.-\AA. Nyman\inst{1,3,4}
          \and
          H. Olofsson\inst{1}
          \and
          W. H. T. Vlemmings\inst{1}
          \and
          D. Tafoya\inst{2}
          \and
          M. Maercker\inst{1}
          \and
          S. B. Charnley\inst{5}
          \and
          M.A. Cordiner\inst{6,7}
          \and
          I. de Gregorio\inst{4}
          \and
          E. Humphreys\inst{3,4}
          \and
          T. J. Millar\inst{8}
          \and
          M. G. Rawlings\inst{9,10}
          }

   \institute{Department of Space, Earth and Environment, Chalmers University of Technology, {SE-412 96 Gothenburg}, Sweden\\\email{ramlal.unnikrishnan@chalmers.se}
      \and
      {Department of Space, Earth and Environment, Chalmers University of Technology, Onsala Space Observatory (OSO), SE-439 92, Onsala, Sweden}
      \and
      Joint ALMA Observatory (JAO), Alonso de Córdova 3107, Vitacura 763-0355, Casilla 19001, Santiago, Chile
      \and
      European Southern Observatory (ESO), Alonso de Córdova 3107, Vitacura 763-0355, Santiago, Chile
      \and
      NASA Goddard Space Flight Center, 8800 Greenbelt Road, Greenbelt, MD 20771, USA
      \and
      Solar System Exploration Division, NASA Goddard Space Flight Center, 8800 Greenbelt Road, Greenbelt, MD 20771, USA.
      \and
      Department of Physics, Catholic University of America, Washington, DC 20064, USA
      \and
      Astrophysics Research Centre, School of Mathematics and Physics, Queen’s University Belfast, University Road, Belfast, BT7 1NN, UK
      \and
      Gemini Observatory / NSF’s NOIRLab, 670 N. A’ohoku Place, Hilo, Hawai’i, 96720, USA
      \and
      East Asian Observatory / James Clerk Maxwell Telescope, 660 N. A’ohoku Place, Hilo, HI 96720, USA
      }

   \date{Received 28 February 2023 / Accepted 4 December 2023}

   \abstract
   {Asymptotic giant branch (AGB) stars are major contributors to the chemical enrichment of the interstellar medium (ISM) through nucleosynthesis and extensive mass loss. Direct measures of both processes can be obtained by studying their circumstellar envelopes in molecular line emission. The relation between circumstellar molecular abundances and stellar elemental abundances is determined by atmospheric and circumstellar chemistry. Most of our current knowledge of these, in particular in a C-rich environment, is based on observations of the carbon star IRC~+10\,216.
   }
   {We aim to obtain a more generalised understanding of the chemistry in C-rich AGB circumstellar envelopes by studying a sample of three carbon stars, \object{IRAS~15194$-$5115}, \object{IRAS~15082$-$4808}, and \object{IRAS~07454$-$7112}, and observationally test the archetypal status often attributed to \object{IRC~+10\,216}.
   }
   {We performed spatially resolved, unbiased spectral surveys in ALMA Band 3 ($85-116$\,GHz). We estimated the sizes of the molecular emitting regions using azimuthally-averaged radial profiles of the line brightness distributions. We derived abundance estimates using a population diagram analysis for molecules with multiple detected lines, and using single-line analytical calculations for the others.
   }
   {We identify a total of 132 rotational transitions from 49 molecular species. There are two main morphologies of the brightness distributions: centrally-peaked (CS, SiO, SiS, HCN) and shell-like (CN, HNC, C$_2$H, C$_3$H, C$_4$H, C$_3$N, HC$_5$N, c-C$_3$H$_2$). The brightness distributions of HC$_3$N and SiC$_2$ have both a central and a shell component. The qualitative behaviour of the brightness distributions of all detected molecules, in particular their relative locations with respect to the central star,  is the same for the three stars, and consistent with those observed towards IRC~+10\,216. Of the shell distributions, the cyanopolyynes peak at slightly smaller radii than the hydrocarbons, and CN and HNC show the most extended emission. The emitting regions for each species are the smallest for IRAS~07454$-$7112, consistent with this object having the lowest circumstellar density within our sample. We find that, within the uncertainties of the analysis, the three stars present similar abundances for most species, also compared to IRC~+10\,216. We find that SiO is more abundant in our three stars compared to IRC+10\,216, and that the hydrocarbons are tentatively under-abundant in IRAS~07454$-$7112 compared to the other stars and IRC~+10\,216. Our estimated $^{12}$C/$^{13}$C ratios match well the literature values for the three sources and our estimated silicon and sulphur isotopic ratios are very similar across the three stars and IRC~+10\,216.
   }
 {The observed circumstellar chemistry appears very similar across our sample and compared to that of IRC~+10\,216, both in terms of the relative location of the emitting regions and molecular abundances. This implies that, to a first approximation, the chemical models tailored to IRC~+10\,216 are, at least, able to reproduce the observed chemistry in C-rich envelopes across roughly an order of magnitude in wind density.}


   \titlerunning{ALMA 3\,mm spectral surveys of C-rich AGB CSEs}
   \authorrunning{R. Unnikrishnan et al.}

   \maketitle

\section{Introduction}
\label{sec:Introduction}
Stars of low-to-intermediate zero-age-main-sequence mass ($\sim$0.8$M_\odot$ < $M$ < 8$M_\odot$) populate the asymptotic giant branch (AGB) during their late evolution, after the end of helium fusion in their cores. In this phase, stars are characterised by an inert carbon-oxygen core surrounded by shells of helium and hydrogen. Contraction of the core causes the outer layers of the star to expand, resulting in cool, luminous giants \citep[$T_\mathrm{eff} \lesssim3000$\,K, $L_{*}\sim10^{3}-10^{4}\,L_\odot$, $R_\mathrm{*} \sim 100 R_\odot$;][]{Herwig_2005}. Microscopic dust grains \citep[{with radius $a$ in the range $0.1\,\mu{\rm m} \leq a \leq {0.5}\,\mu$m}, e.g.][]{Groenewegen_1997, Hofner_2008, Yasuda_and_Kozasa_2012} formed in the cool, dense upper {stellar} atmosphere, are levitated by pulsations, accelerated by radiation pressure, and push gas radially outwards through dust-gas momentum exchange \citep{Hofner_2015}. This causes extensive mass loss {from the stellar surface. AGB mass-loss rates (hereafter MLR) range from} $10^{-8}$ {to} $10^{-4}\,M_\odot\,\mathrm{yr}^{-1}$, {dominating the evolution on the AGB \citep[e.g.][]{Hofner_and_Olofsson_2018}}. The winds of AGB stars are major sources of gas and dust feedback into the interstellar medium \citep[ISM, e.g.][]{Matsuura_et_al_2009, Tielens_2005}.

The mass loss creates an expanding envelope of chemically rich material around the star, called the circumstellar envelope (CSE). The radial density distribution of material in the CSE is a tracer of the temporal variations in the mass loss. The {elemental} compositions of gas and dust in the CSE {reflect} the composition of the atmosphere of the star at the time of the ejection of the circumstellar material. 

Based on their atmospheric carbon-to-oxygen elemental abundance ratio, AGB stars are classified into two main groups: C-type (C/O $>$ 1) and M-type (C/O $<$ 1). AGB stars in the mass range $1.5 - 4 M_\odot$ (at solar composition) evolve into C-type stars in the absence of hot-bottom burning \citep{Lattanzio_et_al_1997}, as the carbon produced in the helium-burning layer is brought up to the surface by convection \citep[e.g.][]{Di_Criscienzo_et_al_2016} during the third dredge-up \citep{Herwig_2005}. The excess of carbon over oxygen for {C-type stars} allows a wide range of C-bearing species to form in their CSEs, including long carbon chains and possibly also PAHs \citep[e.g.][]{Cherchneff_et_al_1992, Allain_et_al_1997}, {as well as} carbonaceous dust \citep[e.g.][]{kraemer2019,sloan1998,zijlstra2006}. {With 96 molecular species detected in carbon star CSEs so far, compared to only around 34 species found in oxygen-rich circumstellar environments, the CSEs of C-type stars are generally richer in chemistry than those of their M-type counterparts \citep{Agundez_2022, McGuire_2022, Cernicharo_et_al_2023_Mg_arXiv, Cernicharo_et_al_2023}}.

\begin{table*}[t]
   \caption{Source properties}
   \label{tab:source_properties}
   \centering
      \begin{tabular}{l l r r c c c c}
      \hline\hline & \\[-2ex]
      Source & \makecell{{Variable}\\Name} & \makecell{Distance$^{(a)}$\\(pc)} & \makecell{MLR$^{(b)}$\\($M_\odot$ yr$^{-1}$)} & $^{12}$C/$^{13}$C$^{(c)}$ & \makecell{$\varv_\mathrm{sys}$$^{(d)}$\\(km/s)} & \makecell{$\varv_\mathrm{exp}$$^{(e)}$\\(km/s)} & \makecell{{Period}\\{(days)}}\\
      \hline & \\[-2ex]
      IRAS 15194$-$5115  & II Lup    & {690 (+130/$-$90)}  & 1.6$\times$10$^{-5}$ & 6  & $-$15.0          & 21.5 & {575$^{(g)}$}         \\
      IRAS 15082$-$4808  & V358 Lup  & 1050 ($\pm$60) & 2.7$\times$10$^{-5}$ & 35 & $-$3.3           & 19.5  & {632$^{(h)}$}       \\
      IRAS 07454$-$7112  & AI Vol    & {580 (+70/$-$60})  & 3.4$\times$10$^{-6}$ & 17 & $-$38.7          & 13.0 & {511$^{(i)}$}        \\
      IRC~+10\,216        & CW Leo   & 190 ($\pm$20)   & 3.0$\times$10$^{-5}$ & 45$^{(f)}$ & $-$26.5$^{(f)}$    & 14.5$^{(f)}$ & {630$^{(j)}$} \\
      \hline
      \end{tabular}
      \tablefoot{$^{(a)}$distance estimates from \citet{Andriantsaralaza_et_al_2022}; 
      $^{(b)}$mass loss rates from \citet{Woods_et_al_2003}, scaled to the listed distance estimates from \citet{Andriantsaralaza_et_al_2022}; 
      $^{(c)}$circumstellar $^{12}$CO/$^{13}$CO ratios derived by \citet{Woods_et_al_2003} from CO radiative transfer modelling; 
      $^{(d)}$systemic velocities of the sources, from \citet{Woods_et_al_2003}; 
      $^{(e)}$expansion velocities of the source CSEs, from \citet{Woods_et_al_2003}; 
      $^{(f)}$The listed $^{12}$C/$^{13}$C ratio, $\varv_\mathrm{sys}$, and $\varv_\mathrm{exp}$ for IRC~+10\,216 are from \citet{Cernicharo_et_al_2000};
      $^{(g)}$ \citet{Feast_et_al_2003};
      $^{(h)}$ \citet{Whitelock_et_al_2006};
      $^{(i)}$ \citet{Whitelock_et_al_2006};
      $^{(j)}$ \citet{Menten_et_al_2012}.}
\end{table*}

The advent of high-angular resolution mm/sub-mm {interferometers} like the Atacama Large Millimeter/submillimeter Array (ALMA) has made it possible to explore the chemical complexity of AGB CSEs in unmatched detail. However, notwithstanding these significant advances in instrumentation, much of our current knowledge of AGB circumstellar chemistry is still based on observational results and modelling focussed on a single object, {the C-type} IRC~+10\,216 (CW~Leonis), owing primarily to its proximity ($120-190$\,pc), high MLR ($\dot{M}=2-4\times10^{-5}\,M_\odot$\,yr$^{-1}$), and molecular richness \citep[e.g.][]{Andriantsaralaza_et_al_2022, Agundez_et_al_2017, He_et_al_2008, Cernicharo_et_al_2000, Cordiner_and_Millar_2009, Groenewegen_et_al_2012, Guelin_et_al_2018, Pardo_et_al_2022, Patel_et_al_2011, Tenenbaum_et_al_2010, Velilla-Prieto_et_al_2019, Van_de_Sande_et_al_2019}.

Studies of the molecular chemistry in the CSE of IRC +10\,216, {from single-dish spectral surveys \citep[e.g.][]{Cernicharo_et_al_2000, Agundez_et_al_2012} to interferometric maps \citep[e.g.][]{Patel_et_al_2011, Velilla-Prieto_et_al_2015, Agundez_et_al_2017}} have revealed a host of information about circumstellar chemical pathways and the physical structure of its CSE, including complex substructure within its molecular emitting shells. Radiative transfer and chemical models of AGB CSEs which take into account the clumps and arcs seen in the emission have also been presented \citep[e.g.][]{Cordiner_and_Millar_2009, Agundez_et_al_2017, Van_de_Sande_et_al_2018, Van_de_Sande_and_Millar_2022}. Chemical modelling of the photochemistry in AGB CSEs has shed light on their diverse chemical pathways \citep[e.g.][]{Li_et_al_2014, Saberi_et_al_2019}, and {the effects of} potential binarity \citep[e.g.][]{Siebert_et_al_2022, Van_de_Sande_and_Millar_2022}.

There have only been very few detailed investigations into the circumstellar chemistry of C-rich AGB stars other than IRC~+10\,216. {\citet{Nyman_et_al_1993, Woods_et_al_2003} and \citet{Smith_et_al_2015} performed single-dish spectral surveys in mm/sub-mm wavelengths towards selected C-type AGB stars. A few} studies of larger source samples in fewer molecules have also been performed \citep[e.g.][]{Schoier_et_al_2006, Schoier_et_al_2007, Schoier_et_al_2013, Massalkhi_et_al_2019}.

A large variety of molecules present in AGB CSEs have rotational transitions at mm/sub-mm wavelengths. Analysis of these lines can help constrain the physics and chemistry of these sources. Therefore, spectral surveys in {this} wavelength regime offer an ideal method to study both the chemical composition and the kinematic and thermodynamic properties of AGB CSEs.

The single-dish spectral line survey by \citet{Woods_et_al_2003} observed the stars in our sample (Sect.~\ref{sec:The_Sources}) in the frequency range 85\,$-$\,266 GHz. However, to determine accurate molecular abundances, spatial information on the emitting regions is required. \citet{Woods_et_al_2003} used indirect methods such as modelling of photodissociation radii and scaling of the known emitting-region sizes for IRC~+10\,216 with the MLRs and expansion velocities of the respective sources, to estimate the sizes of the molecular emitting regions of their stars. It is in this context that the study of spatially-resolved data of a subset of C-type AGB stars, as presented in this work, {is highly} relevant. Our interferometric observations make it possible to directly obtain the spatial extents of the emitting regions for individual stars. Thus, along with mapping the physical conditions like density and temperature across the CSEs, the observations presented will also help provide better estimates of molecular abundances and isotopic ratios.

This paper is the first in a series of planned publications, which together aim to advance our understanding of the molecular chemistry in carbon-rich AGB CSEs by studying a sample of stars other than IRC +10\,216. The primary aim of this paper is to present ALMA spectral surveys of three carbon stars, showcasing the large number of detected spectral lines and their emission maps. Detailed non-LTE radiative transfer modelling is required to constrain molecular abundances to the best possible accuracy, and will be presented in {a subsequent} paper. In {the current} work, we adopt a qualitative approach using simplistic LTE models to obtain order-of-magnitude estimates of the abundances.

This paper is organised as follows. The sample is described in Sect.~\ref{sec:The_Sources}. The details of the observations and data processing procedures employed are described in Sect.~\ref{sec:Observations_and_Data_Reduction}. We present the line identifications and abundance estimates in Sect.~\ref{sec:Analysis_and_Results} and end with a {qualitative} discussion of the chemical and morphological similarities and differences between the three stars, also in comparison to IRC~+10\,216, in Sect.~\ref{sec:Discussion}.

\section{The sources}
\label{sec:The_Sources}
We have observed a sample of three C-type AGB stars, IRAS 15194$-$5115, IRAS 15082$-$4808, and IRAS 07454$-$7112, using ALMA. Owing to their high MLRs, they are expected to present strong emission from a large variety of molecules. The stars were chosen for their similar physical outflow properties (see Table~\ref{tab:source_properties}), both among themselves and in comparison to IRC~+10\,216. They possess expanding, largely spherical CSEs with similar expansion velocities, sampling roughly an order of magnitude in MLRs. They {sample a broad range of} $^{12}$C/$^{13}$C ratios (Table~\ref{tab:source_properties}), which {may be due to} differences in their nucleosynthetic histories. These properties make them ideal candidates for a comparative study of C-type CSEs. {The MLRs listed were obtained by scaling the values from \citet{Woods_et_al_2003} to the new distance estimates from \citet{Andriantsaralaza_et_al_2022}}.

IRAS 15194$-$5115 (II~Lup) is a J-type \citep{Smith_et_al_2015} Mira variable with a pulsation period of 575 days \citep{Feast_et_al_2003}. It is the third brightest carbon star at 12\,$\mu$m, surpassed only by IRC~+10\,216 and CIT~6 \citep{Nyman_et_al_1993}. IRAS 15082$-$4808 (V358~Lup) is a Mira variable with a period of 632 days \citep{Whitelock_et_al_2006}. It has an MLR comparable to that of IRAS 15194$-$5115 and IRC$+$10126. IRAS 07454$-$7112 (AI~Vol) is also a Mira variable with a pulsational period of 511 days \citep{Whitelock_et_al_2006}. It has a lower MLR than the other two stars, by approximately an order of magnitude. Based on their MLRs, all three stars in our sample are currently expected to be in the high-{MLR} phase at the end of their AGB evolution \citep{Vassiliadis_and_Wood_1993}.

\section{Observations and data reduction}
\label{sec:Observations_and_Data_Reduction}
\subsection{Observations}
\label{subsec:Observations}
Interferometric spectral survey observations in the frequency range $85 - 116$\,GHz towards the three sources were carried out {with} angular resolutions {in the range} 0\farcs7 - 1\farcs7 using the band 3 receiver \citep{Claude_et_al_2005} of the ALMA 12 m array in Cycle 2 (project code: 2013.1.00070.S; PI: Nyman, L. \AA.). {A detailed list of the observational and technical parameters is given in Table~\ref{tab:observational_details}}. The maximum recoverable scale (MRS) of the observations ranges from $4.3\arcsec-9.3\arcsec$, and the field of view from $54\arcsec-63\arcsec$. Each source was observed in five tunings, each with four 1.875\,GHz wide spectral windows, together covering the entire Band 3 frequency range {(Table~\ref{tab:observational_setup})}.  {Based on the expected line widths (Table~\ref{tab:source_properties}), a spectral resolution of 0.49 MHz ($\sim$1.6 km/s) was chosen for IRAS 07454-7112, while 0.98 MHz ($\sim$3.2 km/s) was used for the other two stars, in order} to resolve structure in the line profiles to a similar detail for the three sources.

\begin{table*}[t]
  \caption{Observational details}
  \label{tab:observational_details}
  \centering
     \begin{adjustbox}{width=18cm}
        \begin{tabular}{@{\extracolsep{5pt}}lcccccccccc@{}}
        \hline\hline & \\[-2ex]
        \makecell{Source\\(IRAS)} & \makecell{Date(s) of\\Observation} & \makecell{$N_\mathrm{ant}$$^{(a)}$} & \makecell{$B_\mathrm{min}$$^{(b)}$\\(m)} & \makecell{$B_\mathrm{max}$$^{(c)}$\\(m)}  &  \makecell{$\theta$$^{(d)}$\\(\arcsec)} & \makecell{MRS$^{(e)}$\\(\arcsec)} & \makecell{$t_\mathrm{on}$$^{(f)}$ (min)\\(per tuning)} & \makecell{PWV$^{(g)}$\\(mm)} & \makecell{$\Delta$v$^{(h)}$\\(km/s)} \\
        \hline & \\[-2ex]

        15194$-$5115 & 2015 September 15,16                        & 39                & 15.1 & 1574.4 & $0.7 - 0.9$ & 4.9 & 9.6 & 1.4              & 3.2 \\\\
        15194$-$5115$^{(i)}$ & 2016 September 22                & 37                & 15.1 & 3144.0   & 0.4 & 8.0 & 10.1 & 2.5                 & 3.3 \\\\
        15194$-$5115$^{(j)}$ & 2016 May 14                     & 41                & 15.1 & 639.6  & 1.3 & 13.3 & 10.1 & 2.7             & 3.3 \\\\

        15082$-$4808 & \makecell{2015 August 17$^{(k)}$\\2015 September 9$^{(l)}$} & \makecell{37\\36} & \makecell{43.3\\15.1} & 1574.4 & $0.7 - 1.2$ & \makecell{4.3\\5.0} & \makecell{9.6\\9.6}                  & \makecell{1.1\\3.8} & 3.2 \\\\

        15082$-$4808$^{(i)}$ & 2016 September 22                   & 37                & 15.1 & 3144.0  & 0.4 & 8.0 & 31.3 & 2.4                 & 3.3 \\\\
        15082$-$4808$^{(j)}$ & 2016 May 14                        & 41                & 15.1 & 639.6 & 1.3 & 13.3 & 15.6 & 2.8             & 3.3 \\\\

        07454$-$7112 & \makecell{2015 June 7$^{(m)}$\\2015 June 13$^{(n)}$}        & \makecell{36\\35} & 21.2 & 783.1 & $1.3 - 1.7$ & \makecell{9.3\\7.7} & 9.6                  & \makecell{0.5\\0.6} & {1.6} \\
        \hline
        \end{tabular}
     \end{adjustbox}{}
     \tablefoot{$^{(a)}$$N_\mathrm{ant}$: Number of antennae used; 
     $^{(b)}$$B_\mathrm{min}$: minimum baseline; 
     $^{(c)}$$B_\mathrm{max}$: maximum baseline; 
     $^{(d)}$$\theta$: Major axis (FWHM) of median synthesised beam; 
     $^{(e)}$MRS: Maximum Recoverable Scale; 
     $^{(f)}$$t_\mathrm{on}$: time on source per tuning; 
     $^{(g)}$PWV: Precipitable water vapour; 
     $^{(h)}$$\Delta$v: velocity resolution; 
     $^{(i)}$Extended configuration from 2015.1.01271.S; 
     $^{(j)}$Compact configuration from 2015.1.01271.S; 
     $^{(k)}$Tunings a, b, e; 
     $^{(l)}$Tunings c, d; 
     $^{(m)}$Tunings a, b, c; 
     $^{(n)}$Tunings d, e.}
\end{table*}

We also observed the three stars and IRC~+10\,216 using the Atacama Pathfinder Experiment telescope \citep[APEX;][]{gusten2006_apex}. All stars were observed in the full bandwidths of APEX bands 5 ($159-211$\,GHz) and 6 ($200-270$\,GHz). The SEPIA receiver was used for the band 5 observations \citep{billade2012_band5,belitsky2018_sepia}. For band 6, the SHeFI receiver \citep{vassilev2008_shefi} was used for IRAS 15194$-$5115, whereas the PI230 receiver\footnote{PI230 is a collaboration between the European Southern Observatory (ESO) and the Max-Planck-Institut für Radioastronomie (MPIfR).} was used for the other stars. Additionally, IRAS 15194$-$5115 was also observed in APEX band 7 (SHeFI, $272-376$\,GHz). {The Kelvin-to-Jansky conversion factor\footnote{\url{https://www.apex-telescope.org/telescope/efficiency/}} for the APEX lines is 34 (Band 5) and 40 (Bands 6 and 7), with $5-10\%$ uncertainty in each case \citep{Gusten_et_al_2006}}. Several lines from the APEX surveys have been used in this work (Table~\ref{tab:apex_line_detections}, Figs.~\ref{fig:APEX_lines_start}$-$\ref{fig:APEX_B7_lines_15194}). {The integrated intensities of these lines are listed in Table~\ref{tab:apex_line_detections}. The spectra of these lines are shown in Figs.~\ref{fig:APEX_lines_start}$-$\ref{fig:APEX_B7_lines_15194}}. The full APEX surveys will be presented in a forthcoming paper.

\subsection{Data reduction}
\label{subsec:Data_Reduction}
\begin{figure}[h]
   \includegraphics[width=9.2cm]{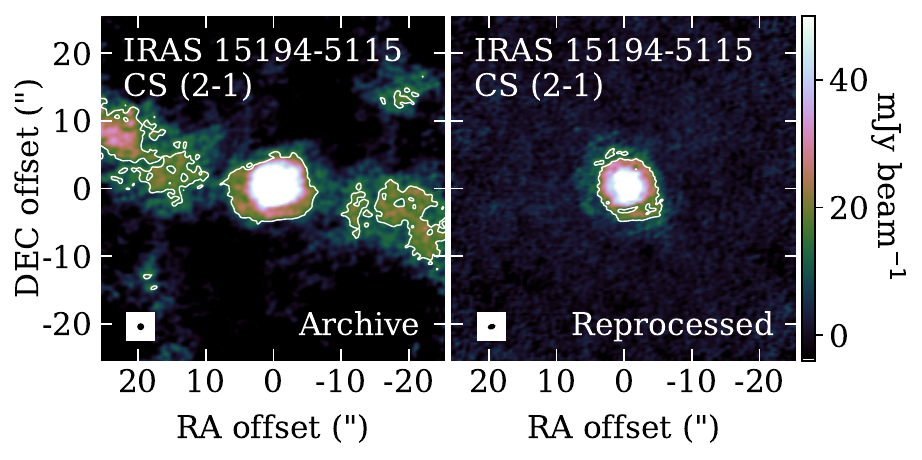}
   \caption{CS $J$\,=\,$2-1$ emission at 97.98 GHz towards IRAS 15194$-$5115, {in a 3.3 km/s wide channel centred on} the systemic velocity. {Left:} archival data, {right:} reprocessed data. The emission shown has been capped at 50\,mJy/beam to make the low-level extended artefacts clearly visible. The white contours are at 5\,mJy/beam. Synthesised beams are shown as filled black ellipses in the bottom left corner of each panel.}
   \label{fig:reprocessing_comparison}
\end{figure}

We retrieved the data stored in the ALMA Science Archive, which were calibrated and imaged using the Common Astronomy Software Applications (CASA) package \citep{McMullin_et_al_2007} versions 4.3 and 4.4. The archival image cubes of IRAS 15194$-$5115 exhibited weak, extended, symmetric artefacts that were not associated with the actual source structure (Fig.~\ref{fig:reprocessing_comparison}). Also, for this source, some of the spectral windows in tunings a and d {(Table~\ref{tab:observational_setup})} showed intensity mismatches for the lines falling within their overlapping frequency ranges. Such issues were partly caused by the fact that the archival images were produced using a now deprecated version of the CASA task {clean}, probably leading to inadequate cleaning at different spatial scales. In addition, we found that the flux scales used for the calibration of the archival data of several tunings of IRAS 15194-5115 and IRAS 15082-4808 were obtained using a model of the solar system object Ceres that was poorly defined \citep[see][and also Appendix C in the CASA User Manual\footnote{\url{https://casa.nrao.edu/docs/UserMan/casa_cookbook014.html}}]{Butler_2012}. This led us to use quasars (Table~\ref{tab:observational_setup}), originally serving as phase calibrators, instead of Ceres to calibrate the flux scales of these tunings. The spectral indices of these quasars were obtained by interpolating the corresponding flux values from the ALMA calibrator source catalogue\footnote{\url{https://almascience.eso.org/sc/}}.

With this change, we manually re-calibrated the whole dataset using CASA version 4.3.1. Water vapour radiometer data was used to correct atmospheric variations at the time of observations. The continuum was estimated using a first order polynomial fit to the line-free channels, and subtracted from the calibrated visibilities. Imaging was performed with the newer CASA version 6.3.0-48, using the Multi-Scale deconvolving parameter of the CASA task {tclean} \citep{Cornwell_2008}, and adopting a Briggs weighting scheme with robust parameter 0.5. The clean masks were set automatically using the {auto-multithresh} algorithm \citep{Kepley_et_al_2020}. This significantly improved the image quality over the archival image cubes (Fig.~\ref{fig:reprocessing_comparison}), and the extended artefacts present in the archival images of IRAS 15194$-$5115 were successfully removed by the reprocessing.

{The use of quasars instead of solar system objects as flux calibrators implies an absolute flux calibration uncertainty of $\sim5-10\%$ \citep[e.g.][]{Goddi_et_al_2019, Guzman_et_al_2019}. In this work, we adopt a conservative estimate of 10\%, also considering the fact that the observation times of our data and the measurements from the calibrator catalogue often differ by several days. With the reprocessing, the intensity mismatches between the lines in the overlapping tunings for IRAS 15194$-$5115 were reduced to within the absolute calibration uncertainty (10\%). In addition}, given that the channel maps in tuning d have more significant residuals after cleaning and the channel maps in tuning a show a better signal-to-noise ratio, we estimated a 20\% calibration uncertainty for the lines from tuning d of IRAS 15194$-$5115.

\begin{figure}[h]
   \includegraphics[width=8.75cm]{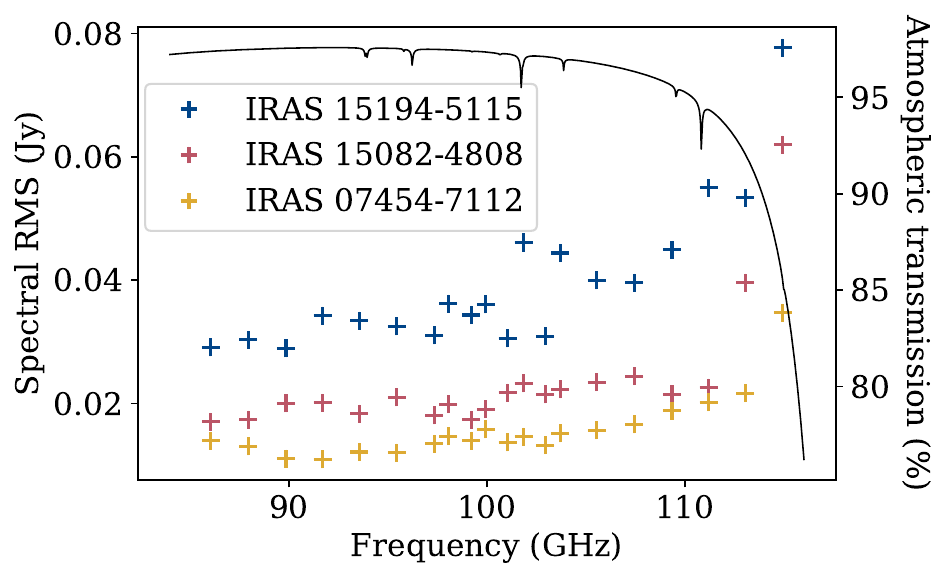}
   \caption{The rms noise in the spectra as a function of frequency. Each point denotes the average rms in a spectral window centred on the corresponding frequency. The black line is the atmospheric transmission at ALMA at 2\,mm PWV.}
   \label{fig:rms_per_tuning}
\end{figure}

Self-calibration was attempted in tunings with bright and compact lines available, but did not result in any significant improvement in signal-to-noise ratio, and was hence not implemented. {The observed continuum emission could not be used for self-calibration due to its low intensity}. The average beam sizes and spectral resolution of the final processed data are listed in Table~\ref{tab:observational_details}. Figure~\ref{fig:rms_per_tuning} shows the variation in the achieved rms in the different spectral windows, as a function of frequency. The atmospheric transmission in ALMA Band 3 is also shown, which corresponds well to the increase in rms at higher frequencies.

\begin{table*}[h]
   \caption{Observational setup and flux calibrators}
   \label{tab:observational_setup}
   \centering
      \begin{tabular}{c c c c c c}
      \hline\hline & \\[-2ex]
      \makecell{Tuning} & \makecell{LSB$^{(a)}$\\(GHz)} & \makecell{USB$^{(b)}$\\(GHz)} & \multicolumn{3}{c}{Flux calibrator$^{(c)}$} \\
      \cline{4-6} & \\[-2ex]
      & & & 15194$-$5115 & 15082$-$4808 & 07454$-$7112 \\
      \hline & \\[-2ex]
      a & $85.1 - 88.9$ & $97.1 - 100.9$ & J1617$-$5848$^{(d)}$  & J1514$-$4748$^{(d)}$  & J0519$-$454  \\
      b & $88.9 - 92.6$ & $100.9 - 104.6$ & J1617$-$5848$^{(d)}$  & J1514$-$4748$^{(d)}$  & Callisto \\
      c & $92.6 - 96.4$ & $104.6 - 108.4$ & J1617$-$5848$^{(d)}$  & Titan & Callisto \\
      d & $96.4 - 100.1$ & $108.4 - 112.1$ & J1617$-$5848$^{(d)}$  & Titan & J0519$-$454  \\
      e & $100.1 - 103.9$ & $112.1 - 115.9$ & J1617$-$5848$^{(d)}$  & J1514$-$4748$^{(d)}$  & J0519$-$454  \\
      \hline
      \end{tabular}
      \tablefoot{$^{(a)}$LSB: Lower side band; $^{(b)}$USB: Upper side band; {$^{(c)}$Flux calibrator used in the manual reprocessing (see Sect.~\ref{subsec:Data_Reduction}) of our dataset (ADS/JAO.ALMA\#2013.1.00070.S)}; $^{(d)}$For IRAS 15194$-$5115, the default flux calibrator Ceres was replaced with the quasar J1617-5848 in all five tunings. For IRAS 15082$-$4808, Ceres was replaced by J1514-4748 in tunings a, b, and e. For IRAS 07454$-$7112, no tunings had Ceres as the default flux calibrator. J1617-5848 and J1514-4748 were originally observed as phase calibrators for the corresponding tunings for IRAS 15194$-$5115 and IRAS 15082$-$4808, respectively.}
\end{table*}

\subsection{Data combination}
\label{subsec:data_combination}
\begin{figure}[h]
   \includegraphics[width=8.75cm]{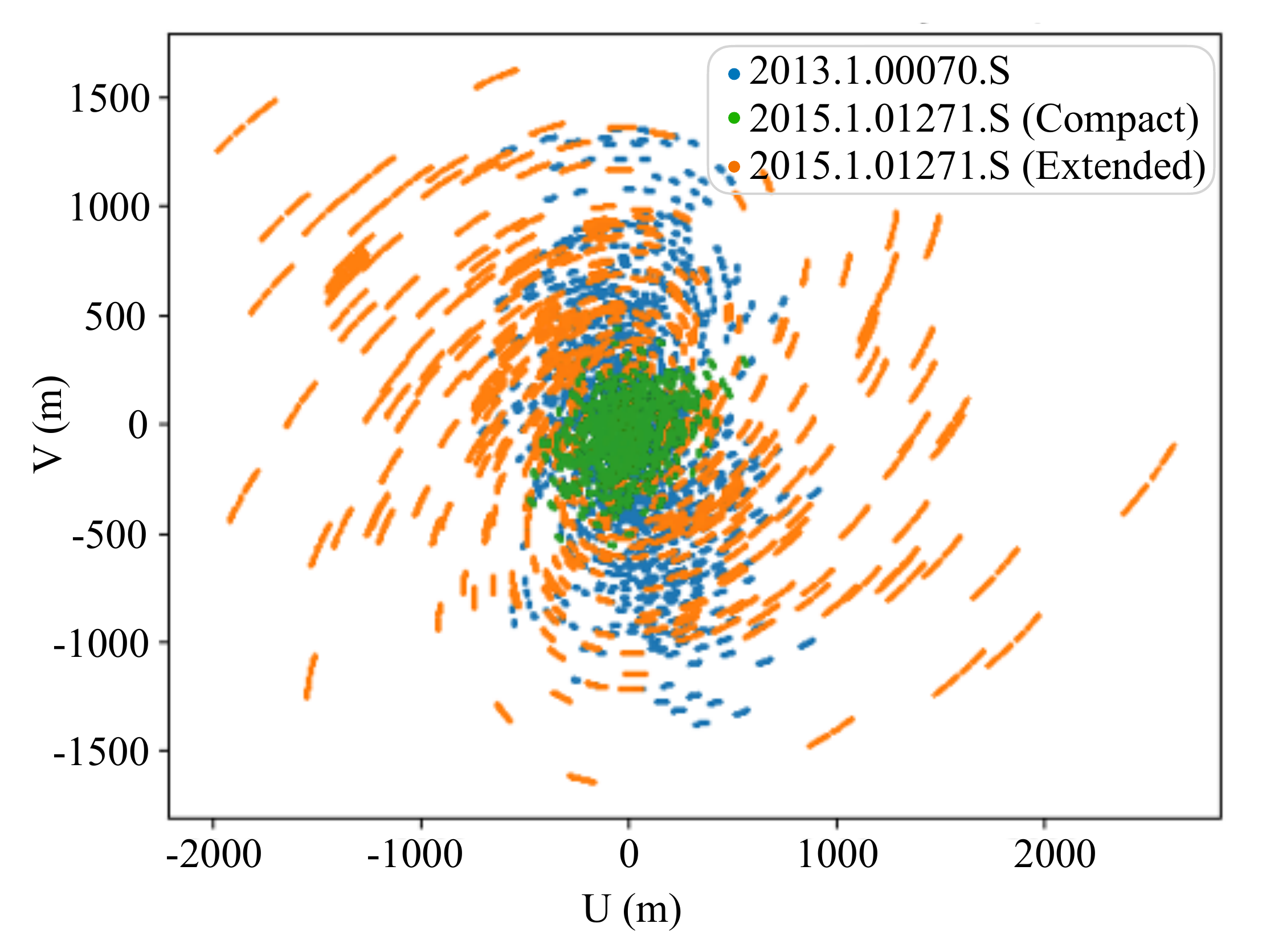}
   \caption{The $uv$ plane of the combined observations towards IRAS 15082$-$4808 in the frequency range $87.2 - 89.1$\,GHz. The data from project 2015.1.01271.S has both long (orange) and short (green) baseline configurations, which help to better recover the compact and extended emission, respectively, compared to our original observations.}
   \label{fig:uv_plane_concat}
\end{figure}

For IRAS 15194$-$5115 and IRAS 15082$-$4808, observations with higher spatial resolution {(0\farcs4)} were available in the ALMA archive (project code: 2015.1.01271.S, PI: Keller, D.), but covering only the frequency ranges $87.27-90.97$\,GHz and $99.14-102.89$\,GHz. The details of this dataset are given in Table~\ref{tab:observational_details}. This dataset includes observations with both more extended {(hereafter high angular resolution dataset)} and more compact {(hereafter low angular resolution dataset) baseline} configurations compared to our data {(hereafter intermediate angular resolution dataset)}. We therefore combined the corresponding visibilities from this project with our observations to increase the sensitivity to the emission {on} compact and extended spatial scales. {This helps to increase the MRS of the observations, enabling us to recover the weak, extended molecular emission on large spatial scales which could have been filtered out due to the lack of very short baselines in the intermediate resolution data before combination, while also lowering the rms noise in the combined maps}. The data combination was performed using the task {concat} in CASA version 6.3.0-48. Figure~\ref{fig:uv_plane_concat} shows the combined visibilities towards IRAS 15082$-$4808, depicting the extended and compact baseline coverages added.

{As these two datasets were observed several months apart, it can be argued that they might not be compatible due to a possible time variation of the sources related to (1) the evolution of the ejected material and (2) changes in the molecular excitation}. We note that any spatial variation in the circumstellar gas distribution occurring during the interval between these observations ($\leq$ 1 yr) will be far below the spatial resolution of both datasets. For example, in the case of IRAS 15194$-$5115, given the expansion velocity of 21.5 km/s at a distance of $\sim$696 pc (Table~\ref{tab:source_properties}), the increase in angular size in 1 year is only 6 mas, around two orders of magnitude less than the angular resolution of the observations. In contrast, by comparing the line intensities from the two epochs used in the data combination, we do detect line variability {likely connected to the variability in the stellar radiation field} in two species, C$_2$H and HC$_3$N, for both IRAS 15194$-$5115 and IRAS 15082$-$4808. The observed variability is quantified in Sect.~\ref{subsec:line_variability}, and its impact on our results is discussed in Sect.~\ref{sec:Discussion}. We still combine the two datasets for these lines, in order to trace the spatial structure in as much detail as possible, while increasing the uncertainty in the line intensities of these species by the observed percentage of variability.

\begin{figure*}
    \begin{subfigure}[t]{0.98\linewidth} \includegraphics[width=\linewidth]{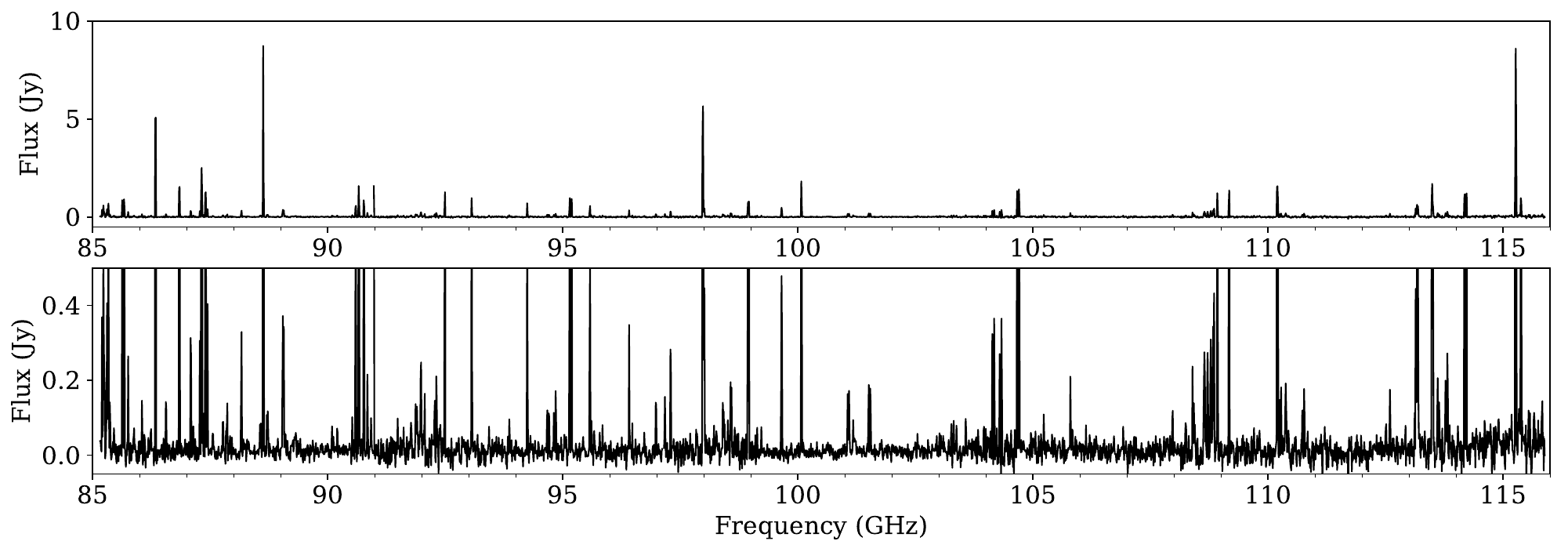}
    \label{subfig:15194_overview_spectrum}
    \end{subfigure}
    \begin{subfigure}[t]{0.98\linewidth} \includegraphics[width=\linewidth]{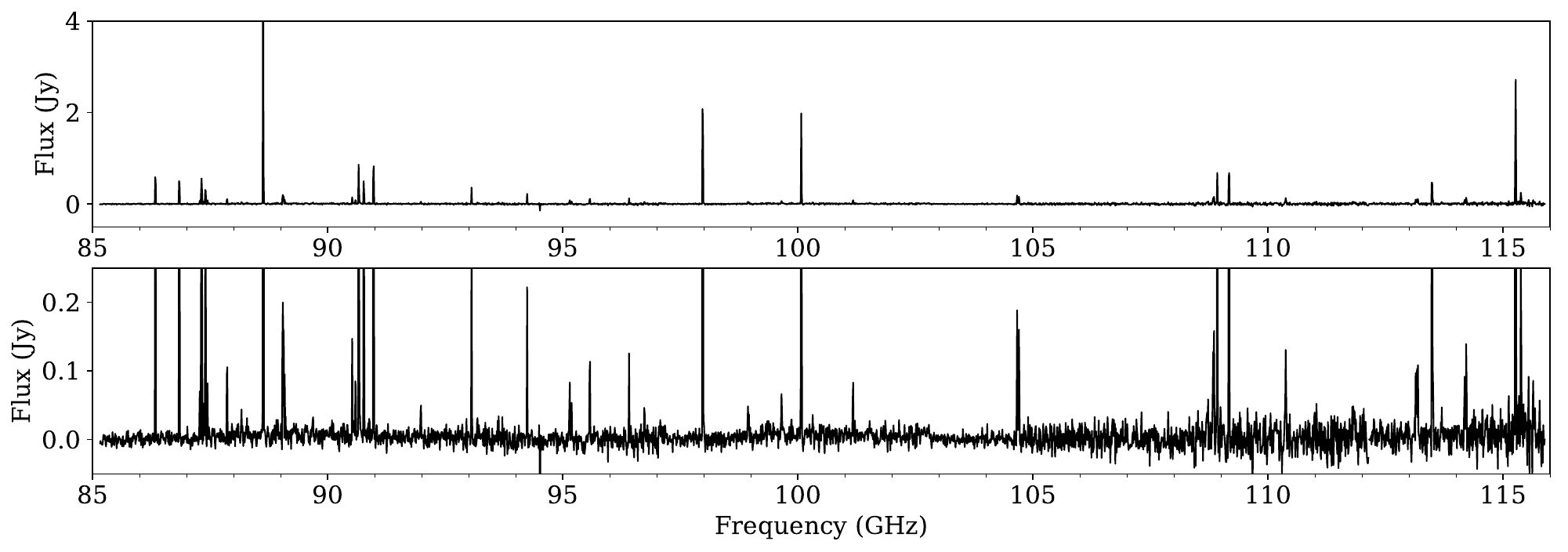}
    \label{subfig:15082_overview_spectrum}
    \end{subfigure}
    \begin{subfigure}[t]{0.98\linewidth} \includegraphics[width=\linewidth]{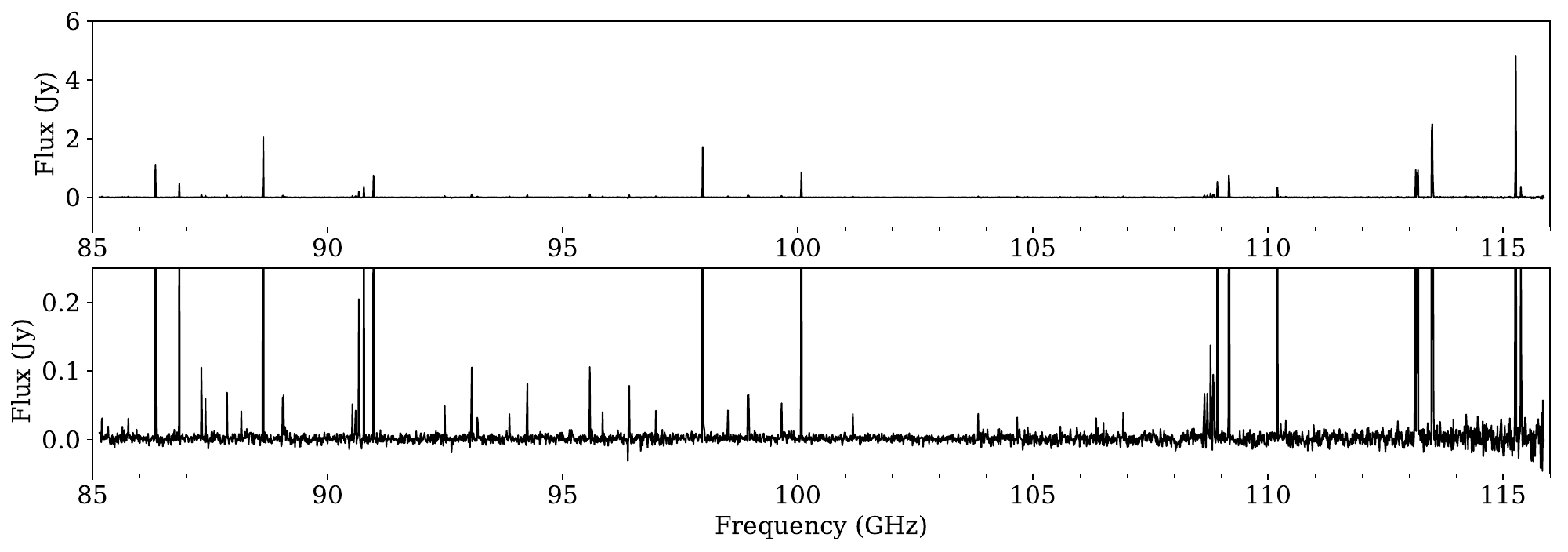}
    \label{subfig:07454_overview_spectrum}
    \end{subfigure}
    \caption{ALMA band 3 spectra extracted using a 12\farcs5 radius aperture centred on the star, for IRAS 15194$-$5115 (top), IRAS 15082$-$4808 (middle), and IRAS 07454$-$7112 (bottom). The spectrum in the lower panel for each star is capped at a low flux density limit to reveal the weaker emission lines. \label{fig:sample_overview_spectra}}
    \end{figure*}

\section{Analysis and results}
\label{sec:Analysis_and_Results}
\subsection{Line detection and identification}
\label{subsec:line_detections_and_identification}
{Continuum maps were produced from the line-free channels across the survey bandwidth (85-116 GHz) using multi-frequency synthesis \citep{Rau_and_Cornwell_2011}. The continuum emission is spatially unresolved for all three stars. The peak positions and total flux of the continuum maps are given in Table~\ref{tab:continuum_details}. The continuum was then removed from the data to obtain the continuum-subtracted line spectra}. 

Figure~\ref{fig:sample_overview_spectra} shows an overview of the ALMA band 3 spectra for all three stars. The full line-labelled ALMA Band 3 spectra of the three sources, extracted by integrating the intensity in a circular aperture of 12.5\arcsec\/radius centred on the respective continuum peaks, so as to recover emission from all molecules, are presented in App.~\ref{app:appendix_A}. We used the Weeds \citep{Maret_et_al_2011} package in GILDAS\footnote{\url{https://iram.fr/IRAMFR/GILDAS}} to identify the emission lines. The Cologne Database for Molecular Spectroscopy \citep[CDMS\footnote{\url{https://cdms.astro.uni-koeln.de}};][]{Muller_et_al_2005} catalogue was referred to for line identification. Splatalogue\footnote{\url{https://splatalogue.online}} \citep{Remijan_et_al_2007} was also used as an interface to obtain spectral line parameters. We detect, above the 3$\sigma$ noise level, a total of 314 spectral emission features across the three stars, out of which 311 could be identified and assigned to {132} known rotational transitions of 49 molecular species, including various isotopologues {(see Fig.~\ref{fig:almaband3_identified_sample} and Table~\ref{tab:alma_line_detections})}. All detected lines, along with the associated spectroscopic parameters and observed integrated intensities towards the three stars are tabulated in Table~\ref{tab:line_detections}. Figures in App.~\ref{app:appendix_C} show the {brightness distributions} at {the} systemic velocity and the spectra of the detected {lines}.

Most of the detected spectral lines are thermal emission lines from transitions in the vibrational ground states. However, some of the {lines} we detect, including HCN $\nu_2$=2, $J$=1$-$0, {$l$=0}, and $\nu_2$=1, $J$=2$-$1, {$l$=1} {are known to} display maser emission towards carbon stars, including IRAS 15082$-$4808 \citep[e.g.][]{Guilloteau_et_al_1987,Smith_et_al_2014,Menten_et_al_2018,Jeste_et_al_2022}. {We find both these lines to be masering towards all three stars in our observations as well, based on their extremely narrow linewidth and significantly different line shapes in comparison to other non-masering lines}. These maser lines are not a target of study in this paper and are not considered for the morphological or abundance analysis presented below. In addition, SiS maser action has been reported for $v$=0 rotational lines towards IRC~+10\,216 using 3D radiative transfer modelling based on mm observations \citep{Fonfria_et_al_2018}. 
{In the SiS lines that we detect in our ALMA/APEX surveys, there are no apparent maser spikes. In this work, we have therefore assumed that all of the observed SiS line emissions are of primarily thermal nature. We will come back on this issue in a forthcoming radiative transfer analysis}. Two of the three unidentified features in our data, at 86.886\,GHz and 100.407\,GHz, present narrow line profiles (Fig. \ref{fig:u_lines_spectra}) indicating that they could be masers and/or originate in the inner parts of the CSE.

Owing to the increased sensitivity and better coverage of spatial scales in the combined data {(see Sect.~\ref{subsec:data_combination})}, we detected several additional lines, including {lines from} C$_5$H, C$_6$H, C$_2$S, and l-C$_4$H$_2$, which were not detected in {the intermediate resolution} data alone. Further, the emission maps of several lines, including those of HC$_3$N, HCN, HNC, C$_3$N, and their $^{13}$C isotopologues, were enhanced by the data combination (e.g. Figs.~\ref{fig:HCN_app_B}, \ref{fig:C3N_cc_appendix_C}, \ref{fig:l-C4H2_cc_appendix_C}, \ref{fig:HNC_app_B}, \ref{fig:HC3N_app_B}), {owing to the same reasons as above}.

We report the first detections of lines of $^{29}$SiO, $^{30}$SiO, $^{29}$SiS, $^{30}$SiS, Si$^{34}$S, and C$^{33}$S towards the three sources. The source with the largest number of detections is IRAS 15194$-$5115, {presenting lines from 119 rotational transitions, compared to the 56 and 67 of IRAS 15082$-$4808, and IRAS 07454$-$7112, respectively (see Table~\ref{tab:line_detections})}. {This is largely due} to the high abundance of $^{13}$C in its CSE which leads to a large number of molecules containing $^{13}$C atoms, most of which are not detected towards the other two stars. We report the first detection of Si$^{13}$CC, and doubly $^{13}$C$-$substituted isotopologues of HC$_3$N towards IRAS 15194$-$5115, along with the long-chain hydrocarbons C$_6$H, C$_8$H, l$-$C$_4$H$_2$, which have not been detected towards this source in previous observations.

\begin{table}
   \caption{Details of continuum emission}
   \label{tab:continuum_details}
   \centering
   \begin{adjustbox}{width=9cm}
      \begin{tabular}{c c c c}
      \hline\hline & \\[-2ex]
      \multirow{2}[1]{*}{IRAS} & \multicolumn{2}{c}{Peak Position} & \multirow{2}[1]{*}{\makecell{{Flux}\\(mJy)}} \\
      \cline{2-3} & \makecell{RA\\ (h:m:s)} & \makecell{DEC\\ (d:m:s)} \\
      \hline & \\[-2ex]
      15194$-$5115 & 15:23:05.048 {$\pm$ 0.004} & $-$51:25:58.95 {$\pm$ 0.06} & 3.25 $\pm$ 0.04 \\
      15082$-$4808 & 15:11:41.453 {$\pm$ 0.005} & $-$48:19:58.95 {$\pm$ 0.08} & 1.84 $\pm$ 0.04 \\
      07454$-$7112 & 07:45:02.456 {$\pm$ 0.008} & $-$71:19:45.79 {$\pm$ 0.12} & 2.17 $\pm$ 0.03 \\
      \hline
      \end{tabular}
   \end{adjustbox}
      \tablefoot{{The listed RA uncertainties are in seconds and DEC uncertainties are in arcseconds}.}
\end{table}

\subsection{Line Variability}
\label{subsec:line_variability}
{The two datasets used for data combination (Sect.~\ref{subsec:data_combination}) were observed several months apart (see Table~\ref{tab:observational_details}). Hence, to check for potential variability, we compared spectra produced from a subset of visibilities that have the same $uv$-coverage in both datasets. This removes any difference in amplitudes caused by the larger $uv$-coverage of one of the datasets, and would reflect only time-dependent amplitude variations, {connected to the variability in the stellar radiation field}.

Owing to the 10\% absolute calibration uncertainty of our {data} (see Sect.~\ref{subsec:Data_Reduction}), and the ALMA band 3 standard calibration uncertainty of 5\%\footnote{See the ALMA Technical Handbook, \url{https://almascience.eso.org/proposing/technical-handbook}} for the {high and low angular resolution datasets} (see Sect.~\ref{subsec:data_combination}), we estimate a relative calibration uncertainty of $\sim$11\% between the two {observations}. For both IRAS 15194$-$5115 and IRAS 15082$-$4808, we find differences in line intensity greater than the relative calibration uncertainty only for two species, C$_2$H and HC$_3$N. For IRAS 15194$-$5115, we find a $\sim15-20\%$ change in the integrated line intensity of the {strongest hyperfine component of the} C$_2$H, $N$=1$-$0 line, and $\sim10-15\%$ in the HC$_3$N, $J$=11$-$10 line. For IRAS 15082$-$4808, we find a $\sim$25-30\% change for C$_2$H, $N$=1$-$0, while for HC$_3$N, $J$=11$-$10, the change is $\sim10-15\%$}. {The phase of the variability is also opposite for C$_2$H and HC$_3$N. See Sect.~\ref{sec:Discussion} for a brief discussion of the detected variability in comparison with that found in IRC~+10\,216}.

\subsection{Morphological diversity}
\label{subsec:Morphological_diversity}
Estimation of the extents of the emitting regions is needed in order to calculate molecular fractional abundances and constrain the chemical models. Fourteen species (see Table~\ref{tab:em_region_sizes}) were found to have non-blended lines with sufficiently high signal-to-noise ratios {($\geq$ 3$\sigma$)} in the channel maps to allow extracting information about the spatial distributions of the emission towards all three stars by producing azimuthally-averaged radial profiles (hereafter AARPs; see Sect.~\ref{subsubsec:AARPs}) of the line brightness distributions.

\begin{figure}[h]
   \centering
   \begin{subfigure}[b]{0.4\textwidth}
       \centering
       \includegraphics[width=\textwidth]{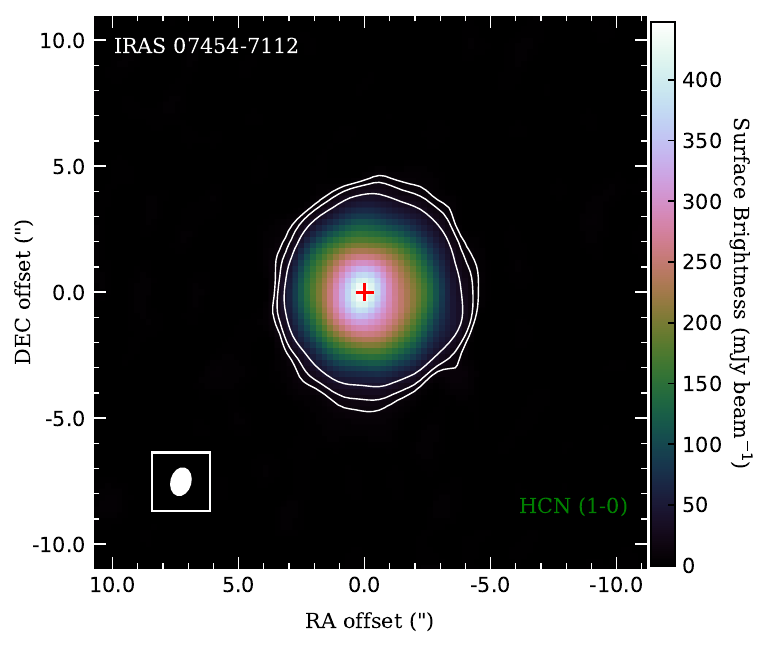}
       \label{subfig:HCN_cp_07454}
   \end{subfigure}
   \hfill
   \begin{subfigure}[b]{0.4\textwidth}
       \centering
       \includegraphics[width=\textwidth]{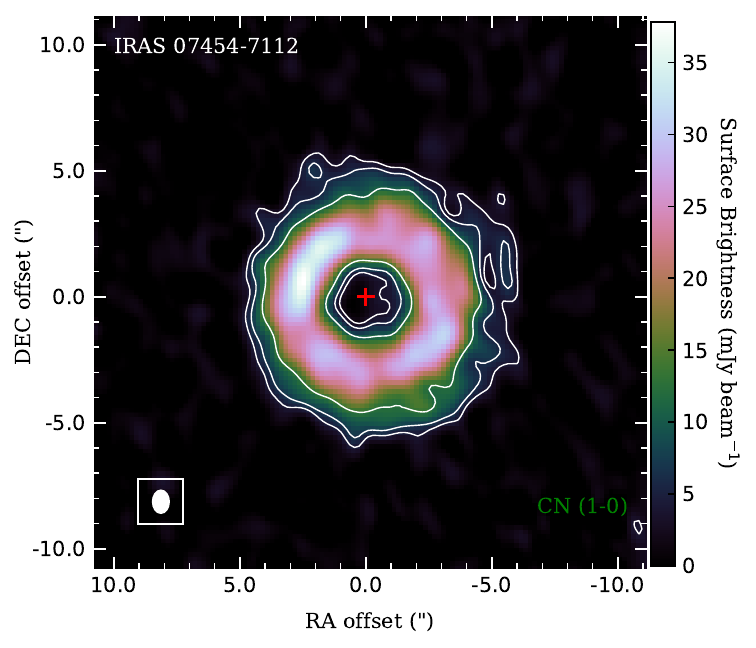}
       \label{subfig:CN_shell_07454}
   \end{subfigure}
      \caption{Examples of centrally peaked (top) and shell (bottom) morphologies. Maps show the emission at the systemic velocity from the HCN ($1-0$) line (top) and the CN ($1-0$, hyperfine components stacked) line (bottom) towards IRAS 07454$-$7112. Contours are at 3, 5, and 10 $\sigma$, where $\sigma$ refers to the rms noise in the emission maps ($\sigma$ = 1.9 mJy/beam (HCN), 1.2 mJy/beam (CN)). {The red $+$ denotes the stellar position (continuum peak)}.}
      \label{fig:morphology_comparison_07454}
\end{figure}

Four of these species, HCN, SiO, SiS, and CS, have line brightness distributions that peak close to the central star and extend radially outwards (Fig.~\ref{fig:morphology_comparison_07454}, top panel and Figs.~\ref{fig:SiO_app_B}, \ref{fig:HCN_app_B}, \ref{fig:SiS_app_B}, \ref{fig:CS_app_B}). These molecules are termed {centrally-peaked} species. {These encompass the so-called parent molecules formed in the stellar photosphere or the extended atmosphere, or molecules formed through circumstellar chemistry in the inner layer of the CSE} \citep[see][]{Millar_and_Herbst_1994, Cordiner_and_Millar_2009, Agundez_et_al_2017, Agundez_et_al_2020}. Eight species, C$_2$H, C$_3$H, C$_4$H, c-C$_3$H$_2$, CN, C$_3$N, HC$_5$N, and HNC, present brightness distributions in the form of rings {at the systemic velocity (e.g. Fig.~\ref{fig:morphology_comparison_07454}, bottom panel and Figs.~\ref{fig:C2H_app_B}, \ref{fig:HC5N_app_B}, \ref{fig:C3N_app_B}, \ref{fig:HNC_app_B}, \ref{fig:C4H_app_B}, \ref{fig:C3H_app_B}, \ref{fig:CN_app_B})}. {Our observations indicate} that these species are distributed around the stars as hollow spherical shells with widths and peak radii that vary between molecules and stars. Hereafter we call these molecules {shell} species. The shell species are {{daughter} molecules formed in the external layers of the CSE \citep[see][]{Agundez_et_al_2020}} from centrally-peaking {parent} molecules, by photodissociation (e.g. C$_2$H and CN) or due to photo-induced chemistry (e.g. HNC and HC$_5$N), {either directly or through series of intermediate steps} \citep[e.g.][]{Daniel_et_al_2012, Agundez_et_al_2017}. Two species, HC$_3$N and SiC$_2$, show {compound} brightness distributions that have both centrally-peaked and shell characteristics (Fig.~\ref{fig:line_em_comp_15194_15082}, top-left panel and Figs.~\ref{fig:SiC2_app_B}, \ref{fig:HC3N_app_B}). Maps of the emission at the systemic velocity for all the above species are shown in App.~\ref{app:appendix_C}.

\subsubsection{Complex morphology in the CSEs of IRAS~15194$-$5115 and IRAS 15082$-$4808}
\label{subsubsec:15194_complex_morphology}
\begin{figure*}[!h]
   \centering
   \begin{subfigure}[t]{0.41\textwidth}
       \includegraphics[width=\linewidth]{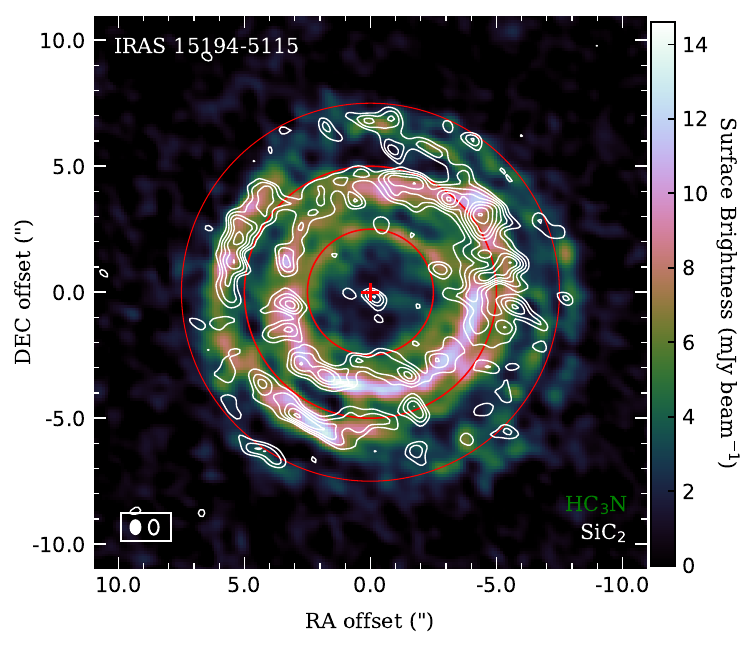}
   \label{subfig:15194_HC3N_SiC2_em_comp}
   \end{subfigure}
    \begin{subfigure}[t]{0.41\textwidth}
       \includegraphics[width=\linewidth]{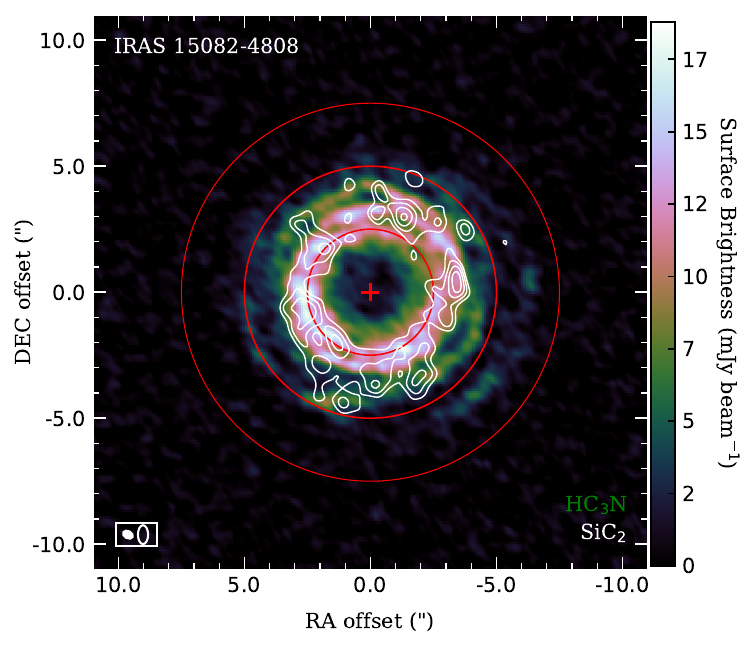}
   \label{subfig:15082_HC3N_SiC2_em_comp}
   \end{subfigure}
   \begin{subfigure}[t]{0.41\textwidth}
      \includegraphics[width=\linewidth]{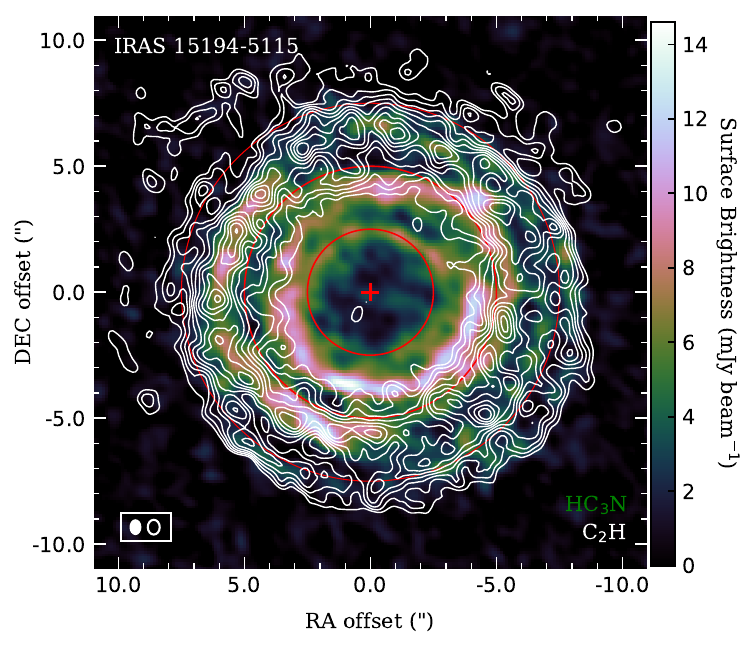}
  \label{subfig:15194_HC3N_C2H_em_comp}
  \end{subfigure}
   \begin{subfigure}[t]{0.41\textwidth}
      \includegraphics[width=\linewidth]{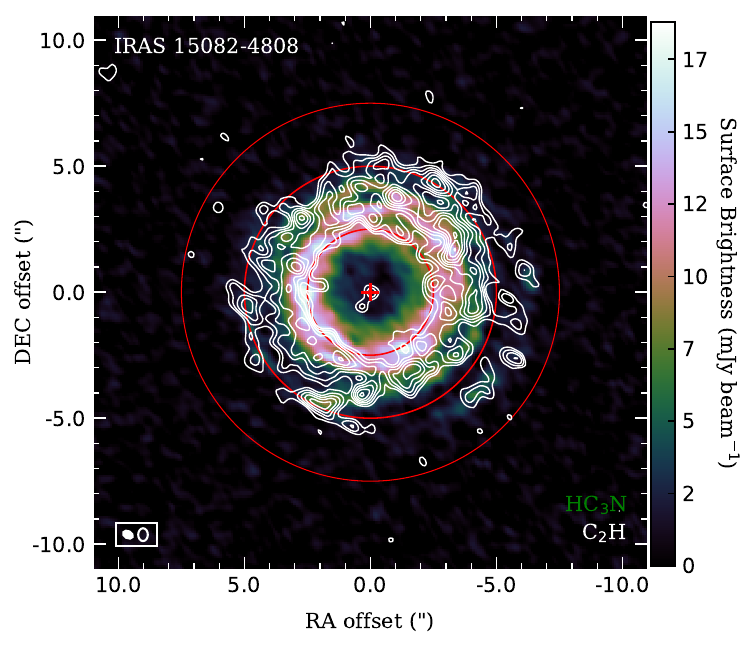}
  \label{subfig:15082_HC3N_C2H_em_comp}
  \end{subfigure}
  \begin{subfigure}[t]{0.41\textwidth}
     \includegraphics[width=\linewidth]{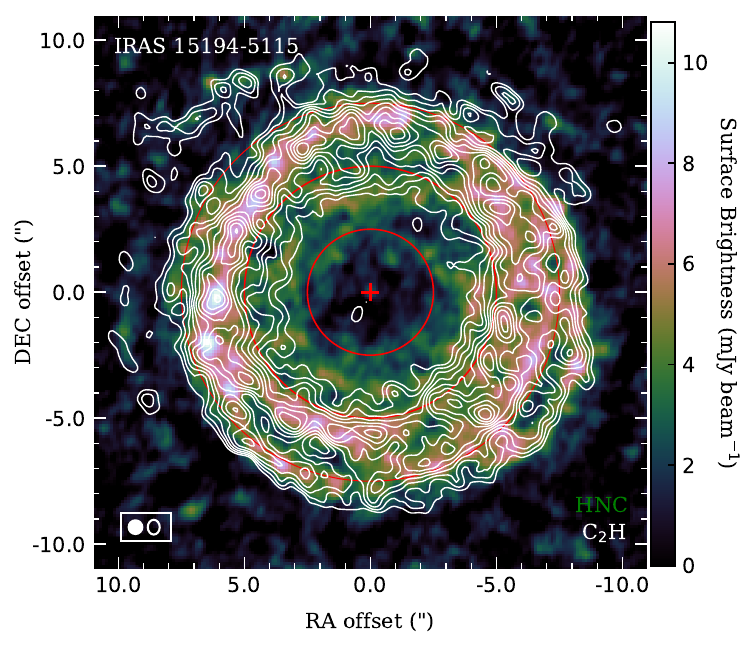}
 \label{subfig:15194_HNC_C2H_em_comp}
 \end{subfigure}
  \begin{subfigure}[t]{0.41\textwidth}
     \includegraphics[width=\linewidth]{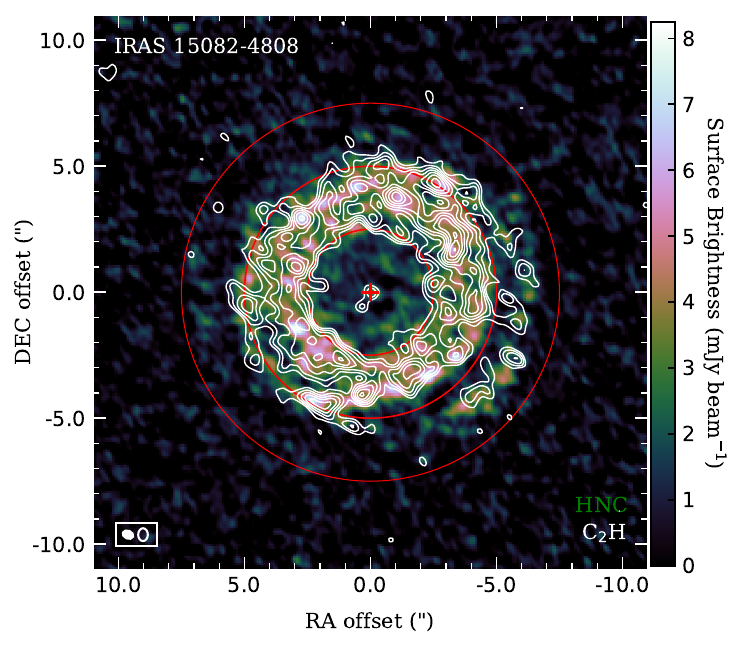}
 \label{subfig:15082_HNC_C2H_em_comp}
 \end{subfigure}
   \caption{Average brightness distributions, {in a 3.3 km/s wide channel centred on} the systemic velocity, of selected molecules towards IRAS 15194$-$5115 (left panels) and IRAS 15082$-$4808 (right panels). The names of the species whose emissions are shown in colour and contours are given at the bottom right corner of each panel in green and white, respectively. The synthesised beams of the colour maps and contours are shown as filled and hollow ellipses, respectively, at the bottom left corner of each panel. Contours are from $3\sigma$ to $15\sigma$ in intervals of $1\sigma$. The $\sigma$ values of the contour maps are, for IRAS 15194$-$5115: SiC$_2$: 0.68 mJy/beam, C$_2$H: 0.50 mJy/beam, and for IRAS 15082$-$4808: SiC$_2$: 1.1 mJy/beam, C$_2$H: 0.31 mJy/beam. The red $+$ denotes the stellar position (continuum peak). Red circles of radius 2\farcs5, 5\farcs0 and 7\farcs5 are drawn on each map to facilitate comparisons.}
   \label{fig:line_em_comp_15194_15082}
\end{figure*}

\begin{figure*}[h]
     \centering
     \begin{subfigure}[b]{0.475\textwidth}
         \centering
         \includegraphics[width=\textwidth]{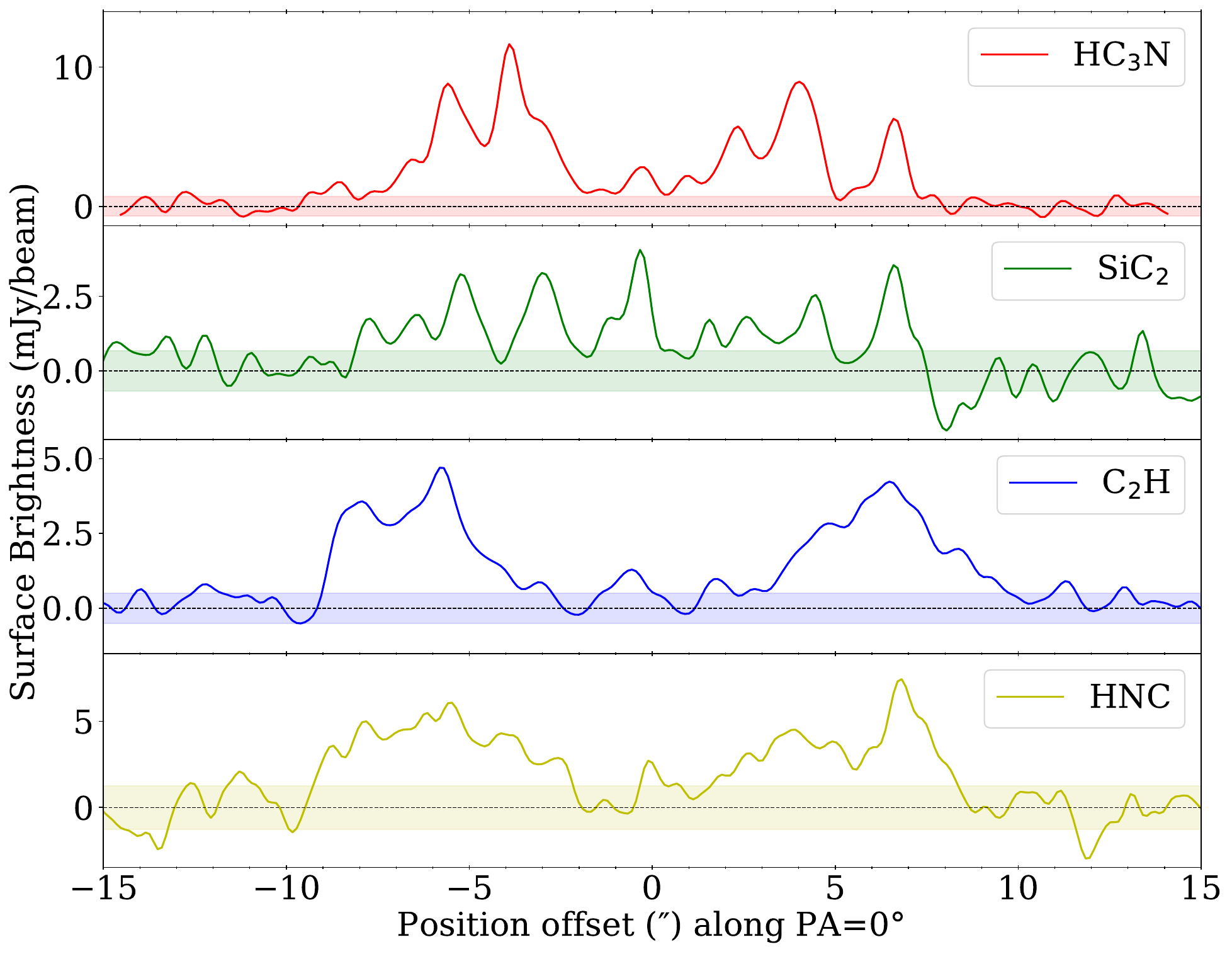}
         \label{subfig:15194_radial_Y_cuts}
     \end{subfigure}
    \hfill
     \begin{subfigure}[b]{0.475\textwidth}
         \centering
         \includegraphics[width=\textwidth]{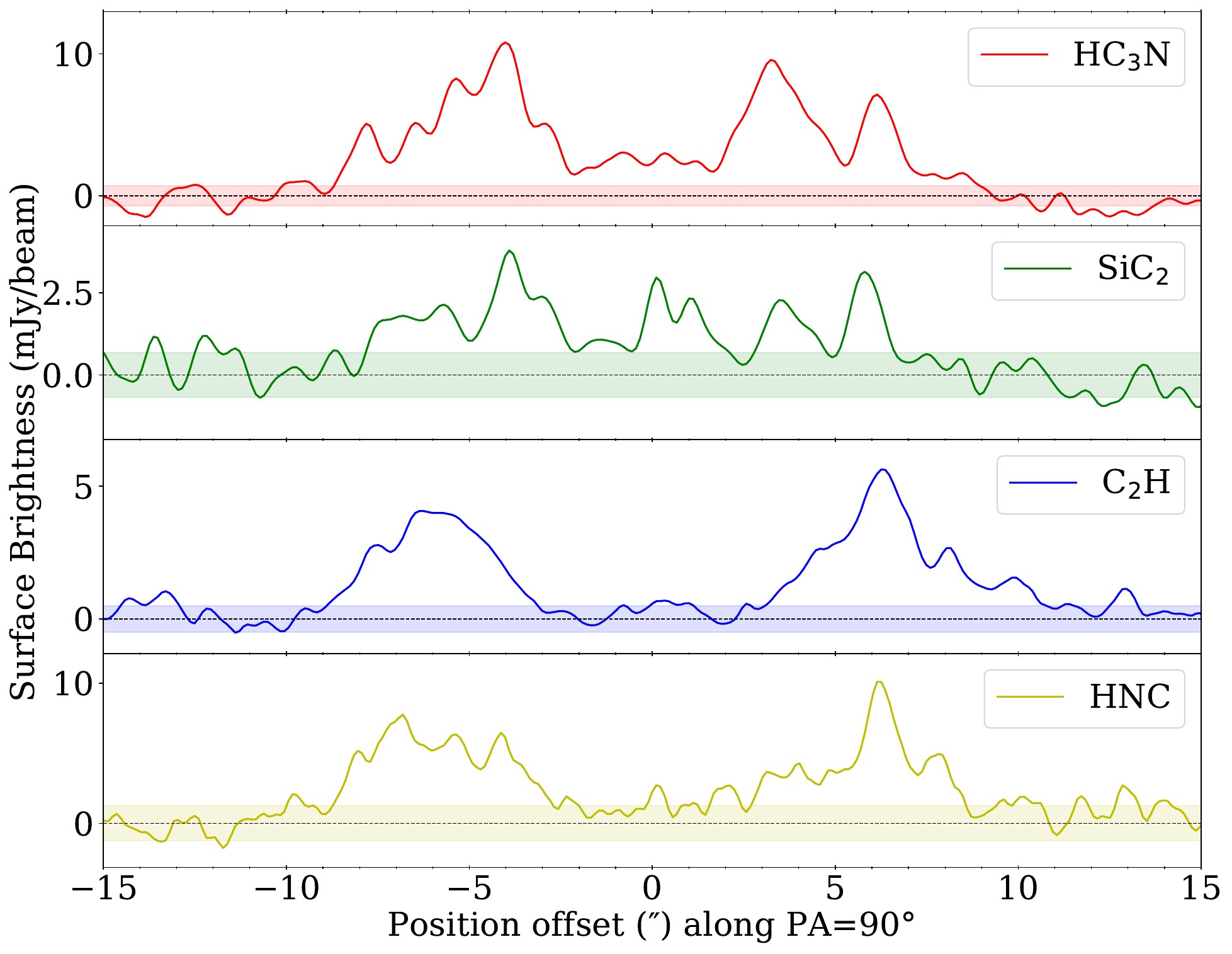}
         \label{subfig:15194_radial_X_cuts}
     \end{subfigure}
    \caption{Brightness distributions of HC$_3$N, SiC$_2$, C$_2$H, and HNC, {in a 3.3 km/s wide channel centred on} the systemic velocity for IRAS 15194$-$5115, along a line that intersects the star at PA=0$\degree$ (left), and PA=90$\degree$ (right). These profiles are extracted from radial cuts {using} the stacked images shown in Fig. \ref{fig:line_em_comp_15194_15082}. The horizontal shaded region for each profile shows the rms noise of the corresponding emission map.}
    \label{fig:15194_radial_cut_profiles}
\end{figure*}

\begin{figure*}
     \centering
     \begin{subfigure}[b]{0.475\textwidth}
         \centering
         \includegraphics[width=\textwidth]{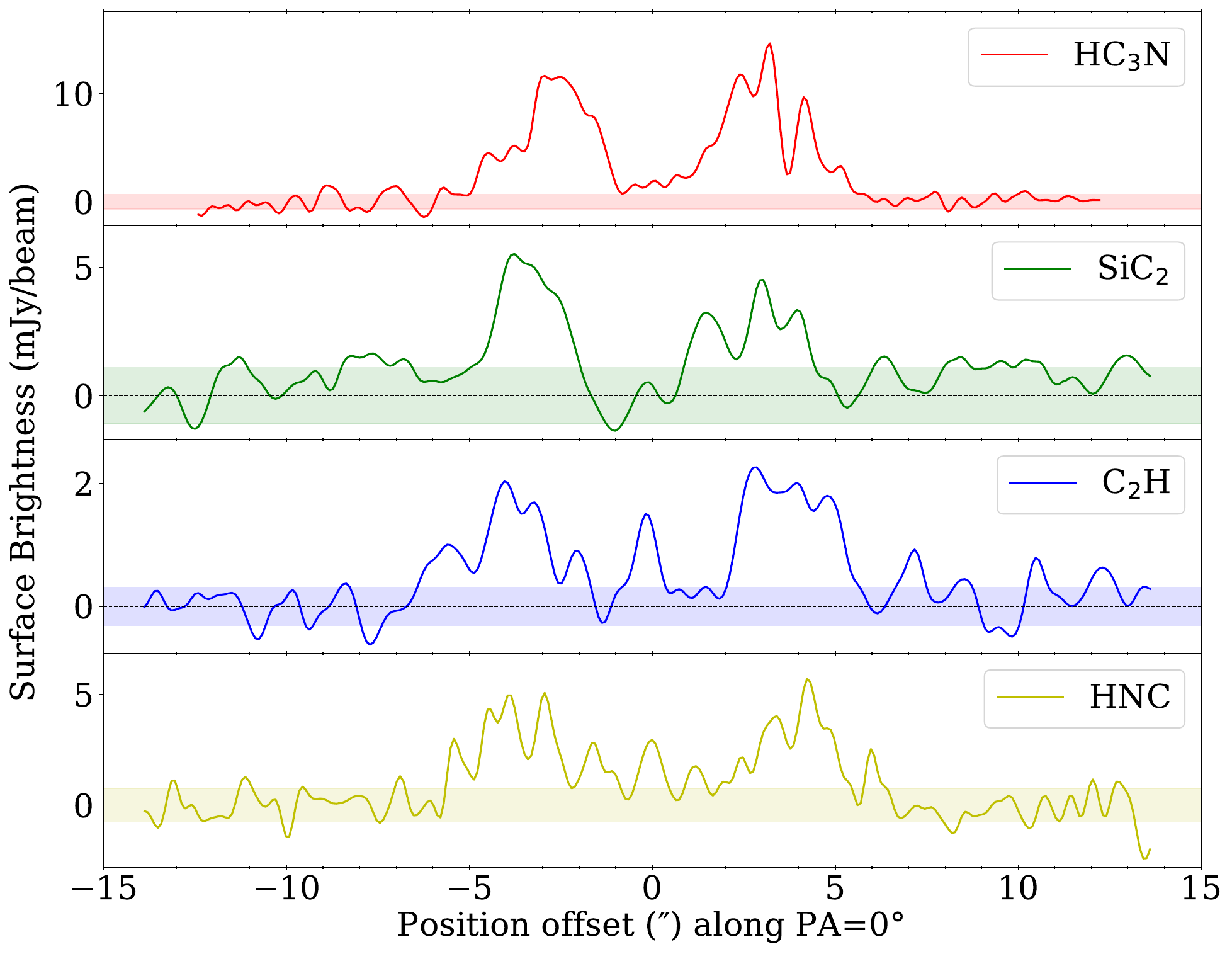}
         \label{subfig:15082_radial_Y_cuts}
     \end{subfigure}
    \hfill
     \begin{subfigure}[b]{0.475\textwidth}
         \centering
         \includegraphics[width=\textwidth]{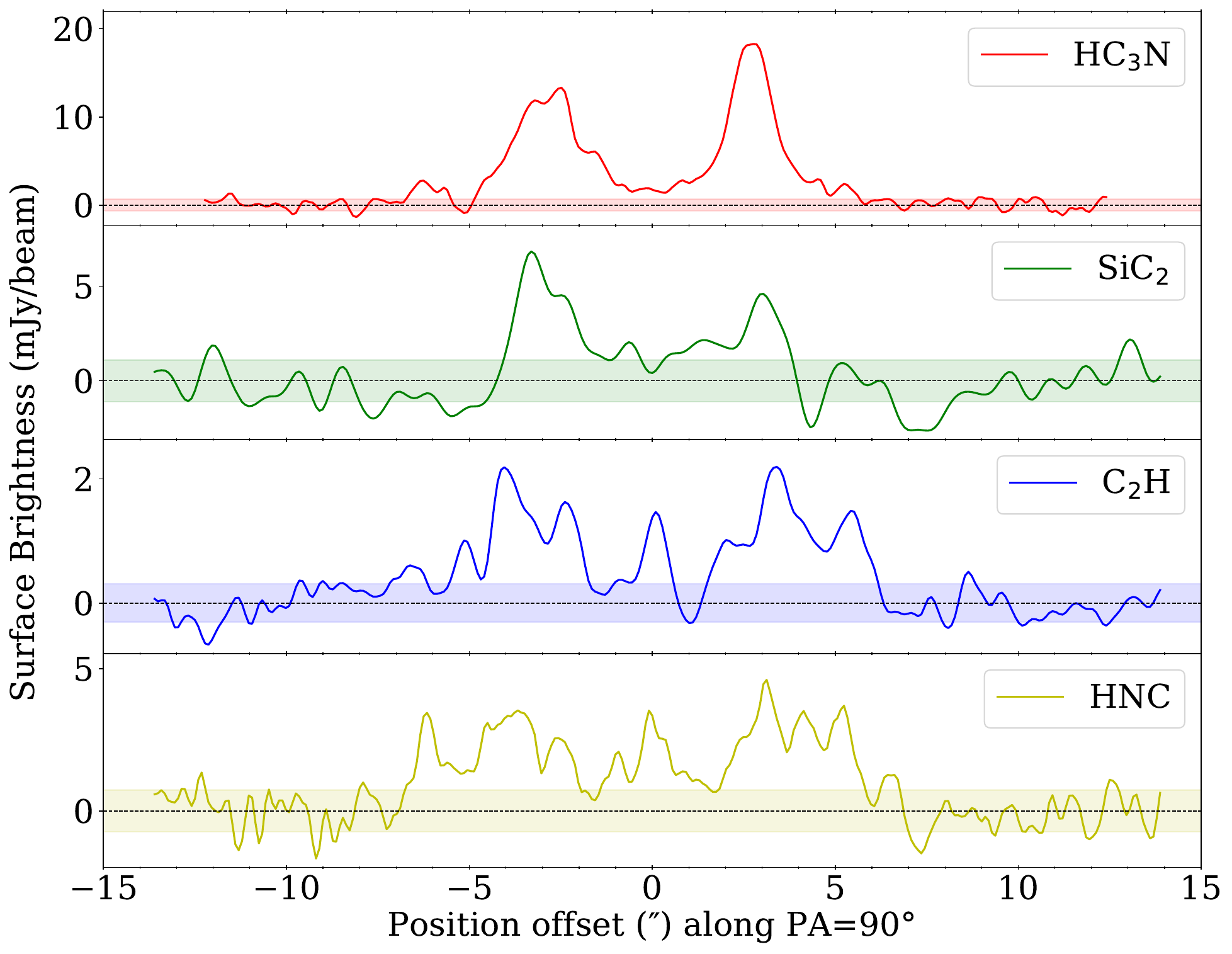}
         \label{subfig:15082_radial_X_cuts}
     \end{subfigure}
    \caption{Same as Fig.~\ref{fig:15194_radial_cut_profiles}, for IRAS 15082$-$4808.}
    \label{fig:15082_radial_cut_profiles}
\end{figure*}

The systemic-velocity brightness distributions of HC$_3$N, SiC$_2$, HNC, and C$_2$H towards IRAS 15194$-$5115 and IRAS 15082$-$4808 are shown in Fig.~\ref{fig:line_em_comp_15194_15082}. The lower spatial resolution {in the maps of IRAS 07454$-$7112} limits {the detection of} any {similar sub-}structure {around this star}. {The maps in Fig.~\ref{fig:line_em_comp_15194_15082}} have been produced by stacking the systemic-velocity emission from all {detected transitions} in the ALMA Band 3 for each of these molecules. {We regridded the line cubes to be stacked to ensure that they have the same spectral resolution (3.3 km/s) and velocity grid, and} used the software LineStacker \citep{Jolly_et_al_2020} to stack the cubes {to} extract the rms-weighted mean emission {of the channels centred on the systemic velocity} for each molecule. {No channel averaging was performed as it would lead to the substructure visible at the systemic velocity to be smeared out}. The spatial profiles obtained through radial cuts at different position angles on these maps are shown in Figs.~\ref{fig:15194_radial_cut_profiles} and \ref{fig:15082_radial_cut_profiles}.

For IRAS 15194$-$5115, the maps in Fig.~\ref{fig:line_em_comp_15194_15082} reveal well-defined arc structures {within the emitting shell} at the same locations in the HC$_3$N and SiC$_2$ maps. The {approximate} radial coincidence of the arcs seen in HC$_3$N and SiC$_2$ emission towards IRAS 15194$-$5115 can also be seen from the spatial profiles in Fig.~\ref{fig:15194_radial_cut_profiles}. However, the SiC$_2$ emission does not appear to fully trace the HC$_3$N emission towards the southwest part of the image, {and also around position angle (PA) $\approx$110$\degree$} (Fig.~\ref{fig:line_em_comp_15194_15082}, top-left panel). This could be due to the higher sensitivity of the HC$_3$N data as compared to those of SiC$_2$, since we have an HC$_3$N line in the combined data set (Sect.~\ref{subsec:data_combination}) with better $uv$-coverage than in the case of SiC$_2$. {Clumpiness cannot be ruled out either as a potential cause of this difference, as there exist significant differences in the chemical pathways of the two species}. {The C$_2$H emission presents a main shell and a faint extended arc at larger radii {($\sim$10$\arcsec$)}, from the north down to the northeast part of the map. The same arc is traced by the HNC emission as well (Fig.~\ref{fig:line_em_comp_15194_15082}, bottom-left panel)}, {though this is somewhat hidden in the figure by the C$_2$H contours}. Notably, there is C$_2$H and HNC emission visible even in areas between the arcs of the HC$_3$N and SiC$_2$ emission, towards IRAS 15194$-$5115 (Fig.~\ref{fig:line_em_comp_15194_15082}, middle-left and  bottom-left panels). 
The {maps of the} lines of C$_2$H and HNC have higher sensitivity in comparison to the SiC$_2$ {maps}, as they are imaged from the combined data {and thus recover emission on more spatial scales}.

In the case of IRAS 15082$-$4808, the SiC$_2$ emission appears confined to only the main shell region of the HC$_3$N map, and does not trace the outer arcs seen in the HC$_3$N emission (Fig.~\ref{fig:line_em_comp_15194_15082}, top-right panel). {This can be due to the lower signal-to-noise ratio of the SiC$_2$ maps, as in the case of IRAS 15194$-$5115}. However, C$_2$H and HNC trace both the main shell and the arcs to the southwest {in} the HC$_3$N image (Fig.~\ref{fig:line_em_comp_15194_15082}, middle-right and bottom-right panels).

\subsubsection{Azimuthally averaged radial brightness profiles}
\label{subsubsec:AARPs}
Despite the complex morphology seen for some of the molecular line emissions, we use AARPs to study the radial structure of the emission for all sources, as discussed below. This is the strategy used also by \citet{Agundez_et_al_2017} in their study of IRC~+10\,216.

To obtain the inner ($R_\mathrm{i}$) and outer ($R_\mathrm{e}$) radii of the line-emitting regions for different species, we constructed the mean AARPs (App.~\ref{app:appendix_D}) for the 14 molecules listed in Table~\ref{tab:em_region_sizes}. Several detected isotopologues of HCN, CS, SiO, and HC$_3$N {(H$^{13}$CN, $^{13}$CS, C$^{34}$S, $^{29}$SiO, H$^{13}$CCCN, HC$^{13}$CCN and HCC$^{13}$CN)} present emission with signal-to-noise ratios $>$3$\sigma$ in their channel maps. {Since the photodissociation of these molecules occurs throughout the broadband continuum \citep{van_Dishoeck_1988}, and is not limited to narrow frequency ranges of discrete UV lines,} their rarer isotopologues are expected to be present in the same spatial range as the main counterparts. 
Hence, to generate the mean AARPs for these species, we have averaged together emission from the detected lines of their non-blended isotopologues {where such averaging increases the signal-to-noise ratio of the resulting profile}.

\begin{figure*}[h]
   \includegraphics[width=\textwidth]{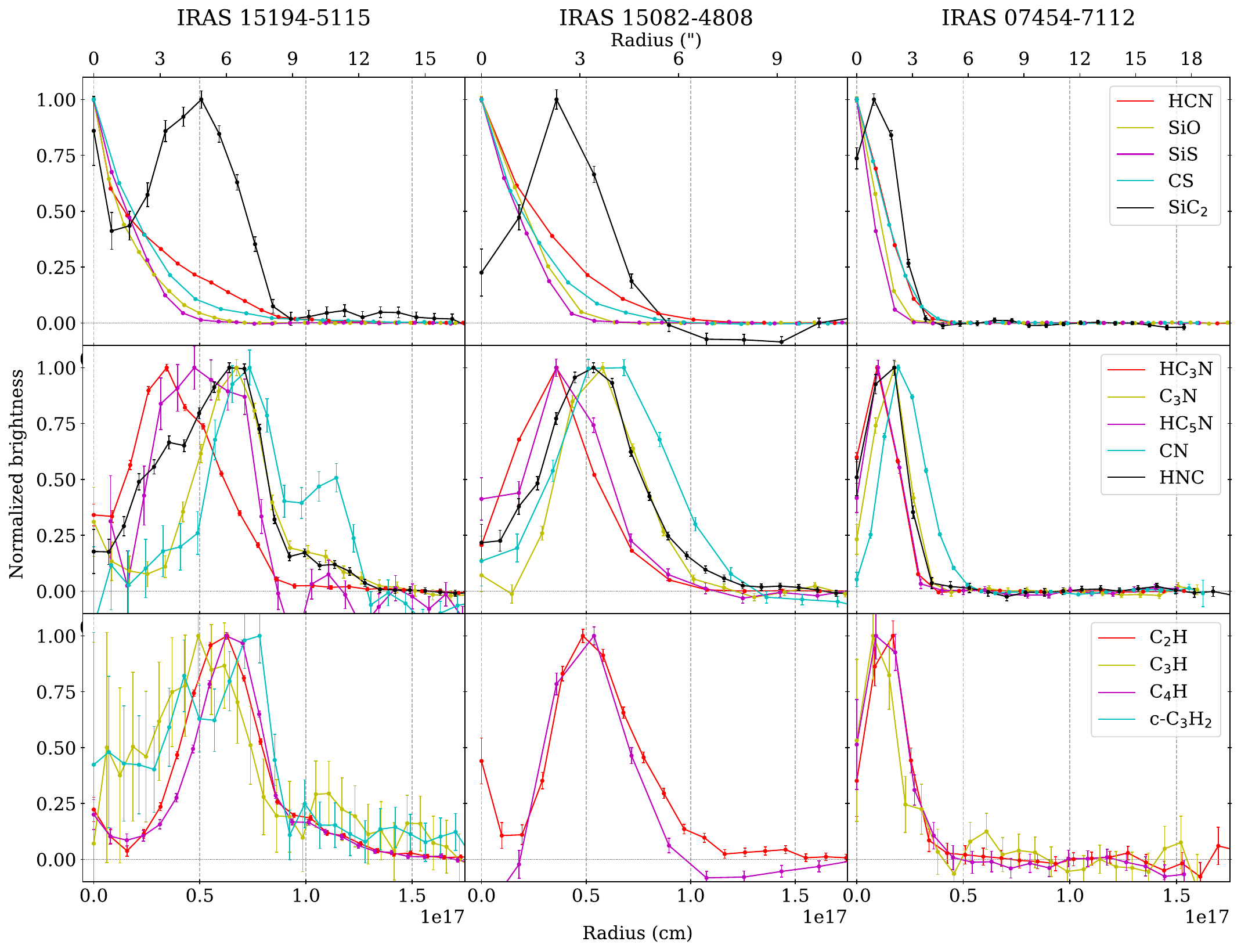}
   \caption{Normalised azimuthally-averaged radial profiles (AARPs) of molecular line brightness distributions for IRAS 15194$-$5115 {(left)}, IRAS 15082$-$4808 {(middle)}, and IRAS 07454$-$7112 {(right)}. The vertical grey gridlines apply to the lower x-axis in cm units.}
   \label{fig:All_AARPs_comparison}
\end{figure*}

Non-blended line cubes for a given species were convolved to a common circular beam with a size equal to the largest major axis among the individual beams. For each of these common-beam cubes, the emission from within {a range of} $\pm20\%$ of the expansion velocity centred on the systemic velocity {was} averaged together to increase the signal-to-noise ratio as we are now focussing on the overall size of the emitting shell, and not the substructure within. {The resulting maps were divided into successive concentric circular {rings} of increasing radii, centred on the continuum peak, with widths equal to the common restoring beam. The emission within each shell was averaged to obtain the azimuthally averaged brightness at the corresponding radius. These average brightnesses are plotted against the radii of the corresponding shells to} produce one AARP per line. The uncertainty in the radial average for each shell corresponds to the rms noise in the channel map divided by the {square root of the} number of beams in the respective shell. The mean of these individual line AARPs was calculated for each species, weighted by the reciprocal of the squared uncertainty, to obtain the mean AARPs of the species shown in Fig.~\ref{fig:All_AARPs_comparison}.

As part of this method, we verified for each species, that the different transitions of all isotopologues are emitted from the same region around the central star. Fewer pixels are used in the azimuthal averaging as we calculate the AARP closer to the centre. This increases the chance that noise at the innermost radii might not be effectively suppressed, and could give rise to artefacts which need not correspond to actual emission. This is the reason why the error bars increase in size towards the inner part of the AARPs.

\begin{figure}[h]
   \centering
   \begin{subfigure}[b]{0.4\textwidth}
       \centering
       \includegraphics[width=\textwidth]{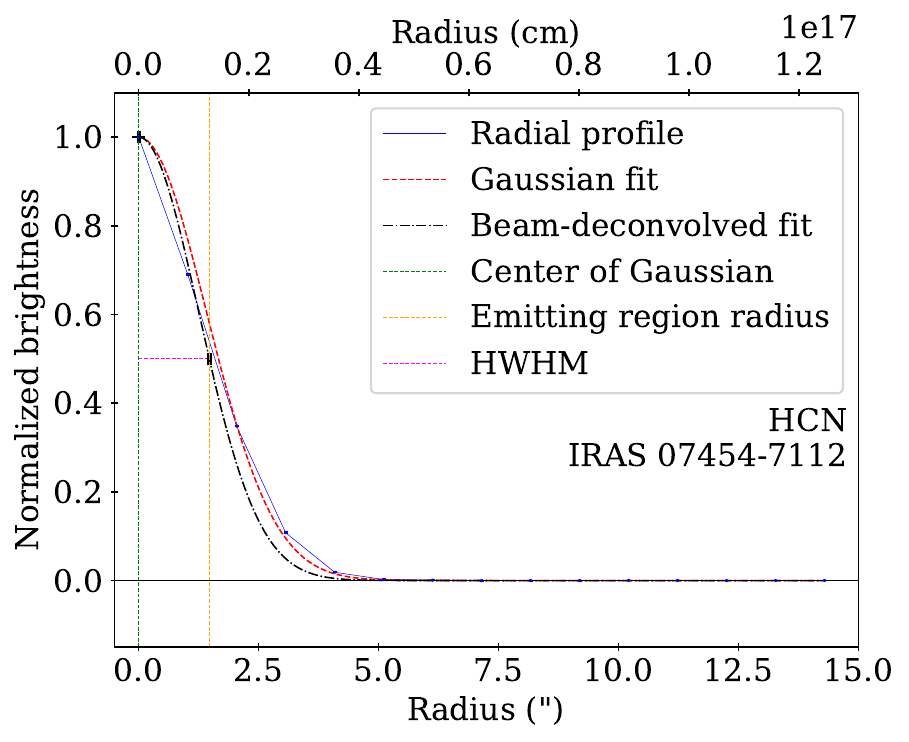}
       \label{subfig:HCN_aarp_and_fit_07454}
   \end{subfigure}
   \hfill
   \begin{subfigure}[b]{0.4\textwidth}
       \centering
       \includegraphics[width=\textwidth]{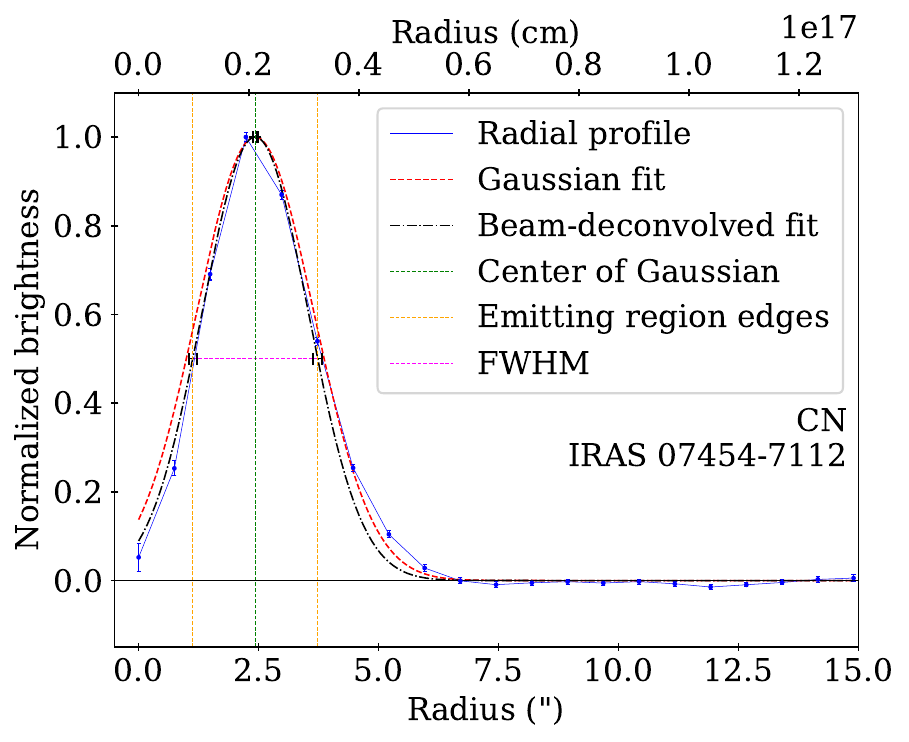}
       \label{subfig:CN_aarp_and_fit_07454}
   \end{subfigure}
      \caption{Examples of Gaussian fits to the AARPs: HCN (centrally-peaked species, top panel), CN (shell species, bottom panel), both towards IRAS 07454$-$7112. The solid blue lines show the mean AARPs and the dashed red lines show the Gaussian fits. The dash-dotted black Gaussians are obtained by deconvolving the beam from the Gaussian fit. The yellow lines show the HWHM distance from the centre of the beam-deconvolved Gaussian on either side of the peak, and the horizontal pink line depicts the HWHM (for centrally-peaked emission, top panel) or the FWHM (for shell emission, bottom panel) of the beam-deconvolved Gaussian.}
      \label{fig:HCN_CN_07454_AARPs_and_fits_example}
\end{figure}

In order to estimate the radial extents of the emitting regions, the mean AARPs of each molecule were fit with Gaussian profiles. {We believe that Gaussian fits to the AARPs provide a reasonable way to systematically estimate the extent of the various emitting regions, to a first approximation}. For instance, Fig.~\ref{fig:HCN_CN_07454_AARPs_and_fits_example} shows such fits for the HCN and CN AARPs for IRAS 07454$-$7112. The Gaussian fits to the AARPs of the 14 selected species for all three sources are shown in App.~\ref{app:appendix_D}. The common restoring beam was used to obtain the beam-deconvolved sizes of the brightness distributions. For centrally-peaked morphologies, the peak {position ($R\mathrm{_p}$)} of the emission, and hence the inner radius ($R_\mathrm{i}$) is set to 0. In this case, the outer radius ($R_\mathrm{e}$) is estimated as the half width at half maximum (HWHM) of the Gaussian fit to the mean AARP. In the case of shell emission, the inner and outer radii of the emission regions are given by the half maxima points on either side of the peak {position} of the Gaussian fit to the AARP. For each molecule, the aperture from which line spectra are extracted is set to a circular region with a radius equal to {the peak position $+$ 3$\sigma$ of the Gaussian fit} to the AARP {(i.e., $R\mathrm{_p}$ + 2.5$\times$HWHM)} of the corresponding molecule, in order to extract the maximum amount of emission while including only a minimum number of pixels {with only noise and no emission}.

\begin{figure}[h]
   \includegraphics[width=8.75cm]{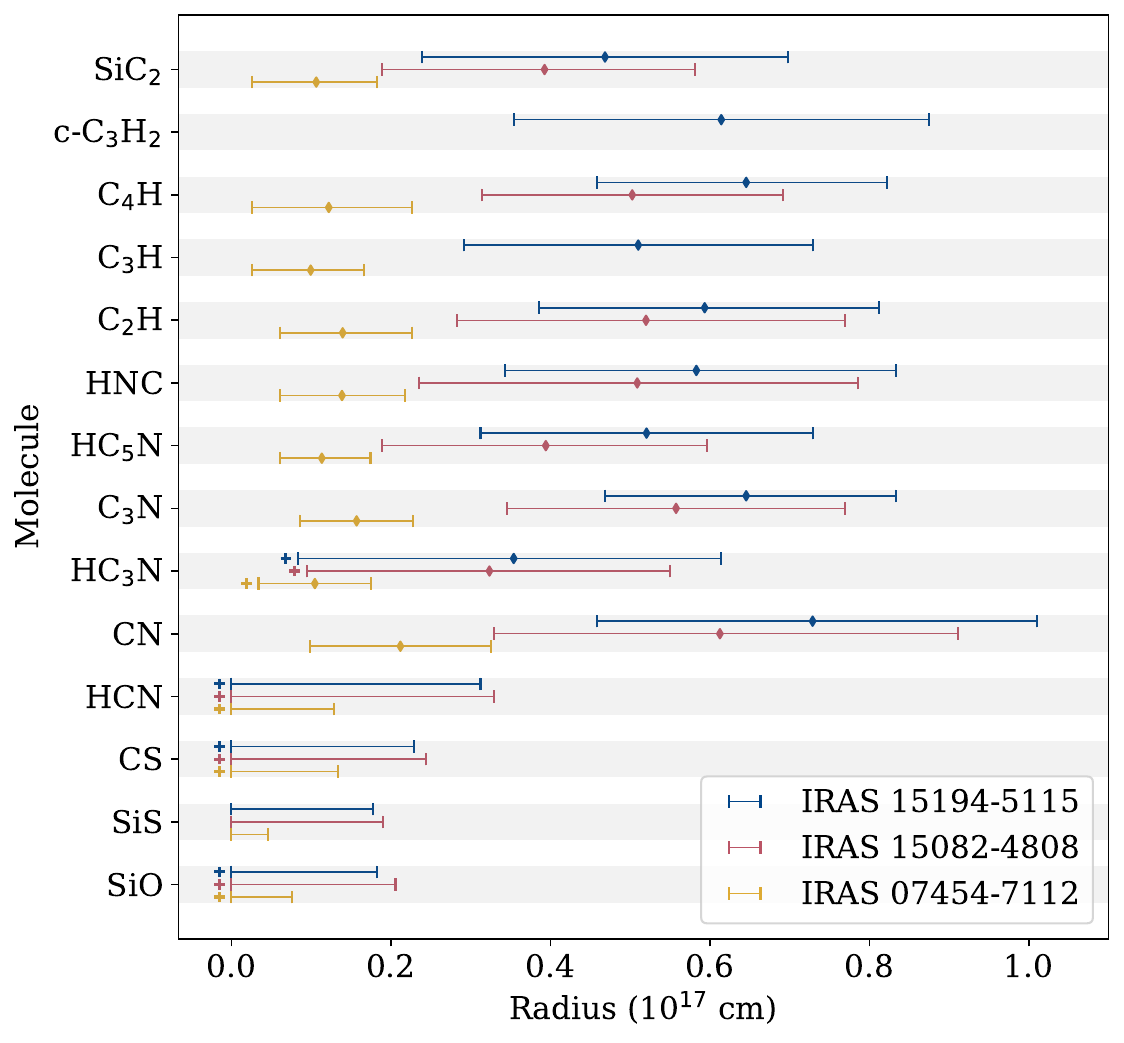}
   \caption{{Estimates of the} extents of the molecular line-emitting regions. The horizontal lines show the radial extent of the emitting regions, {as determined through Gaussian fits}, for each molecule towards IRAS 07454$-$7112 {(yellow)}, IRAS 15082$-$4808 {(red)}, and IRAS 15194$-$5115 {(blue)}.  
   {The $\blacklozenge$ signs denote the positions of the emission peaks}. The $+$ signs indicate species for which the emissions from several isotopologues {(see Sect.~\ref{subsubsec:AARPs})} were averaged to produce the AARP. {The values of the emission extents and their uncertainties are listed in Table~\ref{tab:em_region_sizes}}.}
   \label{fig:all_emission_sizes_comparison}
\end{figure}

Figure~\ref{fig:all_emission_sizes_comparison} shows the radial extent {estimates} of the {average} brightness distributions {of each of the 14 analysed molecules}, calculated using the Gaussian fits to {their mean} AARPs as discussed above. The observed values in angular scale have been converted to a spatial scale using the distances provided in Table~\ref{tab:source_properties}. The values of $R_\mathrm{i}$ and $R_\mathrm{e}$, and the corresponding uncertainties are tabulated in Table~\ref{tab:em_region_sizes}. {We present the qualitative results of the morphological analysis below, taking into account the calculated uncertainties}.

\subsubsection{Centrally-peaked species}
\label{subsubsec:centrally_peaked_species_AARPs}
The AARPs of the centrally-peaked species {(SiO, SiS, CS, HCN)} show clear deviations from a Gaussian profile in both IRAS 15194$-$5115 and IRAS 15082$-$4808, whereas they are fit well using Gaussians for IRAS 07454$-$7112 (an effect of the lower spatial resolution in this case; see Figs.~\ref{fig:SiO_and_SiS_AARPs}, \ref{fig:SiC2_and_CS_AARPs} (right), and \ref{fig:HCN_and_CN_AARPs} (left)). For IRAS 15194$-$5115 and IRAS 15082$-$4808, the emissions from these molecules peak sharply at the centre, and then gradually fall off at larger radii. Due to this, the Gaussian fits underestimate the radii of the actual emitting regions, as can be seen from Figs.~\ref{fig:SiO_and_SiS_AARPs}, \ref{fig:SiC2_and_CS_AARPs} (right), and \ref{fig:HCN_and_CN_AARPs} (left). 

The SiO and SiS emission regions are roughly co-spatial in each source. The CS emission region is more extended than those of SiO and SiS, while the HCN emission is the most extended of the centrally-peaked species for all stars {(see Fig.~\ref{fig:All_AARPs_comparison})}.

\subsubsection{Shell species}
\label{subsubsec:shell_species_AARPs}
{Hydrocarbons}: {Within the calculated uncertainties (see Table~\ref{tab:em_region_sizes})}, the shells of C$_2$H and C$_4$H peak at the same radii (Fig.~\ref{fig:All_AARPs_comparison}) and have the same widths (Fig.~\ref{fig:all_emission_sizes_comparison}) for the three stars, possibly with the exception of IRAS 15082$-$4808, where the C$_2$H emission appears to extend further out than that of C$_4$H {by roughly 10\% (see Table~\ref{tab:em_region_sizes})}. {The main hydrocarbon shells are of roughly the same thickness ($\sim$5500 AU) towards IRAS 15194$-$5115 and IRAS 15082$-$4808, but there are cases of faint extended emission beyond the main shell up to around 9300 AU for IRAS 15194$-$5115 (Fig.~\ref{fig:All_AARPs_comparison}). IRAS 07454-7112 presents much thinner shells for the hydrocarbons, with a thickness of only $\sim$1300 AU (Table~\ref{tab:em_region_sizes})}. C$_5$H, C$_6$H, and C$_8$H have also been detected towards IRAS 15082$-$4808 and/or IRAS 15194$-$5115 in the combined data, but the lines are too weak for any morphological information to be extracted.

\begin{table*}[h]
   \caption{Extents of the emitting regions of selected molecular species.}
   \label{tab:em_region_sizes}
   \centering
   \begin{adjustbox}{width=17.5cm}
   \begin{tabular}{@{\extracolsep{4pt}}lrrrrrrrrrrrr@{}}
   \hline \hline & \\[-1ex]
   \multicolumn{1}{c}{\multirow{3}{*}{Molecule}} & \multicolumn{4}{c}{IRAS 15194$-$5115}                                                                   & \multicolumn{4}{c}{IRAS 15082$-$4808}                                                                   & \multicolumn{4}{c}{IRAS 07454$-$7112}                                                                   \\ \cline{2-5} \cline{6-9} \cline{10-13} & \\[-1ex]

   \multicolumn{1}{c}{}                          & \multicolumn{2}{c}{R$_\mathrm{i}$}                & \multicolumn{2}{c}{R$_\mathrm{e}$}               & \multicolumn{2}{c}{R$_\mathrm{i}$}                & \multicolumn{2}{c}{R$_\mathrm{e}$}               & \multicolumn{2}{c}{R$_\mathrm{i}$}                & \multicolumn{2}{c}{R$_\mathrm{e}$}               \\ \cline{2-3} \cline{4-5} \cline{6-7} \cline{8-9} \cline{10-11} \cline{12-13} & \\[-1ex]
   \multicolumn{1}{c}{}               & \multicolumn{1}{c}{{[}\arcsec{]}} & \multicolumn{1}{c}{{[}cm{]}} & \multicolumn{1}{c}{{[}\arcsec{]}} & \multicolumn{1}{c}{{[}cm{]}} & \multicolumn{1}{c}{{[}\arcsec{]}} & \multicolumn{1}{c}{{[}cm{]}} & \multicolumn{1}{c}{{[}\arcsec{]}} & \multicolumn{1}{c}{{[}cm{]}} & \multicolumn{1}{c}{{[}\arcsec{]}} & \multicolumn{1}{c}{{[}cm{]}} & \multicolumn{1}{c}{{[}\arcsec{]}} & \multicolumn{1}{c}{{[}cm{]}} \\ \hline & \\[-1ex]
    SiO & 0.0 & 0.0 & 1.8(1) & 1.8$\times$10$^{16}$ & 0.0 & 0.0 & 1.3(1) & 2.1$\times$10$^{16}$ & 0.0 & 0.0 & 0.9(1) & 7.6$\times$10$^{15}$ \\ 
    SiS & 0.0 & 0.0 & 1.7(1) & 1.8$\times$10$^{16}$ & 0.0 & 0.0 & 1.2(1) & 1.9$\times$10$^{16}$ & 0.0 & 0.0 & 0.5(1) & 4.6$\times$10$^{15}$ \\ 
    CS & 0.0 & 0.0 & 2.2(1) & 2.3$\times$10$^{16}$ & 0.0 & 0.0 & 1.6(1) & 2.4$\times$10$^{16}$ & 0.0 & 0.0 & 1.5(1) & 1.3$\times$10$^{16}$ \\ 
    HCN & 0.0 & 0.0 & 3.0(2) & 3.1$\times$10$^{16}$ & 0.0 & 0.0 & 2.1(1) & 3.3$\times$10$^{16}$ & 0.0 & 0.0 & 1.5(1) & 1.3$\times$10$^{16}$ \\ 
    CN & 4.4(7) & 4.6$\times$10$^{16}$ & 9.7(7) & 1.0$\times$10$^{17}$ & 2.1(2) & 3.3$\times$10$^{16}$ & 5.8(2) & 9.1$\times$10$^{16}$ & 1.1(1) & 9.9$\times$10$^{15}$ & 3.7(1) & 3.3$\times$10$^{16}$ \\ 
    C$_2$H & 3.7(3) & 3.9$\times$10$^{16}$ & 7.8(3) & 8.1$\times$10$^{16}$ & 1.8(2) & 2.8$\times$10$^{16}$ & 4.9(2) & 7.7$\times$10$^{16}$ & 0.7(2) & 6.1$\times$10$^{15}$ & 2.6(2) & 2.3$\times$10$^{16}$ \\ 
    C$_3$H & 2.8(2) & 2.9$\times$10$^{16}$ & 7.0(2) & 7.3$\times$10$^{16}$ & -- & -- & -- & -- & 0.3(1) & 2.6$\times$10$^{15}$ & 1.9(1) & 1.7$\times$10$^{16}$ \\ 
    C$_4$H & 4.4(2) & 4.6$\times$10$^{16}$ & 7.9(2) & 8.2$\times$10$^{16}$ & 2.0(3) & 3.1$\times$10$^{16}$ & 4.4(3) & 6.9$\times$10$^{16}$ & 0.3(4) & 2.6$\times$10$^{15}$ & 2.6(4) & 2.3$\times$10$^{16}$ \\ 
    c-C$_3$H$_2$ & 3.4(7) & 3.5$\times$10$^{16}$ & 8.4(7) & 8.7$\times$10$^{16}$ & -- & -- & -- & -- & -- & -- & -- & -- \\ 
    SiC$_2$ & 2.3(3) & 2.4$\times$10$^{16}$ & 6.7(3) & 7.0$\times$10$^{16}$ & 1.2(4) & 1.9$\times$10$^{16}$ & 3.7(4) & 5.8$\times$10$^{16}$ & 0.3(1) & 2.6$\times$10$^{15}$ & 2.1(1) & 1.8$\times$10$^{16}$ \\ 
    HC$_3$N & 0.8(3) & 8.3$\times$10$^{15}$ & 5.9(3) & 6.1$\times$10$^{16}$ & 0.6(2) & 9.4$\times$10$^{15}$ & 3.5(2) & 5.5$\times$10$^{16}$ & 0.4(1) & 3.4$\times$10$^{15}$ & 2.0(1) & 1.8$\times$10$^{16}$ \\ 
    C$_3$N & 4.5(2) & 4.7$\times$10$^{16}$ & 8.0(2) & 8.3$\times$10$^{16}$ & 2.2(1) & 3.5$\times$10$^{16}$ & 4.9(1) & 7.7$\times$10$^{16}$ & 0.9(1) & 8.5$\times$10$^{15}$ & 2.6(1) & 2.3$\times$10$^{16}$ \\ 
    HC$_5$N & 3.0(7) & 3.1$\times$10$^{16}$ & 7.0(7) & 7.3$\times$10$^{16}$ & 1.2(2) & 1.9$\times$10$^{16}$ & 3.8(2) & 6.0$\times$10$^{16}$ & 0.7(2) & 6.1$\times$10$^{15}$ & 2.0(2) & 1.7$\times$10$^{16}$ \\ 
    HNC & 3.3(3) & 3.4$\times$10$^{16}$ & 8.0(3) & 8.3$\times$10$^{16}$ & 1.5(1) & 2.4$\times$10$^{16}$ & 5.0(1) & 7.9$\times$10$^{16}$ & 0.7(1) & 6.1$\times$10$^{15}$ & 2.5(1) & 2.2$\times$10$^{16}$ \\  \hline & \\[-1ex]
   \end{tabular}
   \end{adjustbox}
   \tablefoot{R$_\mathrm{i}$ and R$_\mathrm{e}$ denote the inner and outer radii, respectively. For the centrally-peaked species (SiO, SiS, CS, and HCN), we set the inner radii to zero. The values in parentheses in the arcsecond columns show the uncertainty in the last decimal place. These are formal errors from the Gaussian fits from which the radii were obtained. Conversion from angular to spatial distances has been done using the distances in Table~\ref{tab:source_properties}.}
\end{table*}

{Cyanides}: {In addition to the expected shell emission \citep[see e.g.][for the HC$_3$N shell distribution for IRC +10\,216]{Agundez_et_al_2017}, HC$_3$N also shows significant central emission components towards all three stars}. This is discussed in more detail in Sect.~\ref{subsubsec:cen_comps_in_shell_species}. In the cases of IRAS 15082$-$4808 and IRAS 07454$-$7112, the shell-HC$_3$N and HC$_5$N emissions trace {roughly} the same radial range {(Fig.~\ref{fig:all_emission_sizes_comparison} and Table~\ref{tab:em_region_sizes})}, and they peak at {around} the same distance from the centre {(Fig.~\ref{fig:All_AARPs_comparison})}. {In the case of} IRAS 15194$-$5115 the HC$_3$N emission appears to peak at a slightly {smaller radius than} the HC$_5$N emission, {by $\sim$30\%} (Fig.~\ref{fig:All_AARPs_comparison}). However, this could be due to contamination from the central component of the HC$_3$N emission and the low signal-to-noise in the HC$_5$N maps towards this star. Compared to C$_2$H and C$_4$H, the HC$_3$N and the HC$_5$N emissions peak at slightly smaller radii in all three CSEs, by an estimated $10-40\%$ {(Fig.~\ref{fig:All_AARPs_comparison})}.

{In general, CN, HNC, and the hydrocarbons present the most extended emission among the observed molecules}. The peaks of the CN and C$_3$N emissions are located roughly at the outer edges of the HCN and the shell-HC$_3$N emissions, {respectively} (Figs.~\ref{fig:All_AARPs_comparison}, \ref{fig:all_emission_sizes_comparison} and Table~\ref{tab:em_region_sizes}). The CN AARP of IRAS 15194$-$5115 shows a secondary peak at larger radii {($\sim$11$\arcsec$, Fig.~\ref{fig:All_AARPs_comparison})}, which is not present in the other two stars. For IRAS 15194$-$5115, we see weak emission structures at {roughly the same location as the CN secondary peak} also in SiC$_2$, HNC, C$_3$N, C$_2$H, C$_4$H, and possibly also C$_3$H {(Fig.~\ref{fig:All_AARPs_comparison})}. {The fact that these arcs are traced by multiple species suggests that they are most likely the result of density enhancements (see also Sect.~\ref{sec:Discussion})}. The radial range of the HNC emission is roughly coincident with that of the C$_2$H and C$_4$H emission for all three stars {(Fig.~\ref{fig:all_emission_sizes_comparison})}.

{SiC$_2$}: SiC$_2$ is the only silicon-bearing molecule out of the species studied in this paper that emits in a shell, whereas the others (SiO and SiS) are centrally peaked. The peak of the SiC$_2$ emission is close to those of the HC$_3$N and HC$_5$N emissions {for all three stars} (Fig.~\ref{fig:All_AARPs_comparison}). However, SiC$_2$ also presents a central component that is further discussed below. {We note that towards IRAS 15082$-4808$, the SiC$_2$ AARPs show negative dips at larger radii (see Fig.~\ref{fig:All_AARPs_comparison}), which is evidence of interferometric filtering out of extended emission. A similar behaviour is exhibited by C$_4$H towards both IRAS 15194$-$5115 and IRAS 15082$-$4808, and possibly HC$_5$N towards IRAS 15194$-$5115}.

\subsubsection{Central emission components of HC$_3$N and SiC$_2$}
\label{subsubsec:cen_comps_in_shell_species}
For IRAS 15194$-$5115 and IRAS 15082$-$4808, HC$_3$N {and SiC$_2$} show central components {above the respective 3$\sigma$ noise levels} (Fig.~\ref{fig:All_AARPs_comparison}). We stress that what we call {central} emission need not {necessarily originate in the extended atmosphere}, but can instead emanate from roughly anywhere within a synthesised beam {($\sim$700 AU for IRAS 15194$-$5115 and $\sim$1000 AU for IRAS 15082$-$4808, see Table~\ref{tab:observational_details})} centred on the star.  
The central components of these species can be seen in the spatial profile plots shown in Figs.~\ref{fig:15194_radial_cut_profiles} and \ref{fig:15082_radial_cut_profiles} also. {We note that even though C$_2$H appears to have a central component towards IRAS 15082$-$4808 (see Fig.~\ref{fig:line_em_comp_15194_15082}, middle-right panel and Fig.~\ref{fig:15082_radial_cut_profiles}), this is due to the contamination from the blended hyperfine components of C$_2$H, and potentially also weak lines of c-C$_3$H$_2$ and C$_6$H (see Fig.~\ref{fig:almaband3_identified_sample})}. In the case of IRAS 07454$-$7112, the angular resolution compared to the size of the emitting region is too poor to draw any conclusions about the behaviour of the brightness distributions in the inner part of the CSE. {The possible formation mechanisms of these central components are discussed in more detail in Sect.~\ref{sec:Discussion}}.

\begin{table*}[h]
   \caption{{Rotation temperatures and column densities obtained using population diagrams.}}
   \label{tab:N_and_T}
   \centering
      \begin{tabular}{l r r r r r r}
      \hline\hline & \\[-3ex]
      \makecell{\\Molecule} & \multicolumn{2}{c}{IRAS 15194$-$5115} & \multicolumn{2}{c}{IRAS 15082$-$4808} & \multicolumn{2}{c}{IRAS 07454$-$7112} \\
      \cline{2-7} & \\[-1ex]
      & $T_\mathrm{rot}$ (K) & $N_\mathrm{tot}$ (cm$^{-2}$) & $T_\mathrm{rot}$ (K) & $N_\mathrm{tot}$ (cm$^{-2}$) & $T_\mathrm{rot}$ (K) & $N_\mathrm{tot}$ (cm$^{-2}$)\\
      \hline & \\[-1ex]
    SiO        	&	 14(1) 	&	 1.0(1)$\times$10$^{16}$ 	&	 13(1) 	&	 7(1)$\times$10$^{15}$ 	&	 18(2)  	&	 9(2)$\times$10$^{15}$             \\
    $^{29}$SiO 	&	 15(1) 	&	 3.9(6)$\times$10$^{14}$ 	&	 14(1) 	&	 2.1(4)$\times$10$^{14}$ 	&	 18(2)  	&	 2.9(5)$\times$10$^{14}$           \\
    $^{30}$SiO 	&	 28(9) 	&	 9(7)$\times$10$^{13}$   	&	 --    	&	 --                      	&	 --     	&	 --                                \\
    
    SiS        	&	 39(2) 	&	 1.0(1)$\times$10$^{16}$ 	&	 35(3) 	&	 1.3(2)$\times$10$^{16}$ 	&	 91(21)  	&	 8(2)$\times$10$^{16}$             \\
    $^{29}$SiS 	&	 --    	&	 --                      	&	 26(2) 	&	 8(1)$\times$10$^{14}$   	&	 91(22) 	&	 2.5(7)$\times$10$^{15}$           \\
    $^{30}$SiS 	&	 --    	&	 --                      	&	 --    	&	 --                      	&	 85(27) 	&	 1.7(7)$\times$10$^{15}$           \\
    Si$^{34}$S 	&	 --    	&	 --                      	&	 43(6) 	&	 5(1)$\times$10$^{14}$   	&	 --     	&	 --                                \\
    
    SiC$_2$    	&	 23(1) 	&	 4.4(3)$\times$10$^{14}$ 	&	 30(2) 	&	 8(1)$\times$10$^{14}$ 	&	 42(2)  	&	 1.7(1)$\times$10$^{15}$           \\
    
    CS         	&	 12(1) 	&	 3.9(6)$\times$10$^{16}$ &	 11(1) 	&	 6(1)$\times$10$^{16}$ 	&	 --     	&	 --                                \\
    $^{13}$CS  	&	 15(1) 	&	 2.5(5)$\times$10$^{15}$ 	&	 --    	&	 --                      	&	 --     	&	 --                                \\
    C$^{34}$S  	&	 18(2) 	&	 5(1)$\times$10$^{14}$   	&	 15(2) 	&	 7(1)$\times$10$^{14}$   	&	 17(2)  	&	 4.1(8)$\times$10$^{14}$           \\
    
    HCN        	&	 9(1) 		&	 1.4(3)$\times$10$^{16}$ 	&	 8(1)  	&	 4.8(9)$\times$10$^{16}$ 	&	 10(1)   	&	 3.7(6)$\times$10$^{16}$           \\
    H$^{13}$CN 	&	 8(1)  	&	 9(2)$\times$10$^{15}$ 	&	 7(1)  	&	 2.9(5)$\times$10$^{15}$ &	 9(1)   	&	 9(2)$\times$10$^{15}$             \\
    
    HNC        	&	 7(1)  	&	 7(1)$\times$10$^{13}$ 	&	 --    	&	 --                      	&	 --     	&	 --                                \\
    
    HC$_3$N    	&	 15(1) 	&	 5(1)$\times$10$^{14}$   	&	 20(1) 	&	 9(1)$\times$10$^{14}$   	&	 23(1)  	&	 1.6(1)$\times$10$^{15}$           \\
    C$_3$N     	&	 --    	&	 --                      	&	 27(3) 	&	 2.9(6)$\times$10$^{14}$ 	&	 46(2)  	&	 2.7(3)$\times$10$^{14}$           \\
    HC$_5$N    	&	 --    	&	 --                      	&	 20(3) 	&	 8(6)$\times$10$^{14}$   	&	 27(4)  	&	 5(2)$\times$10$^{14}$             \\
    
    C$_2$H     	&	 12(1) 	&	 8(2)$\times$10$^{15}$   	&	 16(6) 	&	 6(3)$\times$10$^{15}$   	&	 --     	&	 --                                \\
    C$_4$H     	&	 35(1) 	&	 8.6(7)$\times$10$^{14}$ 	&	 44(2) 	&	 6.1(7)$\times$10$^{14}$ 	&	 75(10) 	&	 2.6(6)$\times$10$^{14}$ 	          \\ \hline & \\[-1ex]
      \end{tabular}
      \tablefoot{{$T_\mathrm{rot}$ and $N_\mathrm{tot}$ stand for rotational temperature, and total column density, respectively}. The values in parentheses indicate the uncertainty in the immediately preceding digit. This is a formal uncertainty of the fit, and hence provides a measure of the quality of the population diagram. The absolute uncertainty, in particular for the column density, is substantially larger, {as discussed in the text}.}
\end{table*}

\begin{figure*}[h]
   \centering
   \begin{subfigure}[b]{0.475\textwidth}
       \centering
       \includegraphics[width=\textwidth]{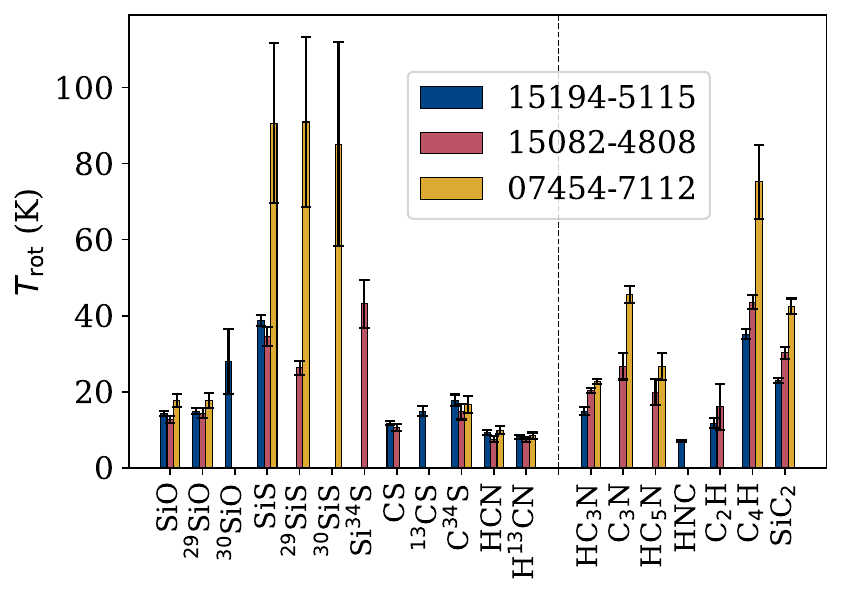}
       \label{subfig:T_rot_comparison}
   \end{subfigure}
   \hfill
   \begin{subfigure}[b]{0.475\textwidth}
       \centering
       \includegraphics[width=\textwidth]{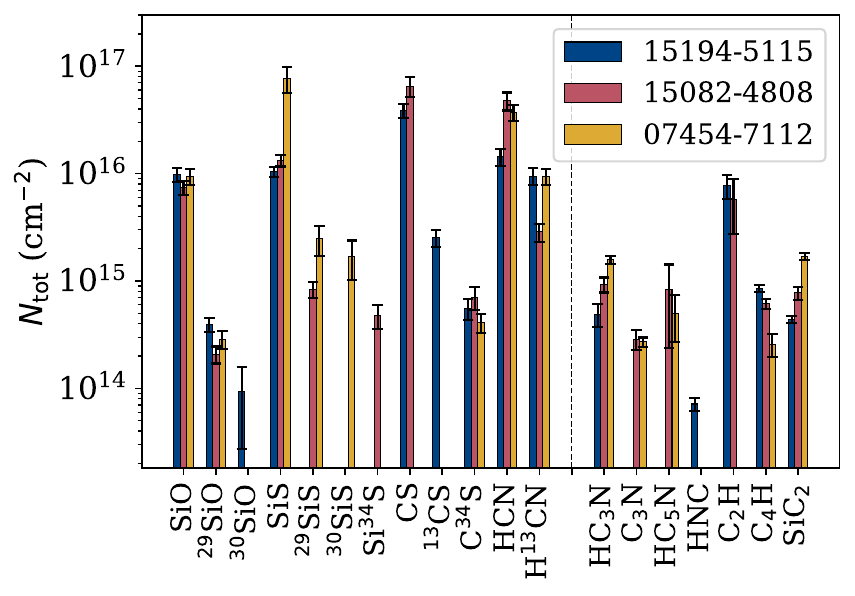}
       \label{subfig:N_tot_comparison}
   \end{subfigure}
      \caption{Rotation temperatures ($T_\mathrm{rot}$, left panel) and total column densities ($N_\mathrm{tot}$, right panel) obtained from population diagrams. The dashed vertical black line separates the shell species and the centrally-peaked species.}
      \label{fig:T_rots_and_N_tots_rot_diags}
\end{figure*}

\begin{table*}[h]
   \caption{{Molecular fractional abundances.}}
   \label{tab:f_As}
   \centering
      \begin{tabular}{l r r r r}
      \hline\hline & \\[-3ex]
      \makecell{\\Molecule} & \multicolumn{4}{c}{X/H$_2$}\\
      \cline{2-5} & \\[-2ex]
      & 15194$-$5115 & 15082$-$4808 & 07454$-$7112 & IRC~+10\,216 \\
      \hline & \\[-2ex]
    SiO                 	&	 4.0(-6)                 	&	 1.8(-6)                 	&	 4.6(-6)         	&	 1.7(-7)$^{(1)}$          \\
    $^{29}$SiO          	&	 1.6(-7)                 	&	 5.2(-8)                 	&	 1.4(-7)         	&	 1.0(-8)$^{(b)}$          \\
    $^{30}$SiO          	&	 3.8(-8)                 	&	 6.5(-9)$^{(a)}$         	&	 2.3(-8)$^{(a)}$ 	&	 7.1(-9)$^{(b)}$          \\
    
    SiS                 	&	 4.2(-6)                 	&	 3.0(-6)                 	&	 2.3(-5)         	&	 2.9(-6)$^{(1)}$          \\
    $^{29}$SiS          	&	 1.7(-7)$^{(a)}$         	&	 1.9(-7)                 	&	 7.3(-7)         	&	 1.7(-7)$^{(b)}$          \\
    $^{30}$SiS          	&	 --                      	&	 3.9(-8)$^{(a)}$         	&	 5.0(-7)         	&	 1.2(-7)$^{(b)}$          \\
    Si$^{34}$S          	&	 8.2(-8)$^{(a)}$         	&	 1.1(-7)                 	&	 1.4(-7)$^{(a)}$ 	&	 1.6(-7)$^{(b)}$          \\
    
    SiC$_2$             	&	 1.1(-6)                 	&	 8.0(-7)                 	&	 2.3(-6)         	&	 3.6(-7)$^{(2)}$          \\
    Si$^{13}$CC         	&	 2.5(-7)$^{(a)}$         	&	 --                      	&	 --              	&	 8.0(-9)$^{(c)}$          \\
    
    CS                  	&	 2.0(-5)                 	&	 1.9(-5)                 	&	 4.8(-6)$^{(a)}$ 	&	 3.9(-6)$^{(1)}$          \\
    $^{13}$CS           	&	 1.3(-6)                 	&	 1.0(-7)$^{(a)}$         	&	 2.2(-7)$^{(a)}$ 	&	 8.7(-8)$^{(c)}$          \\
    C$^{33}$S           	&	 6.9(-8)$^{(a)}$         	&	 1.2(-8)$^{(a)}$         	&	 2.5(-8)$^{(a)}$ 	&	 4.2(-8)$^{(b)}$          \\
    C$^{34}$S           	&	 2.8(-7)                 	&	 2.1(-7)                 	&	 3.5(-7)         	&	 2.0(-7)$^{(b)}$          \\
    
    HCN                 	&	 1.0(-5)                 	&	 1.9(-5)                 	&	 3.1(-5)         	&	 1.9(-5)$^{(1)}$          \\
    H$^{13}$CN          	&	 6.6(-6)                 	&	 1.1(-6)                 	&	 7.7(-6)         	&	 4.3(-7)$^{(c)}$          \\
    
    CN                  	&	 8.5(-7)$^{(a)}$         	&	 1.0(-6)$^{(a)}$         	&	 5.9(-6)$^{(a)}$ 	&	 1.8(-6)$^{(3)}$          \\
    
    HNC                 	&	 2.3(-7)                 	&	 1.2(-7)$^{(a)}$         	&	 5.3(-7)$^{(a)}$ 	&	 6.8(-8)$^{(4)}$          \\
    HN$^{13}$C          	&	 4.6(-8)$^{(a)}$         	&	 --                      	&	 --              	&	 1.6(-9)$^{(c)}$          \\
    
    HC$_3$N             	&	 7.8(-7) 			&	 7.4(-7) 			&	 2.2(-6)         	&	 7.1(-7)$^{(3)}$          \\
    H$^{13}$CCCN        	&	 7.9(-8)$^{(a)}$         	&	 1.4(-8)$^{(a)}$         	&	 5.2(-8)$^{(a)}$ 	&	 5.3(-9)$^{(c)}$          \\
    HC$^{13}$CCN        	&	 8.0(-8)$^{(a)}$         	&	 1.4(-8)$^{(a)}$         	&	 4.5(-8)$^{(a)}$ 	&	 5.3(-9)$^{(c)}$          \\
    HCC$^{13}$CN        	&	 8.0(-8)$^{(a)}$         	&	 1.4(-8)$^{(a)}$         	&	 4.9(-8)$^{(a)}$ 	&	 5.3(-9)$^{(c)}$          \\
    H$^{13}$C$^{13}$CCN 	&	 1.5(-8)$^{(a)}$         	&	 --                      	&	 --              	&	 --                       \\
    H$^{13}$CC$^{13}$CN 	&	 2.2(-8)$^{(a)}$         	&	 --                      	&	 --              	&	 --                       \\
    HC$^{13}$C$^{13}$CN 	&	 1.6(-8)$^{(a)}$         	&	 --                      	&	 --              	&	 --                       \\
    
    C$_3$N              	&	 4.2(-7)$^{(a)}$         	&	 4.9(-7)                 	&	 6.3(-7)         	&	 1.5(-7)$^{(5)}$          \\
    CC$^{13}$CN         	&	 1.1(-7)$^{(a)}$         	&	 --                      	&	 --              	&	 1.1(-9)$^{(c)}$          \\
    
    HC$_5$N             	&	 2.3(-7)$^{(a)}$         	&	 8.7(-7)                 	&	 8.6(-7)         	&	 2.1(-7)$^{(5)}$          \\
    
    C$_2$H              	&	 2.7(-5) 			&	 8.5(-6)  			&	 3.1(-6)$^{(a)}$ 	&	 1.3(-5)$^{(6)}$          \\
    C$_3$H              	&	 5.7(-8)$^{(a)}$         	&	 --                      	&	 3.2(-8)$^{(a)}$ 	&	 2.2(-7)$^{(d)}$          \\
    C$_4$H              	&	 3.6(-6)                 	&	 9.3(-7)                 	&	 4.2(-7)         	&	 1.1(-6)$^{(5)}$          \\
    
    $^{13}$CCCCH        	&	 2.7(-7)$^{(a)}$         	&	 --                      	&	 --              	&	 5.9(-9)$^{(c)}$          \\
    C$^{13}$CCCH        	&	 3.1(-7)$^{(a)}$         	&	 --                      	&	 --              	&	 5.9(-9)$^{(c)}$          \\
    CC$^{13}$CCH        	&	 3.2(-7)$^{(a)}$         	&	 --                      	&	 --              	&	 5.9(-9)$^{(c)}$          \\
    CCC$^{13}$CH        	&	 2.8(-7)$^{(a)}$         	&	 --                      	&	 --              	&	 5.9(-9)$^{(c)}$          \\
    c-C$_3$H$_2$        	&	 1.8(-7)$^{(a)}$         	&	 --                      	&	 --              	&	 5.6(-8)$^{(5)}$          \\ \hline & \\[-1ex]
      \end{tabular}
      \tablefoot{$a(b)$ corresponds to $a\times10^{b}$. We estimate an uncertainty of a factor of {3-5} for the abundances estimated from population diagrams and an order of magnitude for single-line estimates (Sect.~\ref{subsubsec:Summary_of_abundance_estimates}). $^{(a)}$Abundances estimated from single lines. All other abundances (for the three stars in the sample) are from population diagrams. IRC~+10\,216 abundances are from: $^{(1)}$\citet{Agundez_et_al_2012}, $^{(2)}$\citet{Massalkhi_et_al_2018}, $^{(3)}$\citet{Agundez_2009_PhD_thesis}, $^{(4)}$\citet{Daniel_et_al_2012}, $^{(5)}$\citet{Gong_et_al_2015}, and $^{(6)}$\citet{De_Beck_et_al_2012}. $^{(b)}$Calculated by scaling the abundances of the corresponding main isotopologue by the relevant isotopologue ratio from \citet{He_et_al_2008}. $^{(c)}$ Scaled from the corresponding $^{12}$C isotopologue using a $^{12}$C/$^{13}$C ratio of 45 \citep{Cernicharo_et_al_2000}. $^{(d)}$Obtained by scaling the column density from \citet{Agundez_2009_PhD_thesis} by the $f_\mathrm{X}/N_\mathrm{tot}$ ratio of C$_2$H, assuming C$_3$H and C$_2$H to be cospatial.}
\end{table*}

\subsection{{Column density and rotation temperature estimates}}
\label{subsubsec:Results_of_T_rot_estimates}
{We produce population diagrams (also referred to as rotation diagrams) to obtain the rotational temperature ($T_\mathrm{rot}$) and the column density ($N_\mathrm{tot}$) for molecules with detected lines from three or more rotational transitions. {We have made a first-order correction for the optical depths. The method is described in detail in App.~\ref{subsubsec:Rot_Diag_Analysis}. The minimum and maximum line optical depths for each molecule for which we have made population diagrams are listed in Table~\ref{tab:max_optical_depths}}. In this section, we present the column densities and rotation temperatures obtained from the population diagram analysis}. All population diagrams produced are shown in Figs.~\ref{fig:SiO_rot_diags_all_stars} $-$ \ref{fig:C$_4$H_rot_diags_all_stars}. The rotational temperatures and column densities thus obtained are tabulated in Table~\ref{tab:N_and_T} and shown in Fig.~\ref{fig:T_rots_and_N_tots_rot_diags}.

As the centrally-peaked species are located close to the star, they are expected to have relatively high rotational temperatures. However, for all centrally-peaked species, except SiS, the population diagrams yield much lower temperatures than expected ($\sim$$10-20$\,K; Fig.~\ref{fig:T_rots_and_N_tots_rot_diags}, left panel). This is most likely due to three effects: (1) the limited range of energy levels covered {(e.g., the highest energy level observed for HCN is at $\sim$42 K, see Table~\ref{tab:apex_line_detections})}, where, in particular, high-excitation transitions are missing, yielding only a very limited sampling of the excitation properties, (2) the contribution from different lines may come from different regions based on their excitation conditions, and thus may not sample regions with the same temperature, and (3) {the assumption of optically thin lines is not always fulfilled}.
For the SiS emission, where the energy range covered is larger, we obtain a {somewhat higher} rotational temperature of $\sim$40\,K.

The shell species often cover a large range of energy levels {(e.g. 20 - 187 K for C$_4$H, see Tables~\ref{tab:alma_line_detections} and \ref{tab:apex_line_detections})}, {and the emission from different lines of a species come from roughly the same region. They are} therefore are expected to produce reliable estimates of rotation temperatures. HC$_3$N and HC$_5$N yield rotation temperatures in the range $15-30$\,K, whereas C$_4$H and C$_3$N give slightly higher temperatures, around $30-45$\,K. Based on these results, we use rotational temperatures of 40\,K and 20\,K for centrally-peaked and shell species, respectively, in the abundance estimates based on single lines in Sect.~\ref{subsubsec:Single_Line_Analysis}. We also note that, in general, the obtained rotational temperatures are highest for IRAS 07454$-$7112 (Fig.~\ref{fig:T_rots_and_N_tots_rot_diags}, left panel).

\subsection{{Abundance estimates}}
\label{subsubsec:Summary_of_abundance_estimates}
{For the species for which we could produce population diagrams (see Table~\ref{tab:N_and_T} and App.~\ref{app:appendix_E}), we calculated fractional abundances from the corresponding column densities, using Eq.~\ref{eqn:frac_abund_from_N_tot}. For all the other species, we obtain a rough estimate of the abundance analytically using Eq.~\ref{eqn:frac_abund_analytical}. These two methods are described in App.~\ref{app:appendix_E}}.

\begin{figure*}[!h]
   \includegraphics[width=\textwidth]{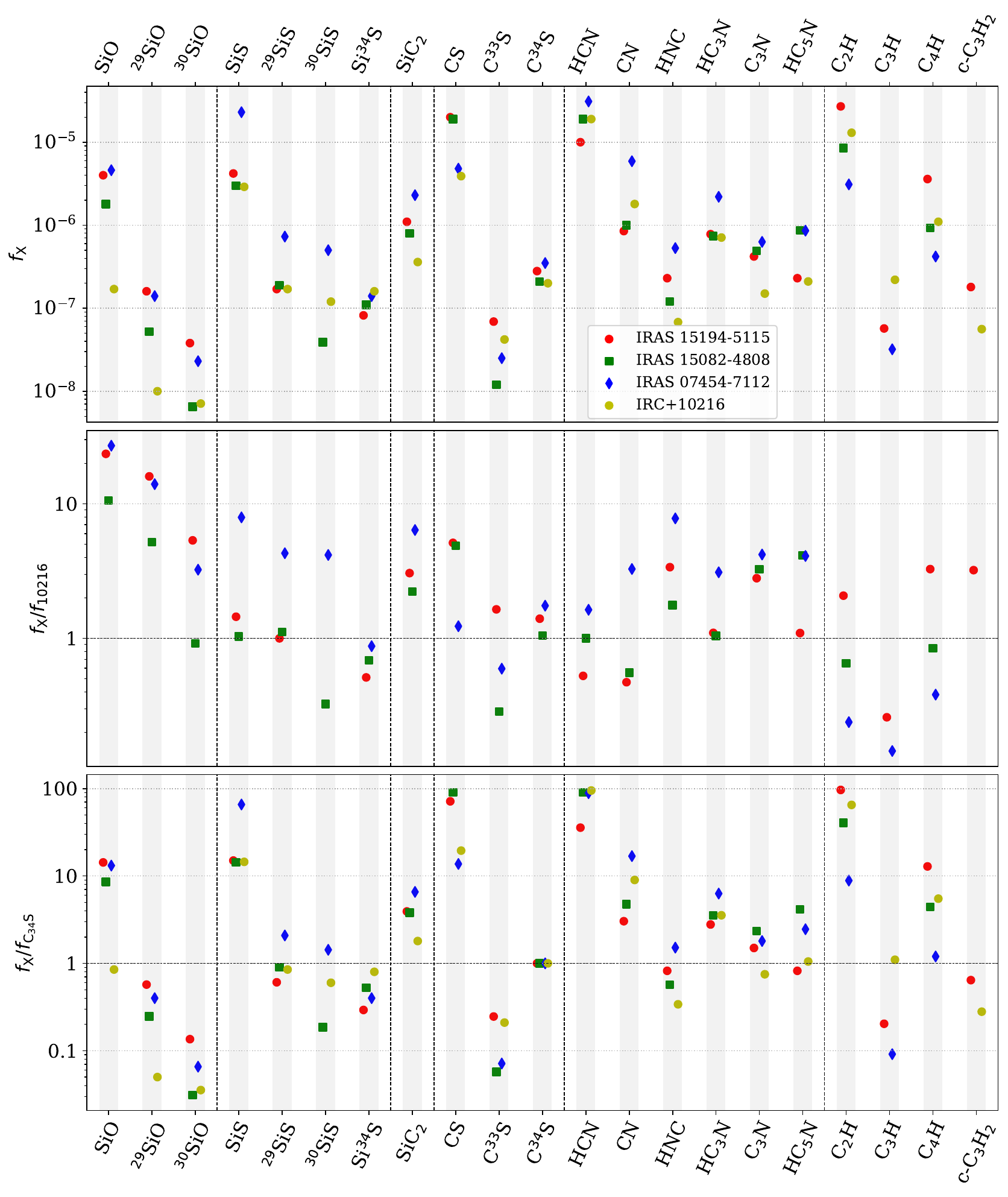}
   \caption{Molecular fractional abundances for the three stars (this work) and IRC~+10\,216 (from the literature, see Table~\ref{tab:f_As}). The absolute abundances are shown in the top panel. The middle panel shows the abundances normalised to IRC~+10\,216, and the bottom panel shows the abundances normalised to each star's own C$^{34}$S abundance. {Uncertainties on these estimates are discussed in Sect.~\ref{subsubsec:Uncertainties_of_abundance_estimates}}.}
   \label{fig:abunds_plots}
\end{figure*}

The fractional abundances estimated towards our stars are listed in Table~\ref{tab:f_As} and shown in Fig.~\ref{fig:abunds_plots}, along with the corresponding abundances for IRC~+10\,216 obtained from the literature. We also show the abundances normalised with those of IRC~+10\,216, for an easier source-to-source comparison, and normalised to each star's C$^{34}$S abundance, to eliminate source-specific uncertainties on distance and MLR. We chose C$^{34}$S in particular, since the S and $^{34}$S abundances are not expected to vary between the sample stars, {nor with respect to} IRC~+10\,216 \citep{henry2012_sulfur}. Whereas we find similar CS abundances for all of the stars, we note that \citet{Massalkhi_et_al_2019} have reported a decreasing CS peak abundance for increasing density in carbon-rich CSEs.

{We have made first-order corrections for optical depths, that are particularly significant for HCN, SiO and CS. However, the estimated rotational temperatures are much lower than expected for these centrally-peaked species (see Table~\ref{tab:N_and_T} and Fig.~\ref{fig:T_rots_and_N_tots_rot_diags}, left panel), and we therefore believe that for these species the reported abundances are lower limits.}

The abundances of the shell species are in general comparable, within the uncertainties, in the three stars. {Since extended emission appears to have been resolved out for SiC$_2$ and C$_4$H towards IRAS 15194$-$5115 and IRAS 15082$-$4808 (see Sect.~\ref{subsubsec:shell_species_AARPs}), the ALMA points in the population diagrams of these species for these sources could be slightly underestimated. We do not expect this to change the abundances beyond the estimated uncertainties}.

\subsubsection{Uncertainties of abundance estimates}
\label{subsubsec:Uncertainties_of_abundance_estimates}
The accuracy of a determined absolute abundance depends on the method used to estimate it, the validity of the assumptions that went into the calculations, and the uncertainties of input parameters. We have assumed a simple circumstellar model (including the shape of the radial abundance distributions), and excitation in LTE of the molecular energy levels. Among the input parameters, we have the adopted distances and the mass-loss rates. In addition, the amount and quality (including calibration uncertainties) of the observational data contribute to the uncertainty, through the measured fluxes and {estimated} emitting-region sizes. This makes it difficult to estimate formal errors of the abundances.

{The arcs seen in the molecular emission towards IRAS 15194$-$5115 and IRAS 15082$-$4808 suggest that there are density enhancements (see Sect.~\ref{subsubsec:shell_species_AARPs}). These are not considered in the assumed smooth H$_2$ flow, as we do not have a well-constrained quantitative estimate of these enhancements. This also adds to the uncertainty of the abundance estimates. Attempts to address this demand detailed modelling which is outside the scope of the present work.

In general, we estimate that the {overall} uncertainty, {including all parameters and model assumptions}, is about a factor of $3-5$ for the absolute abundances estimated using population diagrams, depending on the quality of the population diagram. The single-line estimates are, at best, order-of-magnitude estimates. The uncertainty decreases to a factor of $2-3$ {and a factor of 5 in the cases of population diagram and single-line estimates, respectively}, when comparing abundances that are normalised individually {to the respective C$^{34}$S abundance} for each source (Sect~\ref{subsubsec:Summary_of_abundance_estimates}), since this means that the influences of distance, mass-loss rate, and circumstellar model to a large extent are removed.
{The comparison abundances adopted for IRC~+10\,216, which are obtained using different methods, have an uncertainty of at least a factor of 2-3}.

\subsection{Isotopic ratios}
\label{subsec:isotopic_ratios}
Stellar elemental isotopic ratios are measures of the nucleosynthetic history (including dredge-up) of a star, and are, as such, important observational results. {Spectral lines from our data can be used to estimate circumstellar isotopologue ratios}. In most cases, there is every reason to {think} that these are also good estimates of the corresponding stellar isotope ratios, since the different isotopologues are produced, destroyed, and excited in {a similar way, to first approximation}. This applies, for instance, to silicon and sulphur. For carbon, the situation is special since the photo-dissociation process is isotope selective in the case of CO \citep{Saberi_et_al_2019}. However, the effect of this is estimated to be small for high-mass-loss-rate objects \citep{Saberi_et_al_2020}, that is, the type of stars in this study.

For species for which we could make reliable population diagrams, the isotopic ratios are calculated from the abundance ratios of the relevant molecules. For all other species, the isotopic ratio is determined by calculating the {corresponding} line intensity ratios. This method assumes that the $J-J^{'}$ transition is excited under the same conditions for both isotopologues. {We refrain from performing any frequency corrections on the isotopic ratios, as the uncertainties in the estimated ratios are significantly greater than the magnitude of such corrections}. {We present estimates of the isotopic ratios for the three stars in Table~\ref{tab:isotopic_ratios}, along with isotopic ratios for IRC+10216 and the Sun obtained from the literature for comparison}.

{Carbon}: The $^{12}$C/$^{13}$C ratios calculated from {the} optically thin lines {of} HNC, SiC$_2$, HC$_3$N, C$_4$H, and C$_3$N for all three stars are consistent with their respective values obtained from the CO radiative transfer modelling by \citet{Woods_et_al_2003}, though the C$_4$H ratio is slightly larger than the others. 

{Silicon}: The $^{28}$Si/$^{29}$Si, $^{28}$Si/$^{30}$Si, and $^{29}$Si/$^{30}$Si values in the three stars are similar to the expected values of 20, 30, and 1.5, the solar values \citep{Asplund_et_al_2009}. 

{Sulphur}: The $^{32}$S/$^{33}$S and $^{32}$S/$^{34}$S values, about 100 and 20, respectively, in the three stars, are close to the expected solar values of 125 and 22.

\begin{table*}[h]
   \caption{Isotopic ratios.}
   \label{tab:isotopic_ratios}
   \centering
      \begin{tabular}{l l r r r r r}
      \hline\hline & \\[-2ex]
      \makecell{\\Ratio} & \makecell{\\Species} & \multicolumn{5}{c}{Source}\\
      \cline{3-7} & \\[-2ex]
      & &  15194$-$5115 & 15082$-$4808 & 07454$-$7112 & IRC~+10\,216 & Solar$^{(b)}$ \\
      \hline & \\[-1ex]
    $^{12}$C/$^{13}$C	&	HC$_3$N		&		4.9({0.5})	 &		31.0({3.5})	&			21.0({2.2})	&	  	    &    \\	
    						&	C$_4$H	 	&		10.4({1.5}) 	 &		--			&			--			&	 45(12)$^{(1)}$ 	 	&       \\	
    						&	HNC		&		5.0({0.5})     &		--			&			--			&	 $\geq38$$^{(2)}$ 	 	&       \\						
    						&	SiC$_2$	 	&		5.8({0.6})     &		--			&			--			&	 37(5)$^{(1)}$, 50(7)$^{(1)}$, 41(4)$^{(1)}$ 	 	&       \\	
    						&	C$_3$N	 	&		5.9({0.6})     &		--			&			--			&	35(15)$^{(1)}$  	 	&       \\						
    						&	Average	 	&		6.4({0.6})	 &		31.0({3.5})	&			21.0({2.2})	&	45(3)$^{(1)}$ 	&   89   \\ \hline \\

    $^{32}$S/$^{34}$S	&	SiS		&	19(4)		&	{28(8)}$^{(a)}$&	21(4)		&	18.9(1.3)$^{(3)}$ 	&	22 \\
    $^{34}$S/$^{33}$S	&	CS		&	4.8({1.0})		&	6.1({1.3})			&	5.4({1.0})		&	{5.2(0.4)}$^{(3)}$	&	5.5 \\
    $^{32}$S/$^{33}$S	&	[$^{32}$S/$^{34}$S]$\times$[$^{34}$S/$^{33}$S]	&	93({28})		&	127({44})		&	115({29})	&	93(10)$^{(3)}$ 	&	125 \\ \hline \\

    $^{28}$Si/$^{29}$Si	&	SiS		&	16.8({2.1})		 &	{16(3)}$^{(a)}$	&	{16(7)}		&	17.2(1.1)$^{(3)}$		&	20 \\ \hline \\

    $^{29}$Si/$^{30}$Si	&	SiO		&	1.3({0.2})  		 &	1.4(0.2)	&  	1.4(0.3)			&		&	  \\
    					&	SiS		&		--		     &	2.2({0.5})  	&	1.8(0.8)			&			&	       \\
    					&	Average	&	1.3({0.2})		 &	1.8({0.3})	&	1.6(0.4)			&	1.46(0.11)$^{(3)}$		&	1.5       \\ \hline \\

    $^{28}$Si/$^{30}$Si	&	SiS		&		--		 &	 --		&	{28(12)}	&	24.7(1.8)$^{(3)}$		&	30	\\					
    					&	[$^{28}$Si/$^{29}$Si]$\times$[$^{29}$Si/$^{30}$Si]	&	       21.8({3.2})     &	22(5)	&		21(9)		&			&		\\ \hline & \\[-1ex]
      \end{tabular}
      \tablefoot{The values marked $^{(a)}$ are from abundance ratios, and the others are from line intensity ratios. The multiple number of carbon atoms present in a molecule has been taken into account when calculating the $^{12}$C/$^{13}$C ratio from HC$_3$N, C$_4$H, and C$_3$N. For IRC~+10\,216, isotopic ratios are taken from $^{{(1)}}$\citet{Cernicharo_et_al_2000}, $^{(2)}$\citet{kahane1988}, and $^{(3)}$\citet{He_et_al_2008}. $^{(b)}$Solar values are from \citet{Asplund_et_al_2009}. {The values given in parentheses are the estimated overall uncertainties}.}
\end{table*}

\section{Discussion}
\label{sec:Discussion}
A comparison of molecular radial brightness distributions and abundances between the three stars and with IRC~+10\,216, allows an understanding of whether and how the chemistry in C-type AGB CSEs is affected by variations in properties including the MLR and the stellar $^{12}$C/$^{13}$C ratio. We compare the molecular radial brightness distributions of various molecules focusing on Figs.~\ref{fig:All_AARPs_comparison} and~\ref{fig:all_emission_sizes_comparison} and the molecular abundances based on Fig.~\ref{fig:abunds_plots}. Even though the AARPs show only the radial structure of the emission and average out all azimuthal variations, they still provide valuable information on the chemical stratification in the CSE, and thereby on the formation and destruction mechanisms of various molecules.
 
{We start by analyzing whether the spatial extents of the molecular CSEs follow that expected from photodissociation by an external UV radiation field}. The photodissociation radius of a molecule depends on how well it is shielded from the interstellar radiation field (ISRF). This is determined by the outward column density of dust, which in turn depends on the mass-loss rate, or rather the circumstellar density (proportional to $\dot{M}$/$\varv_\mathrm{exp}$), {and also on the gas-to-dust ratio}. The circumstellar density is lowest for IRAS~07454$-$7112, and about three and five times higher for IRAS~15194$-$5115 and IRAS~15082$-$4808, respectively {(Table~\ref{tab:source_properties})}. {The gas-to-dust ratios for the three sources are likely similar \citep{Groenewegen_1998}}. Thus, we expect the sizes of the molecular shells to be smallest for IRAS~07454$-$7112, larger for IRAS~15194$-$5115, and largest for IRAS~15082$-$4808. Indeed, the shells of IRAS~07454$-$7112 are by far the smallest ones, by about a factor of three {(Figs.~\ref{fig:All_AARPs_comparison}, \ref{fig:all_emission_sizes_comparison}, Table~\ref{tab:em_region_sizes})}. However, those of IRAS~15194$-$5115 and IRAS~15082$-$4808 are of about the same size. They are also comparable in size to the molecular shells seen towards IRC+10216, which has a circumstellar density roughly 3 times higher than that of IRAS~15194$-$5115, and 1.5 times that of IRAS~15082$-$4808. This could be attributed to errors in the distances (which affect mass-loss-rate estimates and spatial extents), differences in the gas-to-dust ratio, differences in the circumstellar dust properties, and/or to the CSEs being exposed to slightly different strengths of the ISRF. Thus, our results, which are corroborated by the radial brightness profiles of also the centrally-peaked species, are not inconsistent with the expected sizes of molecular distributions affected by photo-dissociation by UV light from the outside. {The next paper in the series will address this in a {more} quantitative way (see Sect.~\ref{sec:Introduction})}. {As mentioned above, the lower circumstellar density of IRAS~07454$-$7112 leads to the parent molecules being photodissociated closer to the star than in the other two sources}. This may be reflected in the higher rotation temperatures obtained for {the daughter} species for this object (Fig.~\ref{fig:T_rots_and_N_tots_rot_diags}, left panel). 

When looking at the AARPs of shell species in more detail, we conclude that the cyanopolyynes are located slightly inwards with respect to the hydrocarbons (see Figs. \ref{fig:All_AARPs_comparison}, \ref{fig:all_emission_sizes_comparison} and Table \ref{tab:em_region_sizes}). {The average peak position of the cyanopolyynes falls inside that of the hydrocarbons by around 25-30\% for both IRAS 15194$-$5115 and IRAS 15082$-$4808, and by roughly 10\% for IRAS 07454$-$7112}. We find no {marked} difference in the locations among the species of these respective groups. The largest spatial extents are found for HNC, CN, and possibly C$_3$N, {and the hydrocarbons C$_2$H and C$_4$H}. {This is consistent with the results of \citet{Daniel_et_al_2012} and \citet{Cernicharo_et_al_2013} who found HNC emission extending to more than 30$\arcsec$ towards IRC~+10\,216, and \citet{Agundez_et_al_2017} who found CN to present emission to similar radii, extending beyond that of the other cyanides and cyanopolyynes around this star}. The shell of c-C$_3$H$_2$, detected only towards IRAS 15194$-$5115, appears roughly coincident with the C$_2$H and C$_4$H shells.  In all of these respects, we {reproduce} the findings of \citet{Agundez_et_al_2017} in the case of IRC~+10\,216. {We note that while \citet{Agundez_et_al_2017} found that the CN emission towards IRC~+10\,216 clearly extends beyond that of the C$_2$H and C$_4$H radicals, we observe the faint extended emission from these radicals to have roughly the same radial extent as the CN towards IRAS 15194$-$5115 and IRAS 15082$-$4808 (Fig.~\ref{fig:All_AARPs_comparison})}. {However, this is due to the most extended CN emission being filtered out due to the lack of very short baselines in our observations, in contrast to that of \citet{Agundez_et_al_2017}, who used single dish observations to recover the flux filtered out by ALMA}. 

{From our observations, we find that HC$_3$N is present within one synthesised beam around two of our stars, corresponding to $\sim$700 AU and $\sim$1000 AU for IRAS 15194$-$5115 and IRAS 15082$-$4808, respectively}. For IRC~+10\,216, \citet{Siebert_et_al_2022} found significant emission from high-excitation ($J= 38-37,\, 30-29,\, 28-27$) lines of HC$_3$N within 350 AU from the star. \citet{Herbst_and_Leung1990}, \citet{Howe_and_Millar_1990}, and \citet{Agundez_et_al_2017} suggest that HC$_3$N is formed in {C-type} CSEs mainly through the reaction}
\begin{equation}
   {\mathrm{CN + C_2H_2 \rightarrow HC_3N + H}}
\end{equation}
Hence, it is expected to appear in the inner CSE if there is enough CN available. We cannot confirm this {pathway as the main production channel}, since we do not detect CN emission above the noise level in the innermost part in any of our CSEs. However, this could be a sensitivity effect since the surface brightness of the CN emission is low at the systemic velocity for the two sources where we have sufficient spatial resolution to do this. \citet{Van_de_Sande_and_Millar_2022} {suggested} that a significant inner wind abundance of complex species, including HC$_3$N, can be indicative of the presence of a white dwarf binary companion. However, according to their models, this will also lead to severely reduced SiO and SiS abundances, {an effect not seen in our data}.

We find that the SiC$_2$ abundance distribution has both a central and a shell component. The presence of a central component has been discussed before by, for instance, \citet{Lucas_et_al_1995, Fonfria_et_al_2014, Velilla-Prieto_et_al_2015, Massalkhi_et_al_2018}. It is most likely explained by SiC$_2$ being formed in the extended atmosphere, as already suggested by \citet{MacKay_and_Charnley_1999}, {and predicted by the chemical models of \citet{Agundez_et_al_2020}}. The detection of {hot} SiC$_2$ {ro-vibrational} bands in the optical towards carbon stars \citep{Sarre_et_al_1996} {supports this}. The SiC$_2$ shell, on the other hand, is expected to be formed due to the fast reaction of Si with C$_2$H$_2$ in the low-temperature {($<$100 K)} conditions prevalent in the outer parts of the CSE \citep{Canosa_et_al_2001, Cernicharo_et_al_2010}:
\begin{equation}
   {\mathrm{Si + C_2H_2 \rightarrow Si(CH)_2 \rightarrow SiC_2 + H_2}}
\end{equation}
Further out, it is photo-dissociated to form Si and C$_2$ \citep{Harada_et_al_2010}.

The arcs seen in the HC$_3$N and SiC$_2$ emissions at the systemic velocity for IRAS 15194$-$5115 are clearly co-spatial (Fig.~\ref{fig:line_em_comp_15194_15082}, top left panel), though these molecules themselves do not share any common chemistry. {These arcs appear to be traced by a variety of other molecules as well (Sect.~\ref{subsubsec:15194_complex_morphology}). This suggests that the arc structures seen are, predominantly, the result of density enhancements rather than abundance variations}. Such thin shells in various cyanopolyynes and hydrocarbons are also seen in the CSE of IRC~+10\,216 \citep{Agundez_et_al_2017}. Additionally, for IRC~+10\,216, {\citet{Mauron_and_Huggins_1999, Mauron_and_Huggins_2000, Leao_et_al_2006} and} \citet{Trung_and_Lim_2008} showed that the clumpy shells seen in cyanopolyyne emission coincide spatially with the dust arcs observed in dust-scattered optical light, hence representing the same density enhancements in the CSE. This is in line with the suggestion by \citet{Lykou_et_al_2018}, based on CO, SiO, CS, and HC$_3$N emission and NACO/VLT near-infrared data, that IRAS 15194$-$5115 could be a binary system in which a spiral-like density structure would be imprinted in the CSE by an orbiting companion. Similar substructure was observed in CO{, CN, and C$_4$H} towards IRC~+10\,216 by \citet{Cernicharo_et_al_2015, Decin_et_al_2015} and \citet{Guelin_et_al_2018}, which was attributed to episodic mass loss triggered by the existence of a {putative} binary companion. Towards IRAS 15194$-$5115, we find emission from C$_2$H also in the gaps between the arcs seen in HC$_3$N (Fig.~\ref{fig:line_em_comp_15194_15082}, middle-left panel and Fig.~\ref{fig:15194_radial_cut_profiles}). This need not be contrary to the proposition of density-enhanced shells but can rather be explained by the larger impact of density enhancements on the abundance of HC$_3$N \citep{Cordiner_and_Millar_2009}.

{The rotation temperatures we estimate from our population diagrams match reasonably well with those estimated for IRC~+10\,216. \citet{He_et_al_2008} obtained a rotation temperature of 28 K for HC$_3$N, and \citet{Cernicharo_et_al_2000} reported 35 K for C$_4$H and 20 K for C$_3$N, which align with our estimates. We note that \citet{Pardo_et_al_2022} also present population diagrams of C$_4$H and HC$_5$N, among other species, towards IRC~+10\,216 using lines from their 31.0-50.3 GHz spectral survey. These yield rotation temperatures of 4.9 and 10.1 K respectively, much lower than our estimates for these species (Table~\ref{tab:N_and_T})}.

{We have estimated that the uncertainties of the relative abundances when comparing our three stars are a factor of 2-3 for population diagrams and a factor of 5 for single-line estimates. When comparing with IRC~+10\,216 they increase to a factor of 5 for population diagrams and a factor of 10 for single-line estimates. Considering this, there are no significant differences or trends in the abundances among our three stars}, other than the {possibly} lower hydrocarbon abundances in IRAS~07454$-$7112, compared to the other two stars (Fig. \ref{fig:abunds_plots}). {The latter can be a result of the lower density of the CSE of IRAS 07454$−$7112. When comparing with IRC~+10\,216, all abundances except that of SiO are within a factor of ten, meaning that except for this species, we find no significant differences with the abundances estimated for IRC~+10\,216. The SiO abundances in our stars are higher than that for IRC~+10\,216 by an order of magnitude}. The same discrepancy was observed by \citet{Woods_et_al_2003} {in} their single-dish line survey of these sources. The putative overabundance of SiO {in our stars} needs to be confirmed through a proper radiative transfer analysis. 

Time variability of the intensity of thermal (non-maser) line emission has been reported for several species in the case of IRC~+10\,216 \citep{Cernicharo_et_al_2014, Pardo_et_al_2018, He_et_al_2019}. By comparing the observations from two epochs, we find variability in the 3 mm line emission of C$_2$H and HC$_3$N towards IRAS 15194$-$5115 and IRAS 15082$-$4808. Our findings (Sect.~\ref{subsec:line_variability}) are in line with the conclusions of \citet{Cernicharo_et_al_2014} and \citet{Pardo_et_al_2018}, who show that C$_2$H shows the highest variability at 3 mm, and that the phase of variability of the cyanopolyynes is opposite to that of C$_2$H, for IRC +10\,216 as well. {We note that in addition to C$_2$H and HC$_3$N, \citet{Pardo_et_al_2018} also found time variability in the lines of CN, C$_4$H, C$_5$H, and HC$_5$N towards IRC~+10\,216, which we are unable to ascertain for the stars in our sample from the current data. This can potentially be due to the intensity variation in the lines of these species being lower than the calibration uncertainty}.

The three stars in our sample have significantly different inferred $^{12}$C/$^{13}$C ratios (Table~\ref{tab:isotopic_ratios}), with IRAS~15194$-$5115 standing out with a very low value of 6. For this star, \citet{Ramstedt_and_Olofsson_2014} report a $^{12}$CO/$^{13}$CO ratio of 10, slightly higher than the value of 6 estimated by \citet{Woods_et_al_2003} and us (still within the uncertainties of both estimates). The other two stars, and IRC~+10\,216, have values in the range $20-50$, which is consistent with typical AGB evolution models {\citep[see e.g.][and references therein]{Lattanzio_and_Boothroyd_1997}} where the surface $^{12}$C/$^{13}$C ratio increases as the star evolves along the AGB. {This has also been traced by observations} \citep[e.g.][]{Lambert_et_al_1986, Abia_et_al_2001, Ramstedt_and_Olofsson_2014}. It cannot be concluded with certainty that the low ratio in IRAS 15194$-$5115 is due to hot-bottom burning (HBB) and a subsequent evolution to a carbon star, as the star may not be massive enough for HBB to occur. Whereas HBB efficiently produces N, we find no indications of any N enhancement {in the circumstellar species} for this star. It has been suggested that the presence of a hot binary companion may alter the circumstellar $^{12}$CO/$^{13}$CO {through differential photodissociation} so that it differs from the atmospheric $^{12}$C/$^{13}$C{, whereas this is not the case for the isotopologues of the other observed molecules in this study} \citep[see][]{Vlemmings_et_al_2013, Saberi_et_al_2019}. {They are therefore expected to give good estimates of the stellar  $^{12}$C/$^{13}$C ratio}. The models by \citet{Izzard_and_Tout_2003}, combining nucleosynthesis and binary star evolution, have shown that the atmospheric $^{12}$C/$^{13}$C can be lowered significantly in the presence of binary companions.

{Finally, the observed similarity of the silicon isotopic ratios with the respective solar values is in line with AGB nucleosynthesis in lower-mass stars \citep[$\lesssim$4$M_{\odot}$,][]{karakas2016}. 
In addition, the {sulphur} {isotopic} ratios being consistent with the corresponding solar ratios is {in line with} the fact that these isotopes are not expected to be influenced by nucleosynthesis in lower-mass stars \citep[][and references therein]{shingles2013, humire2020}}.

\section{Summary and conclusions}
\label{sec:Conclusions}
Our unbiased ALMA spectral survey mapped for the first time the 3\,mm molecular rotational transitions in the CSEs of the {C-type} AGB stars IRAS 15194$-$5115, IRAS 15082$-$4808, and IRAS 07454$-$7112 at high spatial resolution ($0\farcs7-1\farcs7$, corresponding to roughly $500-1000$\,AU). Extensive interferometric surveys of this kind have previously been performed only towards IRC~+10\,216. The survey yielded 311 identified lines (including hyperfine splitting components), from {132 rotational} transitions of 49 molecular species, including various isotopologues, and 3 {unidentified} lines. {Data} with sufficient signal-to-noise ratio {(${\gtrsim}3\sigma$ in the emission maps)}, which enabled us to comparatively analyse the morphological characteristics across the three stars {was obtained for 14 species}. We produced azimuthally-averaged radial profiles (AARPs) {to obtain the radial brightness distributions of these species, and estimated the extents of their emitting regions. We used optical-depth-corrected population diagrams or single-line calculations to calculate the fractional abundances of these molecules and their less-abundant isotopologues. Lines from our APEX single-dish surveys of the sources were used in the population diagrams to complement those from the ALMA observations}. We also report several new detections towards these stars, namely $^{29}$SiO, $^{30}$SiO, $^{29}$SiS, $^{30}$SiS, Si$^{34}$S, C$^{33}$S, Si$^{13}$CC, C$_6$H, C$_8$H, l$-$C$_4$H$_2$, and doubly $^{13}$C−substituted isotopologues of HC$_3$N.

We find deviations from a smooth, spherically symmetric CSE towards IRAS 15194$-$5115 and IRAS 15082$-$4808 in several molecules, and detect density enhancements {similar to what has previously been reported in the CSE of IRC~+10\,216}, which {might} indicate the presence of stellar/sub-stellar companions for these stars. {The very low $^{12}$C/$^{13}$C ratio and the presence of a `central' emission component for HC$_3$N may also be indicative of the presence of a binary companion for IRAS 15194$-$5115}.

By comparing our morphological results for the three sources amongst themselves and with results from the literature for IRC~+10\,216, we find that the chemistry in the CSEs, as traced by the radial order in which different molecules exist around the central star, is {very similar} in our sources and IRC~+10\,216. The chemistry does not seem significantly affected by the differences in the nucleosynthetic history of the individual stars, indicated by differences in their $^{12}$C/$^{13}$C ratios, except for the expected increased abundance of singly- and doubly-$^{13}$C-substituted isotopologues in the CSE of IRAS 15194$-$5115, owing to its low $^{12}$C/$^{13}$C ratio. We find that, within the uncertainties of our analysis, most of the estimated fractional abundances are similar across the source sample, and to the corresponding IRC~+10\,216 values. The notable exception is SiO, which appears more abundant in the three stars in our sample in comparison to IRC~+10\,216, {by an order of magnitude}. The calculated abundances of the {shell} species match reasonably well with photochemical models for all sources. We also detect variability in the line emission of C$_2$H and HC$_3$N for two stars in our sample.

We estimate the carbon, silicon, and sulphur isotopic ratios for the three stars. The $^{12}$C/$^{13}$C ratios match the values from the literature for all stars, and the silicon and sulphur ratios of the three stars match reasonably well the corresponding IRC~+10\,216 and solar values.

Overall, we conclude that the chemistry in the {C-type} AGB CSEs studied in this paper, {as judged from radial brightness distributions and estimated abundances}, is consistent across our small sample and with that of IRC~+10\,216. This indicates that IRC~+10\,216 serves reasonably well as an archetypal {C-type} AGB star, {in terms of modelling C-type CSE chemistry, at least for MLRs} within an order of magnitude of that of itself. However, we remind the reader of the possible binary nature of IRC~+10\,216 and potentially also of some of the sources in our study. Binarity might influence the chemical properties differently for different sources, depending on the properties of the objects in the system. 

Non-LTE radiative transfer modelling is necessary to more accurately determine the molecular abundances. This, and the subsequently updated chemical models will be presented in a series of upcoming papers. More extensive observations, both in terms of sources and spectral coverage are required to thoroughly constrain the chemical models and small-scale chemical variations in the CSEs. A detailed investigation of the morphology of the CSEs of the stars in our sample and the possible presence of binary companions will require observations with higher angular resolution and sensitivity.

\begin{acknowledgements}
RU acknowledges data reduction support from the Nordic ALMA Regional Centre (ARC) node based at Onsala Space Observatory (OSO), Sweden. The Nordic ARC node is funded through Swedish Research Council grant No 2017-00648. EDB acknowledges financial support from the Swedish National Space Agency. SBC and MAC were supported by the NASA Planetary Science Division Internal Scientist Funding Program through the Fundamental Laboratory Research work package (FLaRe). IdG acknowledges support from grant PID2020-114461GB-I00, funded by MCIN/AEI/10.13039/501100011033. The work of MGR is supported by NOIRLab, which is managed by the Association of Universities for Research in Astronomy (AURA) under a cooperative agreement with the National Science Foundation, USA. This paper makes use of the following ALMA data: ADS/JAO.ALMA\#2013.1.00070.S, ADS/JAO.ALMA\#2015.1.01271.S, ADS/JAO.ALMA\#2011.0.00001.CAL. ALMA is a partnership of ESO (representing its member states), NSF (USA) and NINS (Japan), together with NRC (Canada), MOST and ASIAA (Taiwan), and KASI (Republic of Korea), in cooperation with the Republic of Chile. The Joint ALMA Observatory is operated by ESO, AUI/NRAO and NAOJ. This paper is based on observations with the Atacama Pathfinder EXperiment (APEX) telescope. APEX is a collaboration between the Max Planck Institute for Radio Astronomy, the European Southern Observatory, and the Onsala Space Observatory. Swedish observations on APEX are supported through Swedish Research Council grant No 2017-00648. The APEX observations were obtained under project numbers O-0107.F-9310 (SEPIA/B5), O-0104.F-9305 (PI230), and O-087.F-9319, O-094.F-9318, O-096.F-9336, and O-098.F-9303 (SHeFI).
\end{acknowledgements}

\bibliography{Charting_Circumstellar_Chemistry_of_Carbon-rich_AGB_Stars}

\begin{appendix}
\onecolumn

\section{ALMA Band 3 spectra}
\label{app:appendix_A}
This appendix shows the entire ALMA Band 3 spectra of the three sources. The spectra have been extracted by integrating the intensity in a circular aperture of 12.5$\arcsec$ radius centred on the respective continuum peaks, so as to recover emission from all molecules. Line detections (Table~\ref{tab:alma_line_detections}) have been labelled in the figure at the corresponding frequencies.

\begin{figure}[h]
\centering
  \includegraphics[width=.9\linewidth, height=0.83\textheight]{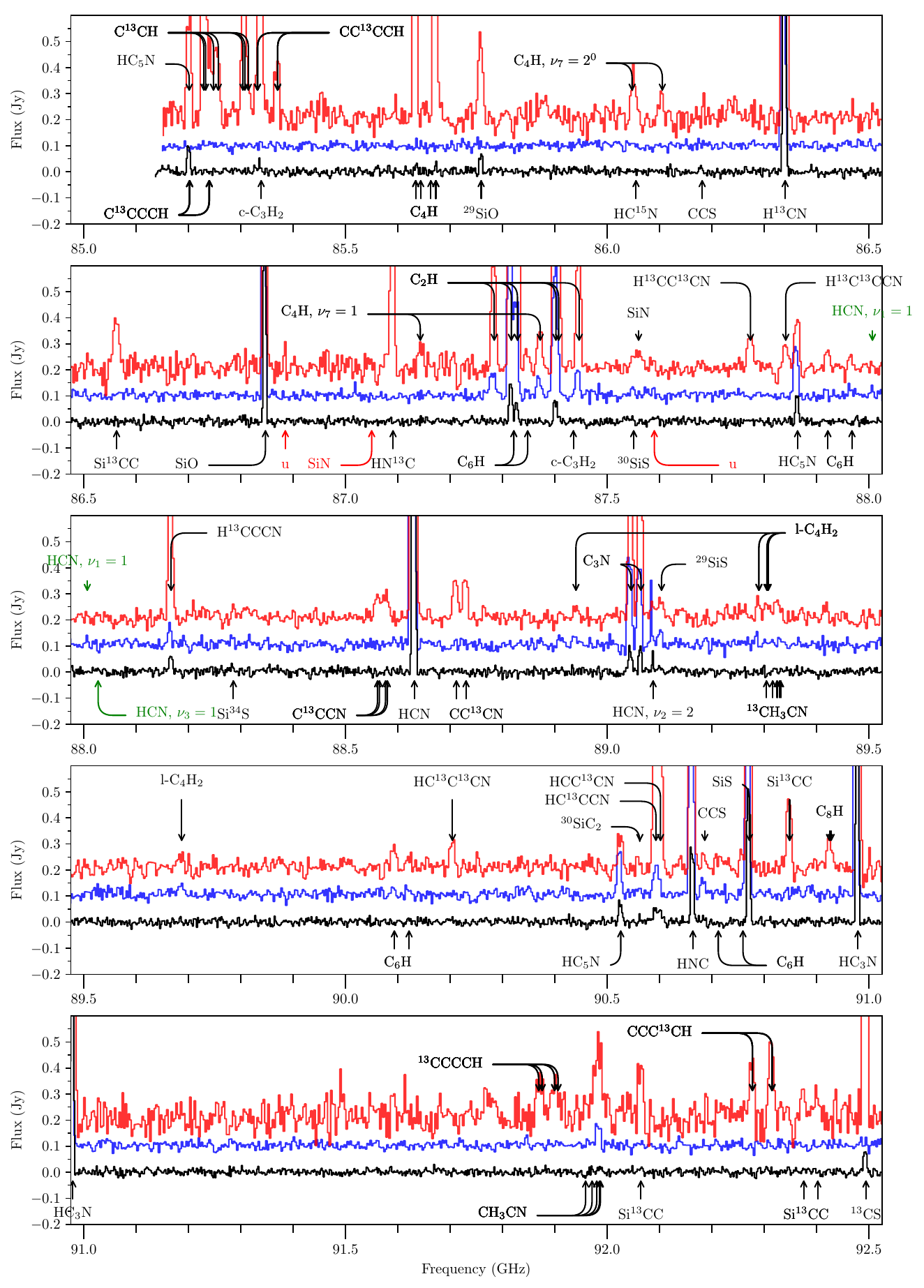}\par
  \caption{ALMA band 3 line survey of IRAS 07454$-$7112 {(black)}, IRAS 15082$-$4808 {(blue; vertical offset of 0.1 Jy)}, and IRAS 15194$-$5115 {(red; vertical offset of 0.2 Jy)}. Labels show the carrier molecule of the indicated emission. Red labels indicate tentative or unidentified detections, and green labels indicate lines that have been identified only in central-pixel spectra (not shown). The spectra from the combined data have been inserted in the respective frequency ranges (see Table~\ref{tab:observational_details}).}
  \label{fig:almaband3_identified_sample}
\end{figure}
\foreach \index in {2, ...,4}
{
\begin{figure*}
\centering
  \includegraphics[width=.9\linewidth]{Figures/Band_3_spectra/SAMPLE_ALMA_\index.pdf}\par
  {\textbf{Fig.~\ref{fig:almaband3_identified_sample}.} continued.}
\end{figure*}
\clearpage
}

\begin{figure}[h]
\centering
\includegraphics[width=.9\linewidth]{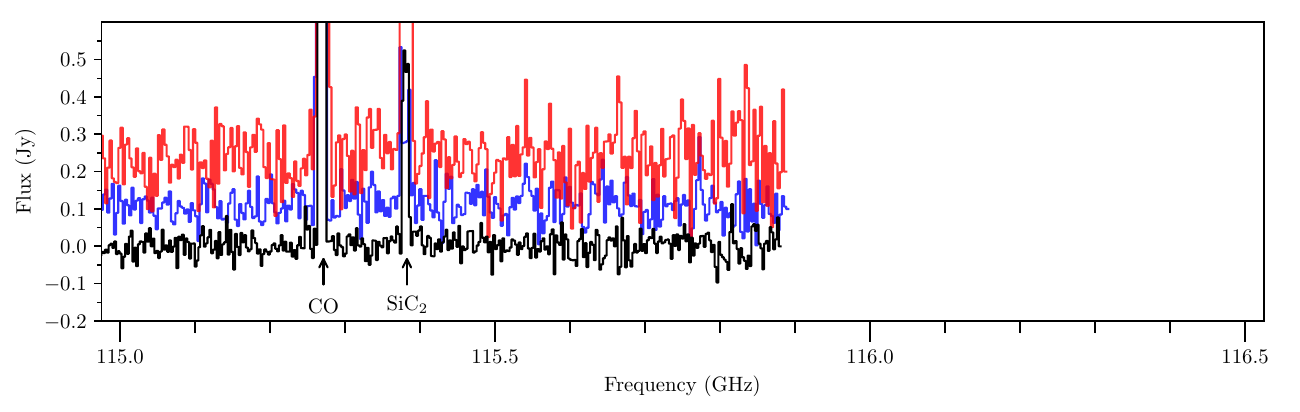}\par
  {\textbf{Fig.~\ref{fig:almaband3_identified_sample}.} continued.}
\end{figure}

\section{Details of detected transitions}
\label{app:appendix_B}
{This appendix lists the detected emission lines along with their upper-level energies and integrated intensities for the three stars. Table~\ref{tab:alma_line_detections} presents the lines detected in the ALMA band 3 spectra, and Table~\ref{tab:apex_line_detections} lists the lines from the APEX surveys (see Sect.~\ref{subsec:Observations}) which are used in this work for abundance calculations.}

\begin{small}
\setlength\LTleft{0pt}
\setlength\LTright{0pt}
\begin{longtable}{@{\extracolsep{\fill}}llrrrrr}
   \caption{\label{tab:line_detections}Details of all detected transitions in the ALMA band 3 survey.}\\
   \hline\hline & \\[-2ex]
   \makecell{\\Molecule} & \makecell{\\Transition} & \makecell{\\Rest Freq.\\(GHz)} & \makecell{\\$E_\mathrm{up}$/$k_\mathrm{B}$ (K)} & \multicolumn{3}{c}{$\int S_\mathrm{ul} d\varv$ (Jy km/s)}\\
   \cline{5-7} & \\[-2ex]
   & & & & 15194 & 15082 & 07454 \\
   \hline & \\[-2ex]
   \endfirsthead
   \caption{continued.}\\
   \hline\hline & \\[-2ex]
   \makecell{\\Molecule} & \makecell{\\Transition} & \makecell{\\Rest Freq. (GHz)} & \makecell{\\$E_\mathrm{up}$/$k_\mathrm{B}$ (K)} & \multicolumn{3}{c}{$\int S_\mathrm{ul} d\varv$ (Jy km/s)}\\
   \cline{5-7} & \\[-2ex]
   & & & & 15194 & 15082 & 07454 \\
   \hline & \\[-2ex]
   \endhead
   \hline & \\[-2ex]
   \endfoot
    HC$_5$N  &  32-31  &85.201340&  67.5 &--&--&2.2(0.1)\\
    C$^{13}$CCCH  &   N= 9-8, J=19/2-17/2  &85.201897&  20.4 &9.4(0.4)&--&--\\
    C$^{13}$CH  &  N= 1-0, J=3/2-1/2  &85.229326&  4.1 &{bl.}&--&--\\
    C$^{13}$CCCH  &   N= 9-8, J=17/2-15/2  &85.239907&  20.5 &{bl.}&--&--\\
    C$^{13}$CH  &  N= 1-0, J=1/2-1/2  &85.314378&  4.1 &{bl.}&--&--\\
    CC$^{13}$CCH  &  N= 9-8, J=19/2-17/2  &85.331904&  20.5 &{bl.}&--&--\\
    c-C$_3$H$_2$  &  2$_{1,2}$-1$_{0,1}$  &85.338894&  6.4 &19.7(0.5)&2.9(0.2)&0.6(0.1)\\
    CC$^{13}$CCH  &  N= 9-8, J=17/2-15/2  &85.370342&  20.5 &5.9(0.4)&--&--\\
    C$_4$H  &  N = 9-8, J = 19/2-17/2  &85.634004&  20.5 &33.8	(0.4)&5.9(0.2)&0.8(0.1)\\
    C$_4$H  &  N = 9-8, J = 17/2-15/2  &85.672582&  20.6 &32.9(0.4)&6.6(0.2)&0.7(0.1)\\
    $^{29}$SiO  &  2-1  &85.759194&  6.2 &9.5(0.4)&2.5(0.2)&1.3(0.1)\\
    C$_4$H, $\nu_7=2^0$  &	 N=9-8, J=19/2-17/2  &86.047530&  397.6& {bl.} &--&--\\
    HC$^{15}$N  &  1-0  &86.054966&  4.1 &{bl.}&--&--\\
    C$_4$H, $\nu_7=2^0$  & 	N=9-8, J=17/2-15/2  &86.105294&  397.6& 3.0(0.7)&--&--\\
    CCS  &  7$_6$-6$_5$  &86.181391&  23.3 & 0.8(0.3) &--& 0.3(0.1)\\
    H$^{13}$CN  &  1-0  &86.339921&  4.1 &249.5(0.4)&48.7(0.2)&37.8(0.1)\\
    Si$^{13}$CC  &  4$_{1,4}$-3$_{1,3}$  &86.562636&  12.3 &6.0(0.4)&--&--\\
    SiO  &  2-1  &86.846985&  6.3 &95.7(0.4)&35.8(0.2)&15.3(0.1)\\
    u  &--&86.886000&--&1.0(0.3)&--&--\\
    SiN?  &  N=2-1, J=3/2-1/2  &87.049857& 6.3 &--&--&--\\
    HN$^{13}$C  &  1-0  &87.090825&  4.2 &13.2(0.4)&--&--\\
    C$_4$H, $\nu_7=1$  &  J=19/2-17/2,$\Omega$=3/2, l=e  &87.142857&  265.5 &2.9(0.7)&--&--\\
    C$_2$H  &  N=1-0$^{(a)}$&87.316898&  4.2 &264.3({1.0})&58.5({0.6})&7.6(0.2)\\
    C$_6$H  &  J=63/2-61/2, $\Omega$=3/2, l=e  &87.321504&  67.9 &{bl.}&{bl.}&--\\
    C$_6$H  &  J=63/2-61/2, $\Omega$=3/2, l=f  &87.347994&  67.9 &2.1(0.5)&0.6(0.2)&--\\
    C$_4$H, $\nu_7=1$  &  J=19/2-17/2,$\Omega$=3/2, l=f  &87.371999&  265.5 &5.0(0.6)&2.0(0.3)&--\\
    c-C$_3$H$_2$  &  5$_{4,2}$-5$_{3,3}$  &87.435318&  45.3 &{bl.}&{bl.}&--\\
    $^{30}$SiS  &  5-4  &87.550558&  12.6 &{bl.}&{bl.}&{bl.}\\
    SiN  &  N=2-1, J=5/2-3/2  &87.559811& 6.3 &{bl.}&{bl.}&--\\
    u  &--&87.590000&--&1.0(0.3)&--&--\\
    H$^{13}$CC$^{13}$CN  &  10-9  &87.773712&  23.2&4.2(0.3)&--&--\\
    H$^{13}$C$^{13}$CCN  &  10-9  &87.841102&  23.2&2.9(0.3)&--&--\\
    HC$_5$N  &  33-32  &87.863630& 71.7 &4.8(0.2)&6.2(0.1)&2.0(0.1)\\
    C$_6$H  &  J=63/2-61/2, $\Omega$=1/2, l=f  &87.921593&  90.7 &2.5(0.5)&0.6(0.2)&--\\
    C$_6$H  &  J=63/2-61/2, $\Omega$=1/2, l=e  &87.967595&  90.7 &1.7(0.4)&1.1(0.2)&--\\
    HCN, $\nu_1=1$?  &  1-0  &88.006690&  4768.7 &--&--&--\\
    HCN, $\nu_3=1$?  &  1-0  &88.027280&  3021.1 &--&--&--\\
    H$^{13}$CCCN  &  10-9  &88.166832&  23.3 &14.9(0.2)&1.8(0.1)&1.3(0.1)\\
    Si$^{34}$S  &  5-4  &88.285828&  12.7 &--&0.7(0.1)&0.3(0.1)\\
    C$^{13}$CCN  &  N=9-8$^{(a)}$&88.561247&  11.8 &9.0(0.4)&--&--\\
    HCN  &  1-0  &88.631602&  4.3 &294.6(0.2)&224.8(0.1)&81.2(0.1)\\
    CC$^{13}$CN  &  N=9-8, J=19/2-17/2  &88.711791&  11.8 &4.9(0.2)&--&--\\
    CC$^{13}$CN  &  N=9-8, J=17/2-15/2  &88.730291&  11.8 &4.5(0.2)&--&--\\
    l-C$_4$H$_2$  &  10$_{1,10}$-9$_{1,9}$  &88.940236&  37.0 &1.8(0.5)&1.5(0.4)&--\\
    C$_3$N  &  9-8 (A)  &89.045592&  21.4 &18.1	(0.2)&10.4	(0.1)&2.0(0.1)\\
    C$_3$N  &  9-8 (B)  &89.064352&  21.4 &18.1	(0.2)&10.0(0.1)&1.8(0.1)\\
    HCN, $\nu_2=2$  &  1$_0$-0$_0$  &89.087690&  2035.0 &2.5(0.2)&2.9(0.3)&0.7(0.1)\\
    $^{29}$SiS  &  5-4  &89.103749&  12.8 &2.5(0.2)&1.7(0.1)&0.4(0.1)\\
    l-C$_4$H$_2$  &  10$_{3,8}$-9$_{3,7}$  &89.289791&  145.2 &{bl.}&{bl.}&--\\
    l-C$_4$H$_2$  &  10$_{3,7}$-9$_{3,6}$  &89.289793&  145.2 &{bl.}&{bl.}&--\\
    $^{13}$CH$_3$CN  &  5$_4$-4$_4$  &89.304386&  127.4 &{bl.}&--&--\\
    l-C$_4$H$_2$  &  10$_{2,9}$-9$_{2,8}$  &89.305141&  77.6 &{bl.}&{bl.}&--\\
    l-C$_4$H$_2$  &  10$_{2,8}$-9$_{2,7}$  &89.307563&  77.6 &{bl.}&{bl.}&--\\
    $^{13}$CH$_3$CN  &  5$_3$-4$_3$  &89.316149&  77.3 &{bl.}&--&--\\
    $^{13}$CH$_3$CN  &  5$_2$-4$_2$  &89.324555&  41.5 &{bl.}&--&--\\
    $^{13}$CH$_3$CN  &  5$_1$-4$_1$  &89.329599&  20.0 &{bl.}&--&--\\
    $^{13}$CH$_3$CN  &  5$_0$-4$_0$  &89.331282&  12.9 &{bl.}&--&--\\
    l-C$_4$H$_2$  &  10$_{1,9}$-9$_{1,8}$  &89.687047&  37.2 &{bl.}&{bl.}&--\\
    C$_6$H  &  J=65/2-63/2, $\Omega$=3/2, l=e  &90.093295&  72.2&3.5(0.5)&--&--\\
    C$_6$H  &  J=65/2-63/2, $\Omega$=3/2, l=f  &90.121407&  72.2&3.6(0.5)&--&--\\
    HC$^{13}$C$^{13}$CN  &  10-9  &90.204324&  23.8 &3.3(0.3)&--&--\\
    HC$_5$N  &  34-33  &90.525890&  76.0 &3.9(0.2)&5.8(0.1)&1.8(0.1)\\
    $^{30}$SiC$_2$  &  4$_{0,4}$-3$_{0,3}$  &90.562283&  10.9 &1.2({0.4})&0.9({0.3})&--\\
    HC$^{13}$CCN  &  10-9  &90.593059&  23.9 &16.4(0.2) ({bl.})& 2.3(0.4) ({bl.})&1.2(0.1)\\
    HCC$^{13}$CN  &  10-9  &90.601777&  23.9 &16.4(0.2) ({bl.})&2.3(0.4) ({bl.})&1.3(0.1)\\
    HNC  &  1-0  &90.663568&  4.4 &74.9(0.2)&36.5(0.1)&6.7(0.1)\\
    CCS  &  7$_7$-6$_6$  &90.686381&  26.1 &1.8(0.4)&2.1(0.4)&--\\
    C$_6$H  &  J=65/2-63/2, $\Omega$=1/2, l=f  &90.712181&  95.0 &3.3(0.6)&--&--\\
    C$_6$H  &  J=65/2-63/2, $\Omega$=1/2, l=e  &90.759297&  95.1 &{bl.}&{bl.}&--\\
    SiS  &  5-4  &90.771564&  13.1 &44.4(0.2)&22.9(0.1)&9.9(0.1)\\
    Si$^{13}$CC  &  4$_{0,4}$-3$_{0,3}$  &90.848527&  11.0 &6.9(0.2)&--&--\\
    C$_8$H  &  J=155/2-153/2, $\Omega$=3/2bbm  &90.924649&  171.2 &4.0(0.6)&--&--\\
    HC$_3$N  &  10-9  &90.979023&  24.0 &75.6(0.5)&56.1(0.3)&24.7(0.1)\\
    $^{13}$CCCCH  &  N=10-9, J=21/2-19/2  &91.875516&  24.3 &3.6(0.4)&--&--\\
    $^{13}$CCCCH  &  N=10-9, J=19/2-17/2  &91.906057&  24.3 &3.4(0.4)&--&--\\
    CH$_3$CN  &  5-4$^{(a)}$  &91.979995&  41.8 &15.4(0.8)&4.0(0.3)&{bl.}\\
    Si$^{13}$CC  &  4$_{2,3}$-3$_{2,2}$  &92.064397&  18.5 &4.3(0.4)&--&--\\
    CCC$^{13}$CH  &  N=10-9, J=21/2-19/2  &92.277460&  24.4 &4.1(0.4)&--&--\\
    CCC$^{13}$CH  &  N=10-9, J=19/2-17/2  &92.314915&  24.4 &6.0(0.4)&--&--\\
    Si$^{13}$CC  &  4$_{3,2}$-3$_{3,1}$  &92.375409&  27.9 &2.7(0.4)&--&--\\
    Si$^{13}$CC  &  4$_{3,1}$-3$_{3,0}$  &92.402216&  27.9 &2.6(0.4)&--&--\\
    $^{13}$CS  &  2-1  &92.494308&  6.7 &37.2(0.5)&2.9(0.3)&2.1(0.1)\\
    SiC$_2$  &  4$_{0,4}$-3$_{0,3}$  &93.063639&  11.2 &39.7(0.4)&15.0(0.2)&5.5(0.1)\\
    HC$_5$N  &  35-34  &93.188123&  80.5 &--&--&1.6(0.1)\\
    Si$^{13}$CC  &  4$_{2,2}$-3$_{2,1}$  &93.432836&  18.6 &2.6(0.4)&--&--\\
    C$_4$H, $\nu_7=1$  &  J=19/2-17/2,$\Omega$=1/2, l=f  &93.863370&  272.1 &{bl.}&{bl.}&{bl.}\\
    CCS  &  7$_8$-6$_7$  &93.870107&  19.9 &{bl.}&{bl.}&{bl.}\\
    SiC$_2$  &  4$_{2,3}$-3$_{2,2}$  &94.245393&  19.1 &26.4(0.4)&2.4(0.2)&3.8(0.1)\\
    C$^{13}$CCCH  &   N=10-9, J=21/2-19/2  &94.670247&  25.0 &5.4(0.4)&--&--\\
    C$^{13}$CCCH  &   N=10-9, J=19/2-17/2  &94.708239&  25.0 &4.7(0.4)&--&--\\
    CC$^{13}$CCH  &  N=10-9, J=21/2-19/2  &94.814723&  25.0 &4.4(0.4)&--&--\\
    CC$^{13}$CCH  &  N=10-9, J=19/2-17/2  &94.853133&  25.0 &6.4(0.4)&--&--\\
    C$_4$H  &  N = 10-9, J = 21/2-19/2  &95.150389&  25.1 &42.7(0.4)&9.3(0.3)&0.9(0.1)\\
    C$_4$H  &  N = 10-9, J = 19/2-17/2  &95.188949&  25.1 &43.5(0.4)&7.4(0.3)&1.0(0.1)\\
    SiC$_2$  &  4$_{2,2}$-3$_{2,1}$  &95.579381&  19.2 &22.6(0.4)&12.0(0.3)&4.1(0.1)\\
    C$_4$H, $\nu_7=2^0$  &  N=10-9, J=21/2-19/2  &95.611500&  561.5 &2.5(0.8)&--&--\\
    C$_4$H, $\nu_7=2^0$  &  N=10-9, J=19/2-17/2  &95.667900&  561.5 &3.2(0.8)&--&--\\
    HC$_5$N  &  36-35  &95.850335&  85.1 &--&--&1.5(0.1)\\
    C$^{34}$S  &  2-1  &96.412950&  6.9 &13.0(0.4)&6.6(0.2)&3.7(0.1)\\
    C$_4$H, $\nu_7=1$  &  J=21/2-19/2,$\Omega$=3/2, l=e  &96.478780&  270.1 &2.1(0.7)&--&--\\
    H$^{13}$CC$^{13}$CN?  &  11-10  &96.550615&  27.8 &--&--&--\\
    C$_4$H, $\nu_7=1$  &  J=21/2-19/2,$\Omega$=3/2, l=f  &96.741210&  270.2 &2.7(0.9)&--&--\\
    H$^{13}$CCCN  &  11-10  &96.983001&  27.9 &10.3(0.4)&--&1.3(0.1)\\
    C$^{33}$S  &  2-1  &97.172064&  7.0 &2.8(0.5)&--&--\\
    CS, $v=1$?  &  2-1  &97.270980&  1837.4 &--&--&--\\
    Si$^{13}$CC  &  4$_{1,3}$-3$_{1,2}$  &97.295257&  13.6 &7.4(0.5)&--&--\\
    CS  &  2-1  &97.980953&  7.1 &252.5(0.5)&127.7(0.2)&55.6(0.1)\\
    C$_3$H  &  J=9/2-7/2, $\Omega$=1/2$^{(a)}$ &98.012070&  12.5 &20.9(0.7)&--&1.9(0.1)\\
    C$^{13}$CCN  &  N=10-9$^{(a)}$  &98.401860&  14.8 &17.0(2.3)&--&--\\
    HC$_5$N  &  37-36  &98.512524&  89.8 &--&--&1.7(0.1)\\
    CC$^{13}$CN  &  N=10-9, J=21/2-19/2  &98.569079&  14.8 &3.9(0.5)&--&--\\
    CC$^{13}$CN  &  N=10-9, J=19/2-17/2  &98.587484&  14.8 &5.2(0.5)&--&--\\
    C$_3$N  &  10-9 (A)  &98.940005&  26.1 &16.7	(0.5)&10.5(0.2)&3.2(0.1)\\
    C$_3$N  &  10-9 (B)  &98.958774&  26.1 &19.0(0.5)&11.5(0.2)&2.9(0.1)\\
    HC$^{13}$C$^{13}$CN  &  11-10  &99.224259&  28.6 &3.1(0.4)&--&--\\
    HC$^{13}$CCN  &  11-10  &99.651849&  28.7 &13.8(0.1) ({bl.})&2.4(0.4) ({bl.})&1.7(0.1)\\
    HCC$^{13}$CN  &  11-10  &99.661467&  28.7 &13.8(0.1) ({bl.})&2.4(0.4) ({bl.})&1.6(0.1)\\
    CCS  &  8$_7$-7$_6$  &99.866521&  28.1 &3.1(0.8)&1.3(0.4)& 0.5(0.1)\\
    HC$_3$N  &  11-10  &100.076392&  28.8 &71.8(0.2)&65.5(0.1)&27.5(0.1)\\
    u  &--&100.407000&--&1.1(0.3)&--&--\\
    $^{13}$CCCCH  &  N=11-10, J=23/2-21/2  &101.063060&  29.1 &6.5(0.2)&--&--\\
    $^{13}$CCCCH  &  N=11-10, J=21/2-19/2  &101.094030&  29.1 &6.3(0.2)&--&--\\
    HC$_5$N  &  38-37  &101.174677&  94.7 &3.2(0.2)&3.6(0.1)&1.4(0.1)\\
    CCC$^{13}$CH  &  N=11-10, J=23/2-21/2  &101.506343&  29.2 &7.7(0.2)&--&--\\
    CCC$^{13}$CH  &  N=11-10, J=21/2-19/2  &101.543766&  29.2 &7.9(0.2)&--&--\\
    $^{30}$SiS, $v=5$  &  6-5  &102.548508&  5206.0 &3.0(0.5)&--&--\\
    C$_4$H, $\nu_7=1$  &  J=21/2-19/2,$\Omega$=1/2, l=e  &103.266200&  277.0 &3.1(1.0)&--&--\\
    C$_3$H  &  J=9/2-7/2, $\Omega$=3/2$^{(a)}$&103.319530&  32.9 &4.7(0.7)&--&--\\
    C$_4$H, $\nu_7=1$  &  J=21/2-19/2,$\Omega$=1/2, l=f  &103.576590&  277.1 &5.0(1.0)&--&--\\
    CCS  &  8$_8$-7$_7$  &103.640759&  31.1 &3.2(1.0)&--&--\\
    HC$_5$N  &  39-38  &103.836817&  99.7 &--&--&1.3(0.1)\\
    C$^{13}$CCCH  &  N=11-10, J=23/2-21/2  &104.138540&  30.0 &9.8(0.5)&--&--\\
    C$^{13}$CCCH  &  N=11-10, J=21/2-19/2  &104.176220&  30.0 &8.8(0.5)&--&--\\
    CC$^{13}$CCH  &  N=11-10, J=23/2-21/2  &104.297340&  30.0 &8.8(0.5)&--&--\\
    CC$^{13}$CCH  &  N=11-10, J=21/2-19/2  &104.335720&  30.1 &10.3(0.5)&--&--\\
    C$_4$H  &  N = 11-10, J = 23/2-21/2  &104.666566&  30.1 &53.7(0.5)&5.6(0.3)&1.3(0.1)\\
    C$_4$H  &  N = 11-10, J = 21/2-19/2  &104.705110&  30.2 &55.5(0.5)&10.6(0.3)&1.0(0.1)\\
    $^{30}$SiS  &  6-5  &105.059203&  17.6 &--&--&0.5(0.1)\\
    C$_4$H, $\nu_7=2^0$  &  N=11-10, J=23/2-21/2  &105.174600&  566.5 &2.6(0.8)&--&--\\
    C$_4$H, $\nu_7=2^0$  &  N=11-10, J=21/2-19/2  &105.230990&  566.5 &5.0(0.8)&--&--\\
    H$^{13}$CCCN  &  12-11  &105.799113&  33.0 &9.0(0.5)&--&1.2(0.1)\\
    C$_4$H, $\nu_7=1$  & J=23/2-21/2,$\Omega$=3/2, l=e  &105.837997&  219.2&4.6(1.2)&--&--\\
    Si$^{34}$S  &  6-5  &105.941503&  17.8 &2.2(0.5)&--&0.5(0.1)\\
    C$_4$H, $\nu_7=1$  & J=23/2-21/2,$\Omega$=3/2, l=f  &106.133368&  219.2&3.2(0.9)&--&--\\
    CCS  &  8$_9$-7$_8$  &106.347726&  25.0 &3.0(1.0)&--&0.6(0.2)\\
    HC$_5$N  &  40-39  &106.498910&  104.8 &--&--&1.2(0.1)\\
    $^{29}$SiS  &  6-5  &106.922980&  18.0 &3.0(0.5)&--&0.8(0.1)\\
    Si$^{13}$CC  &  5$_{1,5}$-4$_{1,4}$  &107.971477&  17.5 &6.5(0.5)&--&--\\
    C$^{13}$CCN  &  N=11-10$^{(a)}$  &108.242334&  18.1 &{bl.}&--&--\\
    HC$^{13}$C$^{13}$CN  &  12-11  &108.244051&  33.8 &{bl.}&--&--\\
    SiS, $v=1$  &  6-5  &108.394291&  1089.3 &--&--&--\\
    CC$^{13}$CN  &  N=11-10, J=23/2-21/2  &108.426168&  18.1 &{bl.}&--&--\\
    CC$^{13}$CN  &  N=11-10, J=21/2-19/2  &108.444688&  18.1 &3.1(0.5)&--&--\\
    $^{13}$CN  &  N= 1-0&108.780201&  5.2 &{bl.}&{bl.}&{bl.}\\
    HC$^{13}$CCN  &  12-11  &108.710532&  33.9 &8.8(0.1) ({bl.})&--&1.6(0.1)\\
    HCC$^{13}$CN  &  12-11  &108.720999&  33.9 &8.8(0.1) ({bl.})&--&1.4(0.1)\\
    C$_3$N  &  11-10 (A)  &108.834250&  31.3 &10.7(0.6)&4.7(0.2)&3.8(0.1)\\
    C$_3$N  &  11-10 (B)  &108.853005&  31.3 &16.4(0.6)&6.2(0.2)&3.6(0.1)\\
    SiS  &  6-5  &108.924301&  18.3 &46.1(0.6)&26.9(0.2)&11.6(0.1)\\
    HC$_5$N  &  41-40  &109.160973&  110.0 &{bl.}&{bl.}&{bl.}\\
    HC$_3$N  &  12-11  &109.173634&  34.1 &57.8(0.6)&50.1(0.3)&32.5(0.1)\\
    $^{13}$CO  &  1-0  &110.201354&5.3&119.6(1.6)&--&18.4(0.2)\\
    $^{13}$CCCCH  &  N=12-11, J=25/2-23/2  &110.250280&  34.4 &4.4(0.5)&--&--\\
    $^{13}$CCCCH  &  N=12-11, J=23/2-21/2  &110.281796&  34.4 &6.2(0.7)&--&--\\
    CH$_3$CN  &  6-5$^{(a)}$  &110.374989&  47.1 &{bl.}&{bl.}&{bl.}\\
    CCC$^{13}$CH  &  N=12-11, J=25/2-23/2  &110.735027&  34.5 &5.2(0.6)&--&--\\
    CCC$^{13}$CH  &  N=12-11, J=23/2-21/2  &110.772434&  34.6 &7.0(0.6)&--&--\\
    HC$_5$N  &  42-41  &111.823024&  115.4 &--&--&1.2(0.1)\\
    Si$^{13}$CC  &  5$_{0,5}$-4$_{0,4}$  &112.593212&  16.4 &8.0(0.6)&--&--\\
    C$_4$H, $\nu_7=1$?  & J=23/2-21/2,$\Omega$=1/2, l=e  &112.922950&  282.5 &--&--&--\\
    C$_4$H, $\nu_7=1$?  & J=23/2-21/2,$\Omega$=1/2, l=f  &113.266080&  282.5 &--&--&--\\
    CN  &  N= 1-0$^{(a)}$  &113.488120&  5.4 &179.0(2.0)&188.0(1.0)&256.5(0.4)\\
    C$^{13}$CCCH  &  N=12-11, J=25/2-23/2  &113.606341&  35.4 &7.4(0.6)&--&--\\
    C$^{13}$CCCH  &  N=12-11, J=23/2-21/2  &113.644283&  35.5 &6.9(0.6)&--&--\\
    CC$^{13}$CCH  &  N=12-11, J=25/2-23/2  &113.779720&  35.5 &9.1(0.6)&--&--\\
    CC$^{13}$CCH  &  N=12-11, J=23/2-21/2  &113.818080&  35.5 &13.1(0.6)&--&--\\
    C$_4$H  &  N = 12-11, J = 25/2-23/2  &114.182514&  35.6 &59.2(0.9)&14.9(0.7)&1.6(0.2)\\
    C$_4$H  &  N = 12-11, J = 23/2-21/2  &114.221044&  35.6 &66.6(0.9)&17.7(0.7)&2.0(0.2)\\
    H$^{13}$CCCN  &  13-12  &114.614995&  38.5 &6.8(0.9)&--&1.5(0.2)\\
    CO  &  1-0  &115.271202&  5.5 &439.2(2.0)$^{(b)}$&120.8(1.1)&166.7(0.5)\\
    SiC$_2$  &  5$_{0,5}$-4$_{0,4}$  &115.382389&  16.8 &53.1(0.9)&28.2(0.7)&13.0(0.2)\\ & \\[-4ex]
   \label{tab:alma_line_detections}
   \end{longtable}
   \tablefoot{All transitions listed are in the ground vibrational state unless otherwise specified. $^{(a)}$Lines with hyperfine components where only the frequency of one of the components has been listed, but the integrated intensity given has been summed over all detected components. $^{(b)}$The extended CO emission in IRAS 15194-5115 is not fully recovered by the observations. The line intensity given is hence underestimated. {bl.} indicates blends, either between hyperfine components of the same transition, or between different lines. A `?' next to the molecule name indicates a tentative detection. The values in parentheses are the uncertainties obtained from the rms noise.\\}
\end{small}

   \begin{small}
      \setlength\LTleft{0pt}
      \setlength\LTright{0pt}
      \begin{longtable}{@{\extracolsep{\fill}}llrrrrr}
         \caption{\label{tab:apex_line_detections}Details of the lines from APEX observations used in the population diagrams.}\\
         \hline\hline & \\[-2ex]
         \makecell{\\Molecule} & \makecell{\\Transition} & \makecell{\\Rest Freq.\\(GHz)} & \makecell{\\$E_\mathrm{up}$/$k_\mathrm{B}$ (K)} & \multicolumn{3}{c}{$\int S_\mathrm{ul} d\varv$ (K km/s)}\\
         \cline{5-7} & \\[-2ex]
         & & & & 15194 & 15082 & 07454 \\
         \hline & \\[-2ex]
         \endfirsthead
         \caption{continued.}\\
         \hline\hline & \\[-2ex]
         \makecell{\\Molecule} & \makecell{\\Transition} & \makecell{\\Rest Freq. (GHz)} & \makecell{\\$E_\mathrm{up}$/$k_\mathrm{B}$ (K)} & \multicolumn{3}{c}{$\int S_\mathrm{ul} d\varv$ (K km/s)}\\
         \cline{5-7} & \\[-2ex]
         & & & & 15194 & 15082 & 07454 \\
         \hline & \\[-2ex]
         \endhead
         \hline & \\[-2ex]
         \endfoot
        C$_4$H&N=17-16&161.796566&70.5&5.83(1.17)&1.84(0.37)&--\\
        SiS&9-8&163.376785&39.3&4.18(0.84)&2.99(0.60)&1.80(0.36)\\
        HC$_3$N&J=18-17&163.753389&73.9&0.84(0.19)&1.26(0.26)&0.75(0.15)\\
        SiC$_2$&7$_{2,6}$-6$_{2,5}$&164.069091&39.4&1.94(0.40)&1.46(0.29)&0.83(0.17)\\
        C$_3$N&N=17-16&168.213682&72.0&--&1.10(0.24)&--\\
        C$_4$H&N=18-17&171.310707&78.8&5.53(1.11)&1.88(0.38)&--\\
        $^{29}$SiO&4-3&171.512796&20.5&0.96(0.21)&0.41(0.09)&0.21(0.06)\\
        H$^{13}$CN&J=2-1&172.677851&12.4&20.58(4.12)&5.32(1.07)&3.49(0.70)\\
        HC$_3$N&J=19-18&172.849300&82.1&0.54(0.13)&0.87(0.18)&0.65(0.14)\\
        SiO&4-3&173.688238&20.8&9.94(1.99)&4.60(0.92)&1.90(0.38)\\
        C$_2$H&N=2-1&174.663199&12.6&43.70(8.75)&17.10(3.44)&4.20(0.86)\\
        HCN&J=2-1&177.261111&12.8&34.48(6.90)&21.55(4.31)&8.41(1.68)\\
        C$_4$H&N=19-18&180.824472&87.6&5.69(1.14)&--&--\\
        HNC&J=2-1&181.324758&13.0&4.39(0.88)&--&--\\
        SiS&10-9&181.525218&48.0&3.93(0.79)&--&--\\
        $^{13}$CS&J=4-3&184.981772&22.2&3.67(0.74)&--&--\\
        SiC$_2$&8$_{6,3}$-7$_{6,2}$&188.385700&110.9&--&0.38(0.11)&--\\
        C$_4$H&N=20-19&190.337804&96.8&5.06(1.02)&1.97(0.40)&--\\
        HC$_3$N&J=21-20&191.040299&99.8&--&0.62(0.13)&0.49(0.11)\\
        C$^{34}$S&4-3&192.818457&23.1&1.67(0.35)&1.14(0.23)&0.61(0.13)\\
        CS&4-3&195.954211&23.5&17.34(3.47)&12.81(2.56)&--\\
        SiS&11-10&199.672229&57.7&4.11(0.83)&3.32(0.66)&1.91(0.38)\\
        C$_4$H&N=21-20&199.850787&106.4&4.91(0.99)&2.33(0.47)&--\\
        HC$_3$N&J=23-22&209.230234&119.2&--&0.22(0.05)&0.22(0.05)\\
        C$_4$H&N=22-21&209.363302&116.6&3.66(0.74)&1.98(0.40)&0.34(0.07)\\
        SiC$_2$&9$_{2,8}$-8$_{2,7}$&209.892001&58.5&1.32(0.27)&1.59(0.32)&1.04(0.21)\\
        SiC$_2$&9$_{6,4}$-8$_{6,3}$&212.031878&121.1&0.53(0.14)&0.36(0.10)&0.31(0.08)\\
        SiC$_2$&9$_{4,6}$-8$_{4,5}$&213.208032&82.3&1.04(0.21)&0.70(0.14)&--\\
        SiC$_2$&9$_{4,5}$-8$_{4,4}$&213.292337&82.3&1.09(0.22)&0.74(0.15)&0.69(0.14)\\
        $^{29}$SiS&12-11&213.816140&66.6&--&0.16(0.03)&0.18(0.04)\\
        $^{29}$SiO&5-4&214.385752&30.8&1.18(0.24)&0.41(0.08)&0.30(0.06)\\
        SiO&5-4&217.104919&31.2&11.40(2.28)&4.56(0.91)&2.73(0.55)\\
        SiS&12-11&217.817663&68.1&3.92(0.78)&2.90(0.58)&2.34(0.47)\\
        HC$_3$N&J=24-23&218.324723&129.6&--&0.24(0.05)&0.25(0.05)\\
        C$_4$H&N=23-22&218.875369&127.2&2.73(0.55)&1.25(0.26)&0.37(0.08)\\
        SiC$_2$&10$_{0,10}$-9$_{0,9}$&220.773685&59.8&1.51(0.31)&1.13(0.23)&1.23(0.25)\\
        SiC$_2$&9$_{2,7}$-8$_{2,6}$&222.009386&60.2&1.66(0.33)&1.19(0.24)&1.21(0.24)\\
        HC$_3$N&J=25-24&227.418905&140.4&--&0.22(0.05)&0.18(0.04)\\
        C$_3$N&N=23-22&227.563276&129.8&--&--&0.45(0.09)\\
        C$_4$H&N=24-23&228.386962&138.2&3.46(0.70)&1.27(0.26)&0.43(0.09)\\
        Si$^{34}$S&13-12&229.500868&76.9&--&0.19(0.04)&0.18(0.04)\\
        $^{13}$CS&J=5-4&231.220685&33.3&5.37(1.07)&--&--\\
        $^{29}$SiS&13-12&231.626673&77.8&--&0.18(0.04)&0.23(0.05)\\
        SiC$_2$&10$_{2,9}$-9$_{2,8}$&232.534070&69.6&1.60(0.32)&1.07(0.21)&1.20(0.24)\\
        SiC$_2$&10$_{6,5}$-9$_{6,4}$&235.712998&132.4&0.47(0.13)&0.36(0.10)&0.39(0.10)\\
        SiS&13-12&235.961363&79.5&6.50(1.30)&3.46(0.69)&2.94(0.59)\\
        HC$_3$N&J=26-27&236.512789&151.6&--&0.17(0.04)&0.18(0.04)\\
        SiC$_2$&10$_{4,7}$-9$_{4,6}$&237.150018&93.7&0.79(0.16)&0.76(0.15)&0.86(0.17)\\
        SiC$_2$&10$_{4,6}$-9$_{4,5}$&237.331309&93.7&1.01(0.20)&0.71(0.14)&0.76(0.15)\\
        C$_3$N&N=24-23&237.453530&141.1&--&--&0.38(0.08)\\
        C$_4$H&N=25-24&237.898060&149.8&2.76(0.56)&0.96(0.20)&0.37(0.08)\\
        C$^{34}$S&5-4&241.016089&34.7&1.44(0.29)&0.91(0.18)&0.65(0.13)\\
        SiC$_2$&11$_{0,11}$-10$_{0,10}$&241.367708&71.4&1.07(0.22)&0.99(0.20)&1.21(0.24)\\
        CS&5-4&242.913610&35.3&24.27(4.85)&10.98(2.20)&7.98(1.60)\\
        Si$^{34}$S&14-13&244.935556&88.7&--&0.18(0.04)&0.22(0.04)\\
        C$_3$N&N=25-24&247.146242&152.9&--&--&0.31(0.07)\\
        C$_4$H&N=26-25&247.343351&161.7&--&0.81(0.17)&0.34(0.07)\\
        SiC$_2$&10$_{2,8}$-9$_{2,7}$&247.408643&72.1&1.36(0.27)&1.11(0.22)&1.33(0.27)\\
        $^{29}$SiS&14-13&247.529119&89.7&--&0.14(0.03)&0.23(0.05)\\
        SiS&14-13&249.435412&91.7&5.50(1.10)&3.08(0.62)&3.53(0.71)\\
        $^{30}$SiO&6-5&254.103211&42.7&0.77(0.16)&--&--\\
        C$_4$H&N=27-26&254.216656&174.2&--&0.62(0.13)&0.34(0.07)\\
        $^{29}$SiO&6-5&256.918691&43.1&1.11(0.22)&0.40(0.08)&0.31(0.06)\\
        H$^{13}$CN&J=3-2&257.255213&24.9&25.64(5.13)&5.88(1.18)&5.86(1.17)\\
        SiC$_2$&11$_{6,6}$-10$_{6,5}$&259.011798&144.8&--&0.31(0.08)&0.37(0.10)\\
        SiO&6-5&259.433309&43.7&10.23(2.05)&4.53(0.91)&3.16(0.63)\\
        SiC$_2$&11$_{4,8}$-10$_{4,7}$&260.518009&106.2&0.68(0.14)&0.55(0.11)&0.67(0.14)\\
        SiC$_2$&11$_{4,7}$-10$_{4,6}$&261.150695&106.2&0.61(0.13)&0.50(0.10)&0.68(0.14)\\
        C$_2$H&N=3-2&261.509329&25.2&48.90(9.79)&25.30(5.07)&10.60(2.14)\\
        $^{30}$SiS&15-14&262.064986&100.2&--&--&0.19(0.04)\\
        Si$^{34}$S&15-14&262.585033&101.4&--&--&0.21(0.04)\\
        HCN&J=3-2&264.789719&25.5&58.90(11.78)&27.29(5.46)&14.82(2.96)\\
        C$_4$H&N=28-27&265.886434&187.1&--&0.48(0.10)&0.32(0.07)\\
        C$_3$N&N=27-26&266.428184&177.8&--&--&0.27(0.06)\\
        $^{29}$SiS&15-14&267.102873&102.5&--&--&0.30(0.06)\\
        SiS&15-14&267.242218&104.8&5.36(1.08)&--&--\\
        $^{13}$CS&J=6-5&272.243052&46.6&5.61(1.12)&--&--\\
        C$^{34}$S&6-5&277.455405&48.6&1.97(0.40)&--&--\\
        SiS&16-15&289.209068&118.8&6.38(1.28)&--&--\\
        CS&6-5&290.380757&49.4&20.15(4.03)&--&--\\
        $^{30}$SiO&7-6&293.912086&57.0&0.62(0.14)&--&--\\
        $^{29}$SiO&7-6&296.575740&57.5&0.96(0.20)&--&--\\
        SiO&7-6&300.120477&58.3&10.04(2.01)&--&--\\
        SiS&17-16&303.926812&133.7&3.99(0.80)&--&--\\
        C$^{34}$S&7-6&308.516144&64.8&1.82(0.37)&--&--\\
        $^{30}$SiO&8-7&337.396459&73.3&0.82(0.18)&--&--\\
        CS&7-6&338.930044&65.9&14.85(2.97)&--&--\\
        $^{29}$SiO&8-7&342.882850&74.0&0.80(0.17)&--&--\\
        SiS&19-18&342.980847&166.0&4.15(0.83)&--&--\\
        H$^{13}$CN&J=4-3&344.779481&41.5&27.50(5.50)&--&--\\
        SiO&8-7&345.339769&75.0&10.41(2.08)&--&--\\
        C$_2$H&N=4-3&347.330581&42.0&70.90(14.19)&--&--\\
        HCN&J=4-3&349.339066&42.5&38.01(7.60)&--&--\\
        HNC&J=4-3&354.505477&43.5&2.13(0.43)&--&--\\
        SiS&20-19&362.630303&183.5&3.88(0.78)&--&--\\
         \end{longtable}
         \tablefoot{All transitions listed are in the ground vibrational state. For lines with hyperfine components, only the frequency of one of the components has been listed, but the integrated intensity given has been summed over all detected components. The values in parentheses denote the total uncertainty in the integrated intensity, including the 20\% uncertainty in flux calibration.}
      \end{small}

\section{Emission maps and line spectra}
\label{app:appendix_C}
\begin{multicols}{2}
The below figures show the line emission maps (top panels) averaged over $\pm$20\% of the expansion velocity around the systemic velocity, and the line spectra (bottom panels), for IRAS 15194$-$5115 (left), IRAS 15082$-$4808 (centre), and IRAS 07454$-$7112 (right). The figures are ordered in increasing order of line rest frequency, which is given in parentheses at the end of the figure caption. The white contours in the emission maps are at 3, 10 and 30 sigma. The synthesised beams are shown as filled white ellipses at the bottom left of each map. The grey-shaded regions in the line spectra represent the channels used for calculating the integrated intensities {of the lines, reported in Table~\ref{tab:alma_line_detections}}. The dotted vertical line in the line spectra marks the systemic velocity of the source. Missing plots indicate non-detection of the concerned line towards the particular source.
\end{multicols}

\begin{figure}[h]
    \centering
    \includegraphics[width=0.8\linewidth]{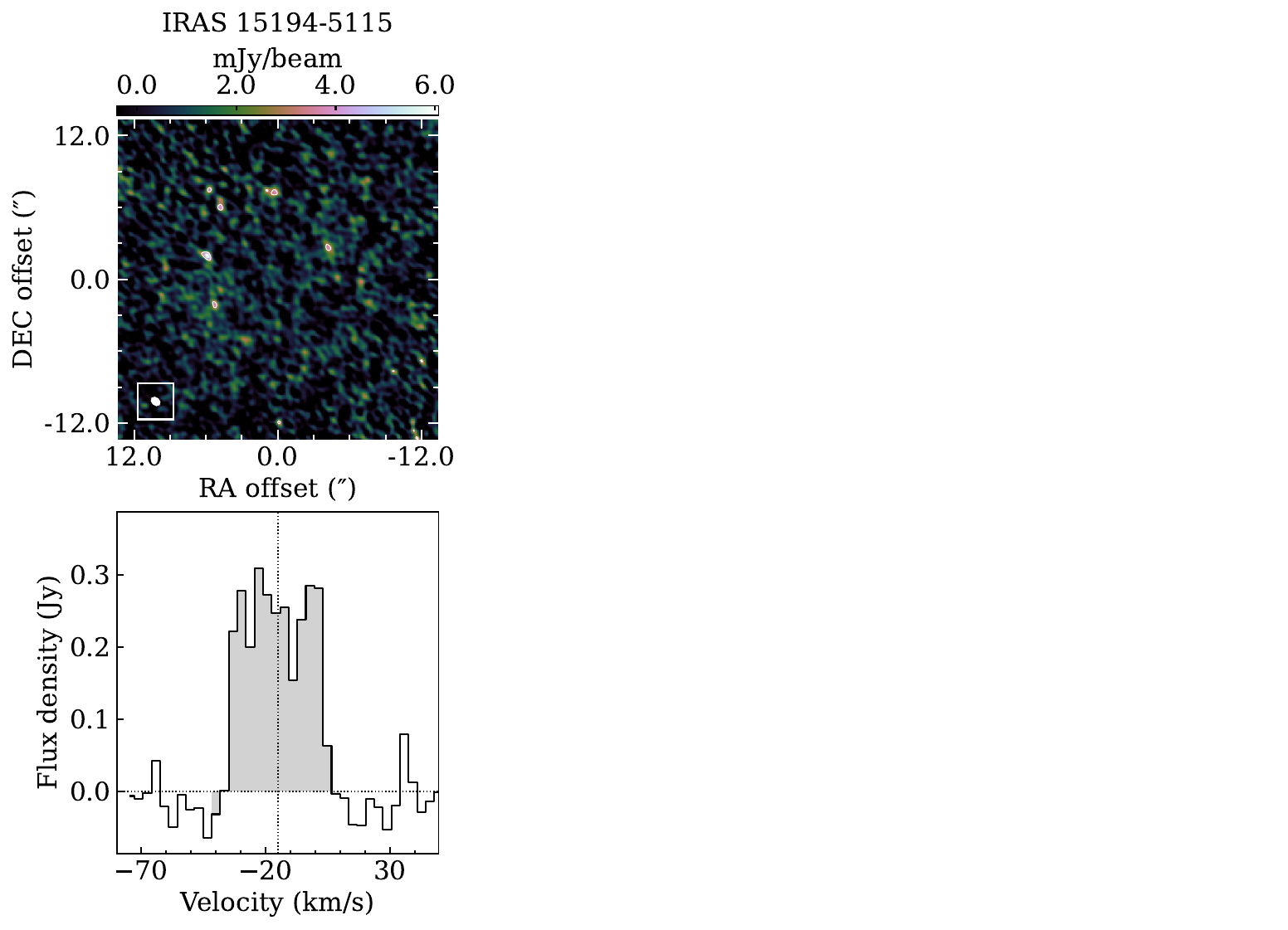}
    \caption{C$^{13}$CCCH, N=9-8, J=19/2-17/2 (85.201906 GHz)}
\end{figure}

\begin{figure}[h]
    \centering
    \includegraphics[width=0.8\linewidth]{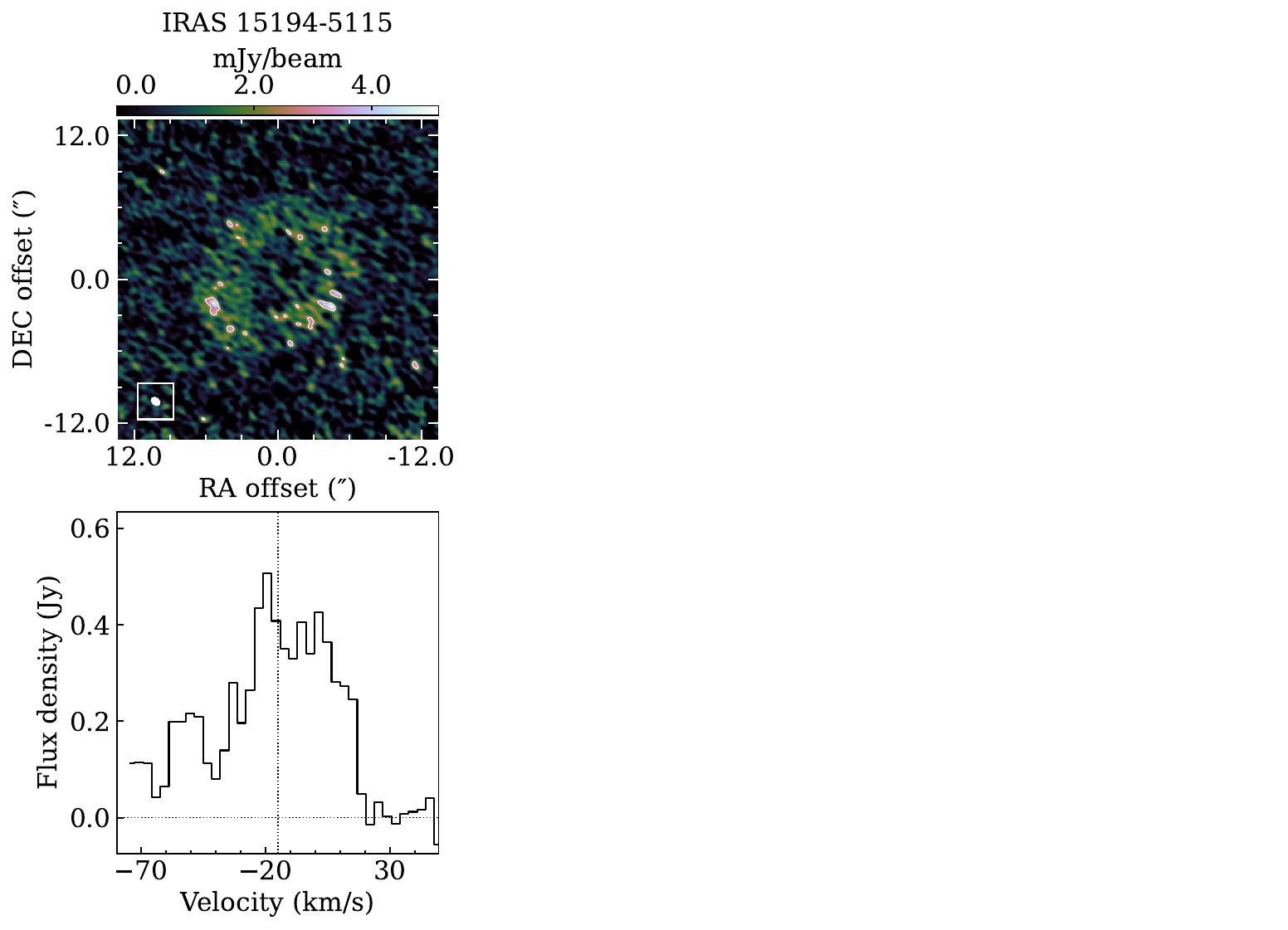}
    \caption{C$^{13}$CH, N=1-0, J=3/2-1/2 (85.232792 GHz)}
\end{figure}

\begin{figure}[h]
    \centering
    \includegraphics[width=0.8\linewidth]{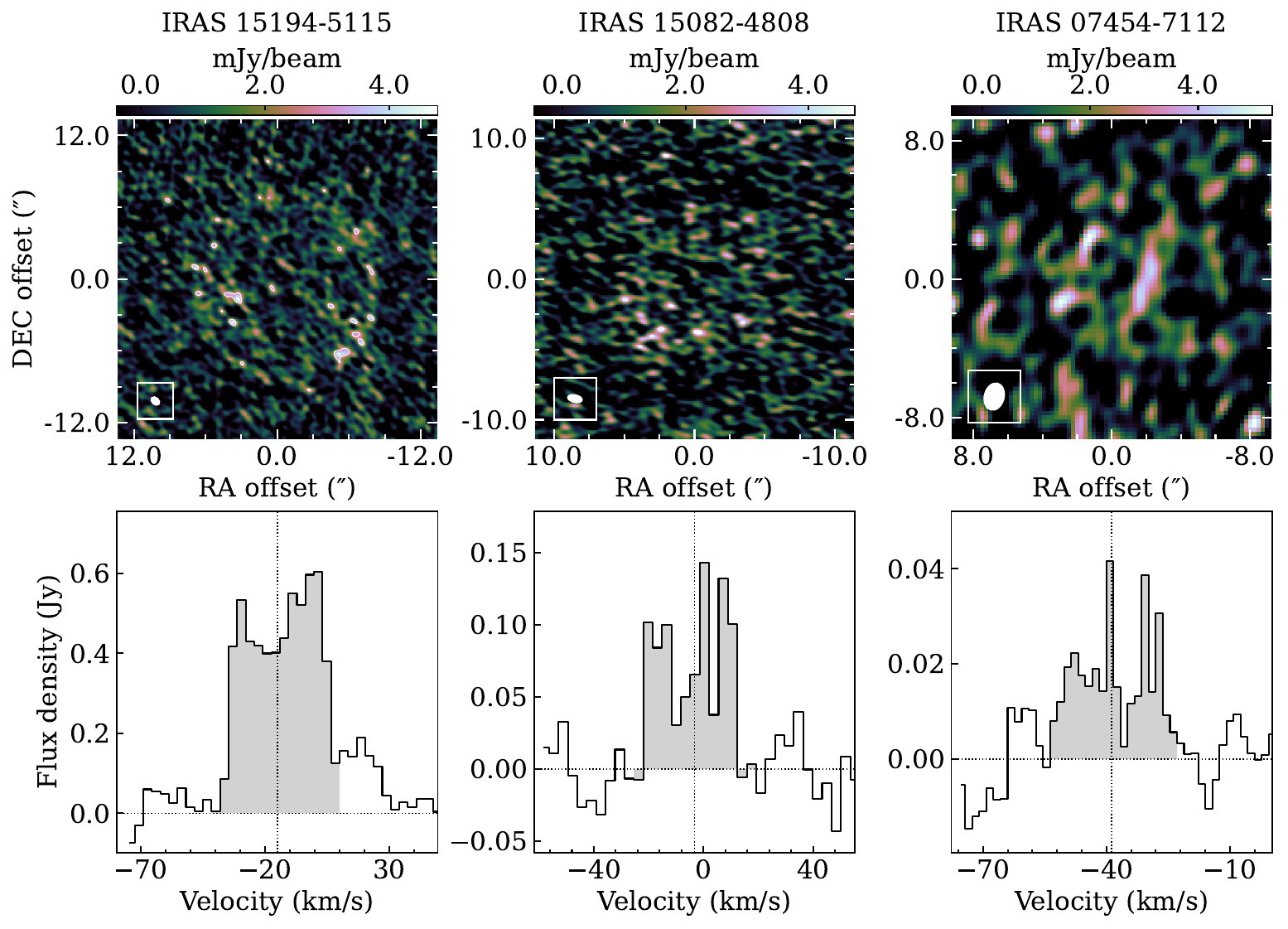}
    \caption{c-C$_3$H$_2$, 2$_{1,2}$-1$_{0,1}$ (85.338894 GHz)}
\end{figure}

\begin{figure}[h]
    \centering
    \includegraphics[width=0.8\linewidth]{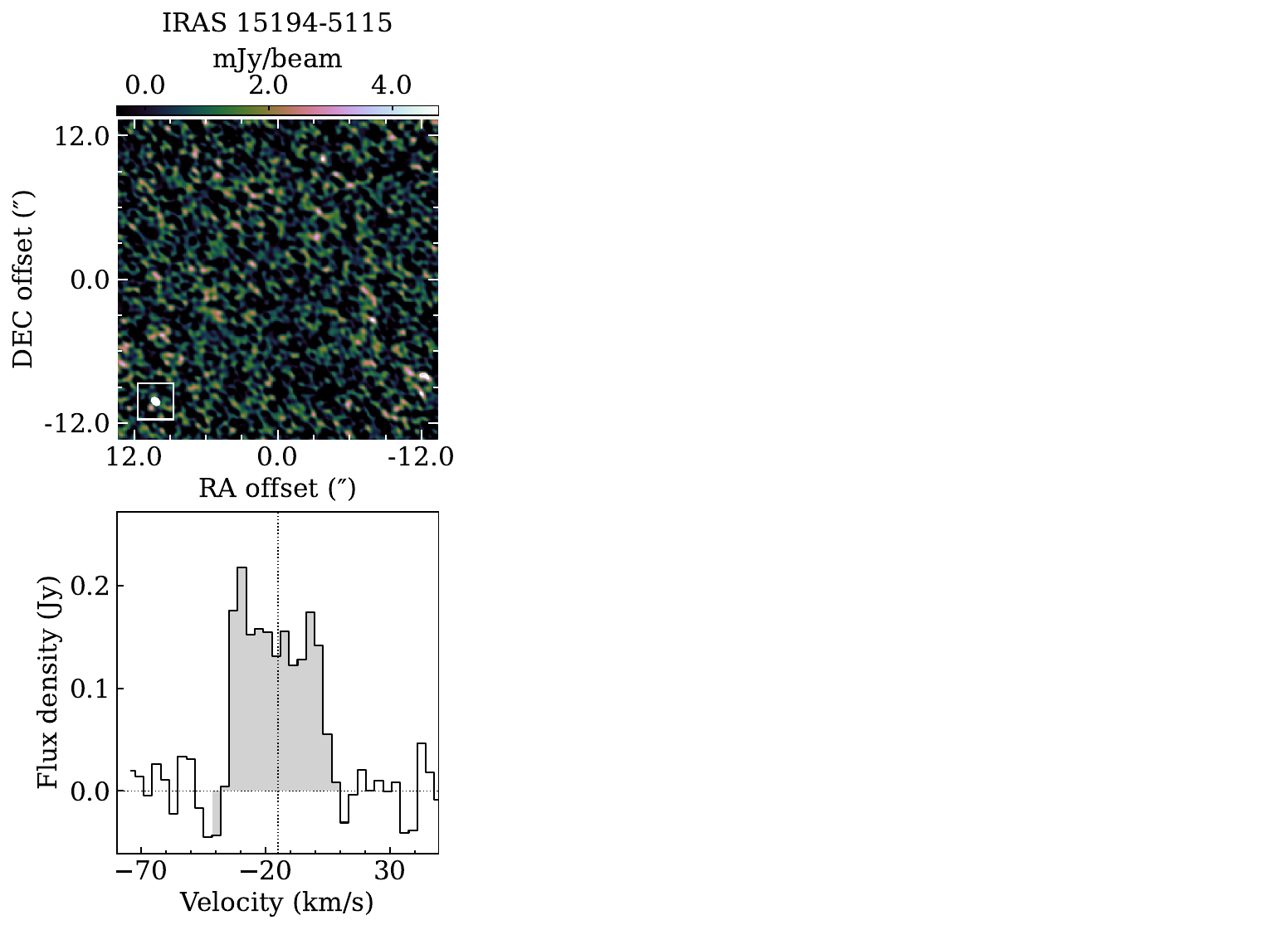}
    \caption{CC$^{13}$CCH, N=9-8, J=17/2-15/2 (85.37033 GHz)}
\end{figure}

\begin{figure}[h]
    \centering
    \includegraphics[width=0.8\linewidth]{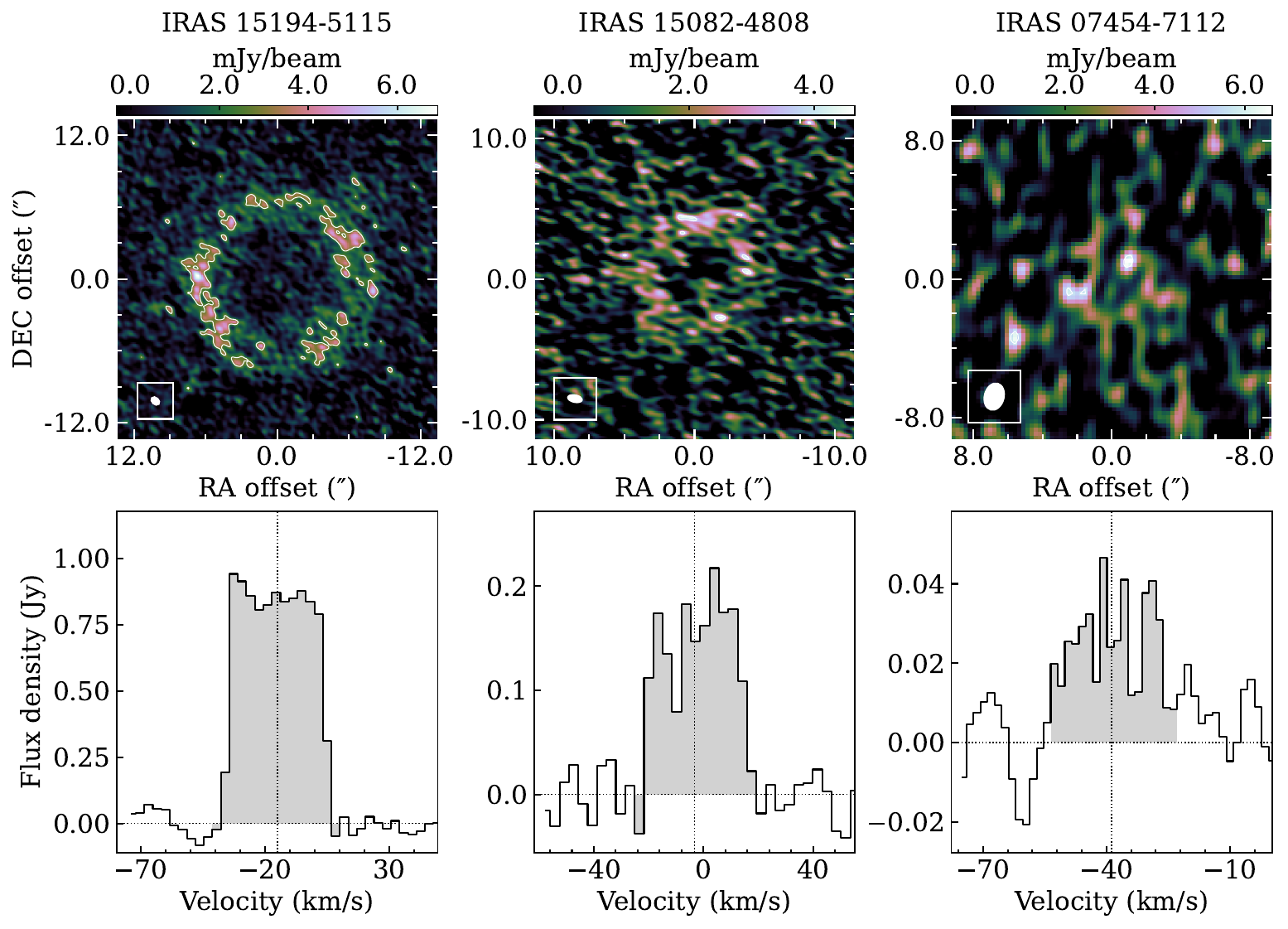}
    \caption{C$_4$H, N=9-8, J=19/2-17/2 (85.634004 GHz)}
\end{figure}

\begin{figure}[h]
    \centering
    \includegraphics[width=0.8\linewidth]{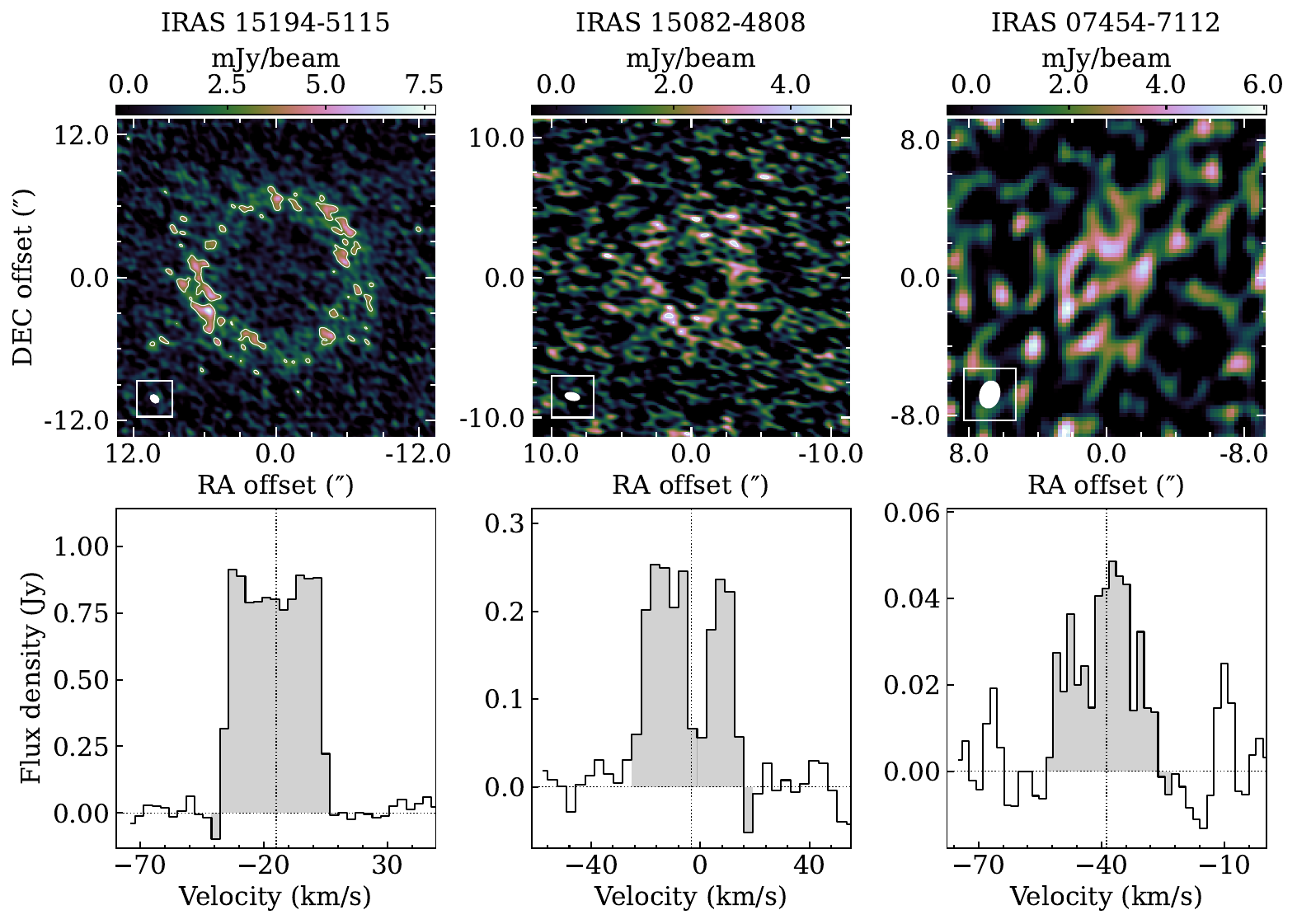}
    \caption{C$_4$H, N=9-8, J=17/2-15/2 (85.672582 GHz)}
\end{figure}

\begin{figure}[h]
    \centering
    \includegraphics[width=0.8\linewidth]{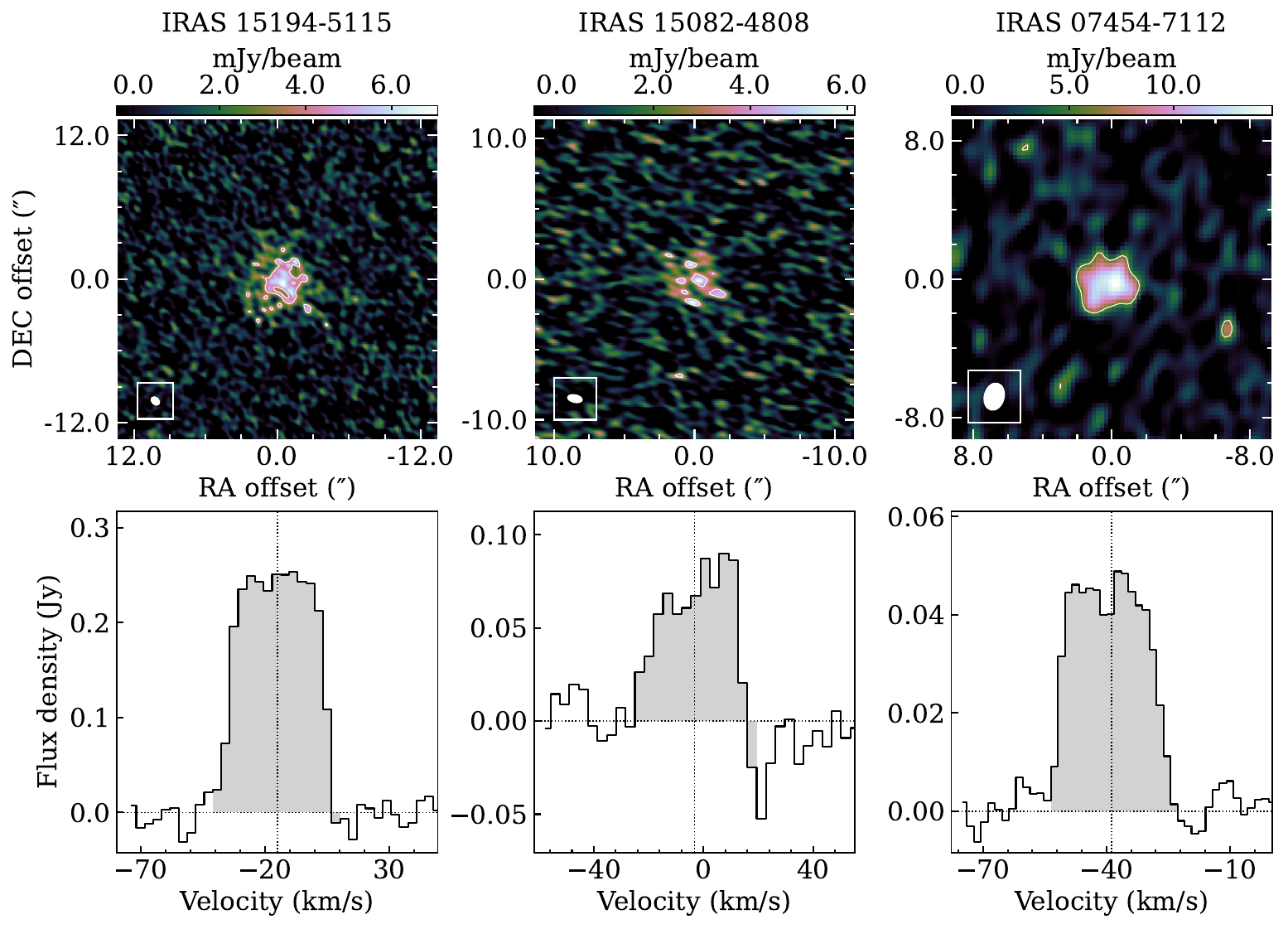}
    \caption{$^{29}$SiO, 2-1 (85.759194 GHz)}
\end{figure}

\begin{figure}[h]
    \centering
    \includegraphics[width=0.8\linewidth]{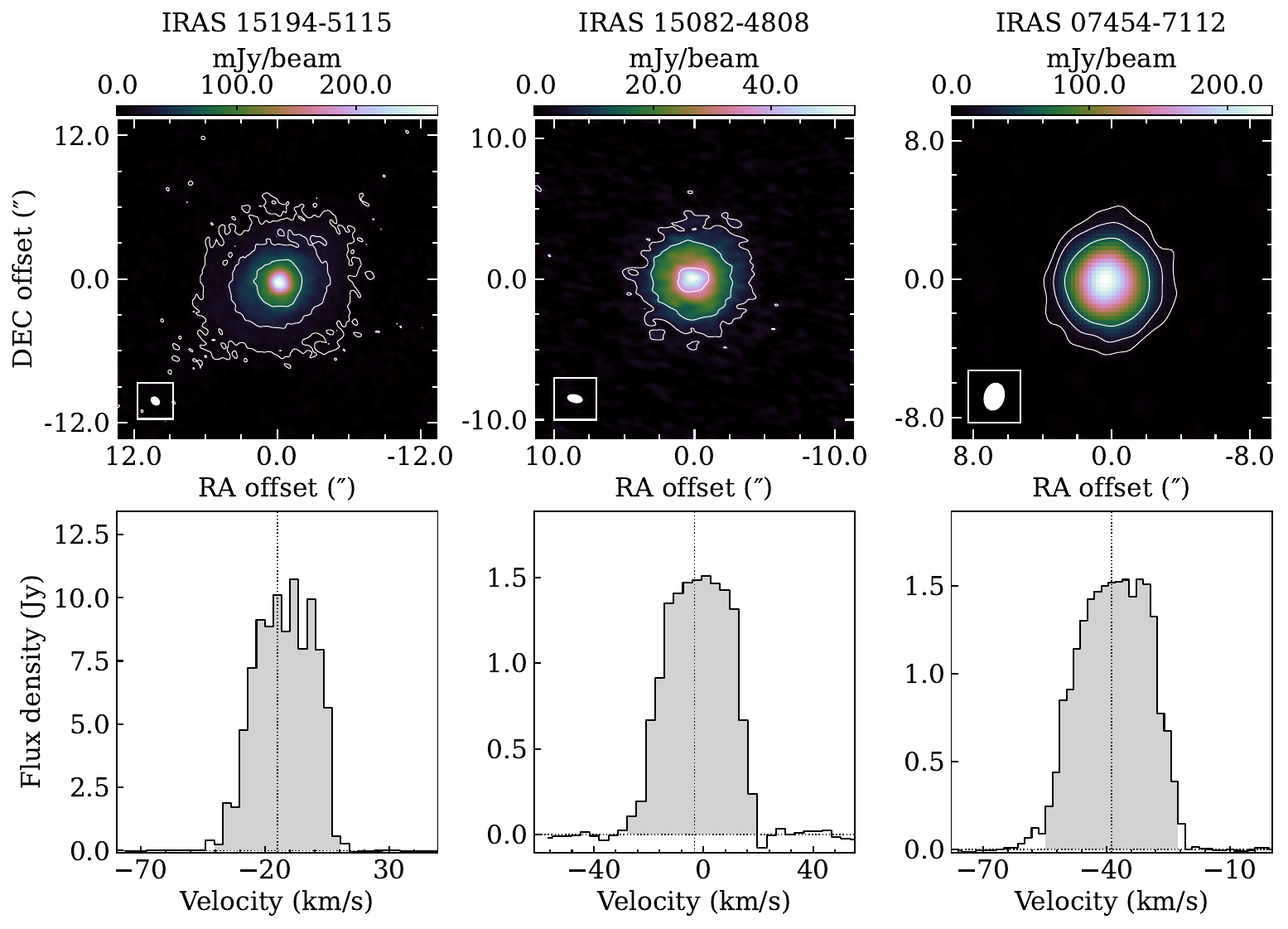}
    \caption{H$^{13}$CN, 1-0 (86.339921 GHz)}
\end{figure}

\begin{figure}[h]
    \centering
    \includegraphics[width=0.8\linewidth]{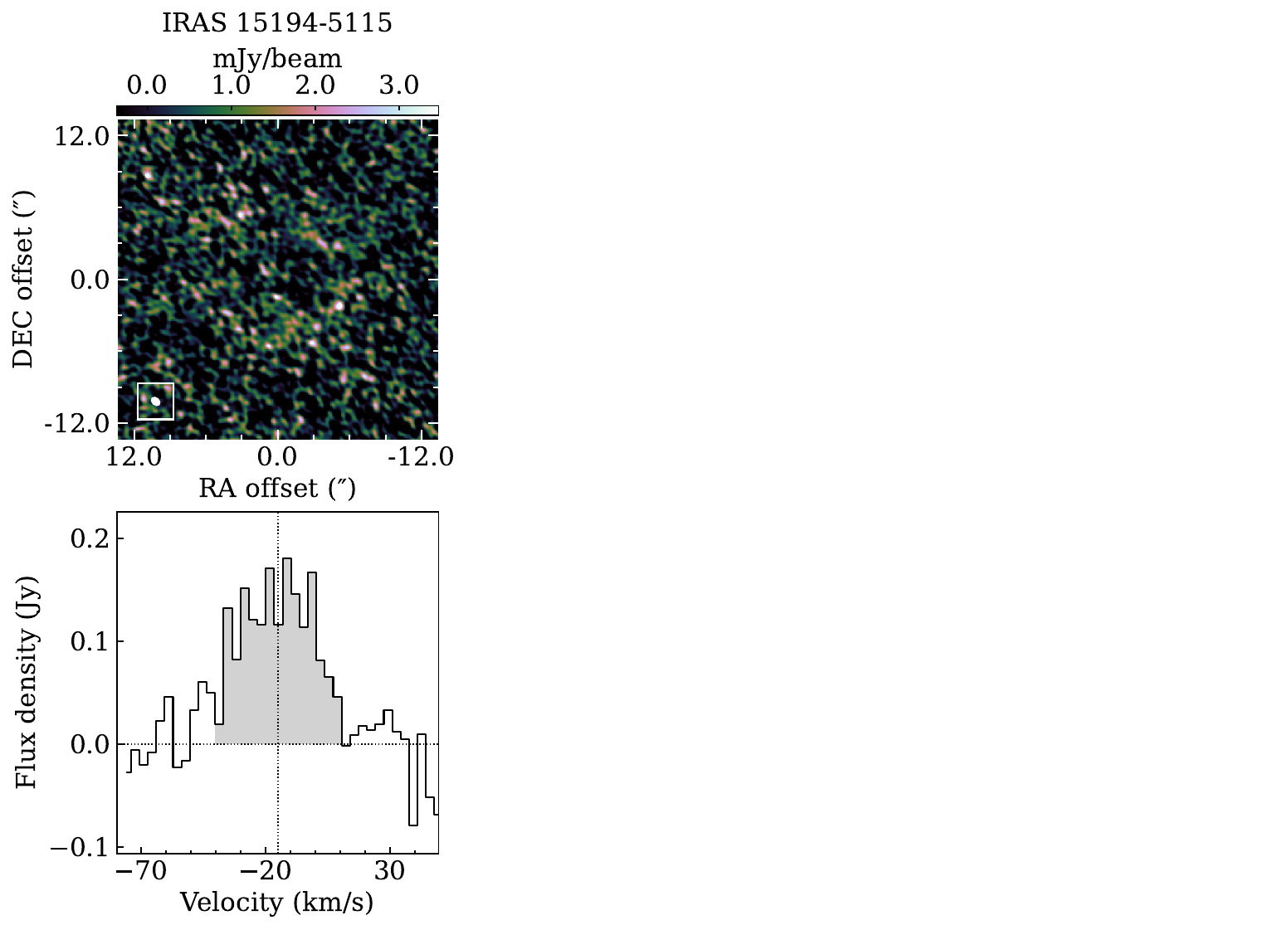}
    \caption{Si$^{13}$CC, 4$_{1,4}$-3$_{1,3}$ (86.562636 GHz)}
\end{figure}

\begin{figure}[h]
    \centering
    \includegraphics[width=0.8\linewidth]{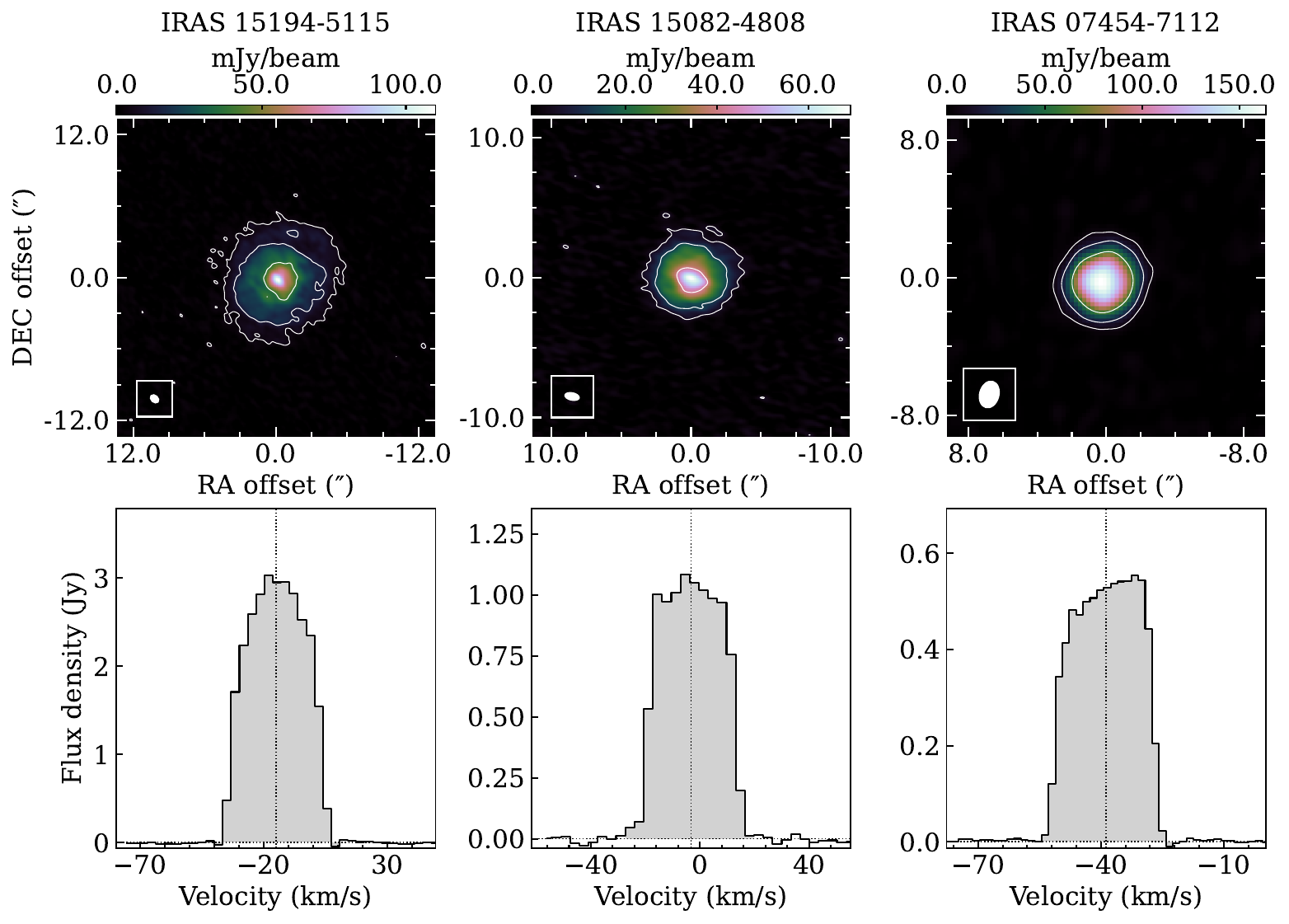}
    \caption{SiO, 2-1 (86.846985 GHz)}
    \label{fig:SiO_app_B}
\end{figure}

\begin{figure}[h]
    \centering
    \includegraphics[width=0.8\linewidth]{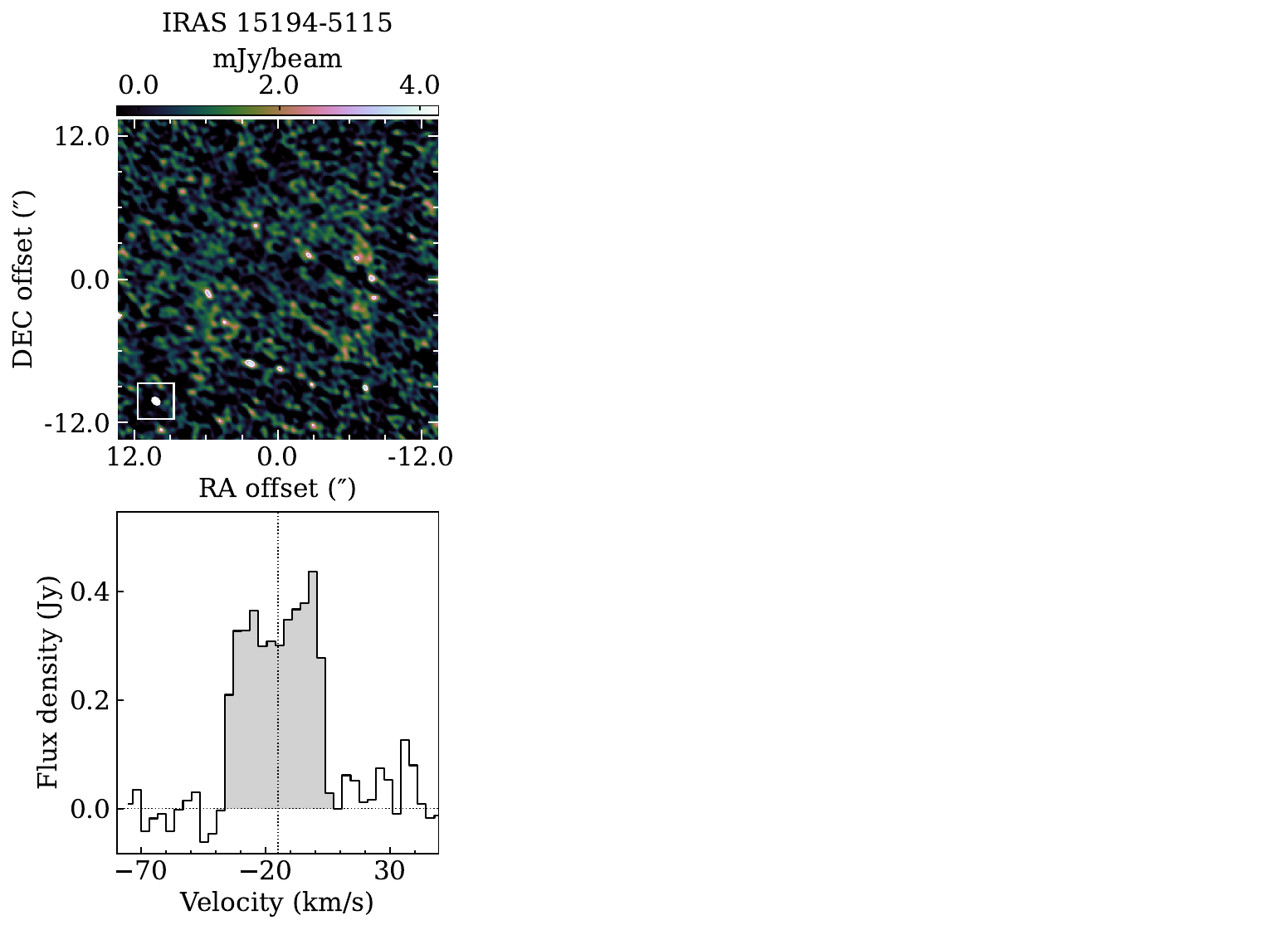}
    \caption{HN$^{13}$C, 1-0 (87.090825 GHz)}
\end{figure}

\begin{figure}[h]
    \centering
    \includegraphics[width=0.8\linewidth]{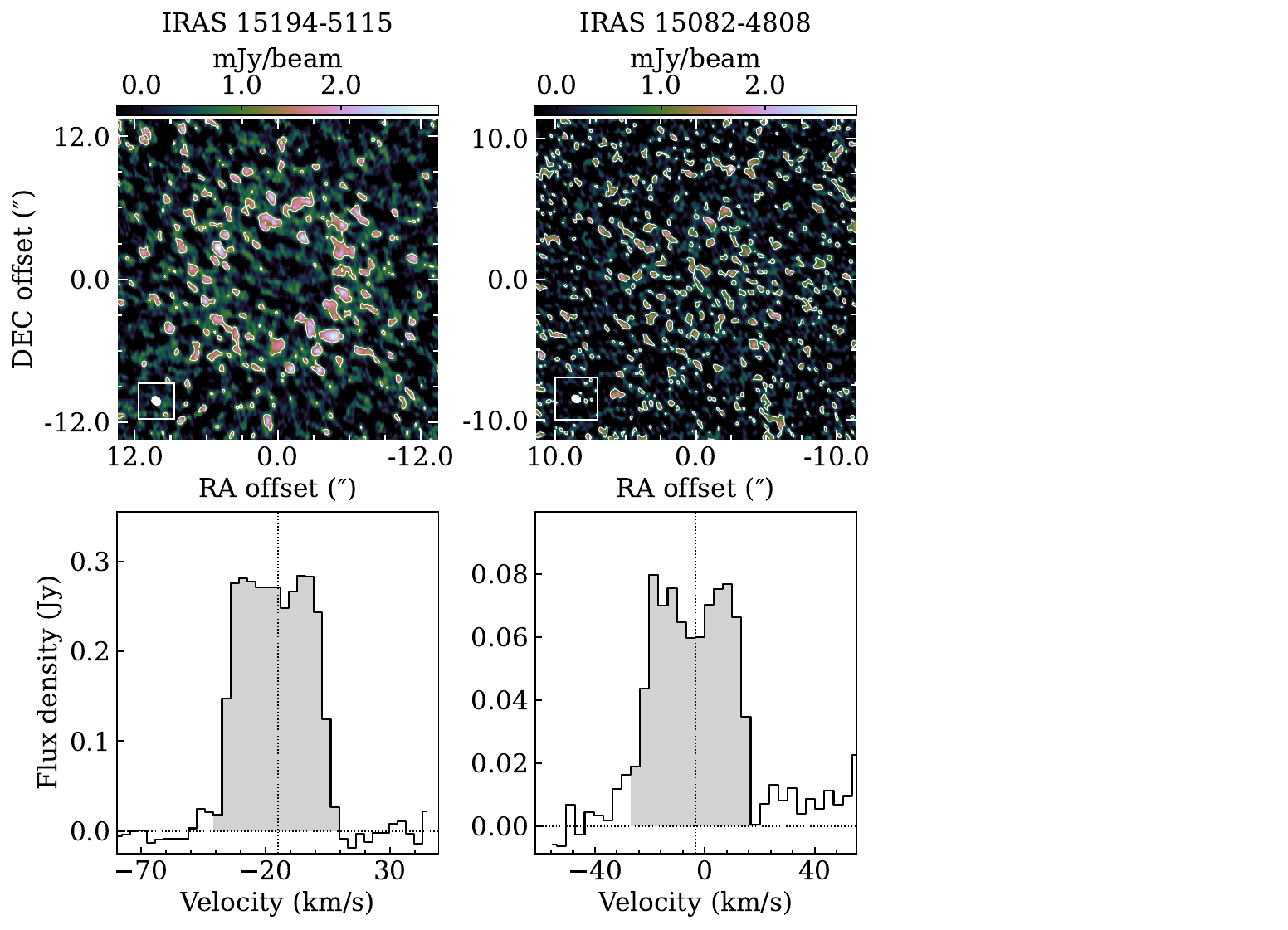}
    \caption{C$_2$H, N=1-0, J=3/2-1/2, F=1-1 (87.284105 GHz)}
\end{figure}

\begin{figure}[h]
    \centering
    \includegraphics[width=0.8\linewidth]{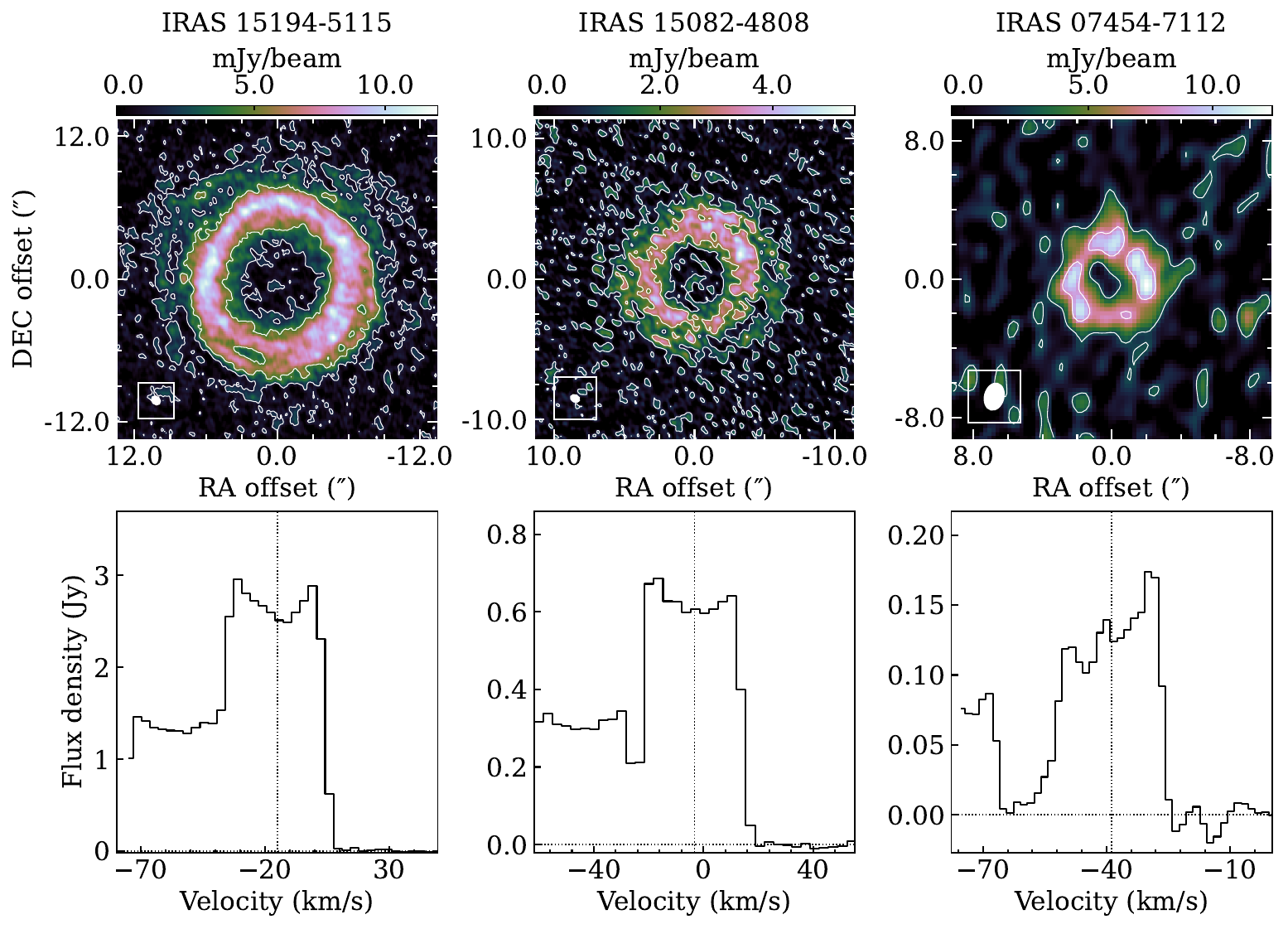}
    \caption{C$_2$H, N=1-0, J=3/2-1/2, F=2-1 (87.316898 GHz)}
\end{figure}

\begin{figure}[h]
    \centering
    \includegraphics[width=0.8\linewidth]{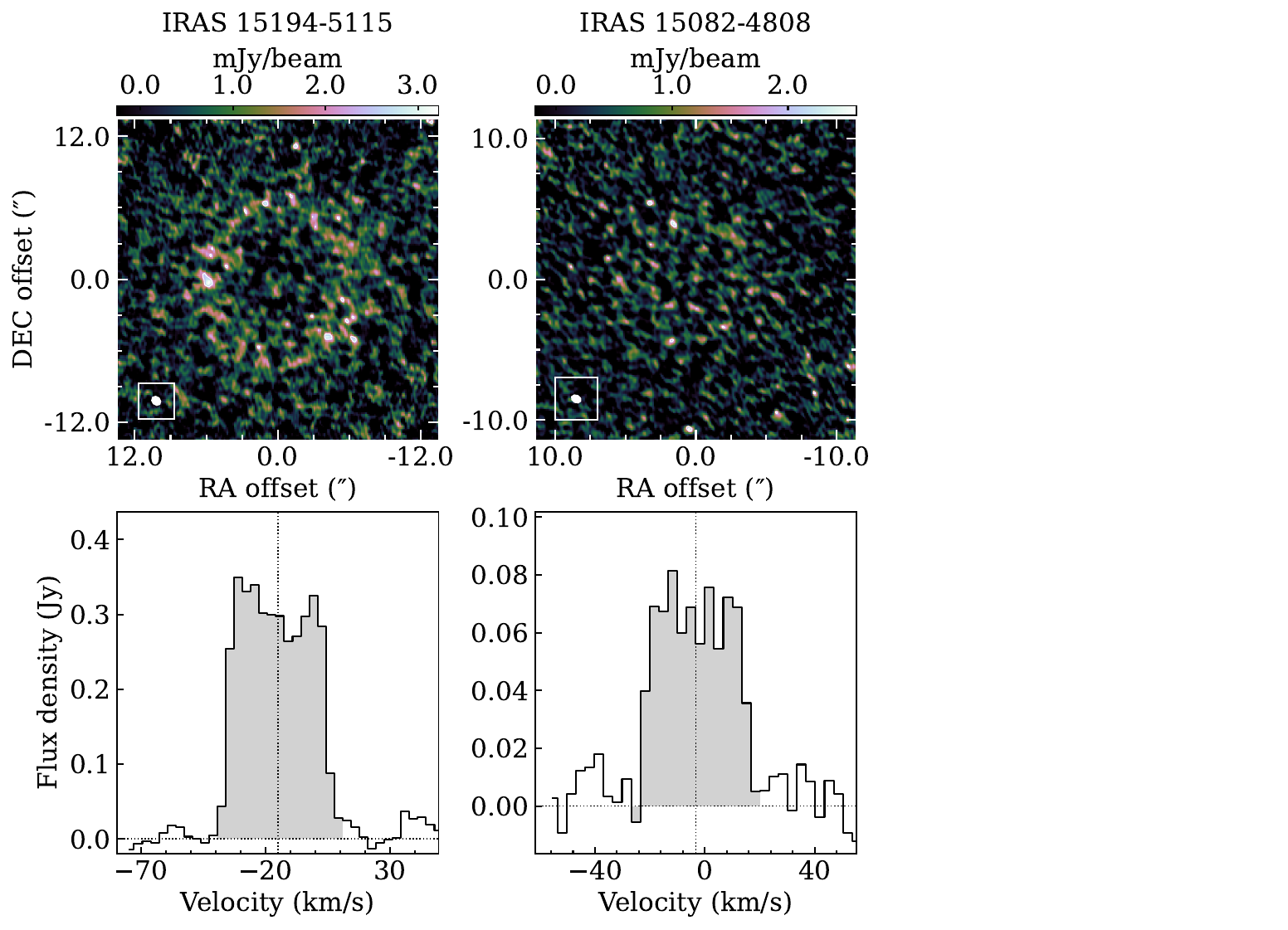}
    \caption{C$_2$H, N=1-0, J=1/2-1/2, F=1-0 (87.44647 GHz)}
    \label{fig:C2H_app_B}
\end{figure}

\begin{figure}[h]
    \centering
    \includegraphics[width=0.8\linewidth]{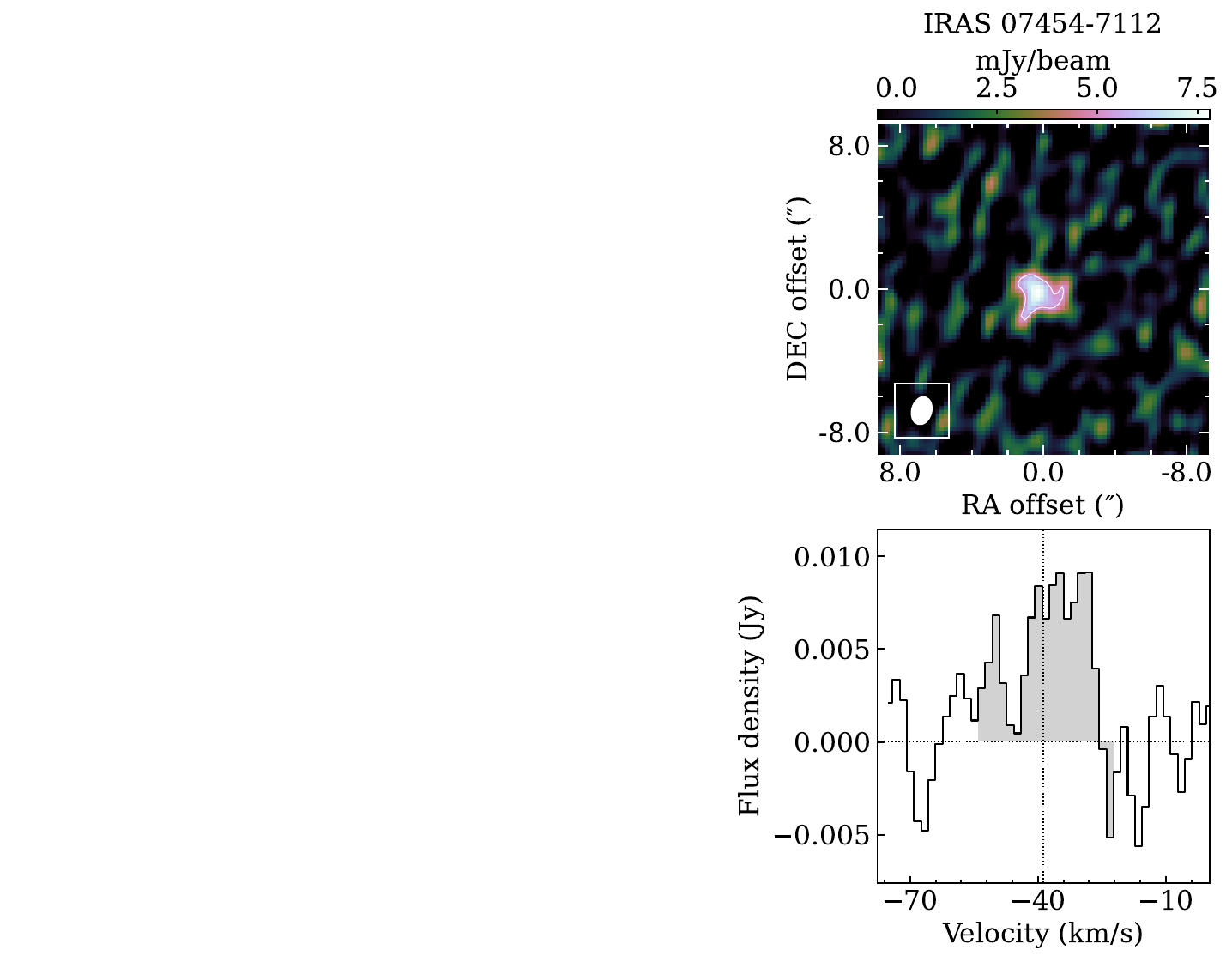}
    \caption{$^{30}$SiS, 5-4 (87.550558 GHz)}
\end{figure}

\begin{figure}[h]
    \centering
    \includegraphics[width=0.8\linewidth]{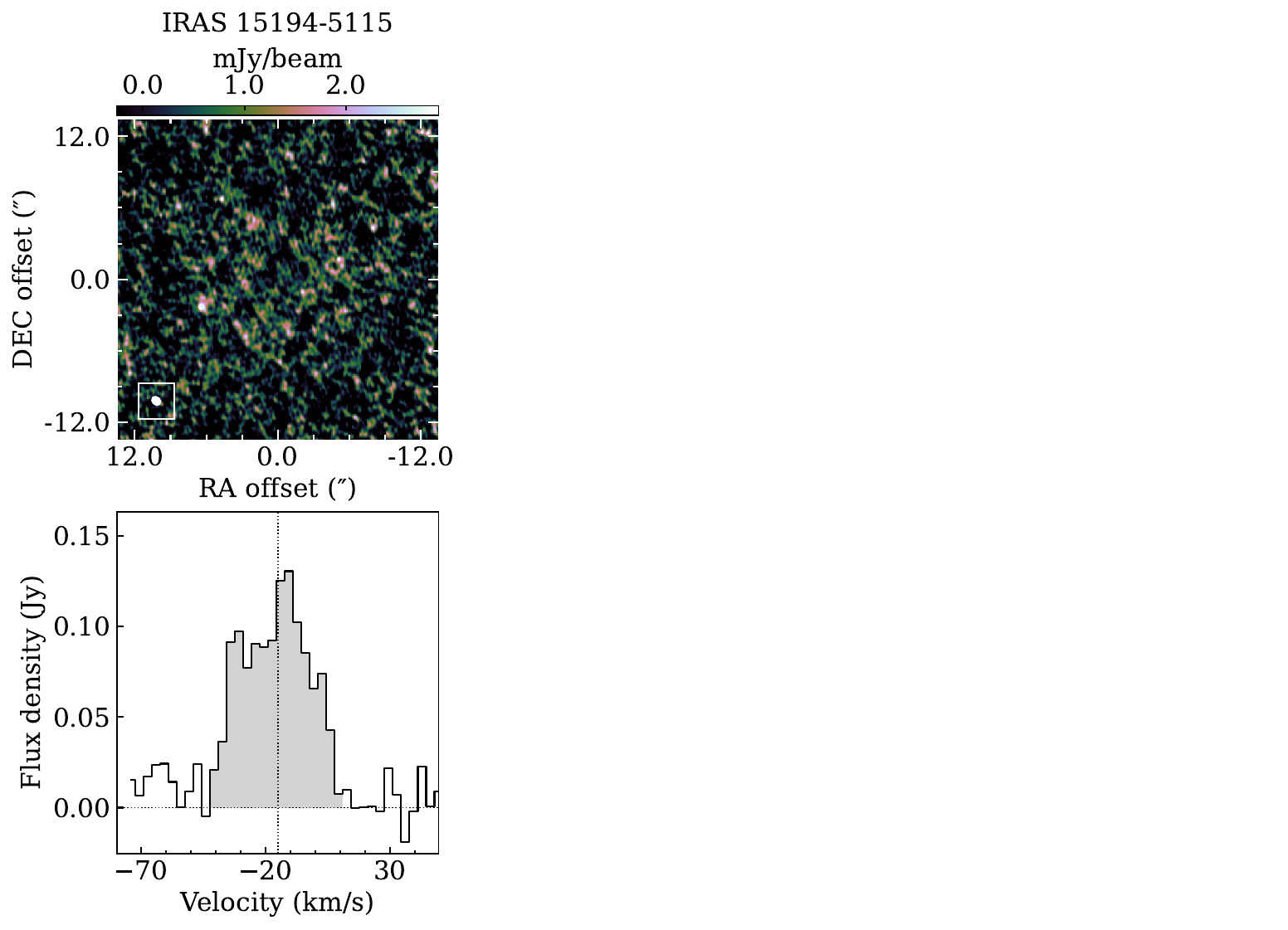}
    \caption{H$^{13}$CC$^{13}$CN, 10-9 (87.773712 GHz)}
\end{figure}

\begin{figure}[h]
    \centering
    \includegraphics[width=0.8\linewidth]{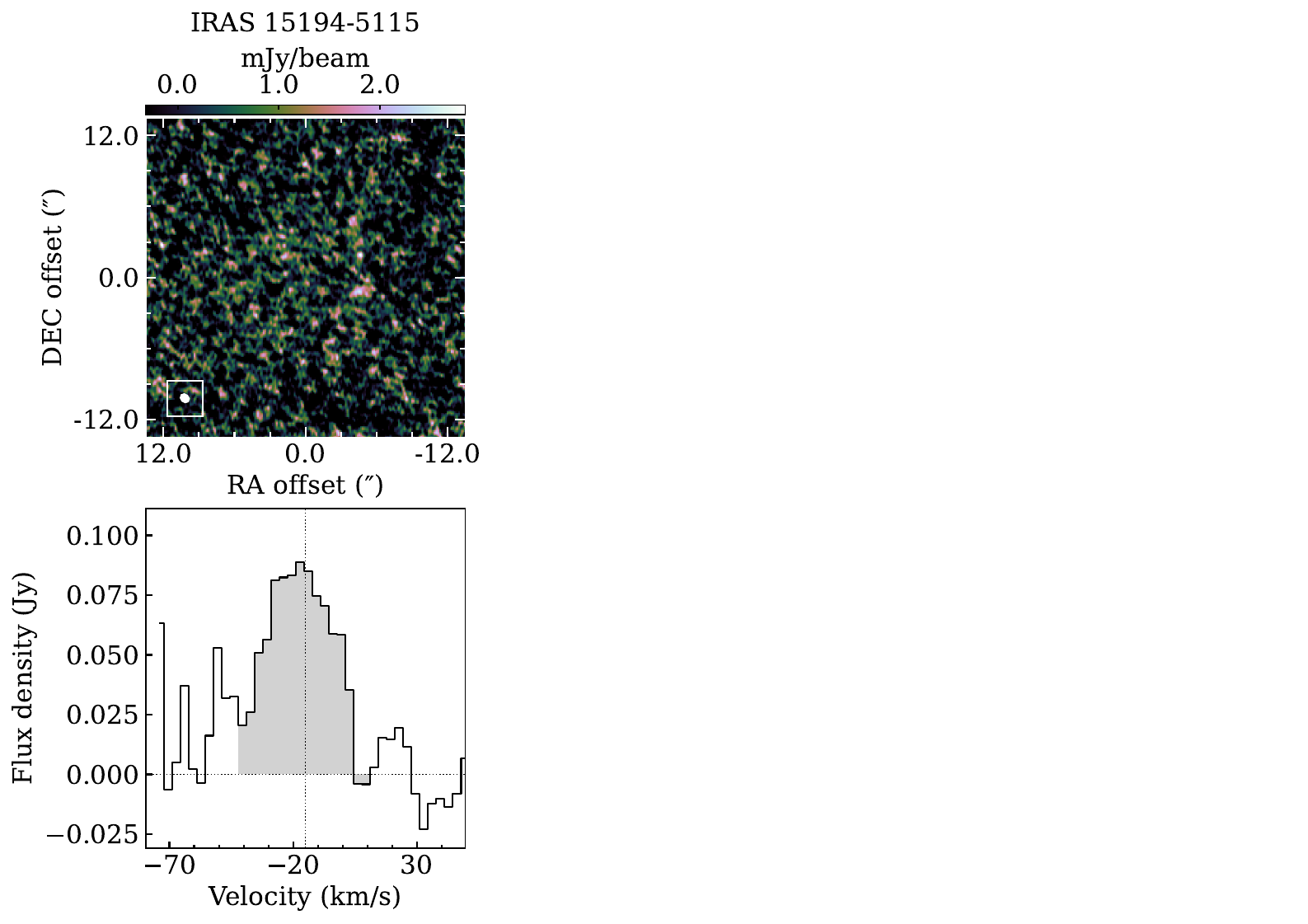}
    \caption{H$^{13}$C$^{13}$CCN, 10-9 (87.841102 GHz)}
\end{figure}

\begin{figure}[h]
    \centering
    \includegraphics[width=0.8\linewidth]{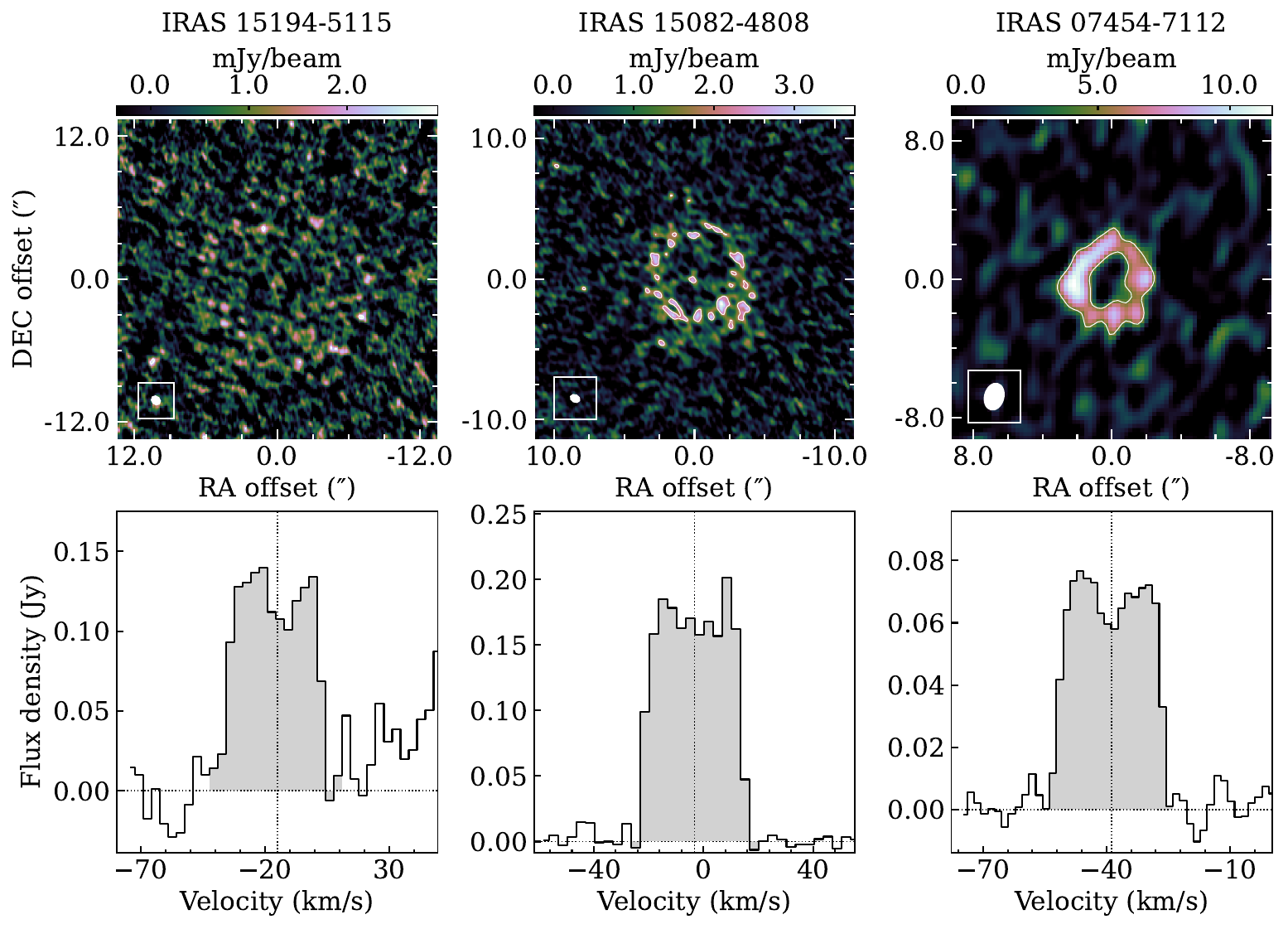}
    \caption{HC$_5$N, 33-32 (87.86363 GHz)}
    \label{fig:HC5N_app_B}
\end{figure}

\begin{figure}[h]
    \centering
    \includegraphics[width=0.8\linewidth]{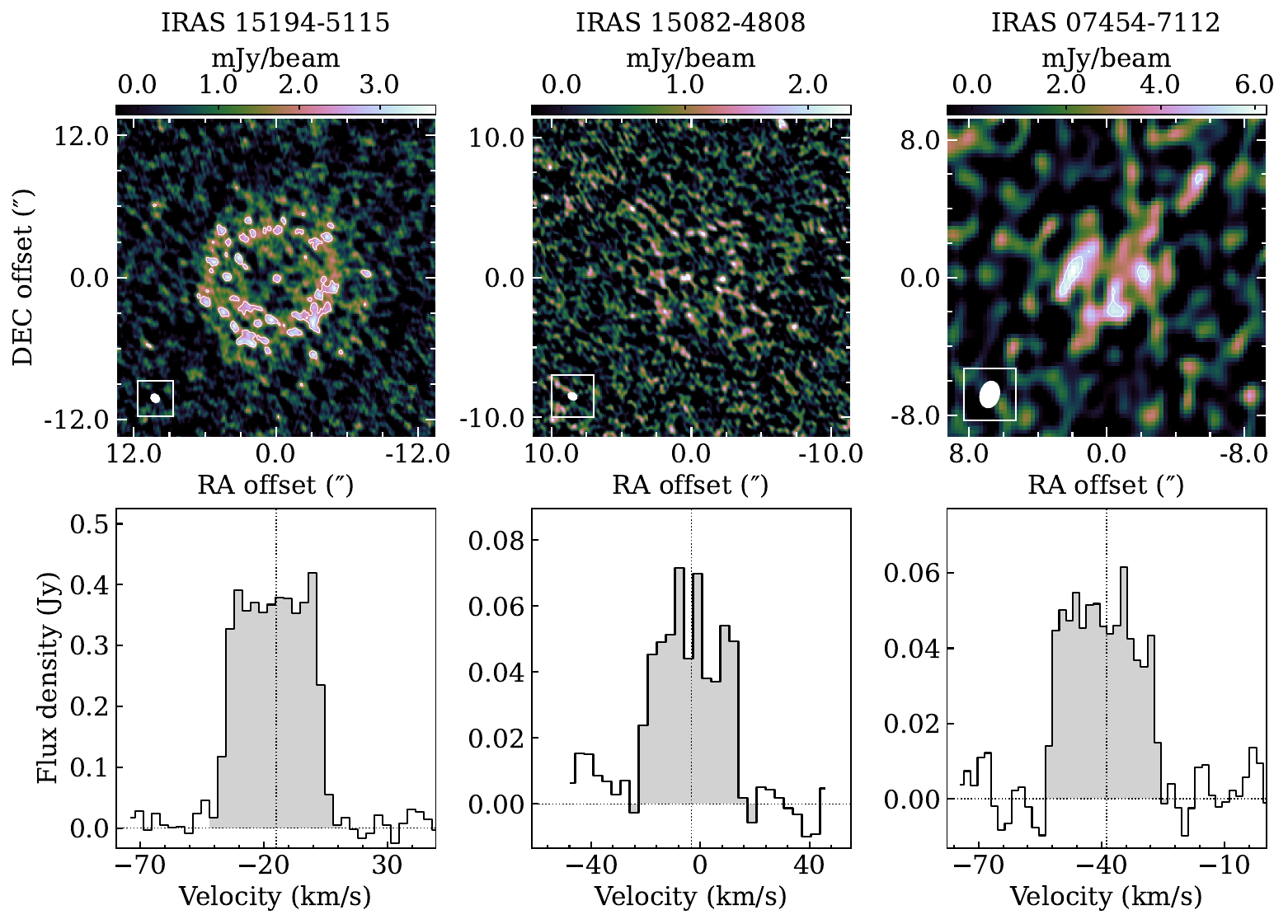}
    \caption{H$^{13}$CCCN, 10-9 (88.166832 GHz)}
\end{figure}

\begin{figure}[h]
    \centering
    \includegraphics[width=0.8\linewidth]{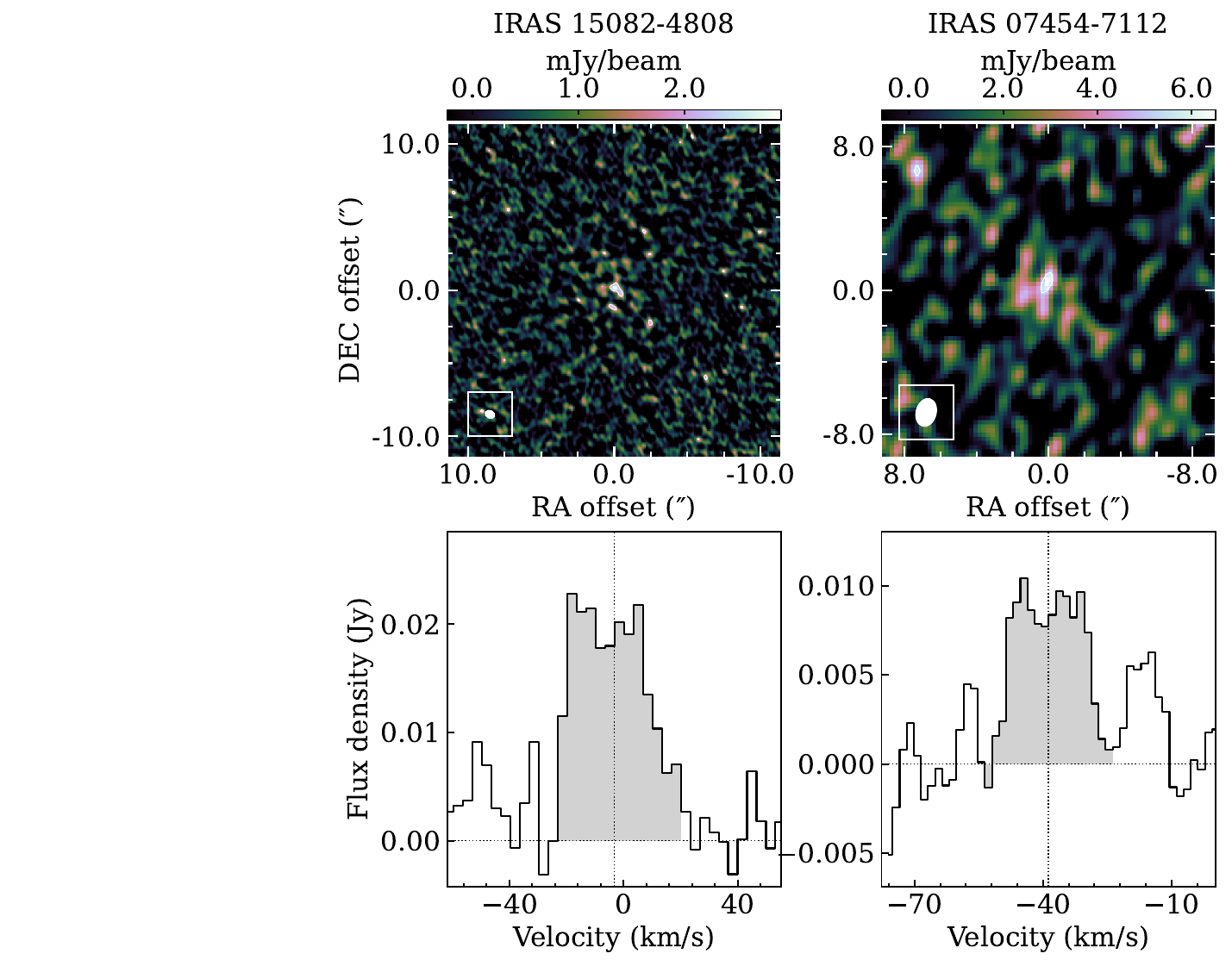}
    \caption{Si$^{34}$S, 5-4 (88.285828 GHz)}
\end{figure}

\begin{figure}[h]
    \centering
    \includegraphics[width=0.8\linewidth]{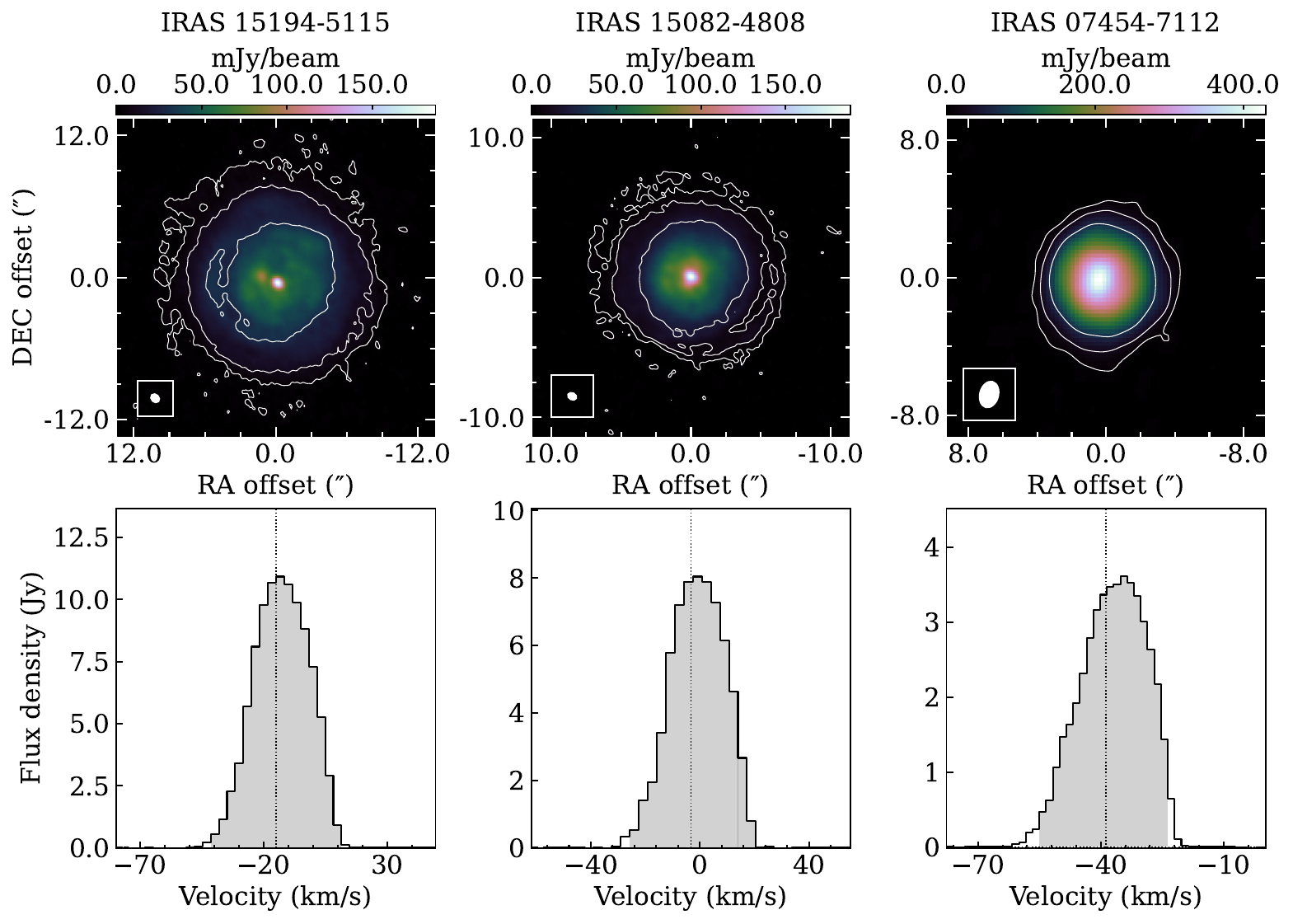}
    \caption{HCN, 1-0 (88.631602 GHz)}
    \label{fig:HCN_app_B}
\end{figure}

\begin{figure}[h]
    \centering
    \includegraphics[width=0.8\linewidth]{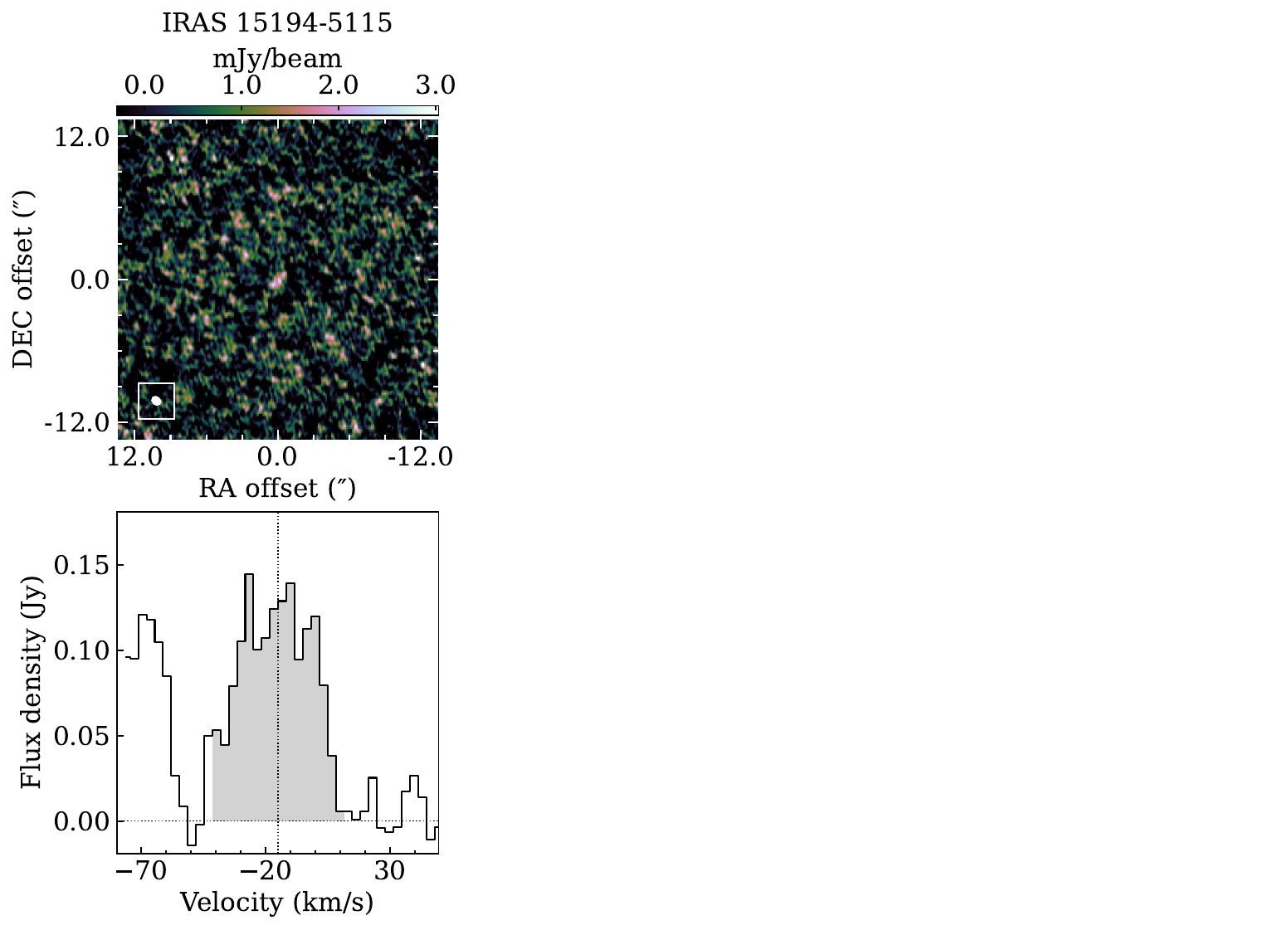}
    \caption{CC$^{13}$CN, N=9-8, J=19/2-17/2 (88.7117 GHz)}
\end{figure}

\begin{figure}[h]
    \centering
    \includegraphics[width=0.8\linewidth]{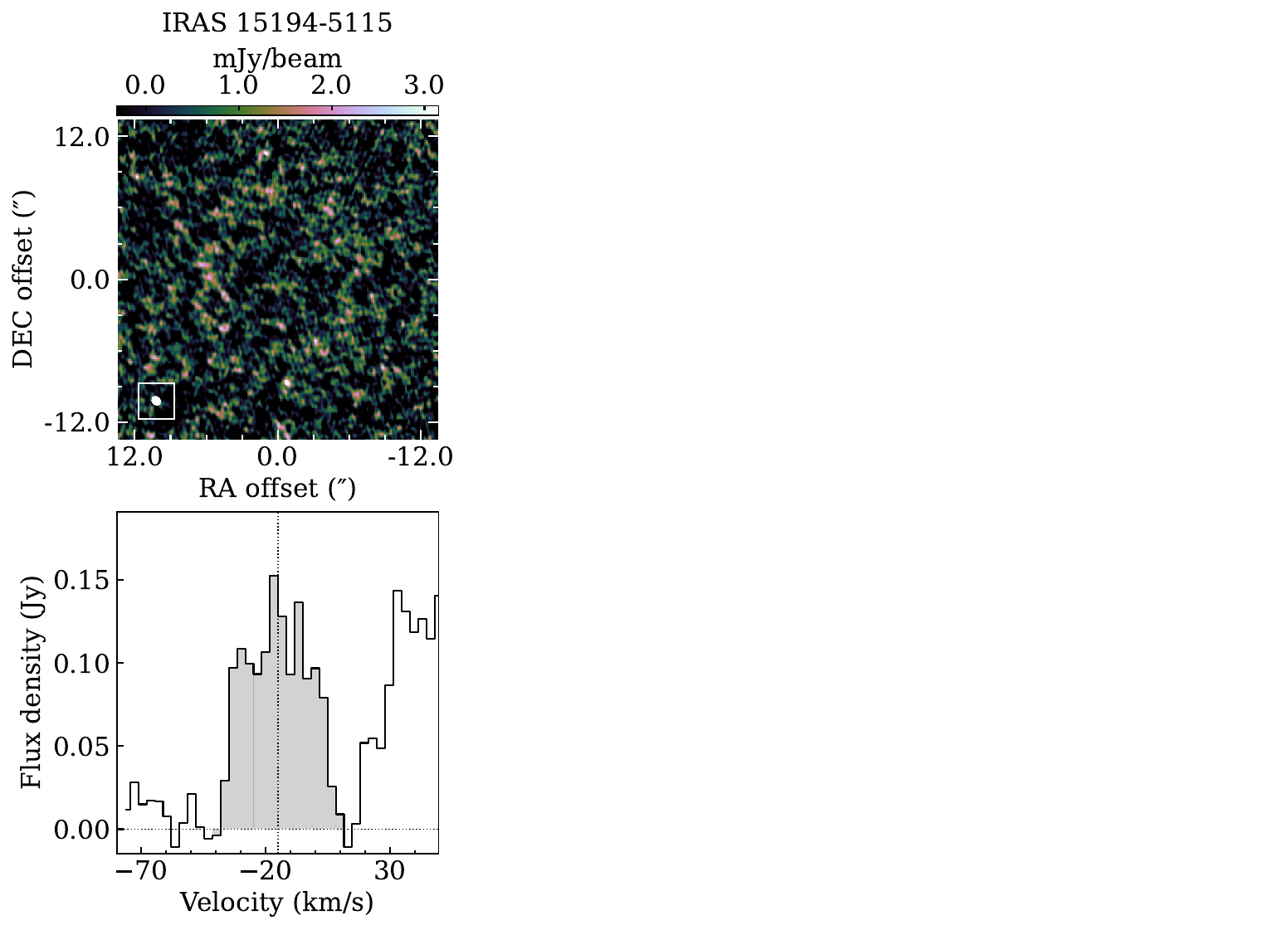}
    \caption{CC$^{13}$CN, N=9-8, J=17/2-15/2 (88.7302 GHz)}
\end{figure}

\begin{figure}[h]
    \centering
    \includegraphics[width=0.8\linewidth]{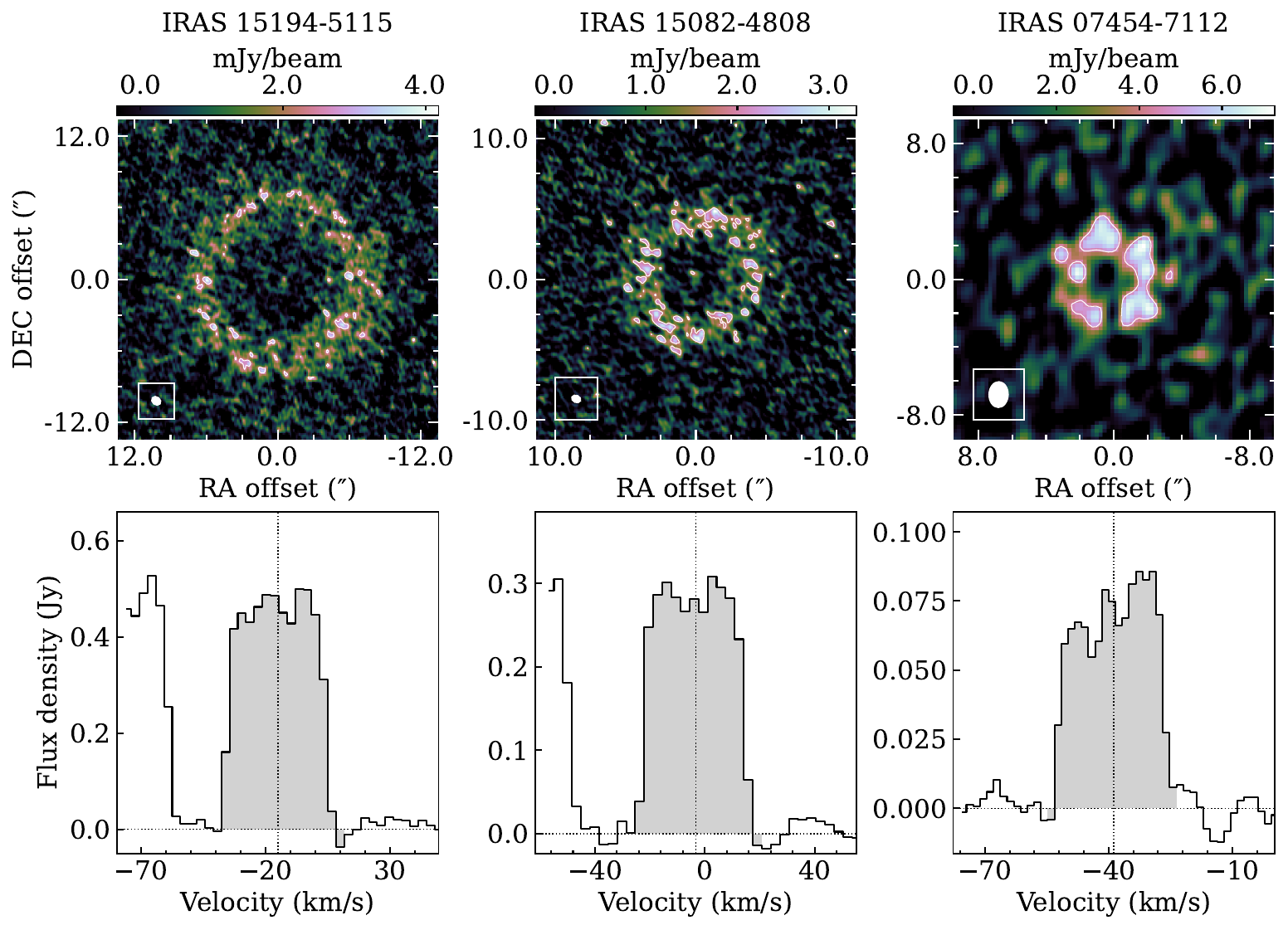}
    \caption{C$_3$N, 9-8 (A) (89.0456 GHz)}
    \label{fig:C3N_app_B}
\end{figure}

\begin{figure}[h]
    \centering
    \includegraphics[width=0.8\linewidth]{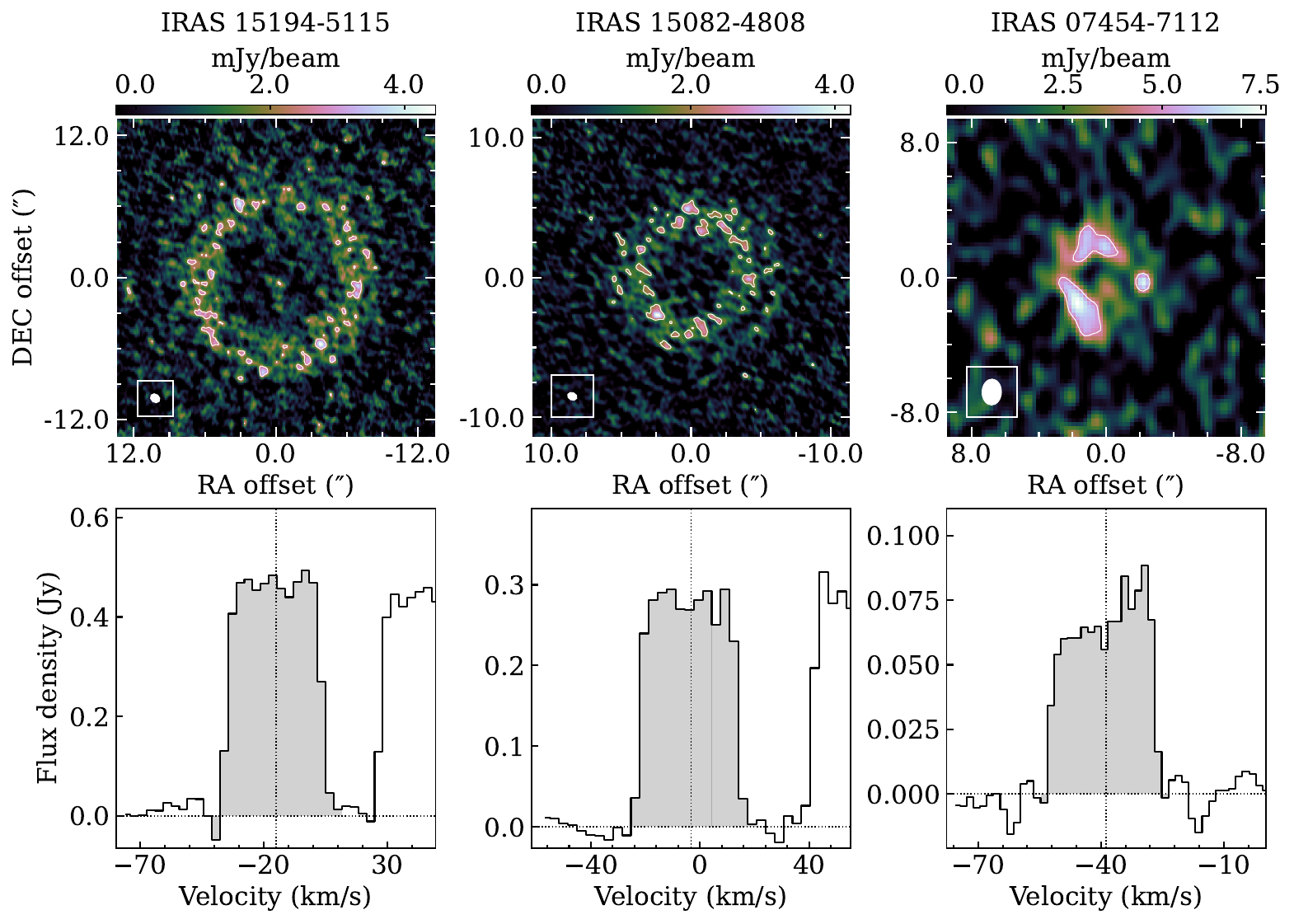}
    \caption{C$_3$N, 9-8 (B) (89.0643 GHz)}
        \label{fig:C3N_cc_appendix_C}
\end{figure}

\begin{figure}[h]
    \centering
    \includegraphics[width=0.8\linewidth]{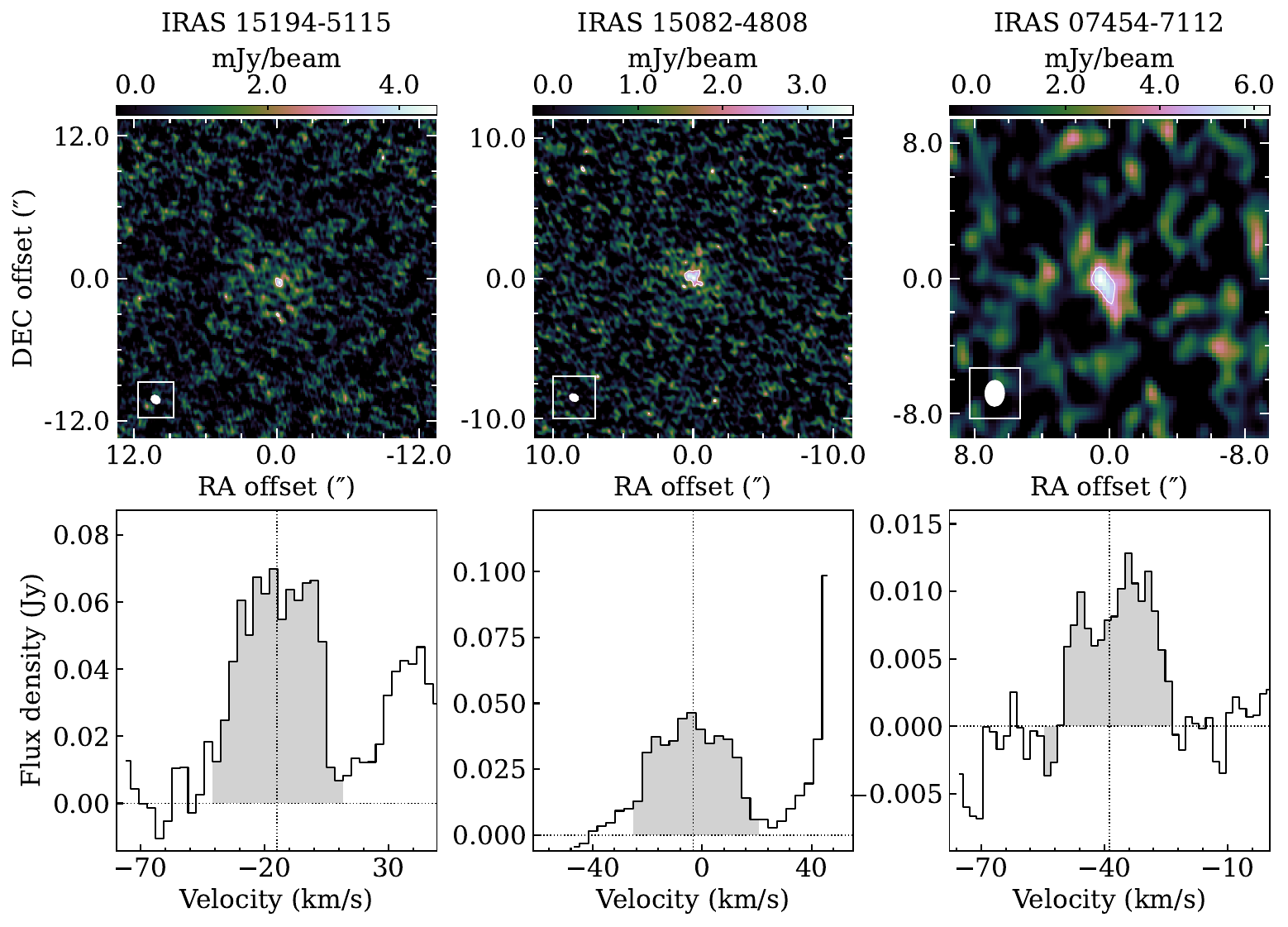}
    \caption{$^{29}$SiS, 5-4 (89.103749 GHz)}
\end{figure}

\begin{figure}[h]
    \centering
    \includegraphics[width=0.8\linewidth]{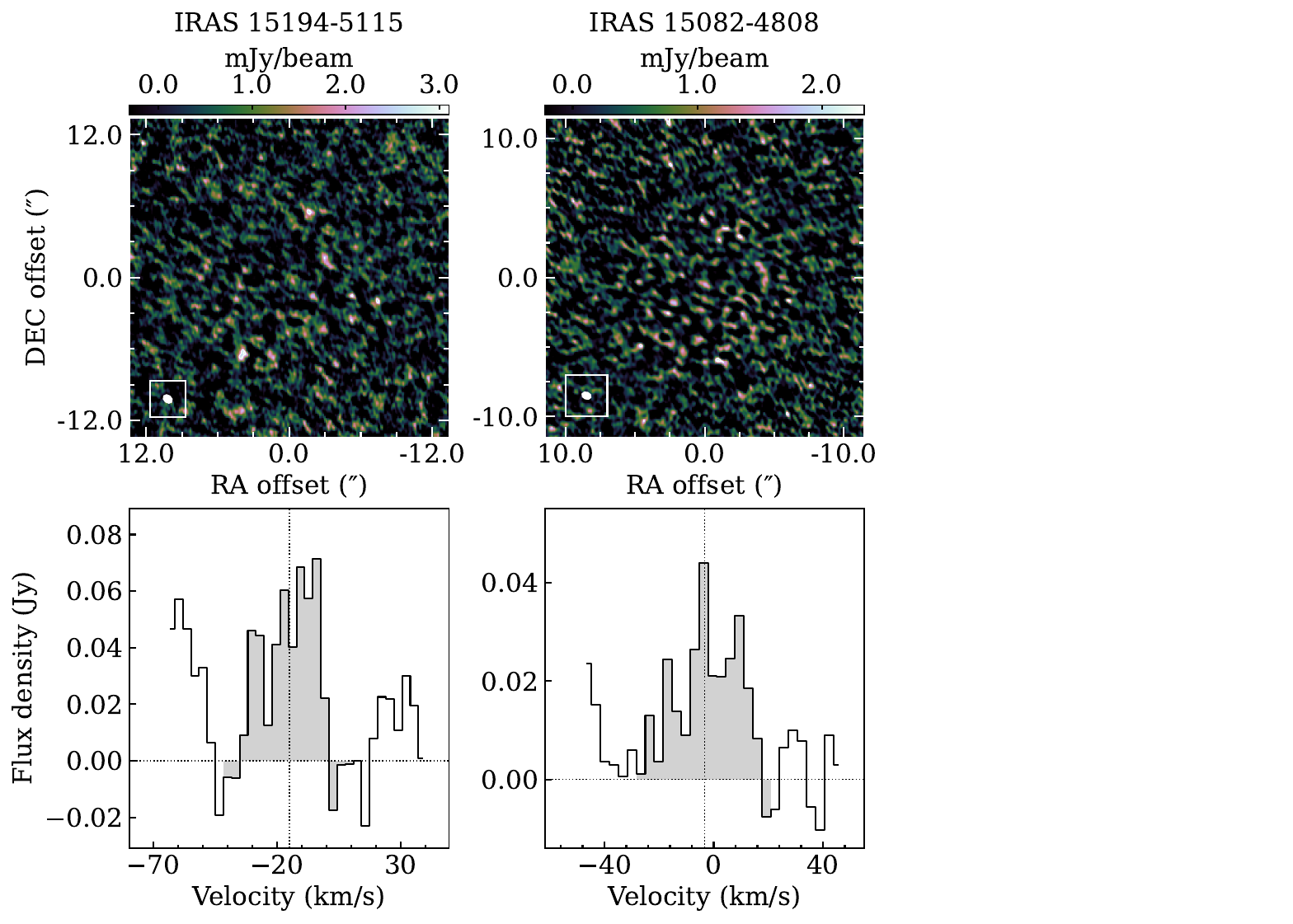}
    \caption{l-C$_4$H$_2$, 10$_{3,8}$-9$_{3,7}$ (89.289791 GHz)}
        \label{fig:l-C4H2_cc_appendix_C}
\end{figure}

\begin{figure}[h]
    \centering
    \includegraphics[width=0.8\linewidth]{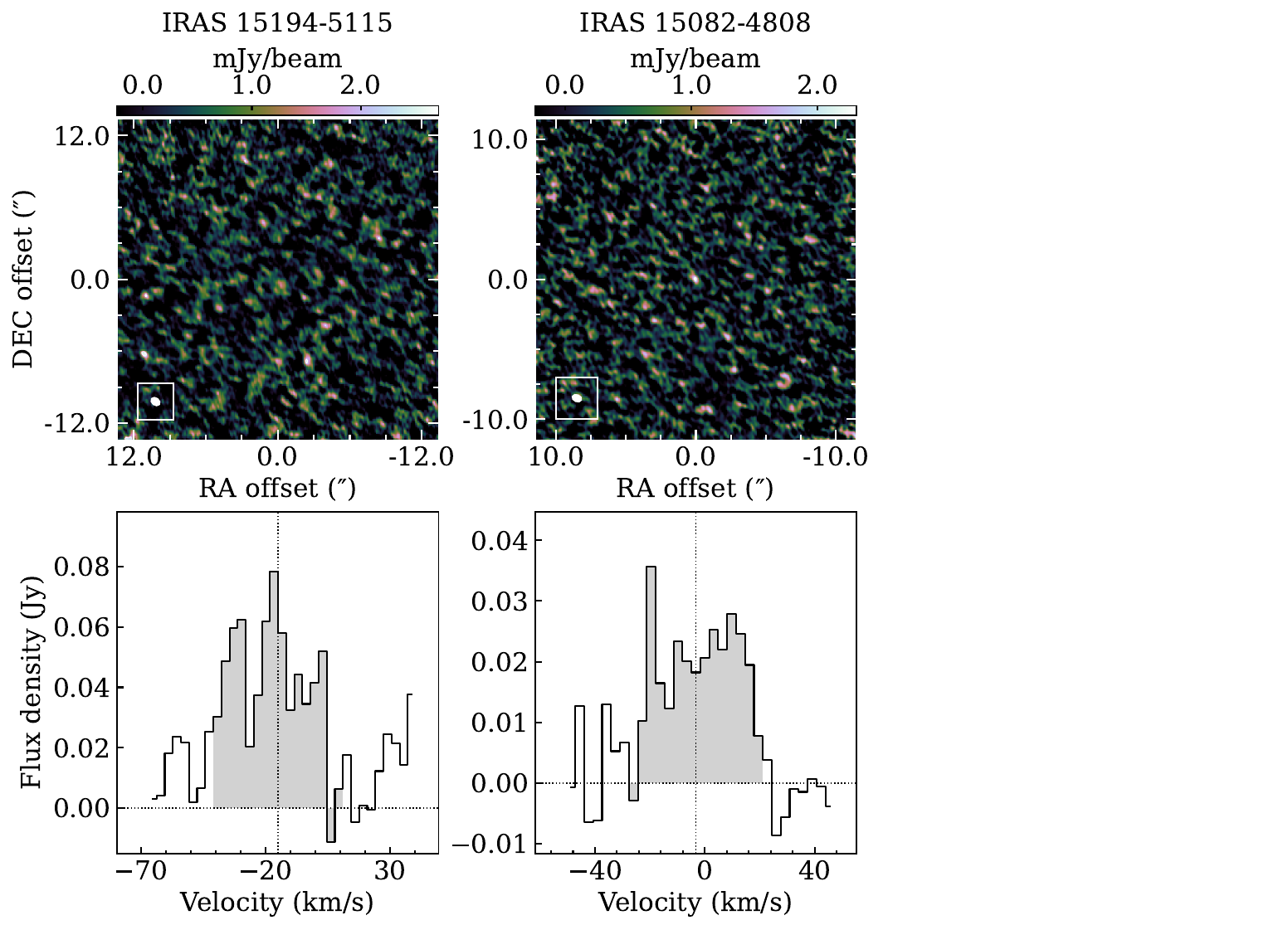}
    \caption{C$_6$H, J=65/2-63/2 (90.093295 GHz)}
        \label{fig:C6H_cc_appendix_C}
\end{figure}

\begin{figure}[h]
    \centering
    \includegraphics[width=0.8\linewidth]{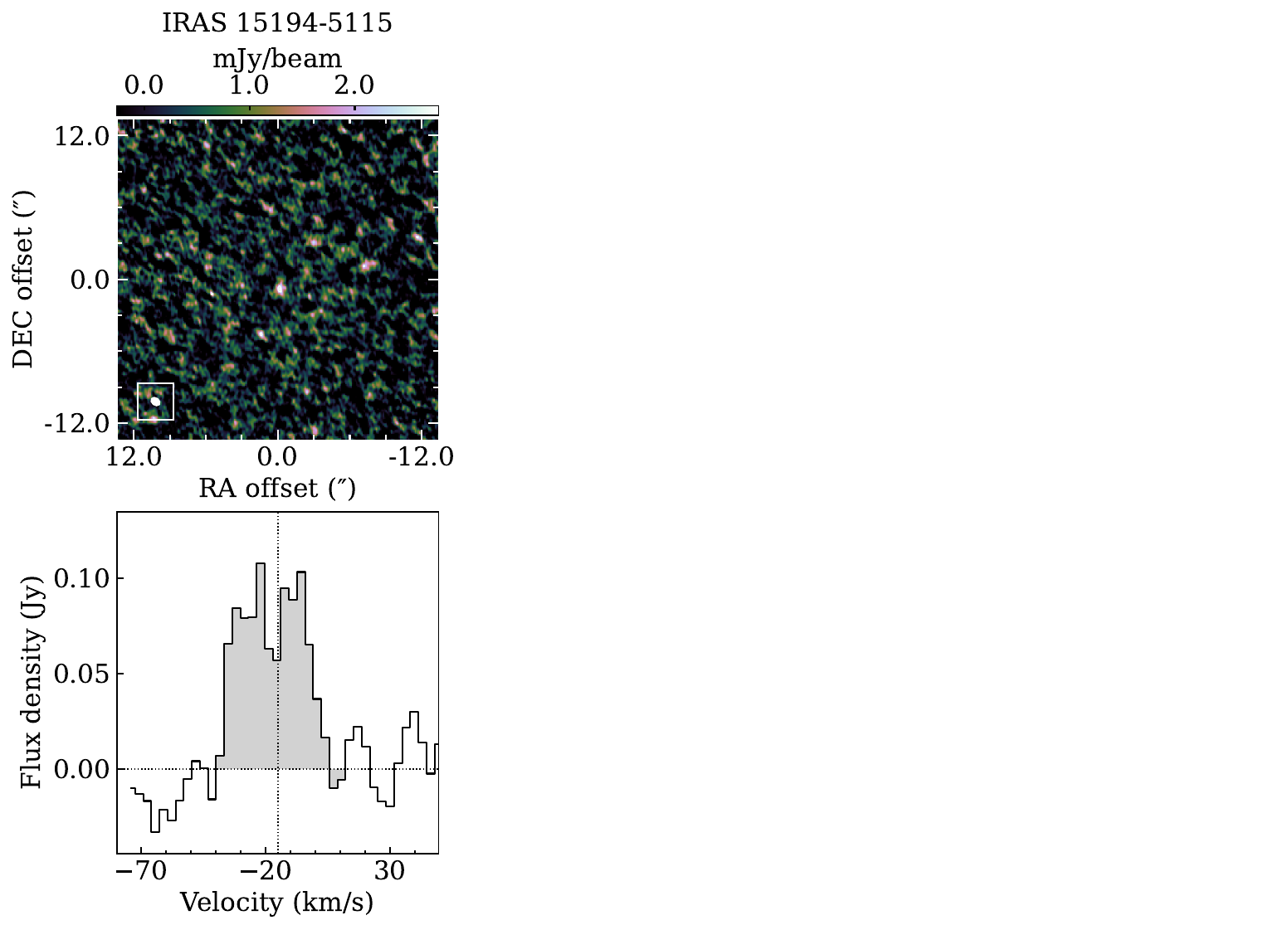}
    \caption{HC$^{13}$C$^{13}$CN, 10-9 (90.204324 GHz)}
\end{figure}

\begin{figure}[h]
    \centering
    \includegraphics[width=0.8\linewidth]{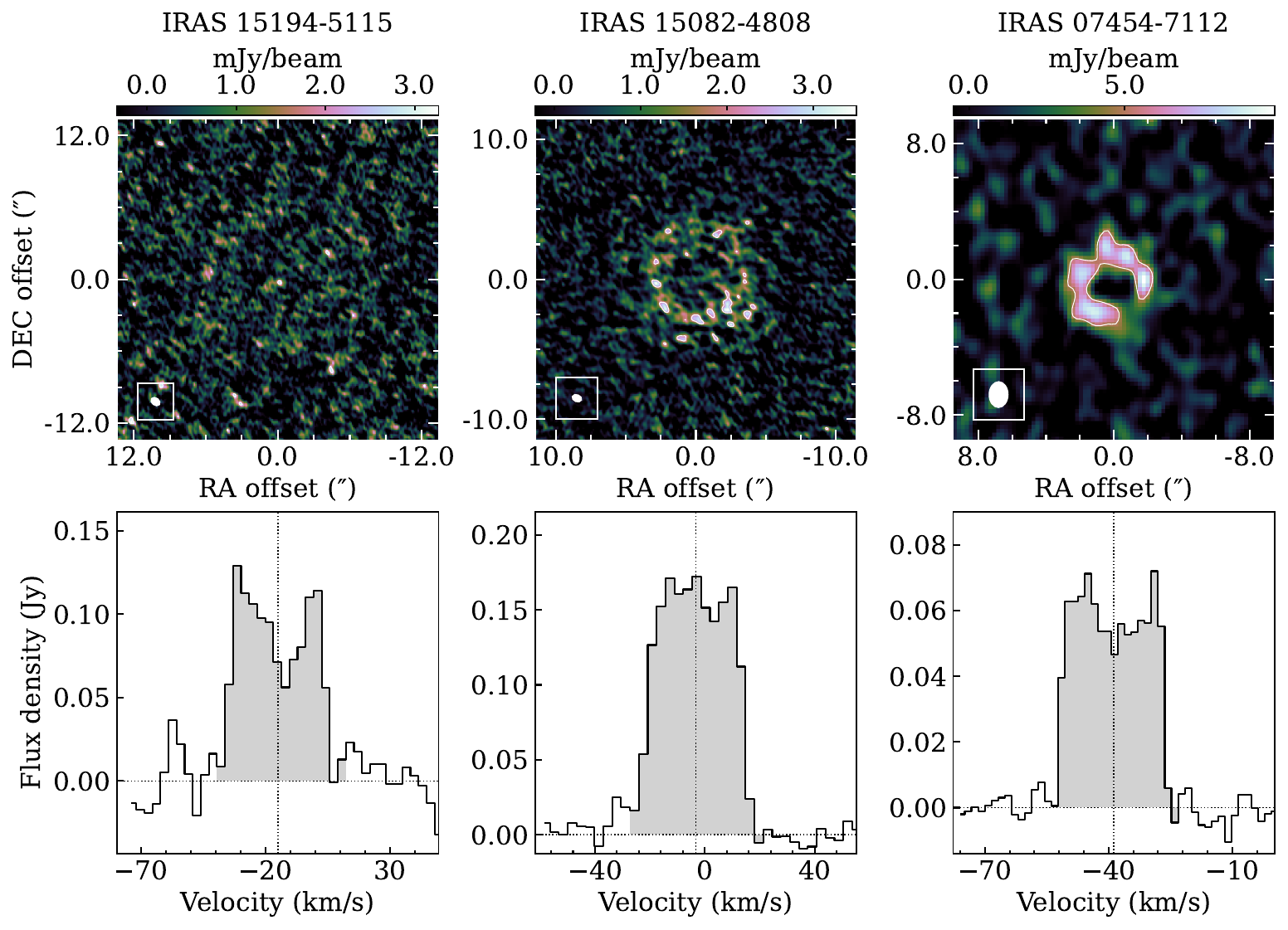}
    \caption{HC$_5$N, 34-33 (90.52589 GHz)}
\end{figure}

\begin{figure}[h]
    \centering
    \includegraphics[width=0.8\linewidth]{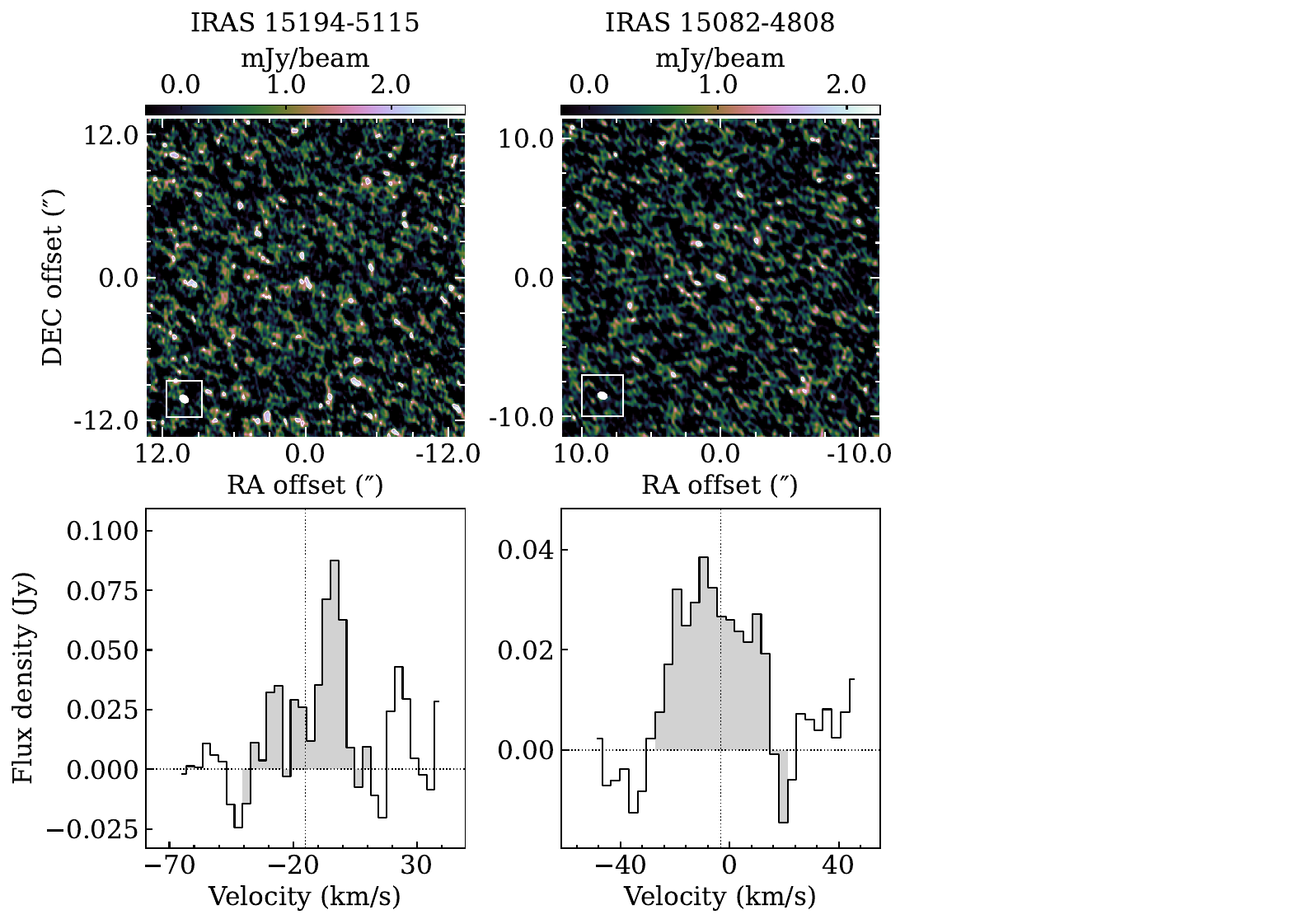}
    \caption{$^{30}$SiC$_2$, 4$_{0,4}$-3$_{0,3}$ (90.562283 GHz)}
\end{figure}

\begin{figure}[h]
    \centering
    \includegraphics[width=0.8\linewidth]{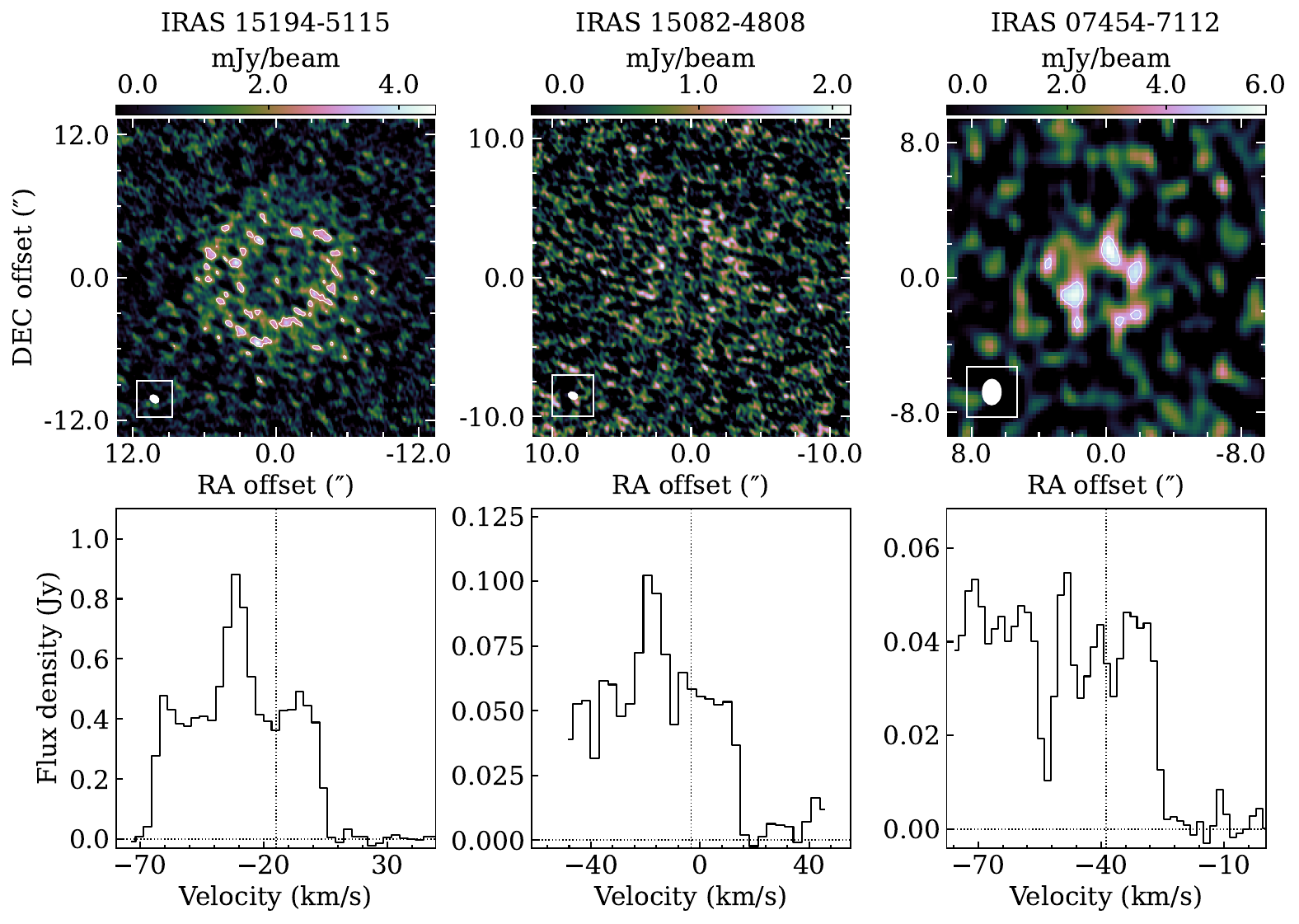}
    \caption{HC$^{13}$CCN, 10-9 (90.593059 GHz)}
        \label{fig:HC13CCN_cc_appendix_C}
\end{figure}

\begin{figure}[h]
    \centering
    \includegraphics[width=0.8\linewidth]{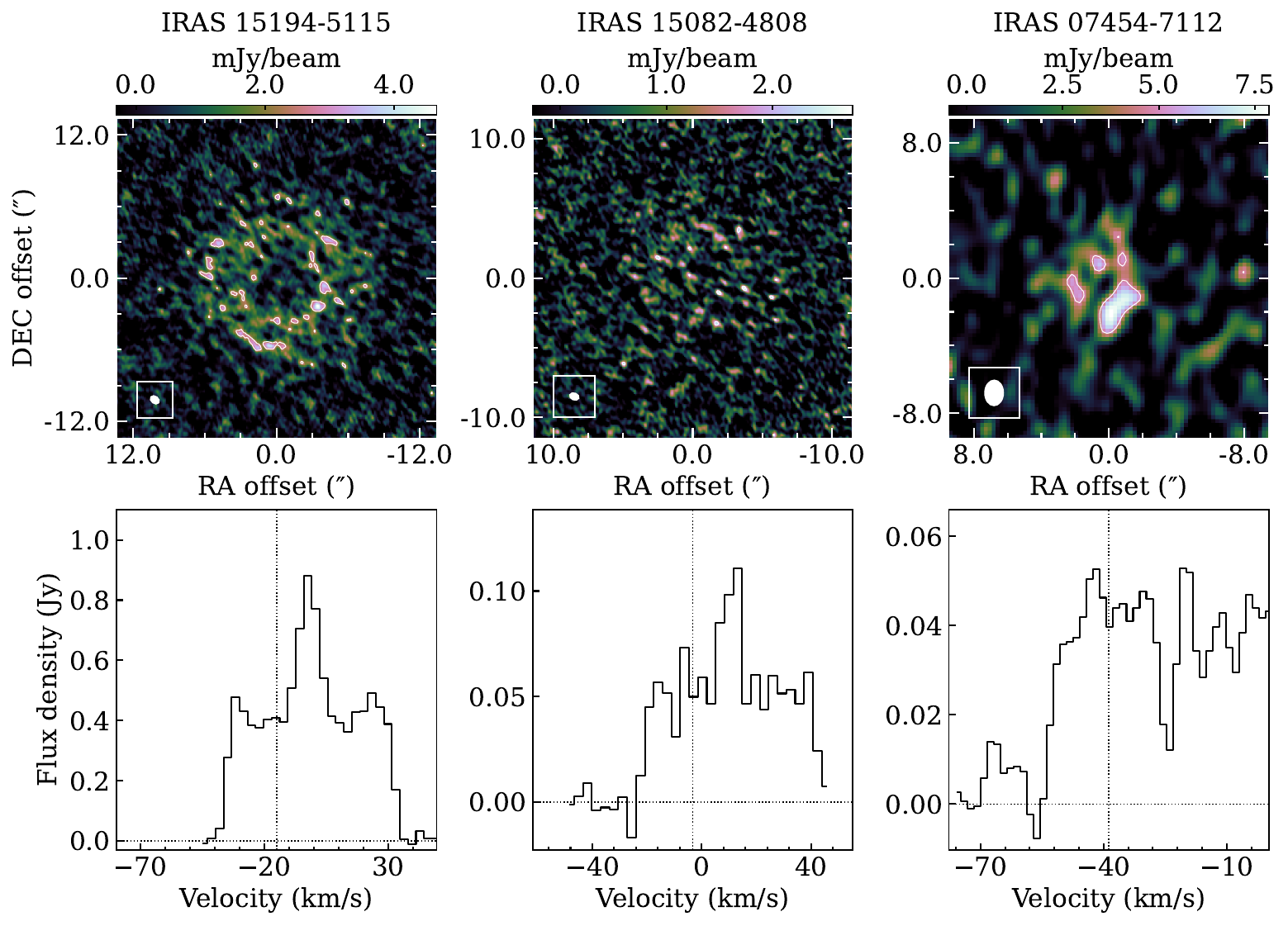}
    \caption{HCC$^{13}$CN, 10-9 (90.601777 GHz)}
    \label{fig:HCC13CN_cc_appendix_C}
\end{figure}

\begin{figure}[h]
    \centering
    \includegraphics[width=0.8\linewidth]{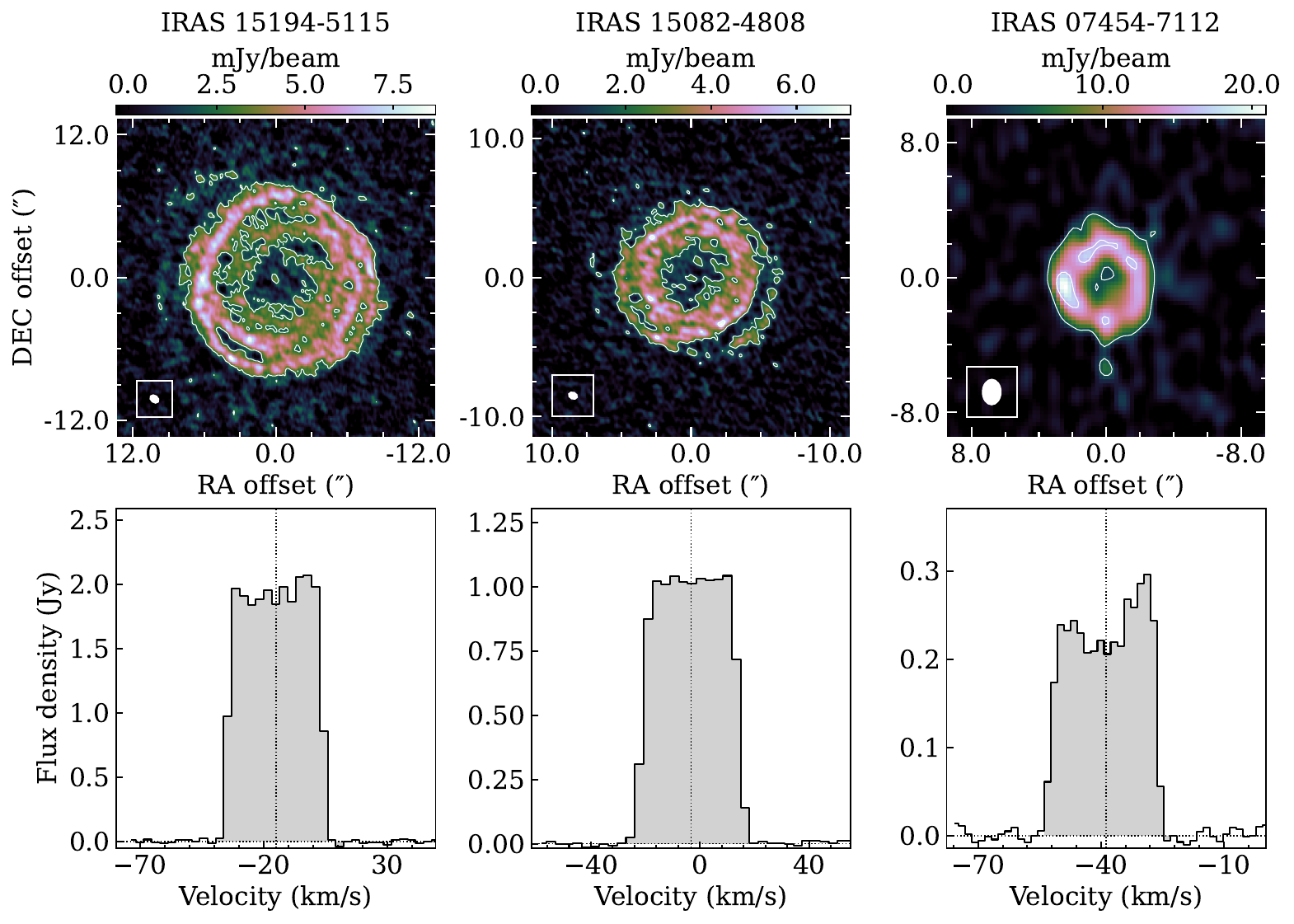}
    \caption{HNC, 1-0 (90.663568 GHz)}
    \label{fig:HNC_app_B}
\end{figure}

\begin{figure}[h]
    \centering
    \includegraphics[width=0.8\linewidth]{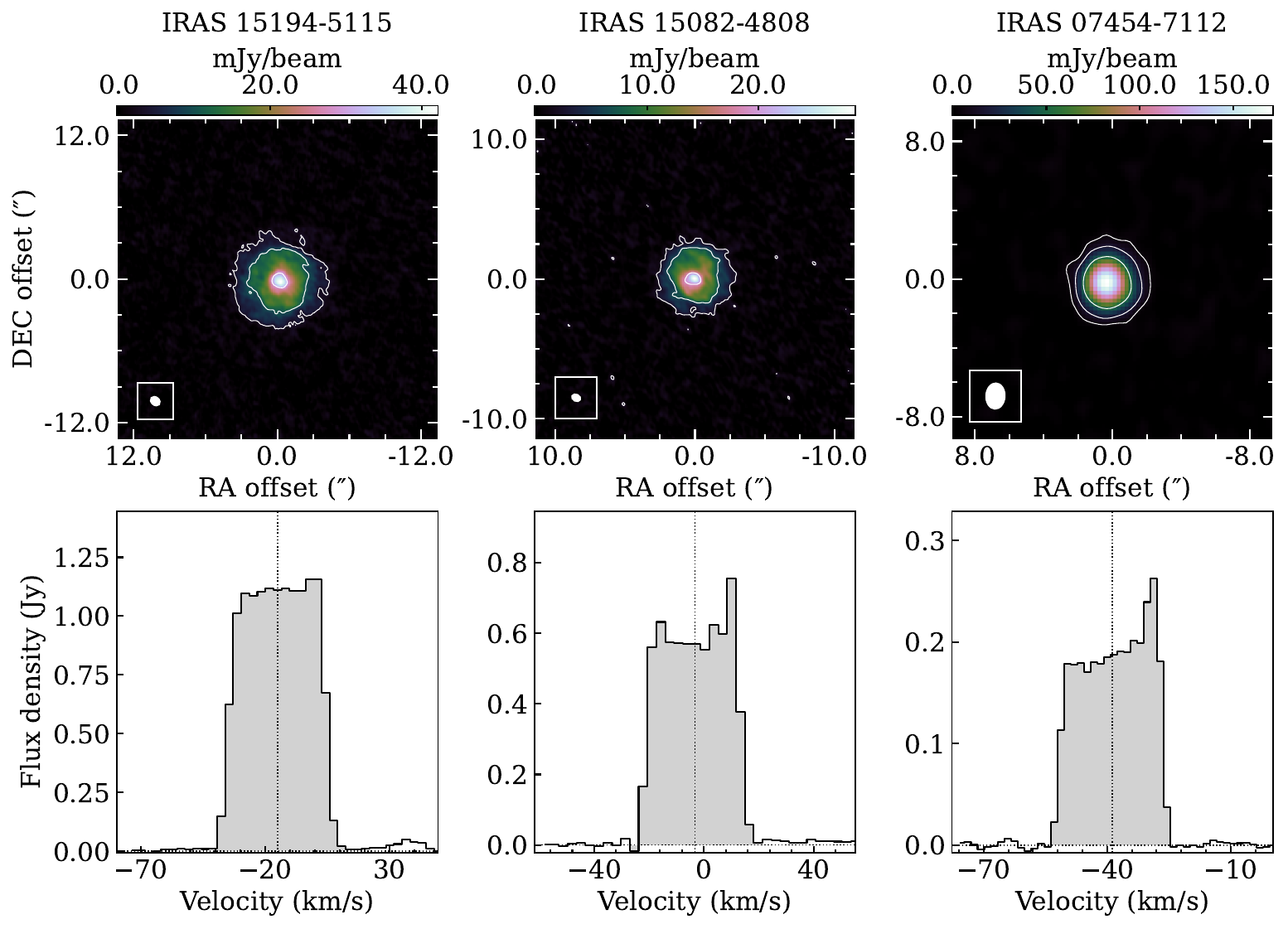}
    \caption{SiS, 5-4 (90.771564 GHz)}
    \label{fig:SiS_app_B}
\end{figure}

\begin{figure}[h]
    \centering
    \includegraphics[width=0.8\linewidth]{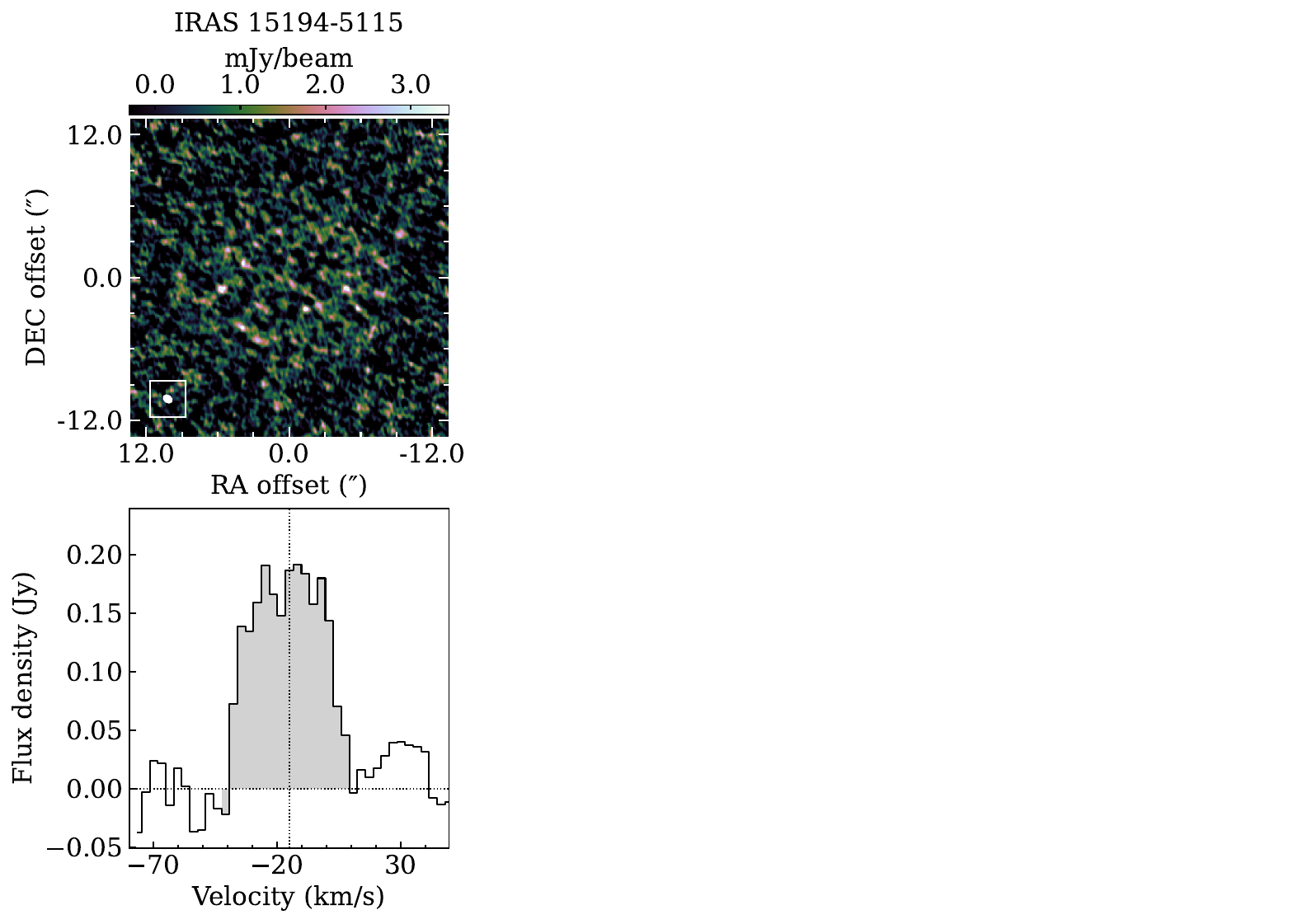}
    \caption{Si$^{13}$CC, 4$_{0,4}$-3$_{0,3}$ (90.848527 GHz)}
\end{figure}

\begin{figure}[h]
    \centering
    \includegraphics[width=0.8\linewidth]{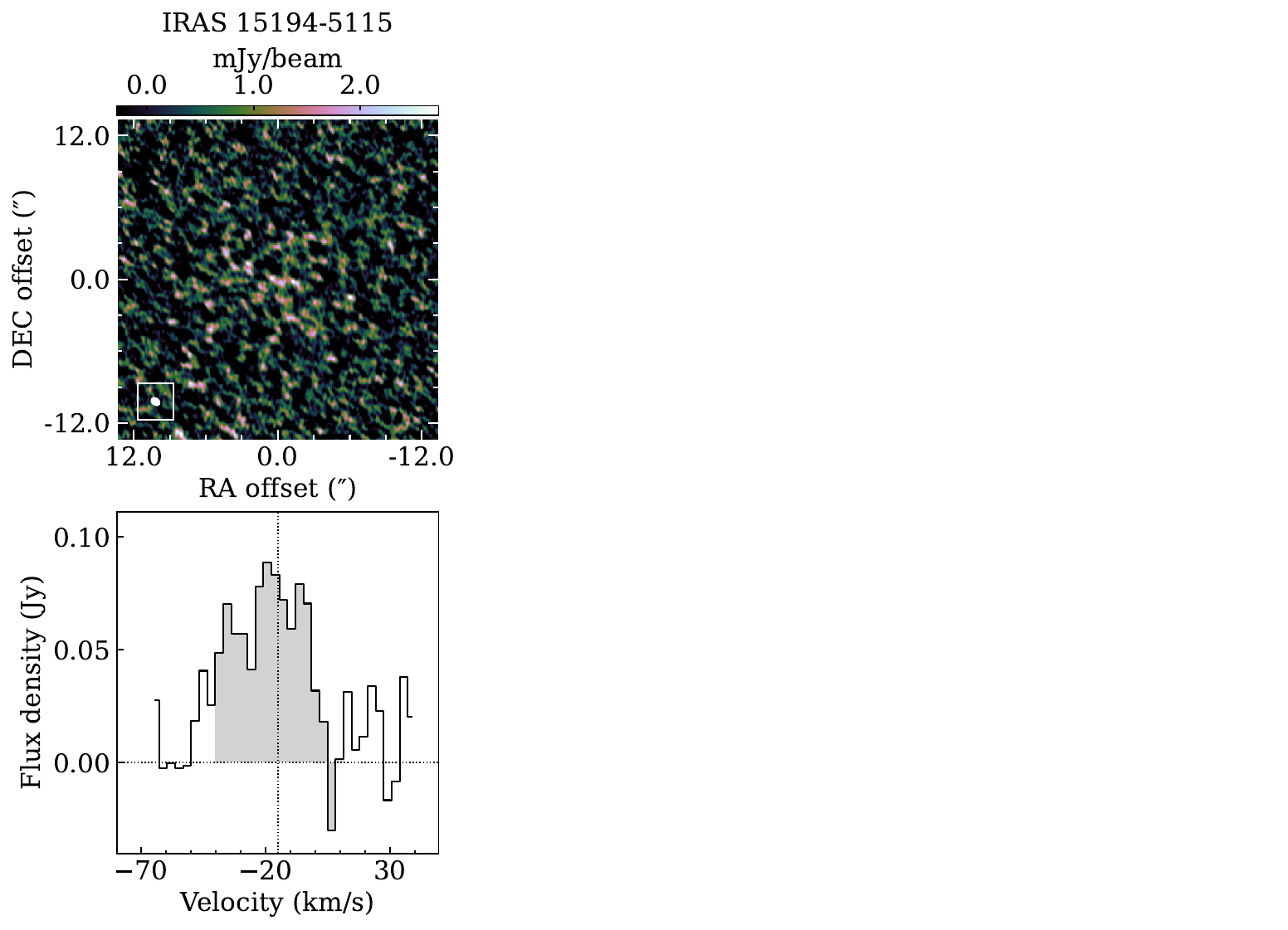}
    \caption{C$_8$H, J=155/2-153/2 (90.92464 GHz)}
\end{figure}

\begin{figure}[h]
    \centering
    \includegraphics[width=0.8\linewidth]{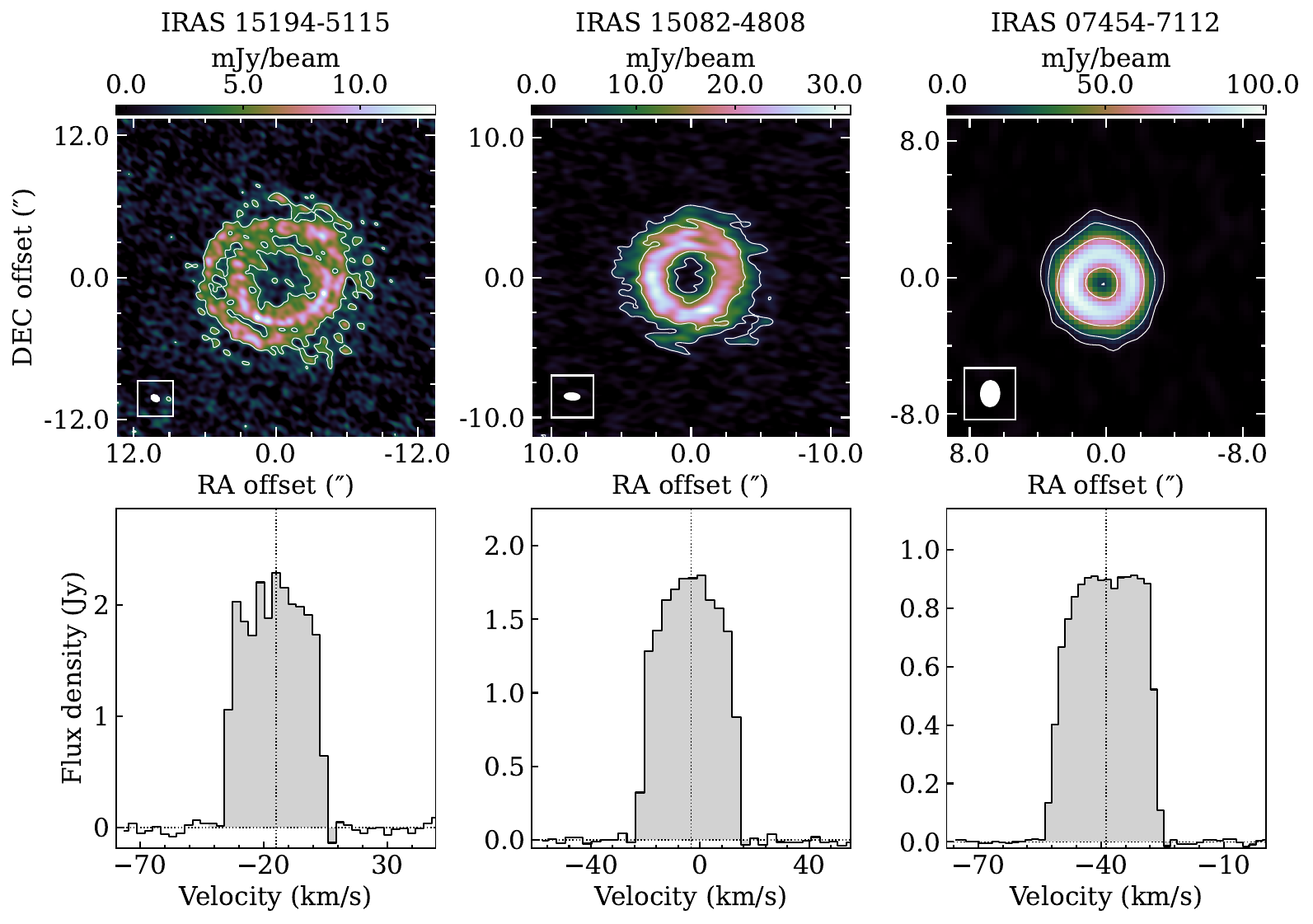}
    \caption{HC$_3$N, 10-9 (90.979023 GHz)}
\end{figure}

\begin{figure}[h]
    \centering
    \includegraphics[width=0.8\linewidth]{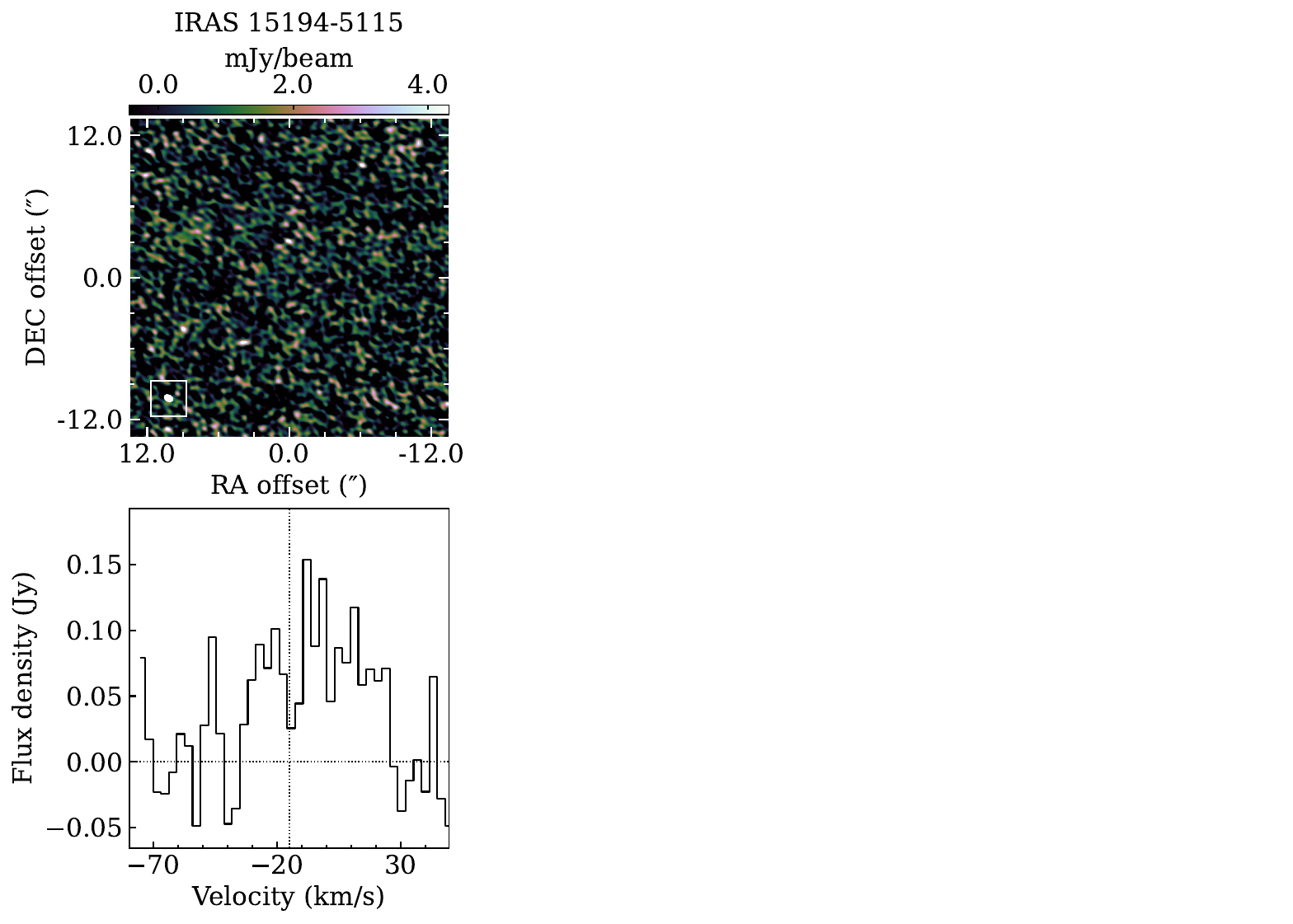}
    \caption{$^{13}$CCCCH, N=10-9, J=21/2-19/2 (91.875516 GHz)}
\end{figure}

\begin{figure}[h]
    \centering
    \includegraphics[width=0.8\linewidth]{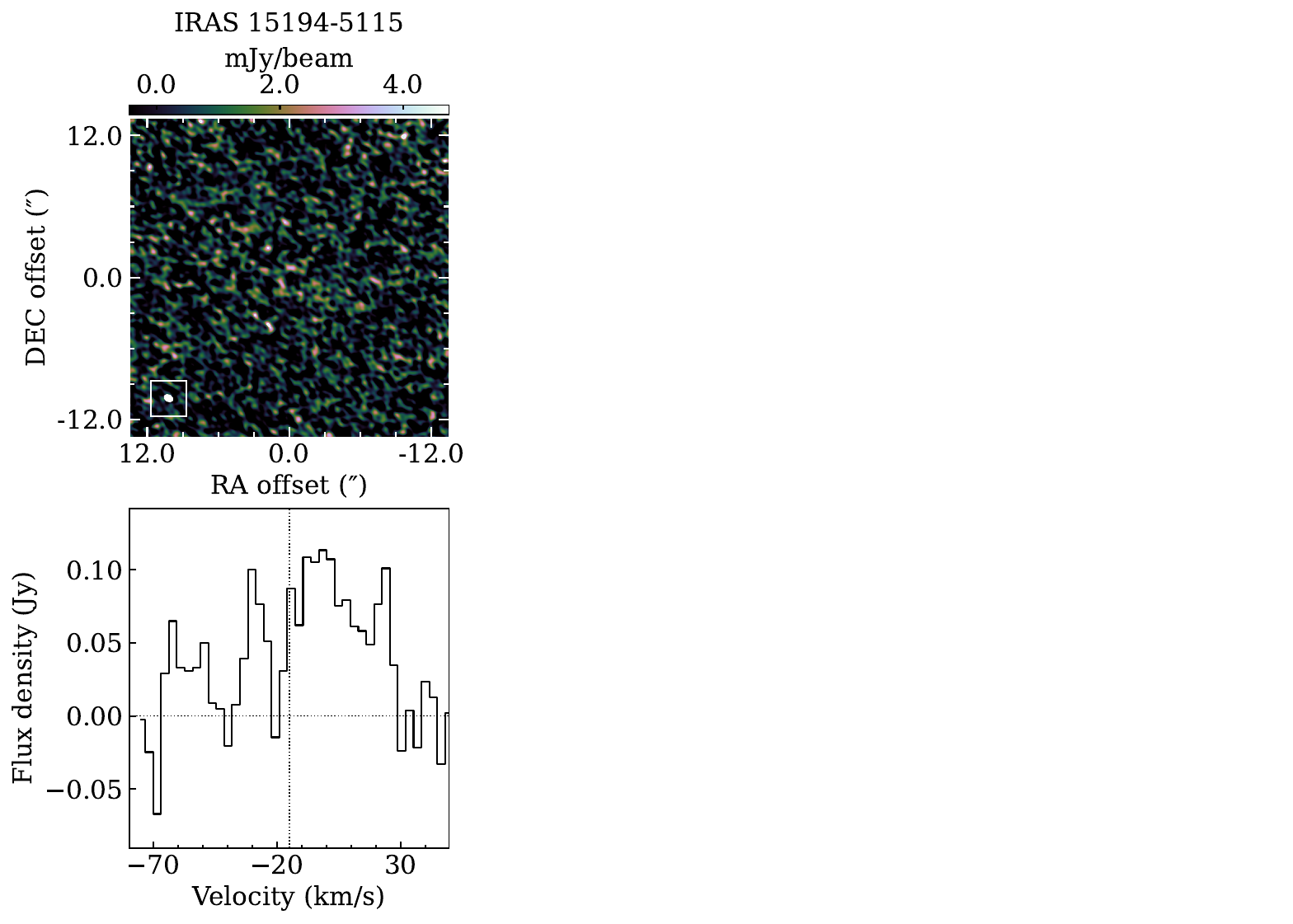}
    \caption{$^{13}$CCCCH, N=10-9, J=19/2-17/2 (91.906057 GHz)}
\end{figure}

\begin{figure}[h]
    \centering
    \includegraphics[width=0.8\linewidth]{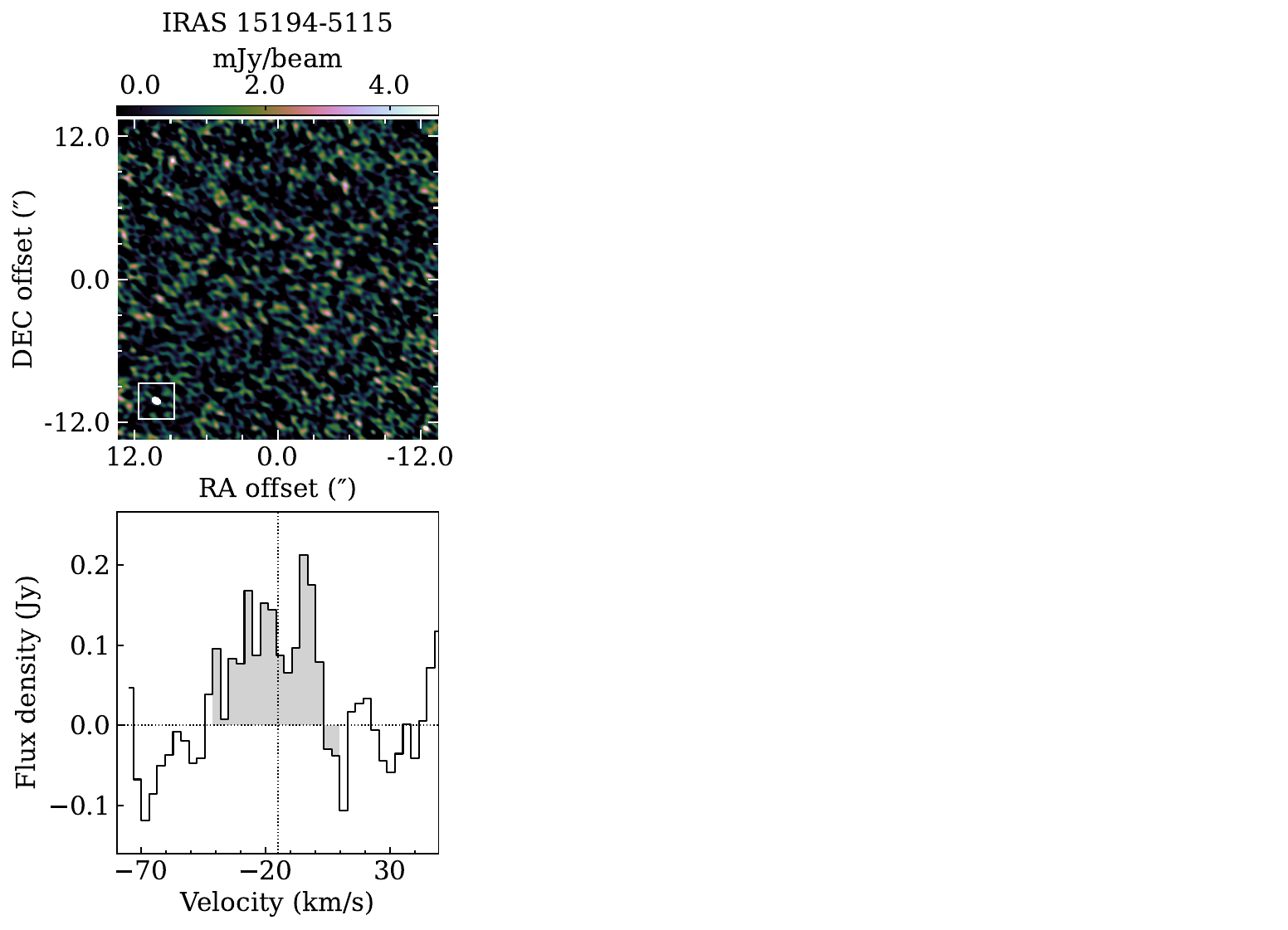}
    \caption{Si$^{13}$CC, 4$_{2,3}$-3$_{2,2}$ (92.064397 GHz)}
\end{figure}

\begin{figure}[h]
    \centering
    \includegraphics[width=0.8\linewidth]{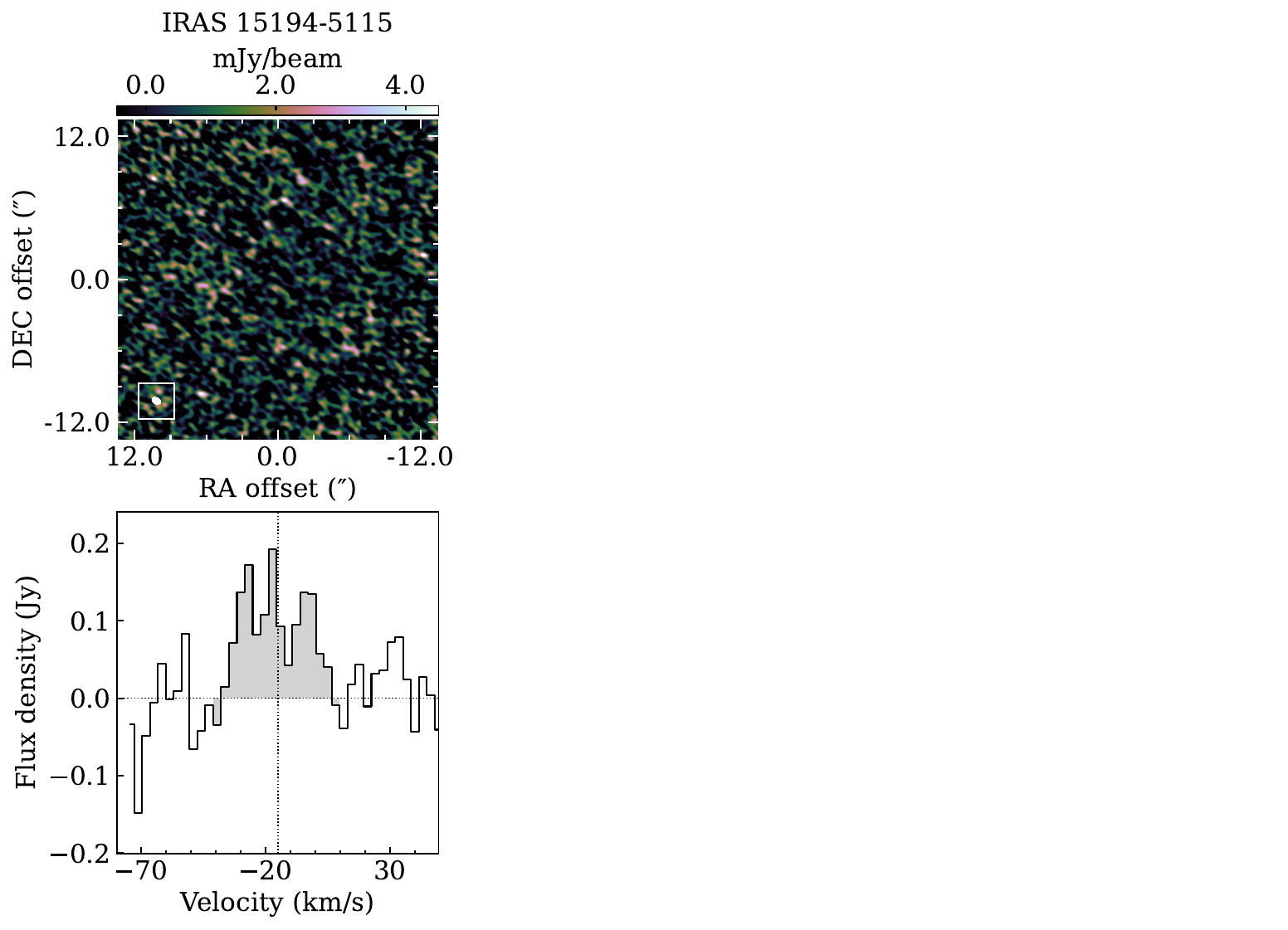}
    \caption{CCC$^{13}$CH, N=10-9, J=21/2-19/2 (92.277495 GHz)}
\end{figure}

\begin{figure}[h]
    \centering
    \includegraphics[width=0.8\linewidth]{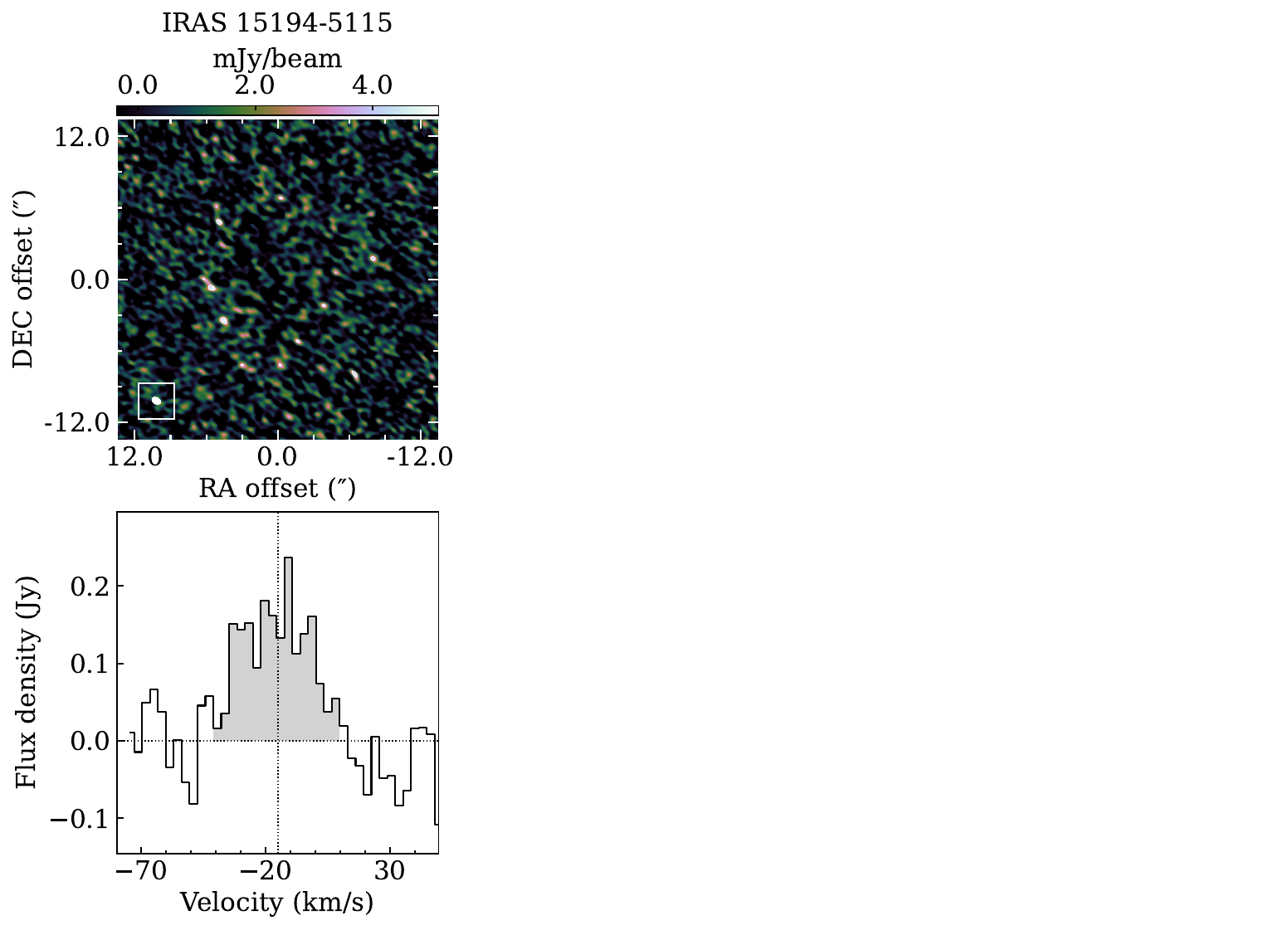}
    \caption{CCC$^{13}$CH, N=10-9, J=19/2-17/2 (92.31491 GHz)}
\end{figure}

\begin{figure}[h]
    \centering
    \includegraphics[width=0.8\linewidth]{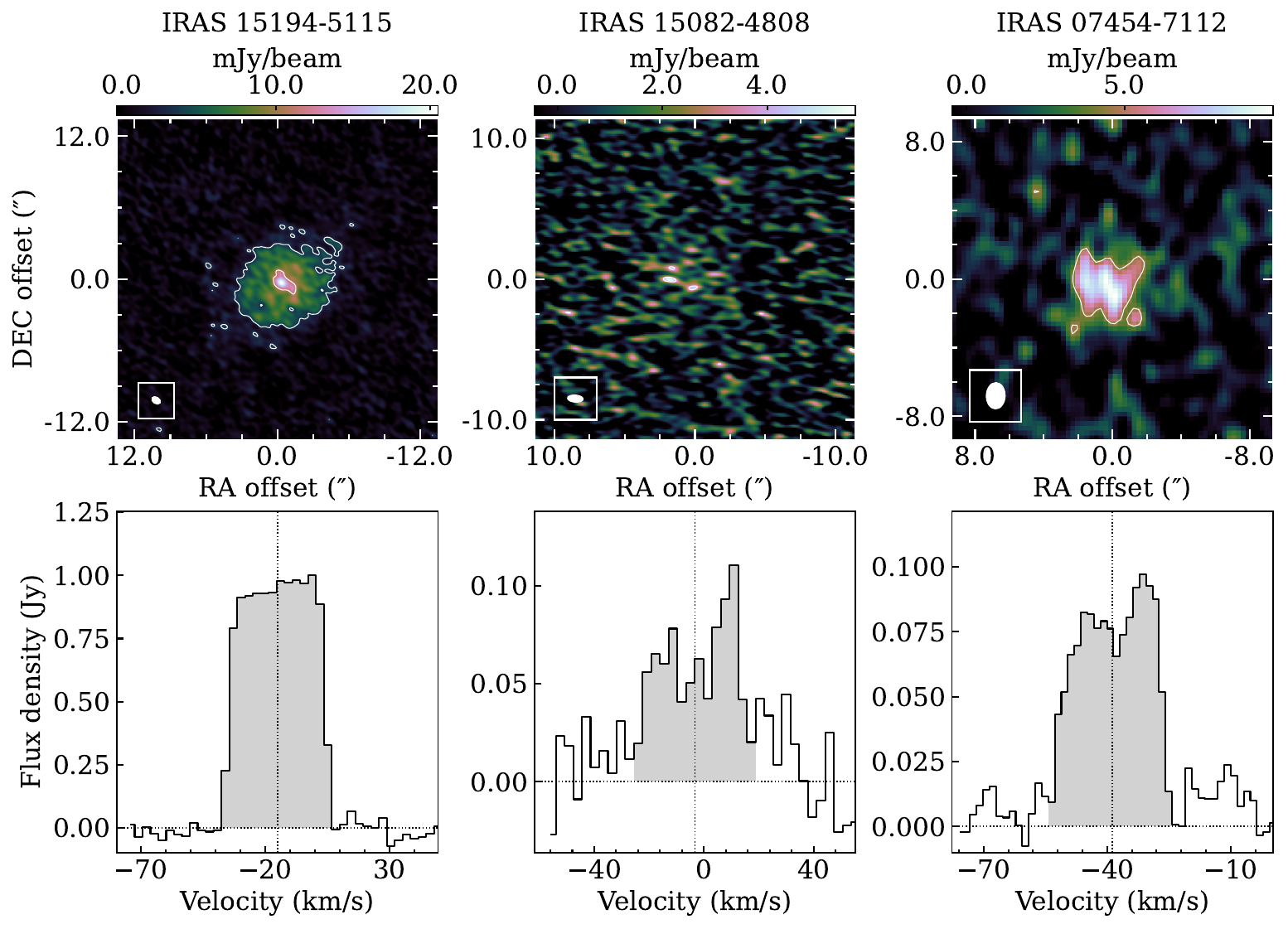}
    \caption{$^{13}$CS, 2-1 (92.494308 GHz)}
\end{figure}

\begin{figure}[h]
    \centering
    \includegraphics[width=0.8\linewidth]{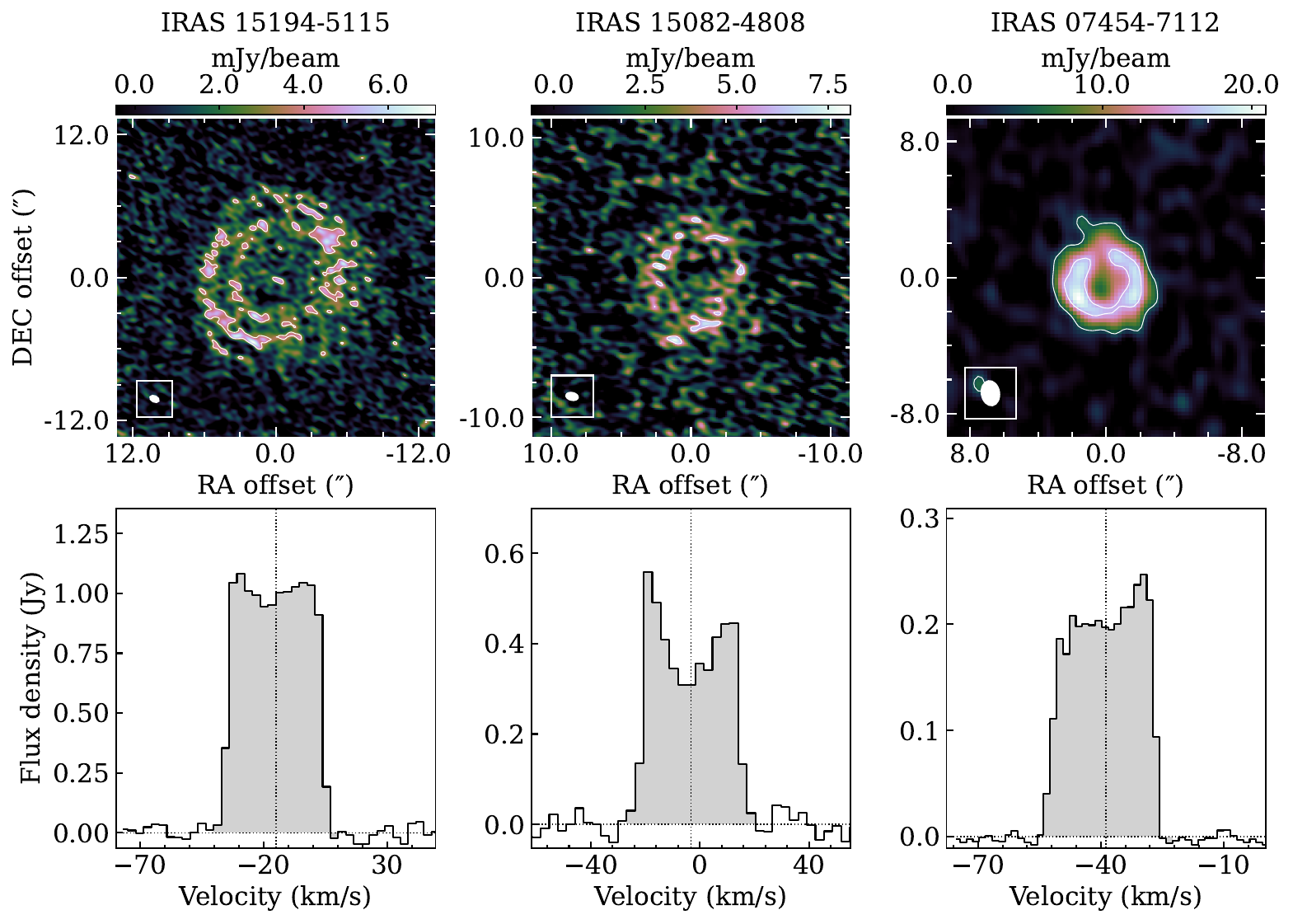}
    \caption{SiC$_2$, 4$_{0,4}$-3$_{0,3}$ (93.063639 GHz)}
    \label{fig:SiC2_app_B}
\end{figure}

\begin{figure}[h]
    \centering
    \includegraphics[width=0.8\linewidth]{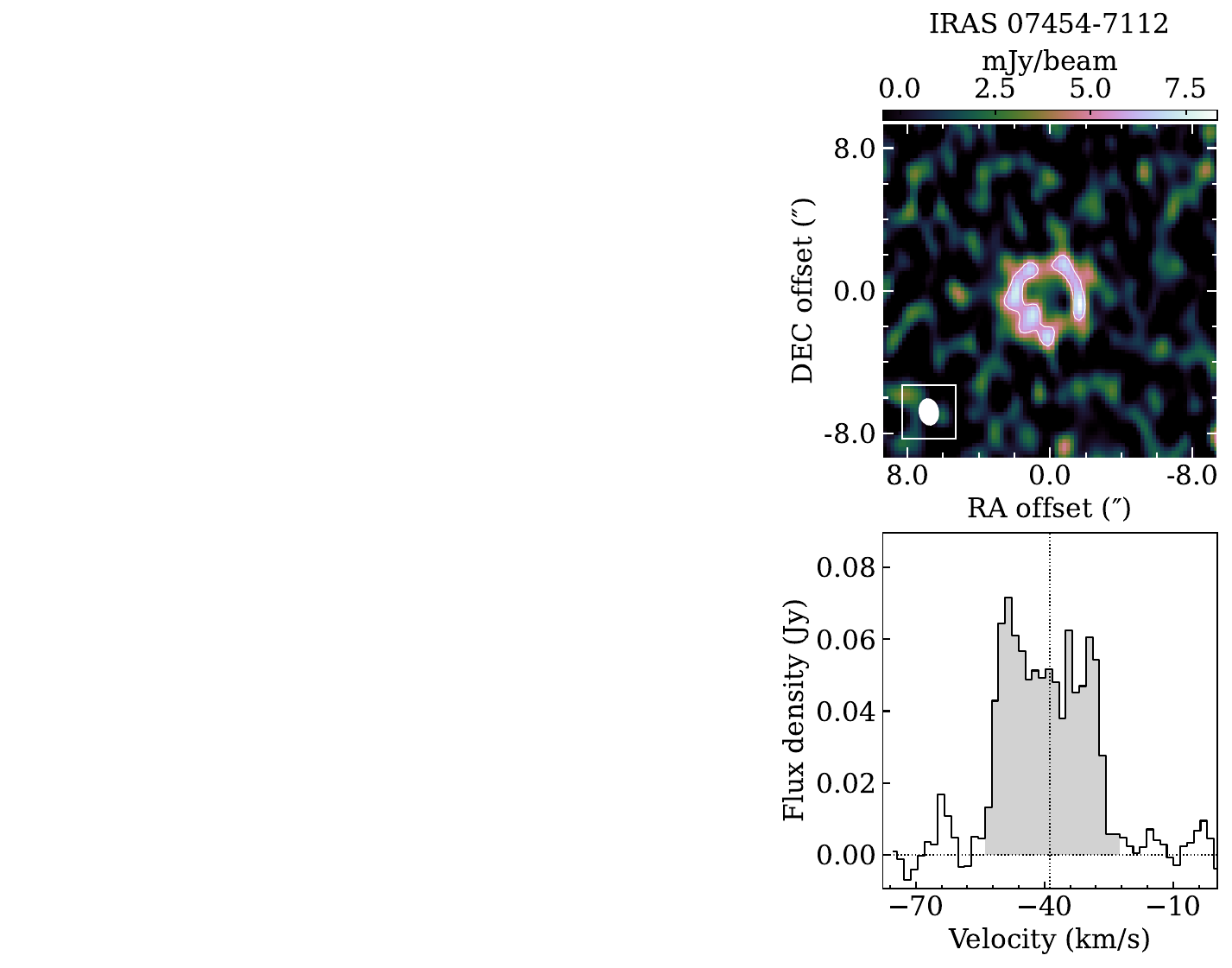}
    \caption{HC$_5$N, 35-34 (93.188123 GHz)}
\end{figure}

\begin{figure}[h]
    \centering
    \includegraphics[width=0.8\linewidth]{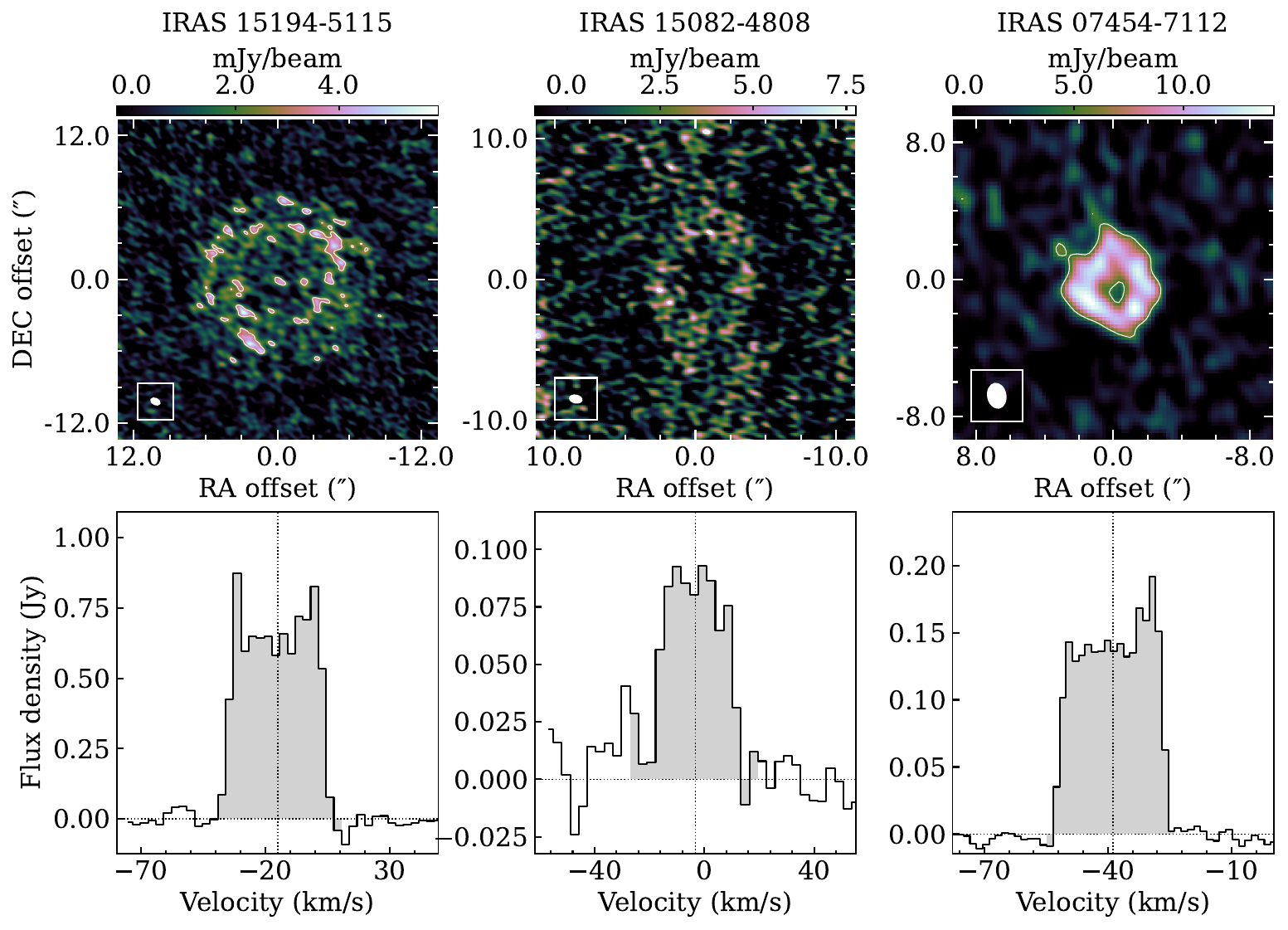}
    \caption{SiC$_2$, 4$_{2,3}$-3$_{2,2}$ (94.245393 GHz)}
\end{figure}

\begin{figure}[h]
    \centering
    \includegraphics[width=0.8\linewidth]{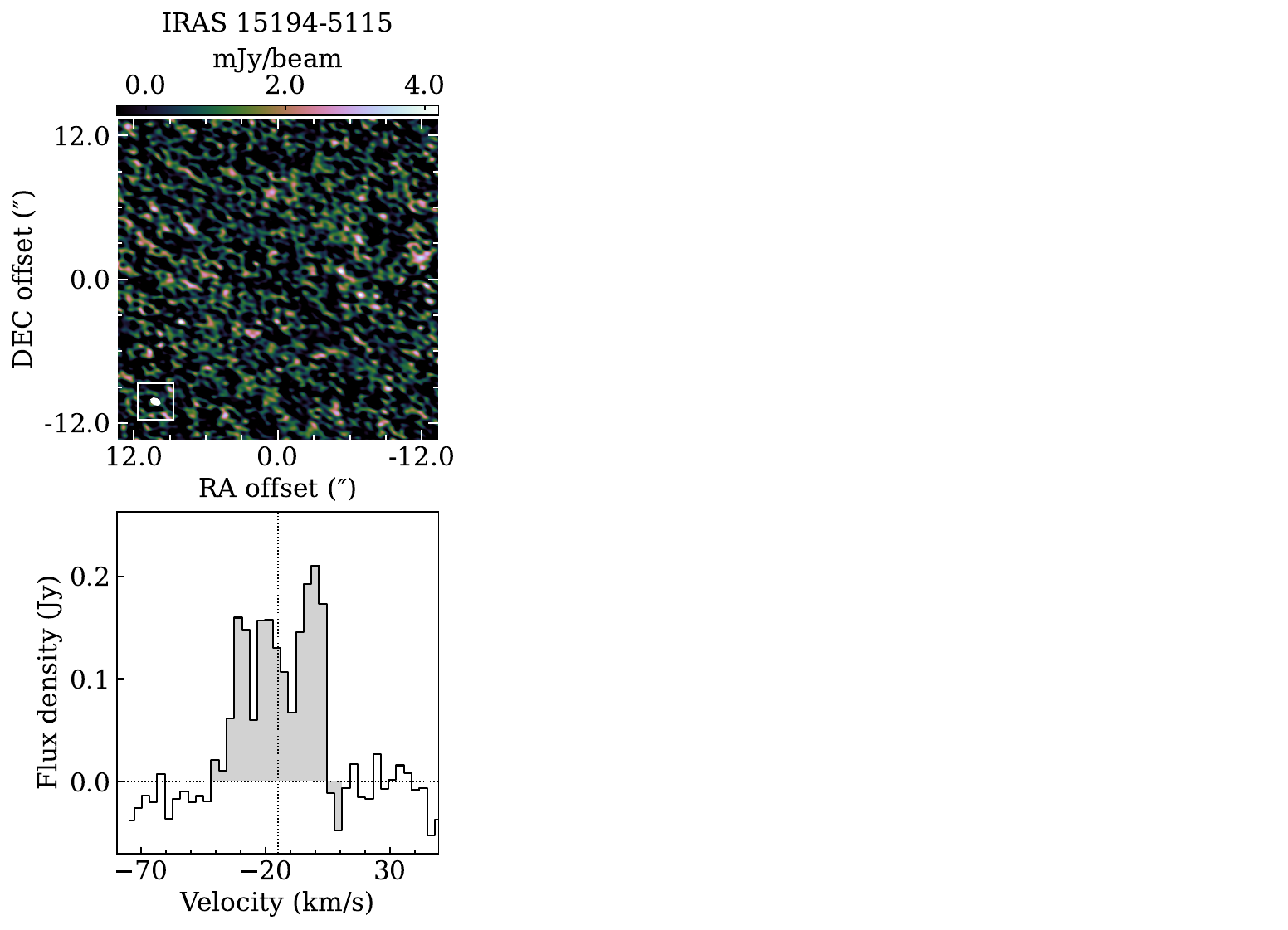}
    \caption{C$^{13}$CCCH, N=10-9, J=21/2-19/2 (94.670583 GHz)}
\end{figure}

\begin{figure}[h]
    \centering
    \includegraphics[width=0.8\linewidth]{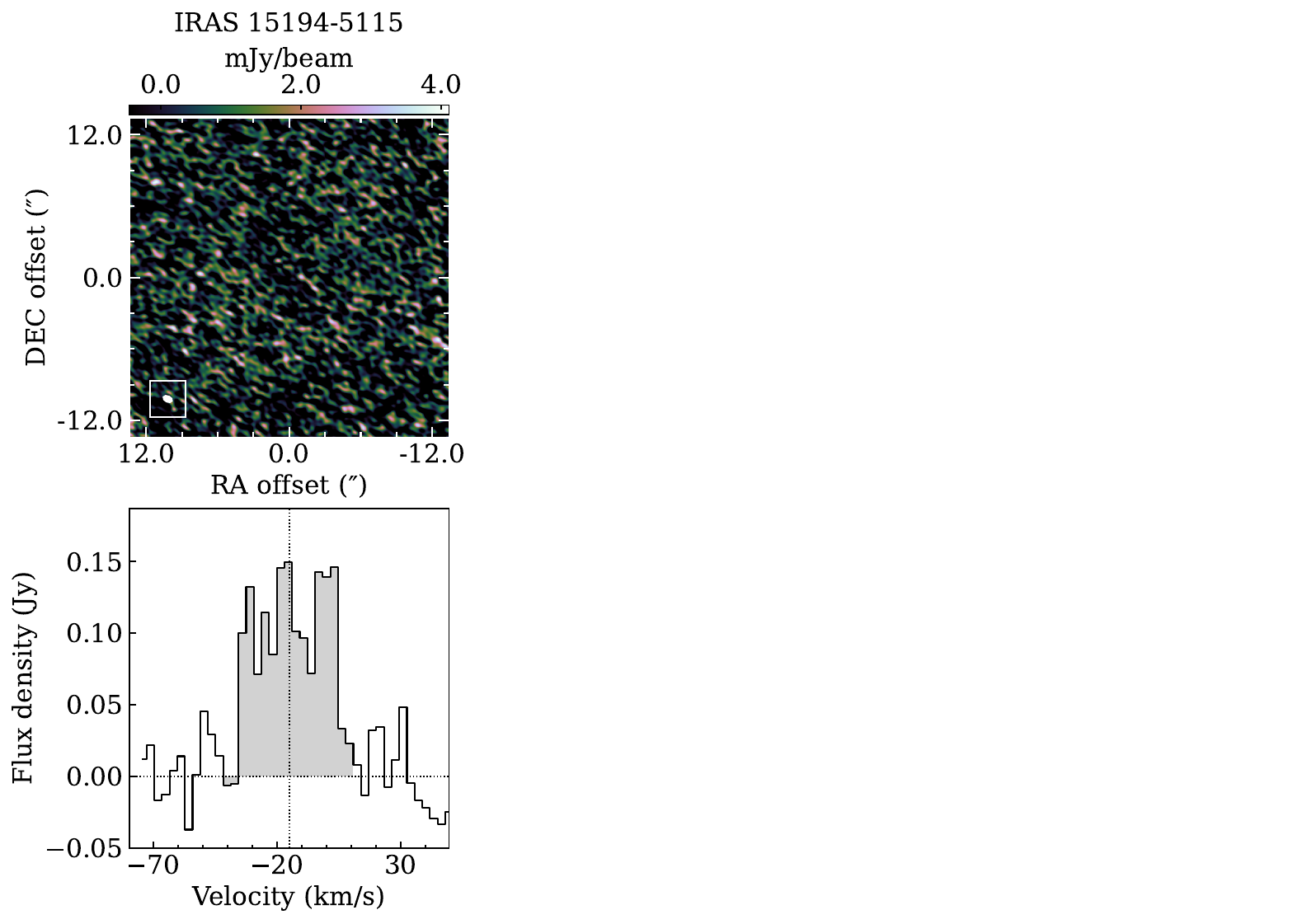}
    \caption{C$^{13}$CCCH, N=10-9, J=19/2-17/2 (94.708235 GHz)}
\end{figure}

\begin{figure}[h]
    \centering
    \includegraphics[width=0.8\linewidth]{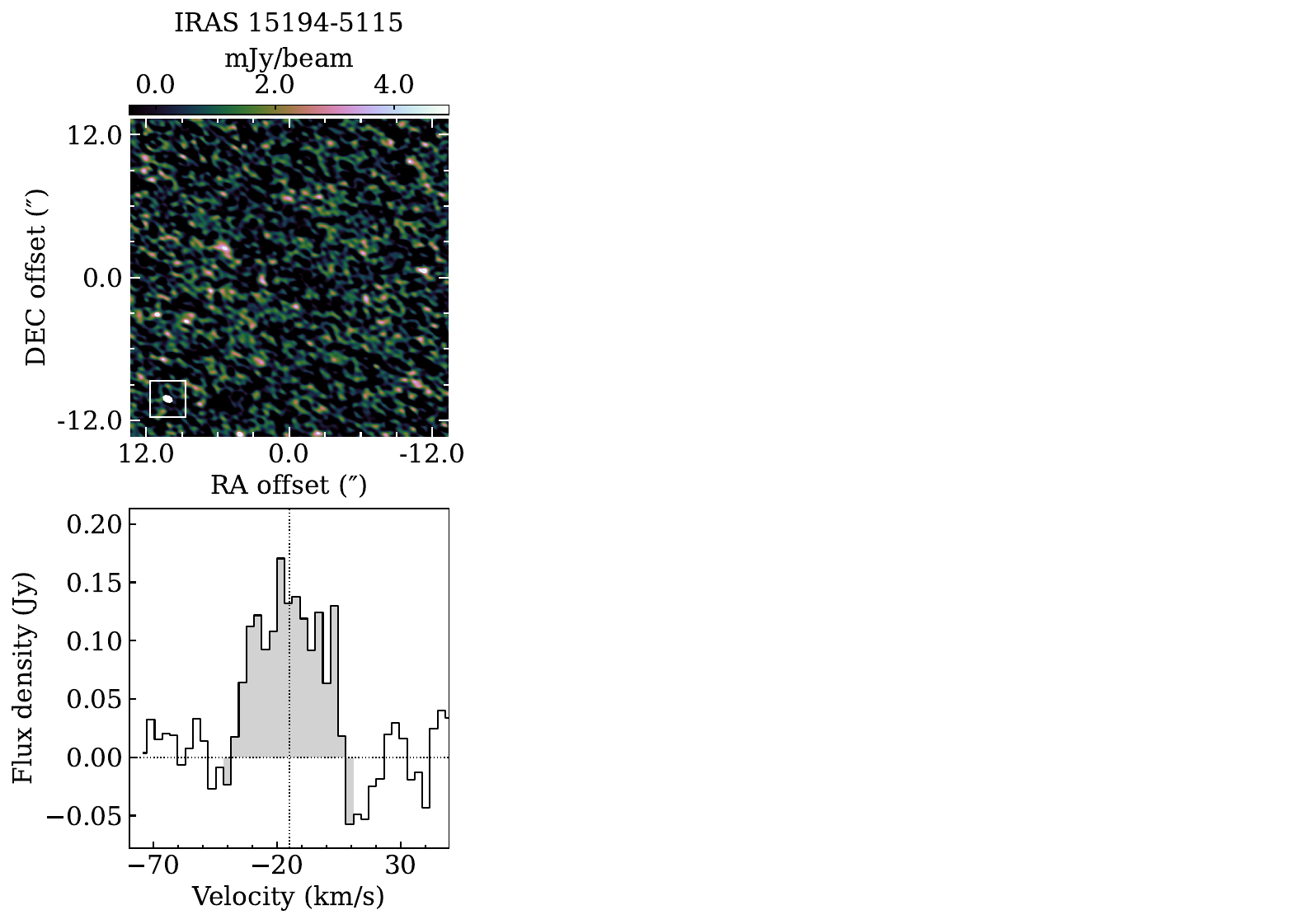}
    \caption{CC$^{13}$CCH, N=10-9, J=21/2-19/2 (94.814724 GHz)}
\end{figure}

\begin{figure}[h]
    \centering
    \includegraphics[width=0.8\linewidth]{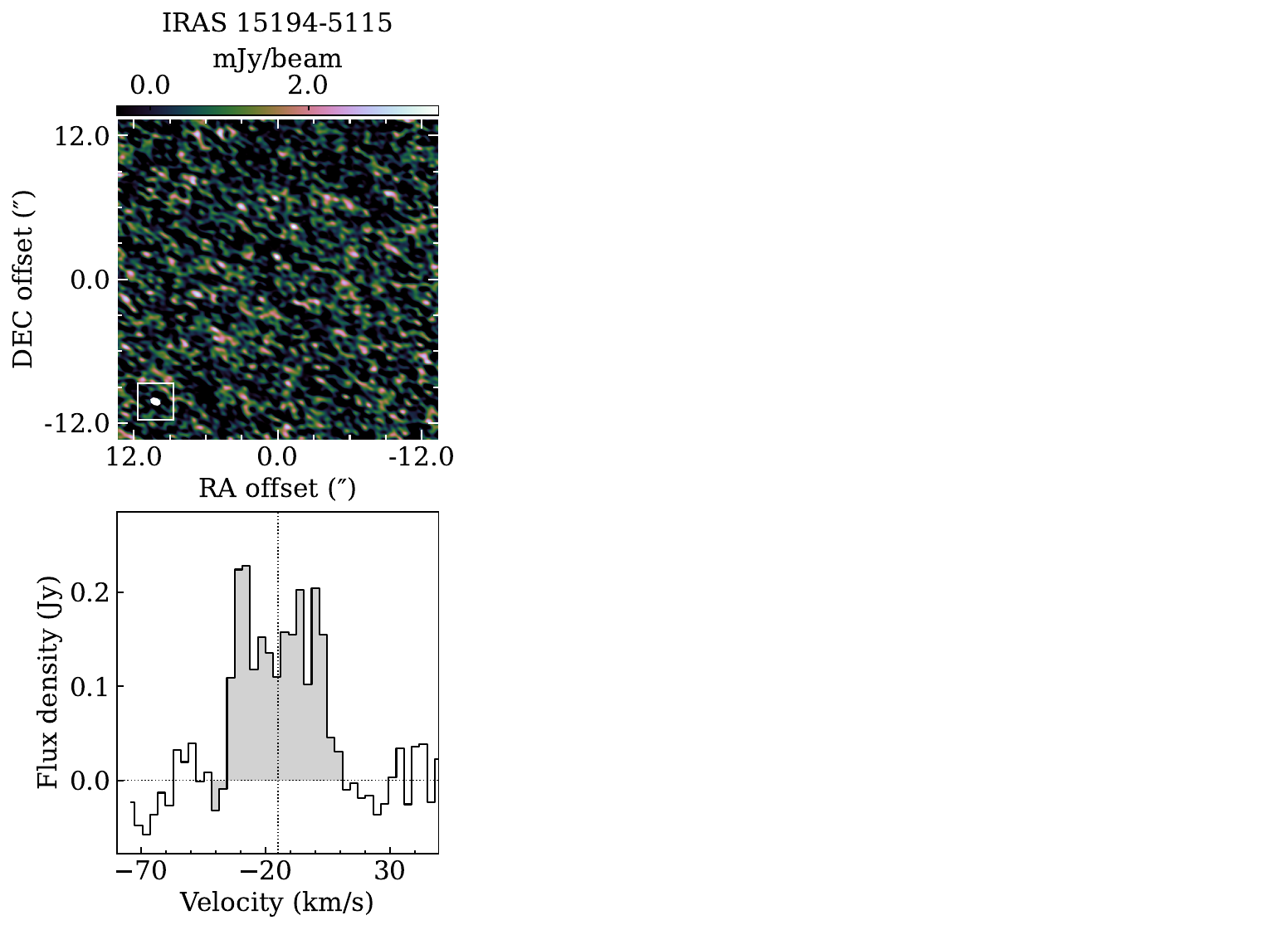}
    \caption{CC$^{13}$CCH, N=10-9, J=19/2-17/2 (94.853126 GHz)}
\end{figure}

\begin{figure}[h]
    \centering
    \includegraphics[width=0.8\linewidth]{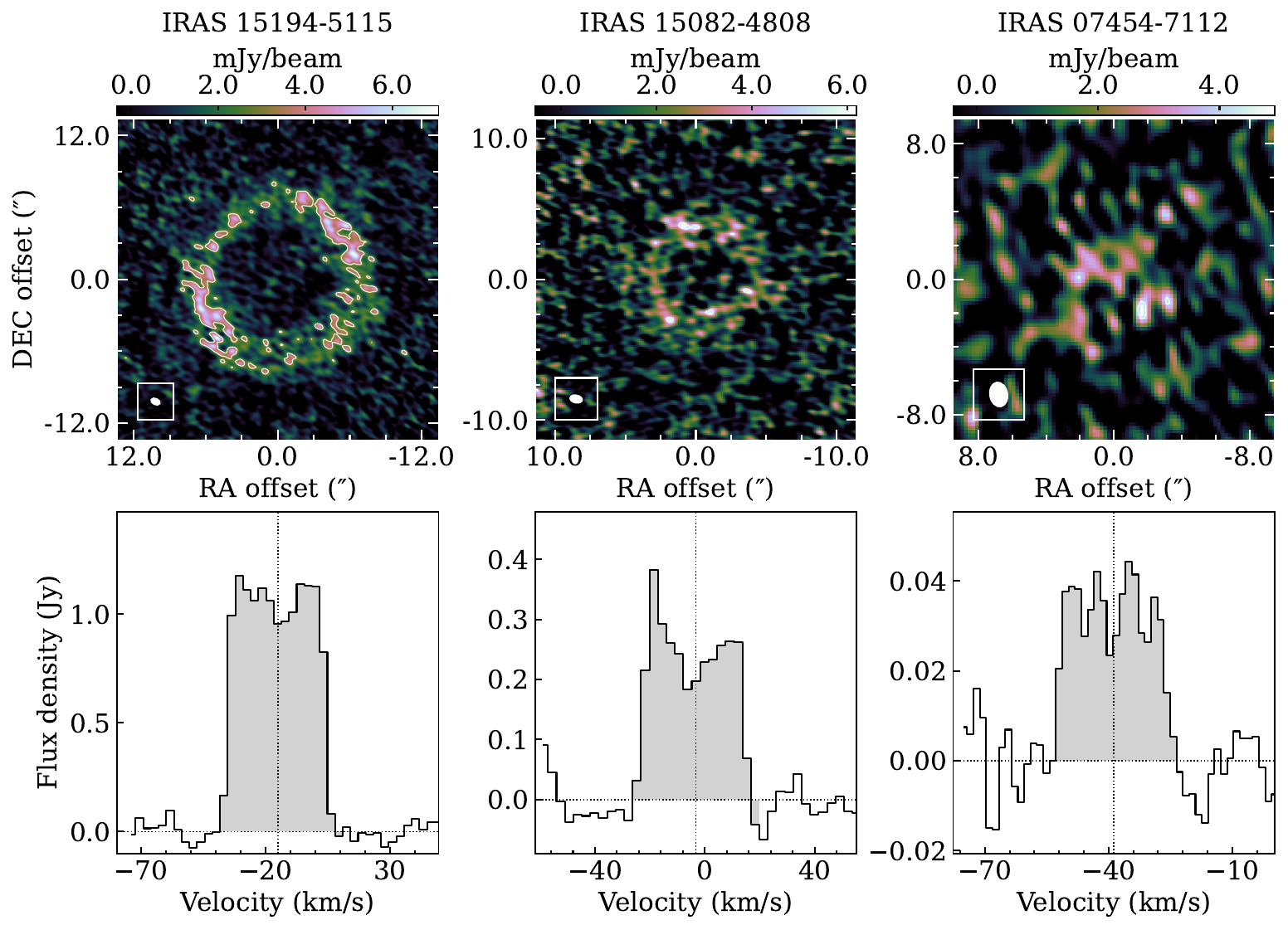}
    \caption{C$_4$H, N=10-9, J=21/2-19/2 (95.150389 GHz)}
\end{figure}

\begin{figure}[h]
    \centering
    \includegraphics[width=0.8\linewidth]{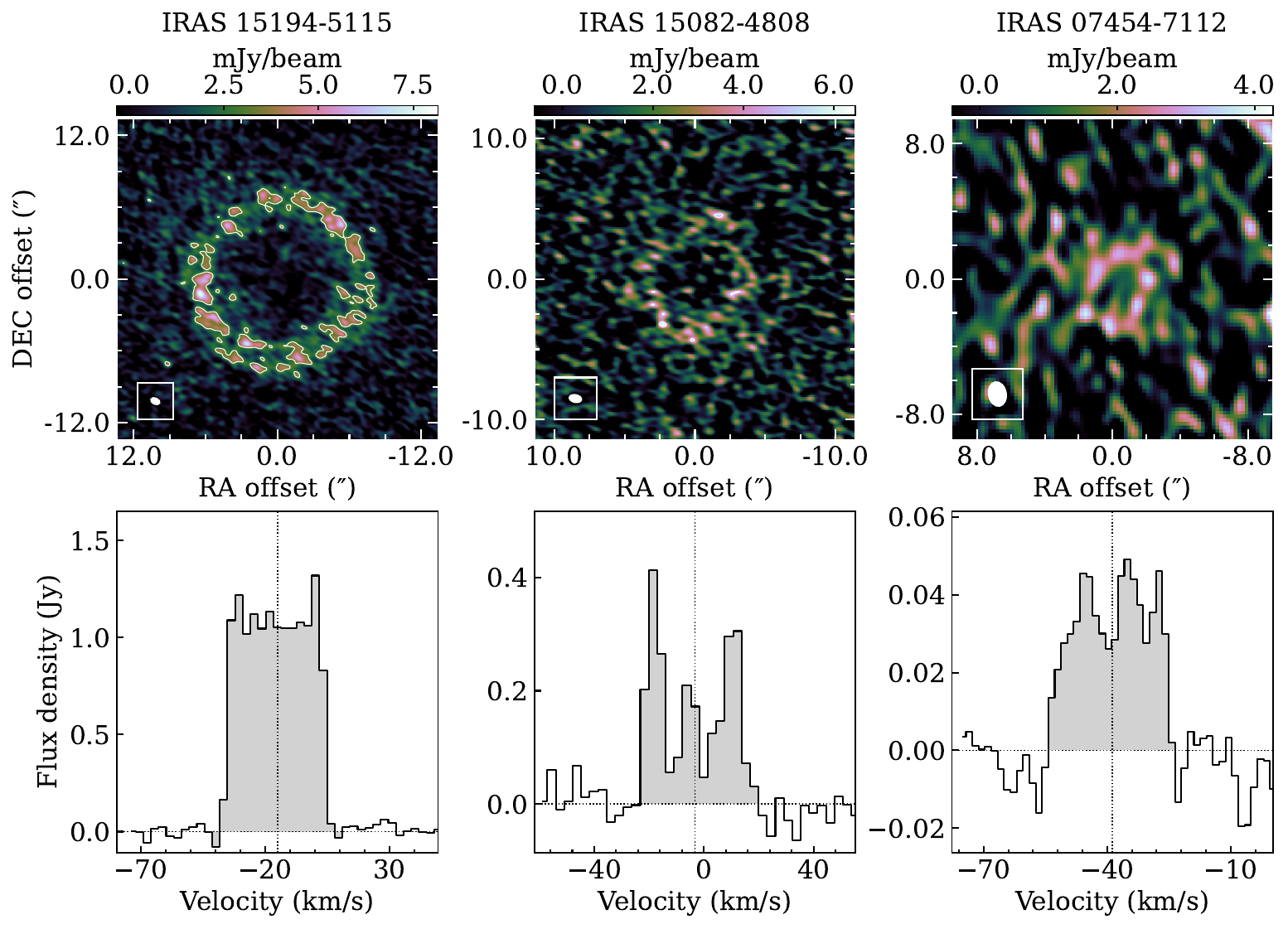}
    \caption{C$_4$H, N=10-9, J=19/2-17/2 (95.188949 GHz)}
    \label{fig:C4H_app_B}
\end{figure}

\begin{figure}[h]
    \centering
    \includegraphics[width=0.8\linewidth]{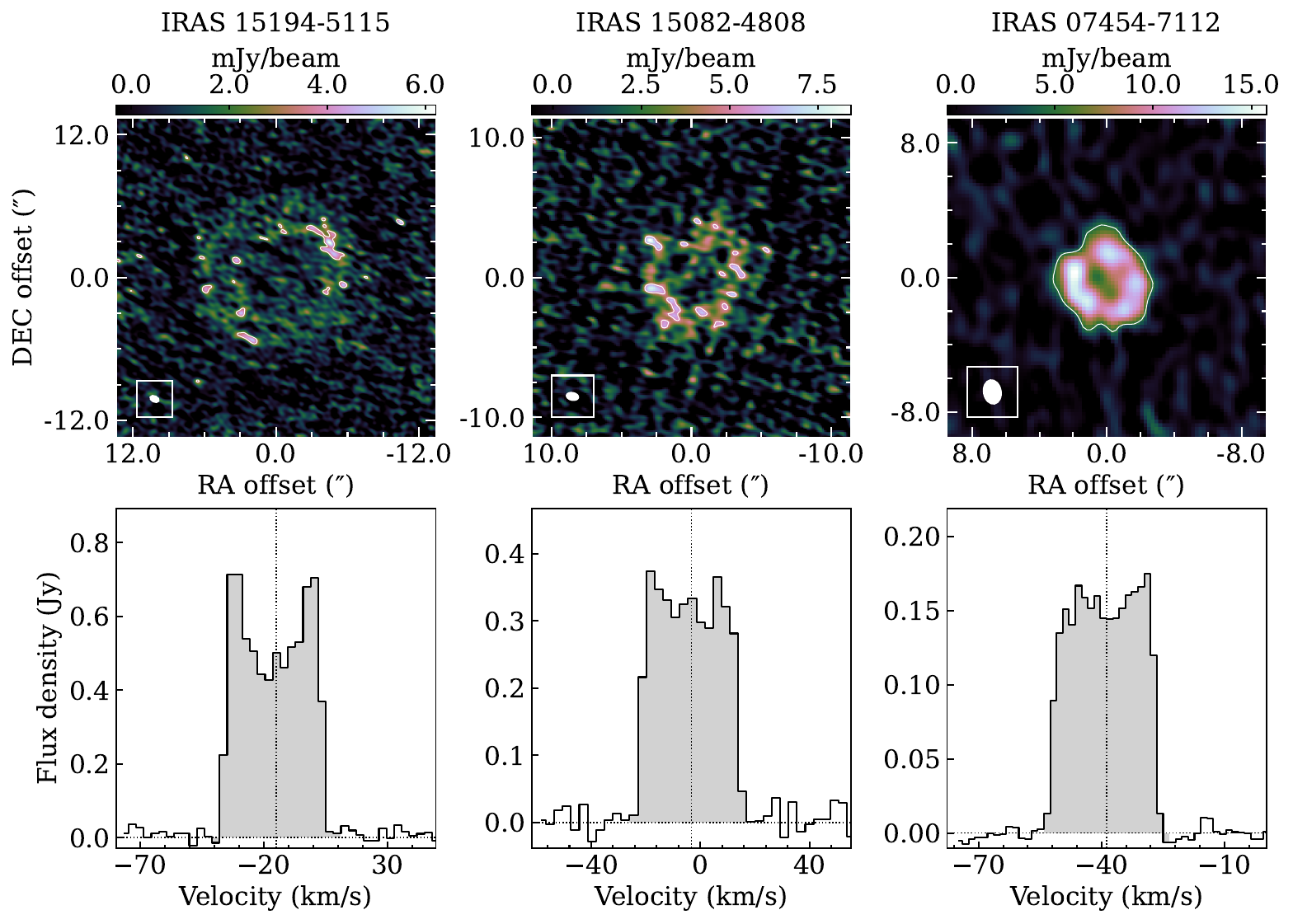}
    \caption{SiC$_2$, 4$_{2,2}$-3$_{2,1}$ (95.579381 GHz)}
\end{figure}

\begin{figure}[h]
    \centering
    \includegraphics[width=0.8\linewidth]{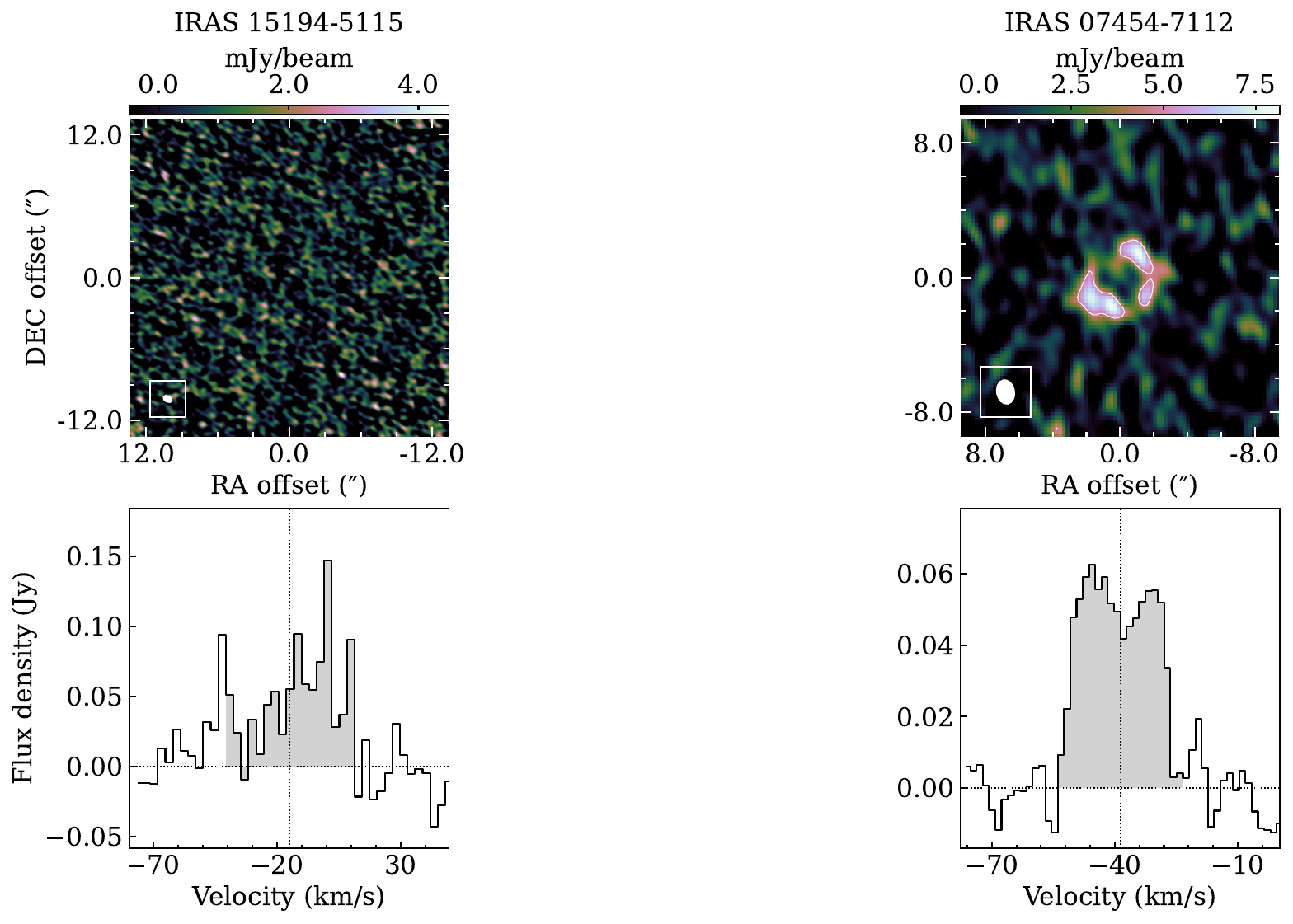}
    \caption{HC$_5$N, 36-35 (95.850335 GHz)}
\end{figure}

\begin{figure}[h]
    \centering
    \includegraphics[width=0.8\linewidth]{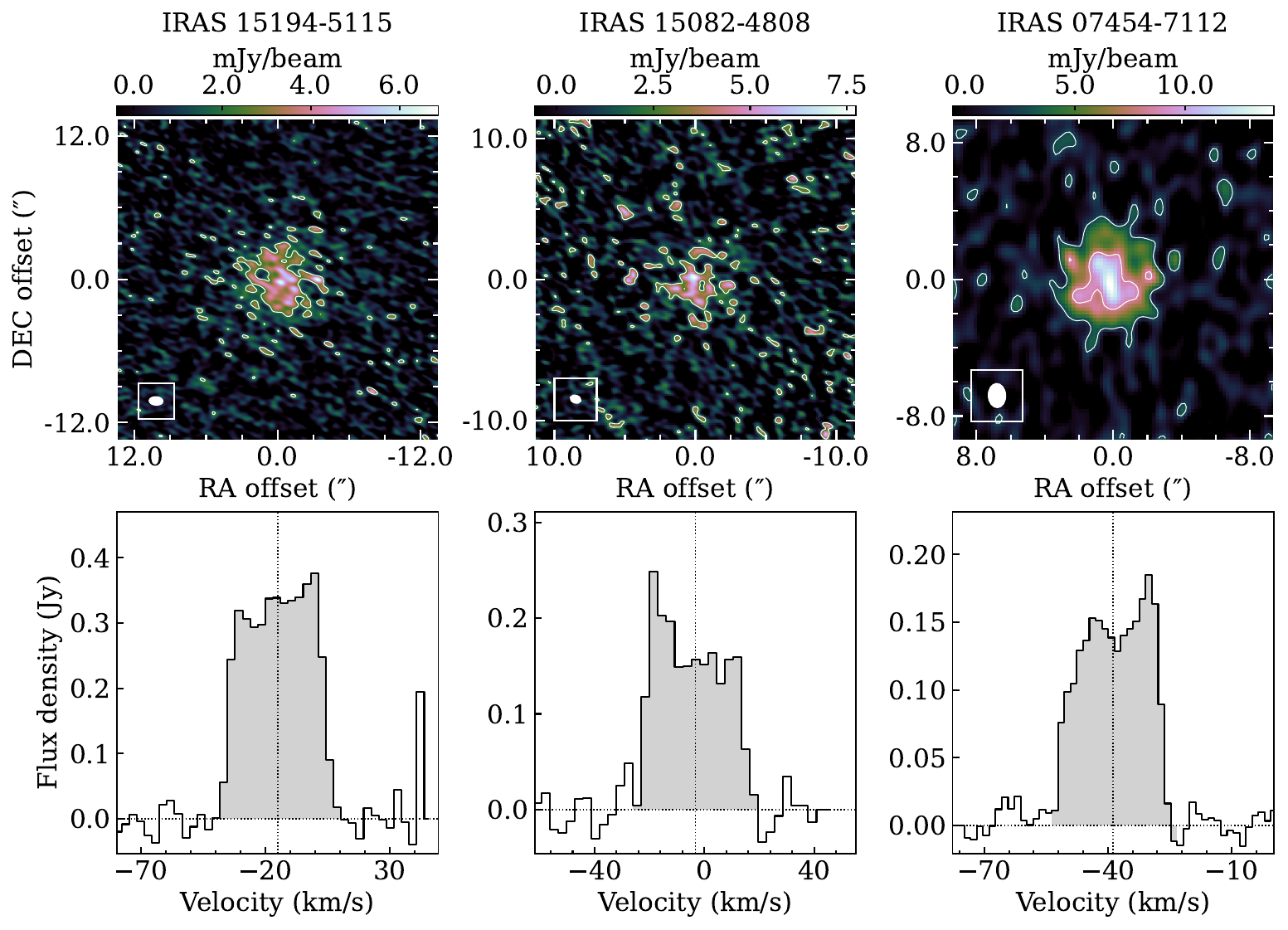}
    \caption{C$^{34}$S, 2-1 (96.41295 GHz)}
\end{figure}

\begin{figure}[h]
    \centering
    \includegraphics[width=0.8\linewidth]{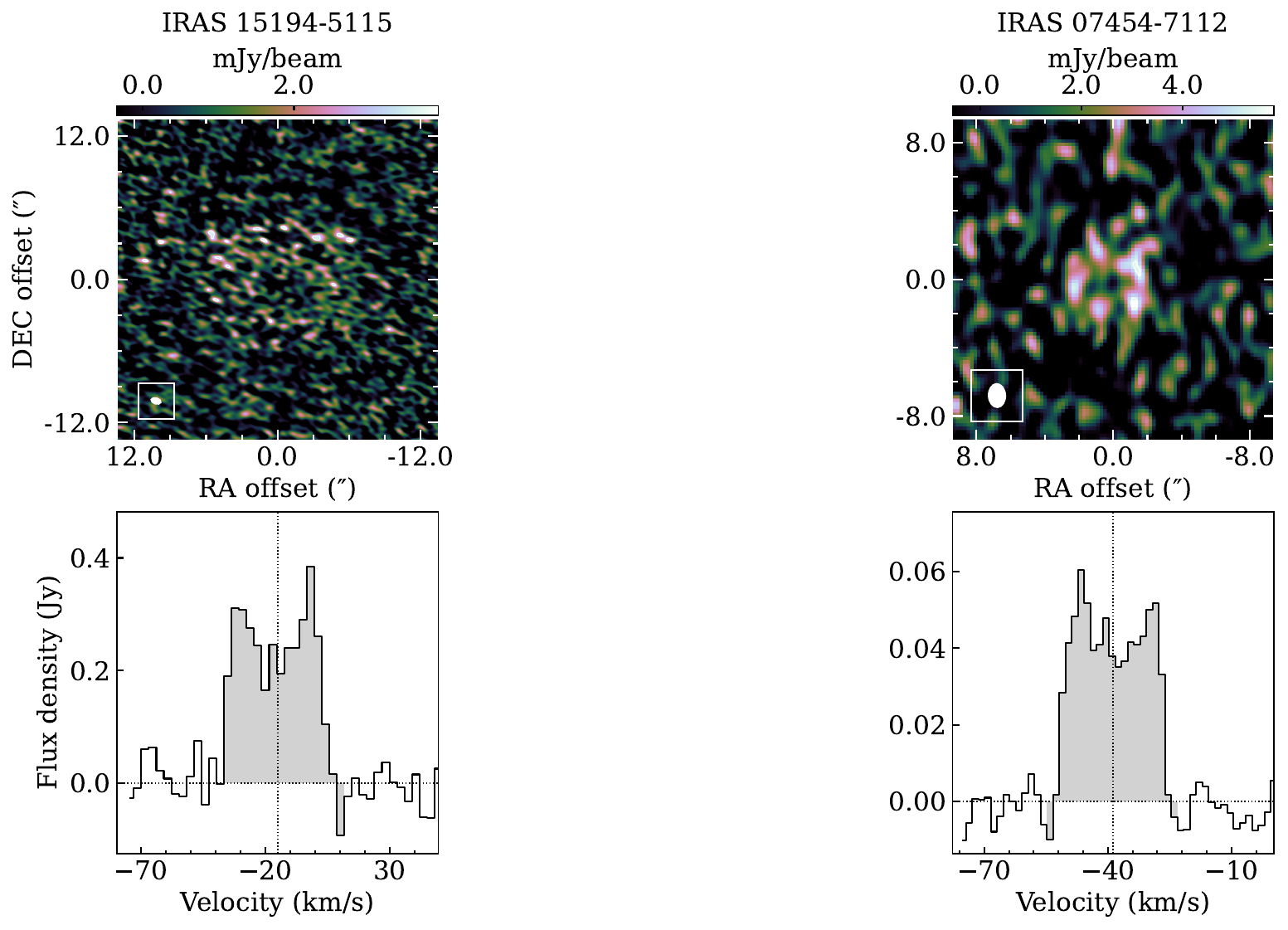}
    \caption{H$^{13}$CCCN, 11-10 (96.983001 GHz)}
\end{figure}

\begin{figure}[h]
    \centering
    \includegraphics[width=0.8\linewidth]{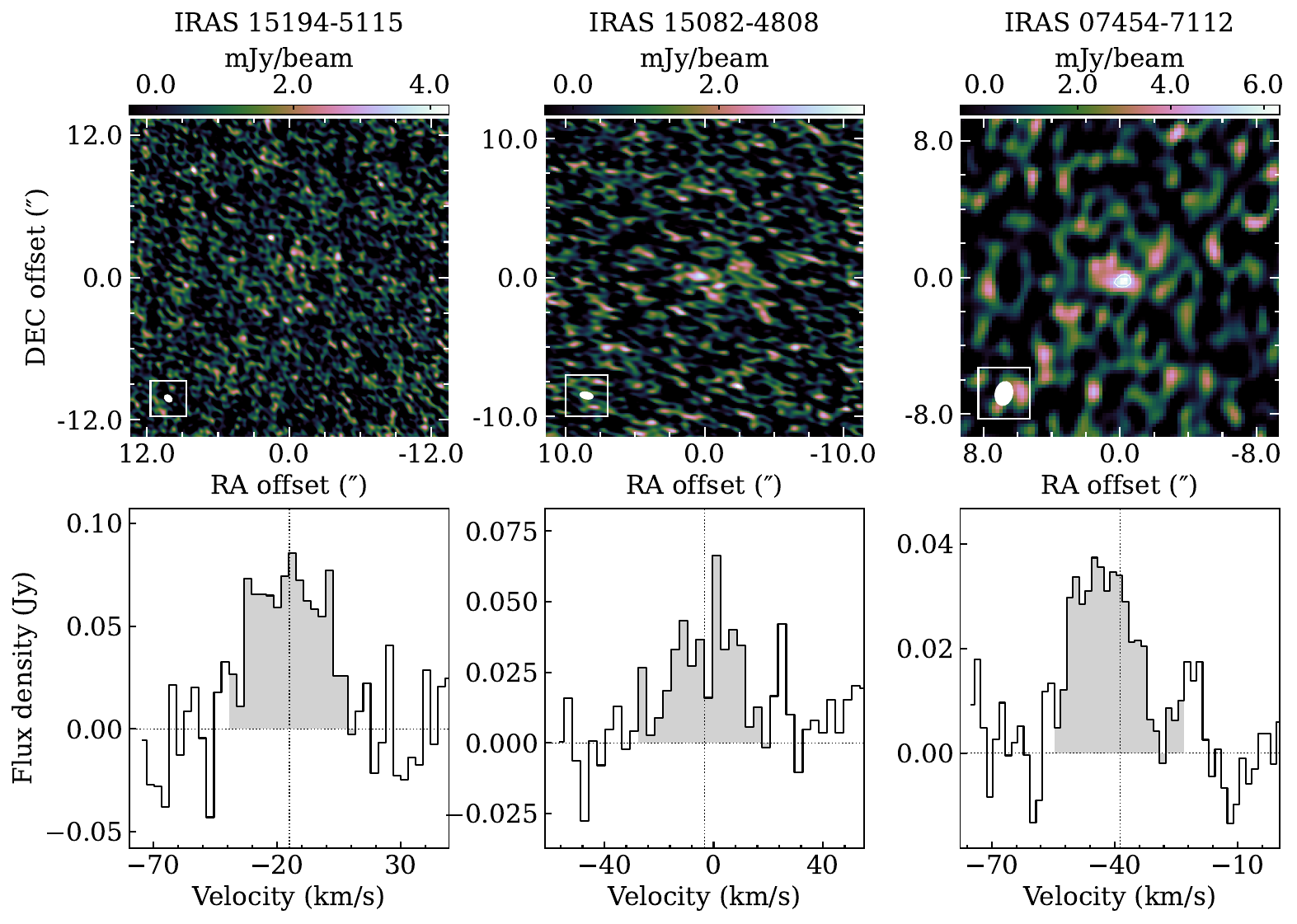}
    \caption{C$^{33}$S, 2-1 (97.172064 GHz)}
\end{figure}

\begin{figure}[h]
    \centering
    \includegraphics[width=0.8\linewidth]{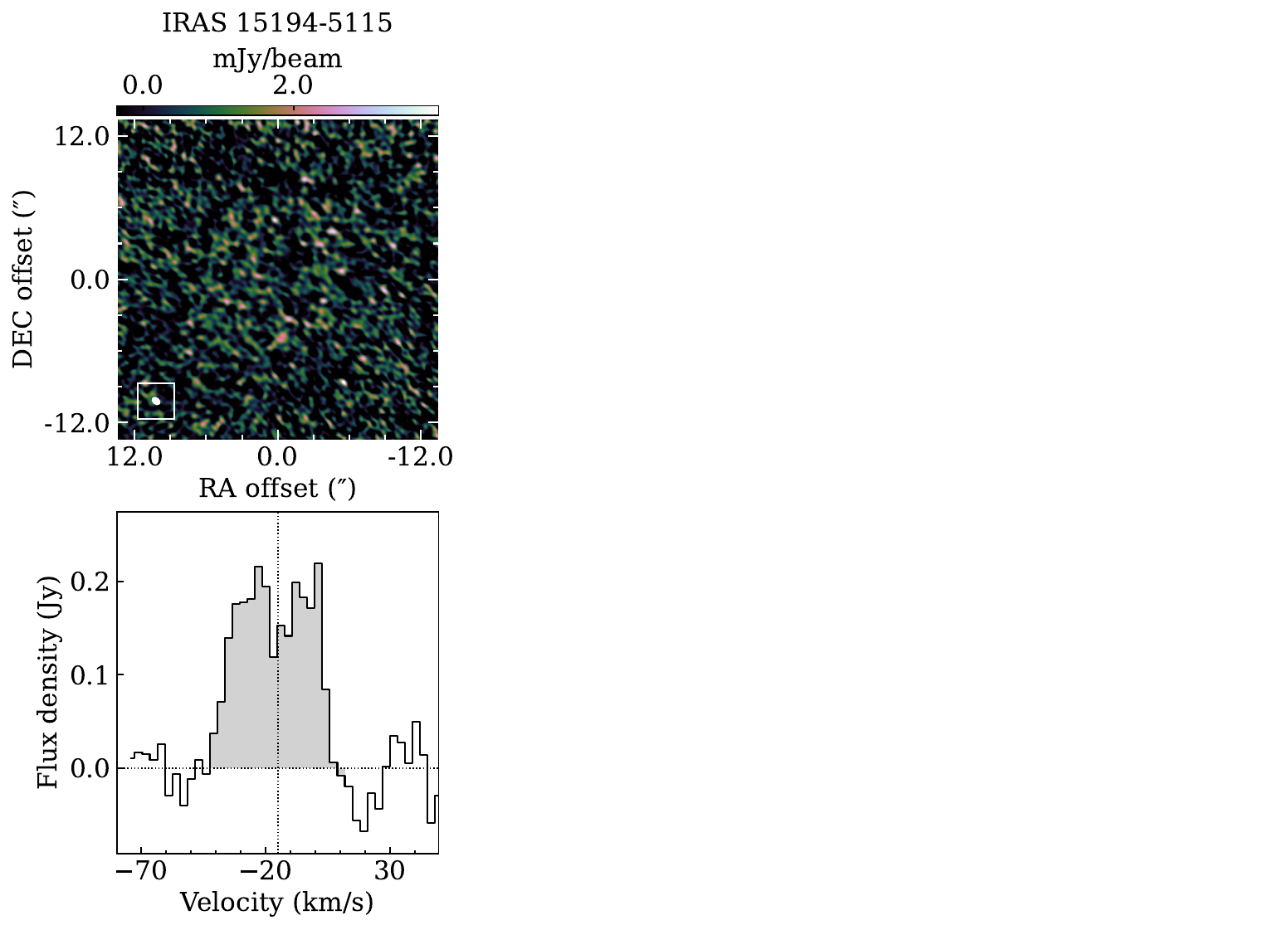}
    \caption{Si$^{13}$CC, 4$_{1,3}$-3$_{1,2}$ (97.295257 GHz)}
\end{figure}

\begin{figure}[h]
    \centering
    \includegraphics[width=0.8\linewidth]{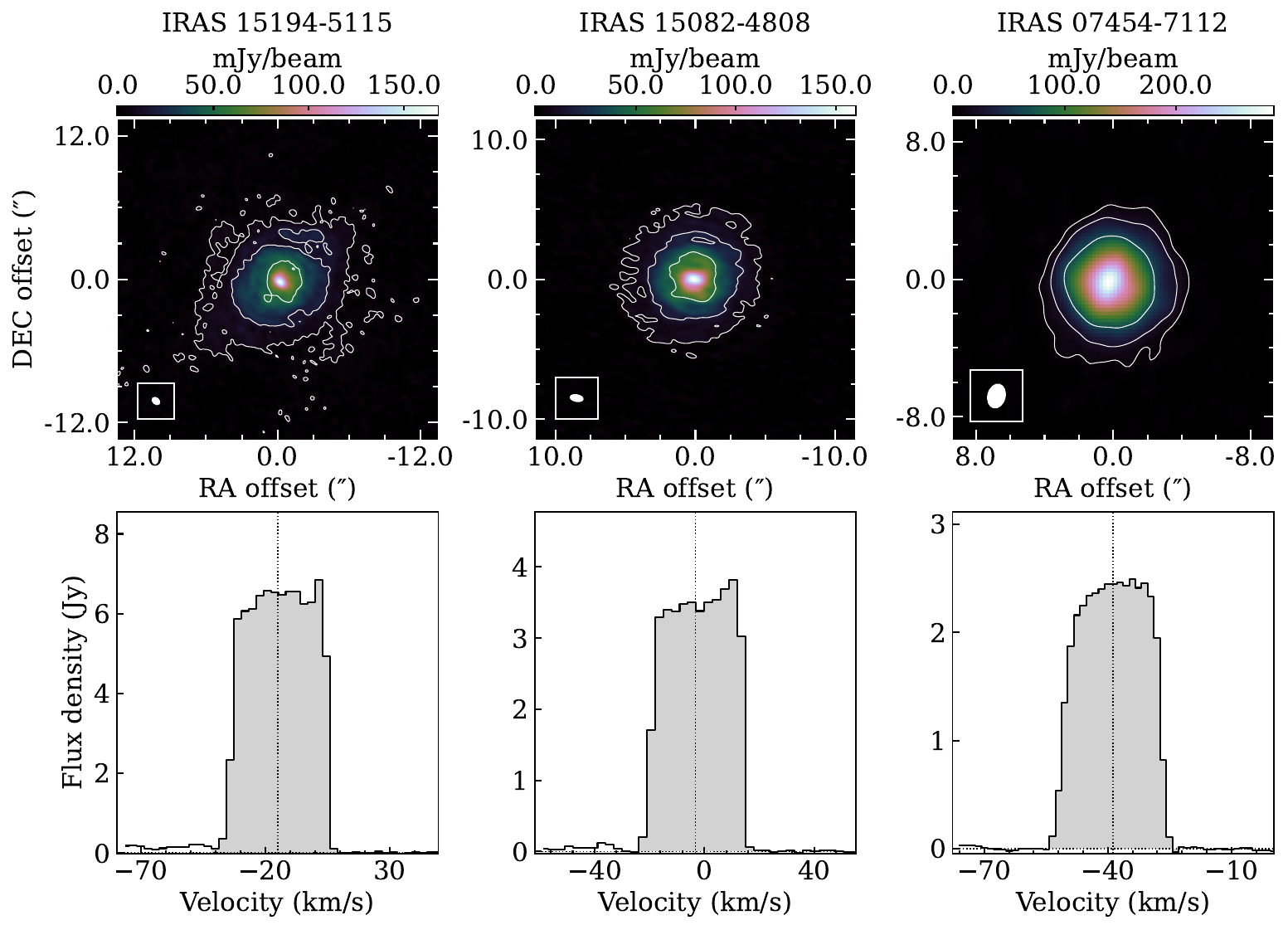}
    \caption{CS, 2-1 (97.980953 GHz)}
    \label{fig:CS_app_B}
\end{figure}

\begin{figure}[h]
    \centering
    \includegraphics[width=0.8\linewidth]{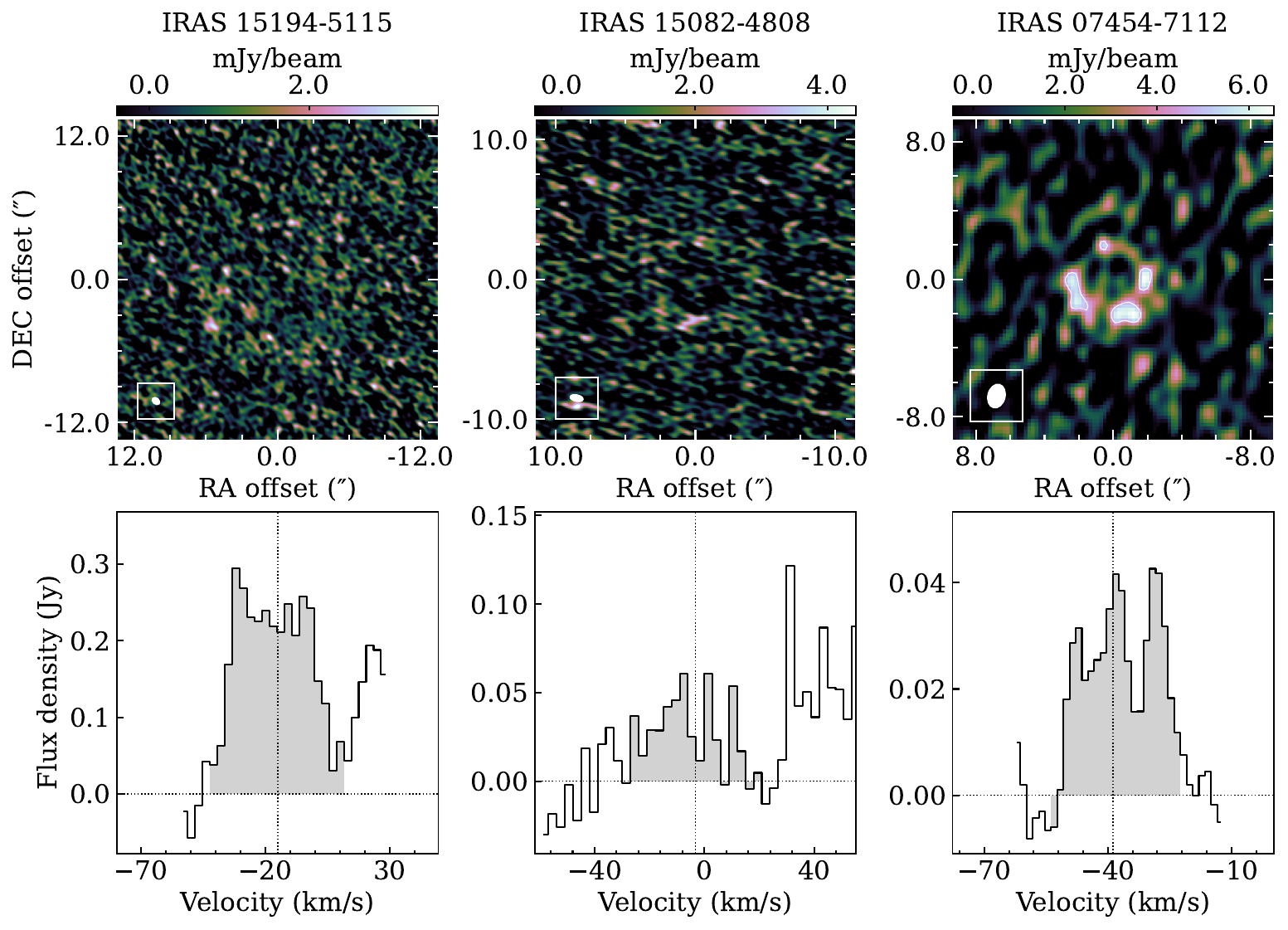}
    \caption{C$_3$H, J=9/2-7/2 (98.01207 GHz)}
    \label{fig:C3H_app_B}
\end{figure}

\begin{figure}[h]
    \centering
    \includegraphics[width=0.8\linewidth]{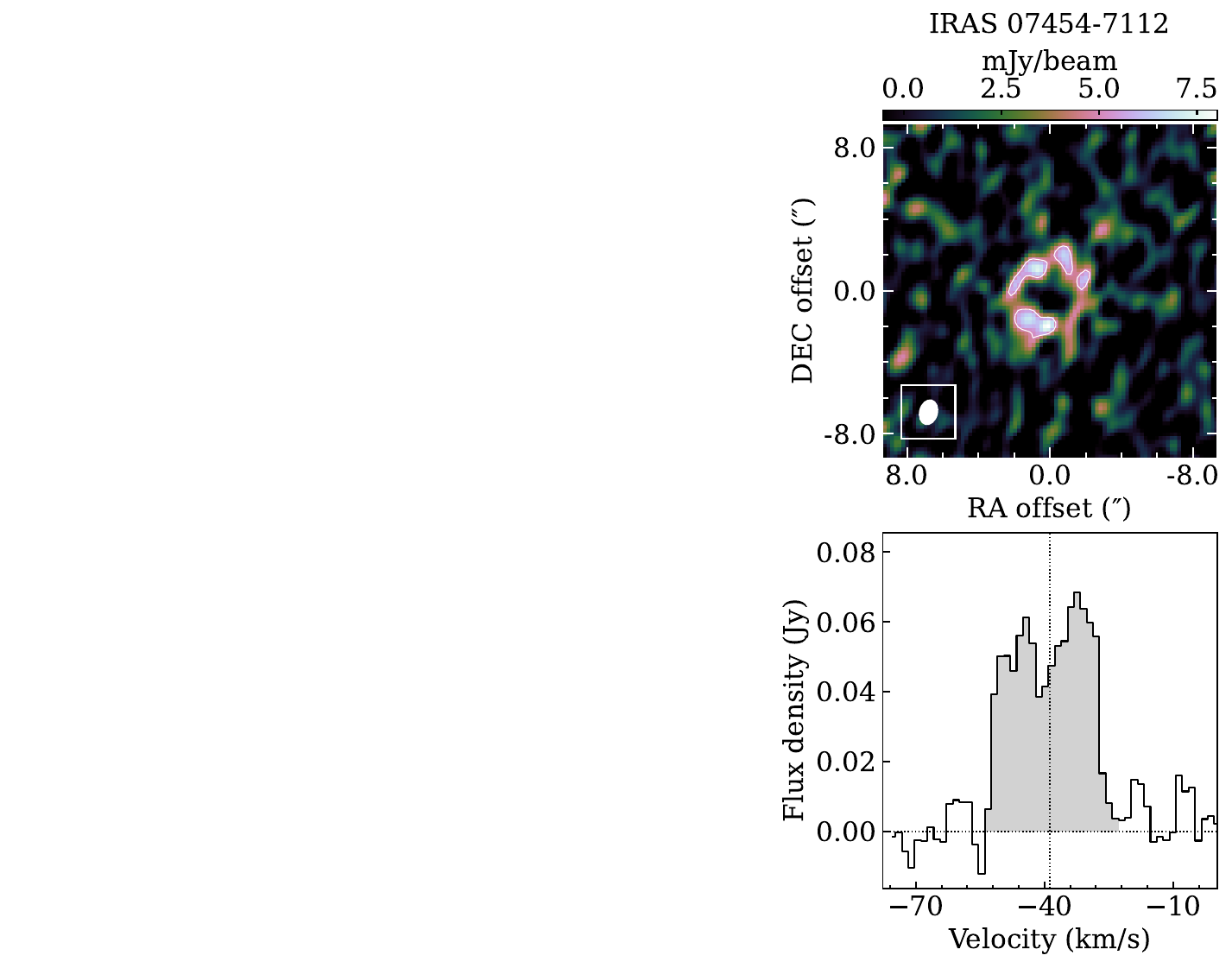}
    \caption{HC$_5$N, 37-36 (98.512524 GHz)}
\end{figure}

\begin{figure}[h]
    \centering
    \includegraphics[width=0.8\linewidth]{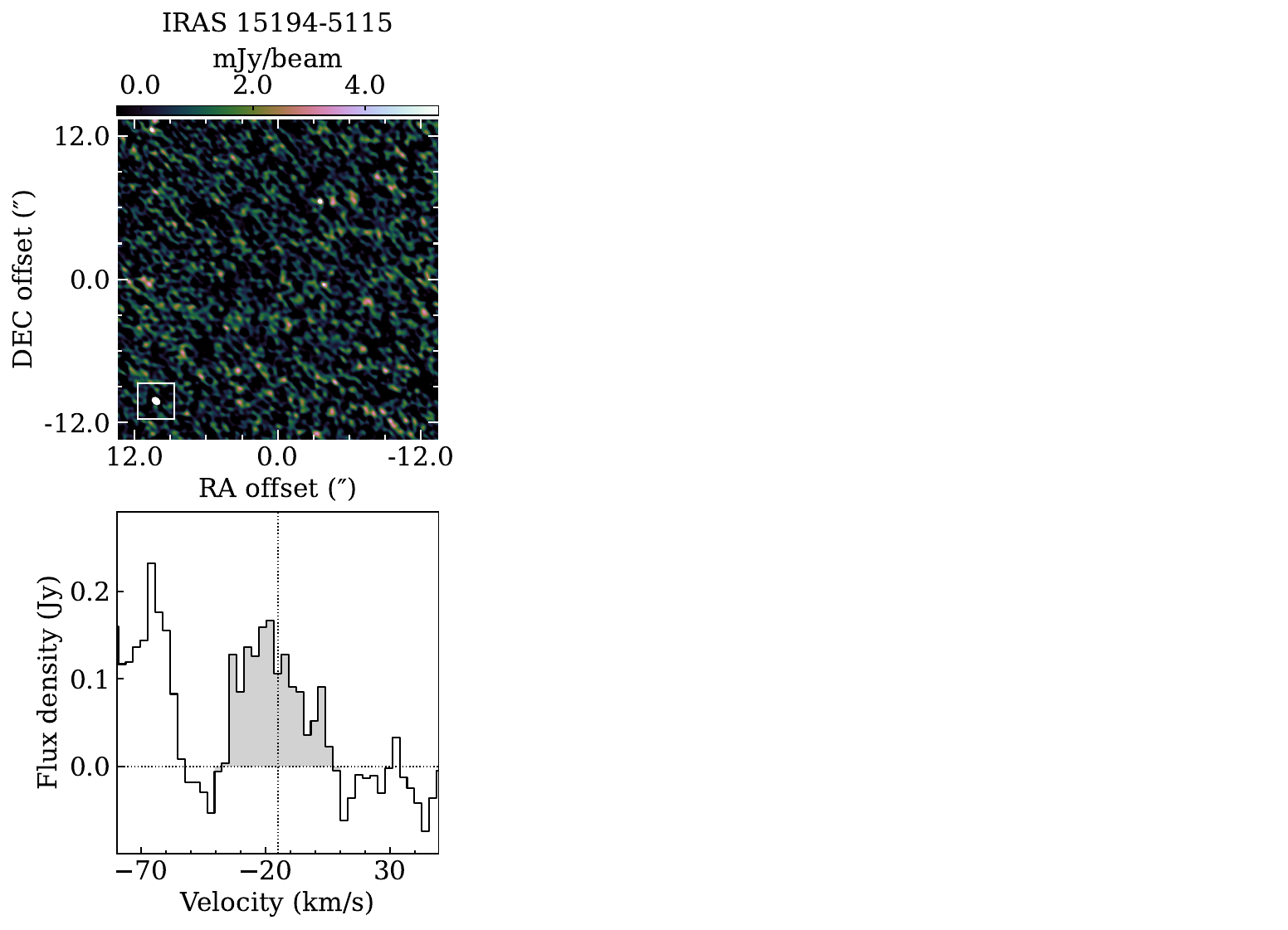}
    \caption{CC$^{13}$CN, N=10-9, J=21/2-19/2 (98.569 GHz)}
\end{figure}

\begin{figure}[h]
    \centering
    \includegraphics[width=0.8\linewidth]{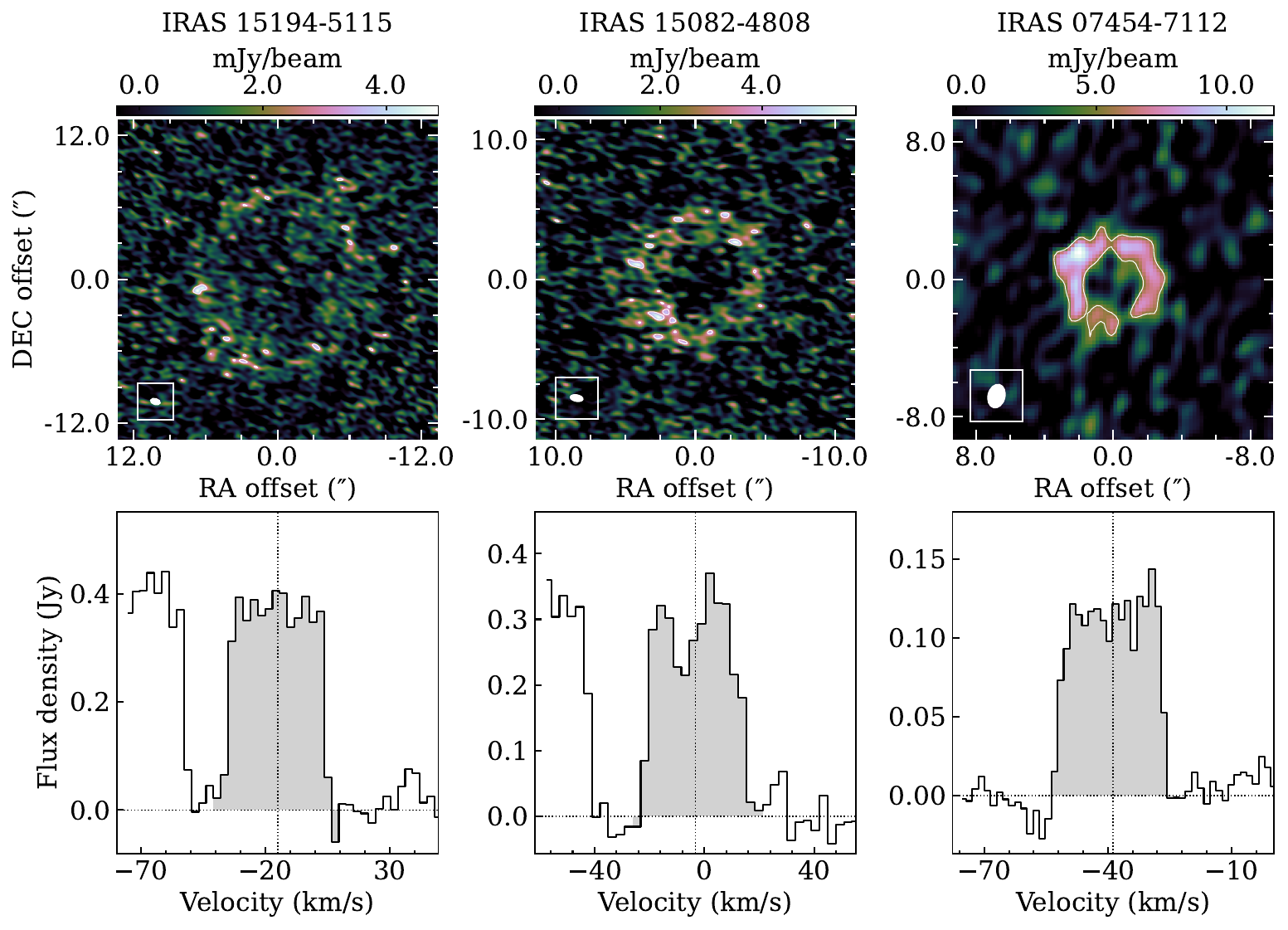}
    \caption{C$_3$N, 10-9 (A) (98.94 GHz)}
\end{figure}

\begin{figure}[h]
    \centering
    \includegraphics[width=0.8\linewidth]{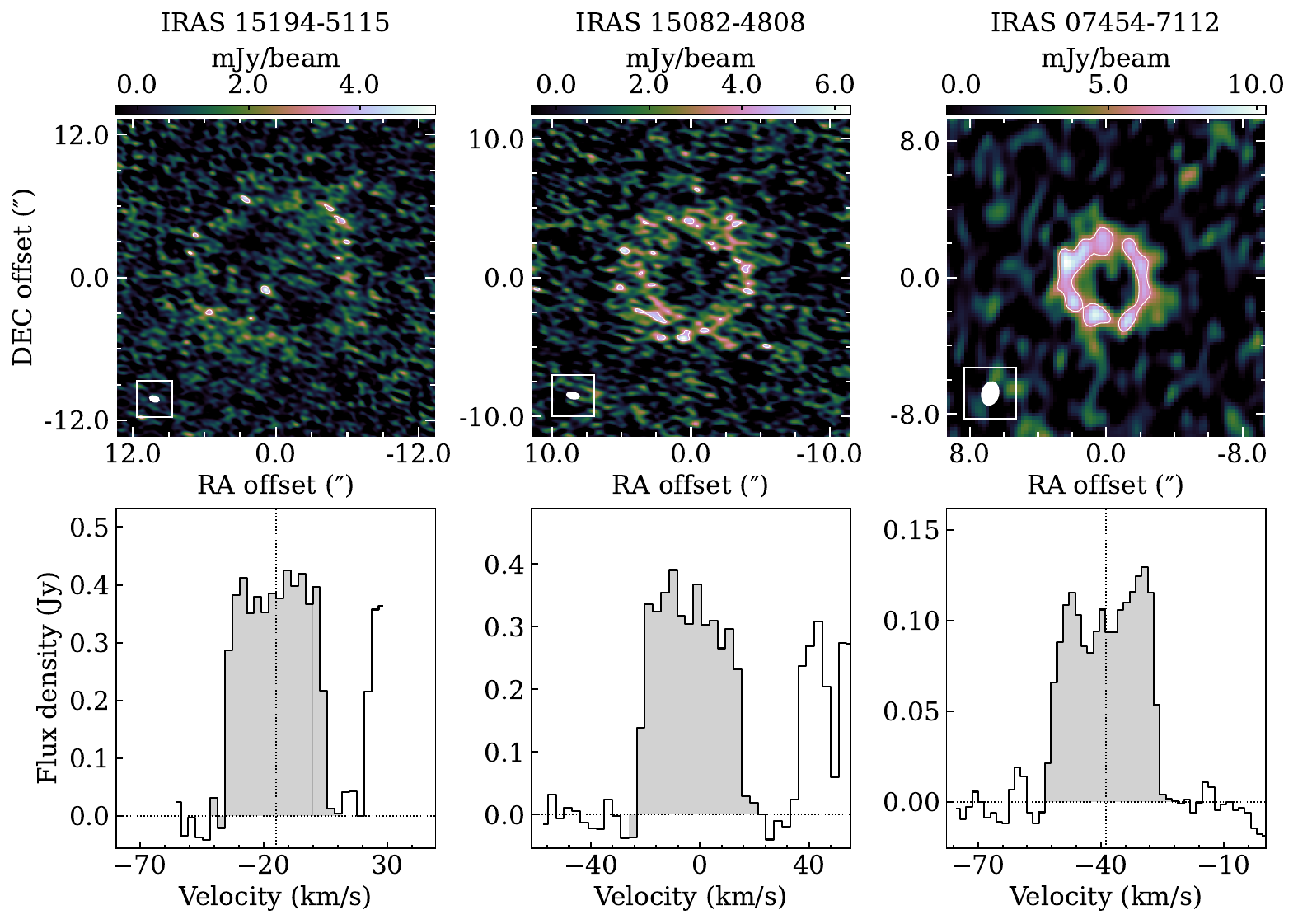}
    \caption{C$_3$N, 10-9 (B) (98.9588 GHz)}
\end{figure}

\begin{figure}[h]
    \centering
    \includegraphics[width=0.8\linewidth]{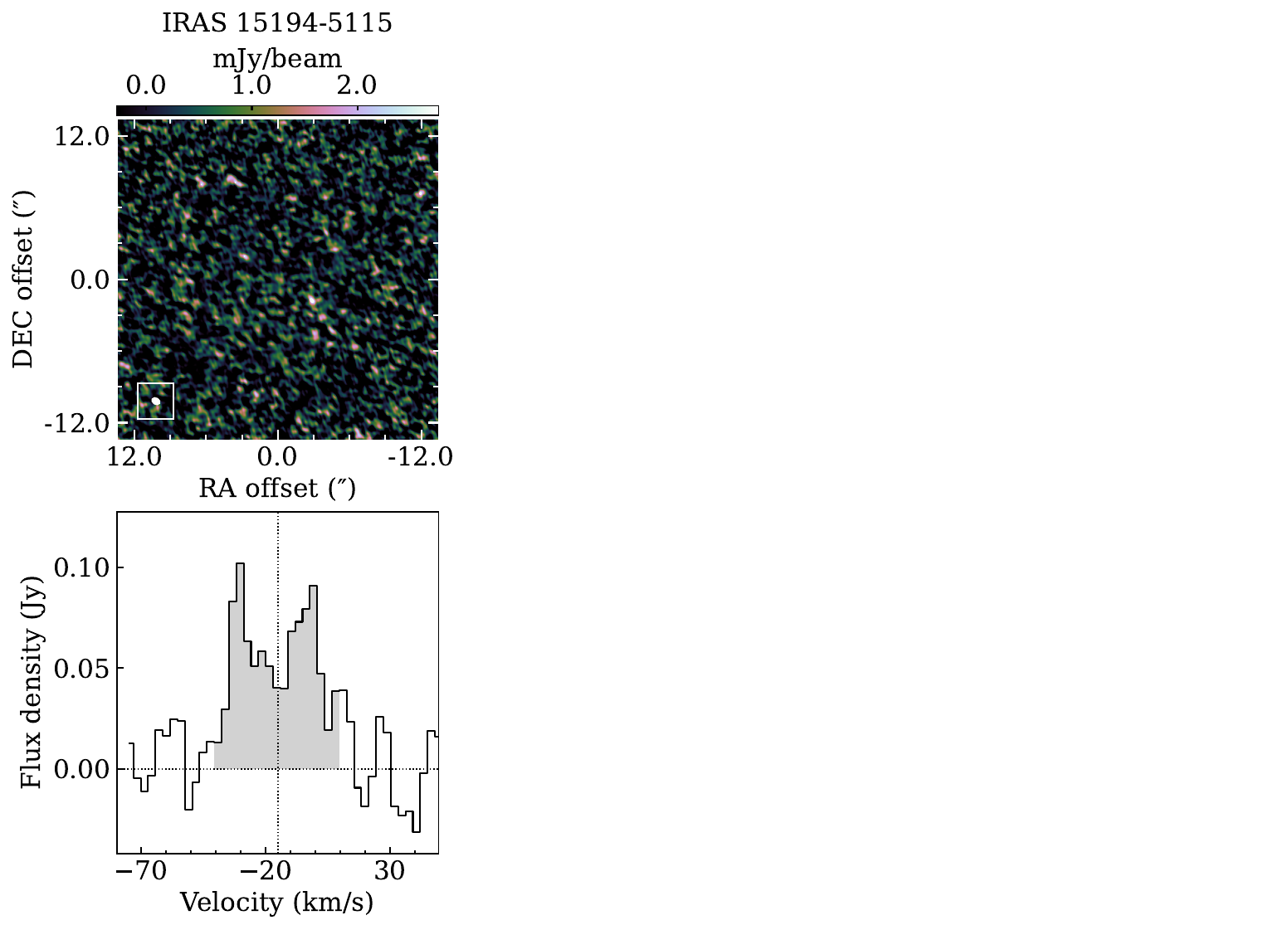}
    \caption{HC$^{13}$C$^{13}$CN, 11-10 (99.224259 GHz)}
\end{figure}

\begin{figure}[h]
    \centering
    \includegraphics[width=0.8\linewidth]{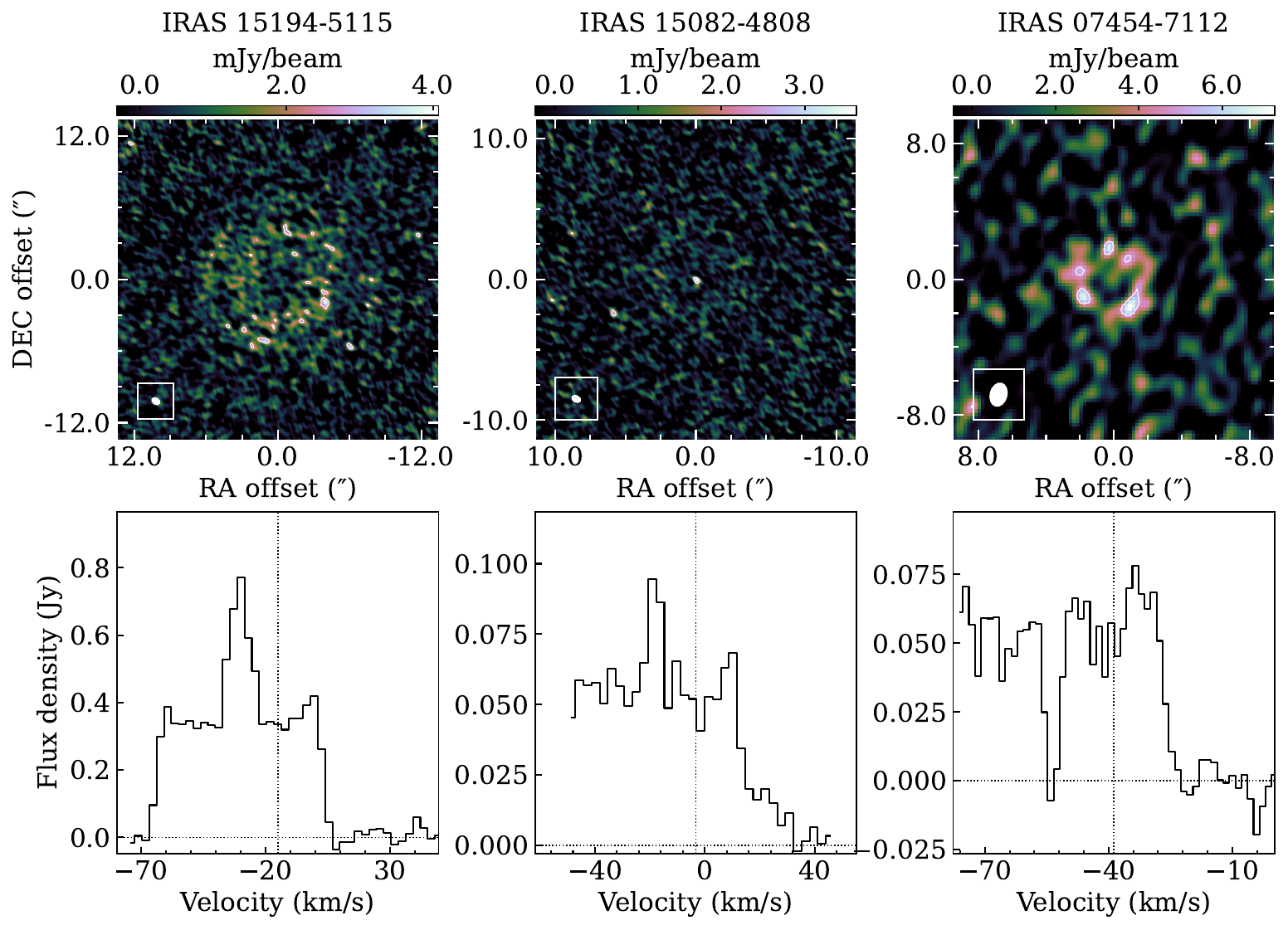}
    \caption{HC$^{13}$CCN, 11-10 (99.651849 GHz)}
\end{figure}

\begin{figure}[h]
    \centering
    \includegraphics[width=0.8\linewidth]{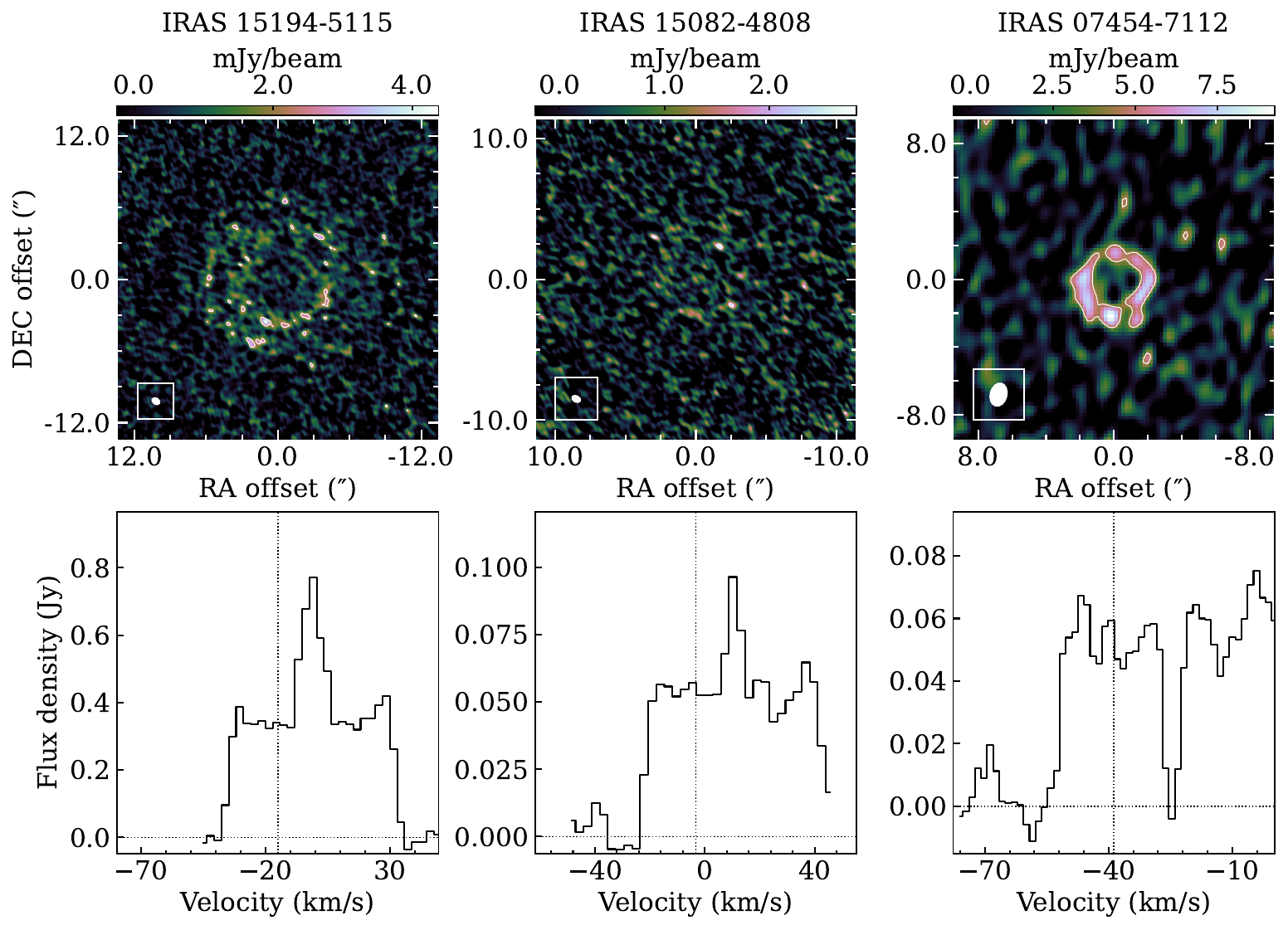}
    \caption{HCC$^{13}$CN, 11-10 (99.661467 GHz)}
\end{figure}

\begin{figure}[h]
    \centering
    \includegraphics[width=0.8\linewidth]{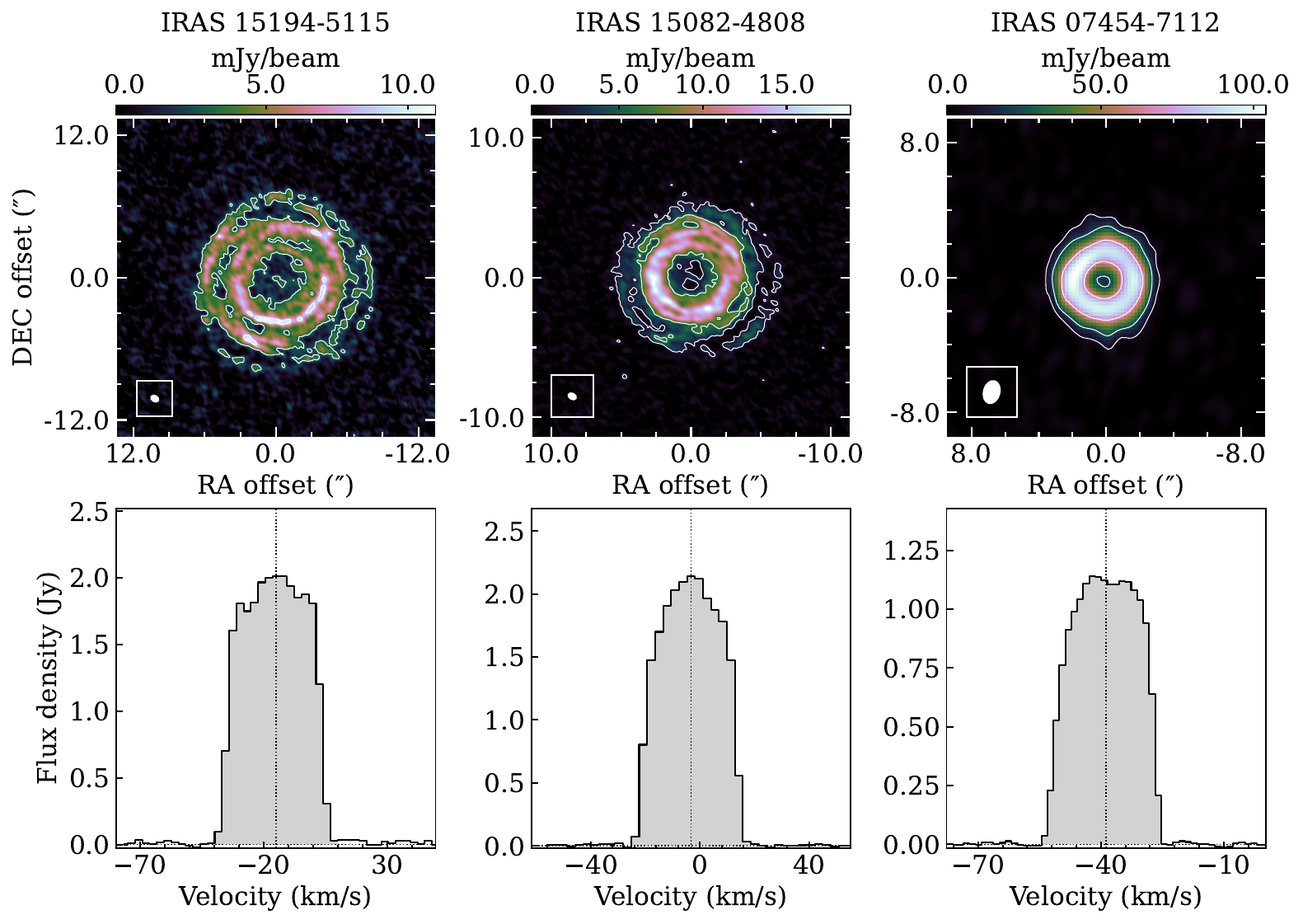}
    \caption{HC$_3$N, 11-10 (100.076392 GHz)}
    \label{fig:HC3N_app_B}
\end{figure}

\begin{figure}[h]
    \centering
    \includegraphics[width=0.8\linewidth]{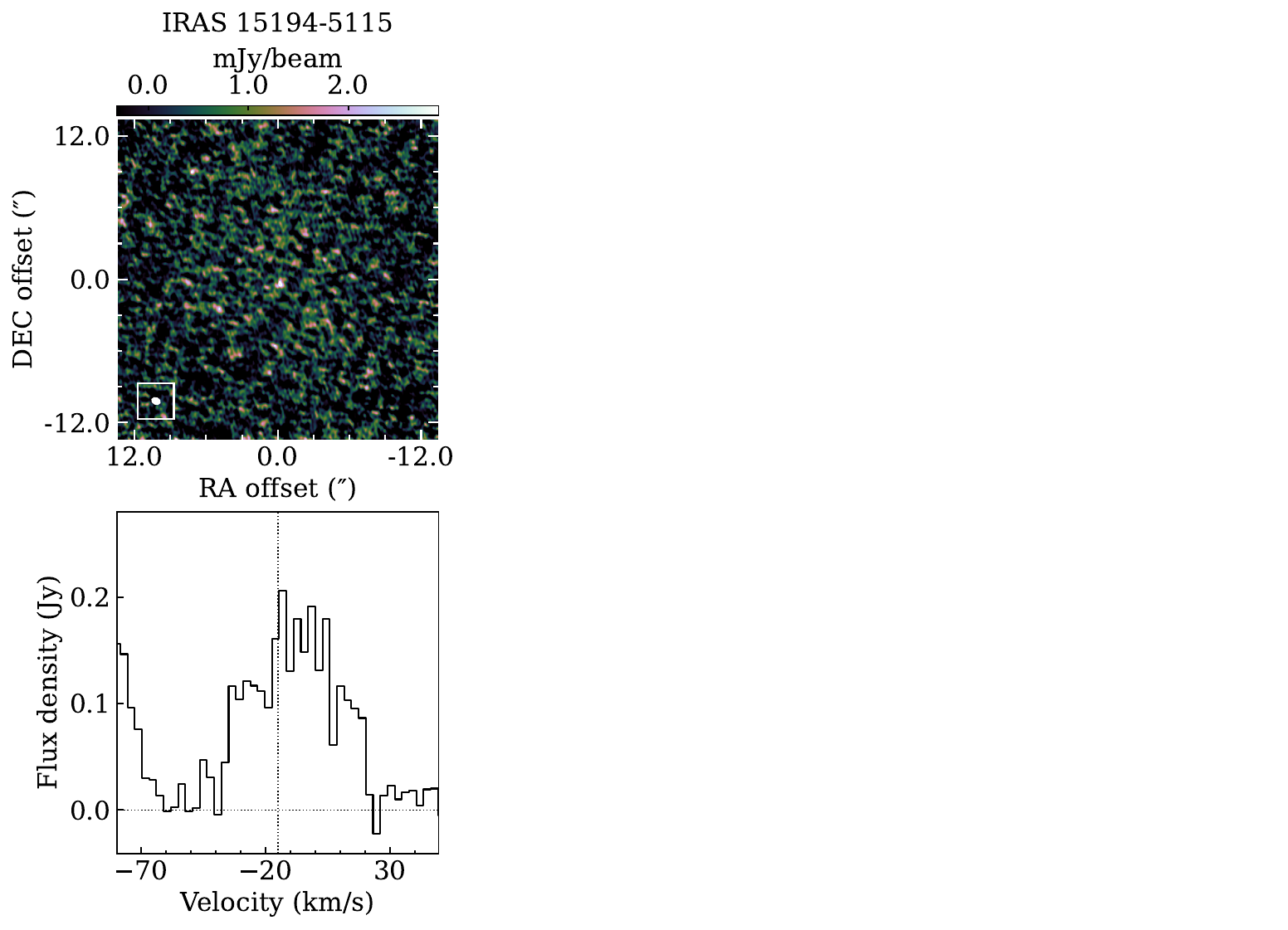}
    \caption{$^{13}$CCCCH, N=11-10, J=23/2-21/2 (101.06306 GHz)}
\end{figure}

\begin{figure}[h]
    \centering
    \includegraphics[width=0.8\linewidth]{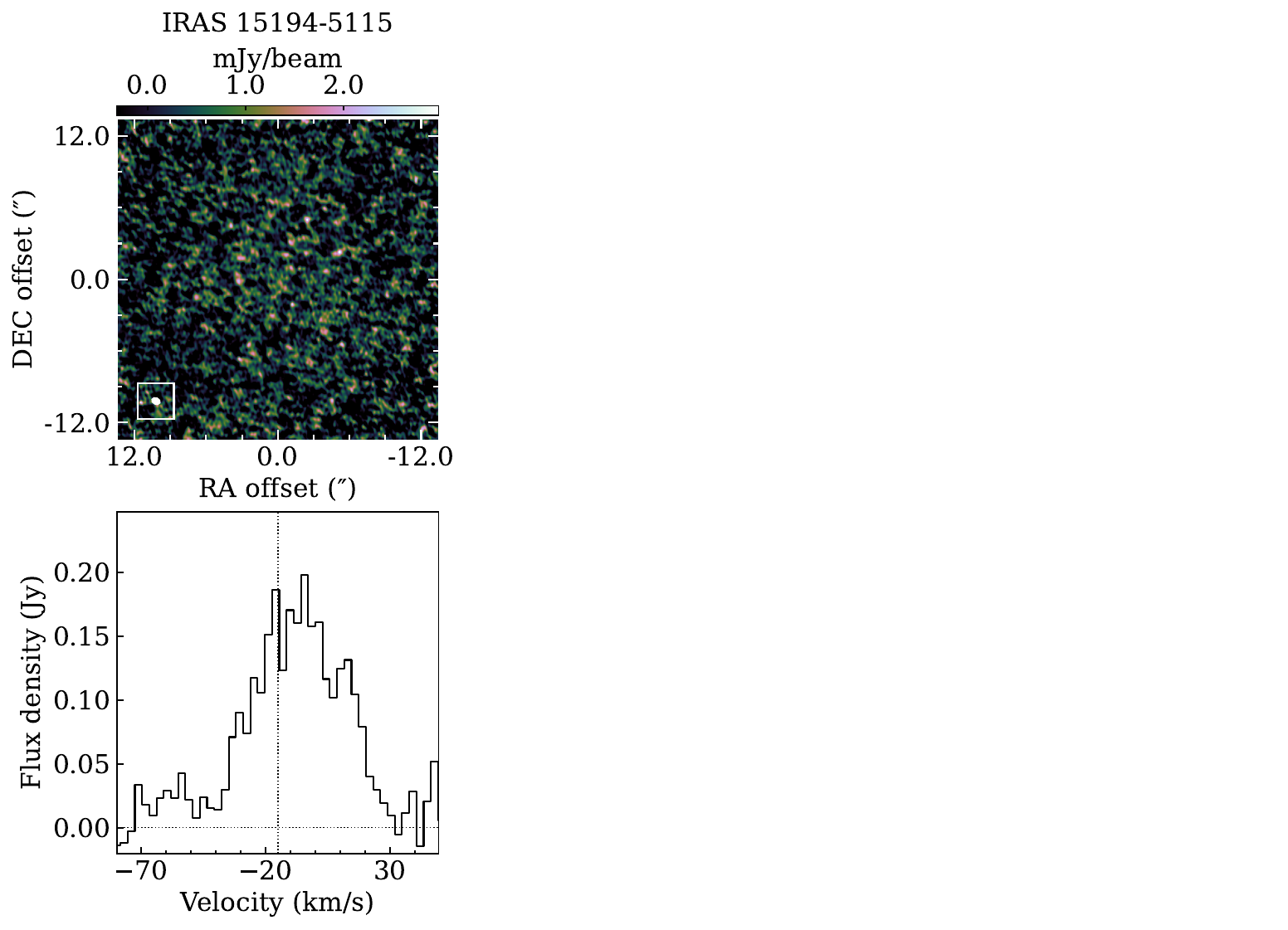}
    \caption{$^{13}$CCCCH, N=11-10, J=21/2-19/2 (101.09403 GHz)}
\end{figure}

\begin{figure}[h]
    \centering
    \includegraphics[width=0.8\linewidth]{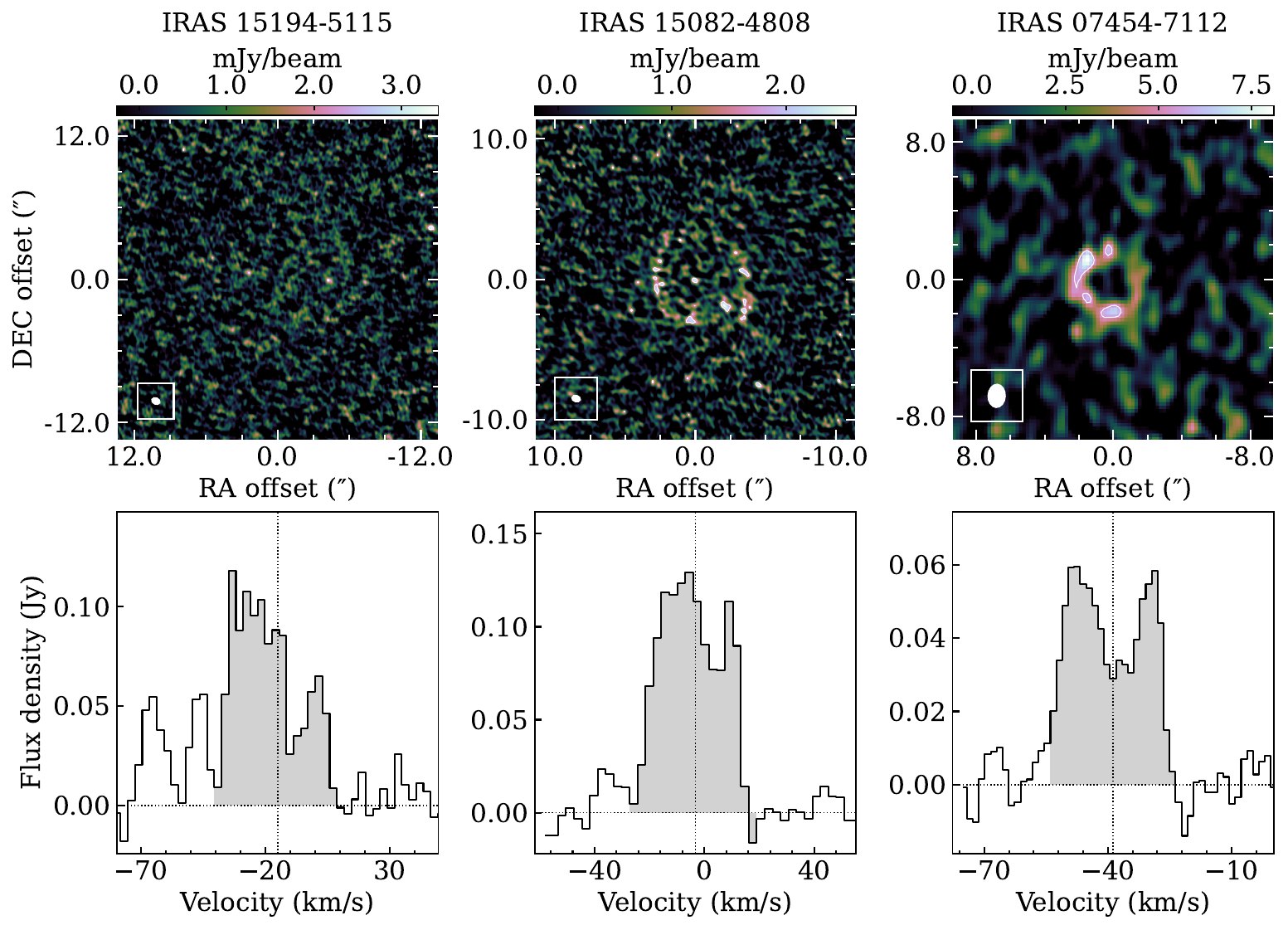}
    \caption{HC$_5$N, 38-37 (101.174677 GHz)}
        \label{fig:HC5N_cc_appendix_C}
\end{figure}

\begin{figure}[h]
    \centering
    \includegraphics[width=0.8\linewidth]{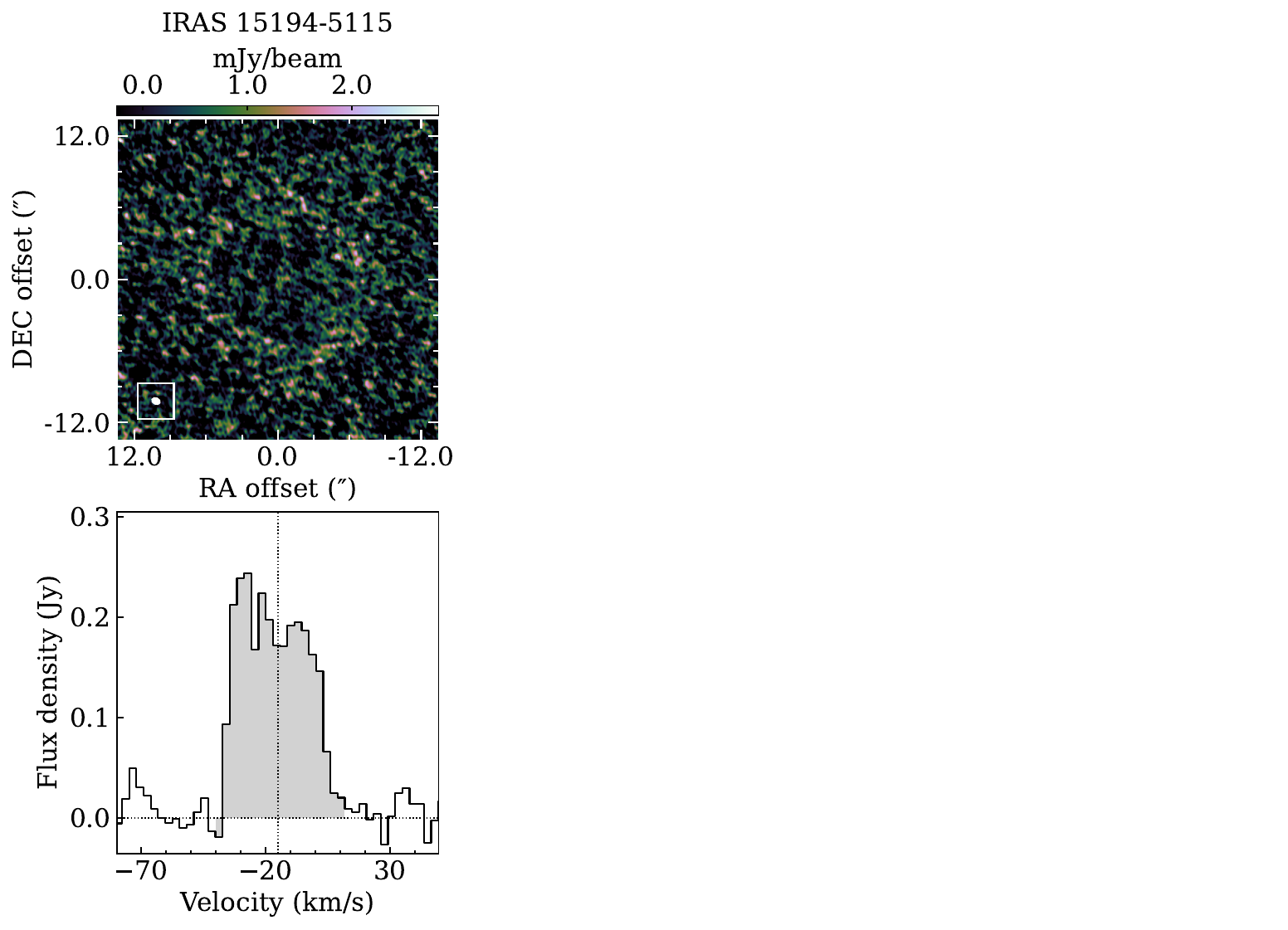}
    \caption{CCC$^{13}$CH, N=11-10, J=23/2-21/2 (101.506343 GHz)}
\end{figure}

\begin{figure}[h]
    \centering
    \includegraphics[width=0.8\linewidth]{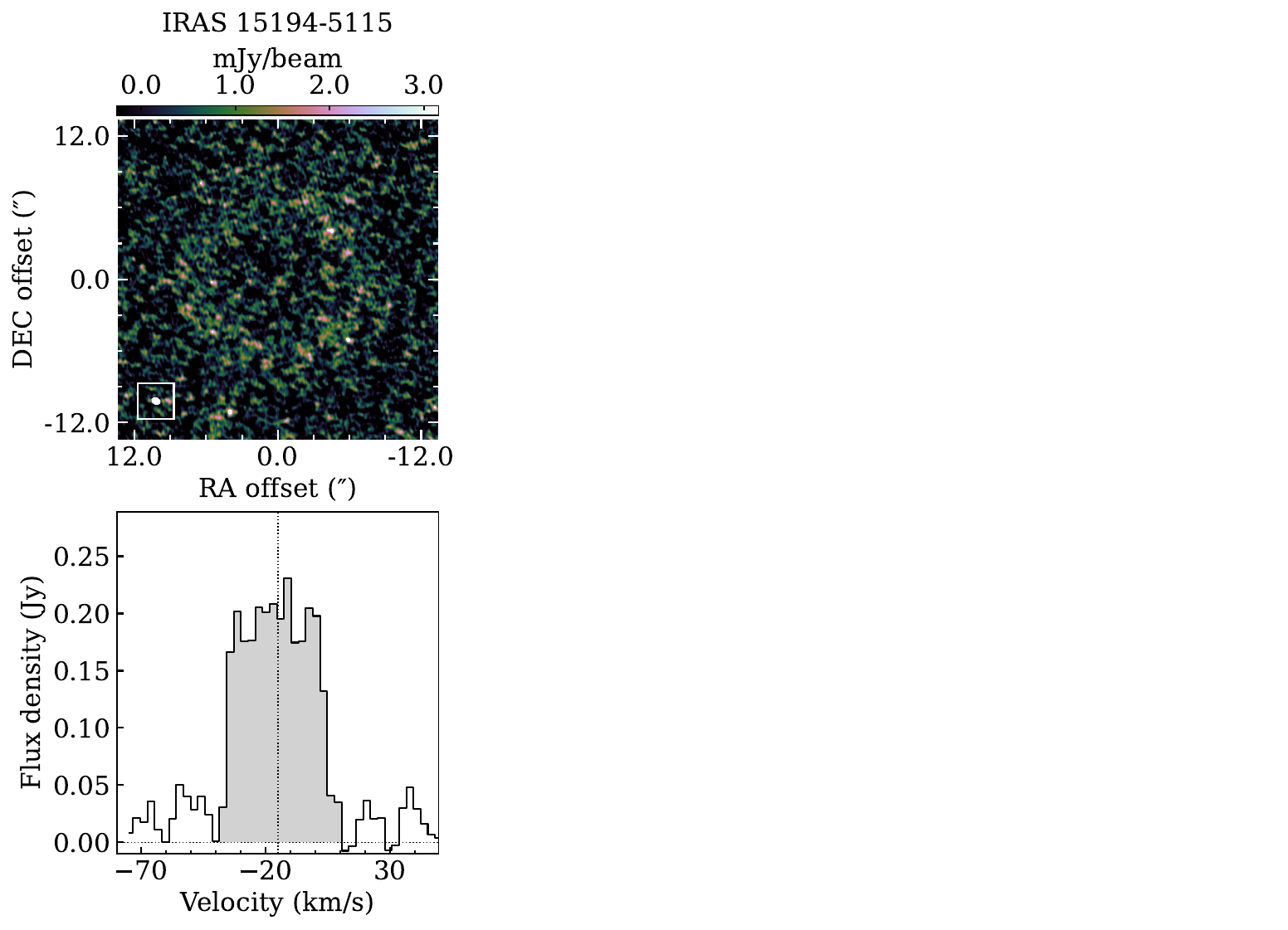}
    \caption{CCC$^{13}$CH, N=11-10, J=21/2-19/2 (101.543766 GHz)}
\end{figure}

\begin{figure}[h]
    \centering
    \includegraphics[width=0.8\linewidth]{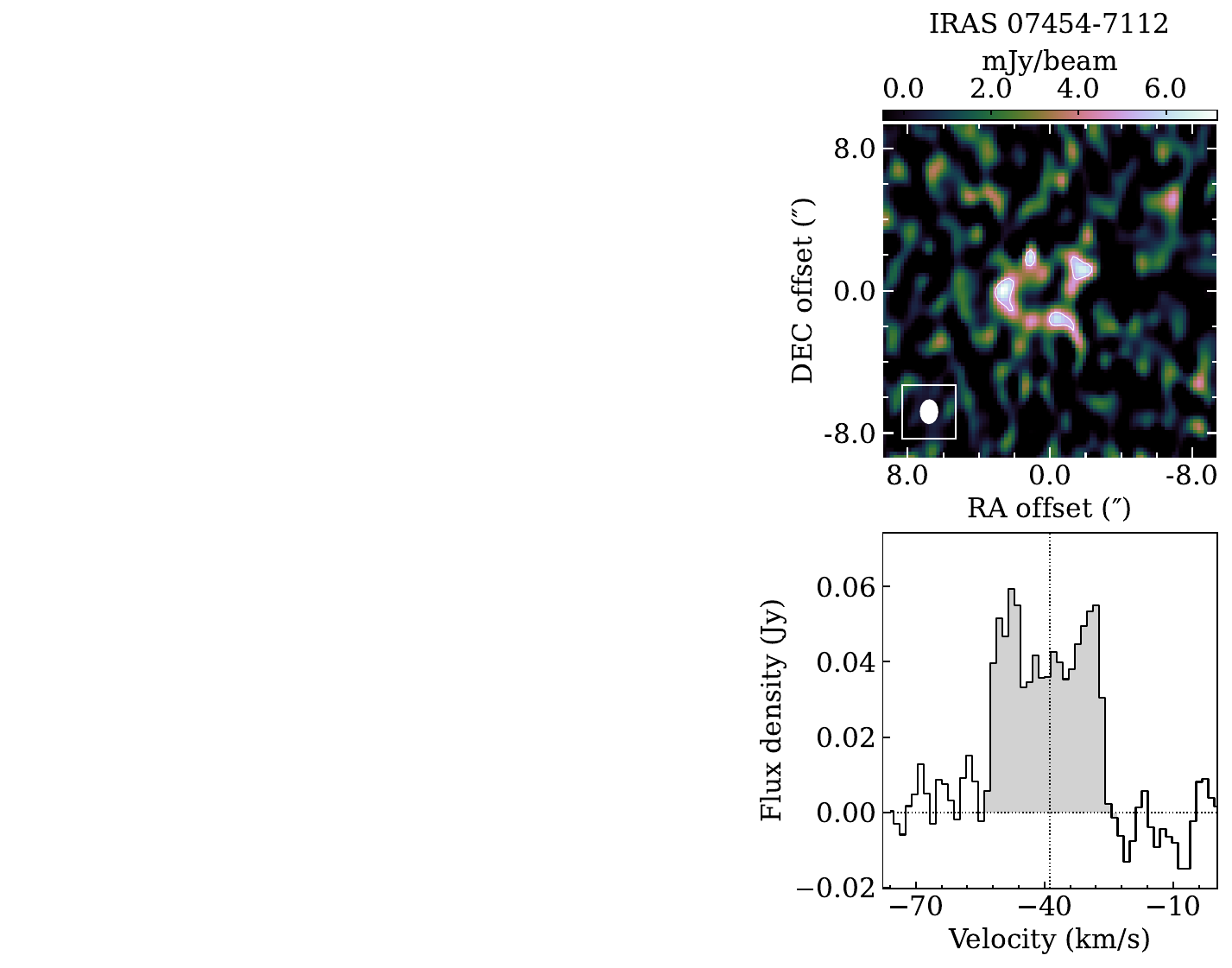}
    \caption{HC$_5$N, 39-38 (103.836817 GHz)}
\end{figure}

\begin{figure}[h]
    \centering
    \includegraphics[width=0.8\linewidth]{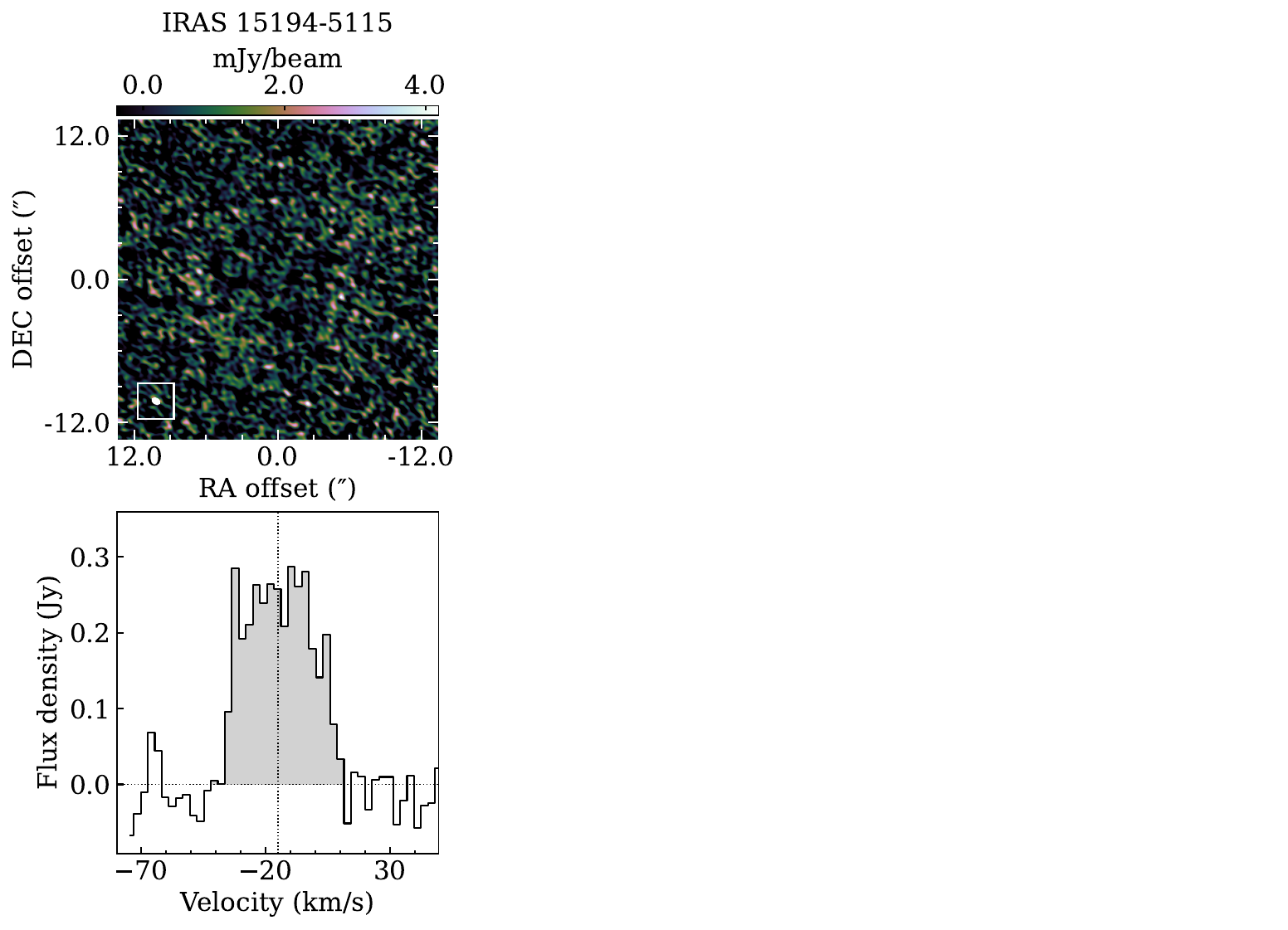}
    \caption{C$^{13}$CCCH, N=11-10, J=23/2-21/2 (104.13854 GHz)}
\end{figure}

\begin{figure}[h]
    \centering
    \includegraphics[width=0.8\linewidth]{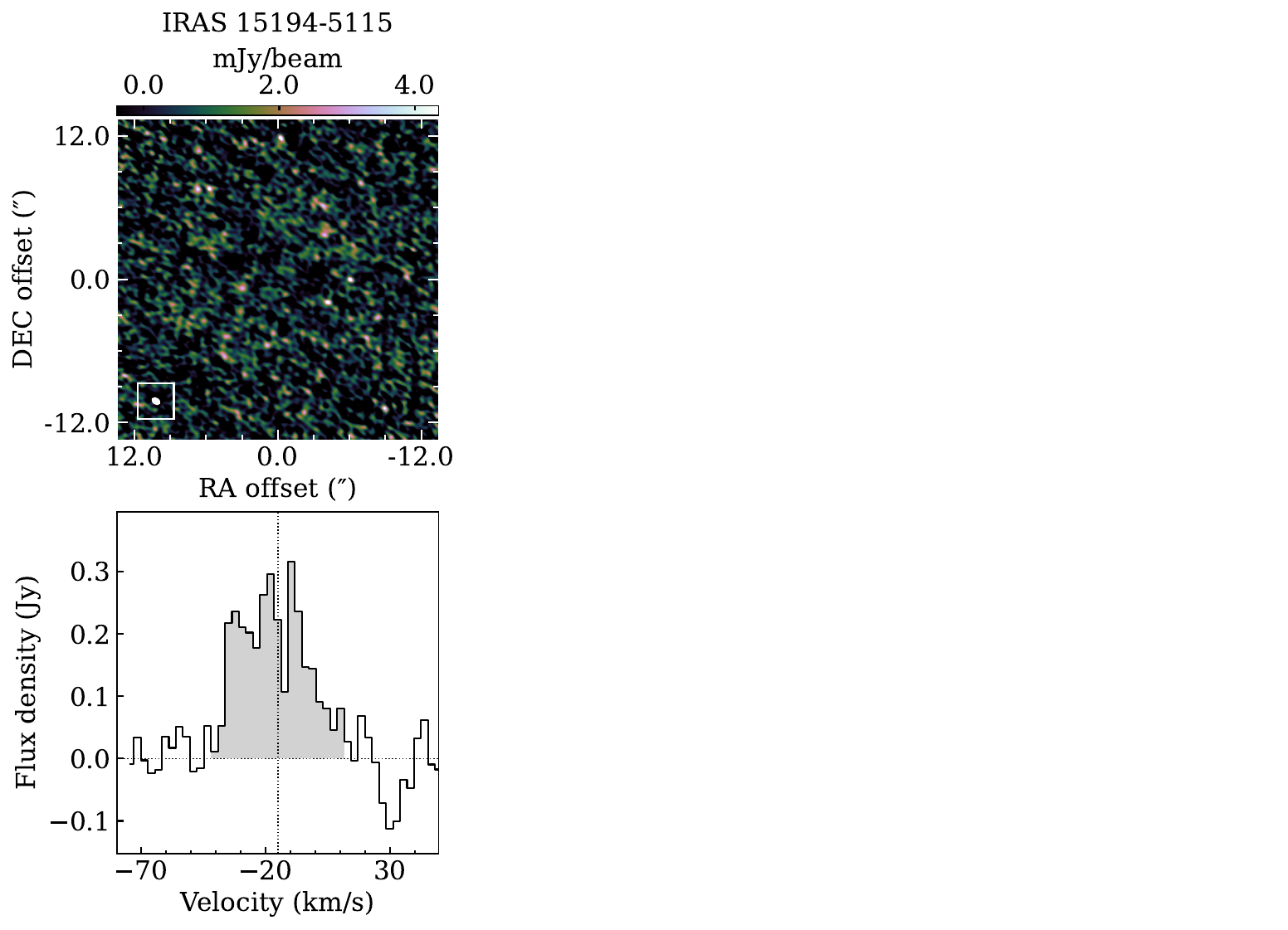}
    \caption{C$^{13}$CCCH, N=11-10, J=21/2-19/2 (104.17622 GHz)}
\end{figure}

\begin{figure}[h]
    \centering
    \includegraphics[width=0.8\linewidth]{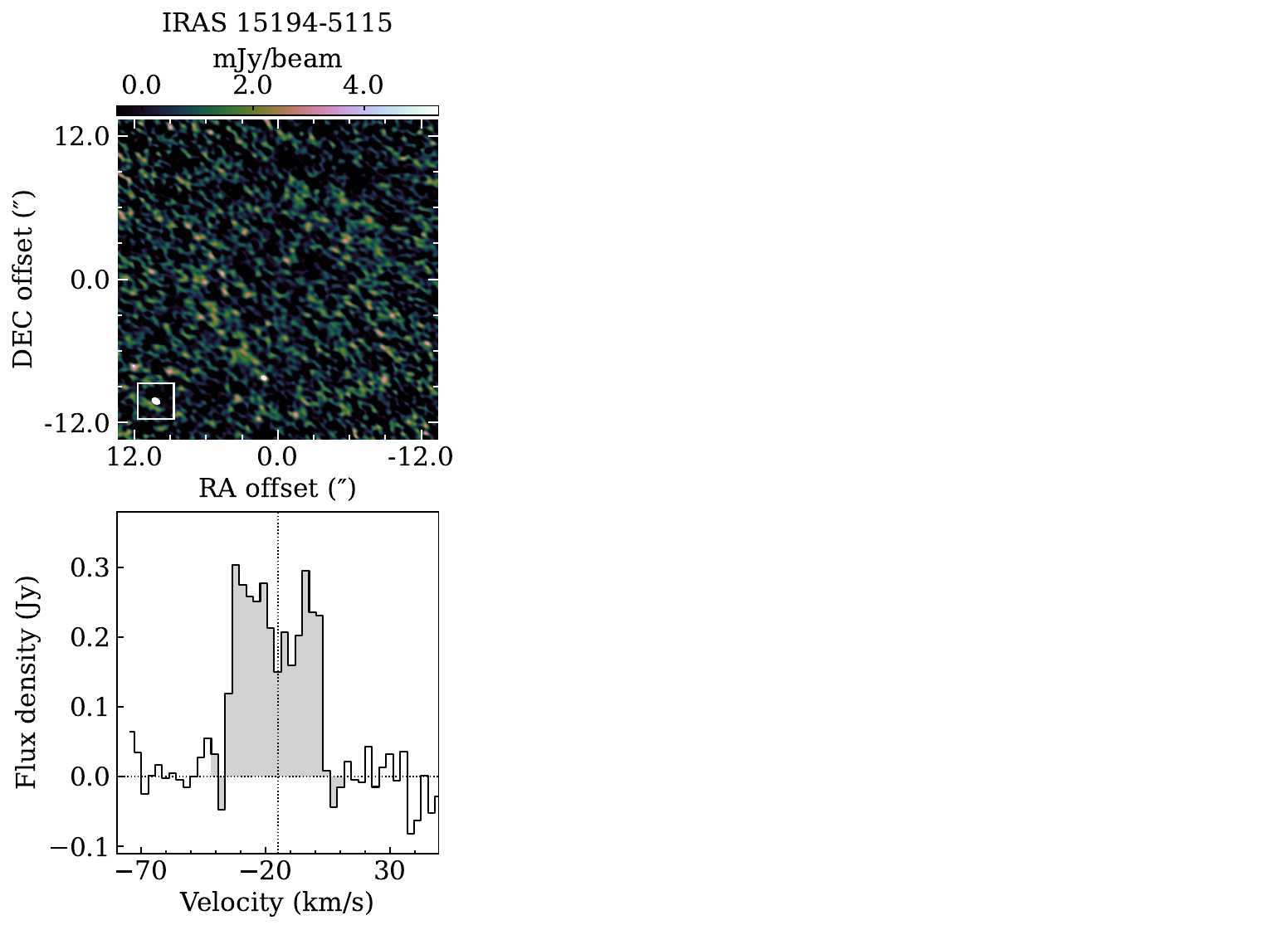}
    \caption{CC$^{13}$CCH, N=11-10, J=23/2-21/2 (104.29734 GHz)}
\end{figure}

\begin{figure}[h]
    \centering
    \includegraphics[width=0.8\linewidth]{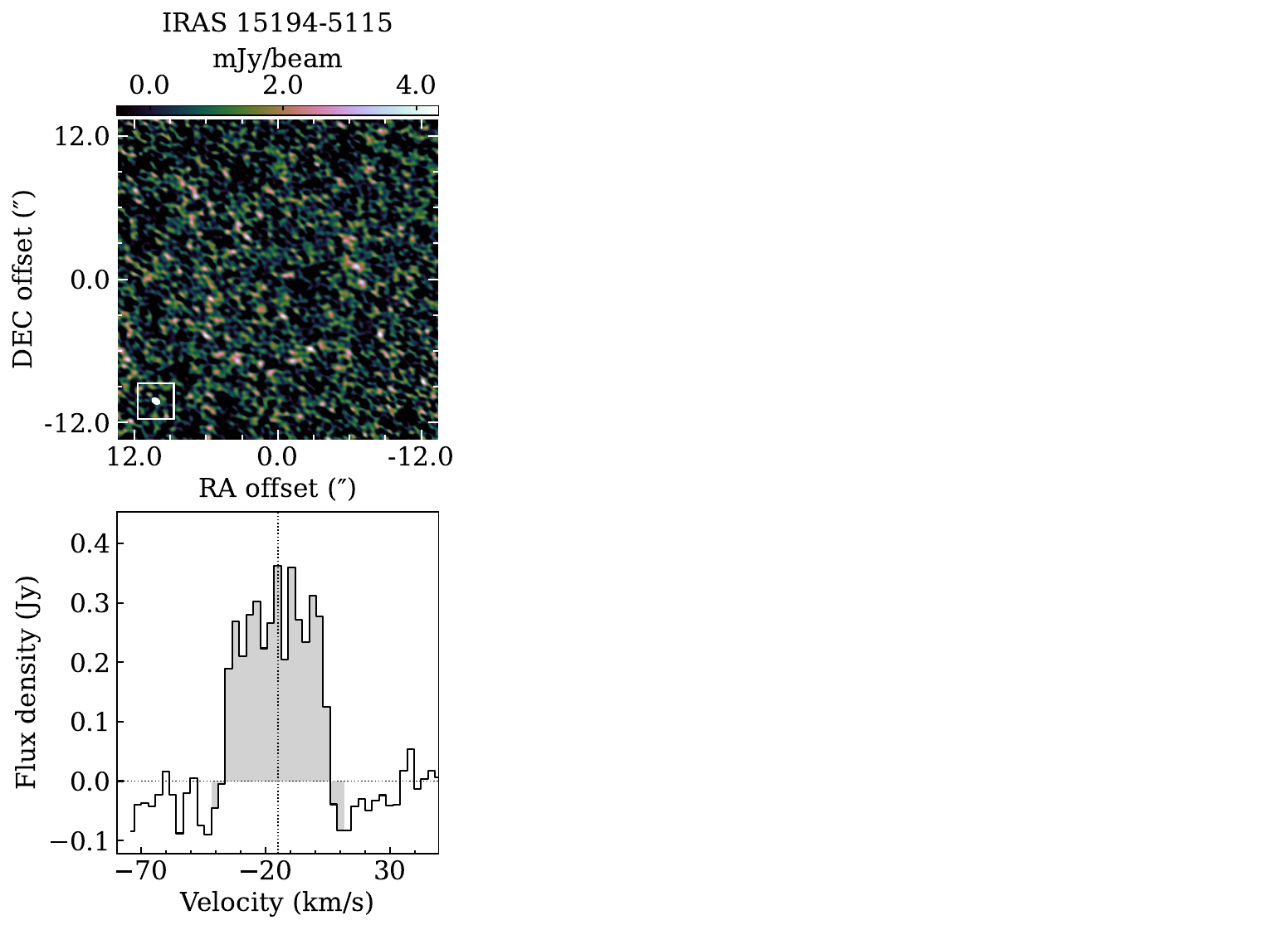}
    \caption{CC$^{13}$CCH, N=11-10, J=21/2-19/2 (104.33572 GHz)}
\end{figure}

\begin{figure}[h]
    \centering
    \includegraphics[width=0.8\linewidth]{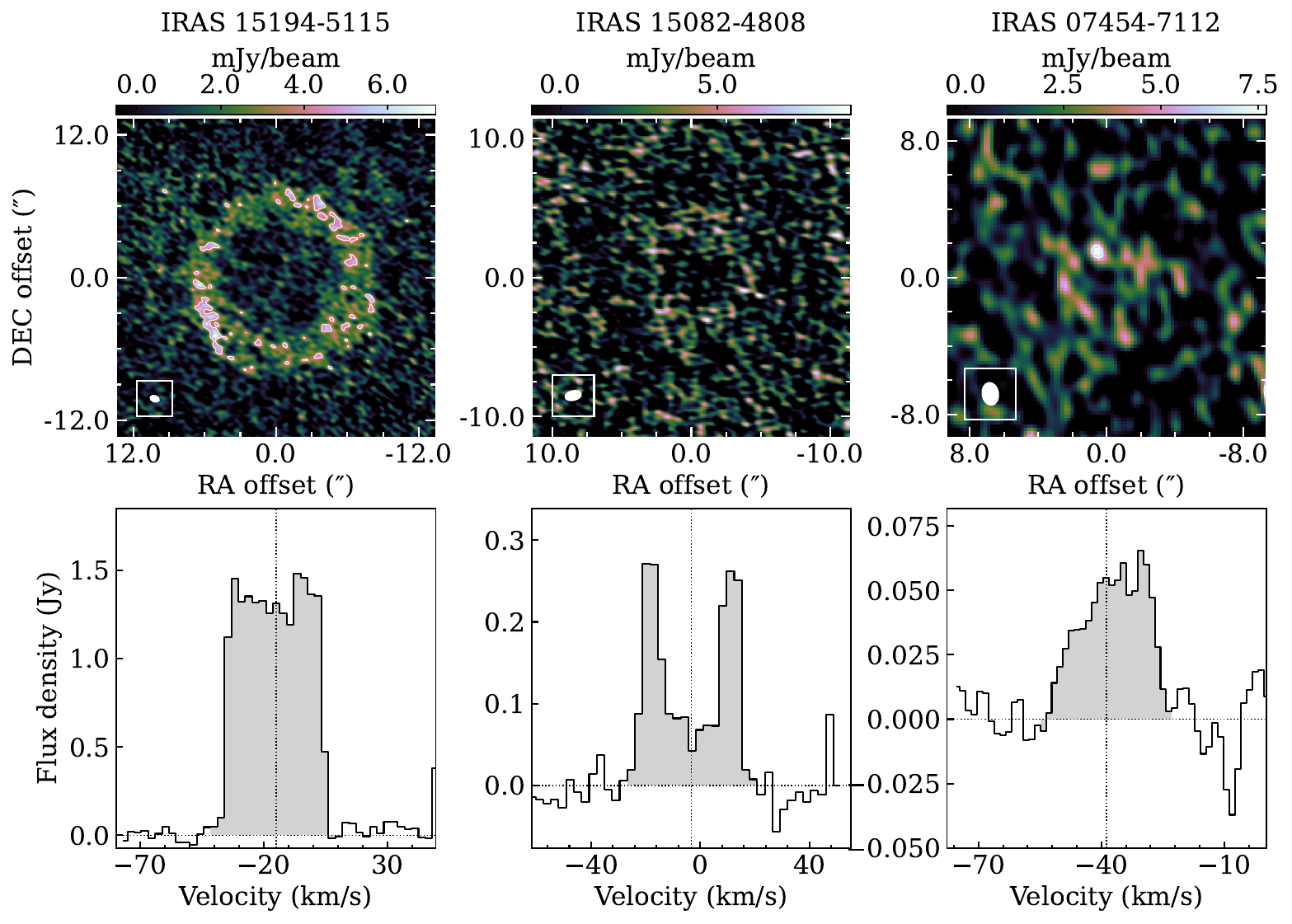}
    \caption{C$_4$H, N=11-10, J=23/2-21/2 (104.666566 GHz)}
\end{figure}

\begin{figure}[h]
    \centering
    \includegraphics[width=0.8\linewidth]{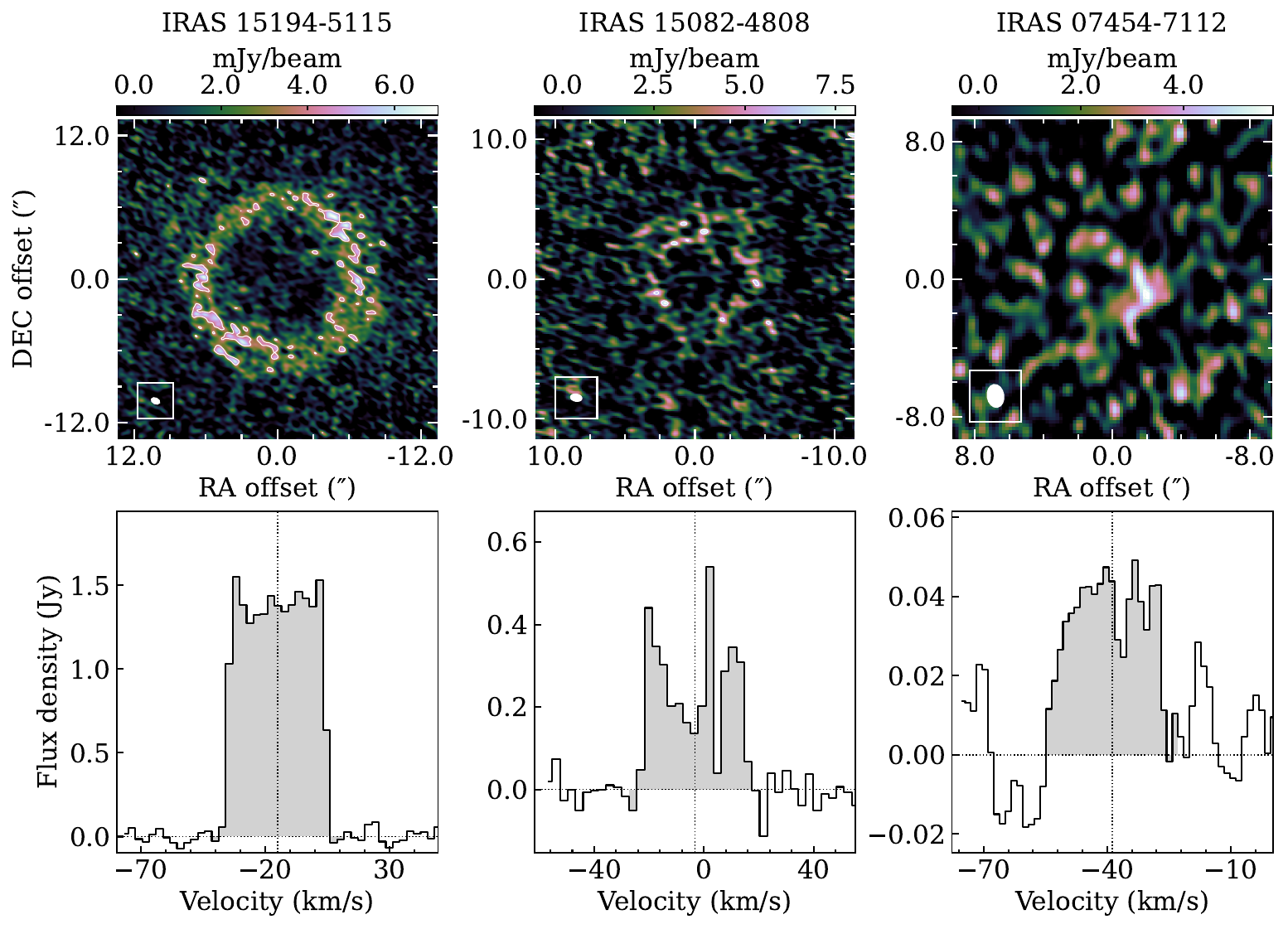}
    \caption{C$_4$H, N=11-10, J=21/2-19/2 (104.70511 GHz)}
\end{figure}

\begin{figure}[h]
    \centering
    \includegraphics[width=0.8\linewidth]{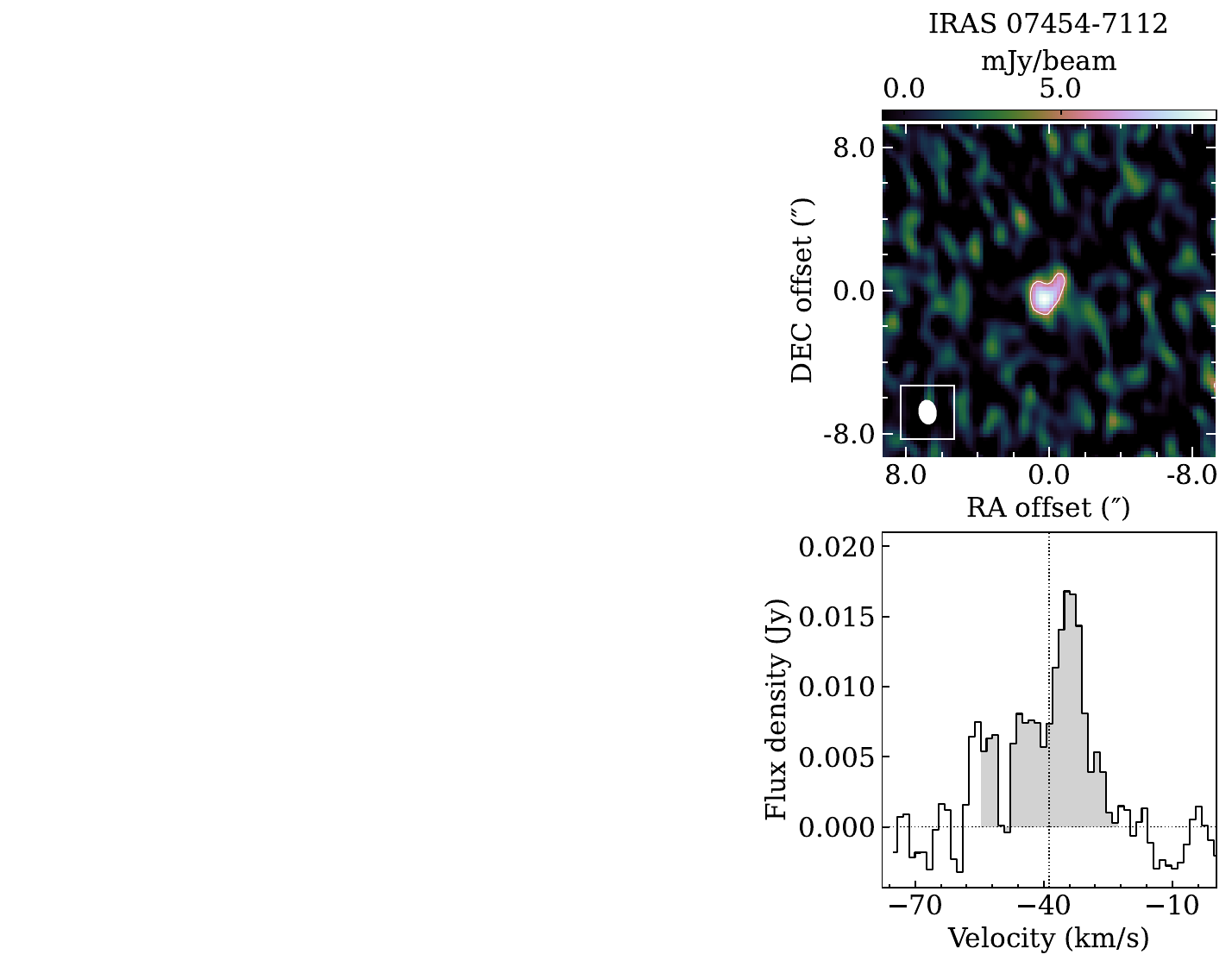}
    \caption{$^{30}$SiS, 6-5 (105.059203 GHz)}
\end{figure}

\begin{figure}[h]
    \centering
    \includegraphics[width=0.8\linewidth]{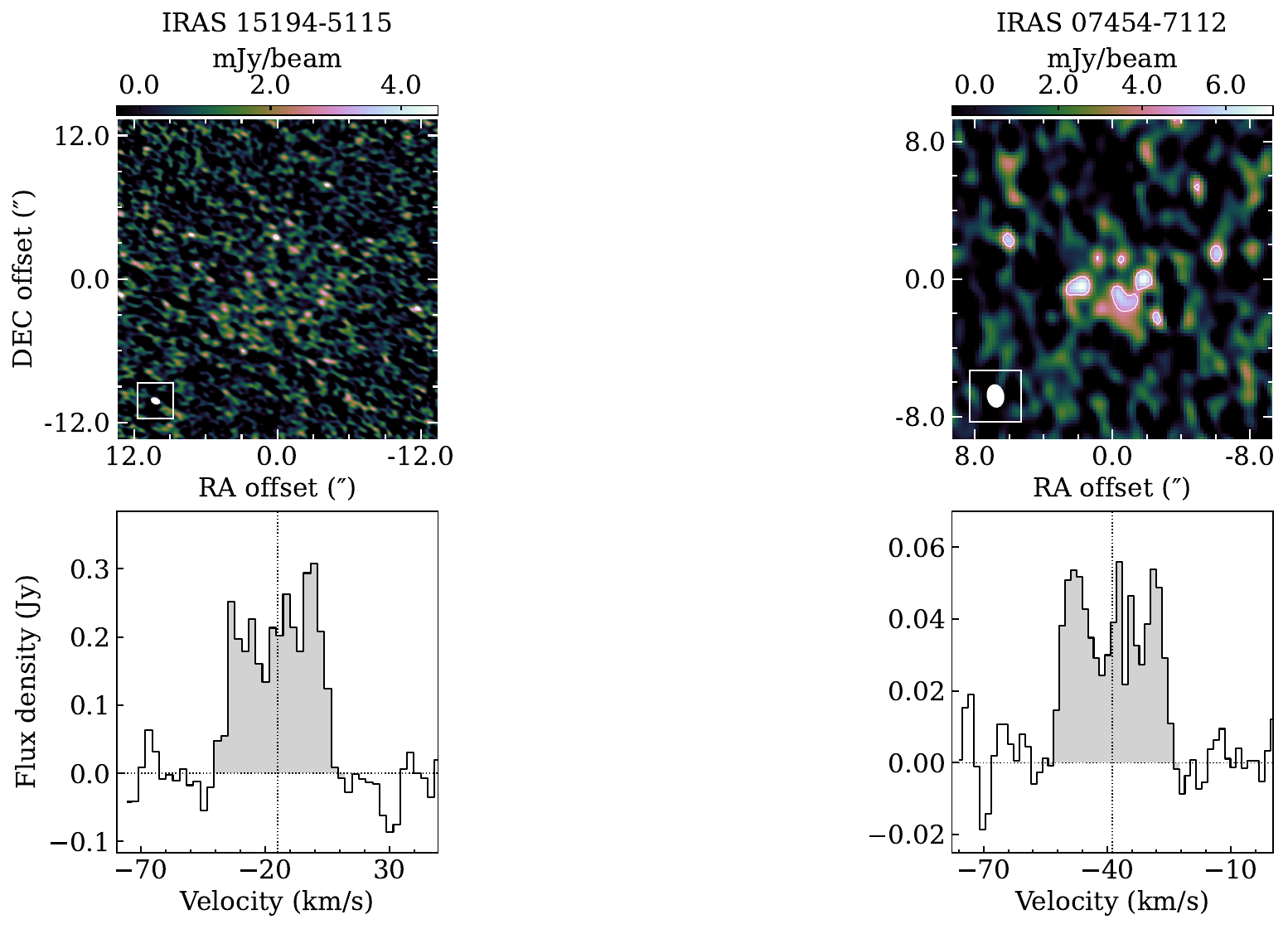}
    \caption{H$^{13}$CCCN, 12-11 (105.799113 GHz)}
\end{figure}

\begin{figure}[h]
    \centering
    \includegraphics[width=0.8\linewidth]{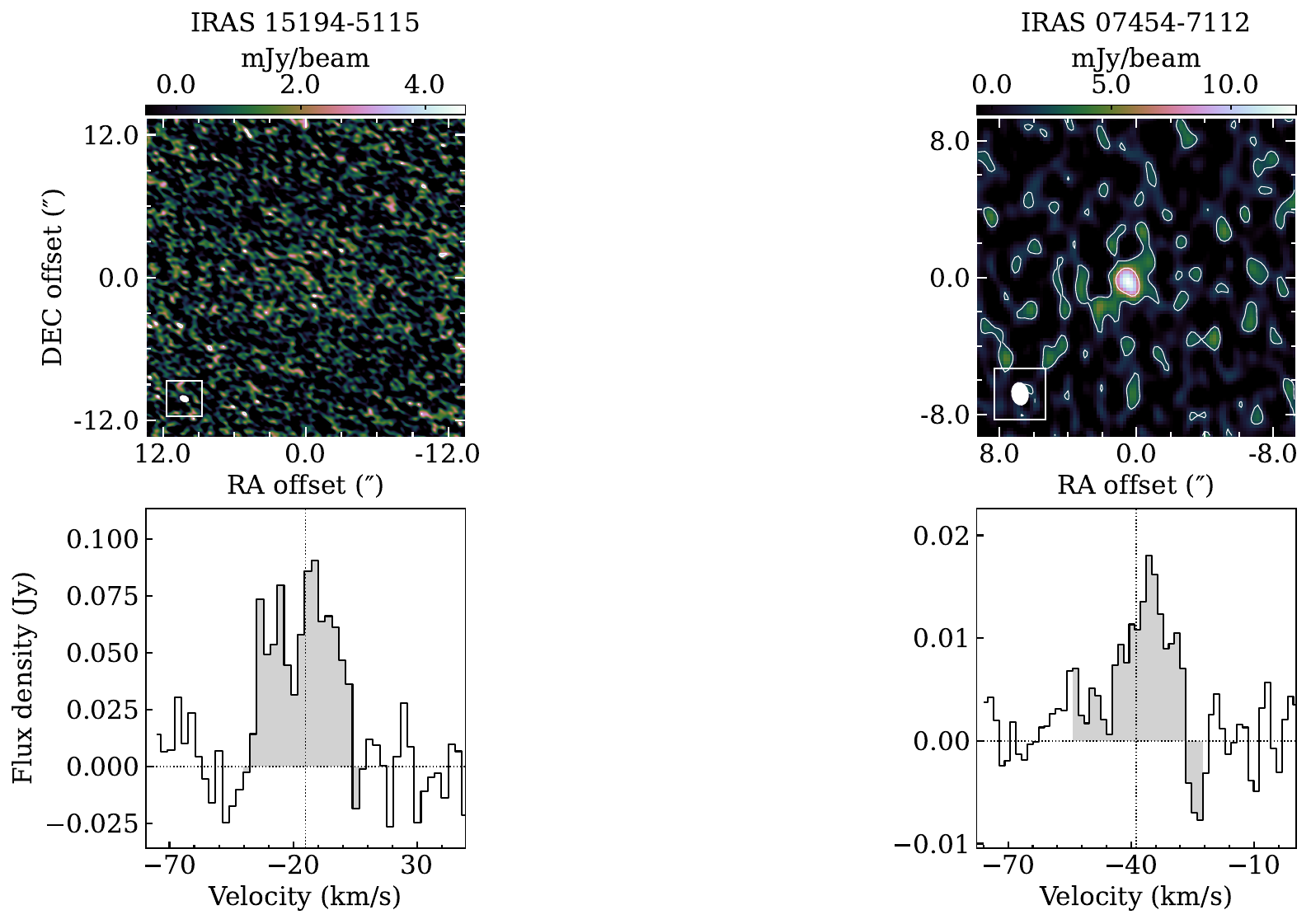}
    \caption{Si$^{34}$S, 6-5 (105.941503 GHz)}
\end{figure}

\begin{figure}[h]
    \centering
    \includegraphics[width=0.75\linewidth]{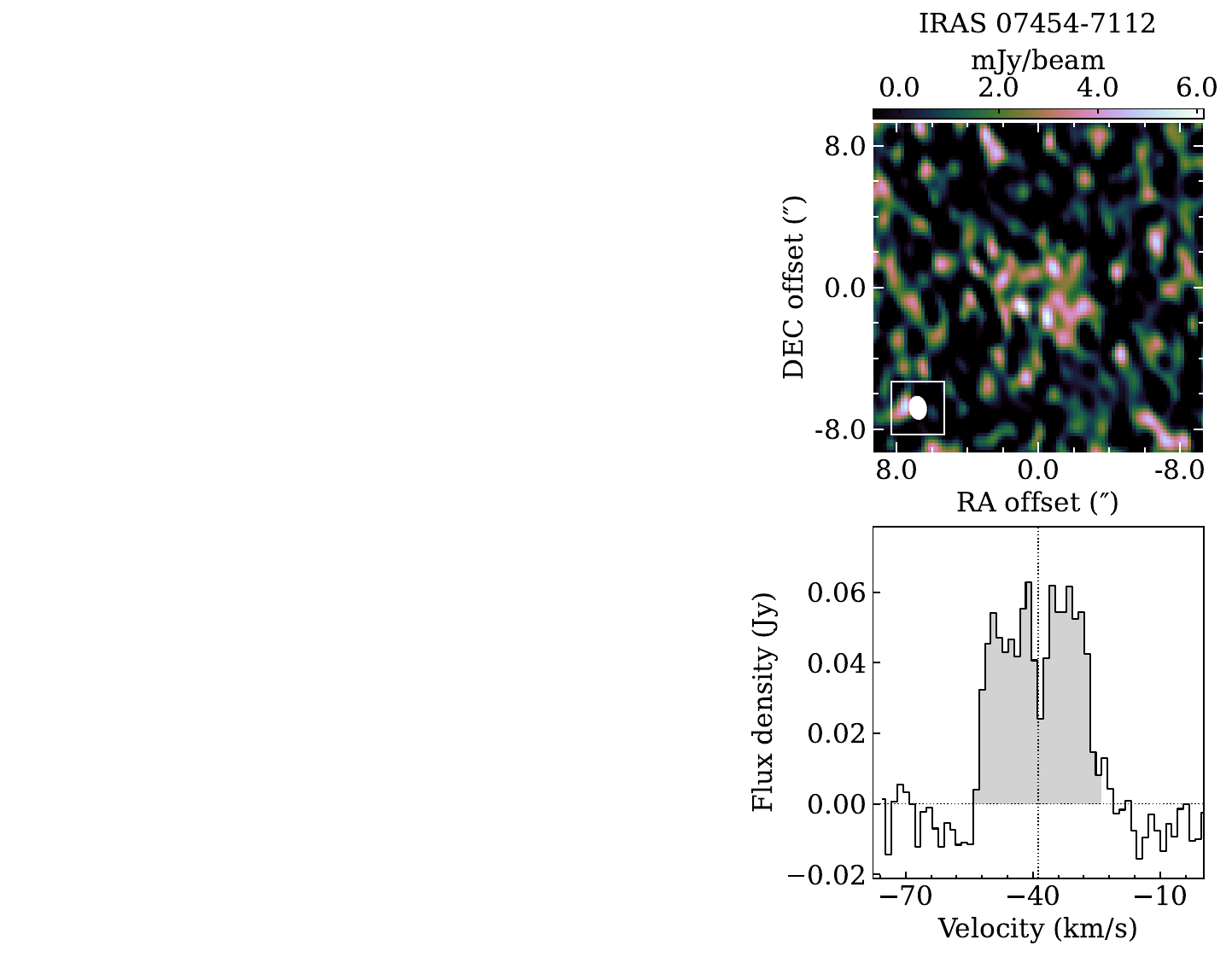}
    \caption{CCS, 8$_9$-7$_8$ (106.347726 GHz)}
\end{figure}

\begin{figure}[h]
    \centering
    \includegraphics[width=0.8\linewidth]{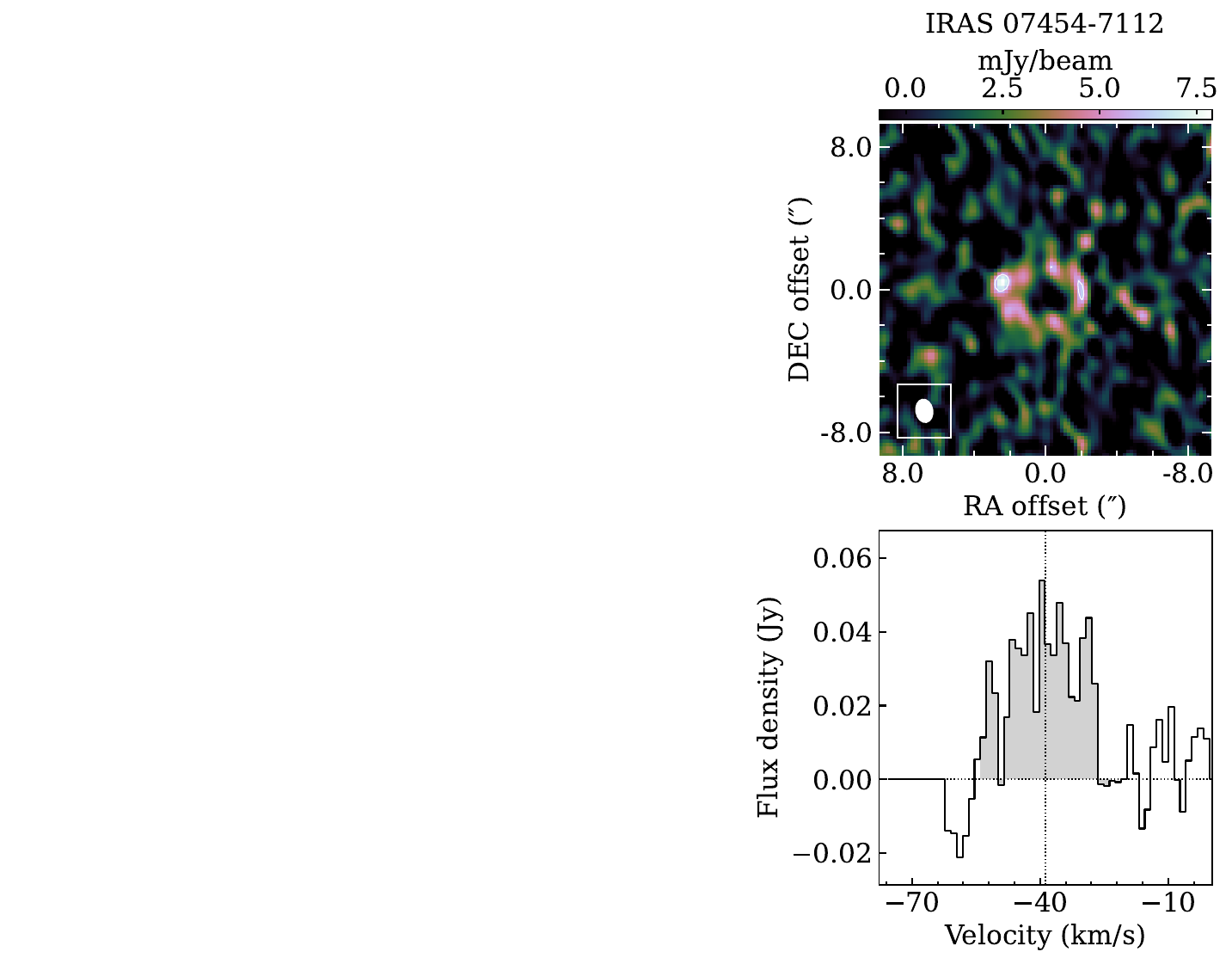}
    \caption{HC$_5$N, 40-39 (106.49891 GHz)}
\end{figure}

\begin{figure}[h]
    \centering
    \includegraphics[width=0.8\linewidth]{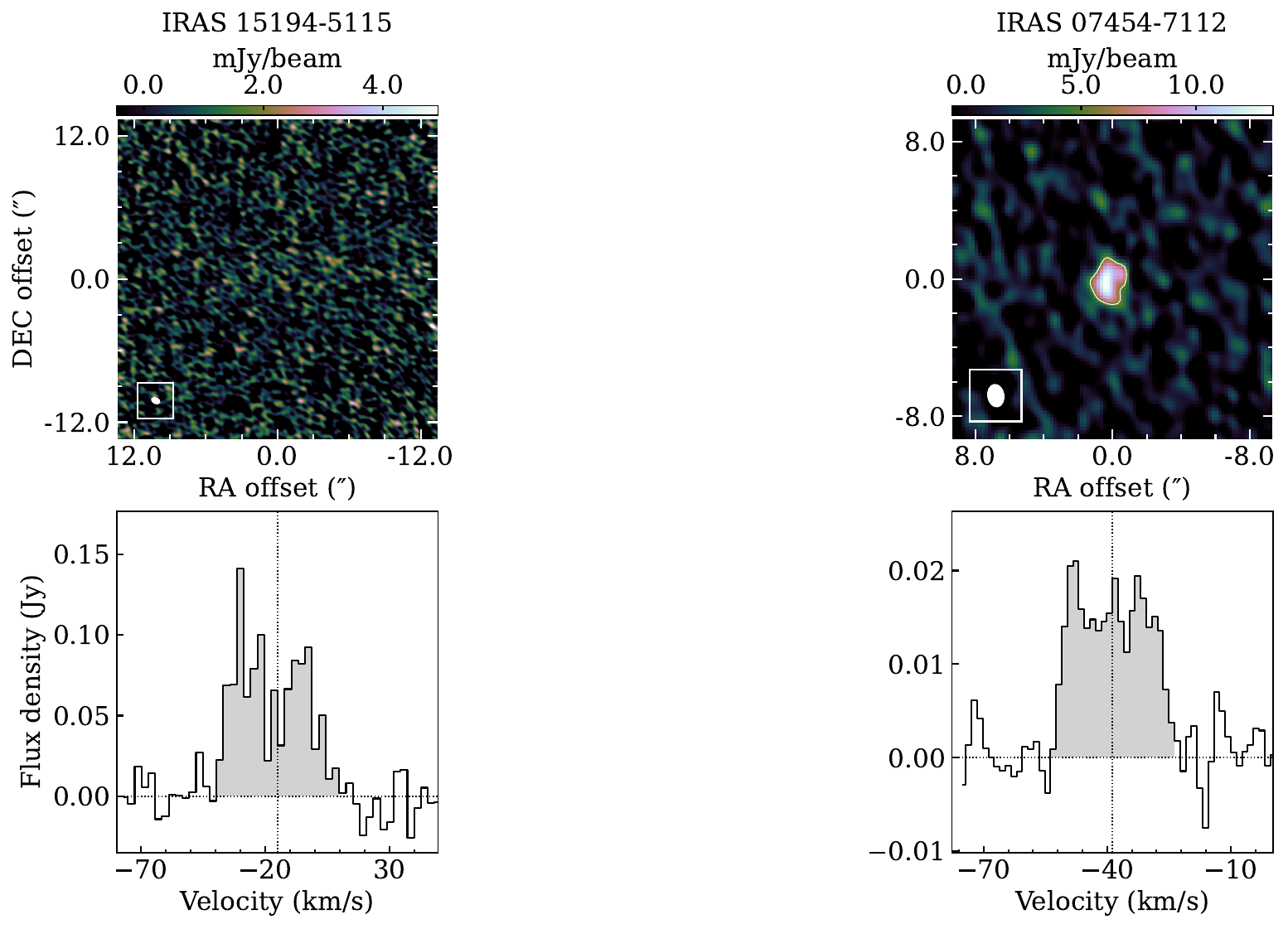}
    \caption{$^{29}$SiS, 6-5 (106.92298 GHz)}
\end{figure}

\begin{figure}[h]
    \centering
    \includegraphics[width=0.8\linewidth]{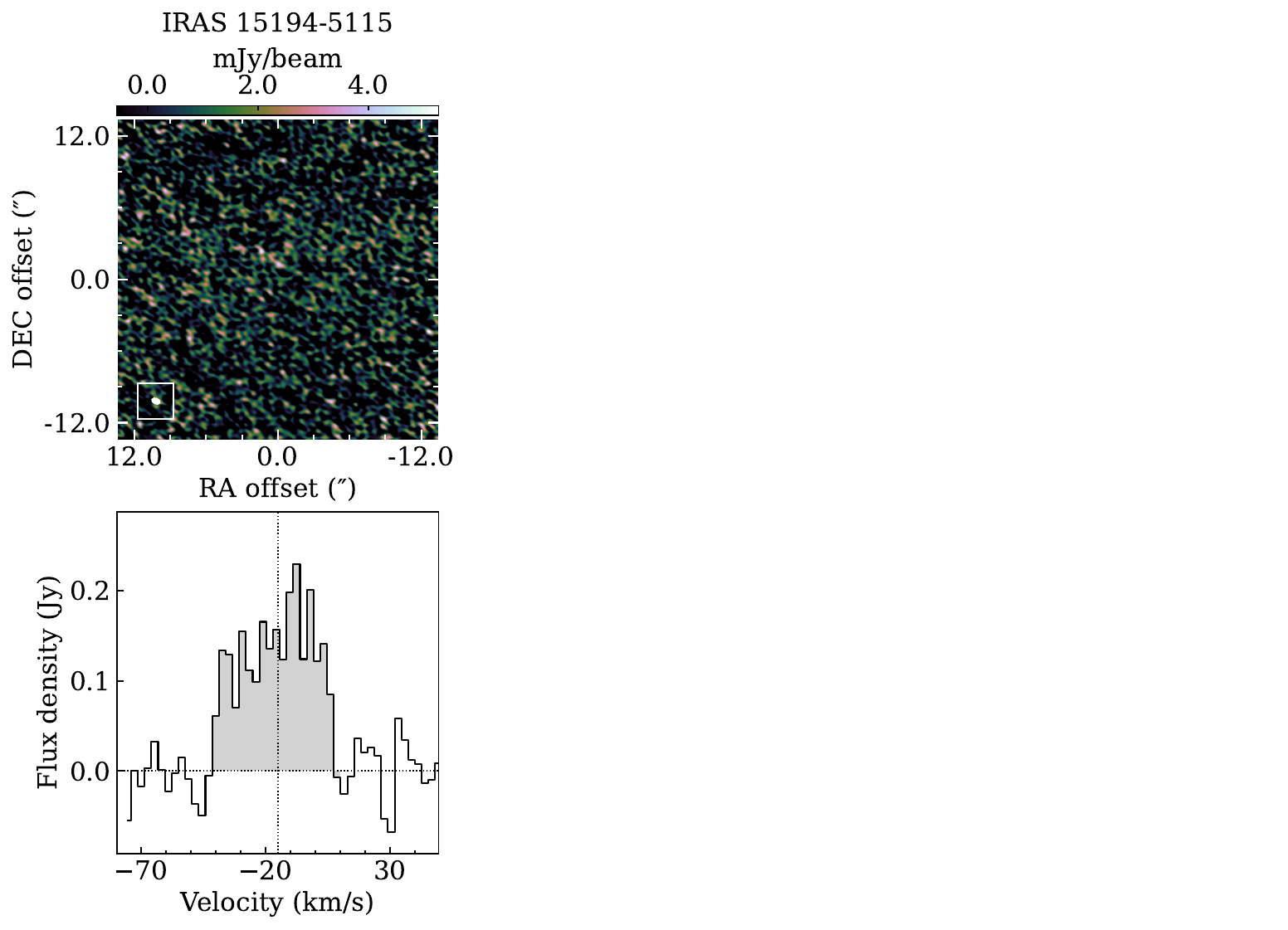}
    \caption{Si$^{13}$CC, 5$_{1,5}$-4$_{1,4}$ (107.971477 GHz)}
\end{figure}

\begin{figure}[h]
    \centering
    \includegraphics[width=0.8\linewidth]{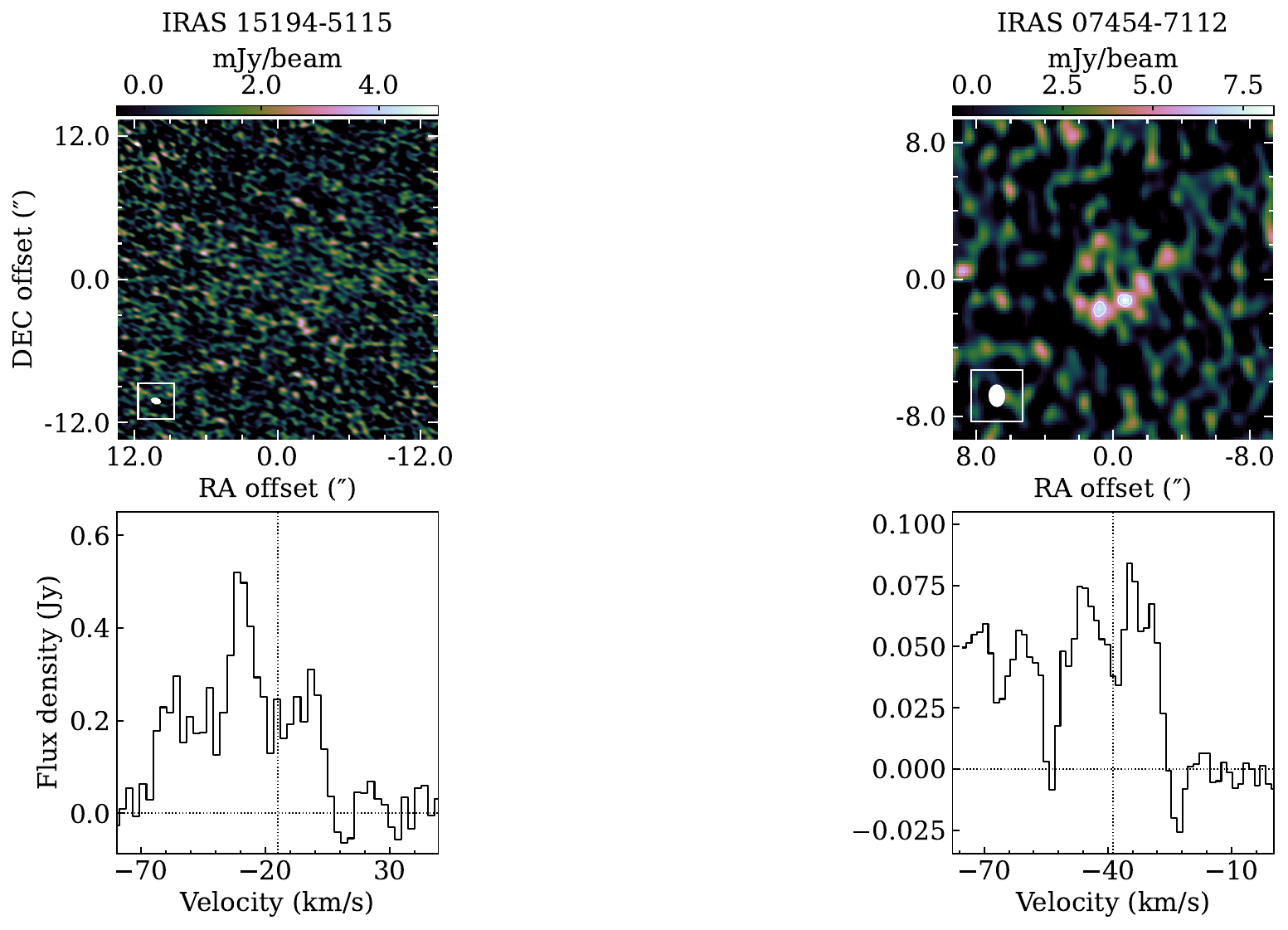}
    \caption{HC$^{13}$CCN, 12-11 (108.710532 GHz)}
\end{figure}

\begin{figure}[h]
    \centering
    \includegraphics[width=0.8\linewidth]{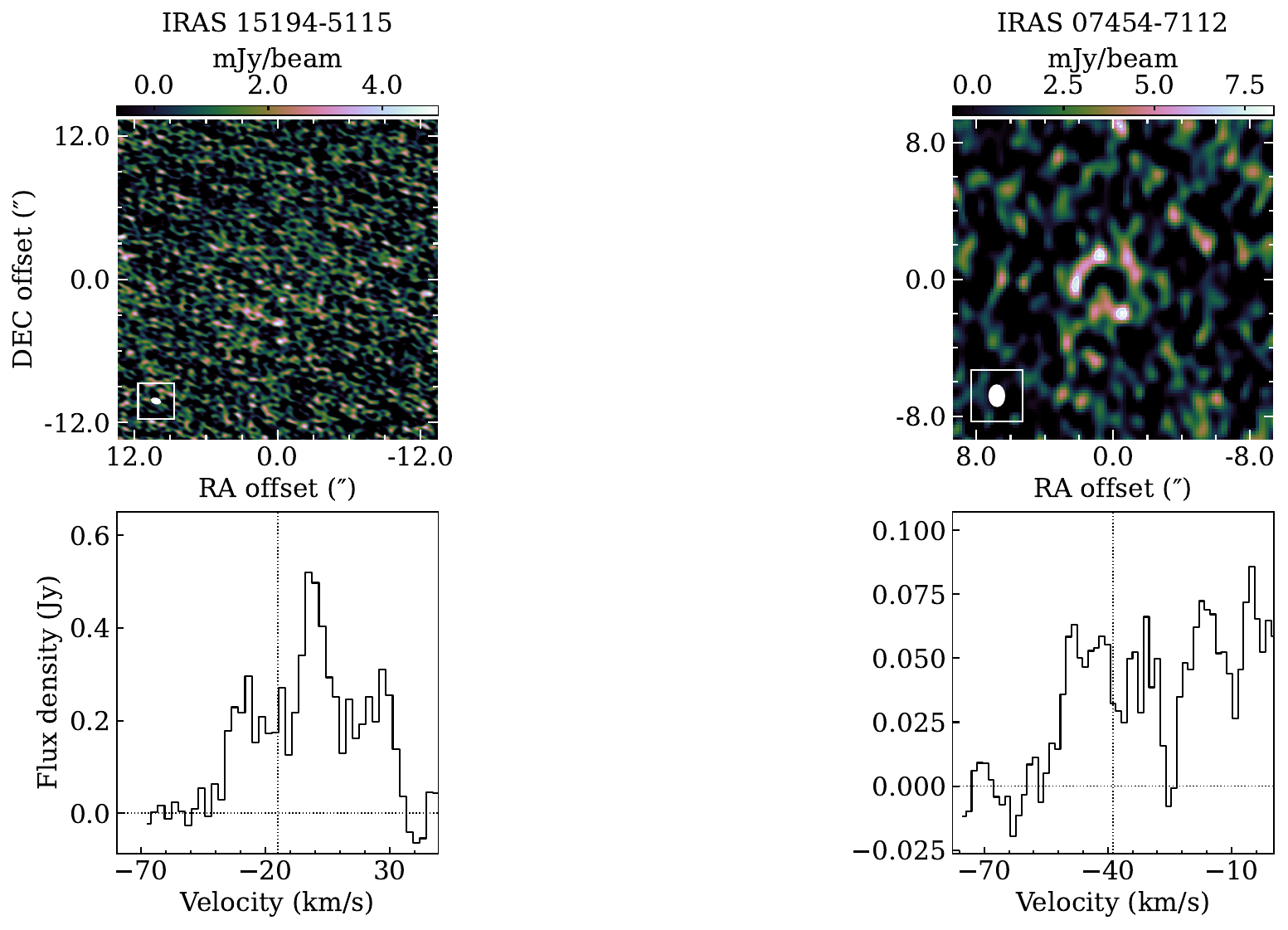}
    \caption{HCC$^{13}$CN, 12-11 (108.720999 GHz)}
\end{figure}

\begin{figure}[h]
    \centering
    \includegraphics[width=0.8\linewidth]{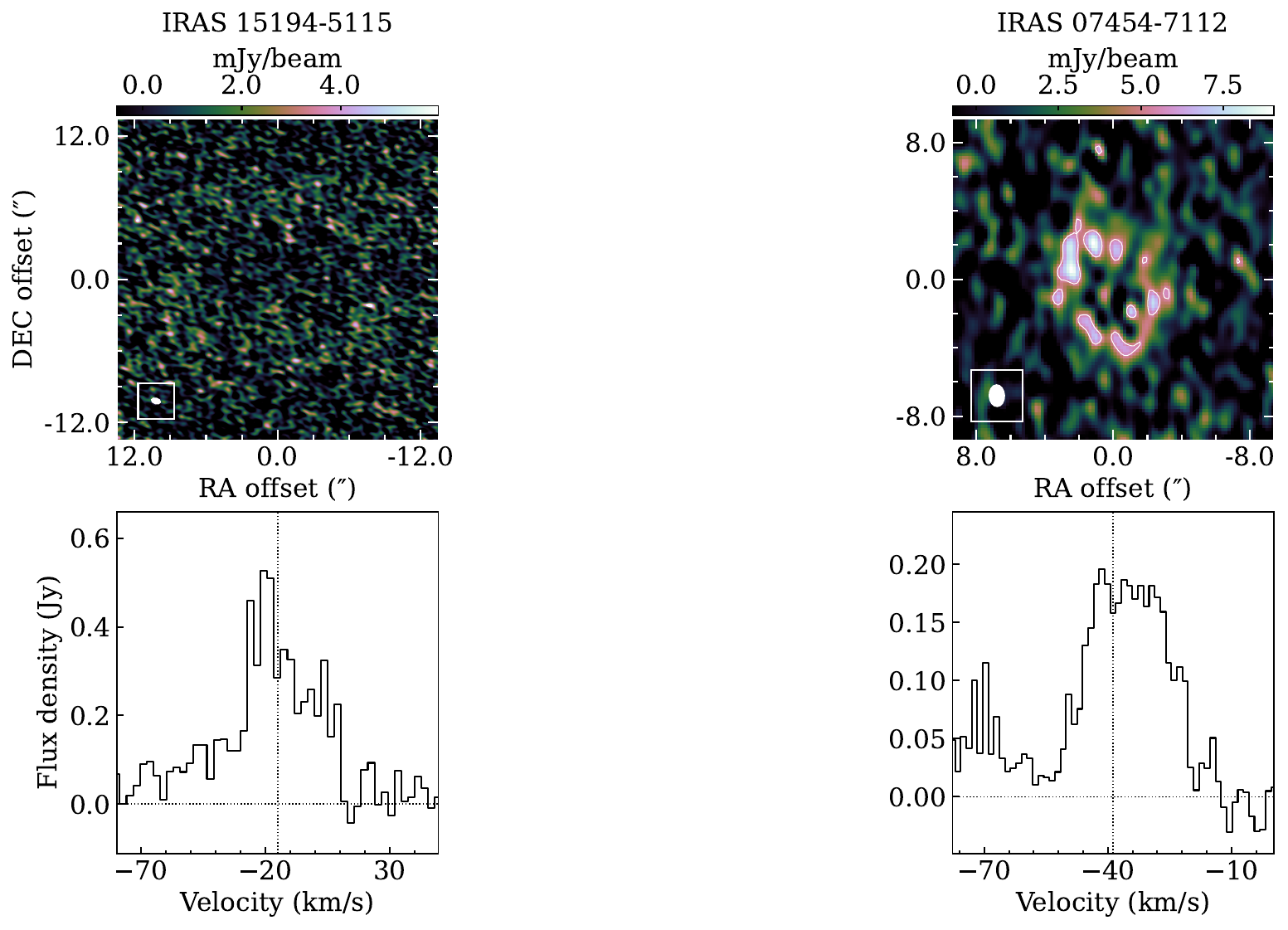}
    \caption{$^{13}$CN, N=1-0, J=3/2-1/2 (108.782374 GHz)}
\end{figure}

\begin{figure}[h]
    \centering
    \includegraphics[width=0.8\linewidth]{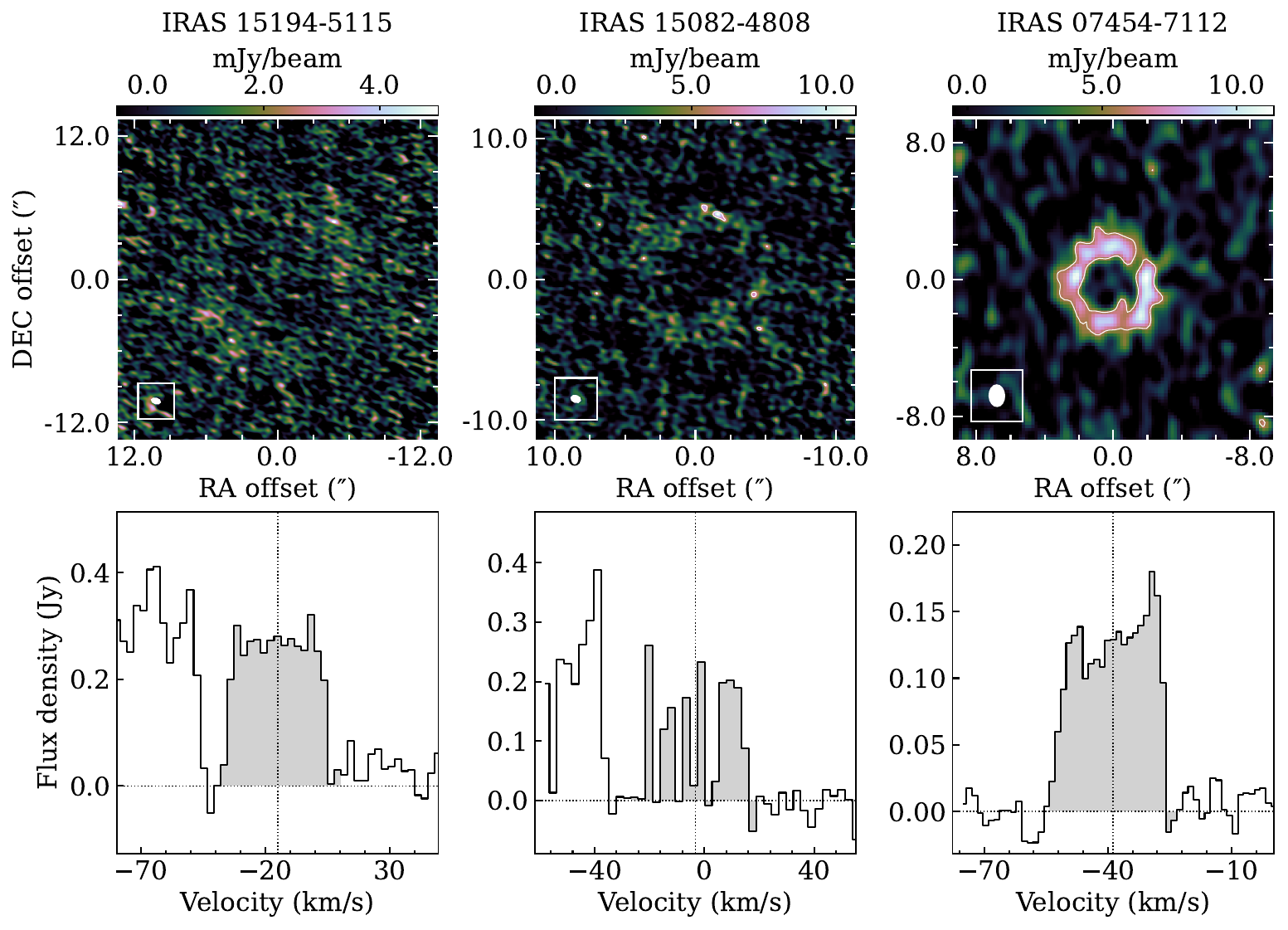}
    \caption{C$_3$N, 11-10 (A) (108.83425 GHz)}
\end{figure}

\begin{figure}[h]
    \centering
    \includegraphics[width=0.8\linewidth]{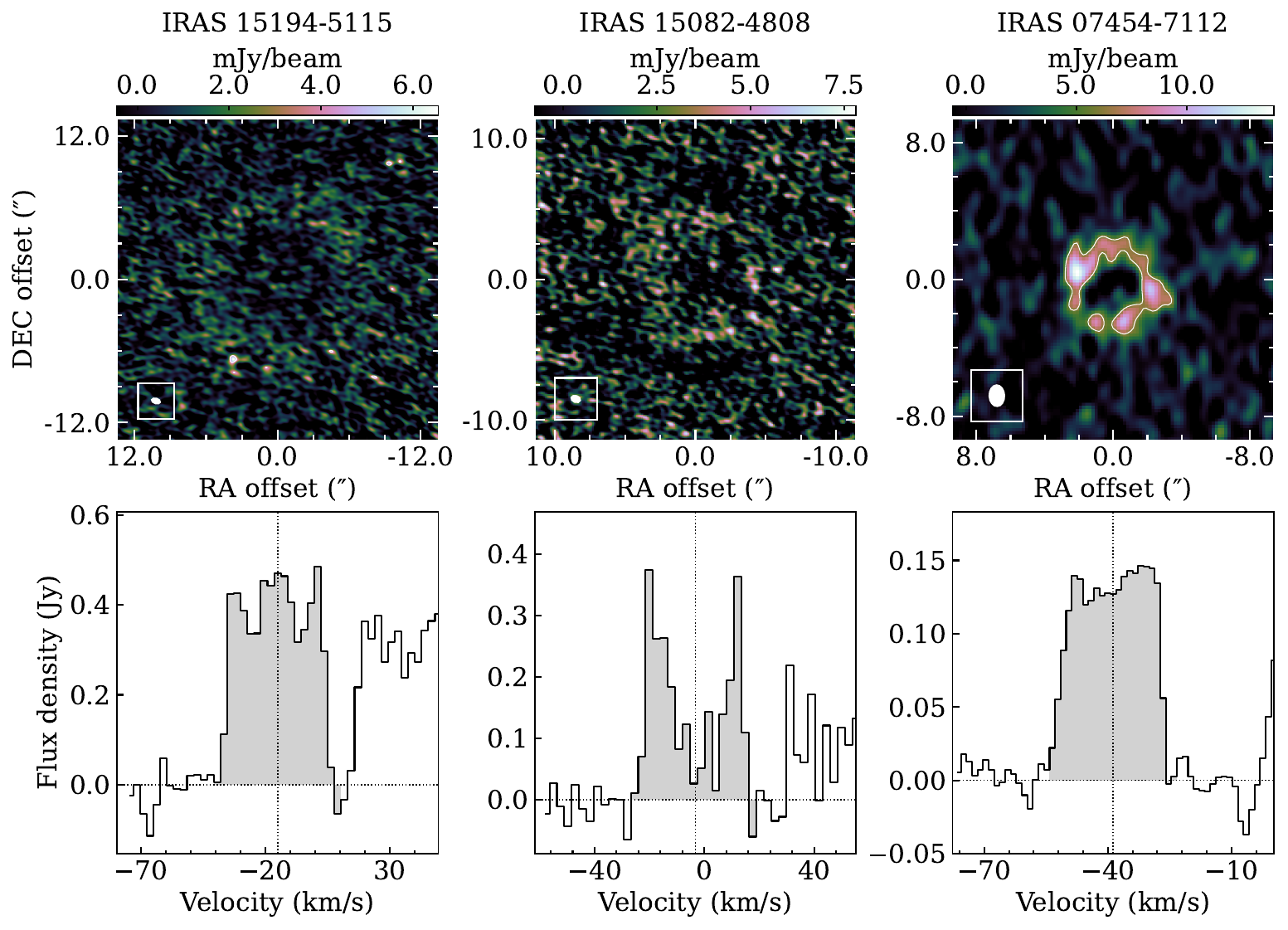}
    \caption{C$_3$N, 11-10 (B) (108.853 GHz)}
\end{figure}

\begin{figure}[h]
    \centering
    \includegraphics[width=0.8\linewidth]{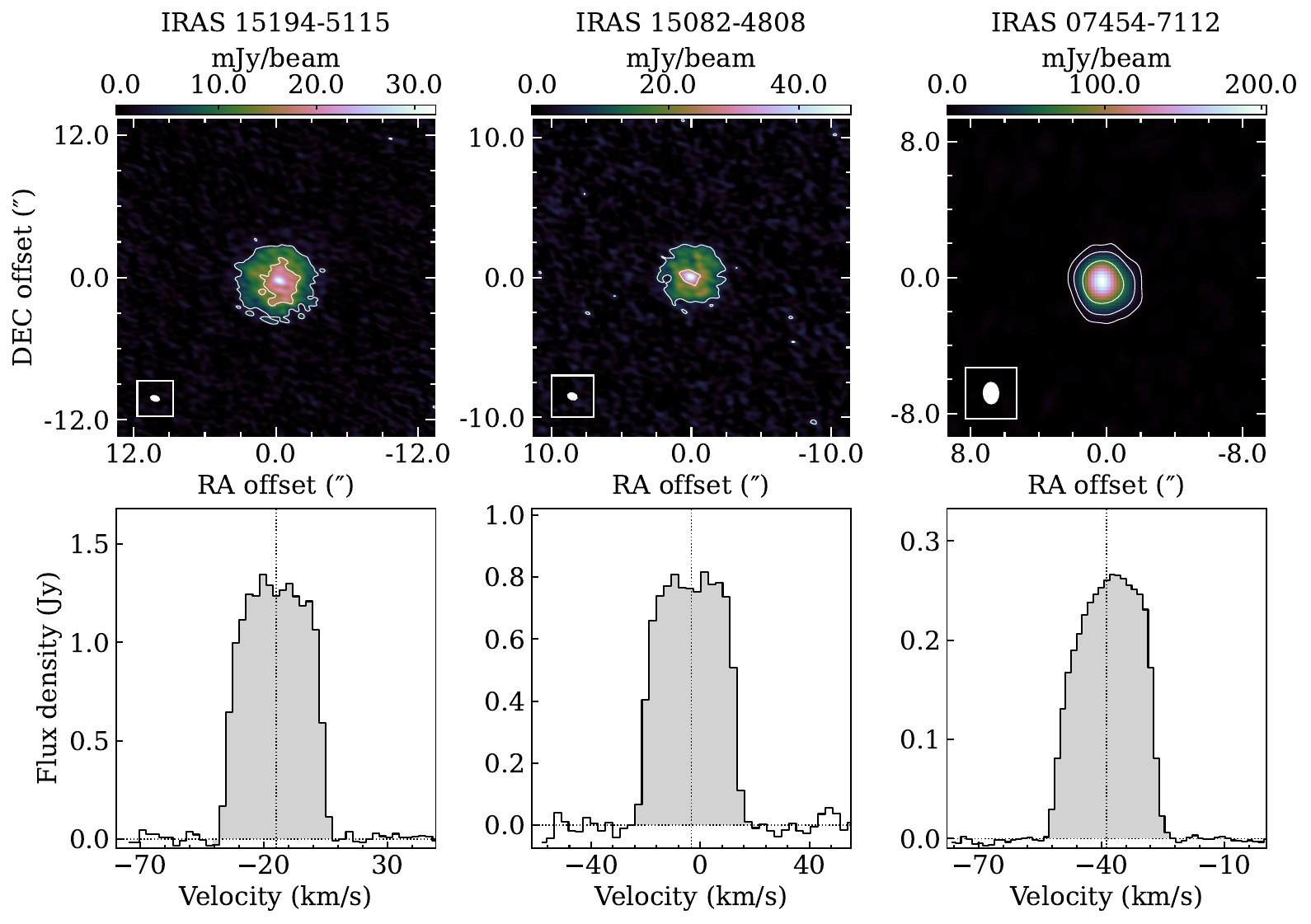}
    \caption{SiS, 6-5 (108.924301 GHz)}
\end{figure}

\begin{figure}[h]
    \centering
    \includegraphics[width=0.8\linewidth]{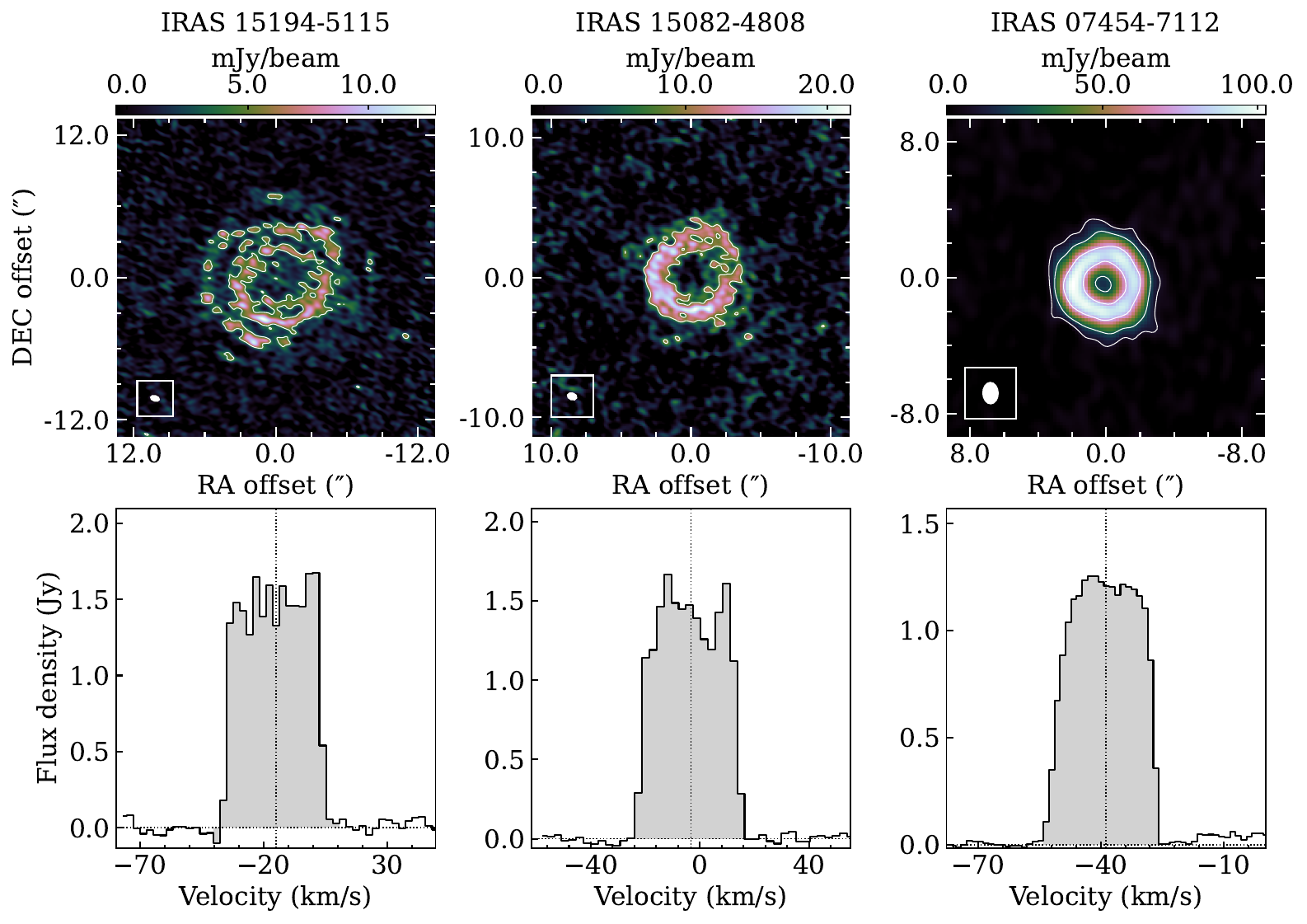}
    \caption{HC$_3$N, 12-11 (109.173634 GHz)}
\end{figure}

\begin{figure}[h]
    \centering
    \includegraphics[width=0.8\linewidth]{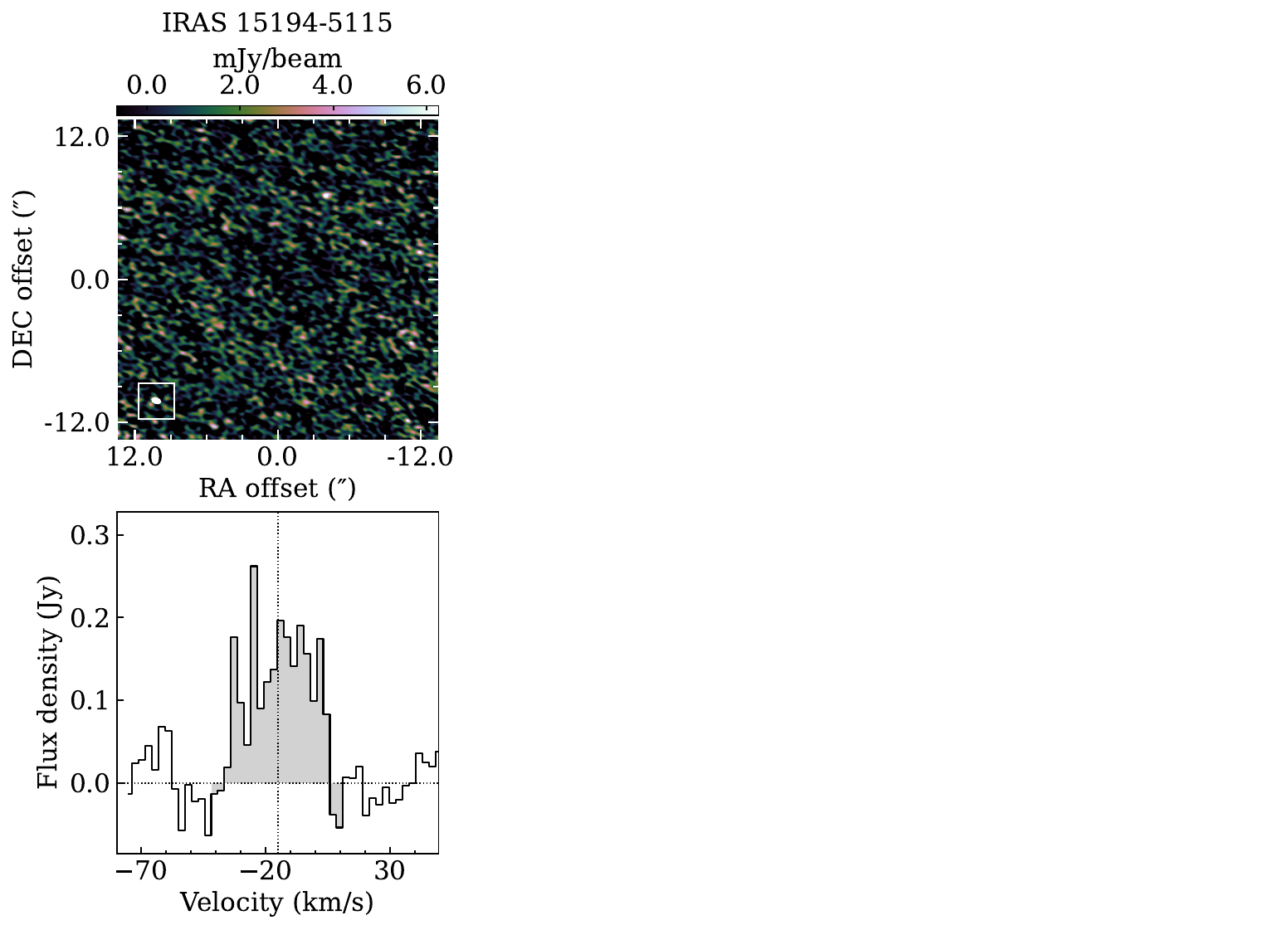}
    \caption{CCC$^{13}$CH, N=12-11, J=25/2-23/2 (110.735027 GHz)}
\end{figure}

\begin{figure}[h]
    \centering
    \includegraphics[width=0.8\linewidth]{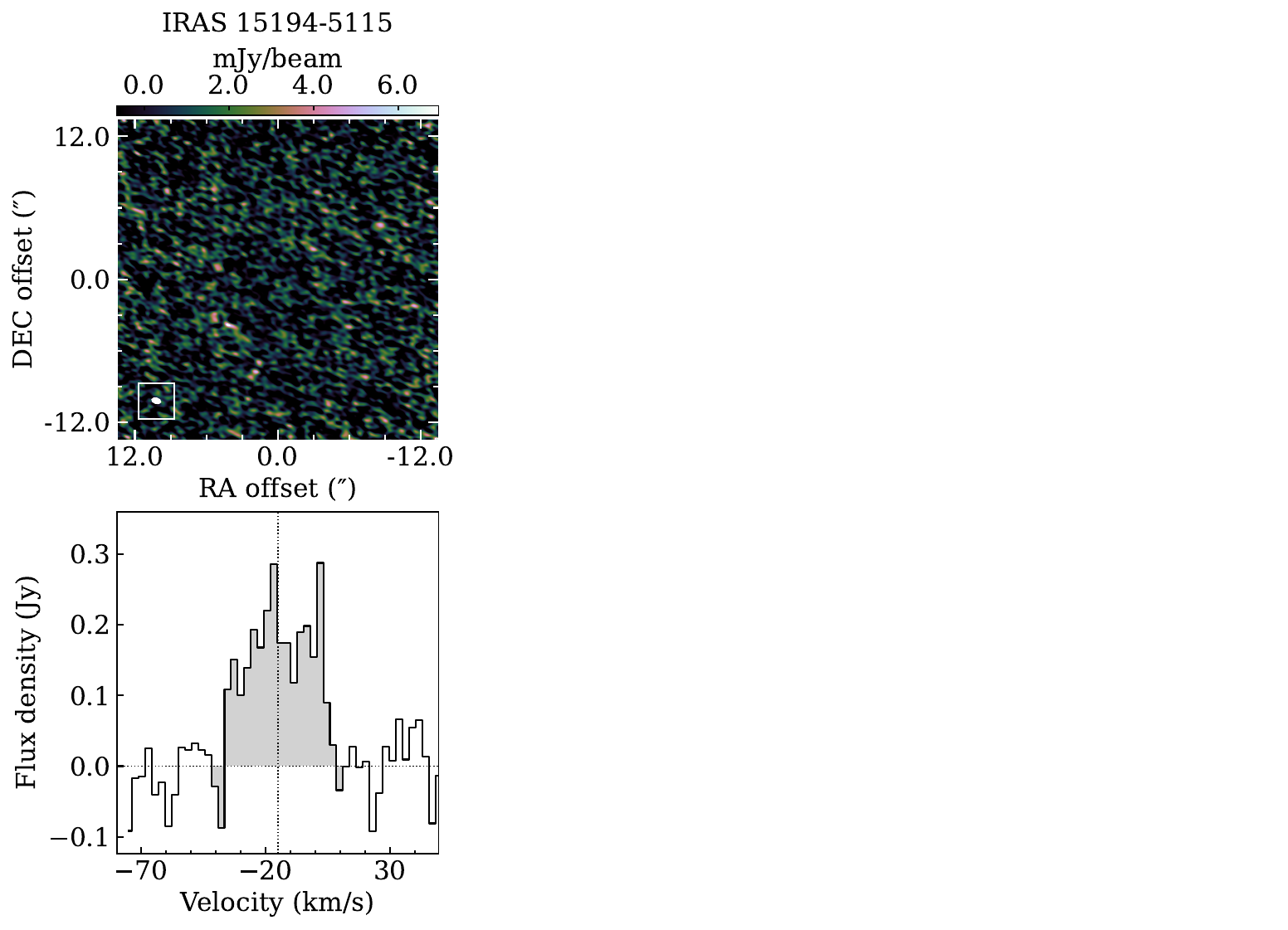}
    \caption{CCC$^{13}$CH, N=12-11, J=23/2-21/2 (110.772434 GHz)}
\end{figure}

\begin{figure}[h]
    \centering
    \includegraphics[width=0.8\linewidth]{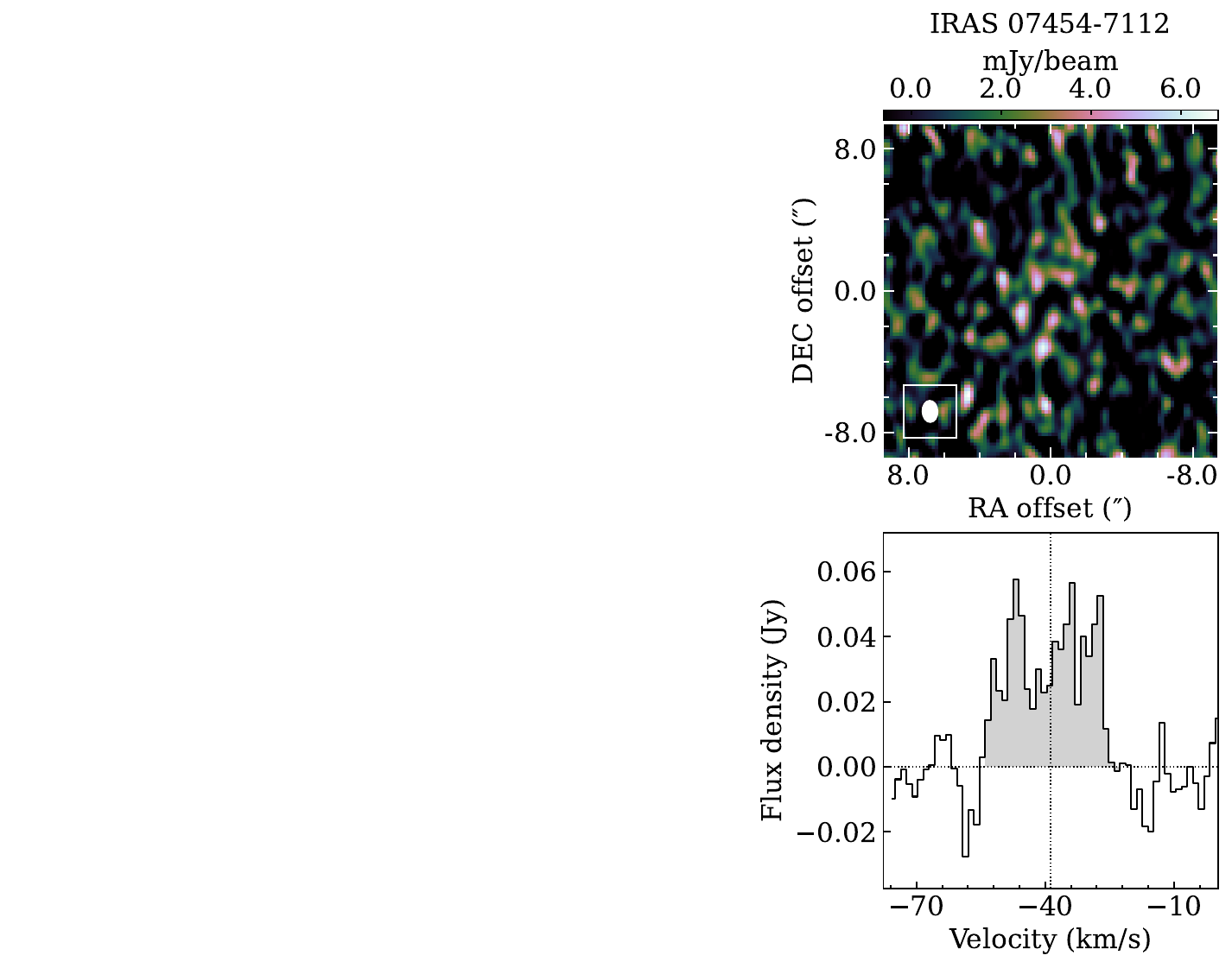}
    \caption{HC$_5$N, 42-41 (111.823024 GHz)}
\end{figure}

\begin{figure}[h]
    \centering
    \includegraphics[width=0.8\linewidth]{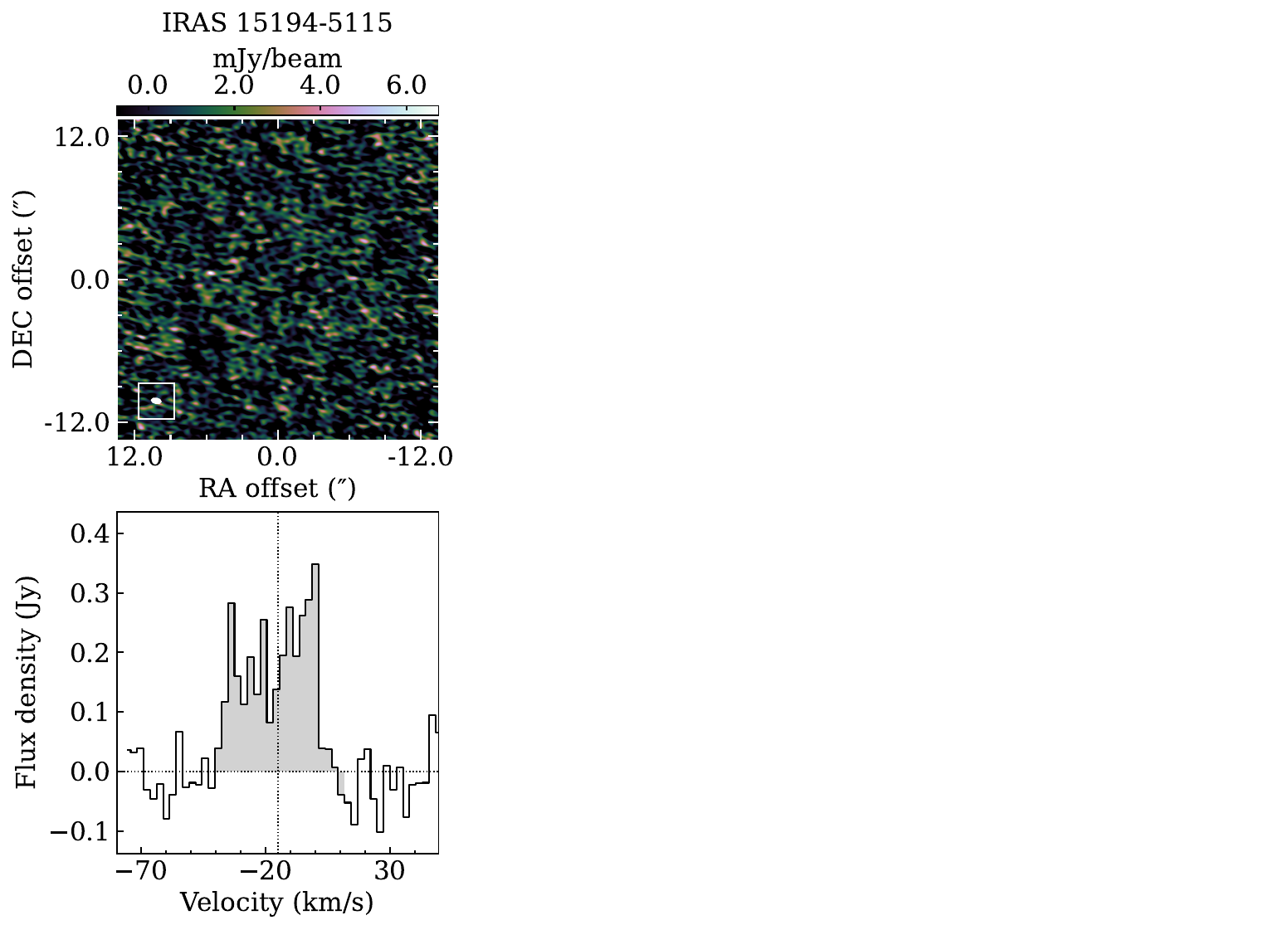}
    \caption{Si$^{13}$CC, 5$_{0,5}$-4$_{0,4}$ (112.593212 GHz)}
\end{figure}

\begin{figure}[h]
    \centering
    \includegraphics[width=0.8\linewidth]{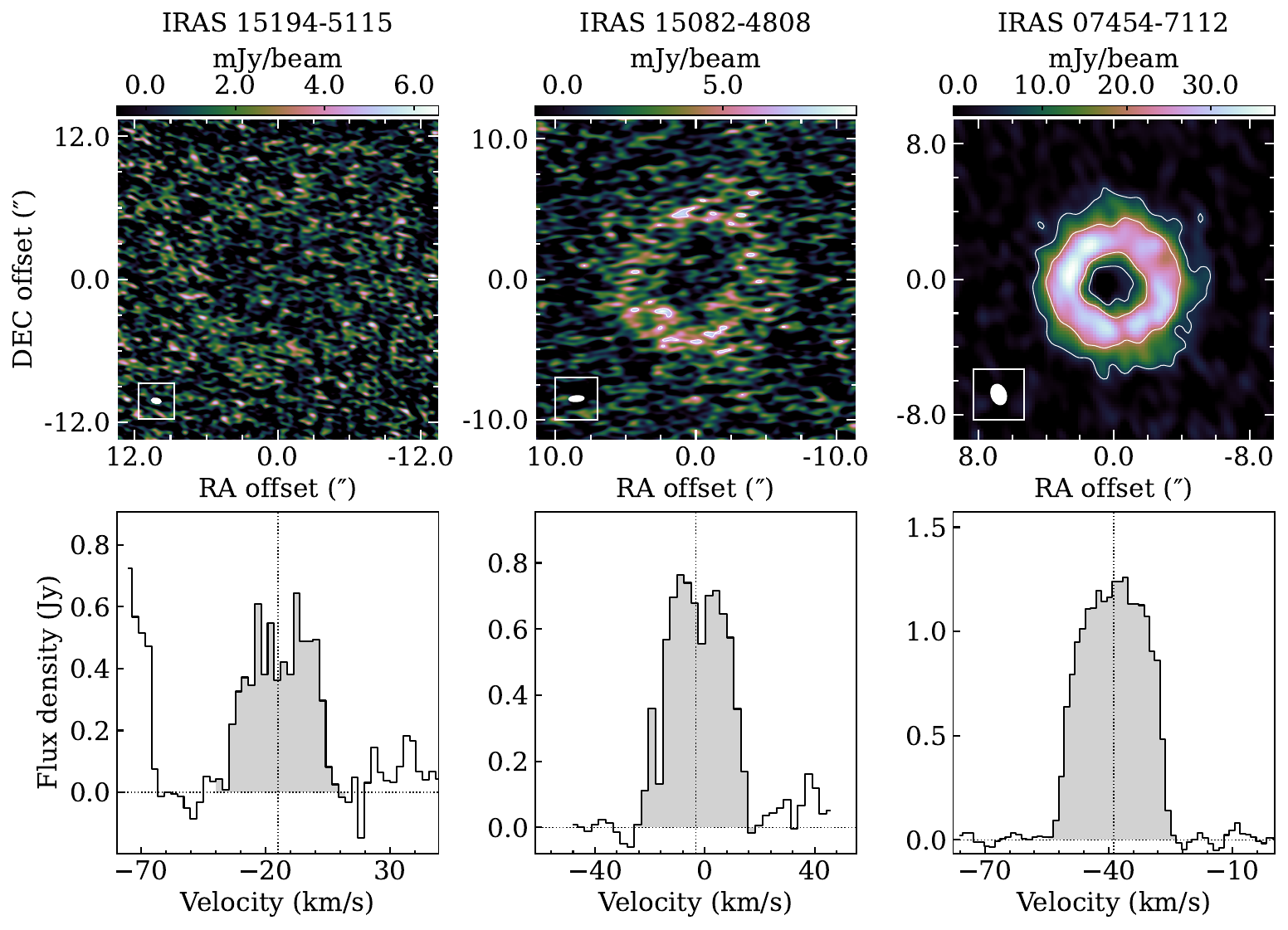}
    \caption{CN, N=1-0, J=1/2-1/2, F=1/2-3/2 (113.144157 GHz)}
\end{figure}

\begin{figure}[h]
    \centering
    \includegraphics[width=0.8\linewidth]{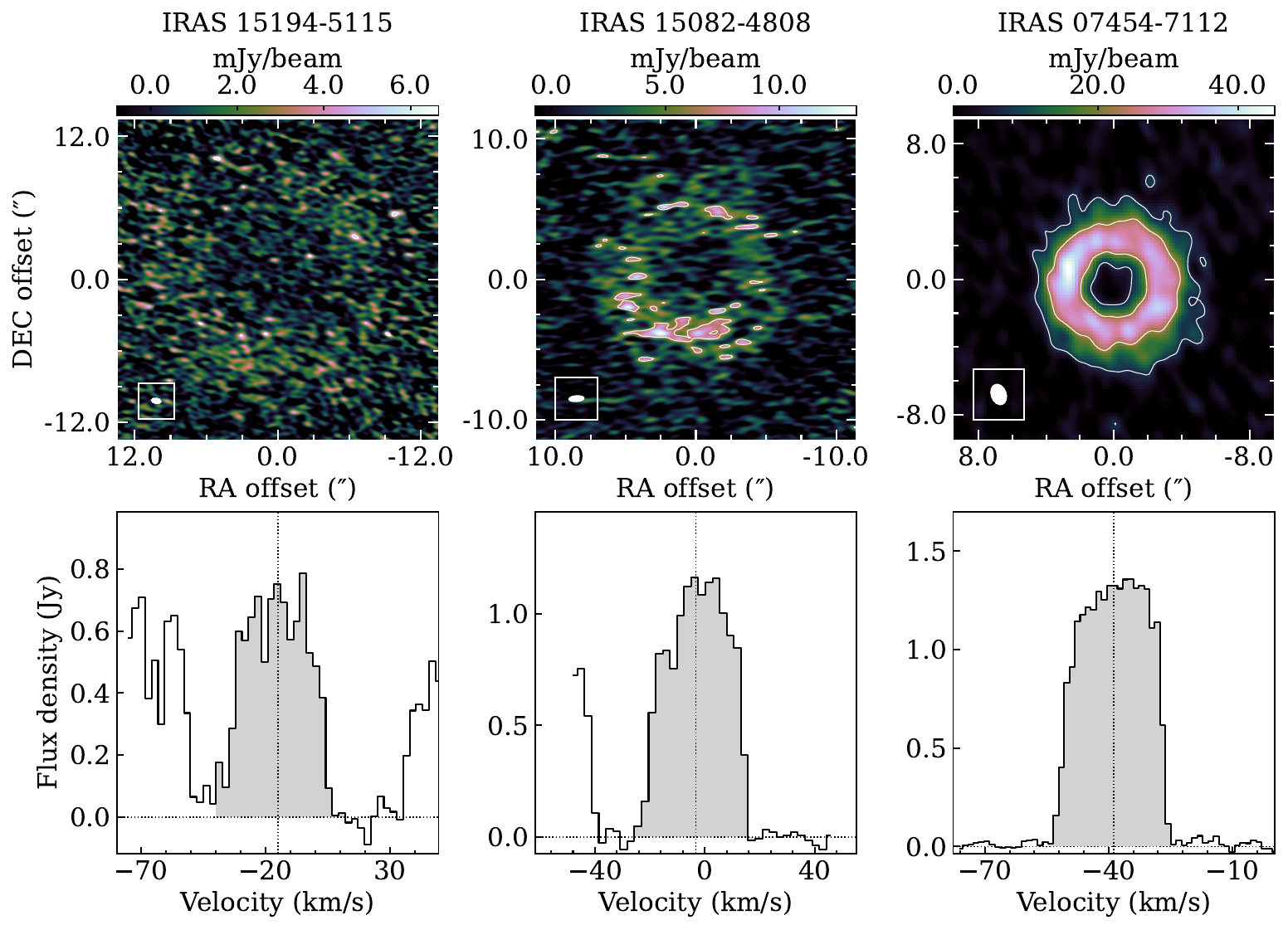}
    \caption{CN, N=1-0, J=1/2-1/2, F=3/2-1/2 (113.170492 GHz)}
    \label{fig:CN_app_B}
\end{figure}

\begin{figure}[h]
    \centering
    \includegraphics[width=0.8\linewidth]{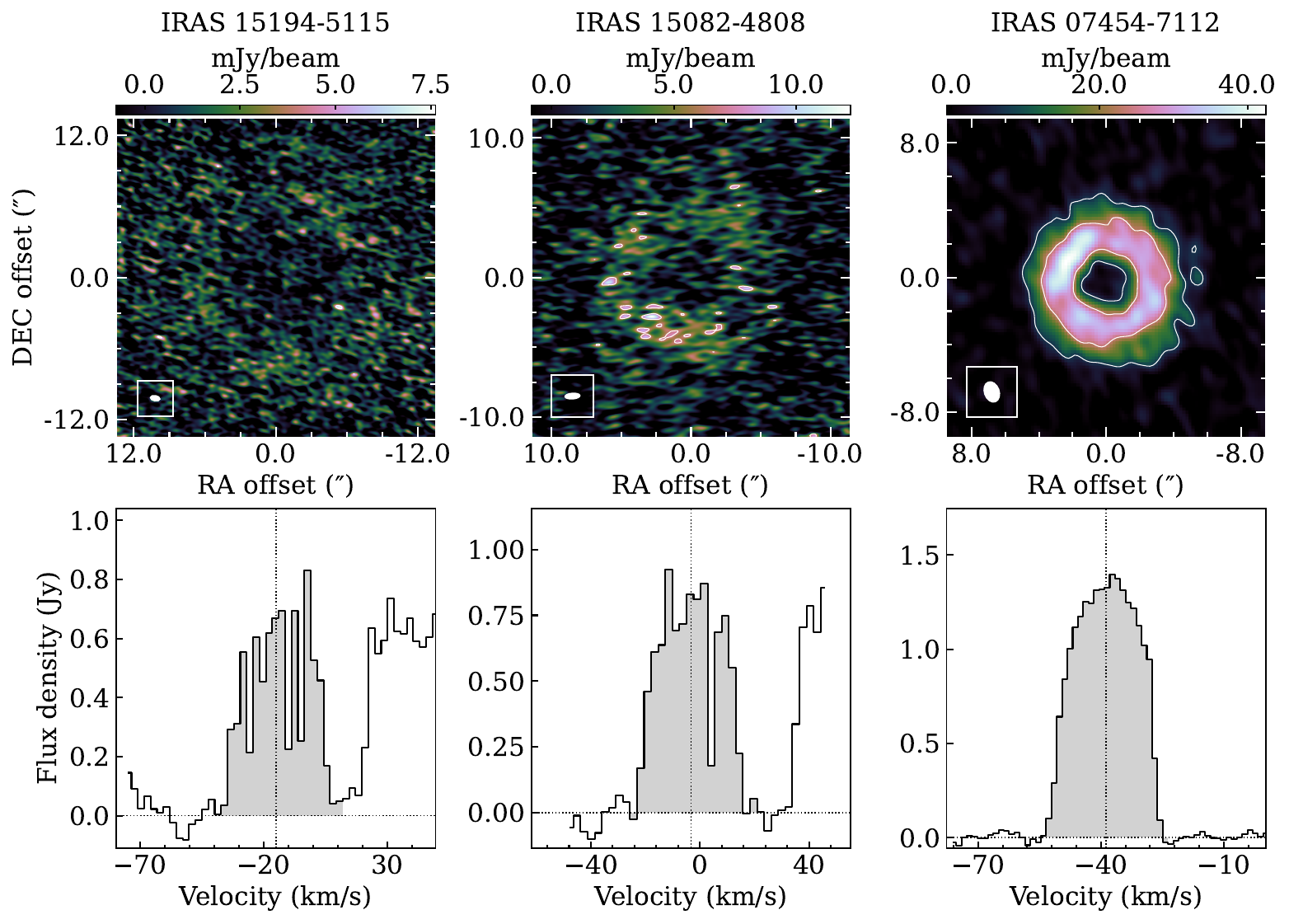}
    \caption{CN, N=1-0, J=1/2-1/2, F=3/2-3/2 (113.191279 GHz)}
\end{figure}

\begin{figure}[h]
    \centering
    \includegraphics[width=0.8\linewidth]{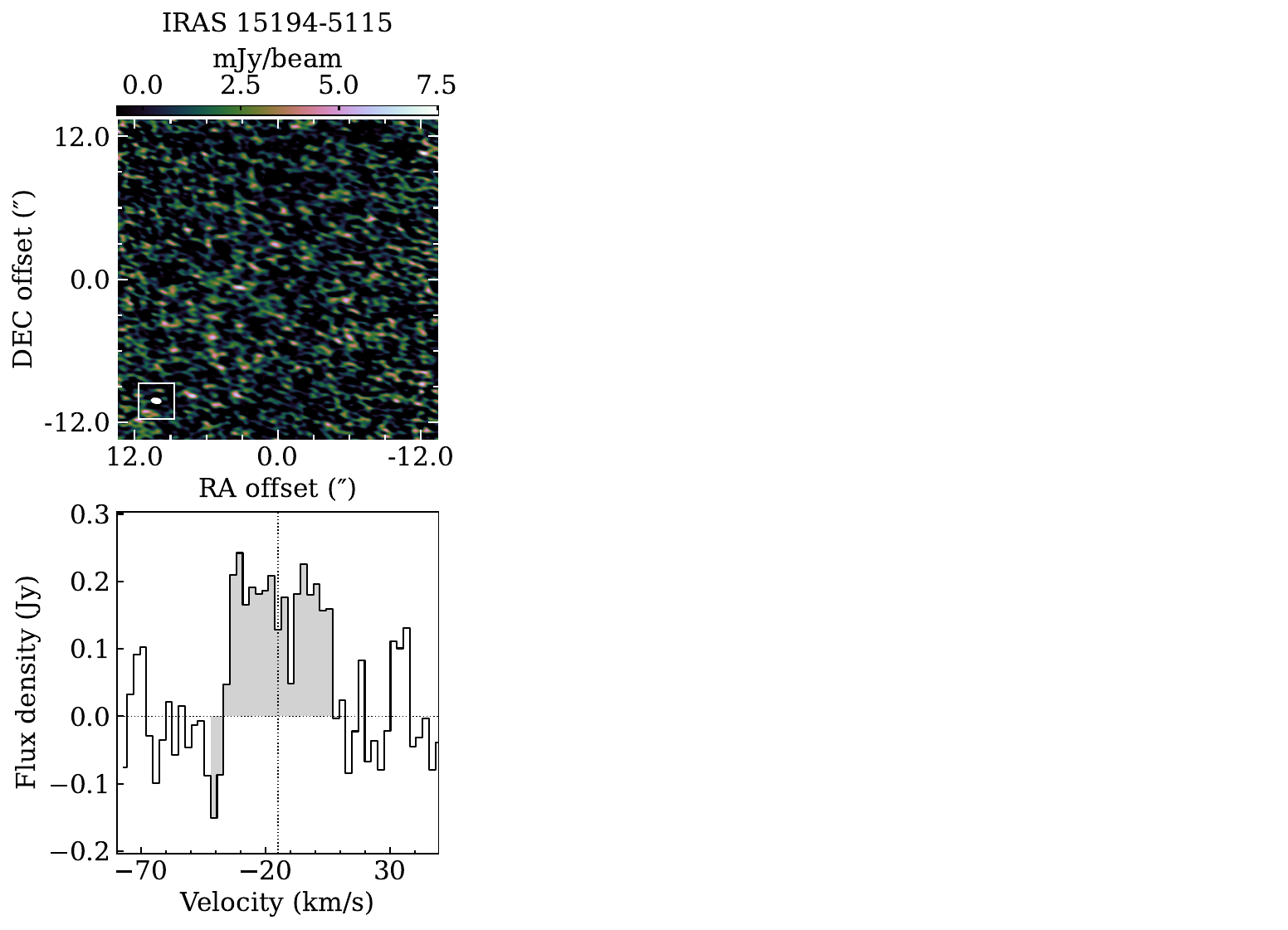}
    \caption{C$^{13}$CCCH, N=12-11, J=23/2-21/2 (113.6442 GHz)}
\end{figure}

\begin{figure}[h]
    \centering
    \includegraphics[width=0.8\linewidth]{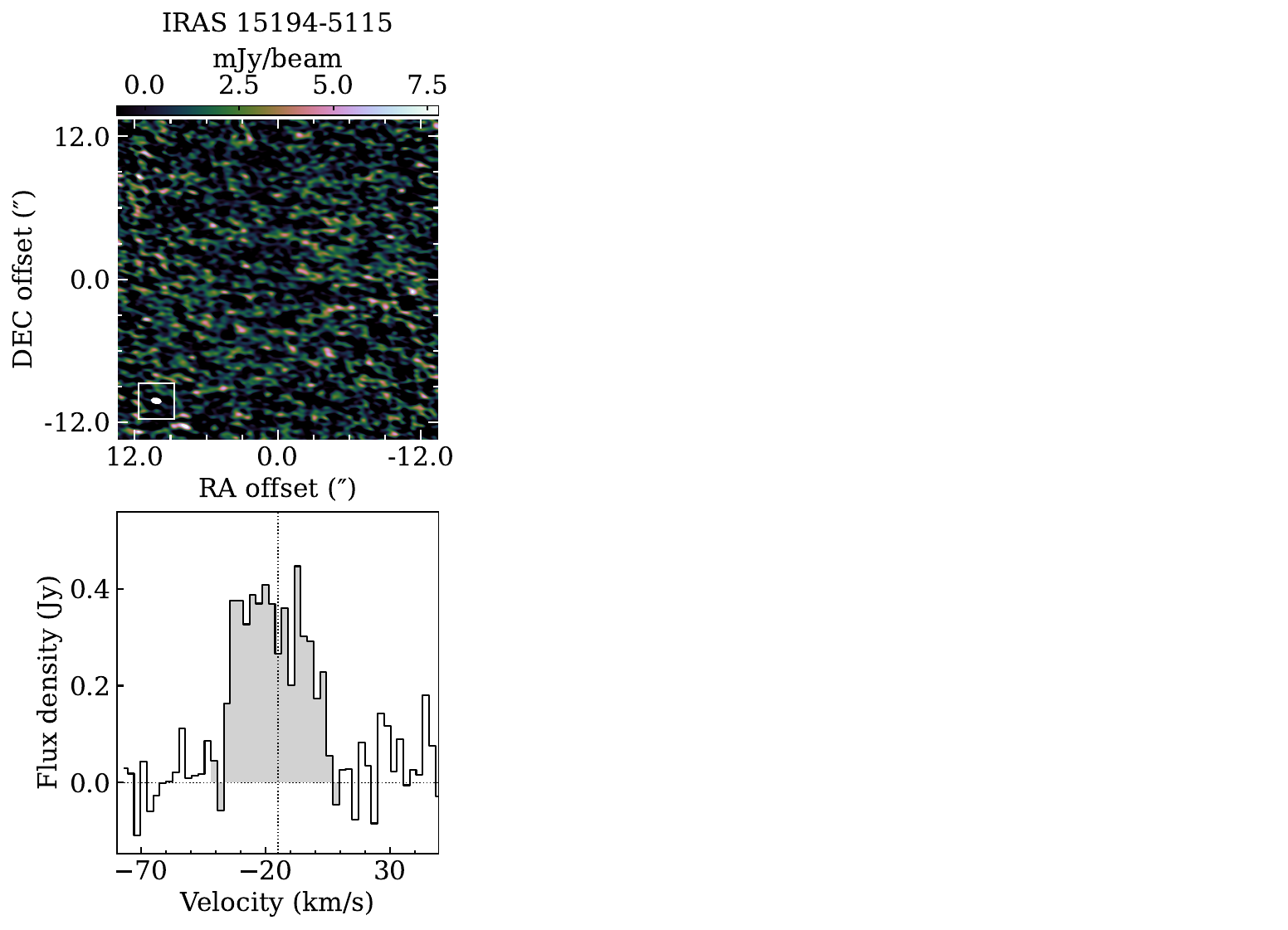}
    \caption{CC$^{13}$CCH, N=12-11, J=23/2-21/2 (113.81808 GHz)}
\end{figure}

\begin{figure}[h]
    \centering
    \includegraphics[width=0.8\linewidth]{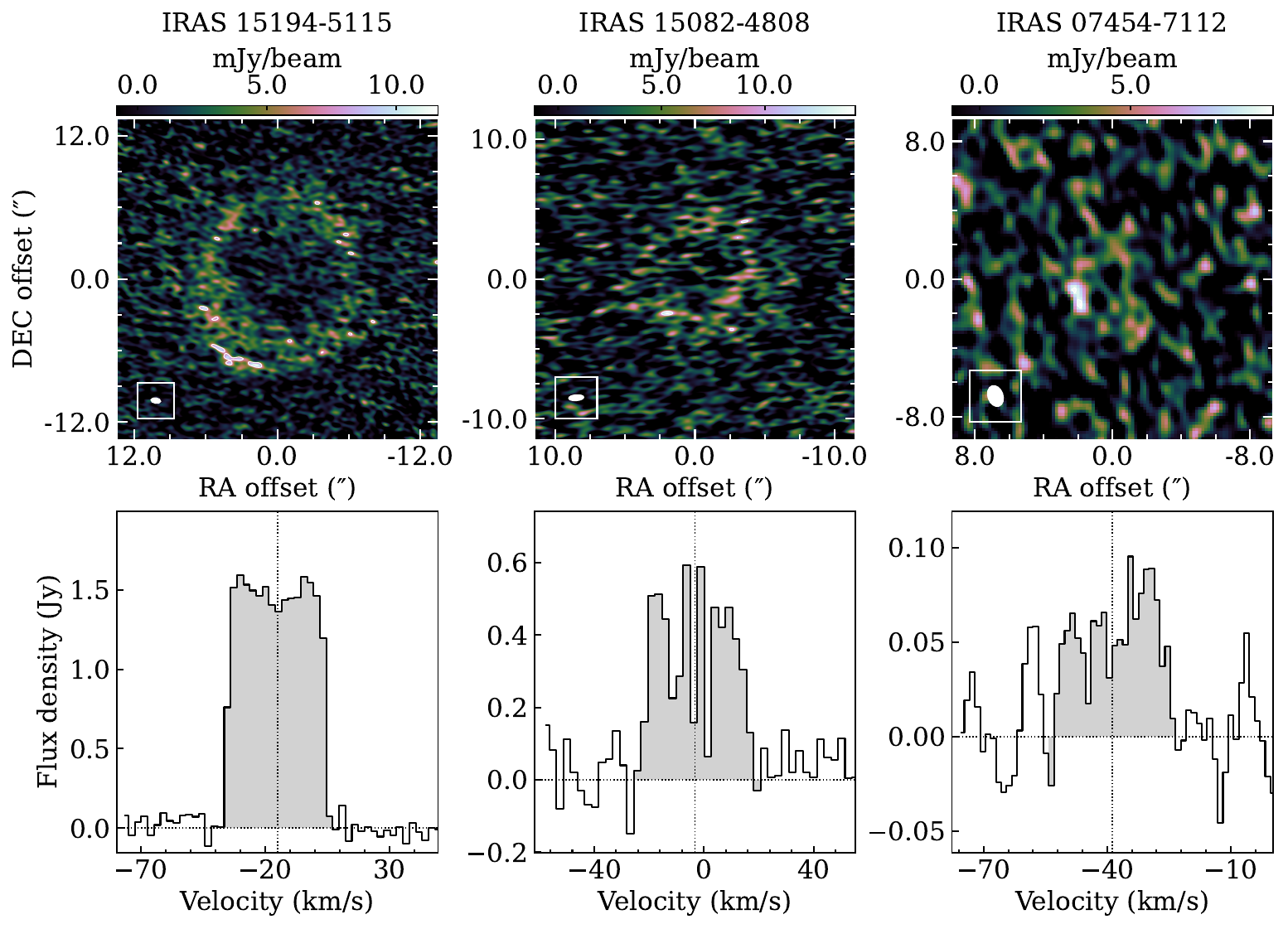}
    \caption{C$_4$H, N=12-11, J=25/2-23/2 (114.182514 GHz)}
\end{figure}

\begin{figure}[h]
    \centering
    \includegraphics[width=0.8\linewidth]{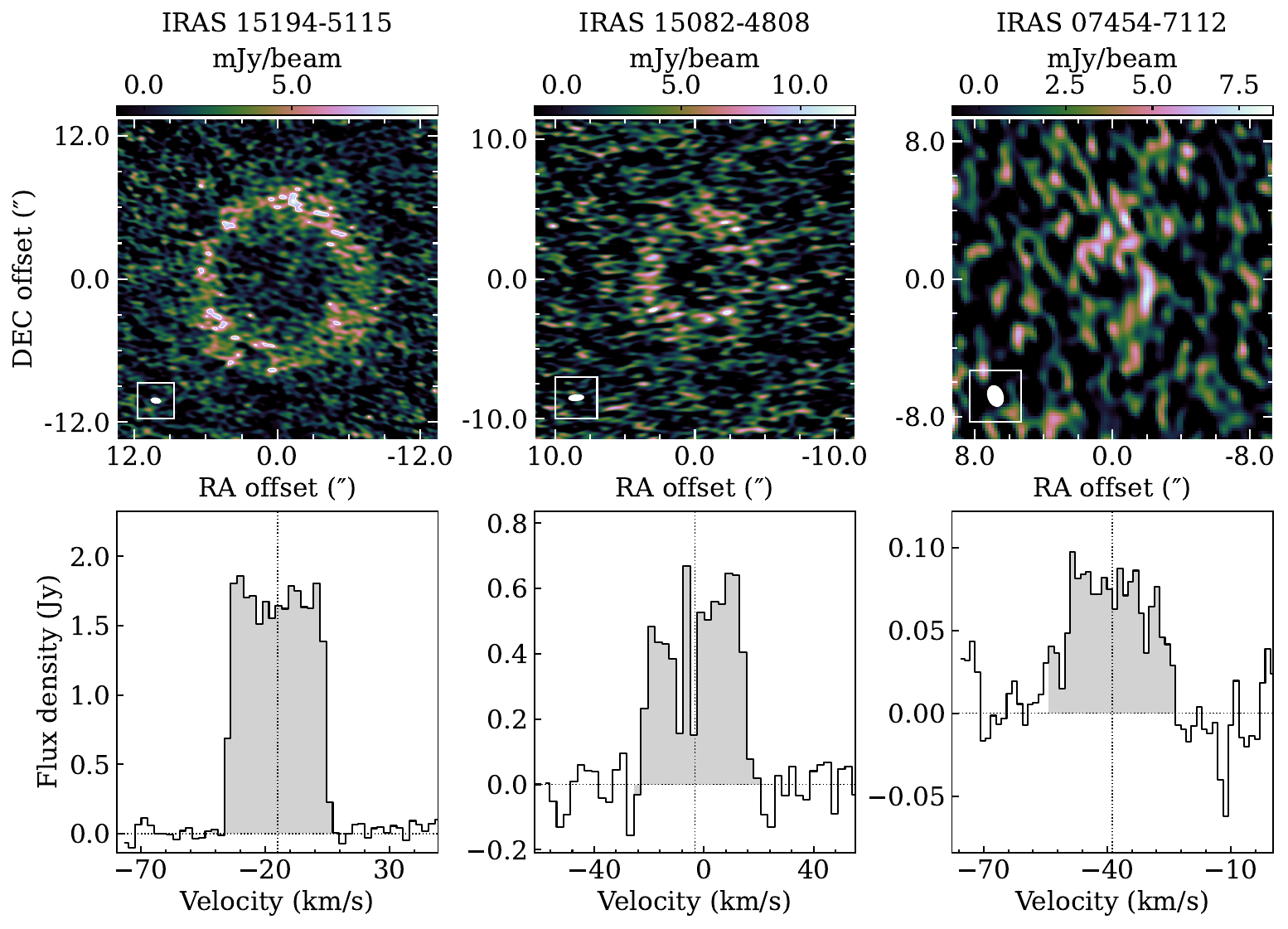}
    \caption{C$_4$H, N=12-11, J=23/2-21/2 (114.221042 GHz)}
\end{figure}

\begin{figure}[h]
    \centering
    \includegraphics[width=0.8\linewidth]{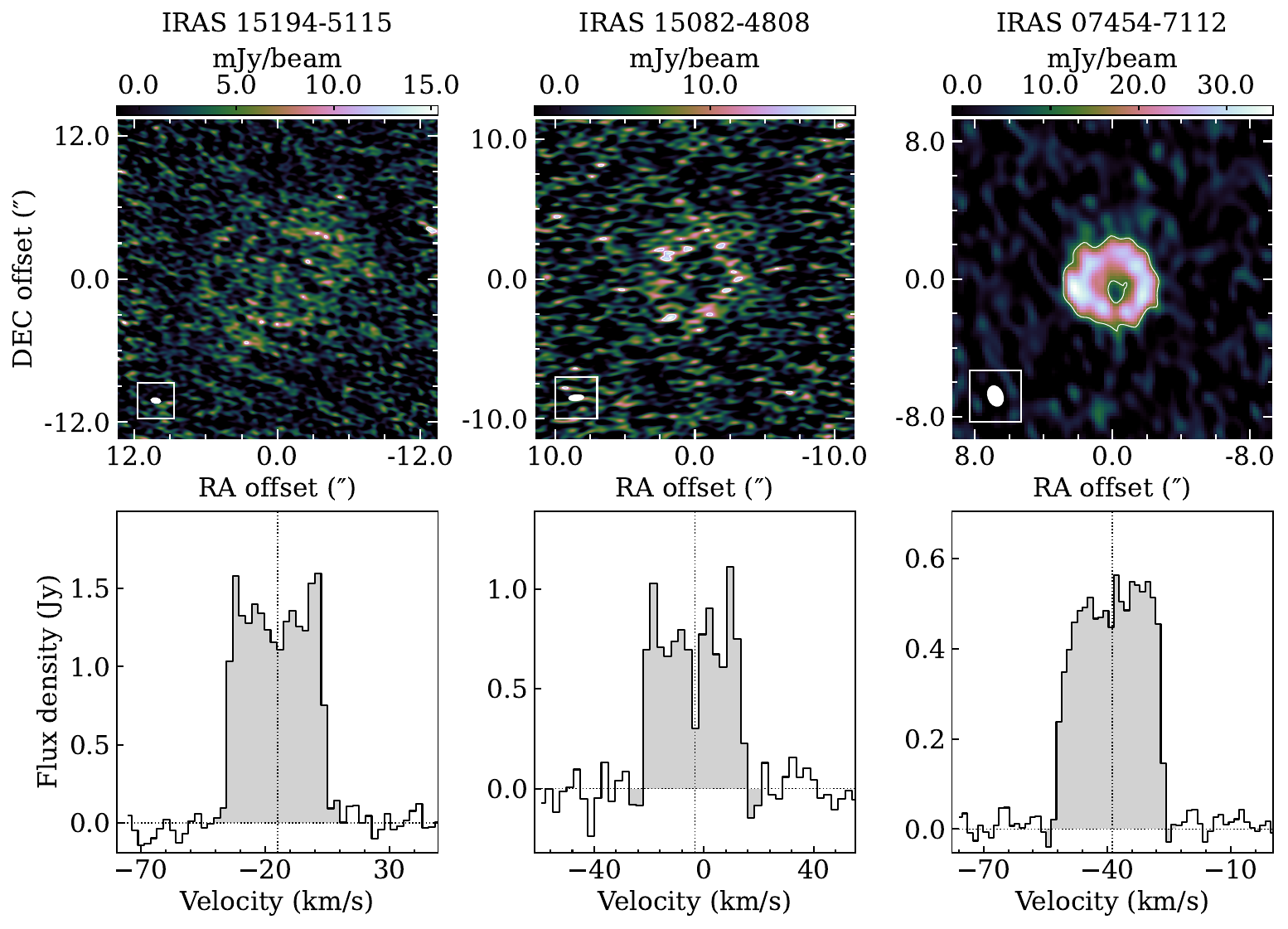}
    \caption{SiC$_2$, 5$_{0,5}$-4$_{0,4}$ (115.382389 GHz)}
\end{figure} 
\FloatBarrier

\begin{figure}[h]
   \centering
   \includegraphics[width=0.8\linewidth]{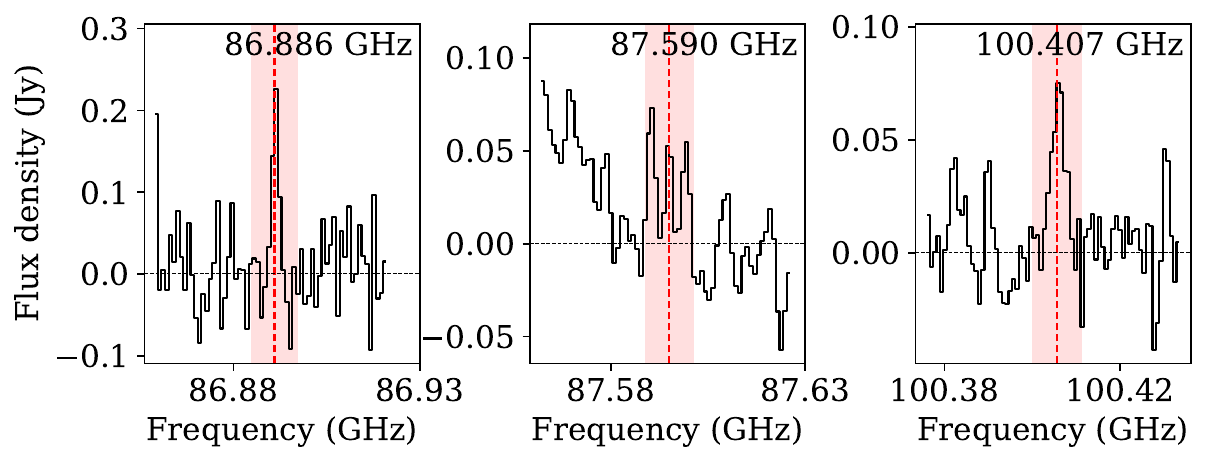}
   \caption{Spectra of unidentified lines towards IRAS 15194$-$5115. The width of the region shaded in pink is twice the expansion velocity of the source CSE. The feature to the left of the 87.59 GHz line is a blend of the $^{30}$SiS 5$-$4 and SiN 2$-$1 lines. The u-lines are too weak to show any structure in the emission maps. They are not detected in the other two sources.}
   \label{fig:u_lines_spectra}
\end{figure}

\begin{figure}[h]
    \centering
    \includegraphics[width=0.65\linewidth]{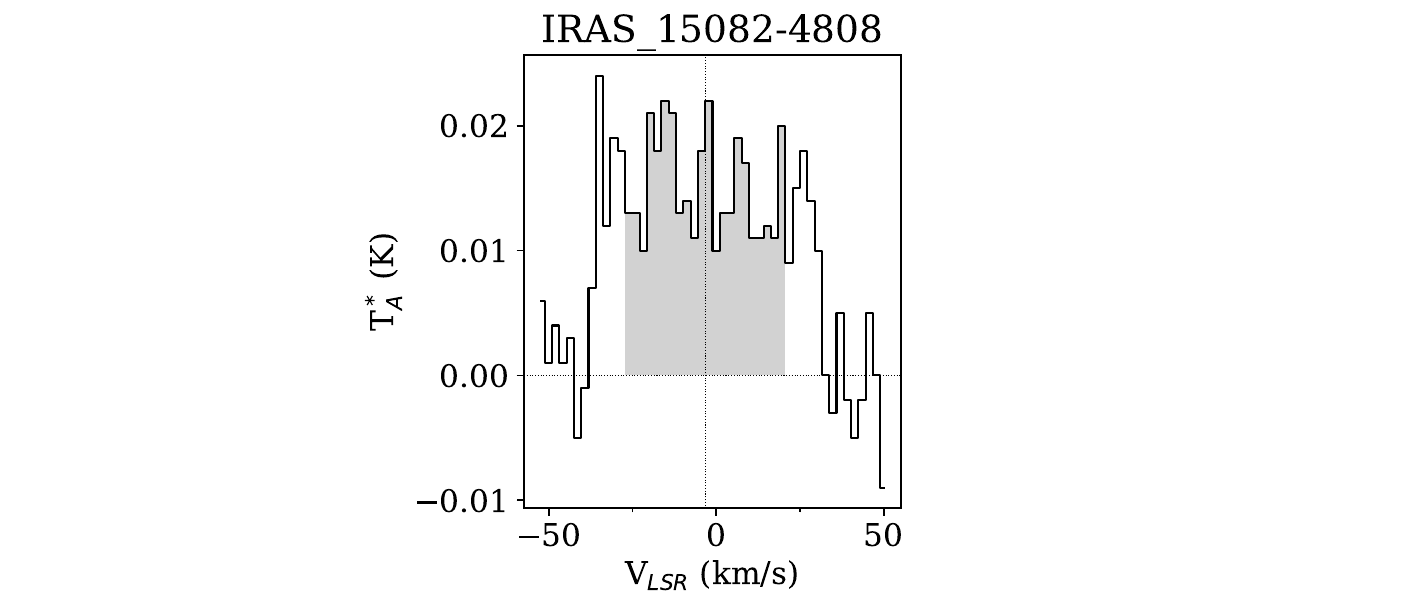}
    \caption{C$_3$N, N=17-16 (168.2043 GHz).}
    \label{fig:C3N_line_1}
\end{figure}

\begin{figure}[h]
    \centering
    \includegraphics[width=0.65\linewidth]{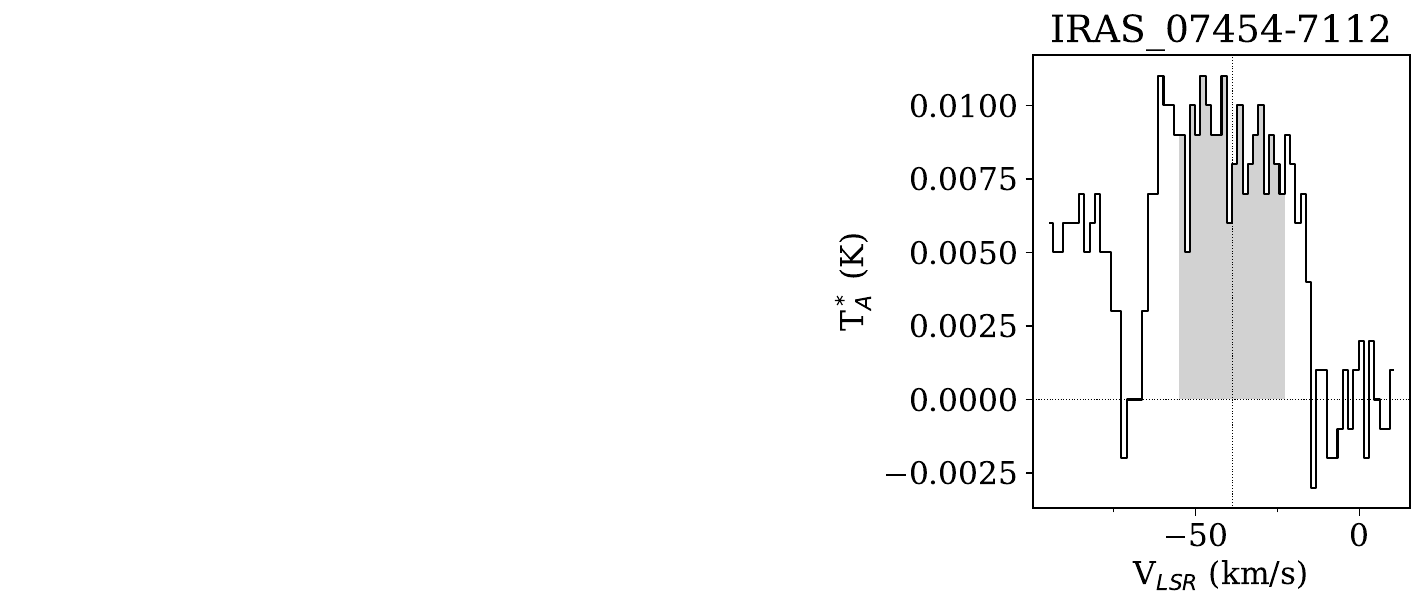}
    \caption{C$_3$N, N=23-22 (227.5539 GHz)}
        \label{fig:C3N_line_2}
\end{figure}

\begin{figure}[h]
    \centering
    \includegraphics[width=0.65\linewidth]{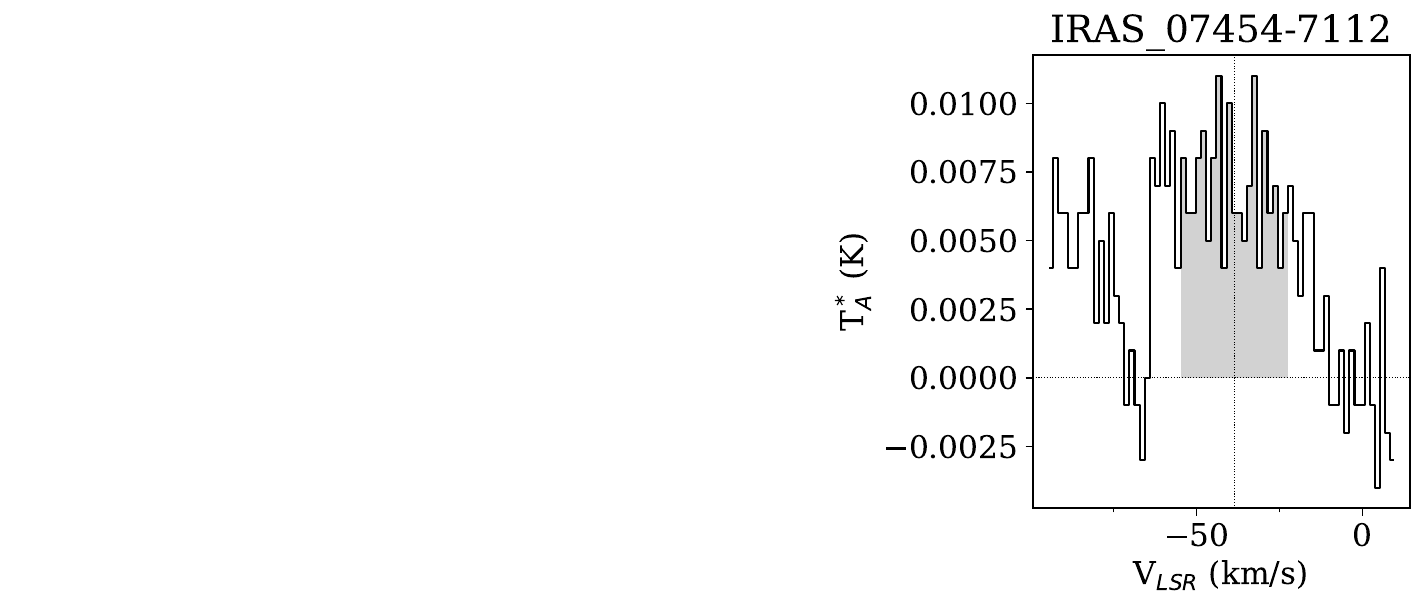}
    \caption{C$_3$N, N=24-23 (237.44415 GHz)}
        \label{fig:C3N_line_3}
\end{figure}

\begin{figure}[h]
    \centering
    \includegraphics[width=0.65\linewidth]{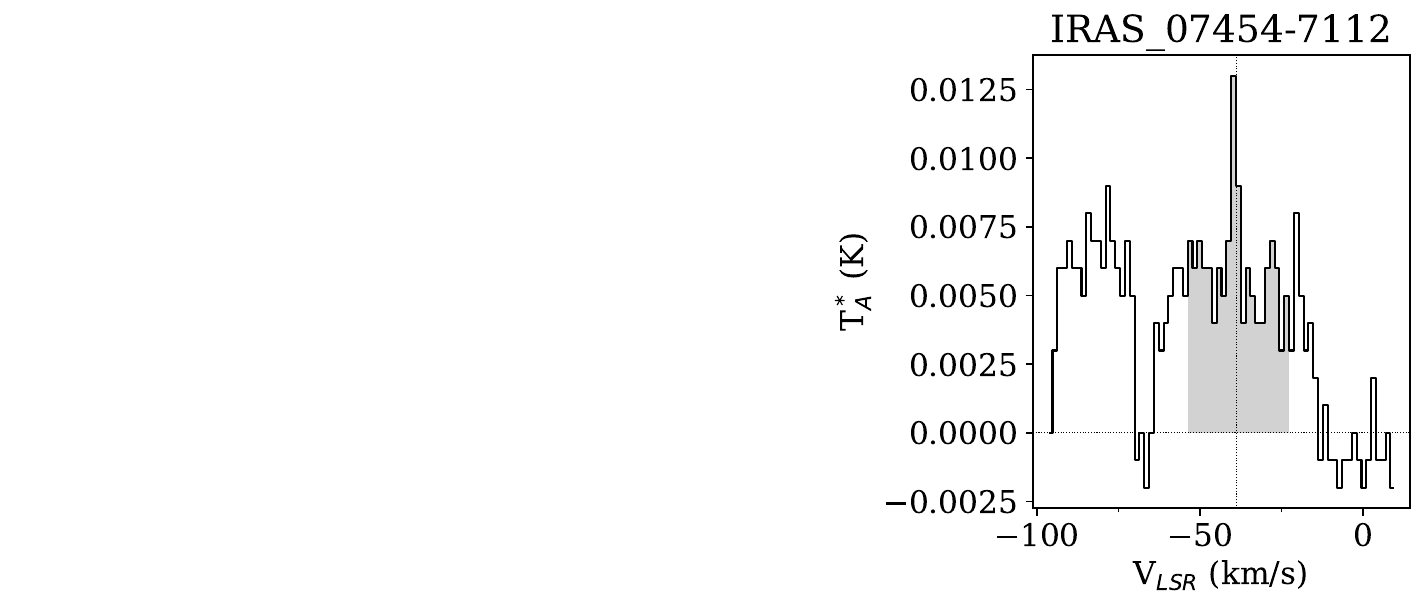}
    \caption{C$_3$N, N=25-24 (247.334 GHz)}
        \label{fig:C3N_line_4}
\end{figure}

\begin{figure}[h]
    \centering
    \includegraphics[width=0.65\linewidth]{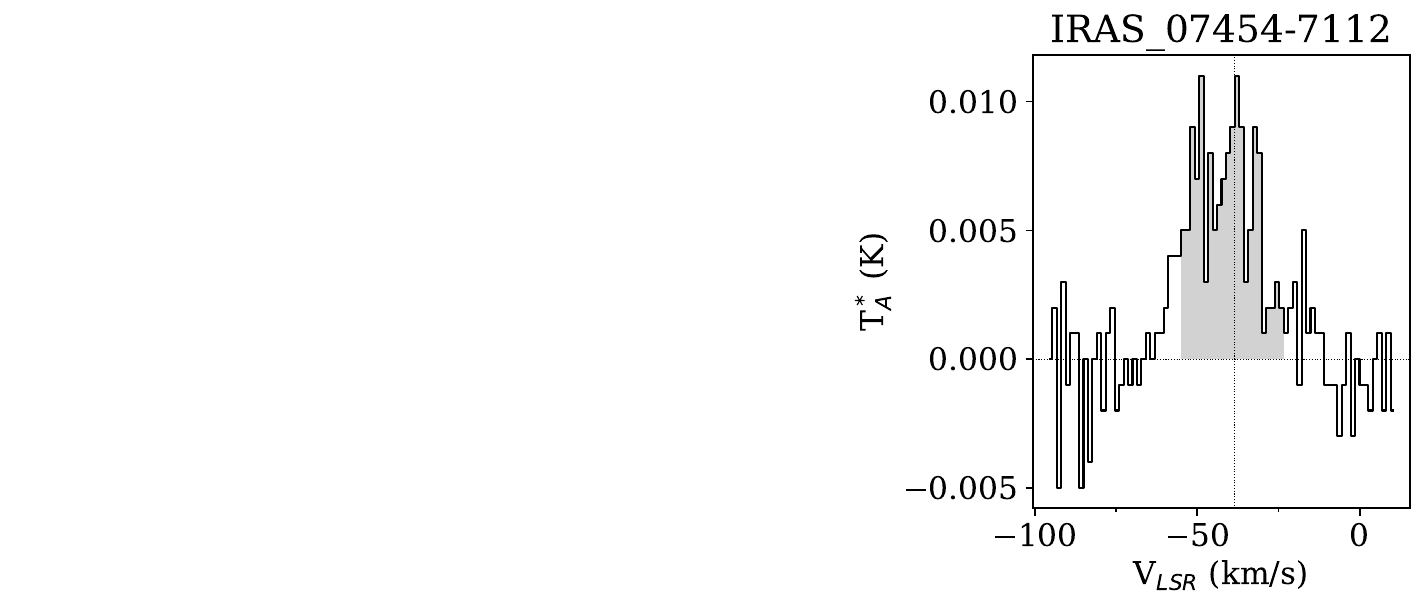}
    \caption{C$_3$N, N=27-26 (267.11225 GHz)}
        \label{fig:C3N_line_5}
\end{figure}

\begin{multicols}{2}
The below figures show the spectra of the lines from the APEX surveys that we have used in this work, for IRAS 15194$-$5115 (left), IRAS 15082$-$4808 (centre), and IRAS 07454$-$7112 (right). The figures are ordered in increasing order of line rest frequency, which is given in parentheses at the end of the figure caption. Missing plots indicate non-detection of the concerned line towards the particular source. The dotted vertical line marks the systemic velocity of the source. The grey-shaded regions cover 1.2 times the expansion velocity on either side of the systemic velocity. For non-blended lines, this region represents the channels used for calculating the integrated line intensities, {reported in Table~\ref{tab:apex_line_detections}}. For the lines of C$_3$N (Figs.~\ref{fig:C3N_line_1}, \ref{fig:C3N_line_2}, \ref{fig:C3N_line_3}, \ref{fig:C3N_line_4}, \ref{fig:C3N_line_5}), where the line profile consists of blended hyperfine components, the whole line profile, not just the grey shaded region, has been integrated to obtain the integrated line intensity. For lines of C$_2$H and C$_4$H, which have hyperfine components, we have shown only the strongest/unblended component, but the total integrated intensity was calculated by summing up all the components per transition.
\end{multicols}

\begin{figure}[h]
    \centering
    \includegraphics[width=0.65\linewidth]{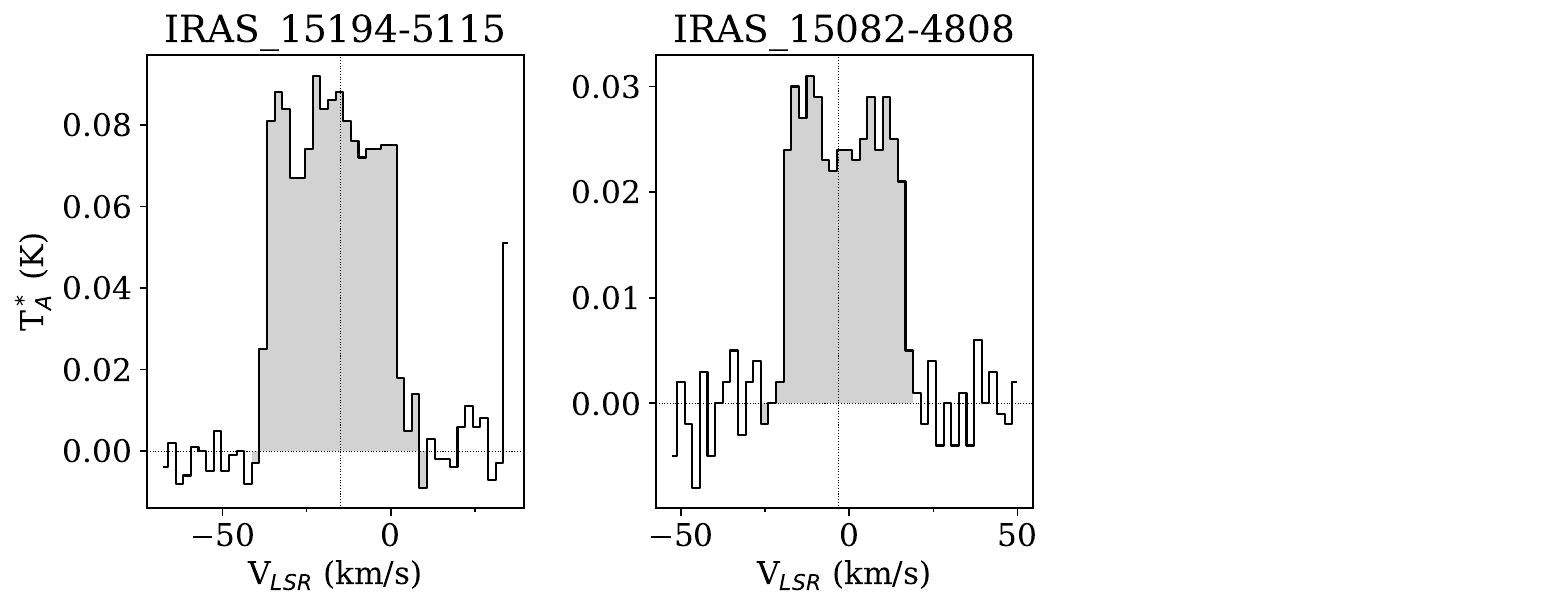}
    \caption{C$_4$H, N=17-16 (161.796566 GHz)}
    \label{fig:APEX_lines_start}
\end{figure}

\begin{figure}[h]
    \centering
    \includegraphics[width=0.65\linewidth]{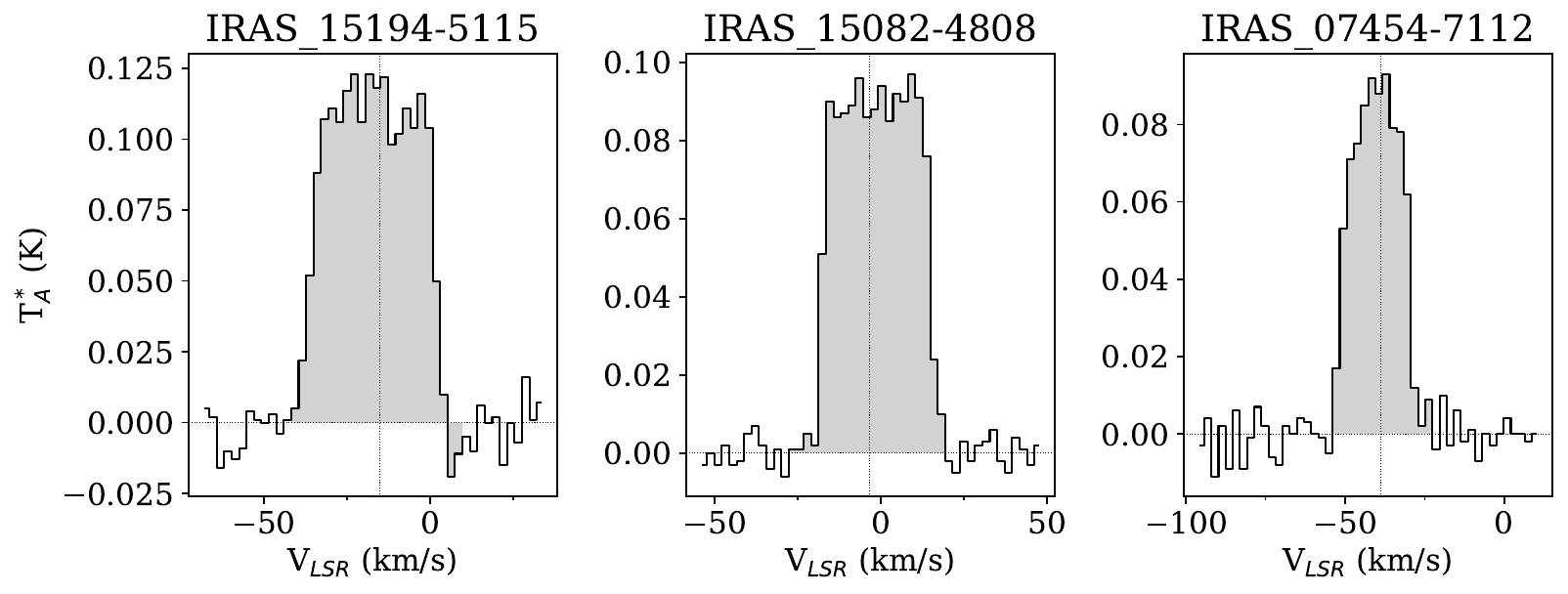}
    \caption{SiS, 9-8 (163.376785 GHz)}
\end{figure}

\begin{figure}[h]
    \centering
    \includegraphics[width=0.65\linewidth]{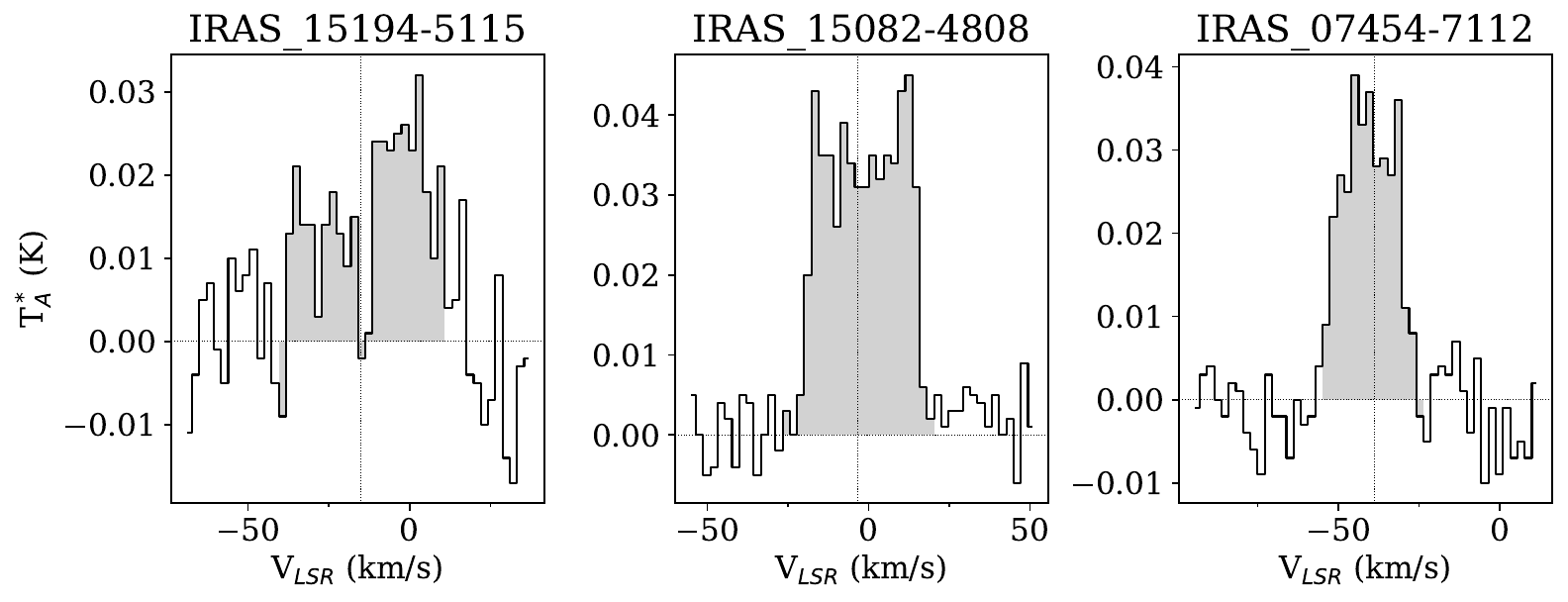}
    \caption{HC$_3$N, J=18-17 (163.753389 GHz)}
\end{figure}

\begin{figure}[h]
    \centering
    \includegraphics[width=0.65\linewidth]{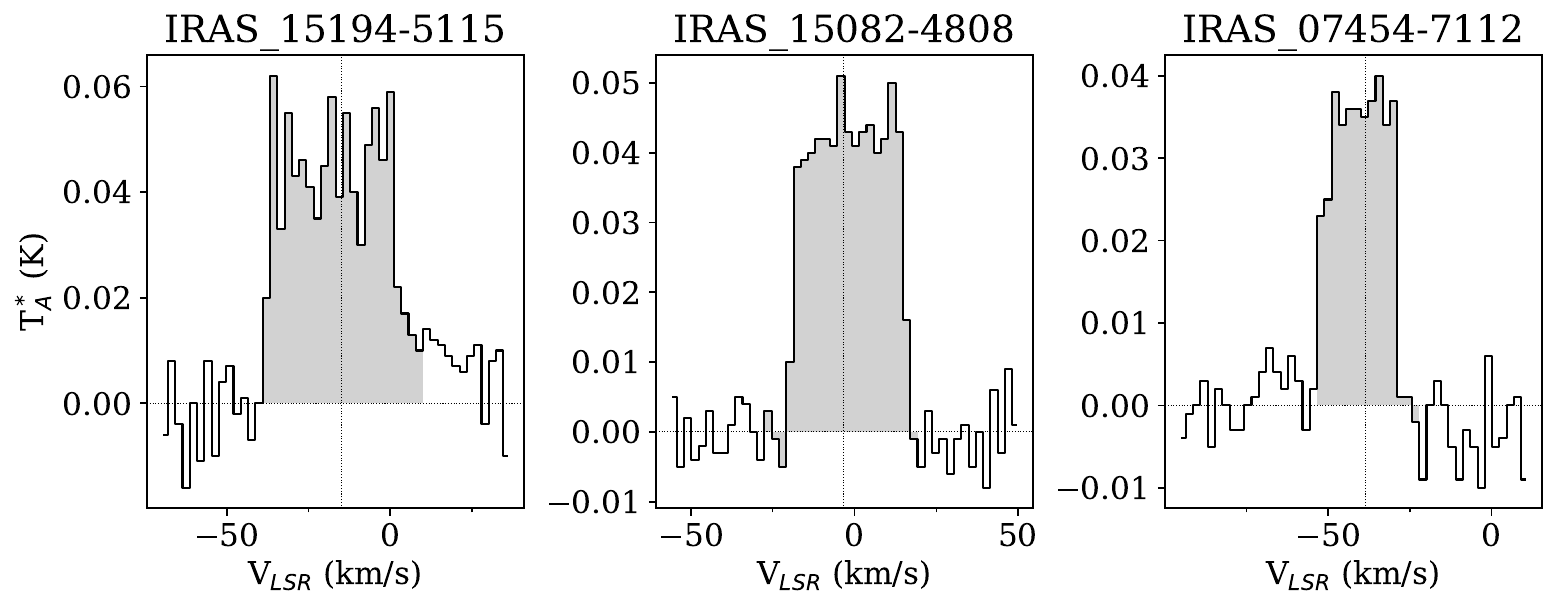}
    \caption{SiC$_2$, 7( 2, 6)- 6( 2, 5) (164.069091 GHz)}
\end{figure}

\begin{figure}[h]
    \centering
    \includegraphics[width=0.65\linewidth]{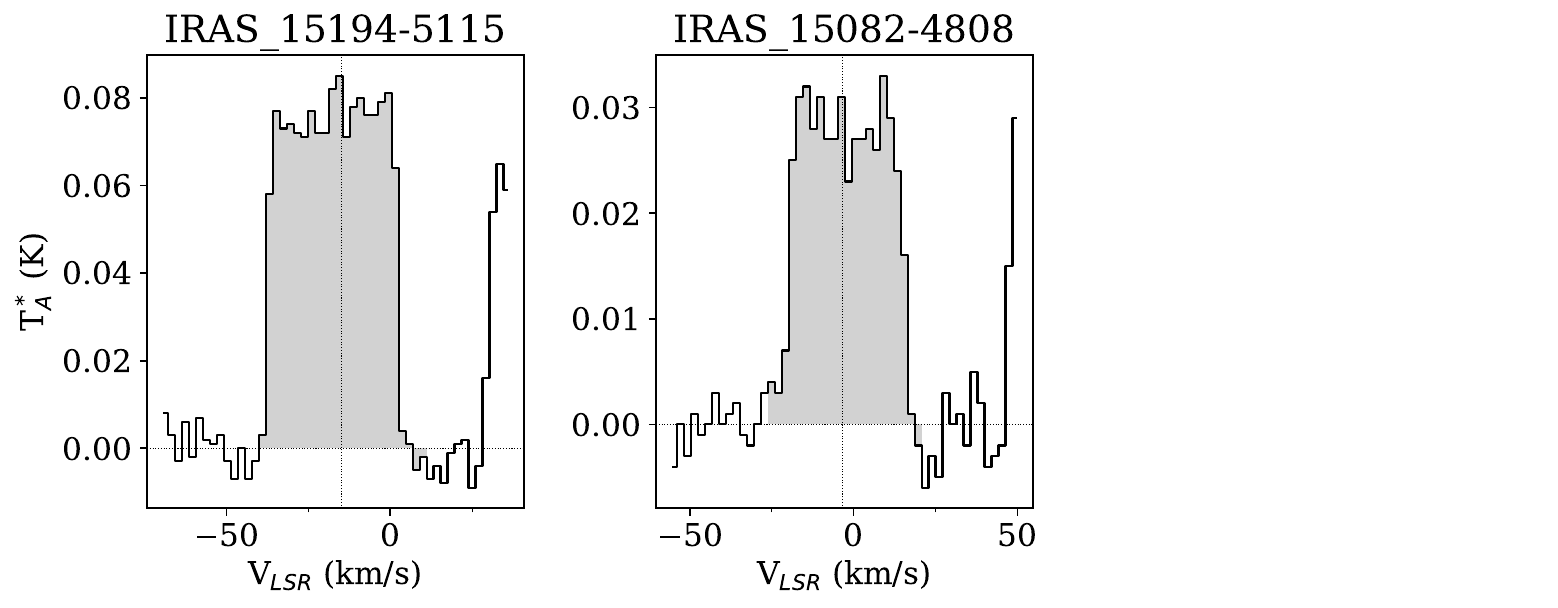}
    \caption{C$_4$H, N=18-17 (171.310707 GHz)}
\end{figure}

\begin{figure}[h]
    \centering
    \includegraphics[width=0.65\linewidth]{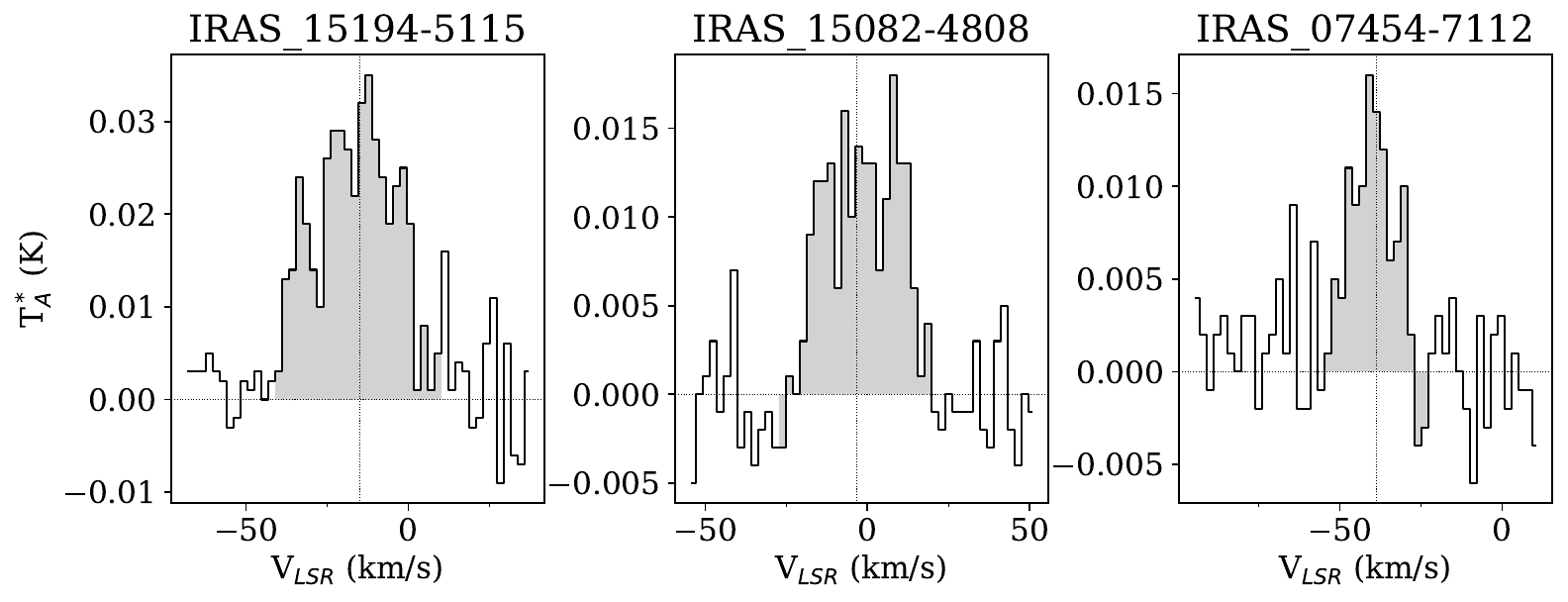}
    \caption{$^{29}$SiO, 4-3 (171.512796 GHz)}
\end{figure}

\begin{figure}[h]
    \centering
    \includegraphics[width=0.65\linewidth]{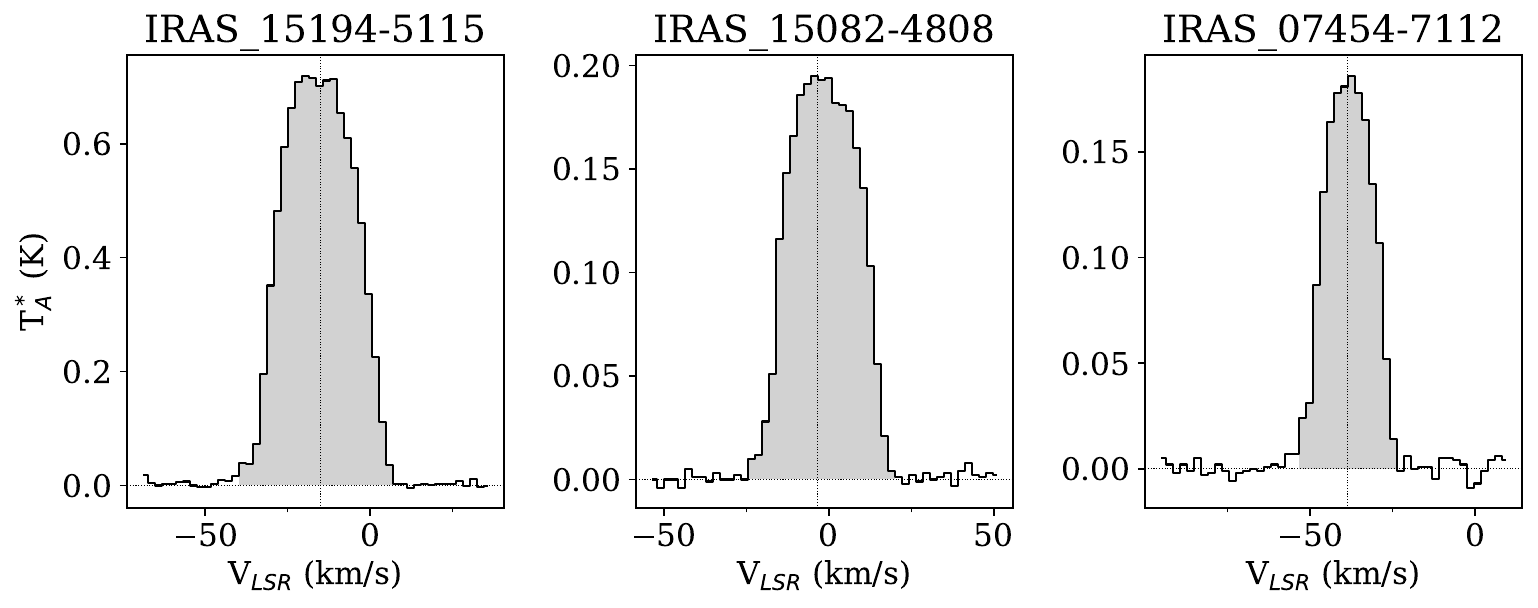}
    \caption{H$^{13}$CN, J=2-1 (172.677851 GHz)}
\end{figure}

\begin{figure}[h]
    \centering
    \includegraphics[width=0.65\linewidth]{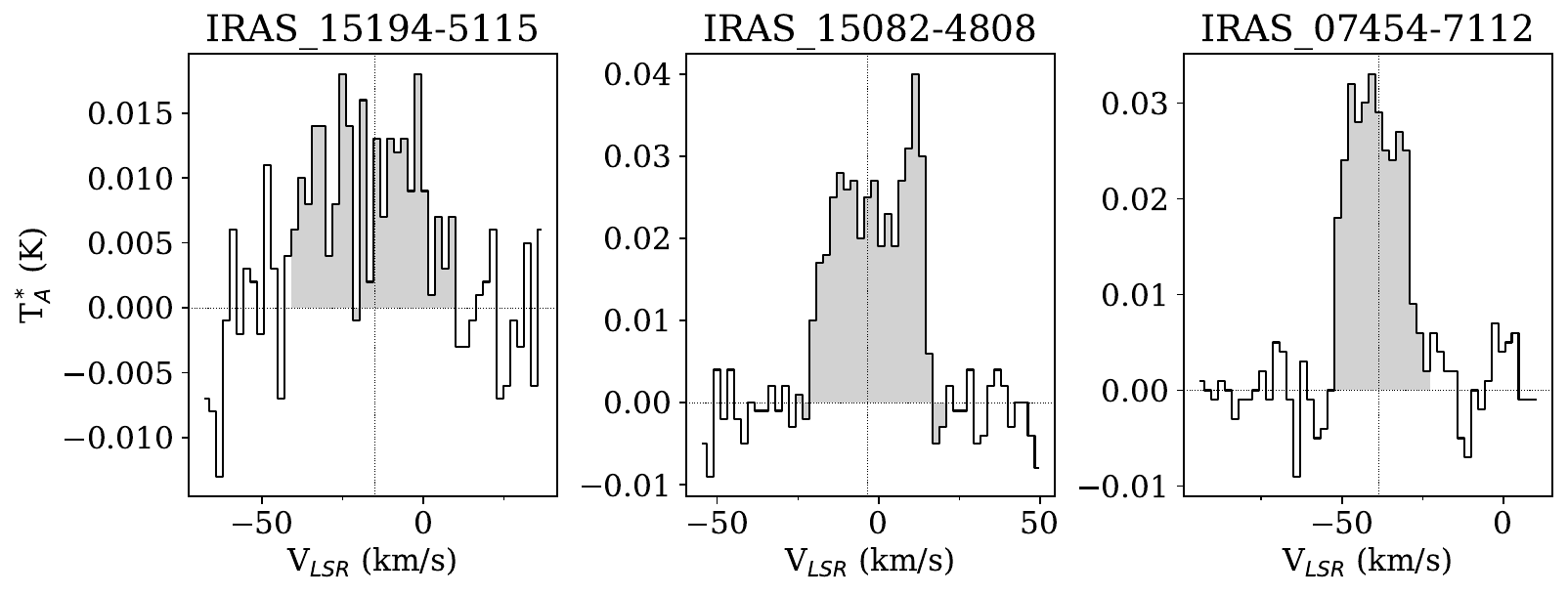}
    \caption{HC$_3$N, J=19-18 (172.8493 GHz)}
\end{figure}

\begin{figure}[h]
    \centering
    \includegraphics[width=0.65\linewidth]{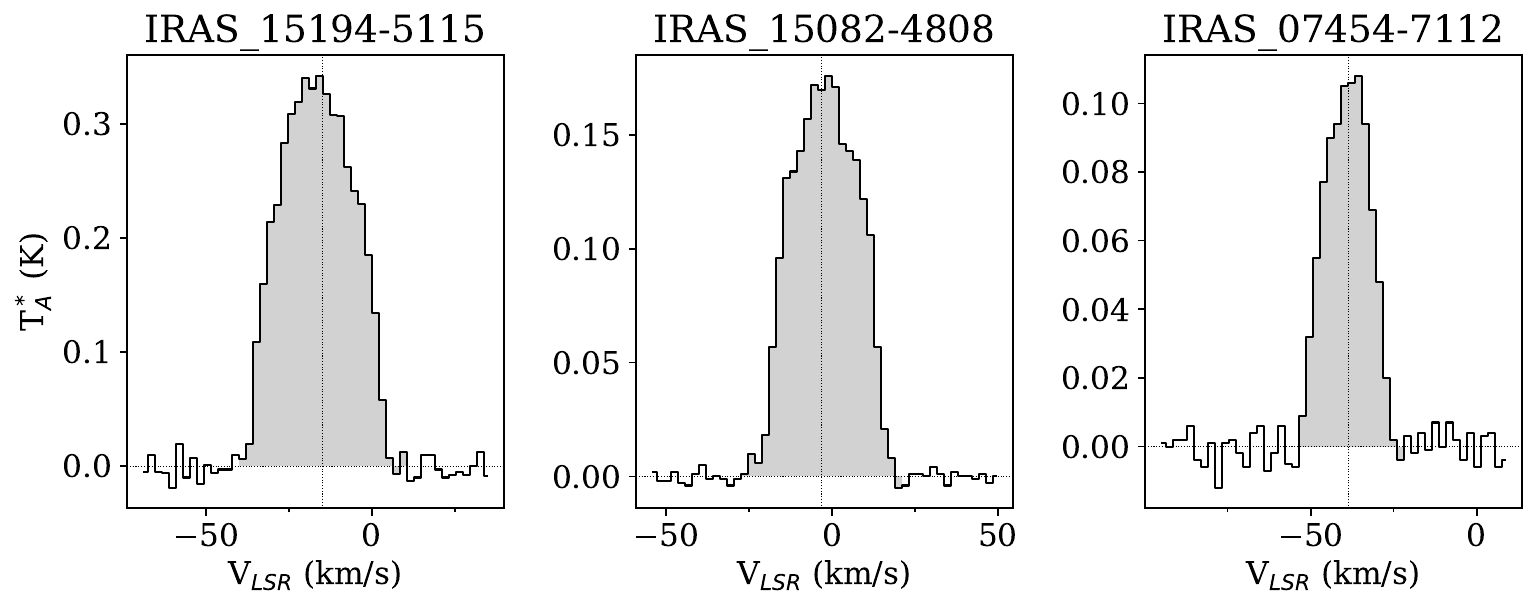}
    \caption{SiO, 4-3 (173.688238 GHz)}
\end{figure}

\begin{figure}[h]
    \centering
    \includegraphics[width=0.65\linewidth]{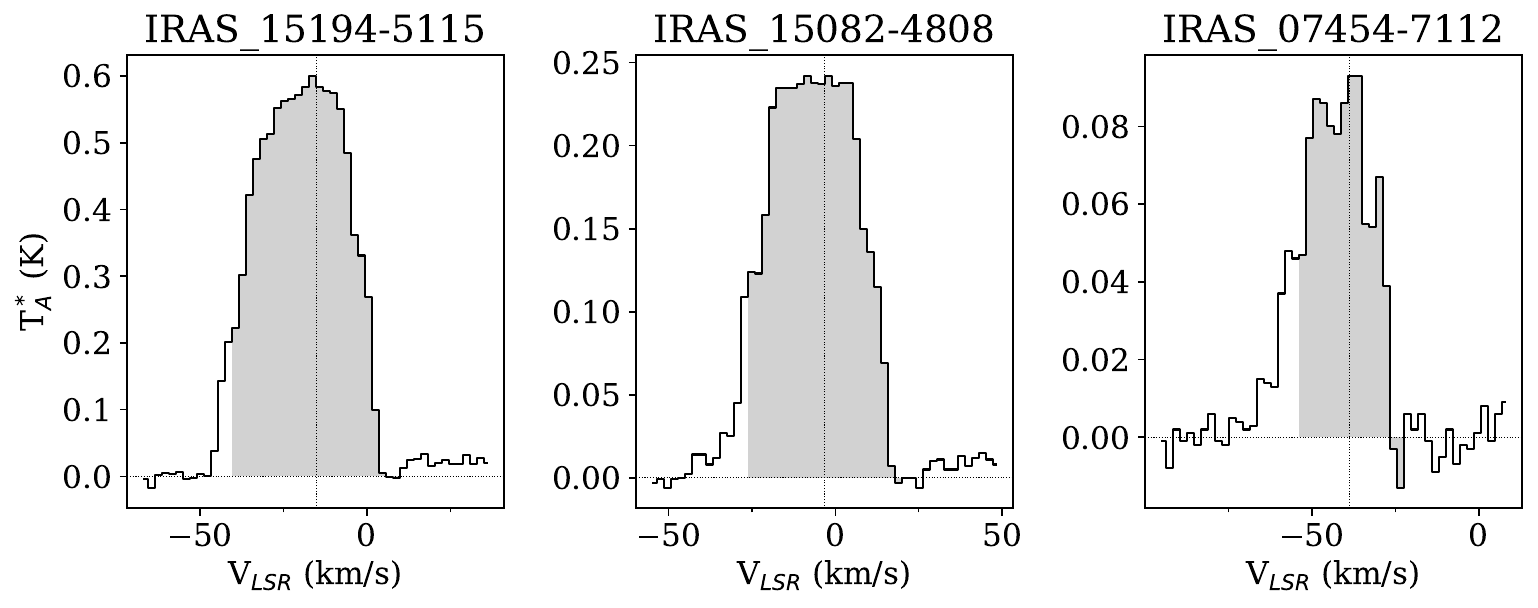}
    \caption{C$_2$H, N=2-1 (174.663199 GHz)}
    \label{fig:C2H_line_5}
\end{figure}

\begin{figure}[h]
    \centering
    \includegraphics[width=0.65\linewidth]{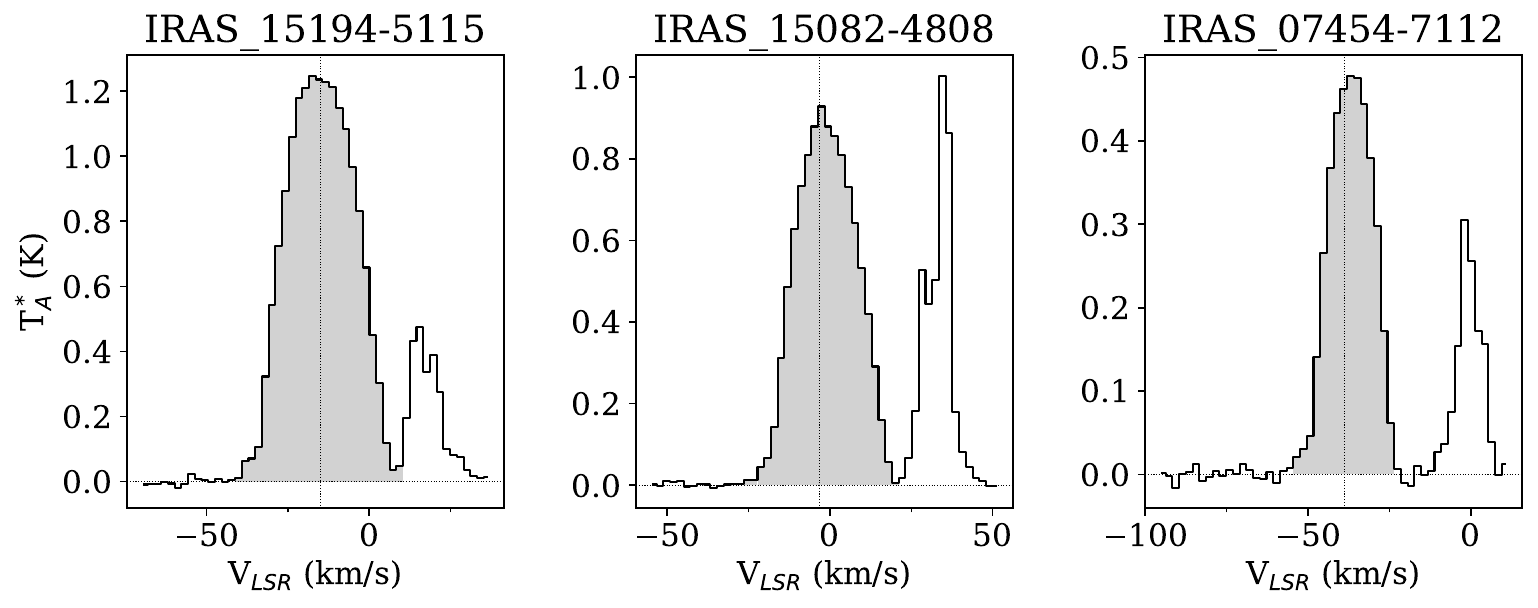}
    \caption{HCN, J=2-1 (177.261111 GHz)}
\end{figure}

\begin{figure}[h]
    \centering
    \includegraphics[width=0.65\linewidth]{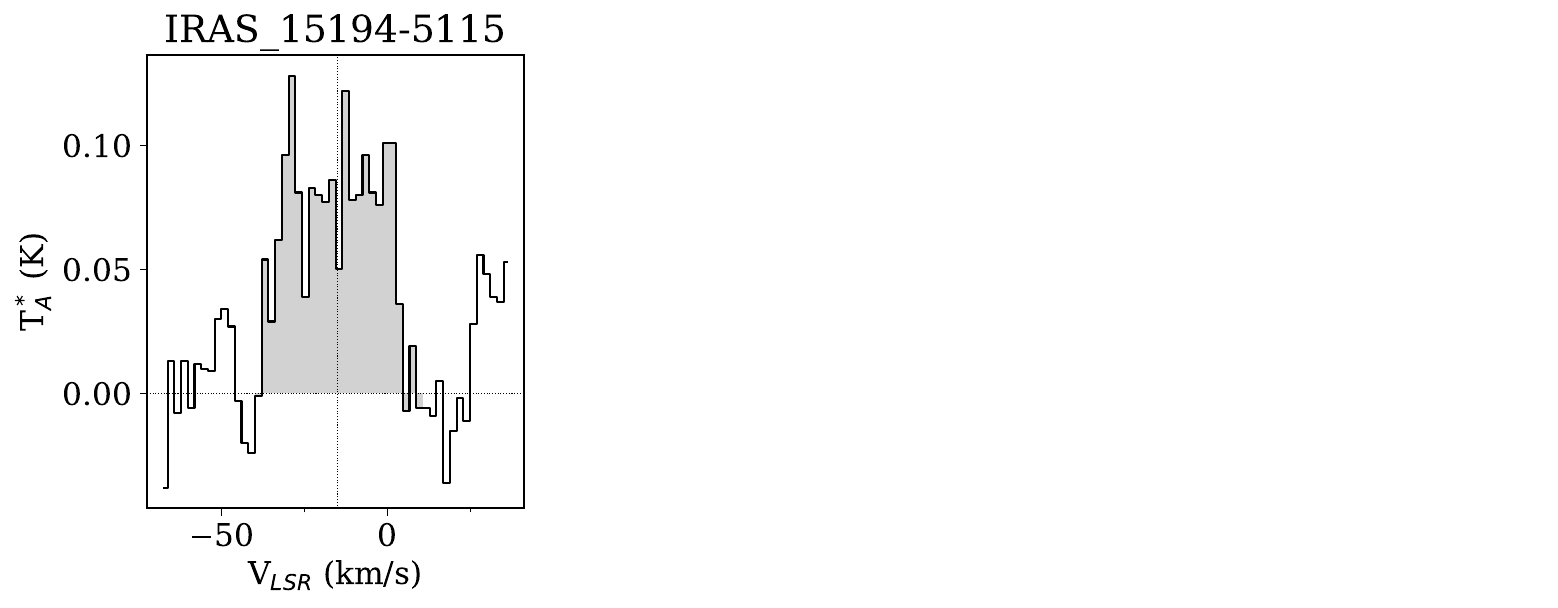}
    \caption{C$_4$H, N=19-18 (180.824472 GHz)}
\end{figure}

\begin{figure}[h]
    \centering
    \includegraphics[width=0.65\linewidth]{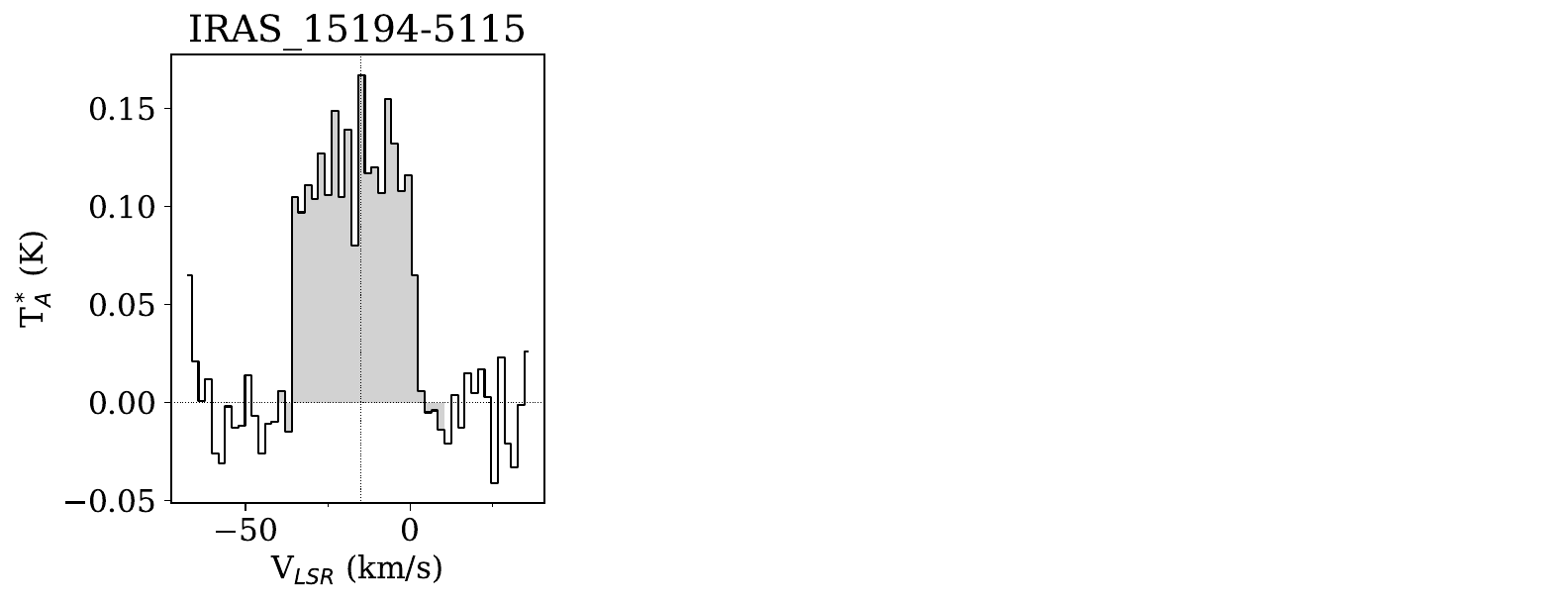}
    \caption{HNC, J=2-1 (181.324758 GHz)}
\end{figure}

\begin{figure}[h]
    \centering
    \includegraphics[width=0.65\linewidth]{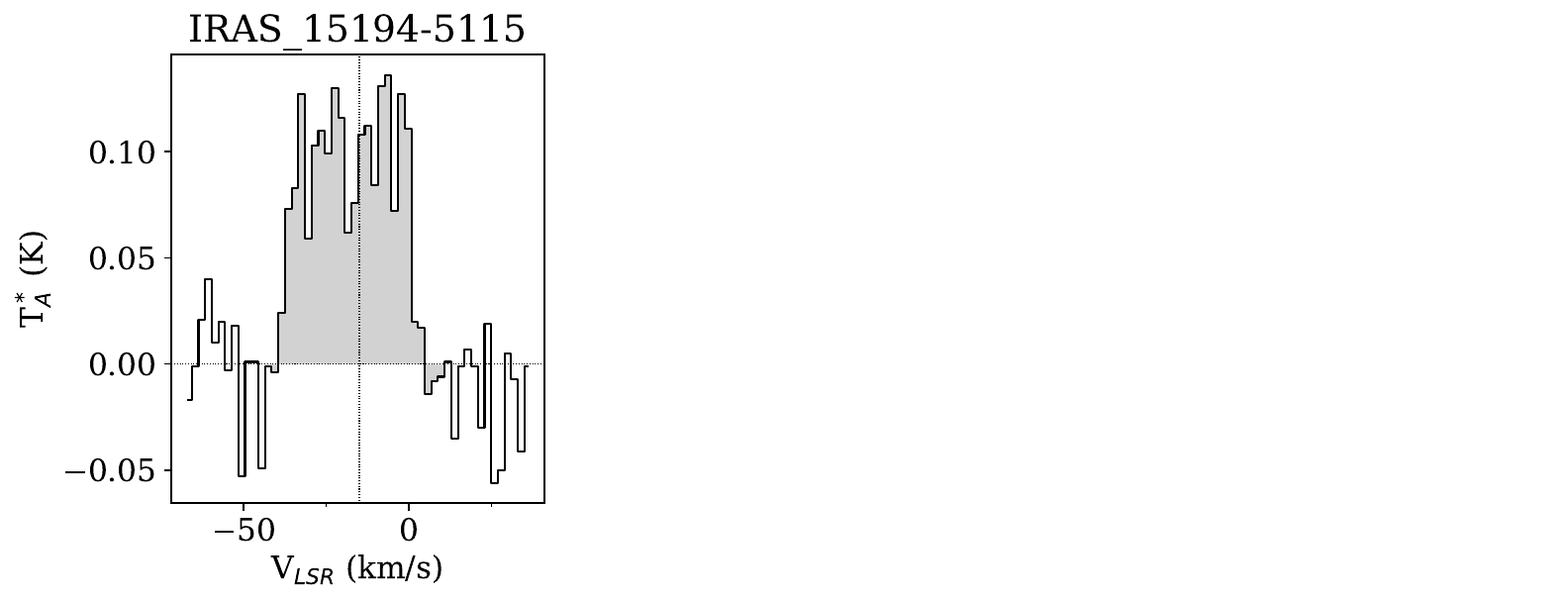}
    \caption{SiS, 10-9 (181.525218 GHz)}
\end{figure}

\begin{figure}[h]
    \centering
    \includegraphics[width=0.65\linewidth]{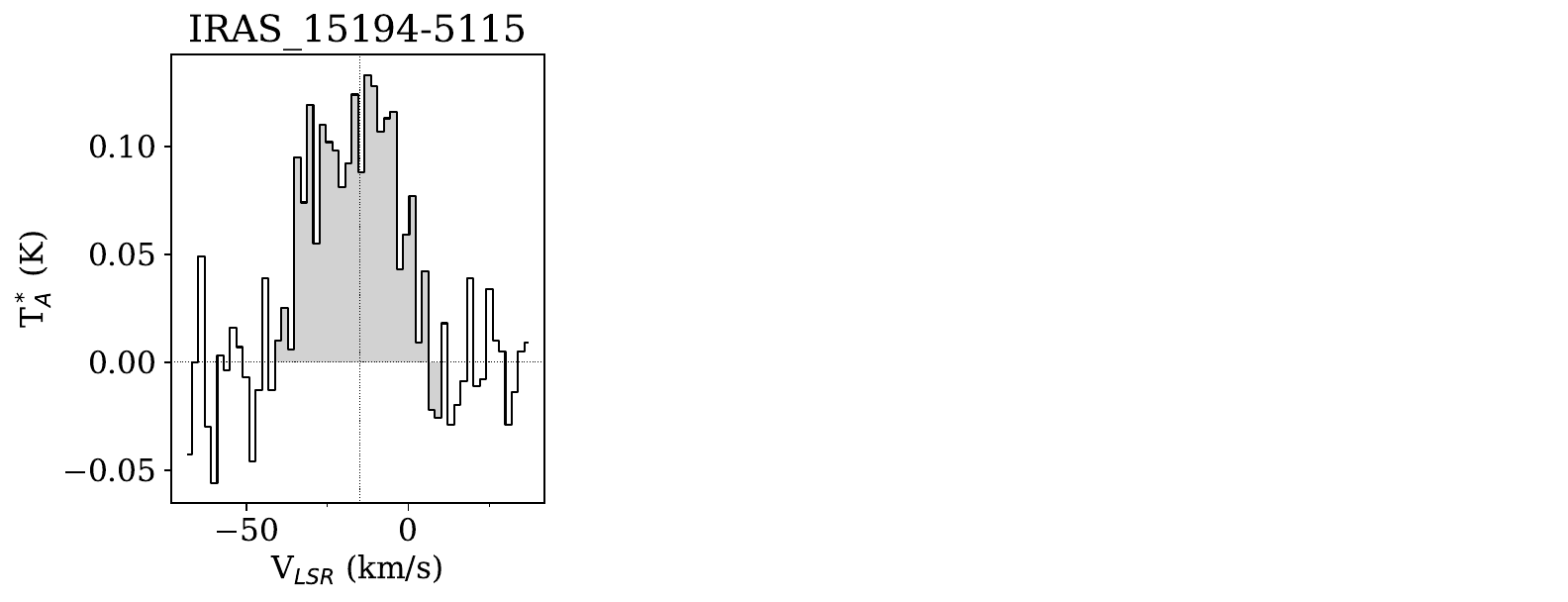}
    \caption{$^{13}$CS, J=4-3 (184.981772 GHz)}
\end{figure}

\begin{figure}[h]
    \centering
    \includegraphics[width=0.65\linewidth]{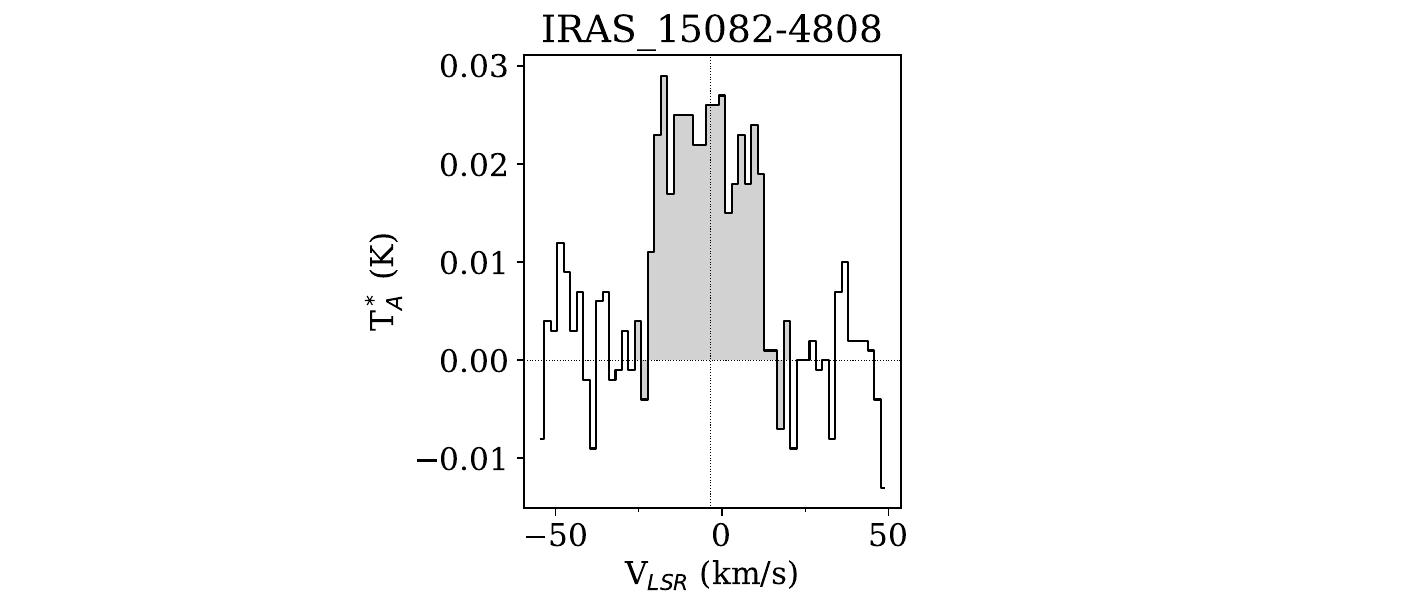}
    \caption{SiC$_2$, 8( 6, 3)- 7( 6, 2) (188.3857 GHz)}
\end{figure}

\begin{figure}[h]
    \centering
    \includegraphics[width=0.65\linewidth]{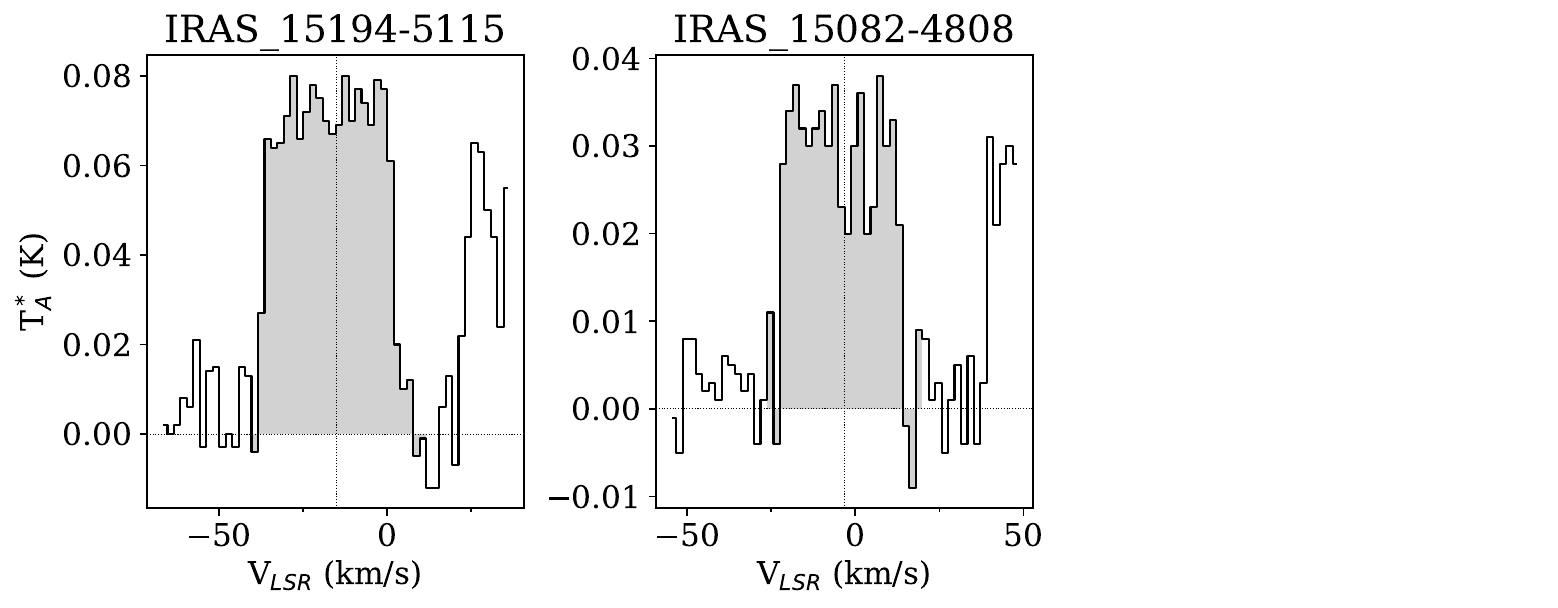}
    \caption{C$_4$H, N=20-19 (190.337804 GHz)}
\end{figure}

\begin{figure}[h]
    \centering
    \includegraphics[width=0.65\linewidth]{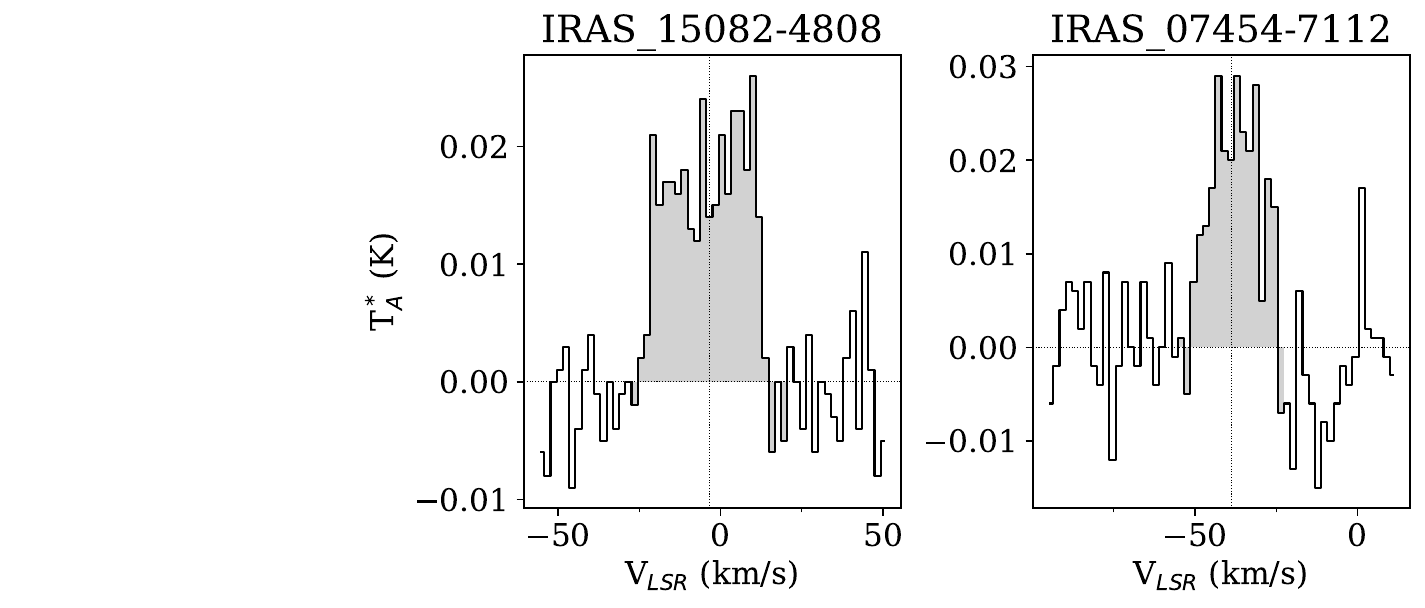}
    \caption{HC$_3$N, J=21-20 (191.040299 GHz)}
\end{figure}

\begin{figure}[h]
    \centering
    \includegraphics[width=0.65\linewidth]{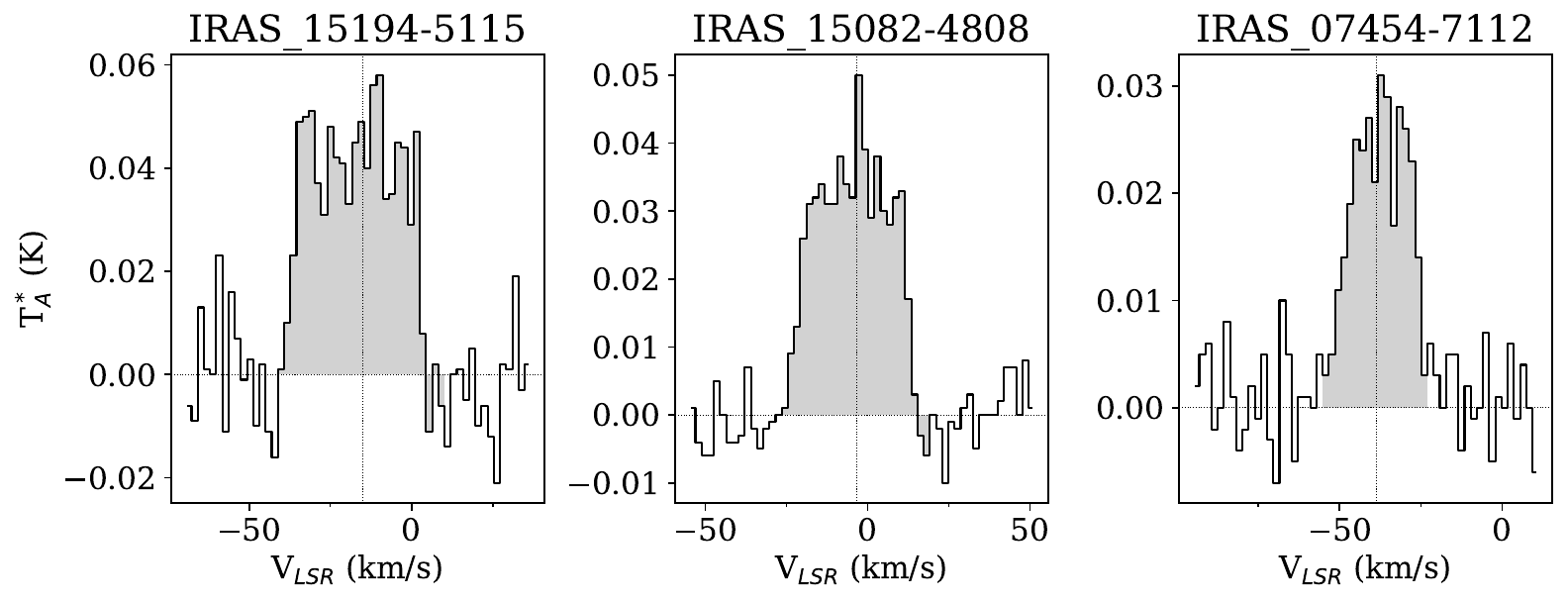}
    \caption{C$^{34}$S, 4-3 (192.818457 GHz)}
\end{figure}

\begin{figure}[h]
    \centering
    \includegraphics[width=0.65\linewidth]{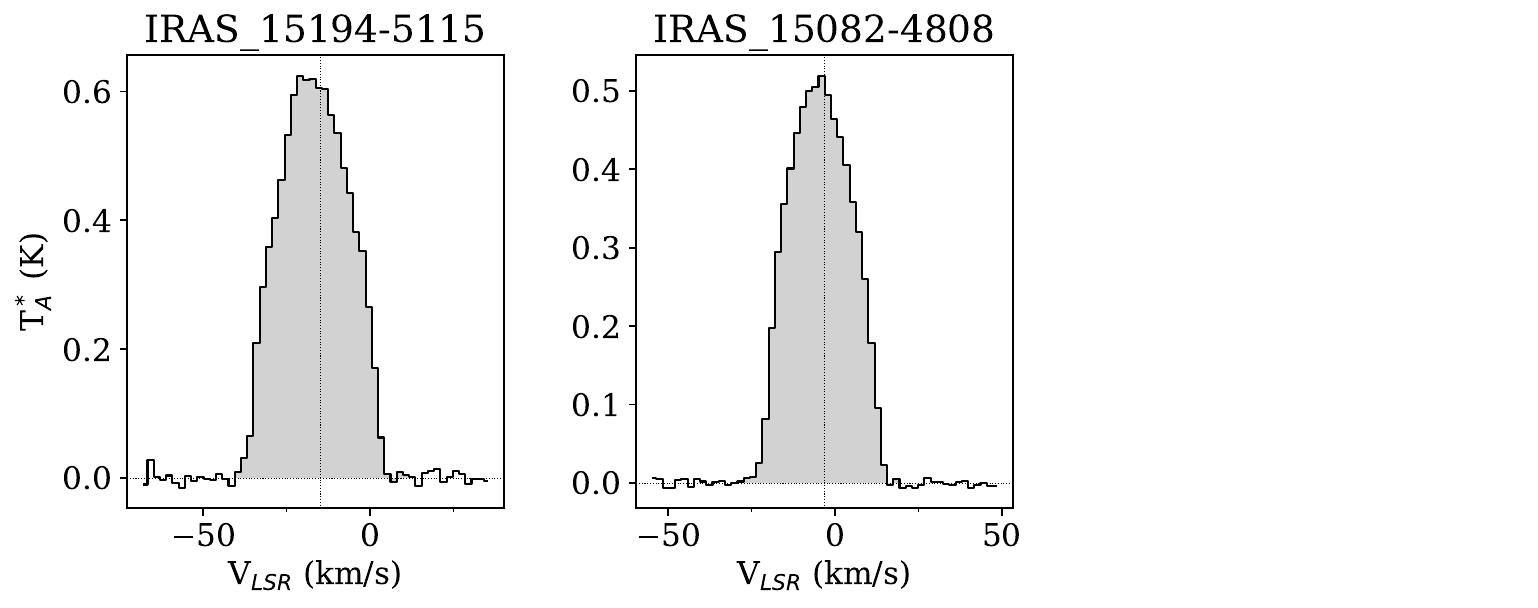}
    \caption{CS, 4-3 (195.954211 GHz)}
\end{figure}

\begin{figure}[h]
    \centering
    \includegraphics[width=0.65\linewidth]{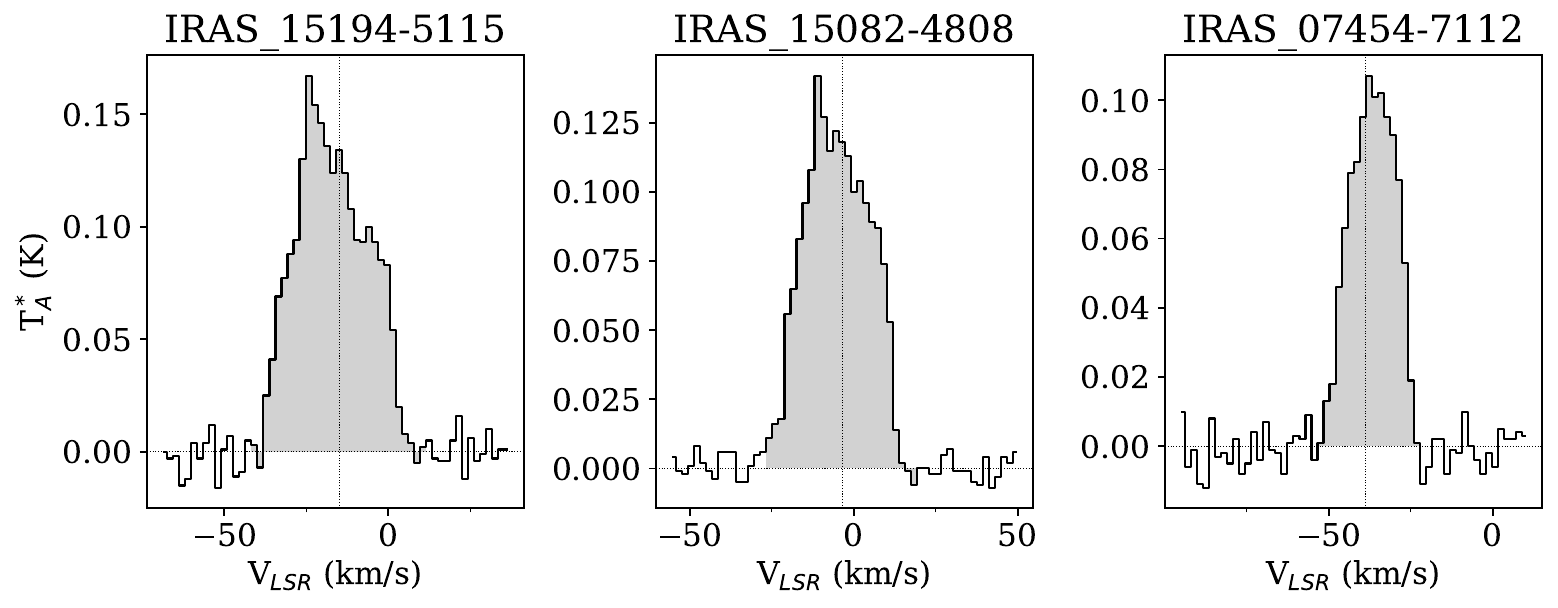}
    \caption{SiS, 11-10 (199.672229 GHz)}
\end{figure}

\begin{figure}[h]
    \centering
    \includegraphics[width=0.65\linewidth]{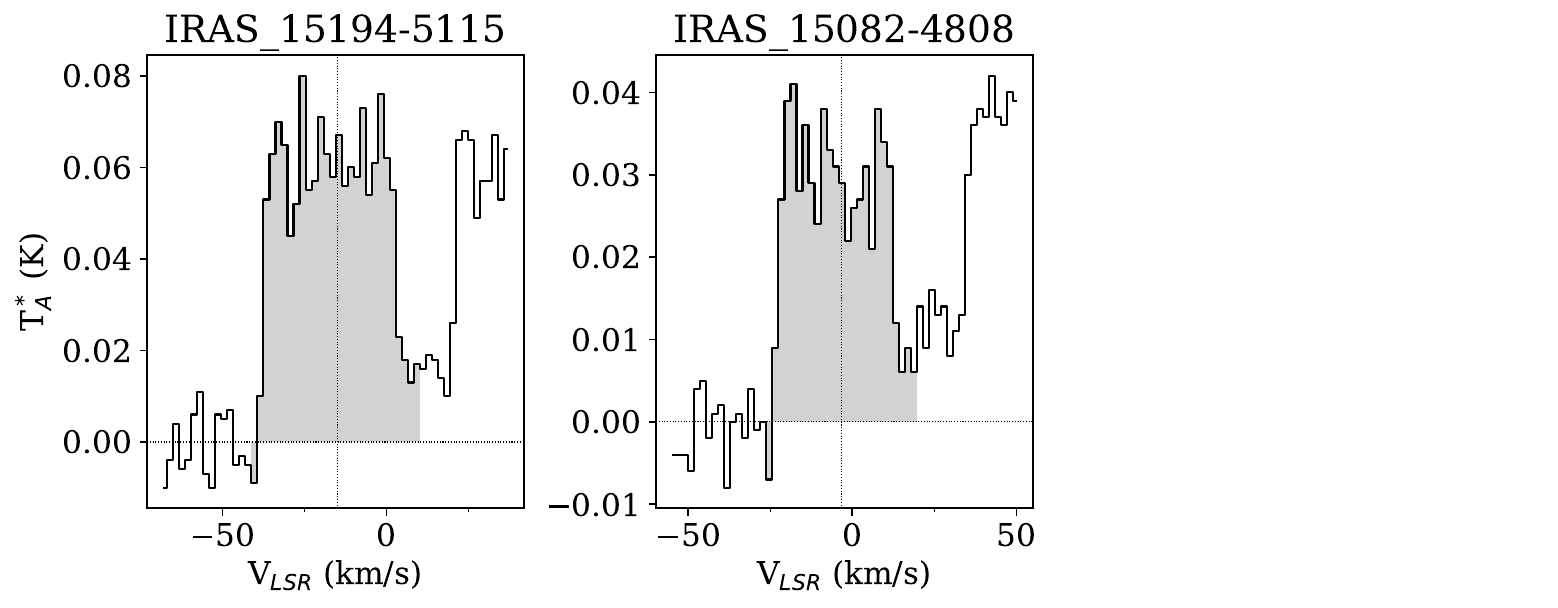}
    \caption{C$_4$H, N=21-20 (199.850787 GHz)}
\end{figure}

\begin{figure}[h]
    \centering
    \includegraphics[width=0.65\linewidth]{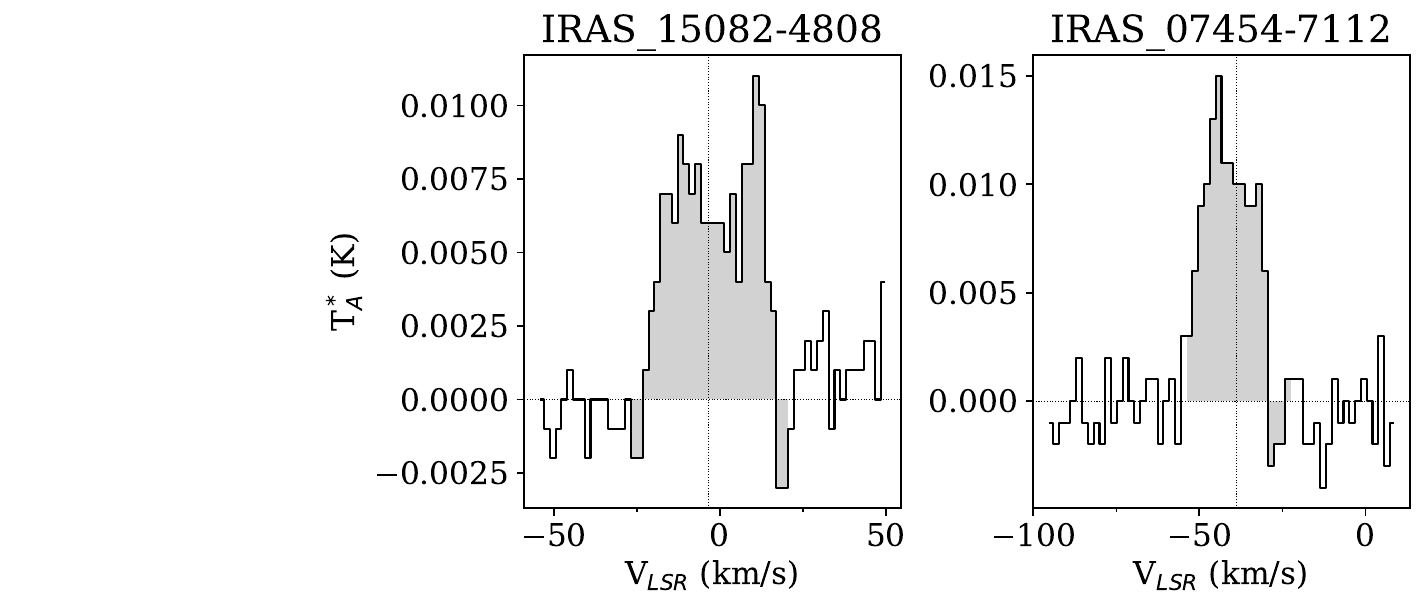}
    \caption{HC$_3$N, J=23-22 (209.230234 GHz)}
\end{figure}

\begin{figure}[h]
    \centering
    \includegraphics[width=0.65\linewidth]{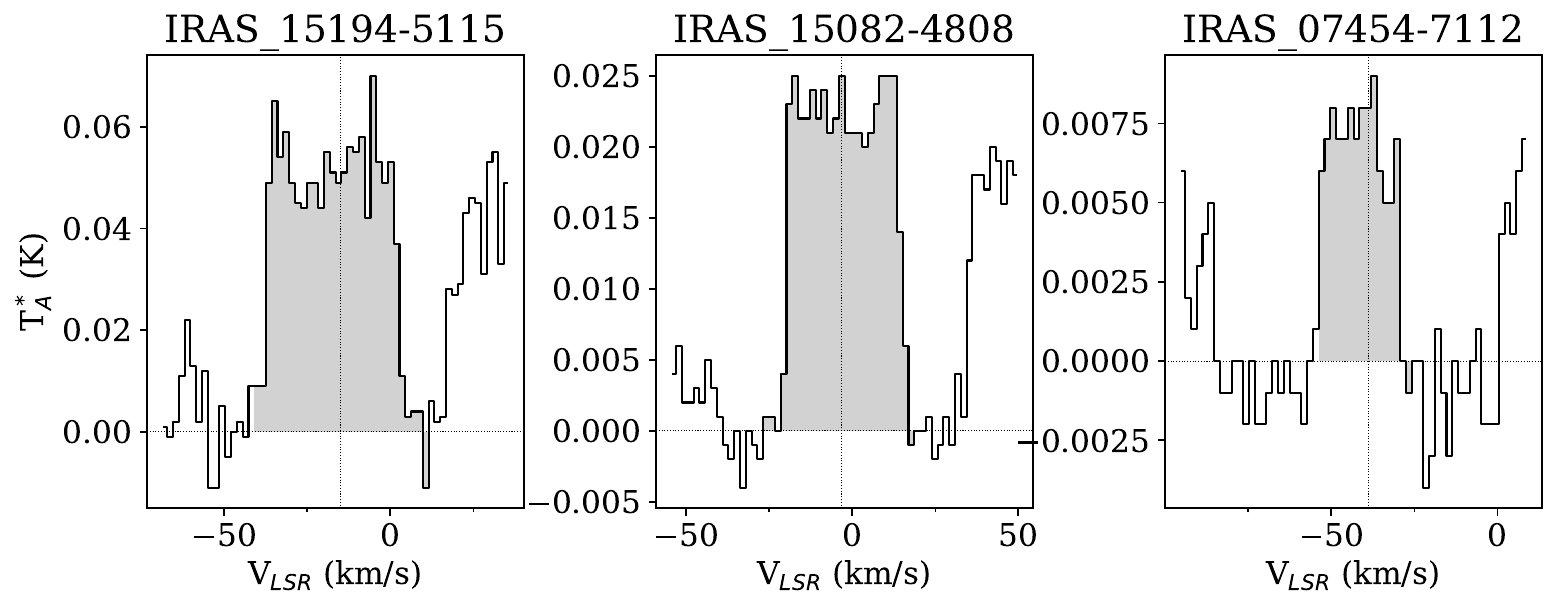}
    \caption{C$_4$H, N=22-21 (209.363302 GHz)}
\end{figure}

\begin{figure}[h]
    \centering
    \includegraphics[width=0.65\linewidth]{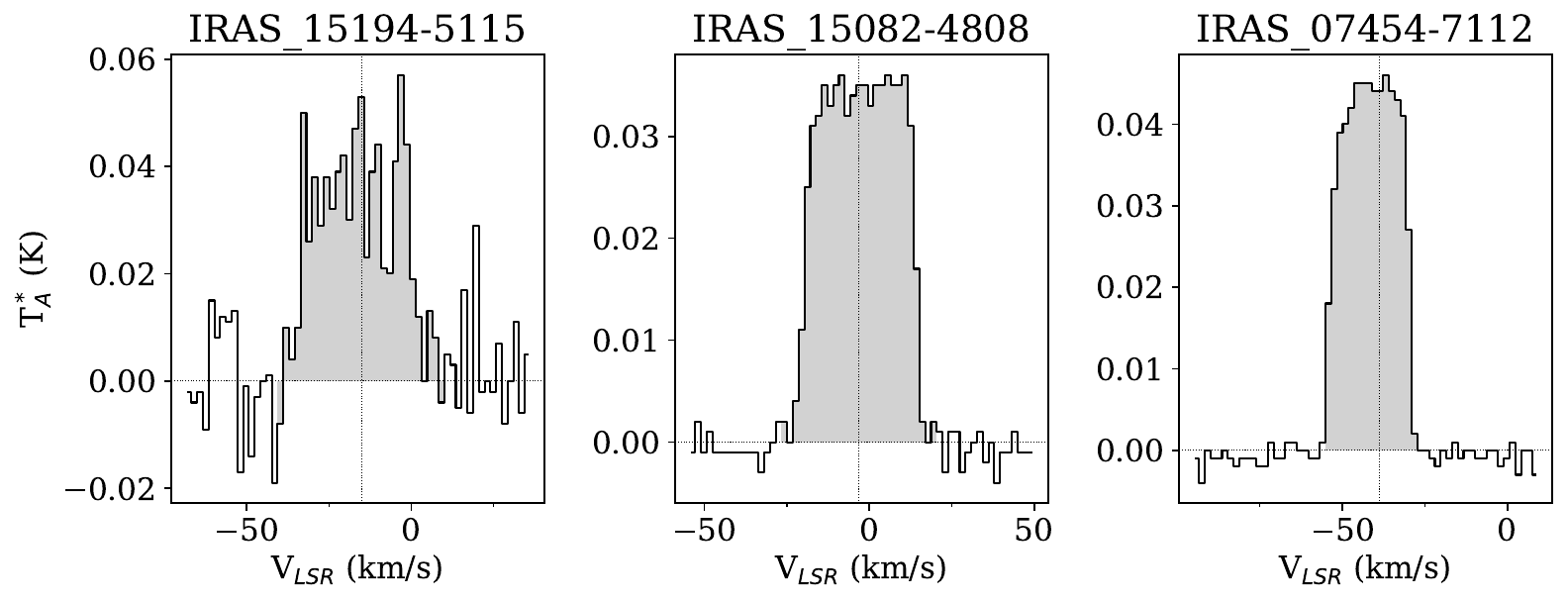}
    \caption{SiC$_2$, 9( 2, 8)- 8( 2, 7) (209.892001 GHz)}
\end{figure}

\begin{figure}[h]
    \centering
    \includegraphics[width=0.65\linewidth]{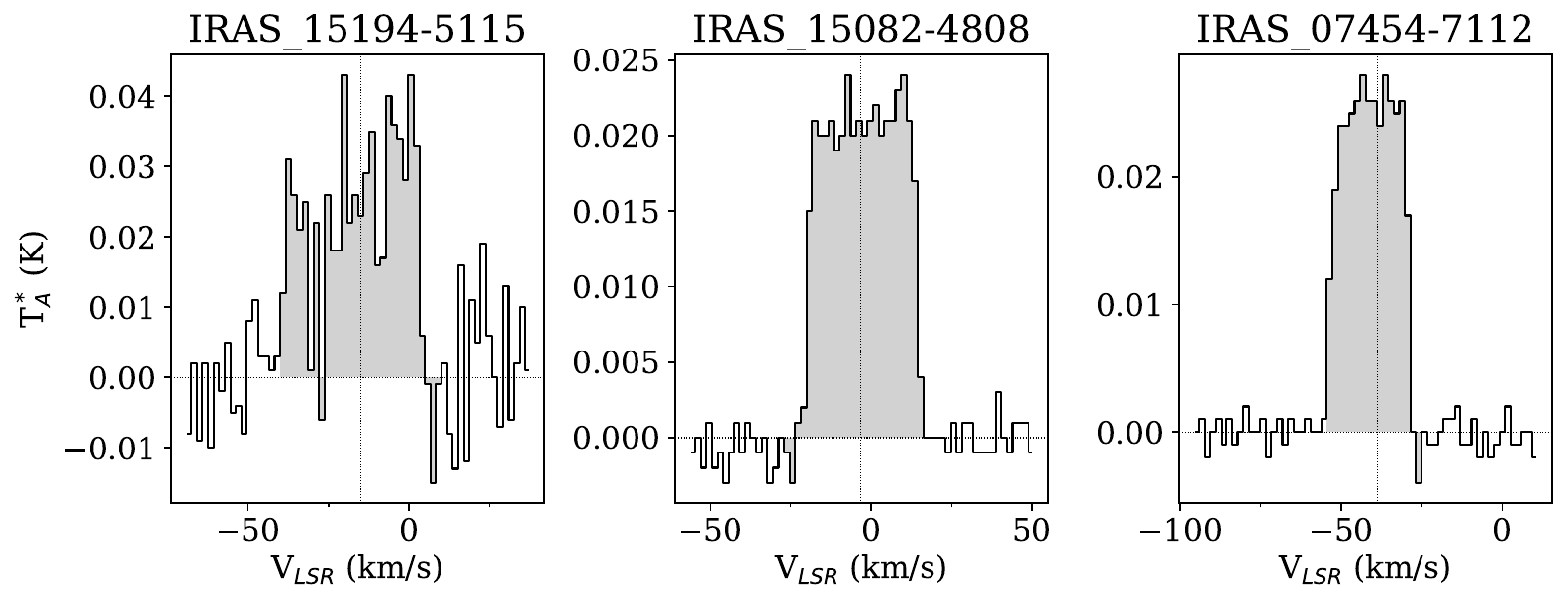}
    \caption{SiC$_2$, 9( 6, 4)- 8( 6, 3) (212.031878 GHz)}
\end{figure}

\begin{figure}[h]
    \centering
    \includegraphics[width=0.65\linewidth]{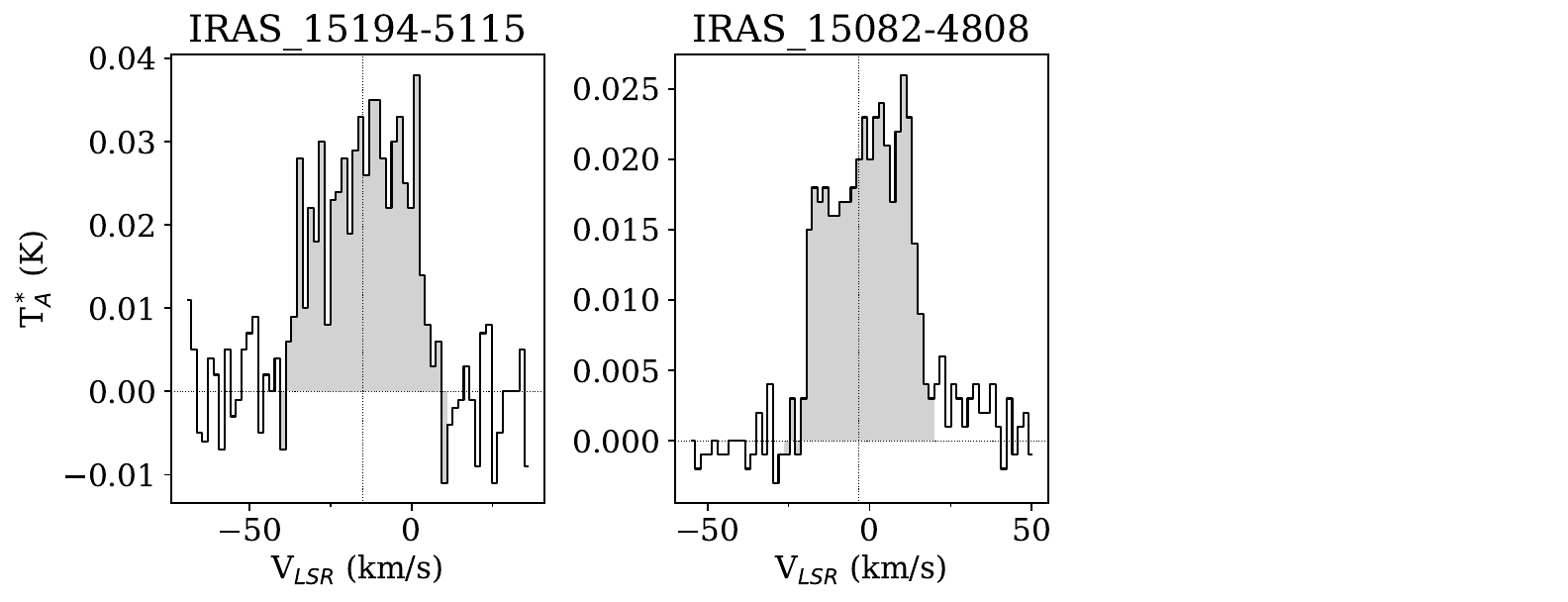}
    \caption{SiC$_2$, 9( 4, 6)- 8( 4, 5) (213.208032 GHz)}
\end{figure}

\begin{figure}[h]
    \centering
    \includegraphics[width=0.65\linewidth]{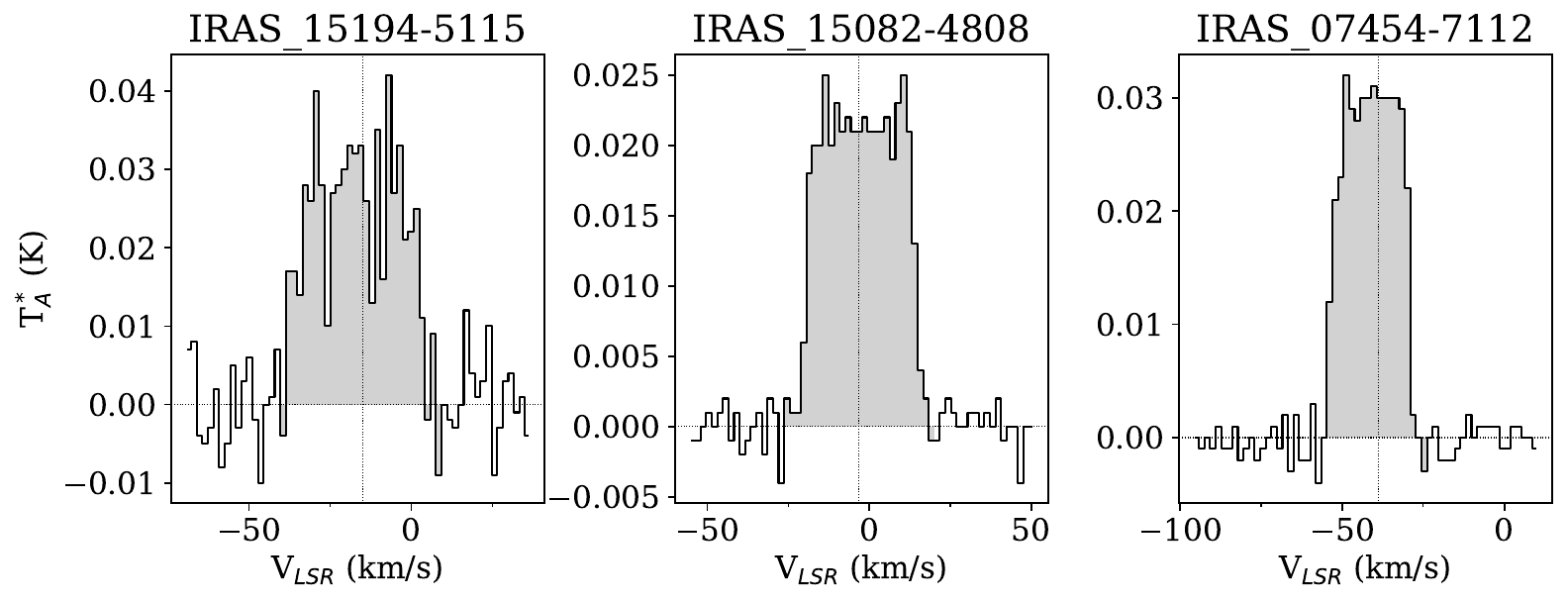}
    \caption{SiC$_2$, 9( 4, 5)- 8( 4, 4) (213.292337 GHz)}
\end{figure}

\begin{figure}[h]
    \centering
    \includegraphics[width=0.65\linewidth]{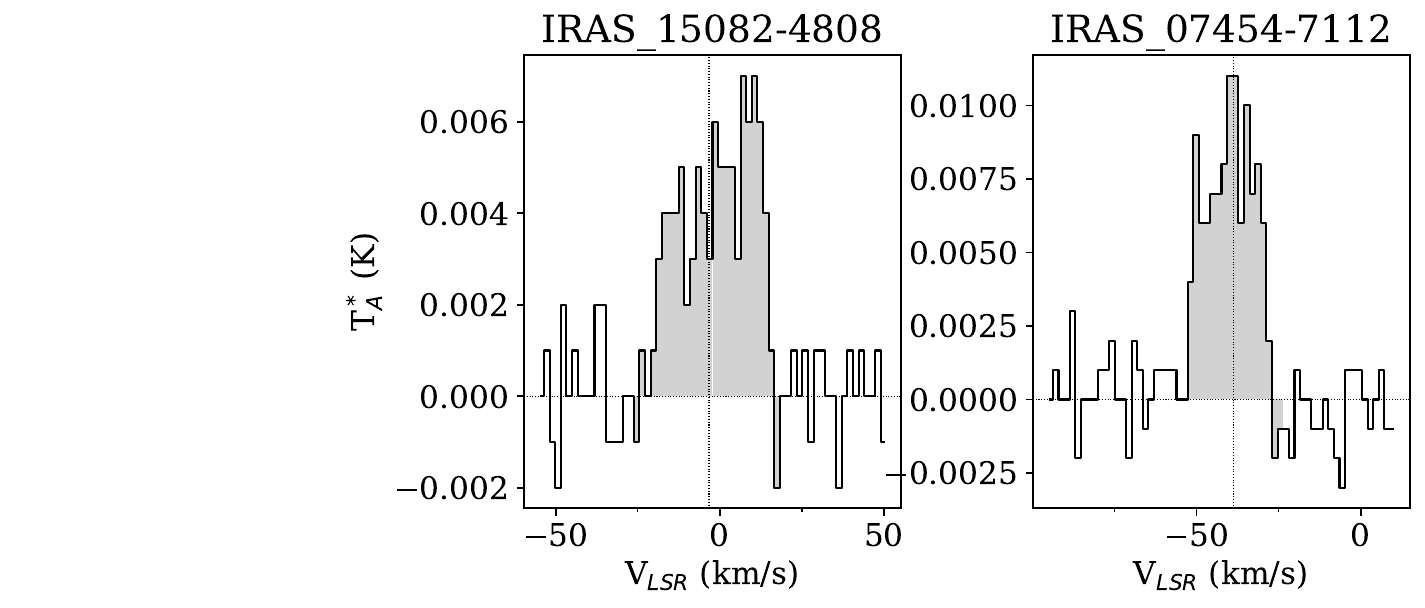}
    \caption{$^{29}$SiS, 12-11 (213.81614 GHz)}
\end{figure}

\begin{figure}[h]
    \centering
    \includegraphics[width=0.65\linewidth]{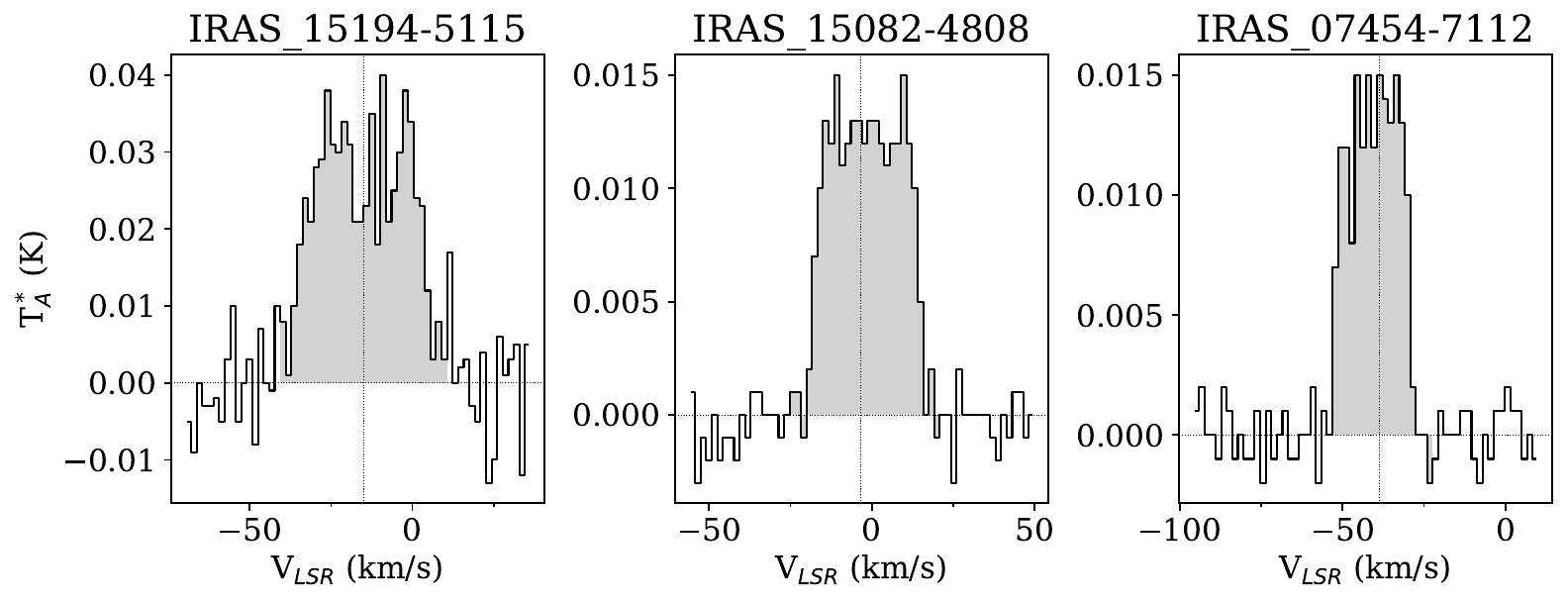}
    \caption{$^{29}$SiO, 5-4 (214.385752 GHz)}
\end{figure}

\begin{figure}[h]
    \centering
    \includegraphics[width=0.65\linewidth]{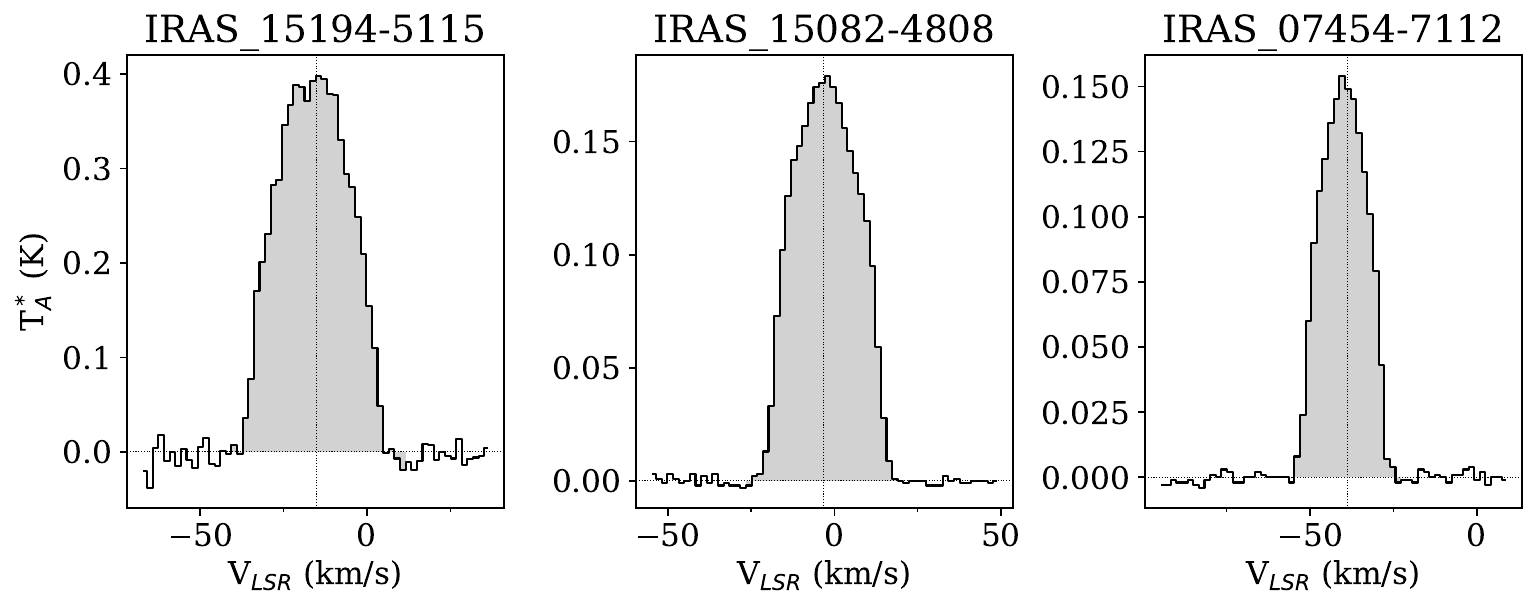}
    \caption{SiO, 5-4 (217.104919 GHz)}
\end{figure}

\begin{figure}[h]
    \centering
    \includegraphics[width=0.65\linewidth]{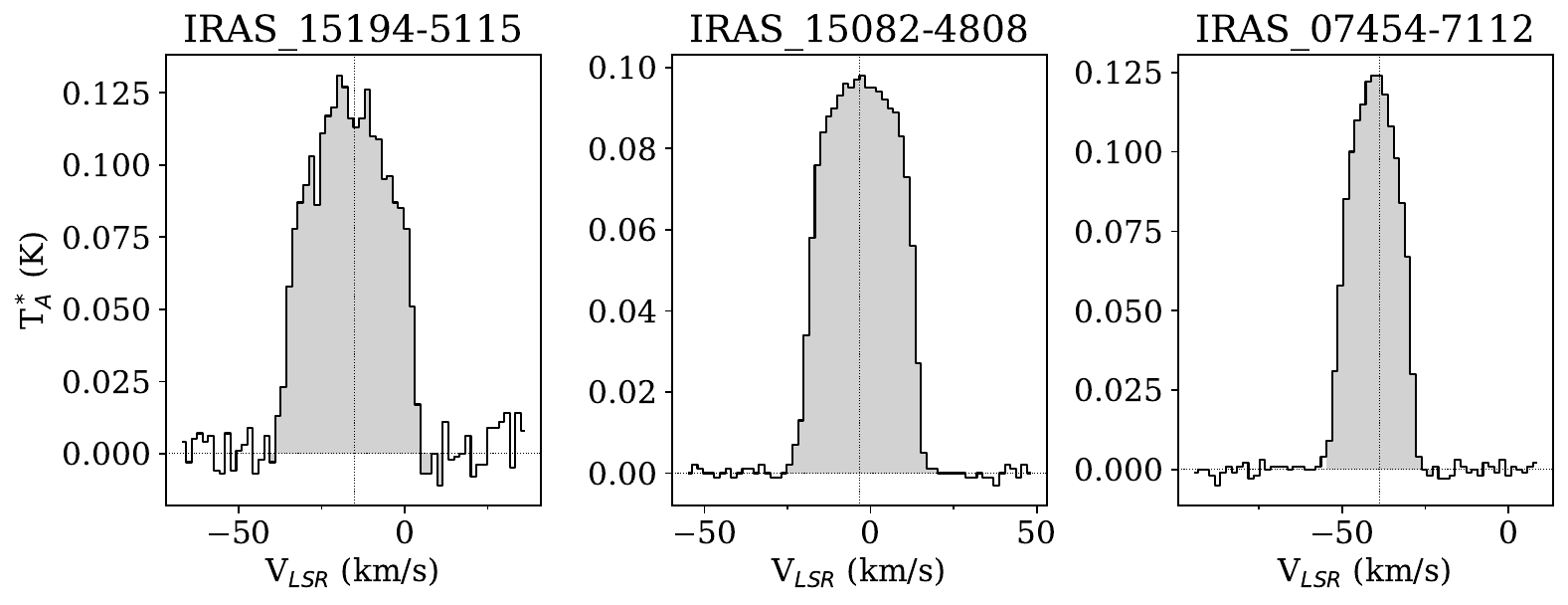}
    \caption{SiS, 12-11 (217.817663 GHz)}
\end{figure}

\begin{figure}[h]
    \centering
    \includegraphics[width=0.65\linewidth]{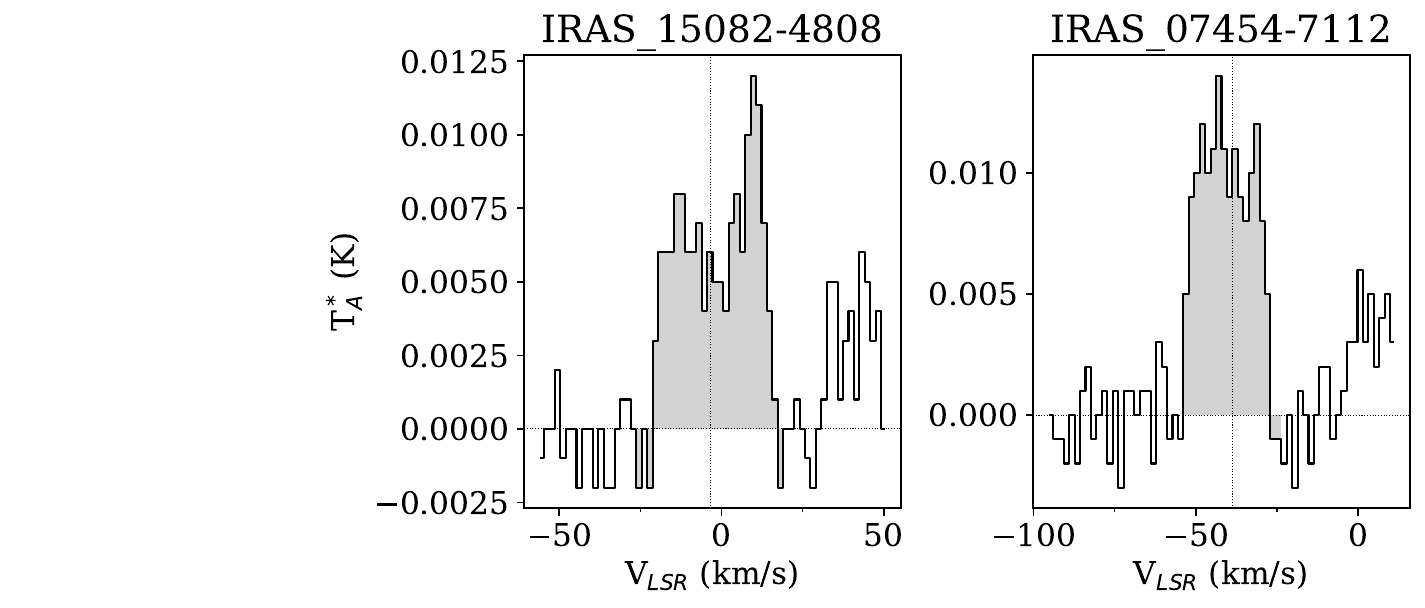}
    \caption{HC$_3$N, J=24-23 (218.324723 GHz)}
\end{figure}

\begin{figure}[h]
    \centering
    \includegraphics[width=0.65\linewidth]{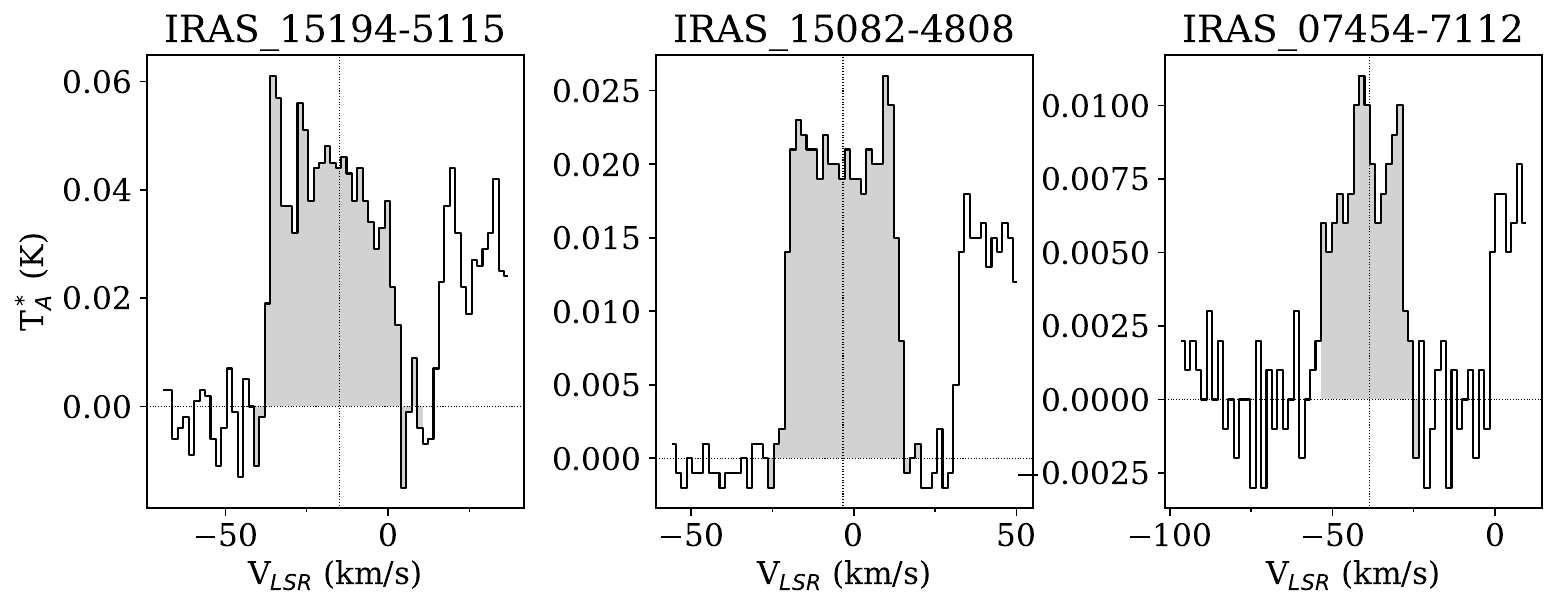}
    \caption{C$_4$H, N=23-22 (218.875369 GHz)}
\end{figure}

\begin{figure}[h]
    \centering
    \includegraphics[width=0.65\linewidth]{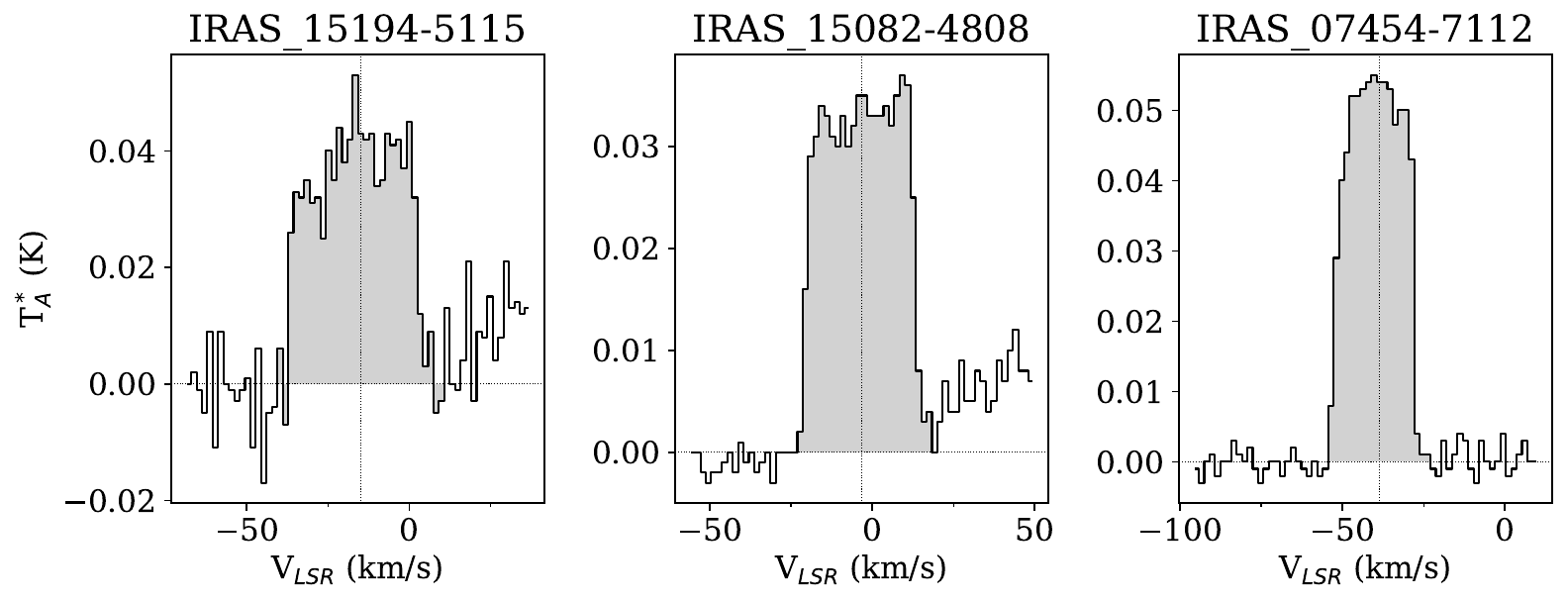}
    \caption{SiC$_2$, 10( 0,10)- 9( 0, 9) (220.773685 GHz)}
\end{figure}

\begin{figure}[h]
    \centering
    \includegraphics[width=0.65\linewidth]{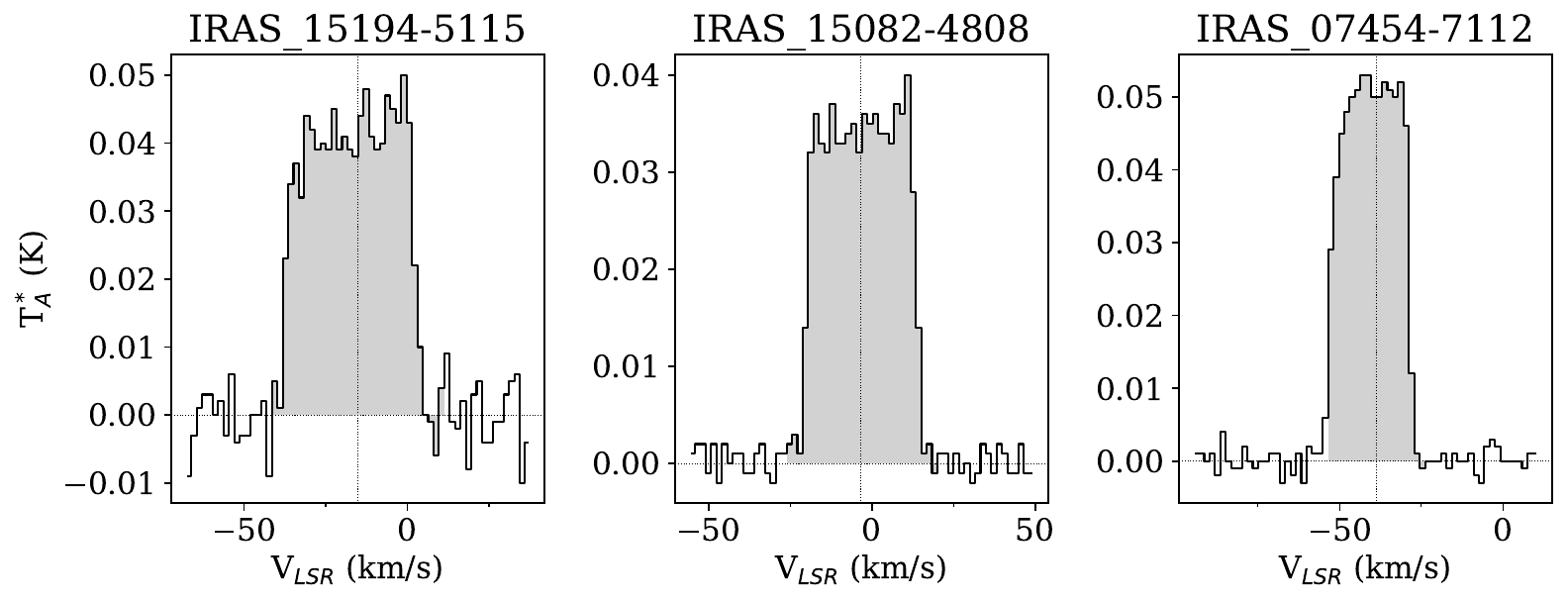}
    \caption{SiC$_2$, 9( 2, 7)- 8( 2, 6) (222.009386 GHz)}
\end{figure}

\begin{figure}[h]
    \centering
    \includegraphics[width=0.65\linewidth]{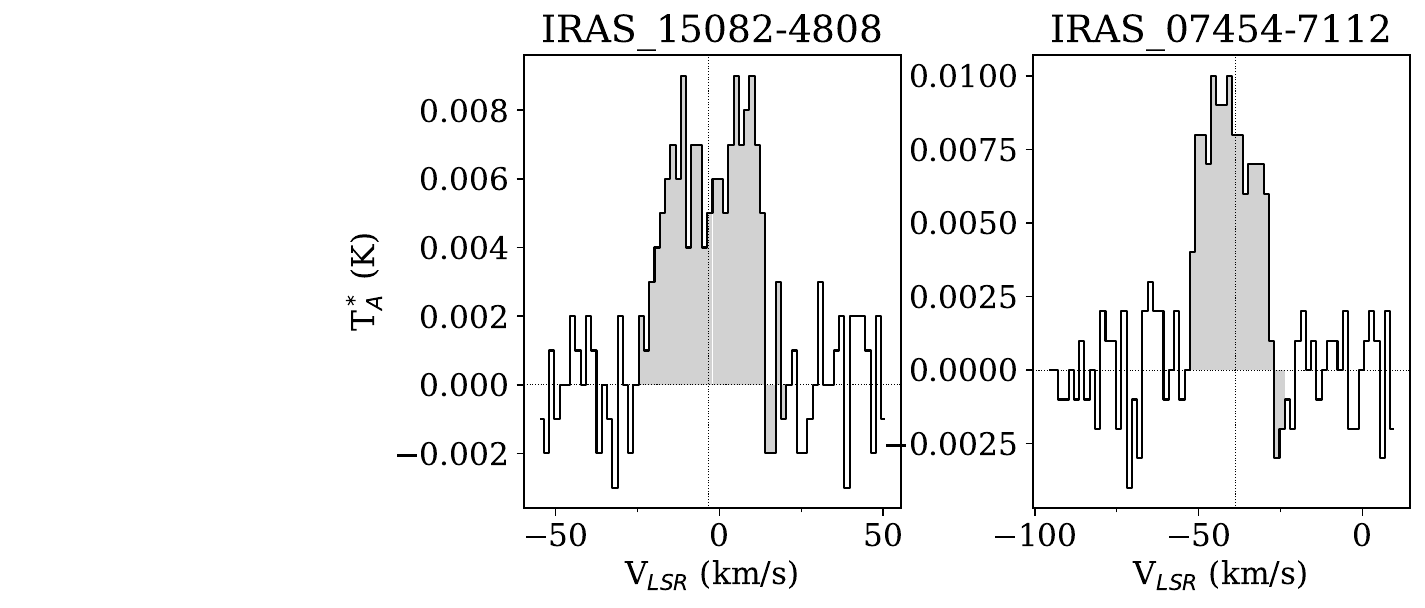}
    \caption{HC$_3$N, J=25-24 (227.418905 GHz)}
\end{figure}

\begin{figure}[h]
    \centering
    \includegraphics[width=0.65\linewidth]{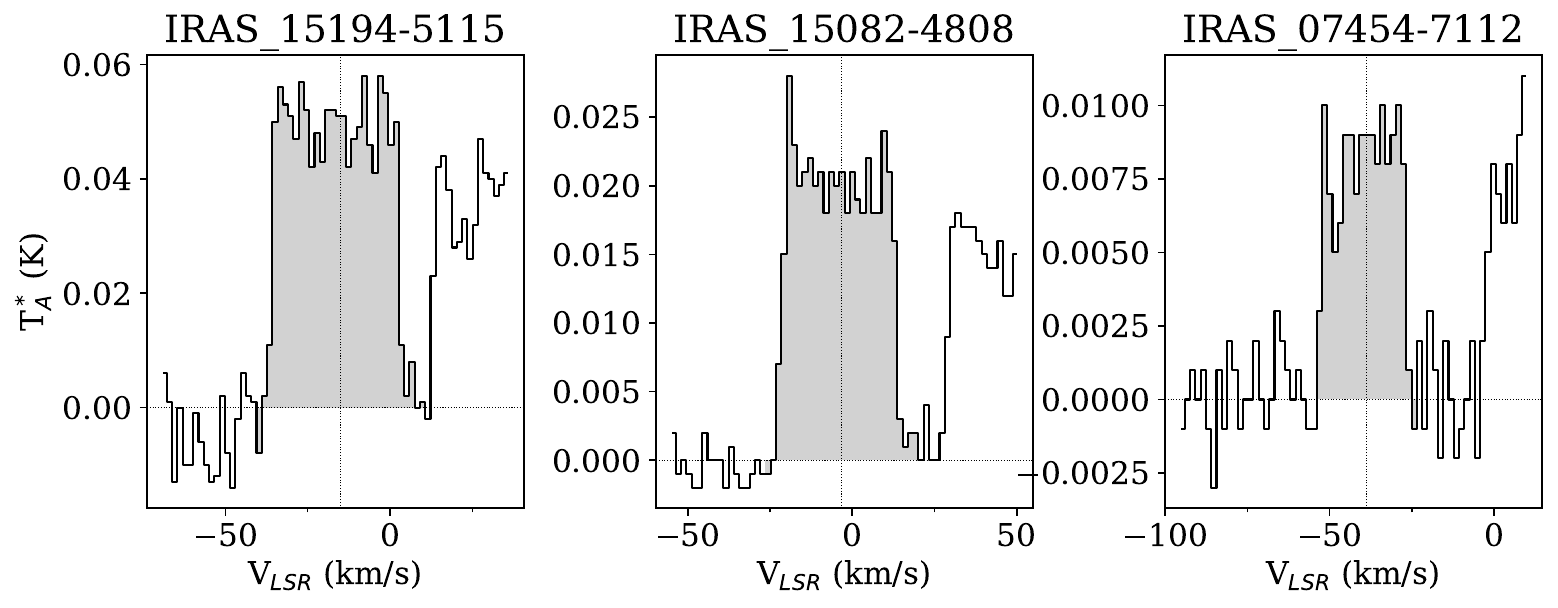}
    \caption{C$_4$H, N=24-23 (228.386962 GHz)}
\end{figure}

\begin{figure}[h]
    \centering
    \includegraphics[width=0.65\linewidth]{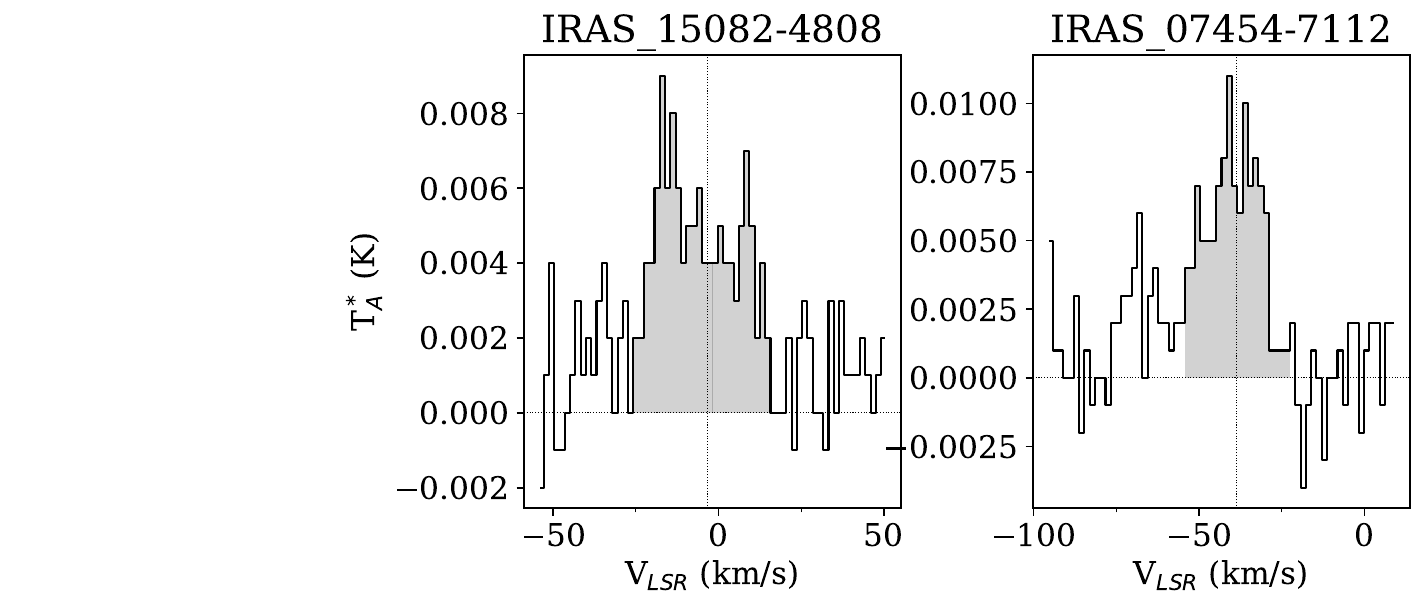}
    \caption{Si$^{34}$S, 13-12 (229.500868 GHz)}
\end{figure}

\begin{figure}[h]
    \centering
    \includegraphics[width=0.65\linewidth]{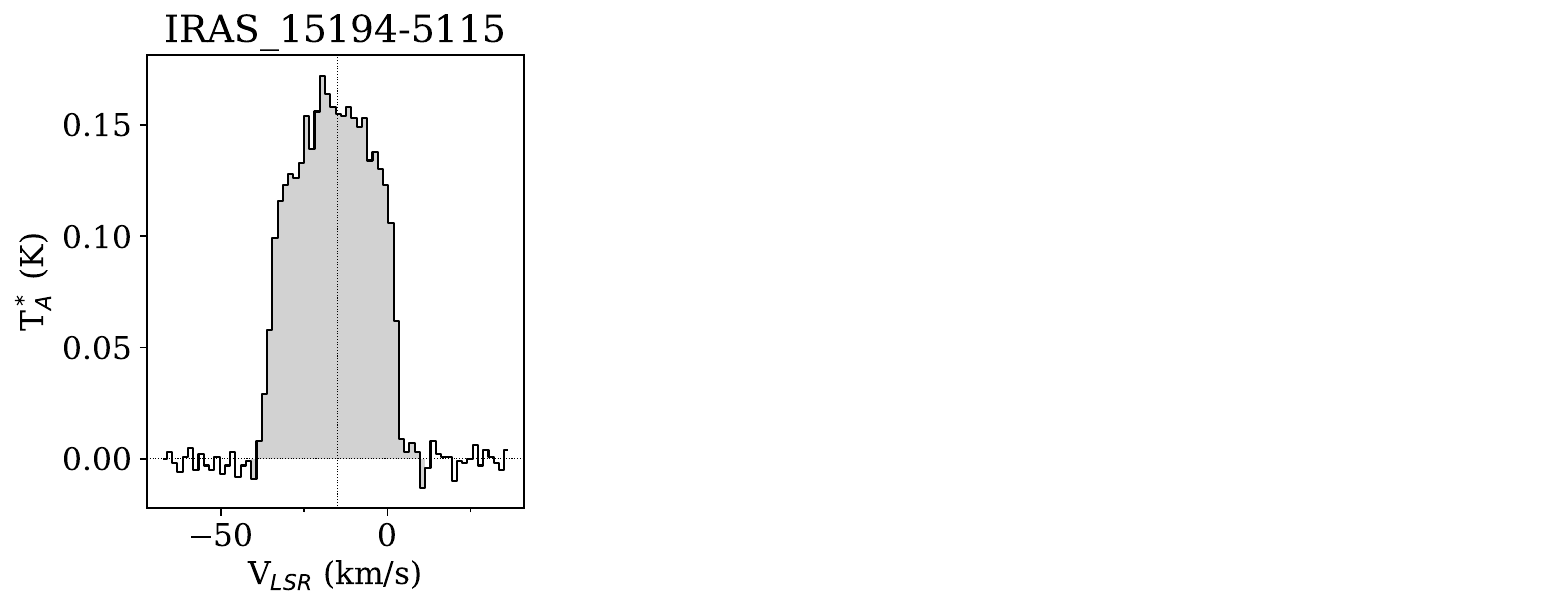}
    \caption{$^{13}$CS, J=5-4 (231.220685 GHz)}
\end{figure}

\begin{figure}[h]
    \centering
    \includegraphics[width=0.65\linewidth]{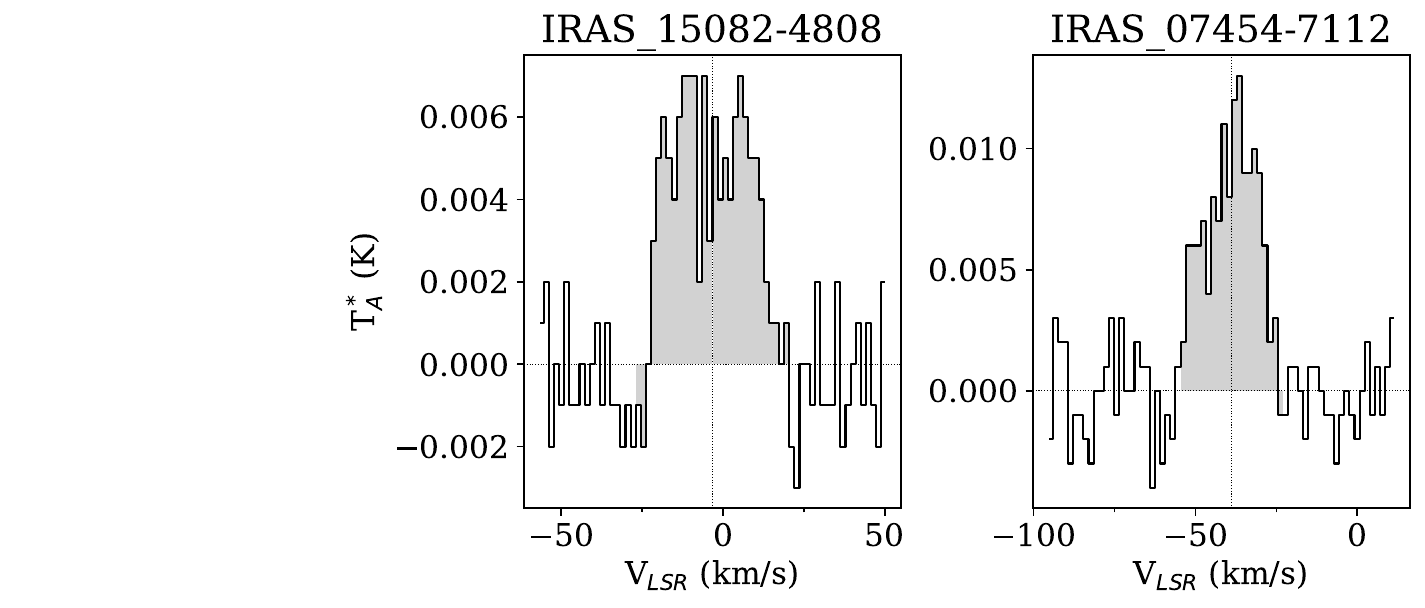}
    \caption{$^{29}$SiS, 13-12 (231.626673 GHz)}
\end{figure}

\begin{figure}[h]
    \centering
    \includegraphics[width=0.65\linewidth]{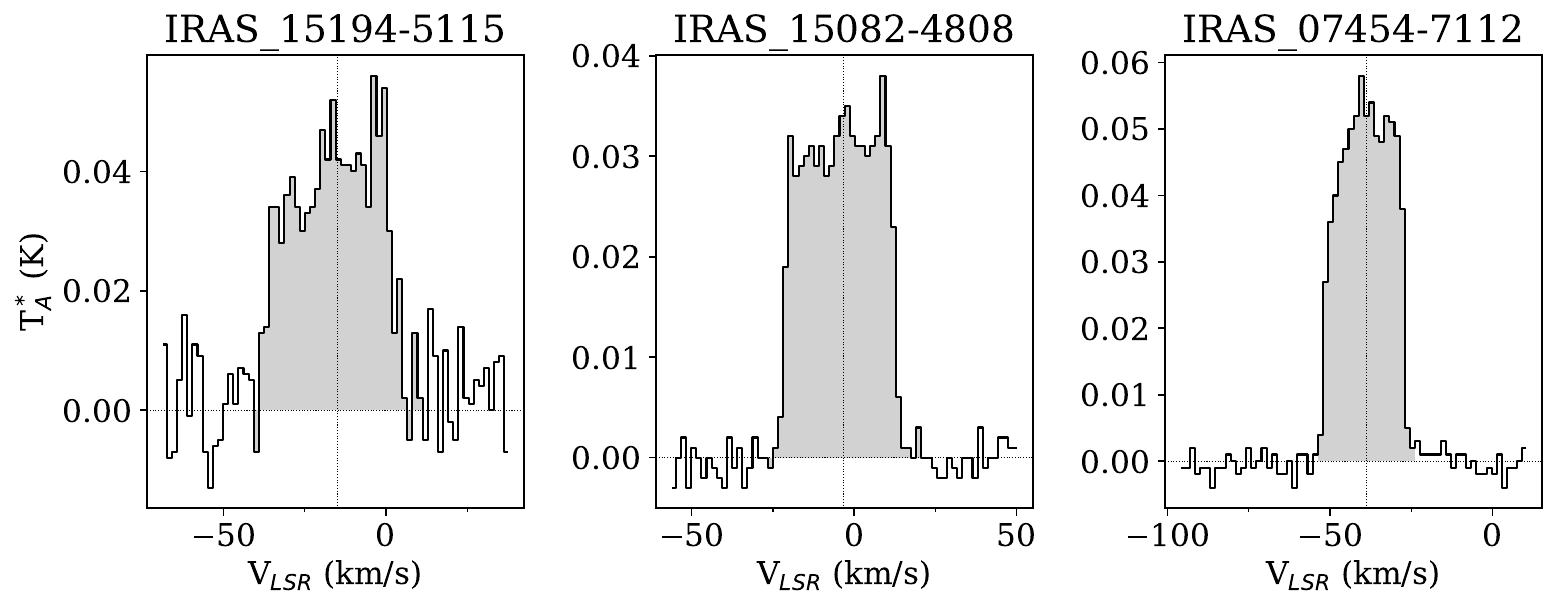}
    \caption{SiC$_2$, 10( 2, 9)- 9( 2, 8) (232.53407 GHz)}
\end{figure}

\begin{figure}[h]
    \centering
    \includegraphics[width=0.65\linewidth]{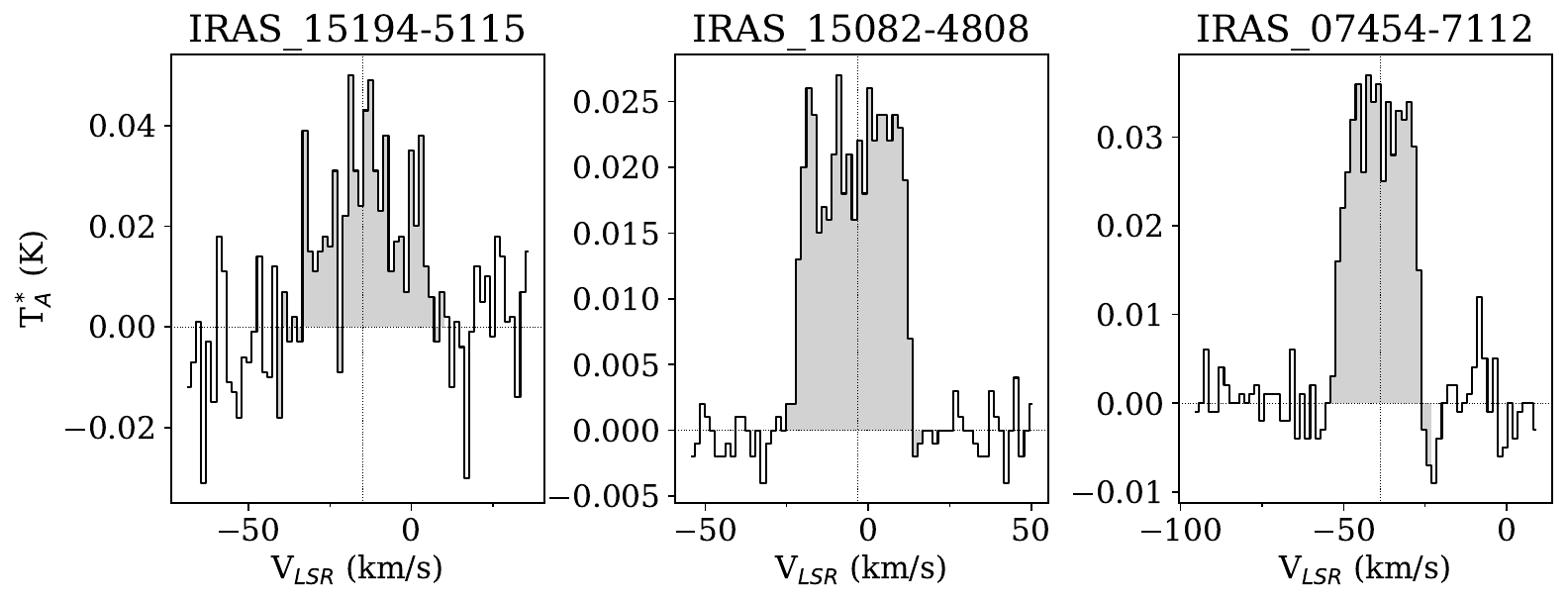}
    \caption{SiC$_2$, 10( 6, 5)- 9( 6, 4) (235.712998 GHz)}
\end{figure}

\begin{figure}[h]
    \centering
    \includegraphics[width=0.65\linewidth]{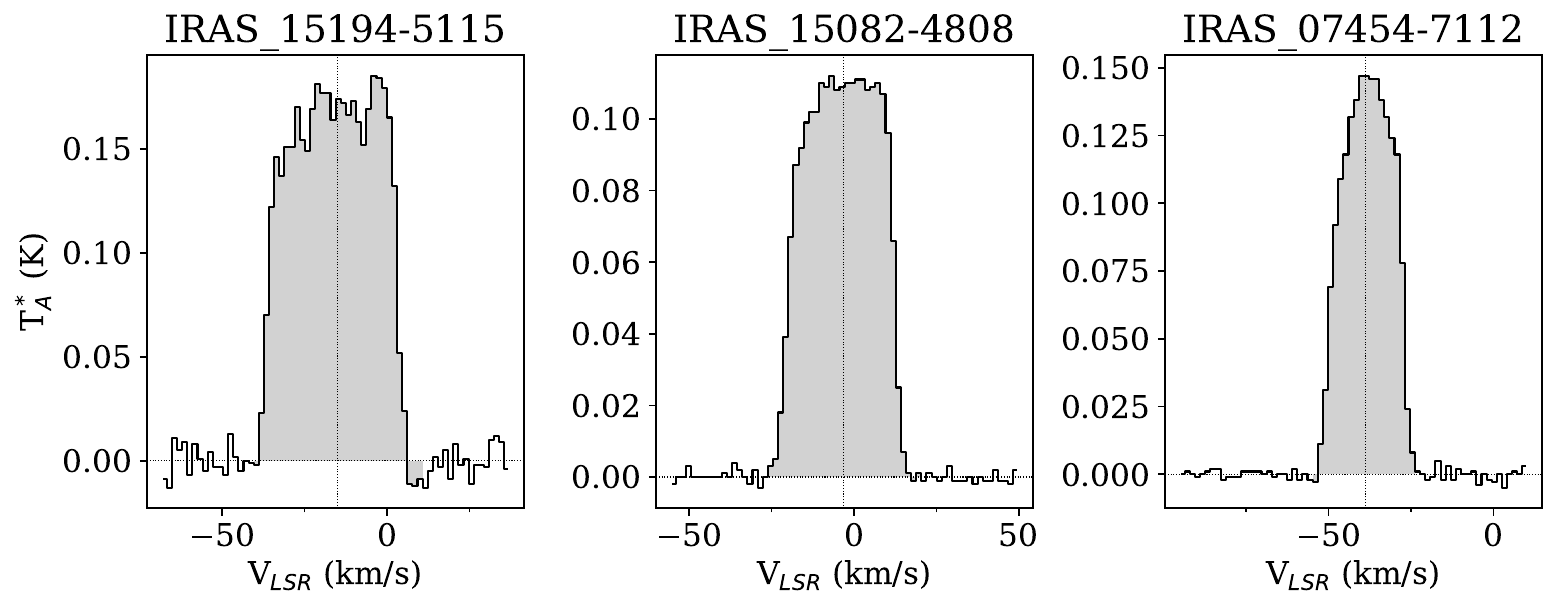}
    \caption{SiS, 13-12 (235.961363 GHz)}
\end{figure}

\begin{figure}[h]
    \centering
    \includegraphics[width=0.65\linewidth]{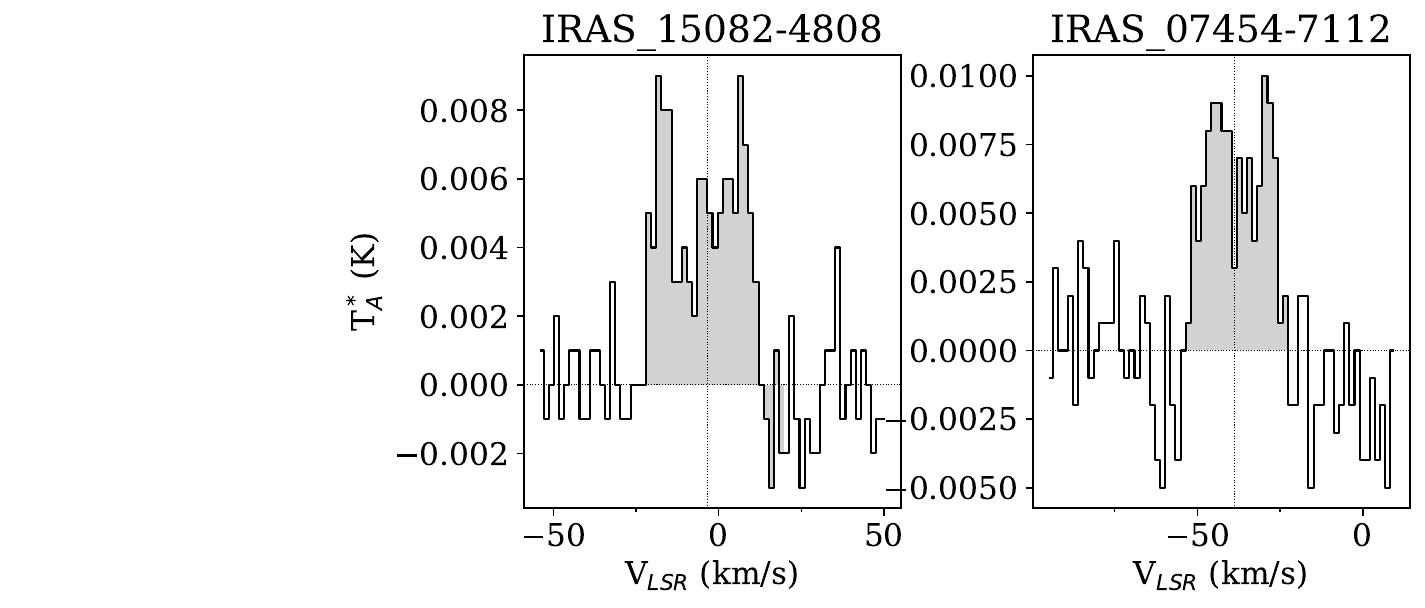}
    \caption{HC$_3$N, J=26-27 (236.512789 GHz)}
\end{figure}

\begin{figure}[h]
    \centering
    \includegraphics[width=0.65\linewidth]{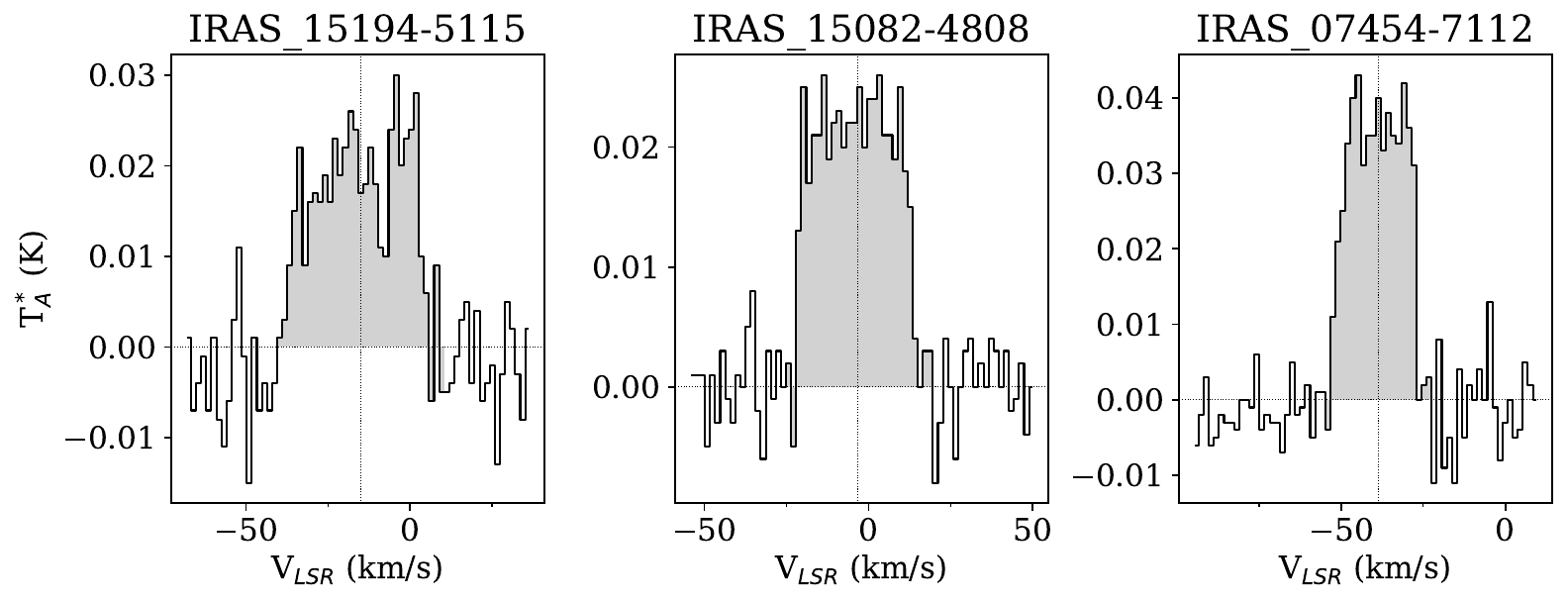}
    \caption{SiC$_2$, 10( 4, 7)- 9( 4, 6) (237.150018 GHz)}
\end{figure}

\begin{figure}[h]
    \centering
    \includegraphics[width=0.65\linewidth]{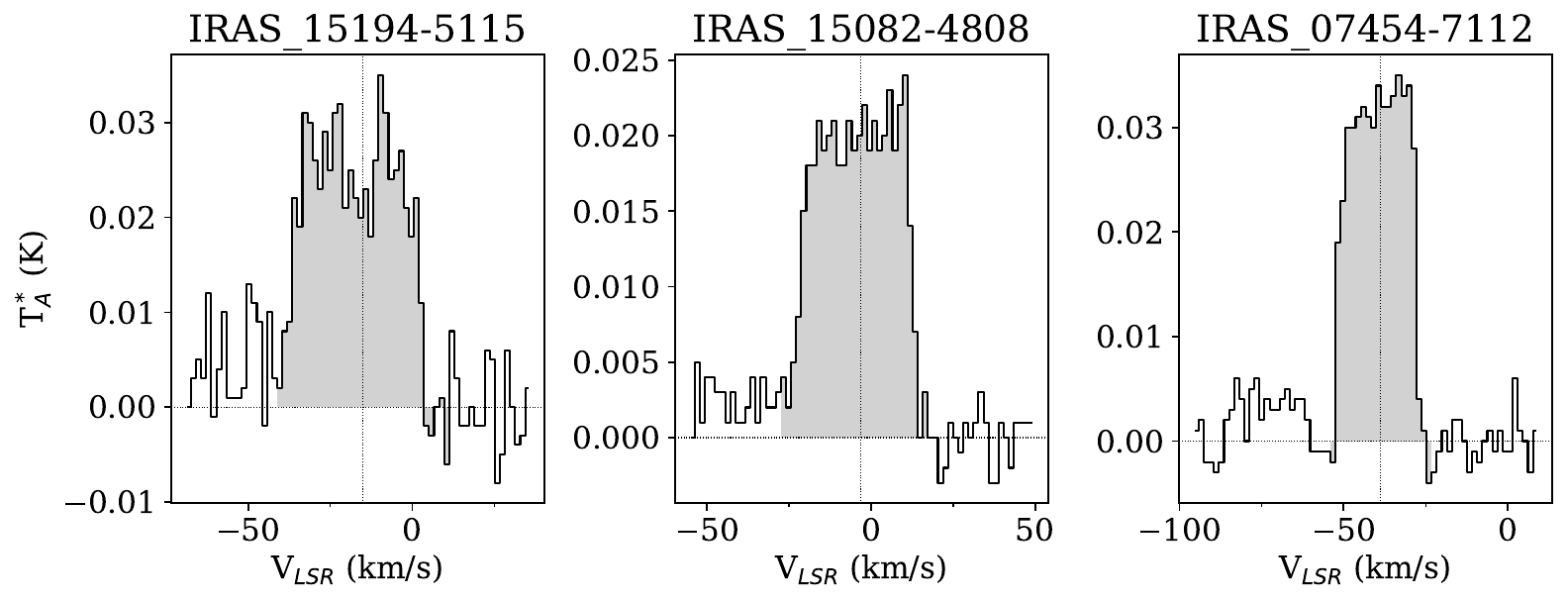}
    \caption{SiC$_2$, 10( 4, 6)- 9( 4, 5) (237.331309 GHz)}
\end{figure}

\begin{figure}[h]
    \centering
    \includegraphics[width=0.65\linewidth]{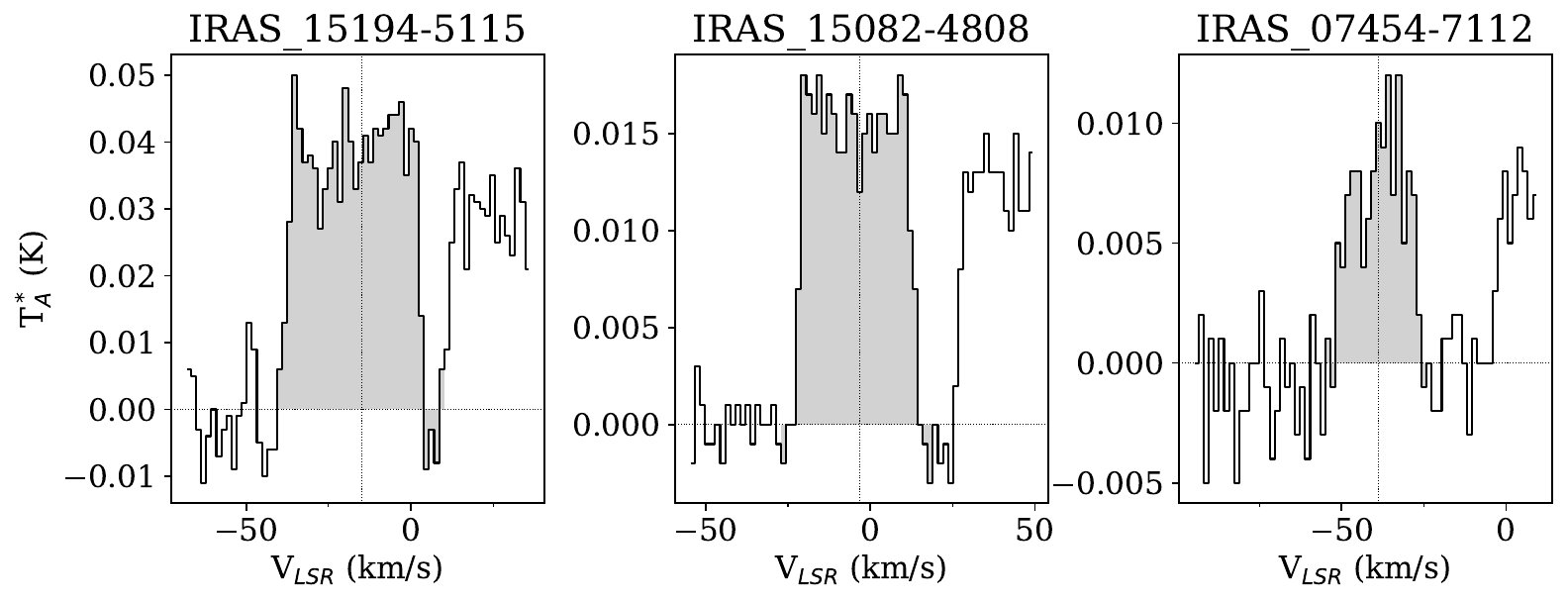}
    \caption{C$_4$H, N=25-24 (237.89806 GHz)}
\end{figure}

\begin{figure}[h]
    \centering
    \includegraphics[width=0.65\linewidth]{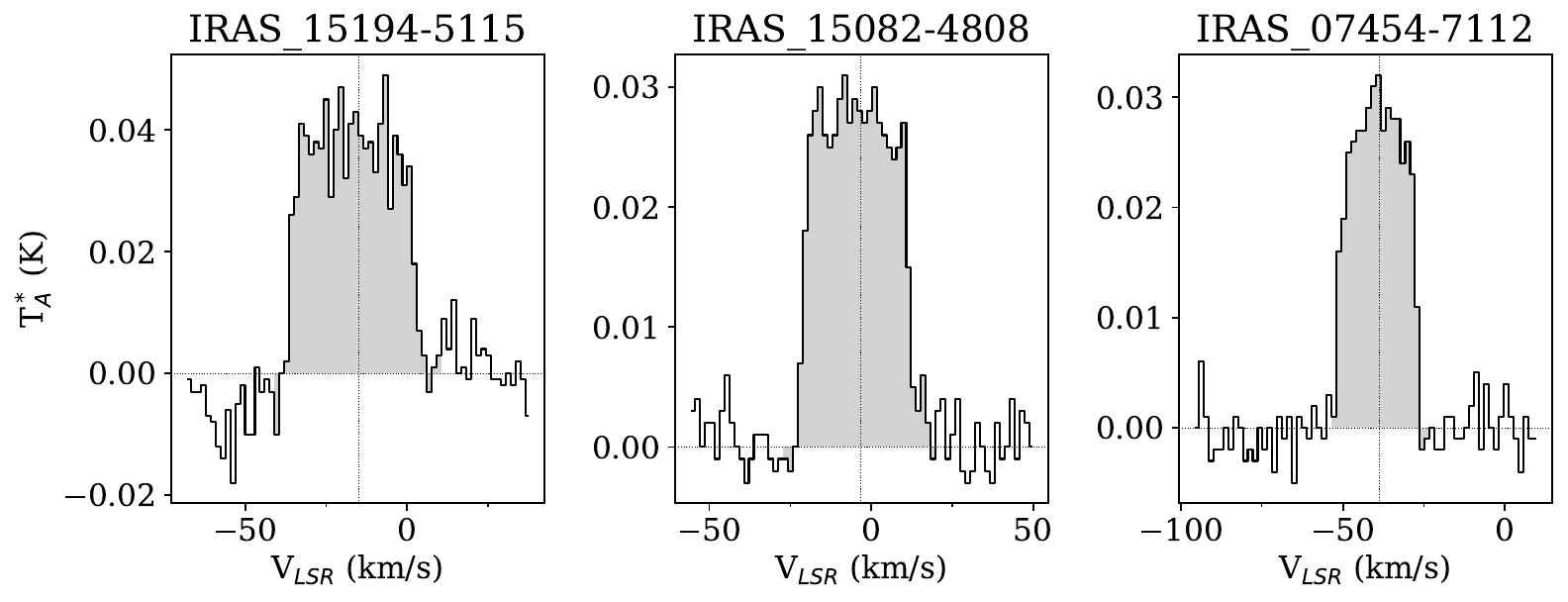}
    \caption{C$^{34}$S, 5-4 (241.016089 GHz)}
\end{figure}

\begin{figure}[h]
    \centering
    \includegraphics[width=0.65\linewidth]{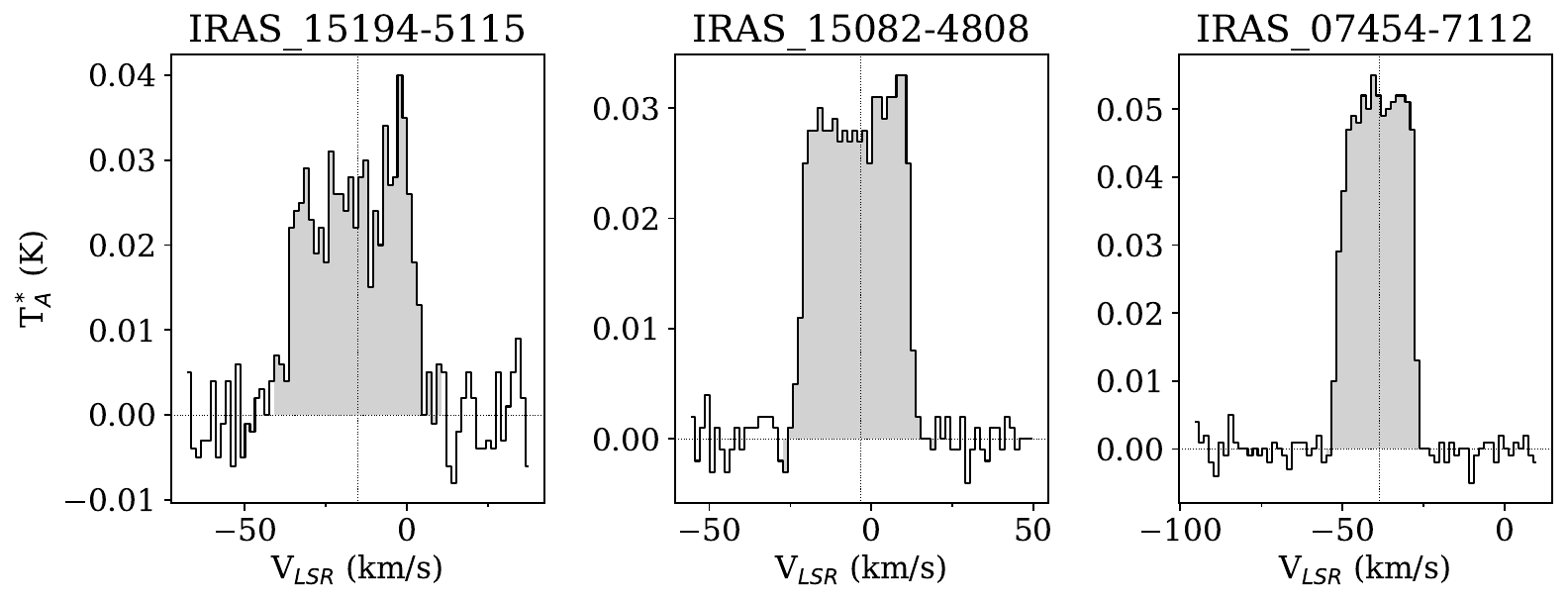}
    \caption{SiC$_2$, 11( 0,11)-10( 0,10) (241.367708 GHz)}
\end{figure}

\begin{figure}[h]
    \centering
    \includegraphics[width=0.65\linewidth]{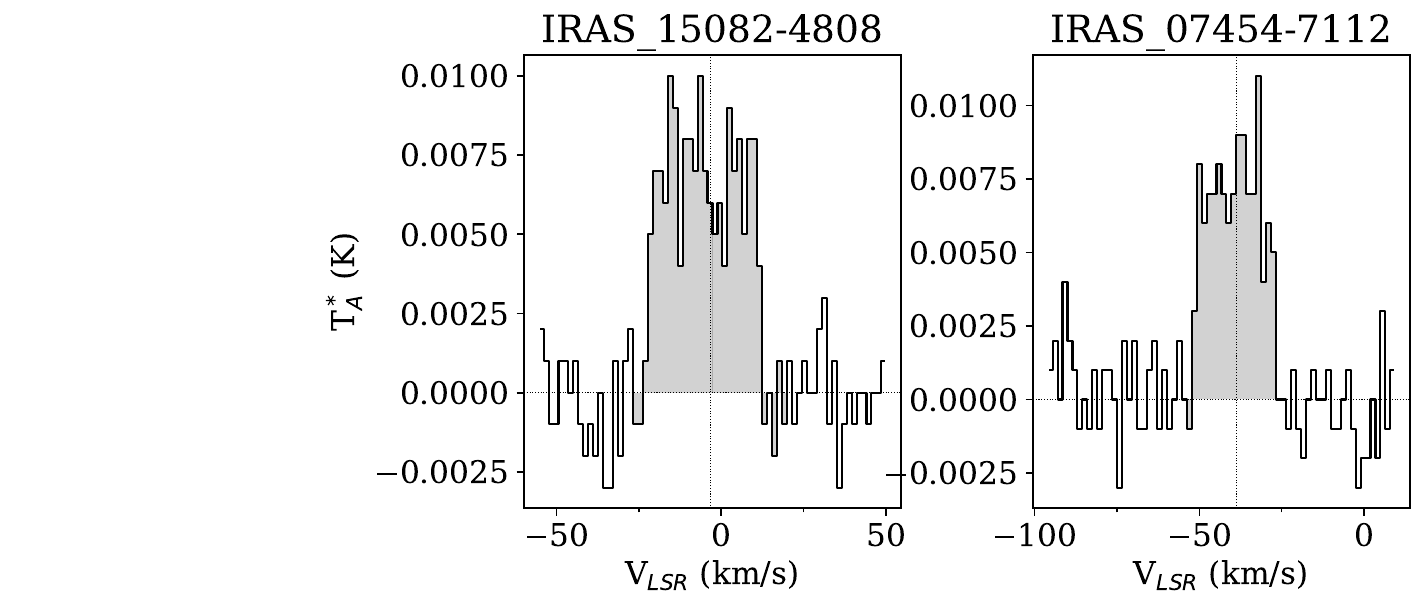}
    \caption{C$^{33}$S, 5-4 (242.91361 GHz)}
\end{figure}

\begin{figure}[h]
    \centering
    \includegraphics[width=0.65\linewidth]{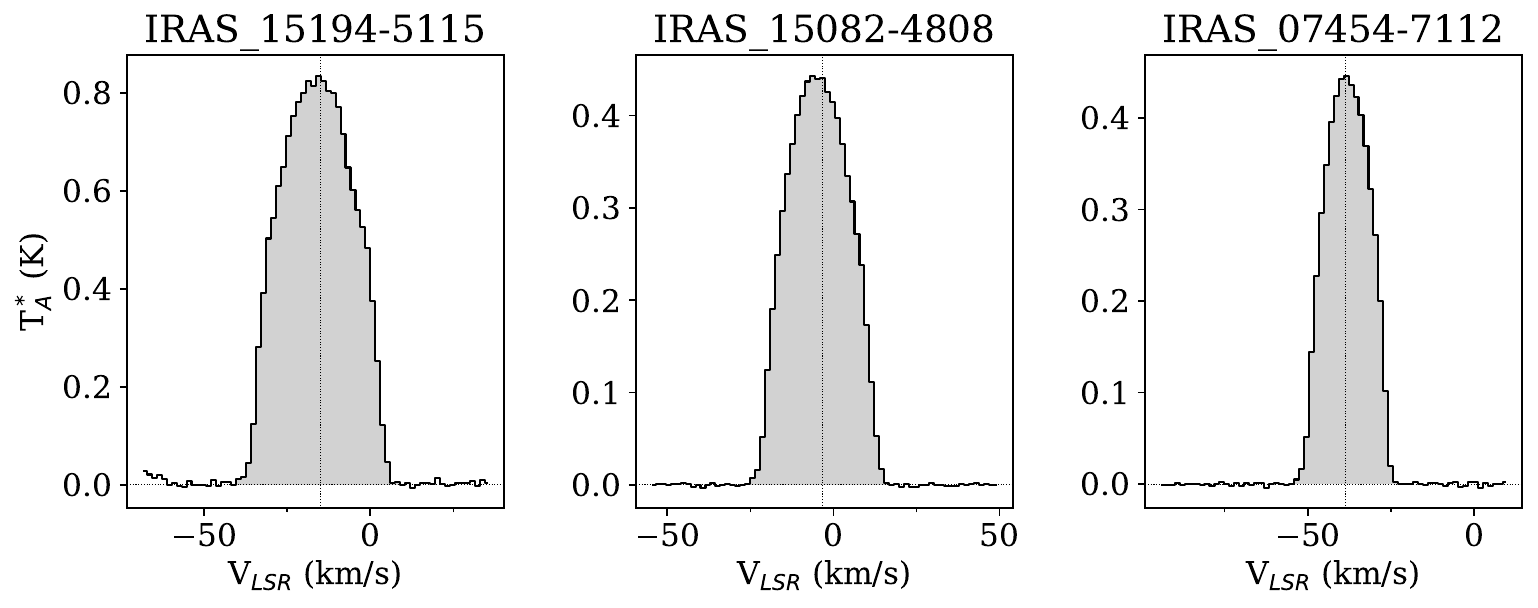}
    \caption{CS, 5-4 (244.935556 GHz)}
\end{figure}

\begin{figure}[h]
    \centering
    \includegraphics[width=0.65\linewidth]{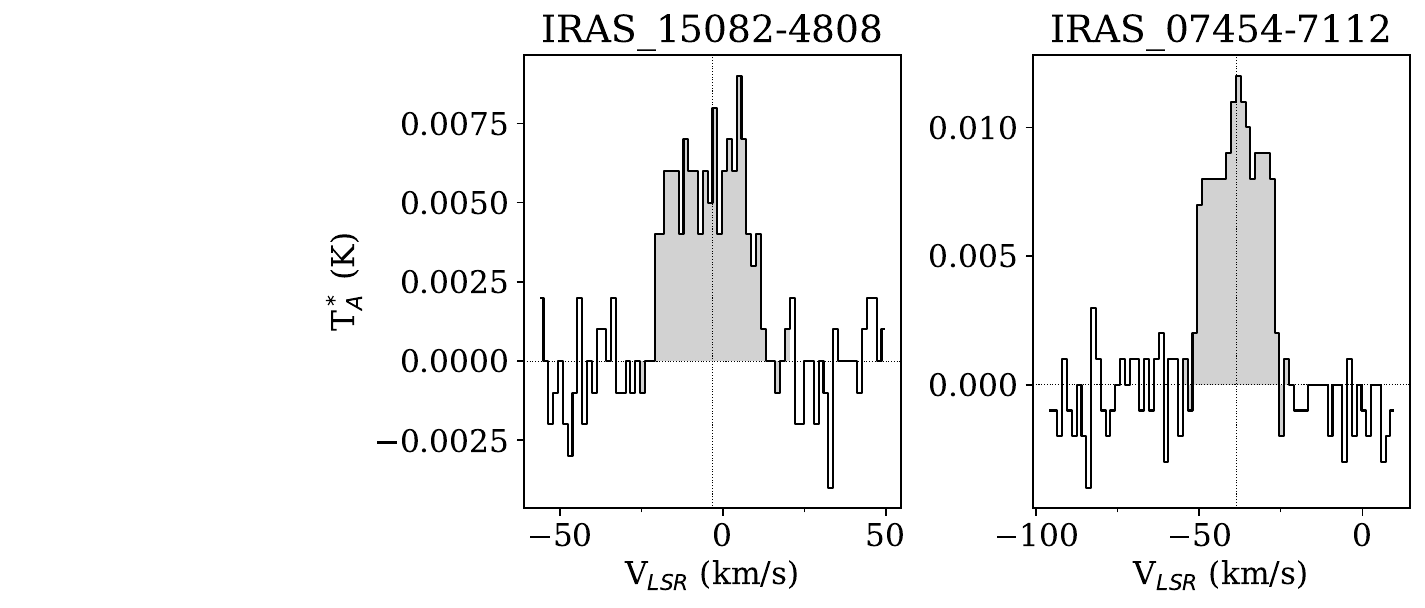}
    \caption{Si$^{34}$S, 14-13 (247.146242 GHz)}
\end{figure}

\begin{figure}[h]
    \centering
    \includegraphics[width=0.65\linewidth]{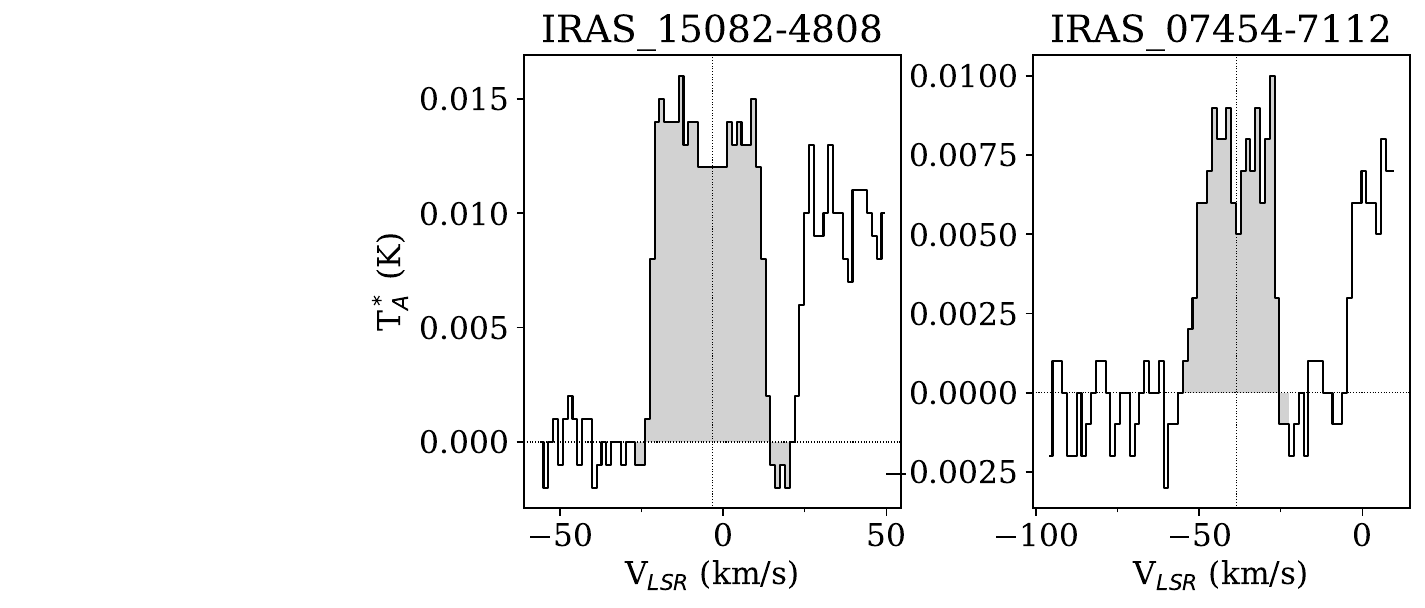}
    \caption{C$_4$H, N=26-25 (247.408643 GHz)}
\end{figure}

\begin{figure}[h]
    \centering
    \includegraphics[width=0.65\linewidth]{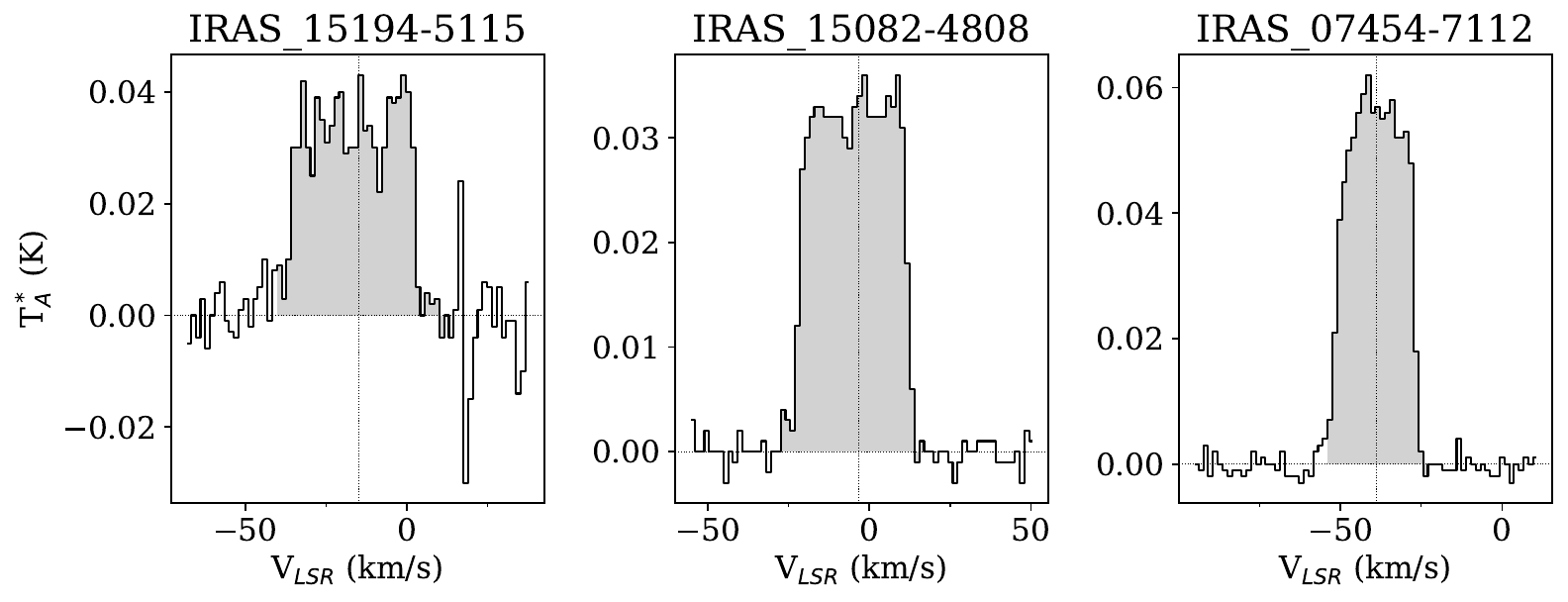}
    \caption{SiC$_2$, 10( 2, 8)- 9( 2, 7) (247.52912 GHz)}
\end{figure}

\begin{figure}[h]
    \centering
    \includegraphics[width=0.65\linewidth]{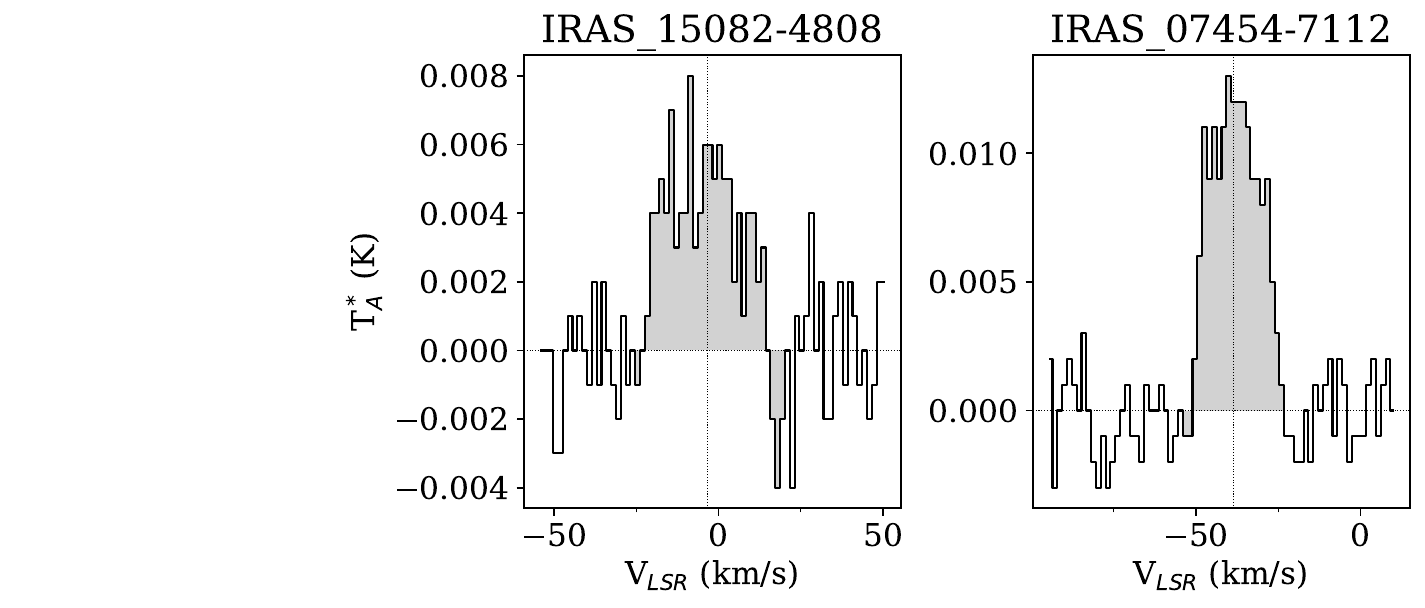}
    \caption{$^{29}$SiS, 14-13 (249.435412 GHz)}
\end{figure}

\begin{figure}[h]
    \centering
    \includegraphics[width=0.65\linewidth]{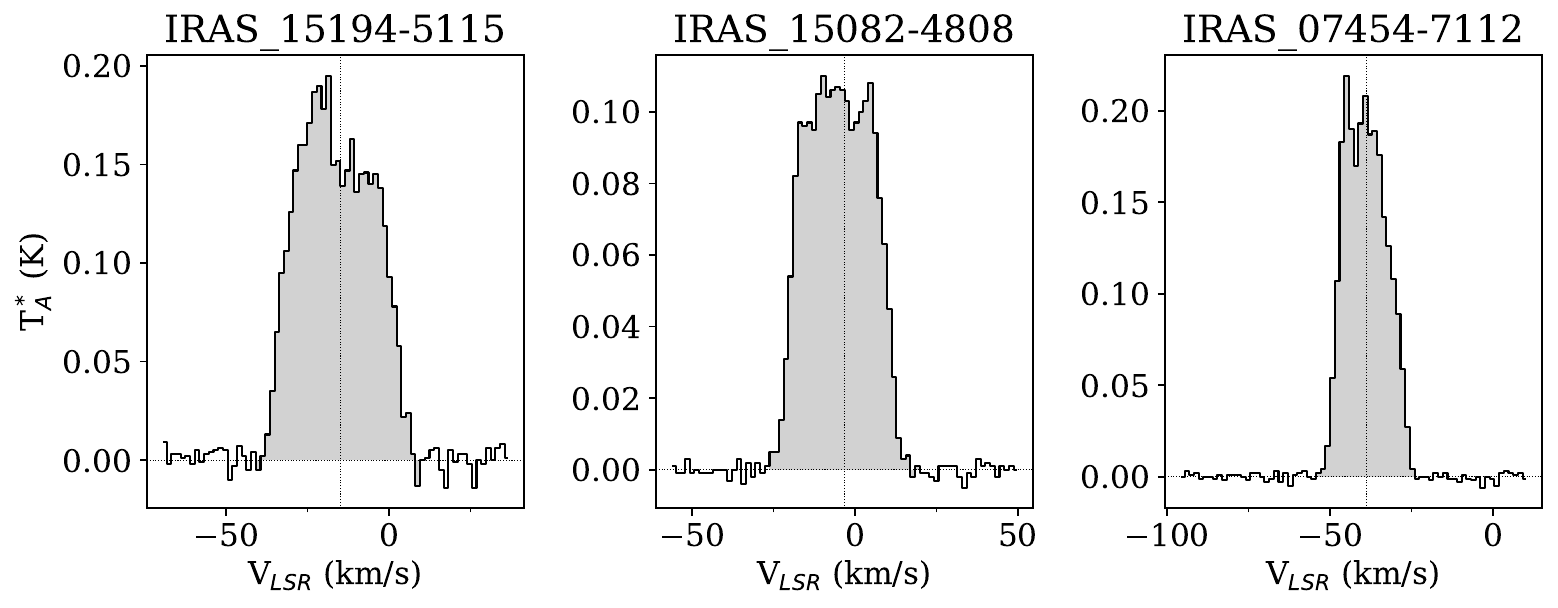}
    \caption{SiS, 14-13 (254.103211 GHz)}
\end{figure}

\begin{figure}[h]
    \centering
    \includegraphics[width=0.65\linewidth]{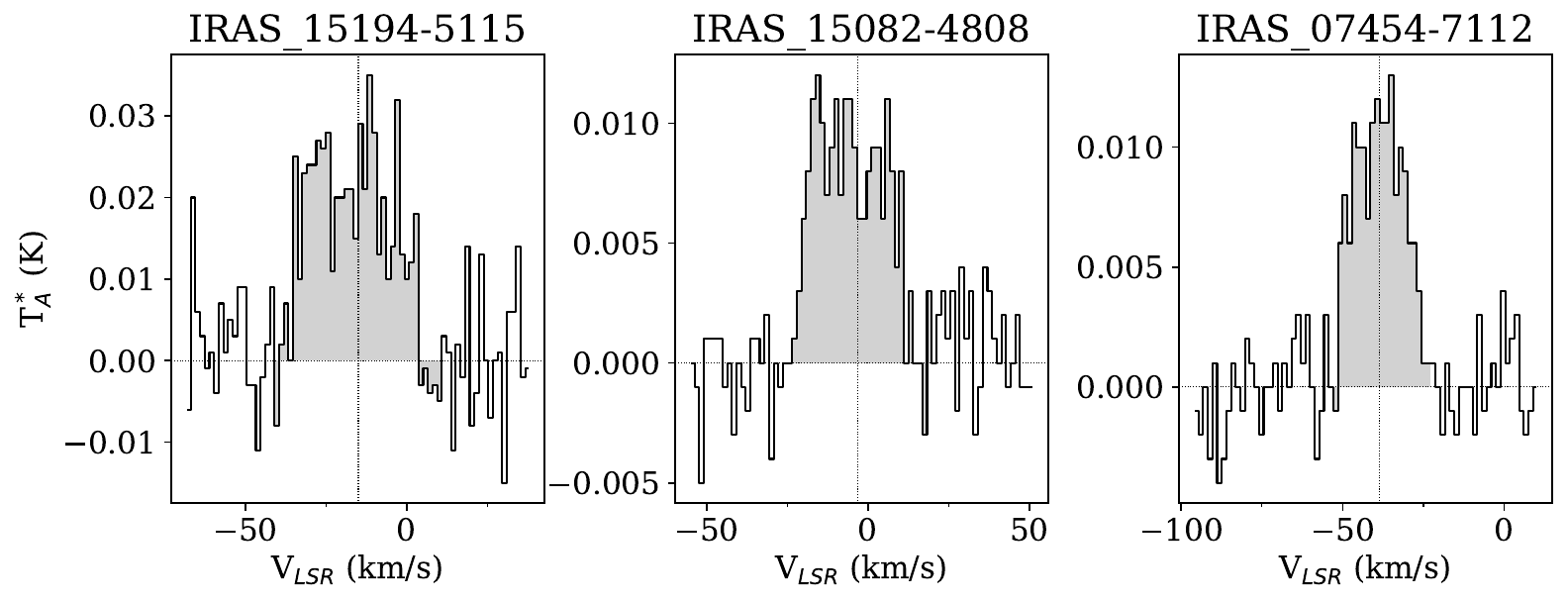}
    \caption{$^{30}$SiO, 6-5 (254.216656 GHz)}
\end{figure}

\begin{figure}[h]
    \centering
    \includegraphics[width=0.65\linewidth]{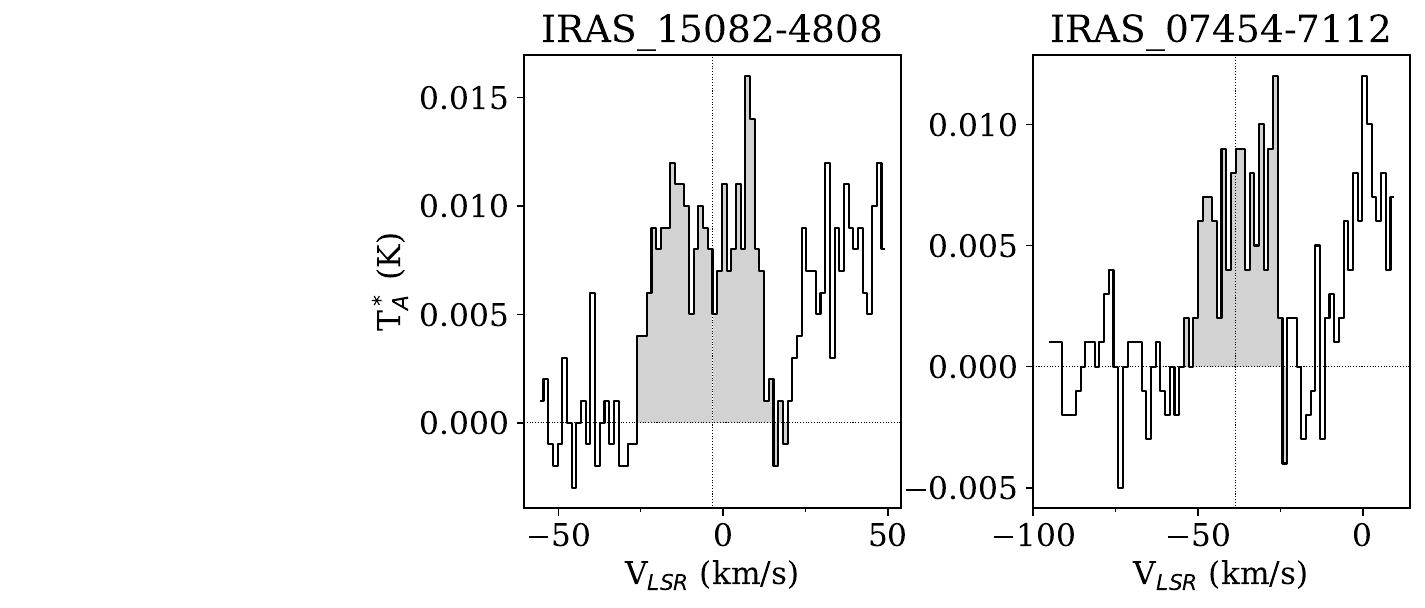}
    \caption{C$_4$H, N=27-26 (256.918691 GHz)}
\end{figure}

\begin{figure}[h]
    \centering
    \includegraphics[width=0.65\linewidth]{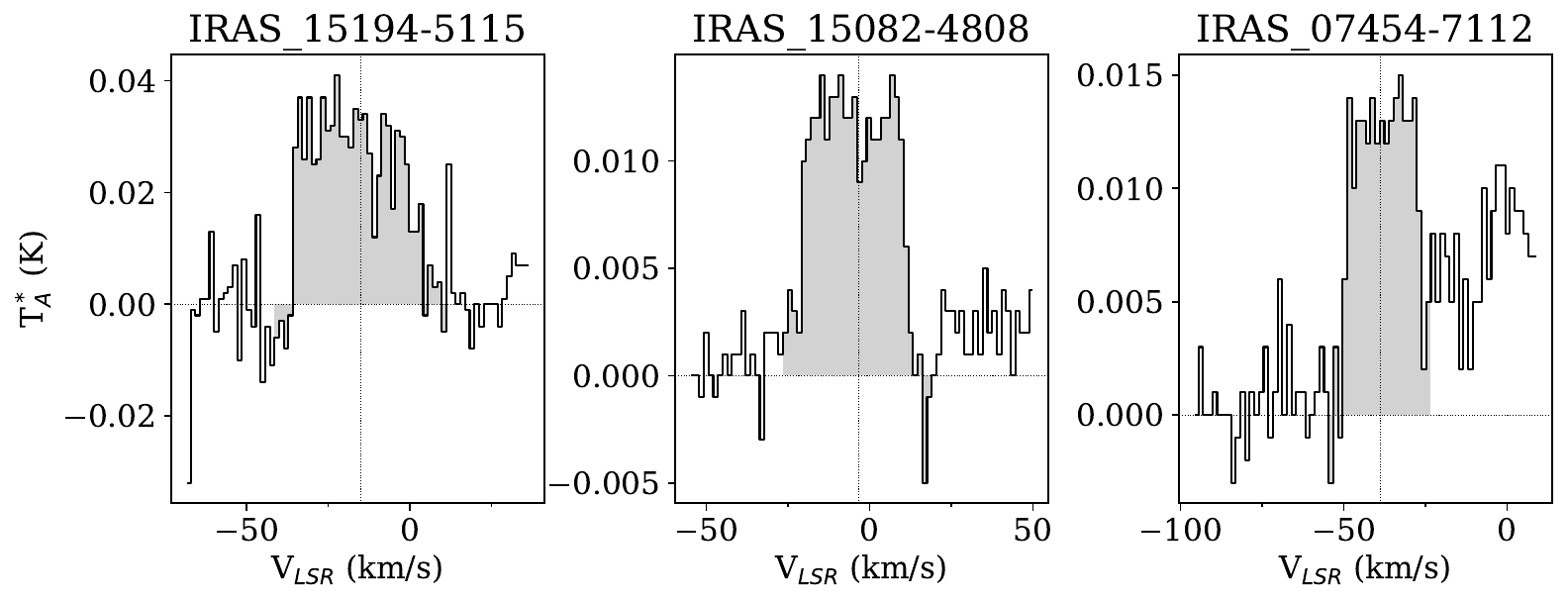}
    \caption{$^{29}$SiO, 6-5 (257.255213 GHz)}
\end{figure}

\begin{figure}[h]
    \centering
    \includegraphics[width=0.65\linewidth]{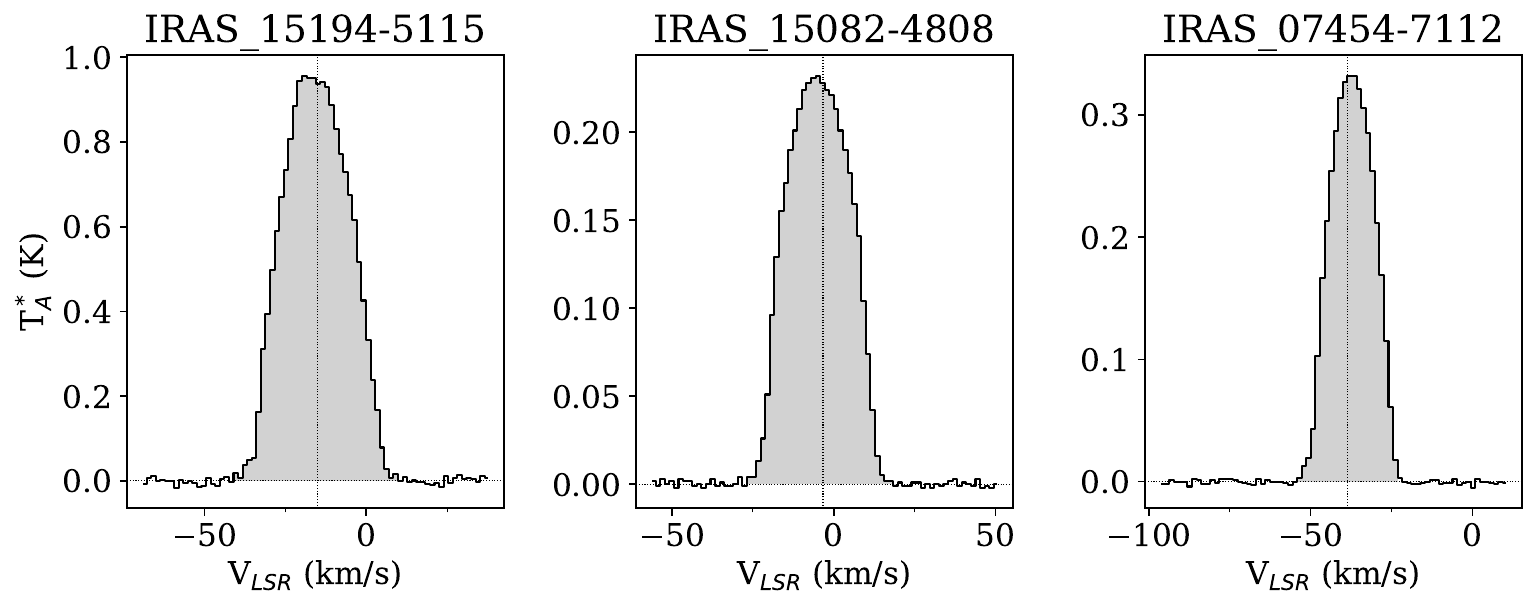}
    \caption{H$^{13}$CN, J=3-2 (259.011798 GHz)}
\end{figure}

\begin{figure}[h]
    \centering
    \includegraphics[width=0.65\linewidth]{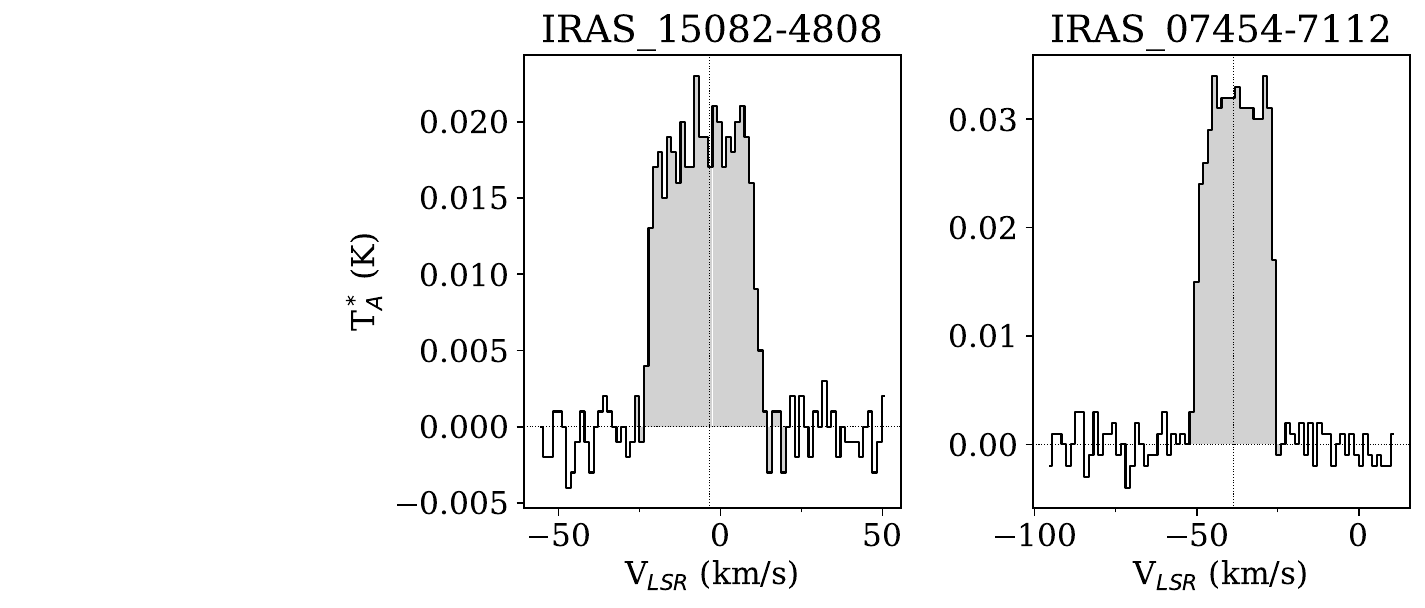}
    \caption{SiC$_2$, 11( 6, 6)-10( 6, 5) (259.433309 GHz)}
\end{figure}

\begin{figure}[h]
    \centering
    \includegraphics[width=0.65\linewidth]{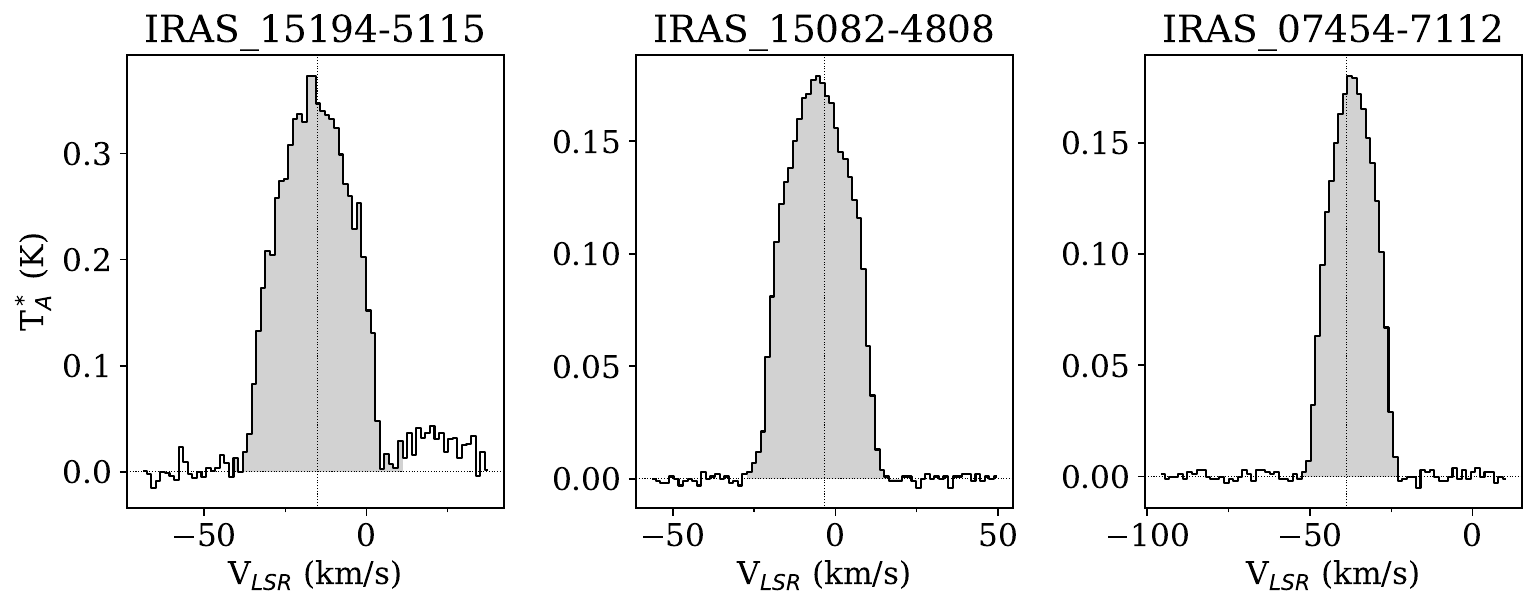}
    \caption{SiO, 6-5 (260.518009 GHz)}
\end{figure}

\begin{figure}[h]
    \centering
    \includegraphics[width=0.65\linewidth]{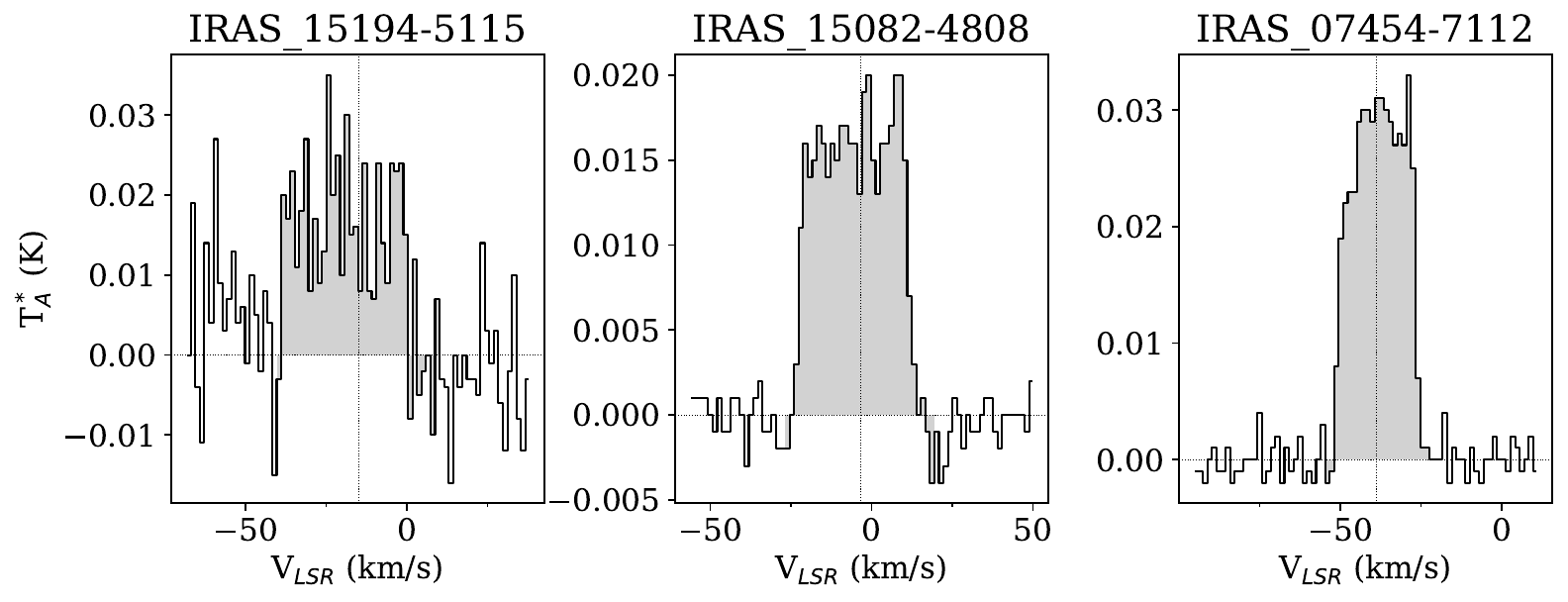}
    \caption{SiC$_2$, 11( 4, 8)-10( 4, 7) (261.150695 GHz)}
\end{figure}

\begin{figure}[h]
    \centering
    \includegraphics[width=0.65\linewidth]{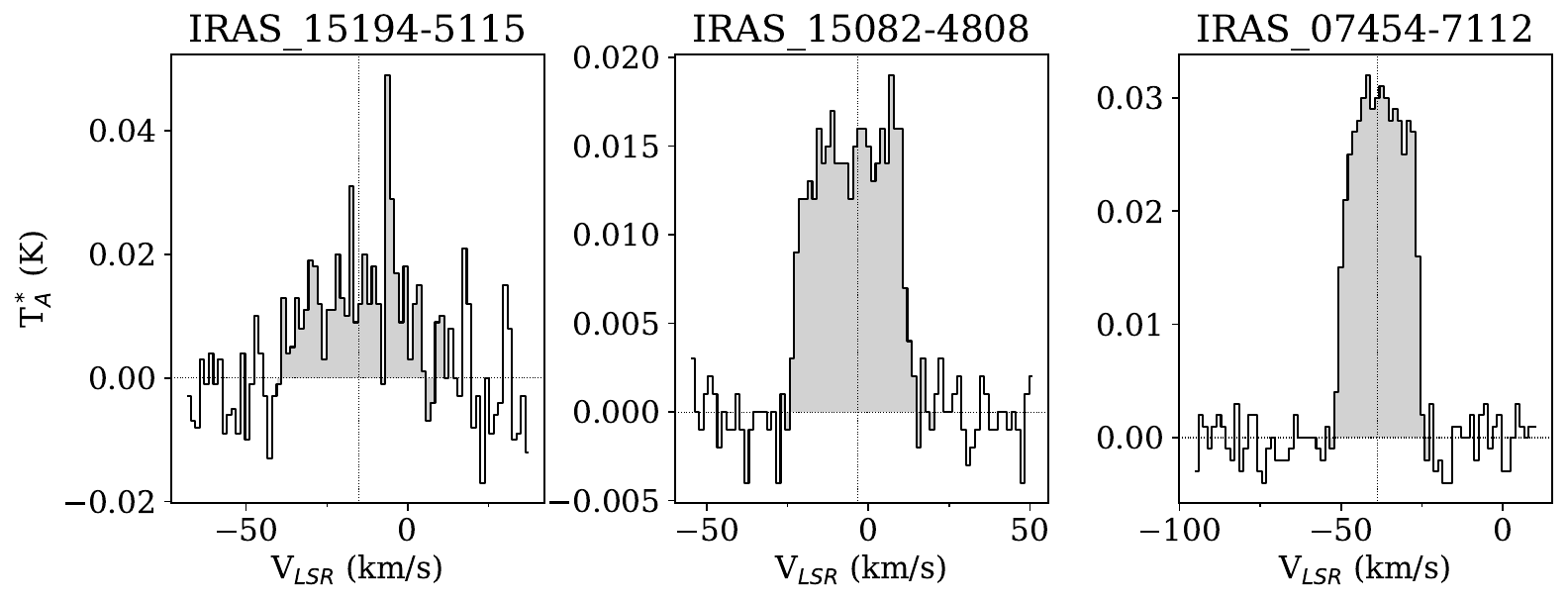}
    \caption{SiC$_2$, 11( 4, 7)-10( 4, 6) (261.509329 GHz)}
\end{figure}

\begin{figure}[h]
    \centering
    \includegraphics[width=0.65\linewidth]{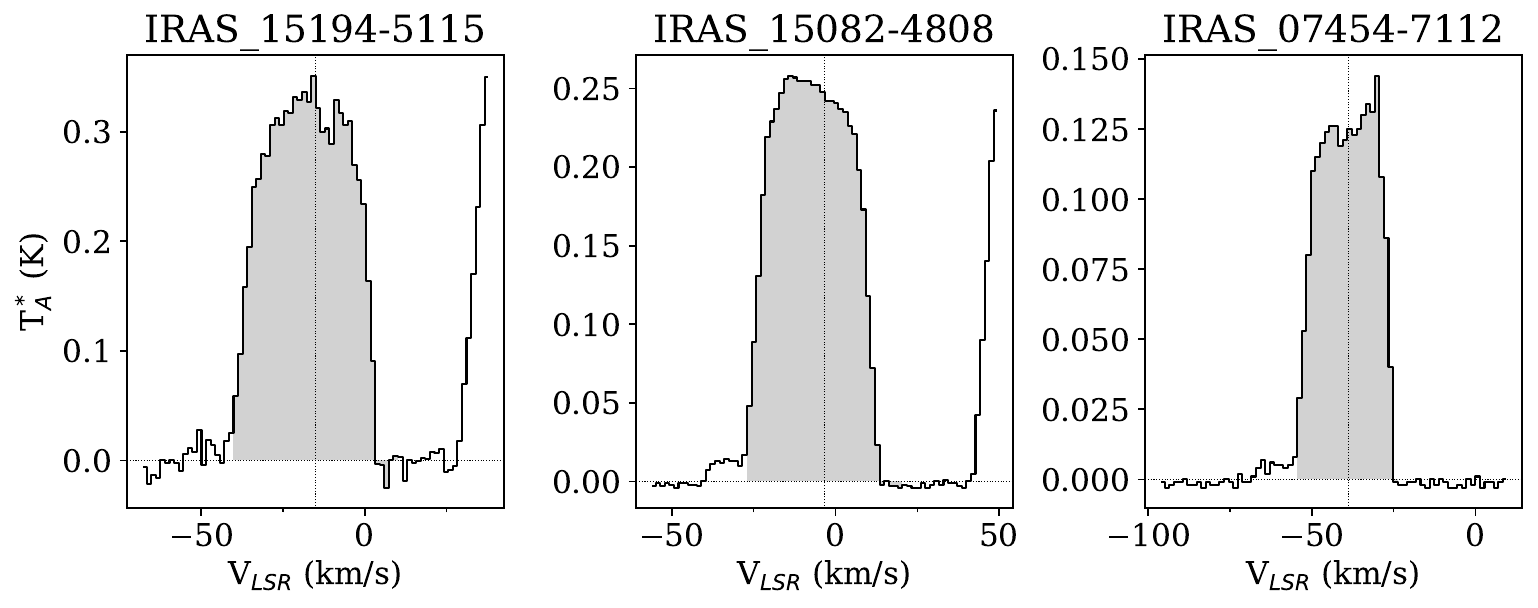}
    \caption{C$_2$H, N=3-2 (262.064986 GHz)}
\end{figure}

\begin{figure}[h]
    \centering
    \includegraphics[width=0.65\linewidth]{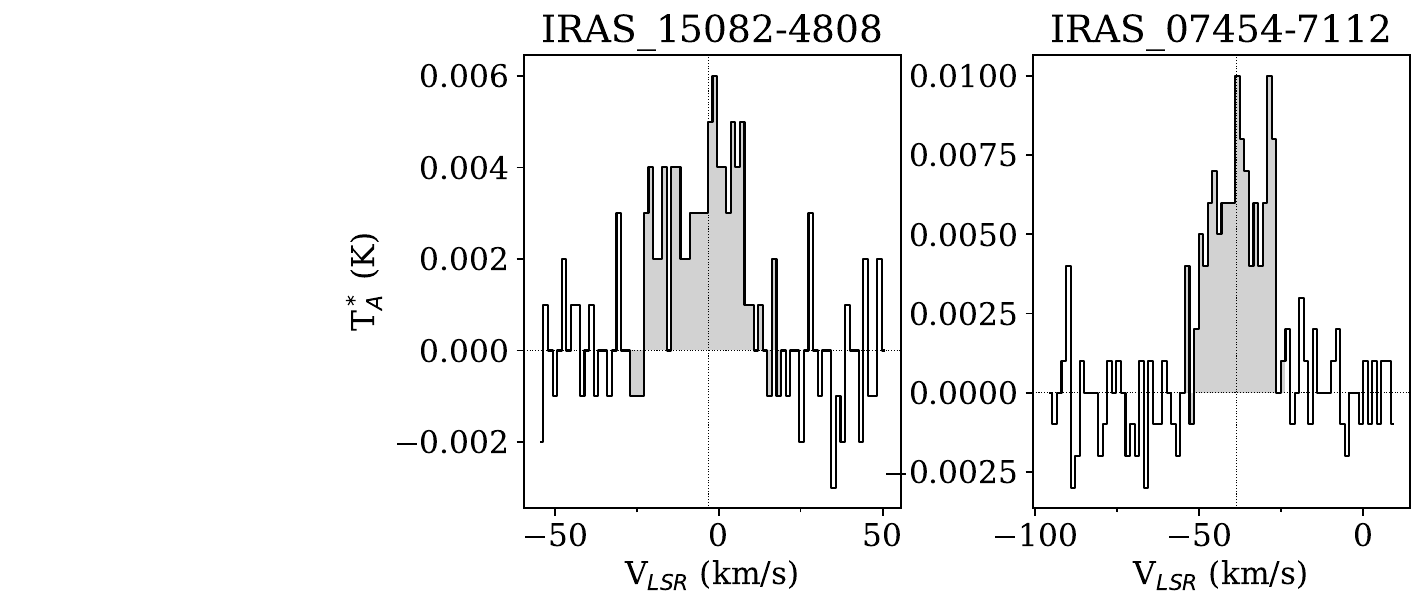}
    \caption{$^{30}$SiS, 15-14 (262.585033 GHz)}
\end{figure}

\begin{figure}[h]
    \centering
    \includegraphics[width=0.65\linewidth]{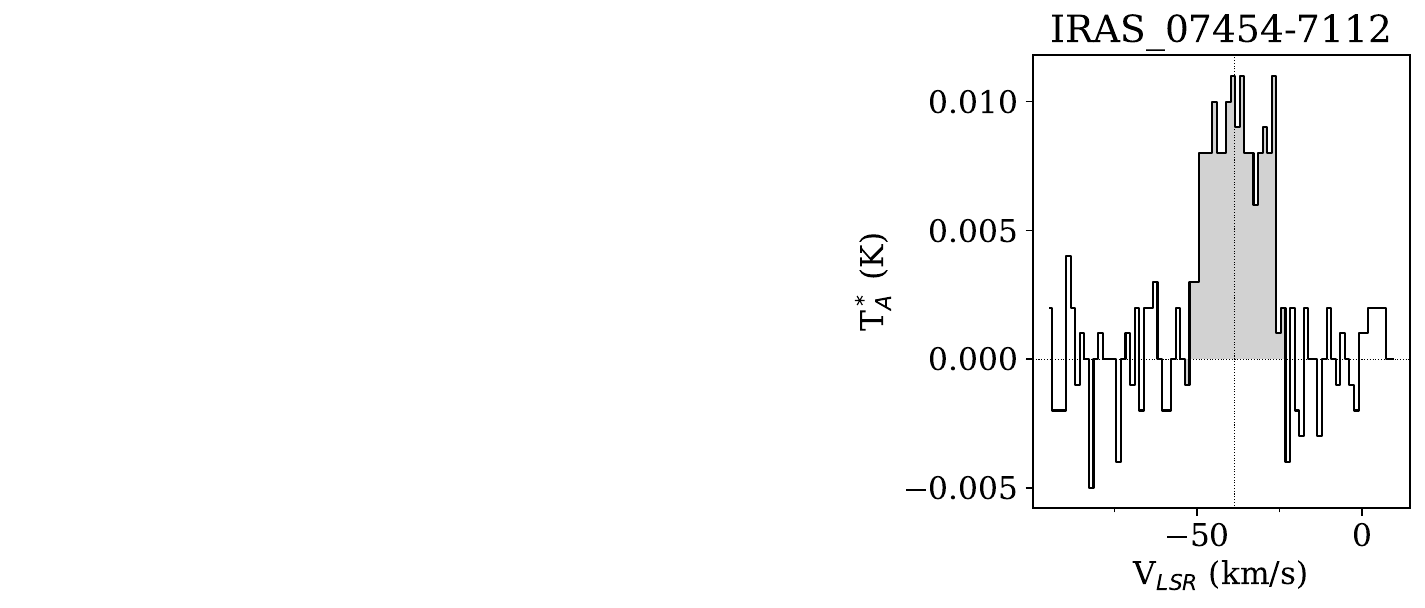}
    \caption{Si$^{34}$S, 15-14 (264.789719 GHz)}
\end{figure}

\begin{figure}[h]
    \centering
    \includegraphics[width=0.65\linewidth]{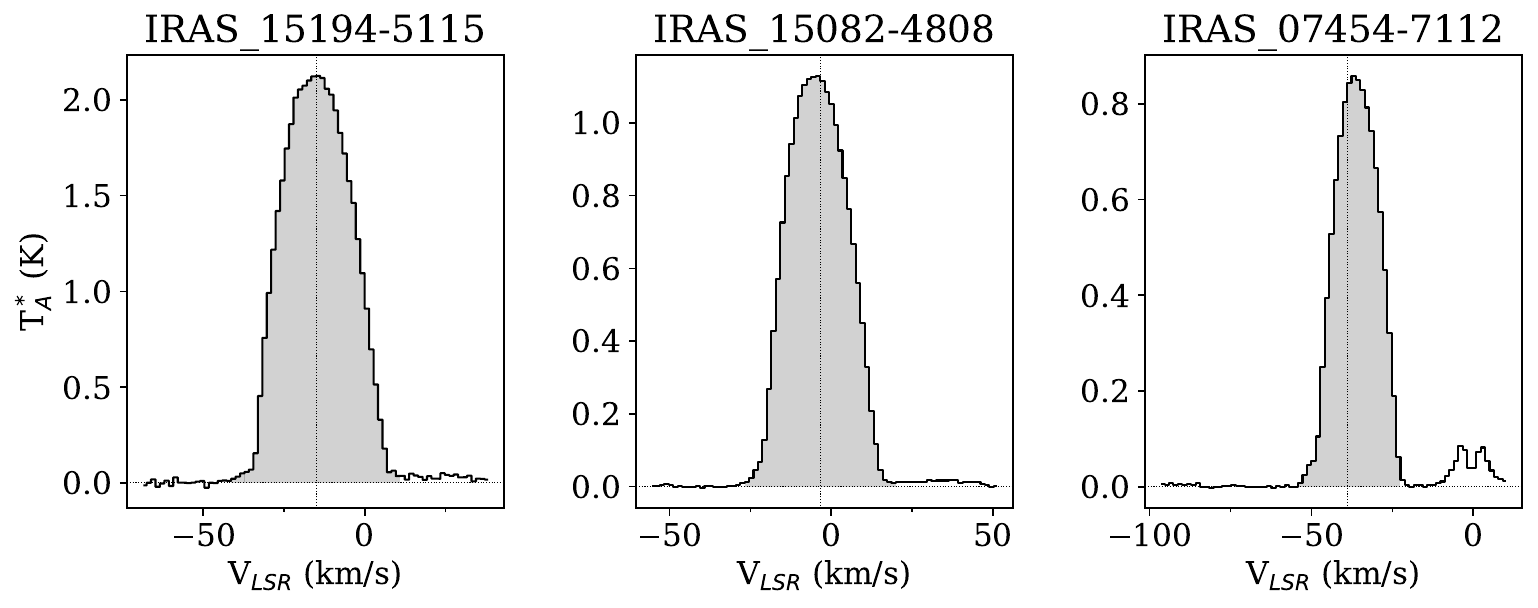}
    \caption{HCN, J=3-2 (265.886434 GHz)}
\end{figure}

\begin{figure}[h]
    \centering
    \includegraphics[width=0.65\linewidth]{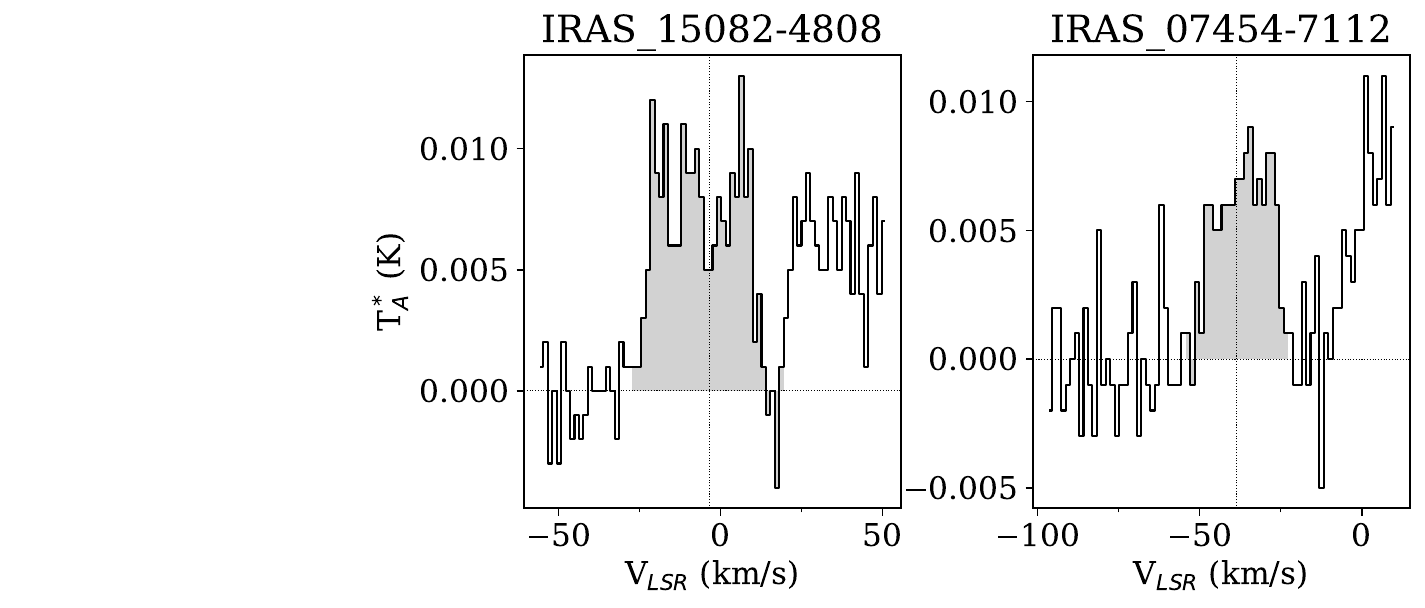}
    \caption{C$_4$H, N=28-27 (266.428184 GHz)}
\end{figure}

\begin{figure}[h]
    \centering
    \includegraphics[width=0.65\linewidth]{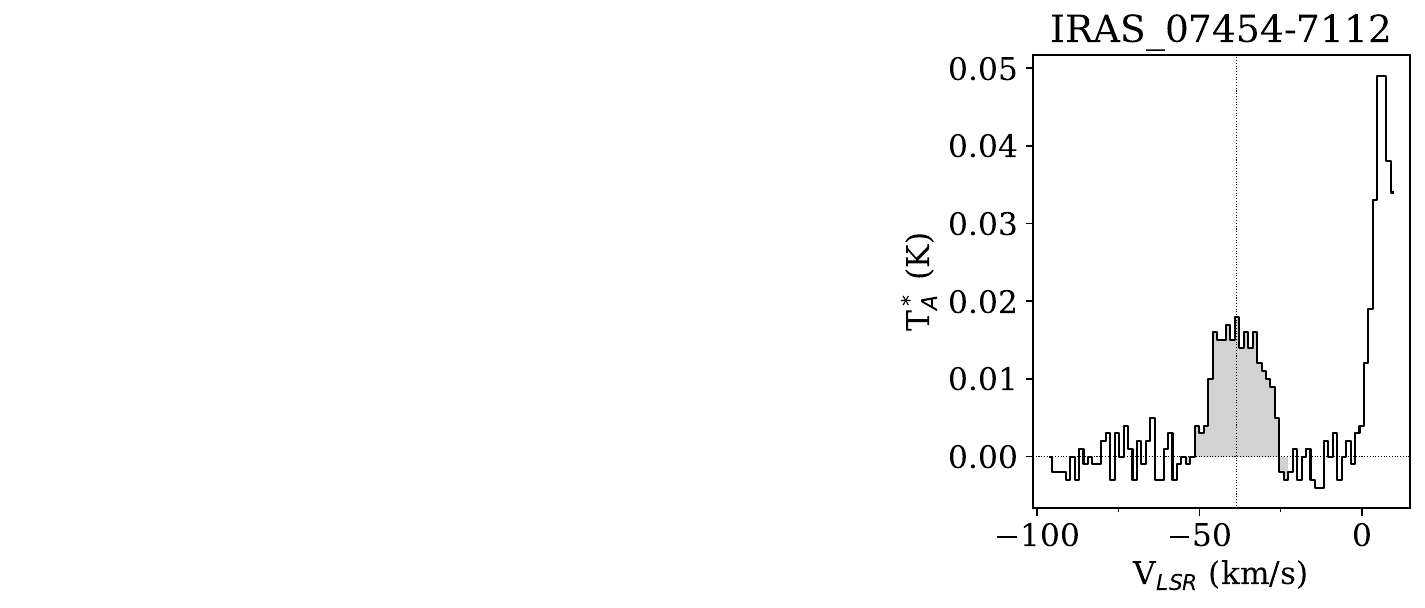}
    \caption{$^{29}$SiS, 15-14 (267.242218 GHz)}
\end{figure}

\begin{landscape}
\begin{figure}[!h]
   \centering
   \includegraphics[width=0.85\linewidth]{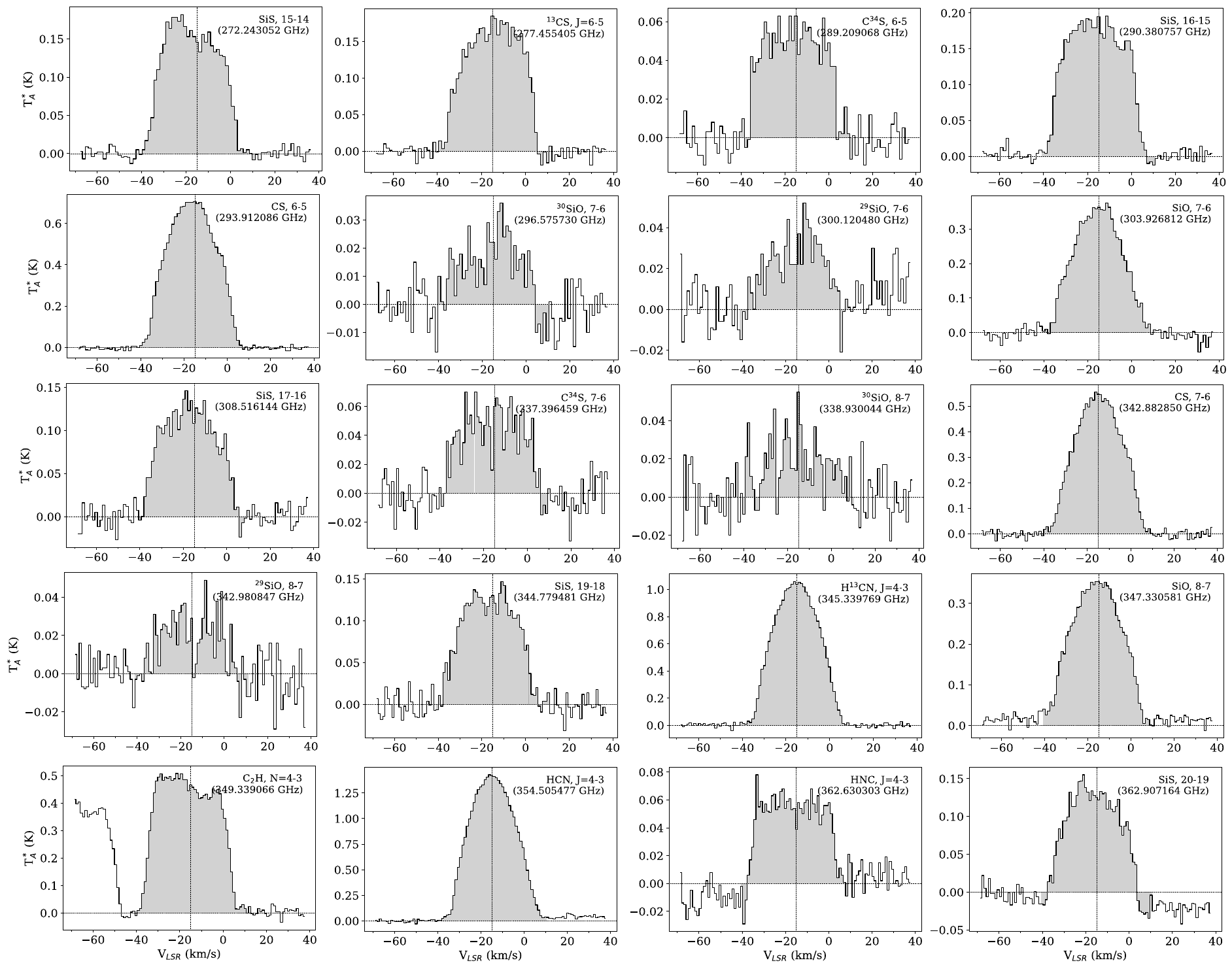}
   \caption{Lines from the APEX Band 7 (272$-$376 GHz) survey of IRAS 15194$-$5115, which have been used in this work. The grey$-$shaded regions represent the channels used for calculating the integrated intensities {of the lines, reported in Table~\ref{tab:apex_line_detections}}. The dotted vertical line marks the systemic velocity of the source.}
   \label{fig:APEX_B7_lines_15194}
\end{figure}
\end{landscape}

\clearpage
\section{Azimuthally-averaged radial profiles (AARPs)}
\label{app:appendix_D}
{The figures below show the Gaussian fits to the azimuthally-averaged radial profiles (AARPs) of the systemic velocity emission maps of the 14 species shown in Fig.~\ref{fig:All_AARPs_comparison}. The details of these profiles and fits are explained in Sect.~\ref{subsubsec:AARPs}.}

\begin{figure}[!h]
   \centering
   \begin{subfigure}[b]{0.365\textwidth}
      \centering
      \includegraphics[width=\textwidth]{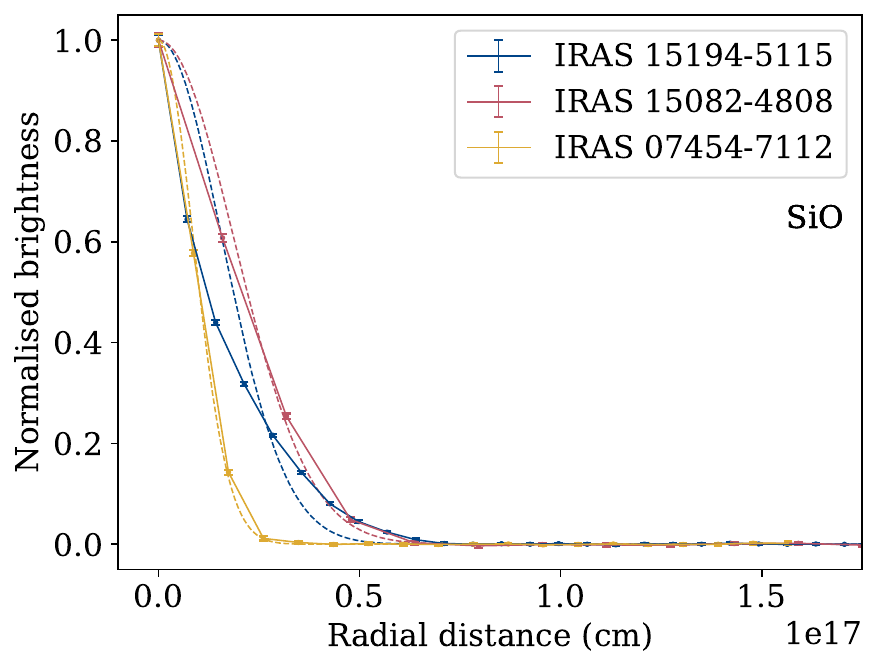}
      \label{subfig:SiO_aarps}
   \end{subfigure}
   \begin{subfigure}[b]{0.365\textwidth}
      \centering
      \includegraphics[width=\textwidth]{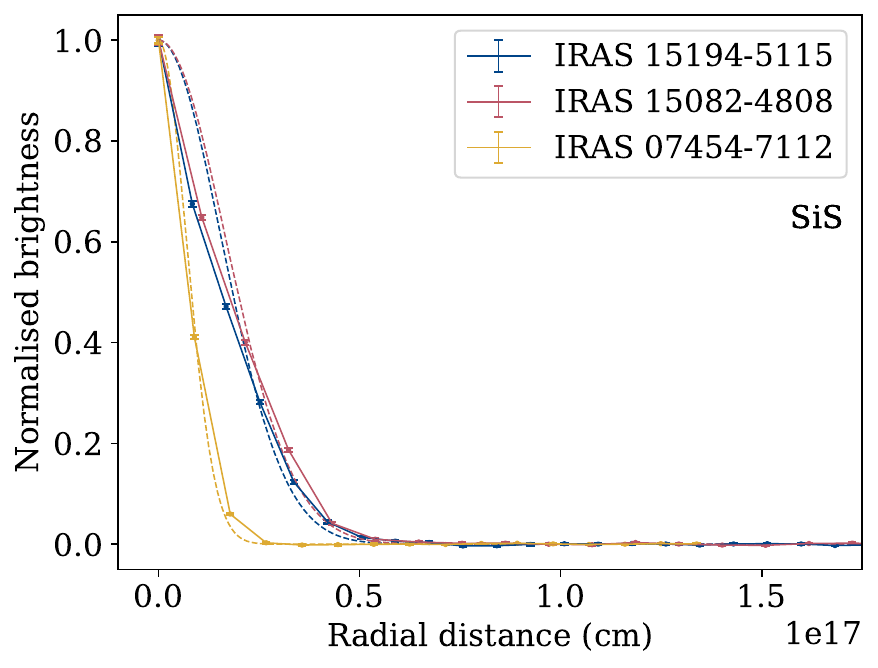}
      \label{subfig:SiS_aarps}
   \end{subfigure}
      \caption{AARPs {(solid)} and Gaussian fits {(dashed)} for SiO (left) and SiS (right) emission.}
      \label{fig:SiO_and_SiS_AARPs}
\end{figure}

\begin{figure}[!h]
   \centering
   \begin{subfigure}[b]{0.365\textwidth}
      \centering
      \includegraphics[width=\textwidth]{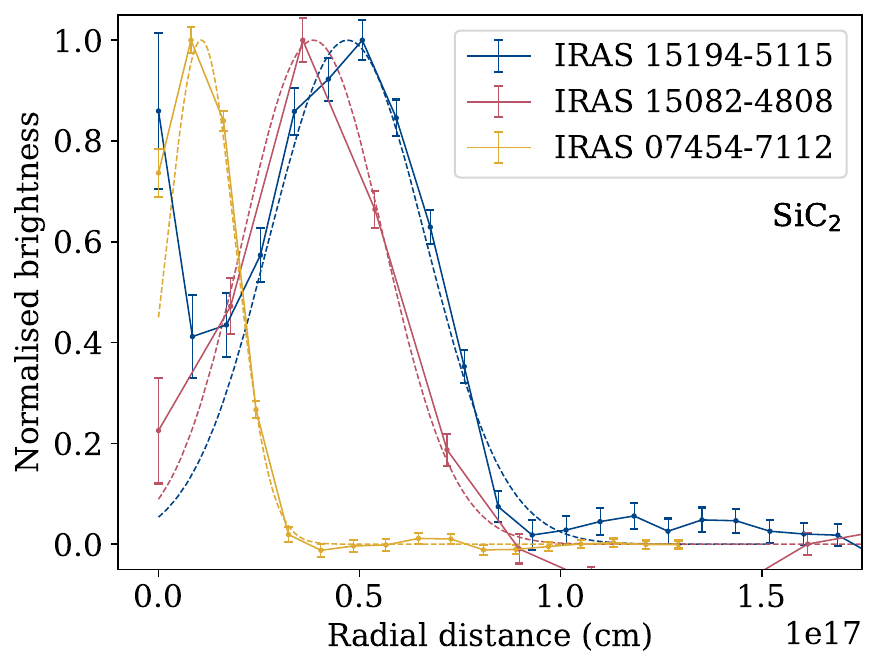}
      \label{subfig:SiC2_AARPs}
   \end{subfigure}
   \begin{subfigure}[b]{0.365\textwidth}
      \centering
      \includegraphics[width=\textwidth]{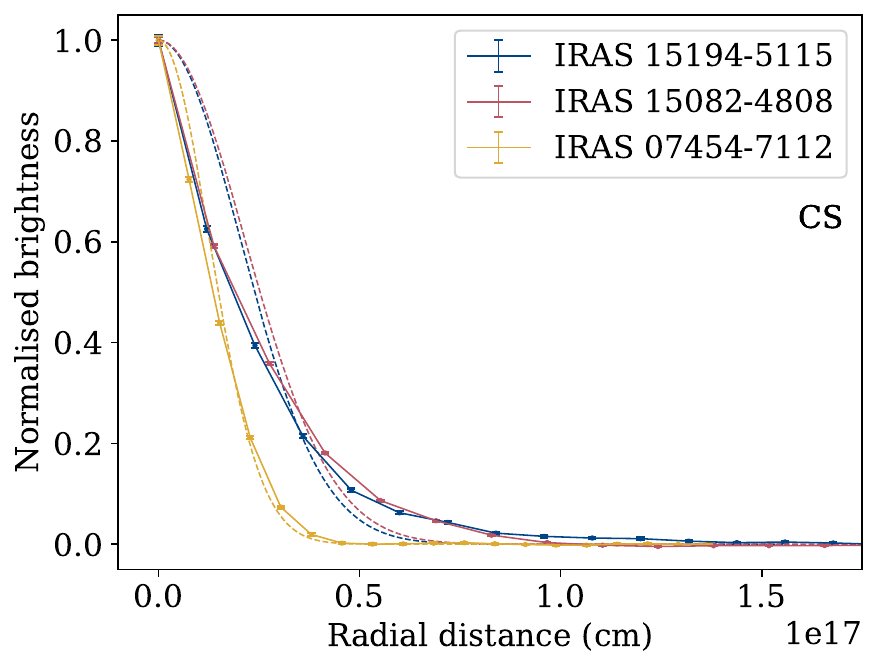}
      \label{subfig:CS_AARPs}
   \end{subfigure}
      \caption{Same as Fig.~\ref{fig:SiO_and_SiS_AARPs} for SiC$_2$ (left) and CS (right) emission.}
      \label{fig:SiC2_and_CS_AARPs}
\end{figure}

\begin{figure}[!h]
   \centering
   \begin{subfigure}[b]{0.365\textwidth}
      \centering
      \includegraphics[width=\textwidth]{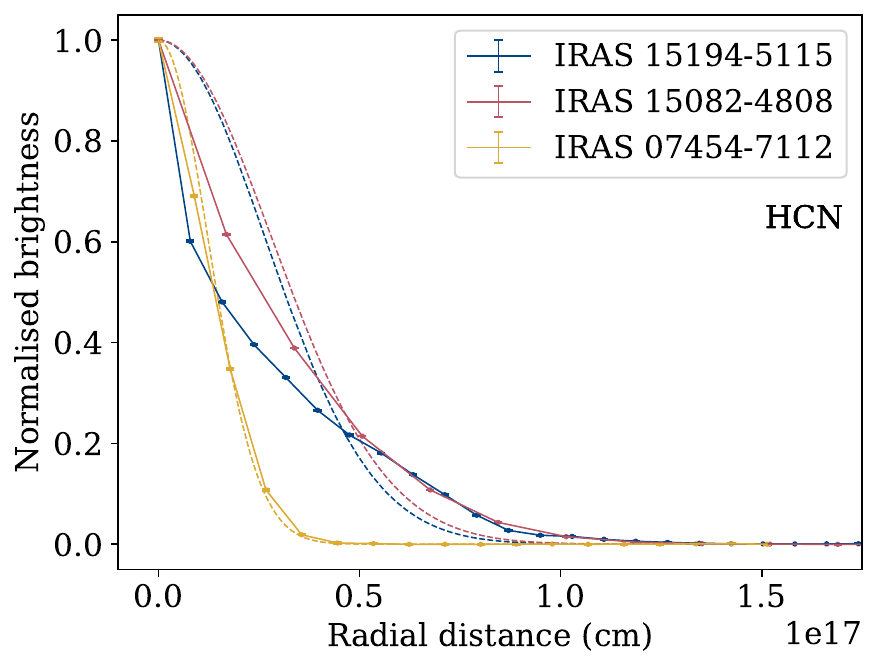}
      \label{subfig:HCN_AARPs}
   \end{subfigure}
   \begin{subfigure}[b]{0.365\textwidth}
      \centering
      \includegraphics[width=\textwidth]{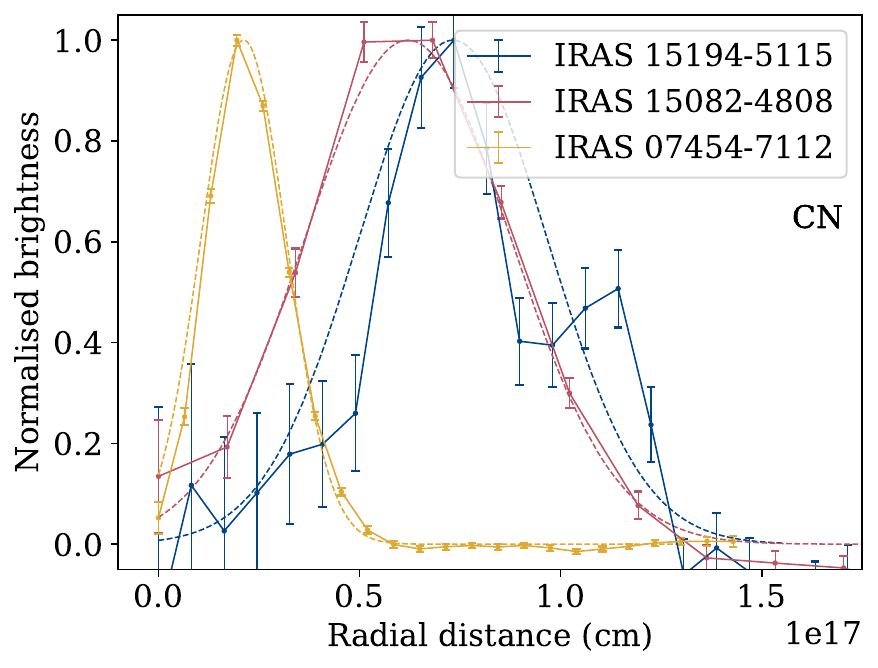}
      \label{subfig:CN_AARPs}
   \end{subfigure}
      \caption{Same as Fig.~\ref{fig:SiO_and_SiS_AARPs}  for HCN (left) and CN (right) emission.}
      \label{fig:HCN_and_CN_AARPs}
\end{figure}

\begin{figure}[!h]
   \centering
   \begin{subfigure}[b]{0.365\textwidth}
      \centering
      \includegraphics[width=\textwidth]{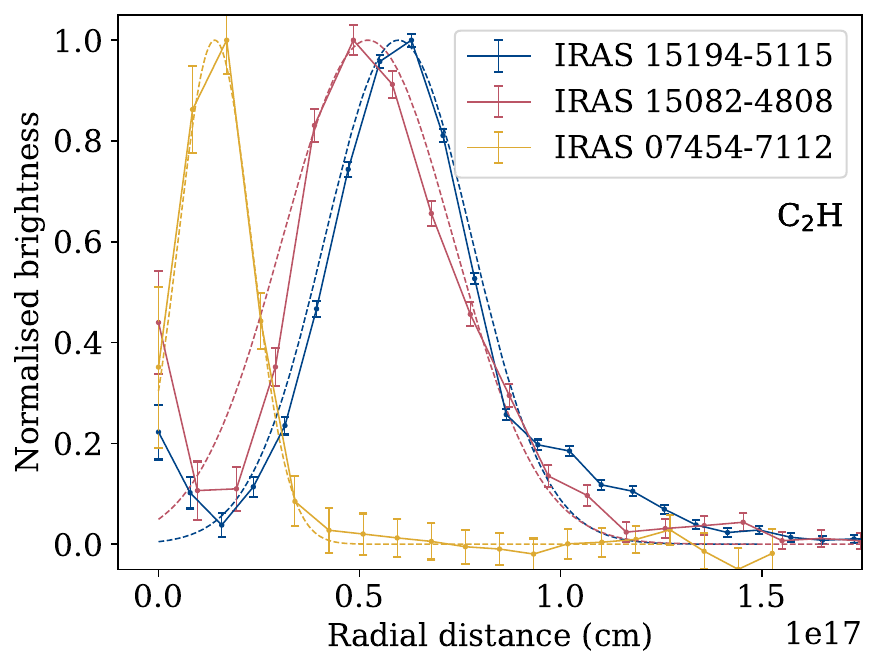}
      \label{subfig:C2H_aarp_and_fit}
   \end{subfigure}
   \begin{subfigure}[b]{0.365\textwidth}
      \centering
      \includegraphics[width=\textwidth]{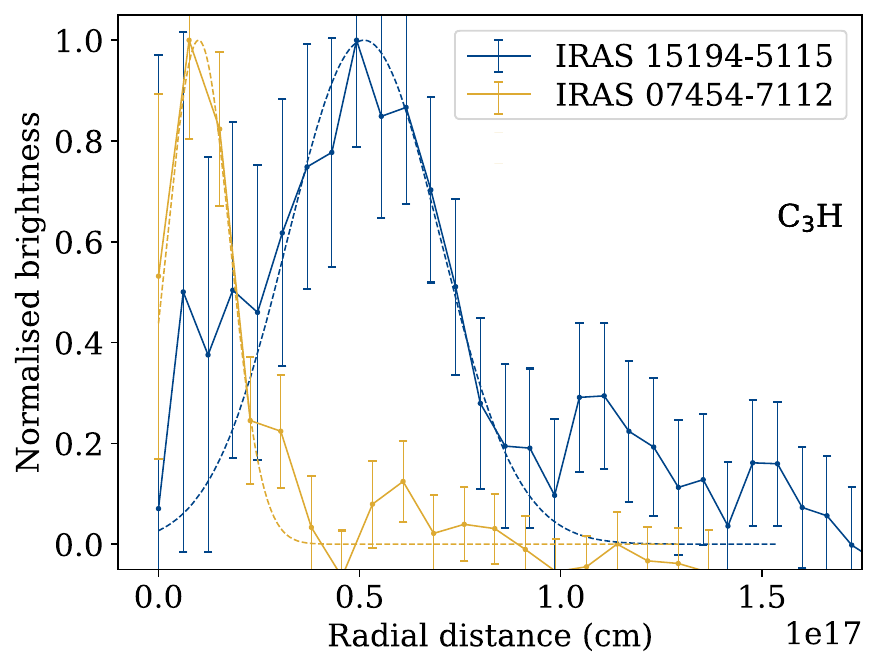}
      \label{subfig:C3H_aarp_and_fit}
   \end{subfigure}
   \begin{subfigure}[b]{0.365\textwidth}
      \centering
      \includegraphics[width=\textwidth]{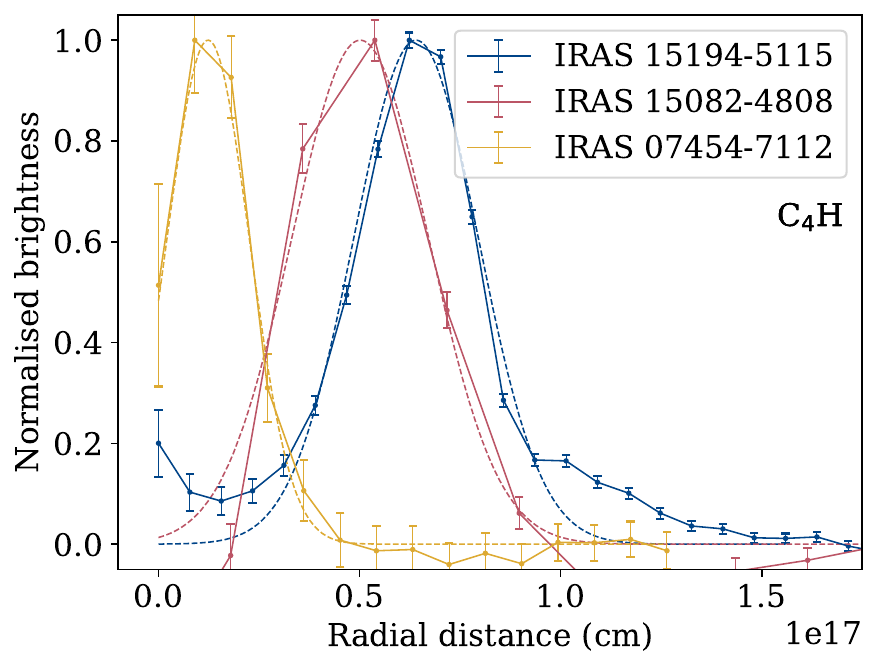}
      \label{subfig:C4H_aarp_and_fit}
   \end{subfigure}
   \begin{subfigure}[b]{0.365\textwidth}
      \centering
      \includegraphics[width=\textwidth]{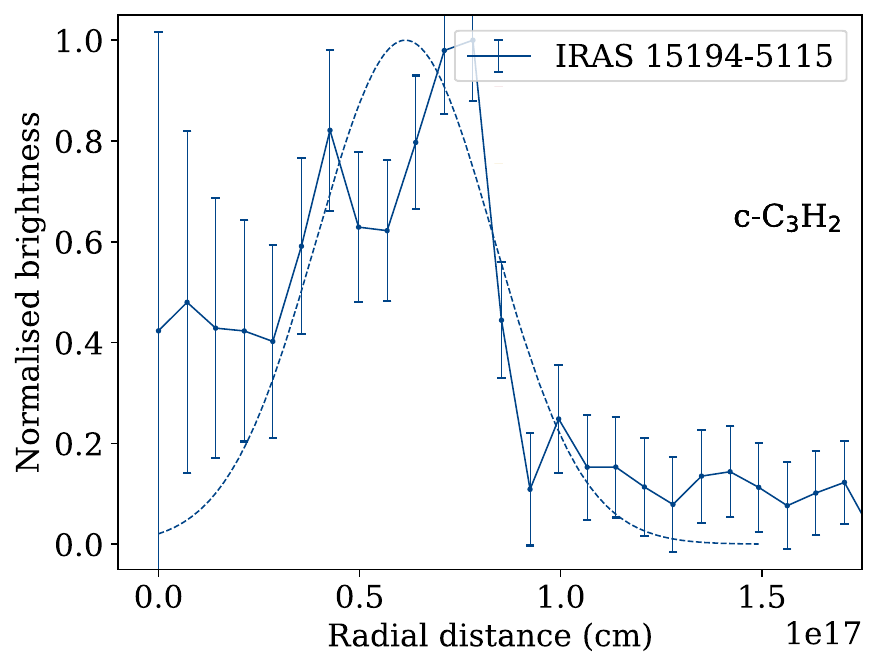}
      \label{subfig:c-C3H2_aarp_and_fit}
   \end{subfigure}
      \caption{Same as Fig.~\ref{fig:SiO_and_SiS_AARPs}  for C$_2$H (top left), C$_3$H (top right), C$_4$H (bottom left), and c-C$_3$H$_2$ (bottom right) emission}
      \label{fig:carbon_chains_AARPs}
\end{figure}

\begin{figure}[!h]
   \centering
   \begin{subfigure}[b]{0.365\textwidth}
      \centering
      \includegraphics[width=\textwidth]{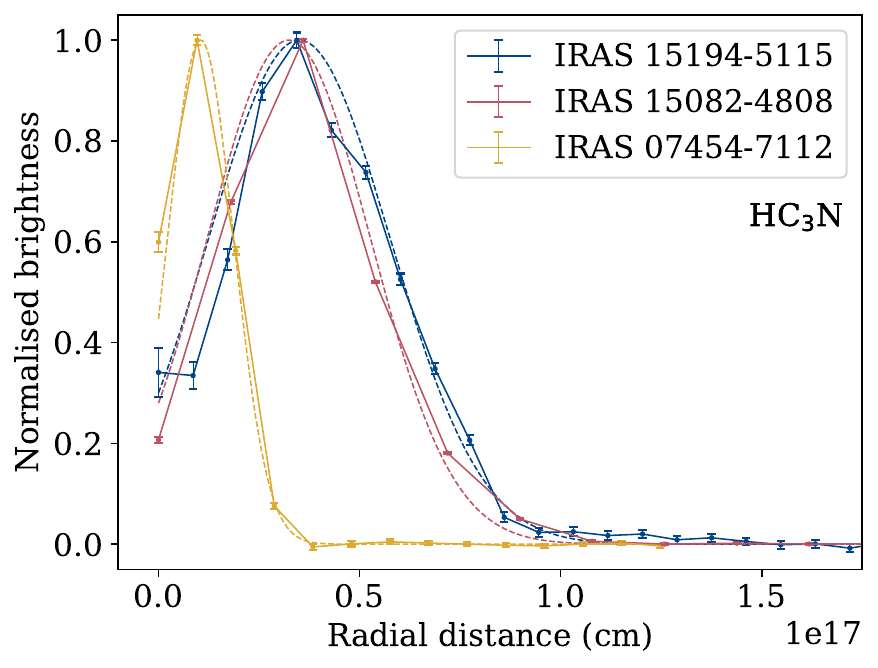}
      \label{subfig:HC3N_aarp_and_fit}
   \end{subfigure}
   \begin{subfigure}[b]{0.365\textwidth}
      \centering
      \includegraphics[width=\textwidth]{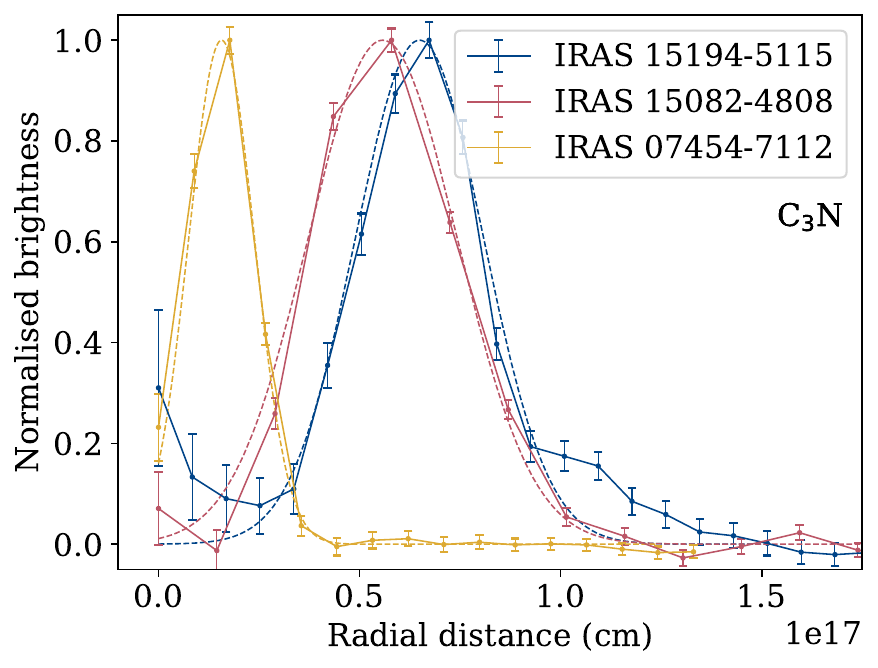}
      \label{subfig:C3N_aarp_and_fit}
   \end{subfigure}
   \begin{subfigure}[b]{0.365\textwidth}
      \centering
      \includegraphics[width=\textwidth]{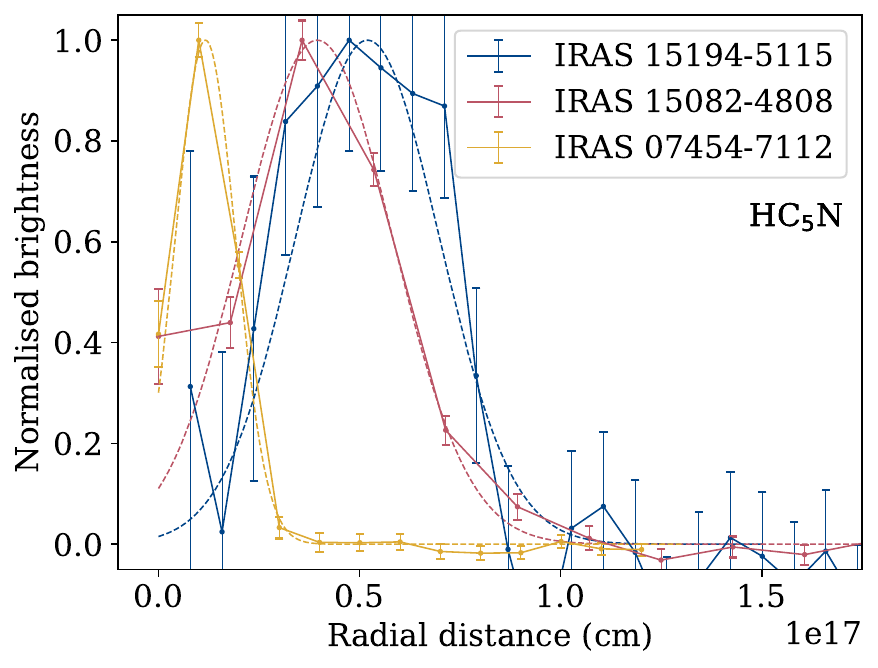}
      \label{subfig:HC5N_aarp_and_fit}
   \end{subfigure}
      \begin{subfigure}[b]{0.365\textwidth}
         \centering
         \includegraphics[width=\textwidth]{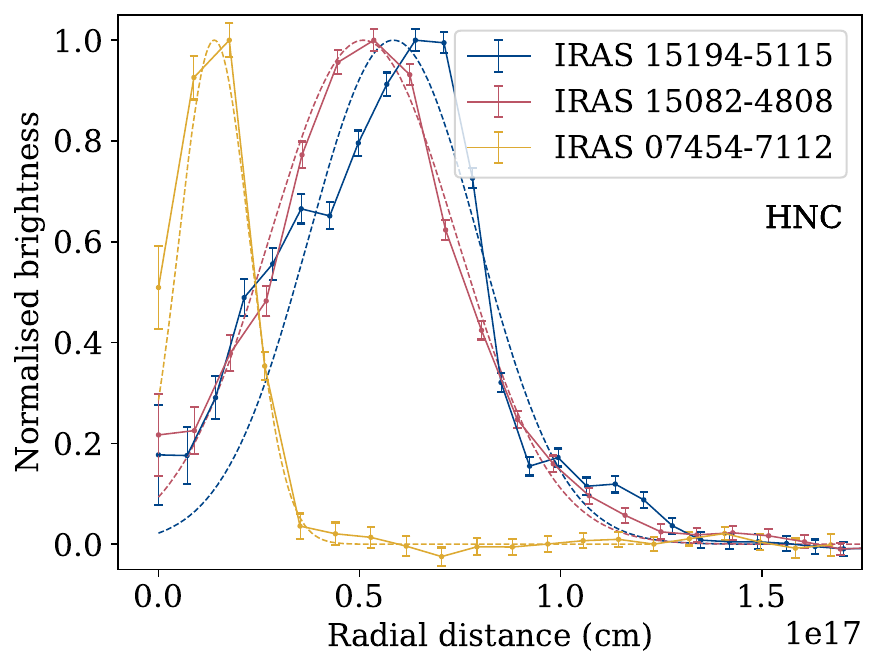}
         \label{subfig:HNC_aarp_and_fit}
      \end{subfigure}
      \caption{Same as Fig.~\ref{fig:SiO_and_SiS_AARPs}  for HC$_3$N (top left), C$_3$N (top right), HC$_5$N (bottom left), and HNC (bottom right) emission}
      \label{fig:cyanopolyynes_AARPs}
\end{figure}

\clearpage
\section{{Column densities and abundances: methods}}
\label{app:appendix_E}
\begin{multicols}{2}
The results in this section are based on a simple model for the CSE, assuming an isotropic, constant mass-loss rate and the gas expanding at its terminal velocity. Nevertheless, there are clear indications of deviations from such a simple model, particularly in terms of the morphology, as discussed in Sect.~\ref{subsubsec:15194_complex_morphology}.

We express molecular abundances as fractional abundances relative to that of molecular hydrogen, H$_2$, which has the abundance distribution with the largest spatial extent {\citep{Huggins_and_Glassgold_1982}}. The fractional abundance profile of a molecular species X is defined as
\begin{equation}
   f_\mathrm{X}(r) = \frac{n_\mathrm{X}(r)}{n_\mathrm{H_{2}}(r)},
\end{equation}
where $n_\mathrm{X}(r)$ and $n_\mathrm{H_{2}}(r)$ are the number density distributions of the molecule and molecular hydrogen, respectively.

The excitation of a molecule in the CSE is determined by collisions and radiation, in particular the radiation fields of the central star and circumstellar dust. Non-local thermodynamic equilibrium (non-LTE) conditions can be prevalent in CSEs, which calls for a detailed radiative transfer analysis for the proper treatment of molecular excitation. Such {a} non-LTE treatment is outside the scope of the current work, and shall be presented in an upcoming paper. A simpler analysis assuming LTE conditions can provide reasonable estimates of the molecular abundances and offer substantial insights into circumstellar chemistry{, and is regularly employed as a first tool to analyse spectral surveys and other large datasets of evolved stars \citep[e.g.][]{Cernicharo_et_al_2000, Justtanont_et_al_2000, Wesson_et_al_2010, Patel_et_al_2011, Sanchez_Contreras_et_al_2015, Smith_et_al_2015, Velilla-Prieto_et_al_2015_OH_231.8+4.2, Pardo_et_al_2022}}. 

For molecules with multiple rotational transitions detected, the rotational temperature and column density can be estimated using a population diagram (Sect.~\ref{subsubsec:Rot_Diag_Analysis}). The column densities thus obtained, together with the emission region sizes calculated from the AARPs, can be used to estimate molecular fractional abundances. In those cases where the population-diagram method can not be employed, owing to the lack of sufficient number of transitions covering a broad range of upper-level energies, we perform an approximate calculation of the fractional abundance using a single line for each species {(Sect.~\ref{subsubsec:Single_Line_Analysis})}. We describe both methods below.

\subsection{The population diagram {method}}
\label{subsubsec:Rot_Diag_Analysis}
The column density of each rotational energy level is related to the rotational temperature, {$T_\mathrm{rot}$},  and total column density, {$N_\mathrm{tot}$}, by the relation
\begin{equation}
   \label{eqn:rot_diag}
   \ln\frac{N_\mathrm{u}}{g_\mathrm{u}} = \ln\frac{N_\mathrm{tot}}{Z(T_\mathrm{rot})} - \frac{E_\mathrm{u}}{k_\mathrm{B}T_\mathrm{rot}}
\end{equation}
where $N_\mathrm{u}$ is the column density of molecules in the upper energy level having energy $E_\mathrm{u}$ and statistical weight $g_\mathrm{u}$, and $Z$ and $k_\mathrm{B}$ {are} the partition function and the Boltzmann constant, respectively \citep{Goldsmith_and_Langer_1999}. 

For the calculation of partition functions, all molecules used for producing population diagrams (see Table~\ref{tab:N_and_T}) have been regarded as linear rigid rotors, except SiC$_2$ and c-C$_3$H$_2$, which are asymmetric tops. In order to use the integrated intensities summed over the hyperfine components wherever present (e.g. C$_2$H, C$_3$N), we have calculated the partition functions, degeneracy factors, and the Einstein $A$-coefficients under the assumption that the molecules do not possess hyperfine structure. For the asymmetric tops, we interpolated the partition function values from the CDMS \citep{Muller_et_al_2005} catalogue to the rotation temperatures obtained from our analysis.

The source-averaged column density of the upper state, $N_\mathrm{u}$, can be calculated from the observed velocity-integrated flux, $\int S_\mathrm{ul} d\varv$, of the corresponding rotational emission line as
\begin{equation}
   N_\mathrm{u} = \frac{4 \pi \int S_\mathrm{ul} d\varv}{h c A_\mathrm{ul} \Omega_s} {C_{\tau_\nu}}
\end{equation}
where $\nu_\mathrm{ul}$ is the rest frequency of the transition, $A_\mathrm{ul}$ is the Einstein $A$-coefficient, {$\Omega_s$ is the solid angle of the emitting region}, $h$ is the Planck constant, $c$ is the speed of light in vacuum, {and C$_{\tau_\nu}$ is the optical depth correction factor \citep[see][]{Goldsmith_and_Langer_1999}}, given by
\begin{equation}
    {C_{\tau_\nu} = \left( \frac{\tau_\nu}{1-e^{-\tau_\nu}}\right)}
    \label{eqn:optical_depth_correction_factor}
\end{equation}
{where $\tau_\nu$ is the optical depth of the line}.

If the assumptions that the populations of the corresponding energy levels follow a Boltzmann distribution which can be described by a single excitation temperature, $T_\mathrm{rot}$, and that the lines are optically thin are correct, the points in the plot of $\ln(N_\mathrm{u}/g_\mathrm{u})$ versus $E_\mathrm{u}/k_\mathrm{B}$ are expected to lie on a straight line {(see e.g. Fig.~\ref{fig:C$_4$H_rot_diags_all_stars})}. The rotational temperature and column density of each molecule are calculated from the slope and intercept, respectively, of a linear fit to such a plot.

\begin{figure}[!h]
   \centering
   \begin{subfigure}[b]{0.33\textwidth}
      \centering
      \includegraphics[width=\textwidth]{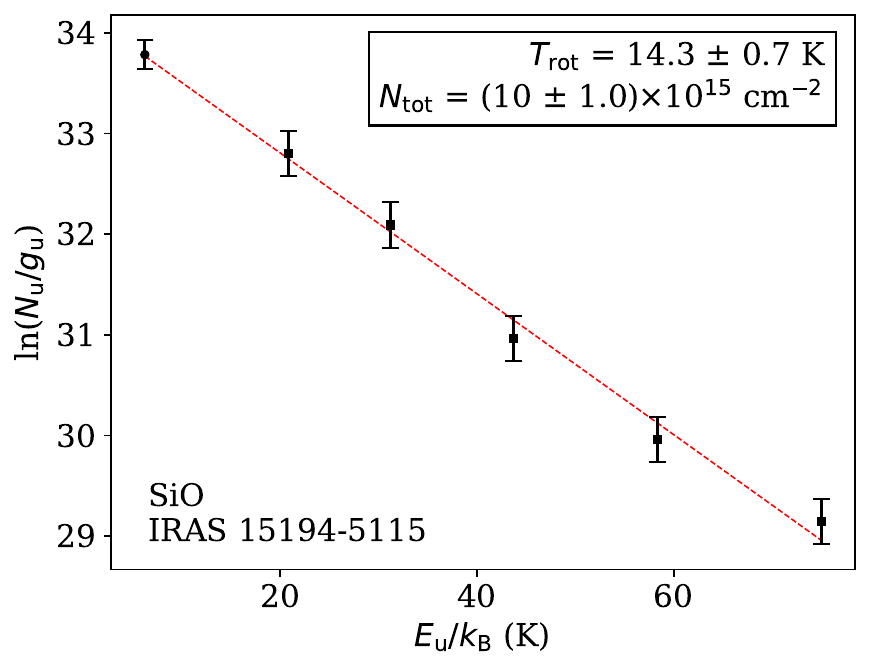}
      \label{subfig:SiO_rot_diag_15194}
   \end{subfigure}
   \begin{subfigure}[b]{0.33\textwidth}
      \centering
      \includegraphics[width=\textwidth]{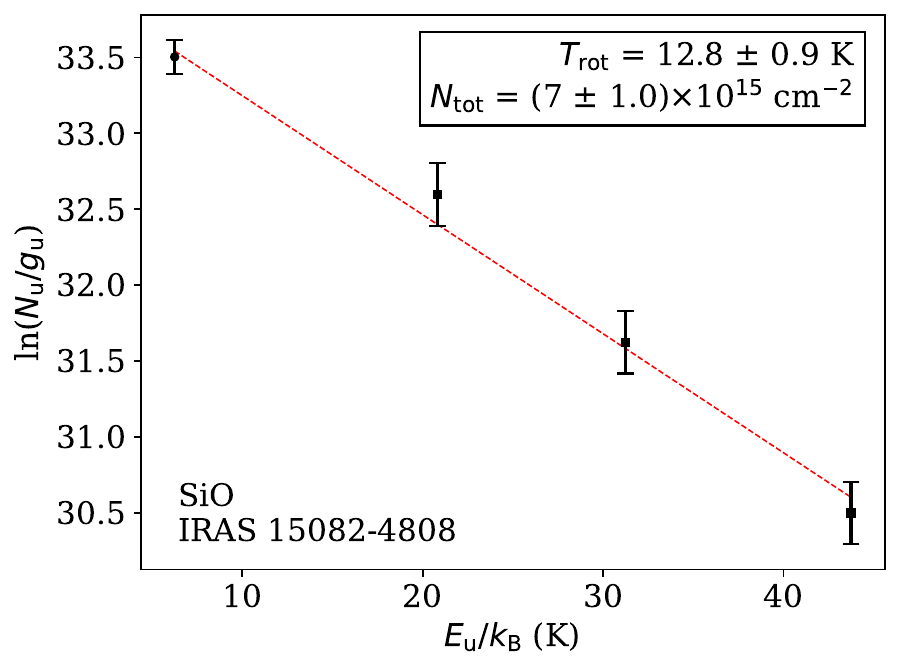}
      \label{subfig:SiO_rot_diag_15082}
   \end{subfigure}
   \begin{subfigure}[b]{0.33\textwidth}
      \centering
      \includegraphics[width=\textwidth]{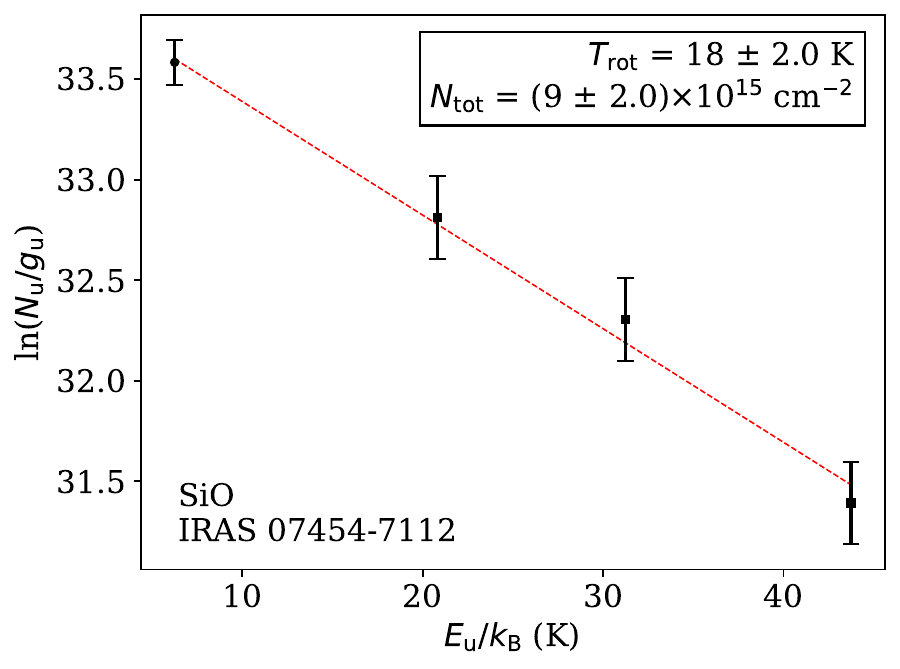}
      \label{subfig:SiO_rot_diag_07454}
   \end{subfigure}
      \caption{Population diagrams for SiO for IRAS 15194$-$5115 (left), IRAS 15082$-$4808 (middle), and IRAS 07454$-$7112 (bottom). The solid line (red) is the best-fit linear model to the data (black). Rotation temperatures $T_{\mathrm{rot}}$ and source-averaged column densities $N_{\mathrm{tot}}$ are listed in the legends. Circle and square points denote lines observed with ALMA and APEX, respectively. Hollow circles denote lines for which data combination (see Sect.~\ref{subsec:data_combination}) has been performed}.
      \label{fig:SiO_rot_diags_all_stars}
\end{figure}

\begin{figure}[!h]
   \centering
   \begin{subfigure}[b]{0.33\textwidth}
      \centering
      \includegraphics[width=\textwidth]{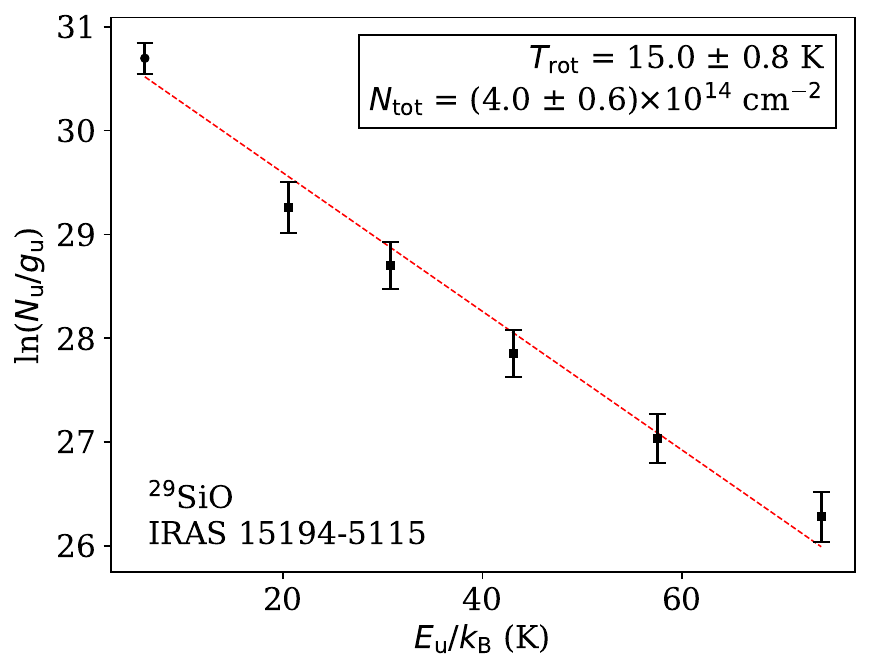}
      \label{subfig:29SiO_rot_diag_15194}
   \end{subfigure}
   \begin{subfigure}[b]{0.33\textwidth}
      \centering
      \includegraphics[width=\textwidth]{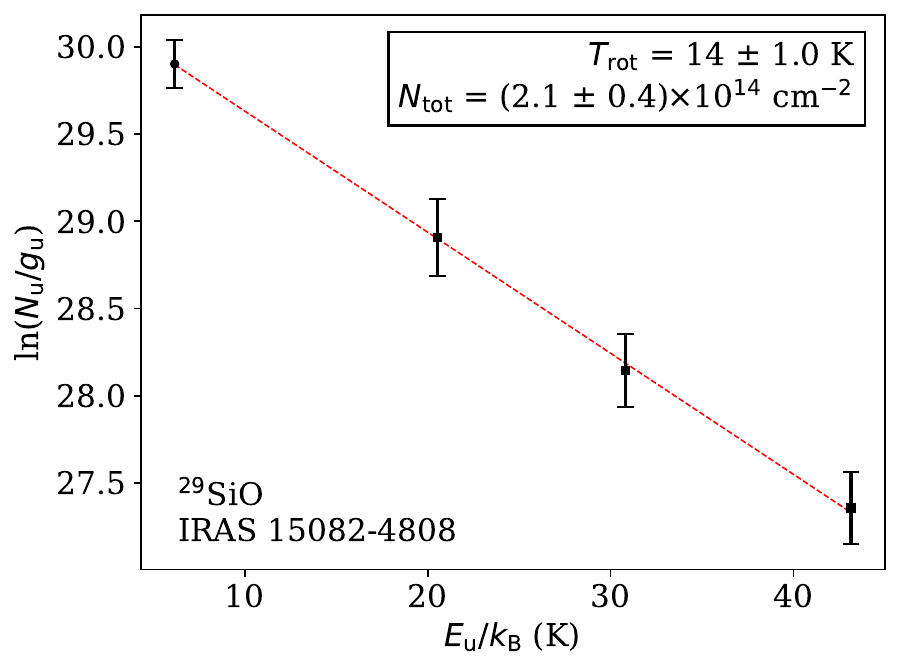}
      \label{subfig:29SiO_rot_diag_15082}
   \end{subfigure}
   \begin{subfigure}[b]{0.33\textwidth}
      \centering
      \includegraphics[width=\textwidth]{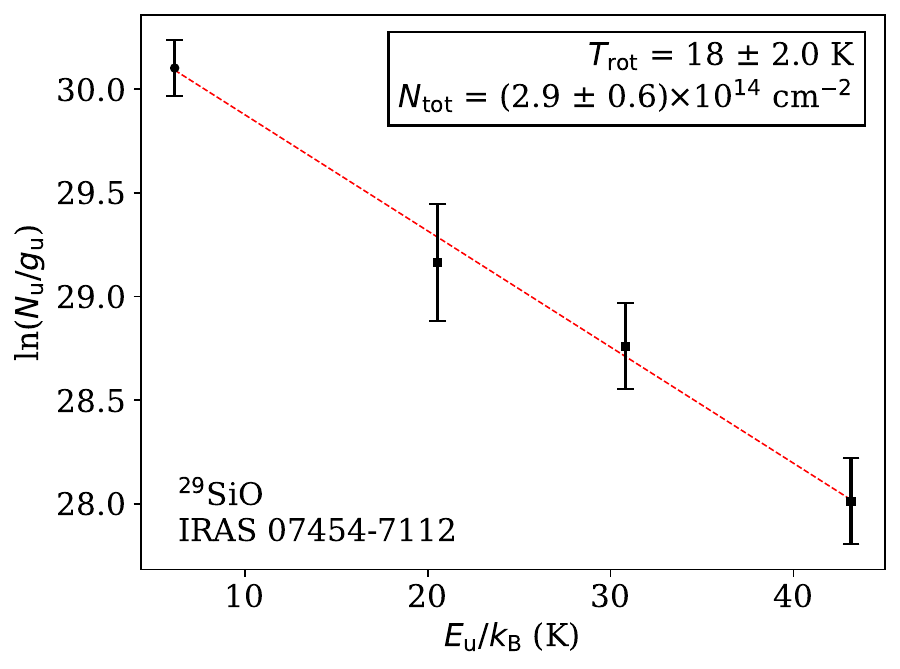}
      \label{subfig:29SiO_rot_diag_07454}
   \end{subfigure}
      \caption{Same as Fig.~\ref{fig:SiO_rot_diags_all_stars} for $^{29}$SiO.}
      \label{fig:29SiO_rot_diags_all_stars}
\end{figure}

\begin{figure}[!h]
   \centering
   \hspace*{-0.69\textwidth}
   \begin{subfigure}[b]{0.33\textwidth}
      \centering
      \includegraphics[width=\textwidth]{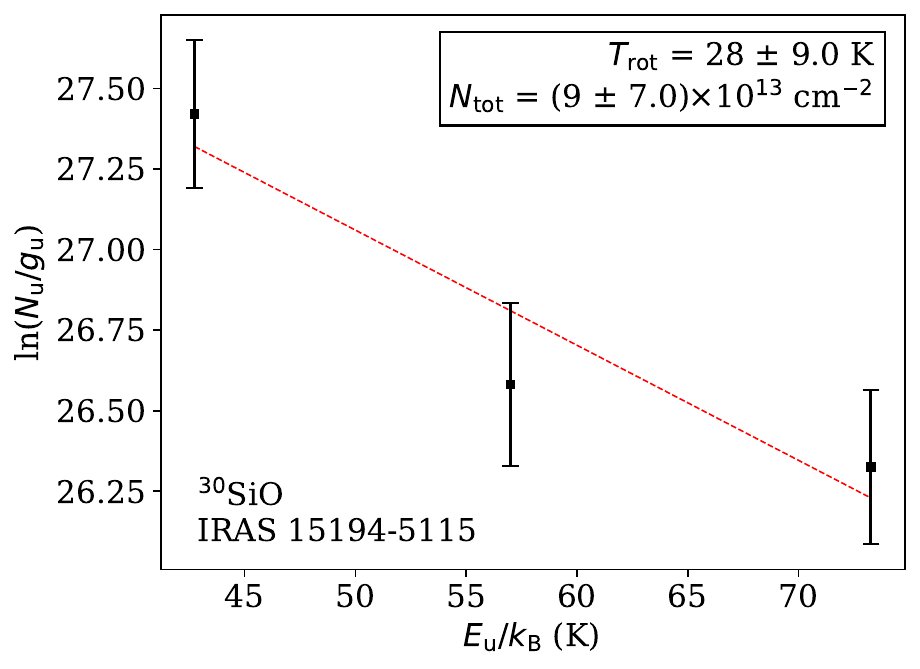}
      \label{subfig:30SiO_rot_diag_15194}
   \end{subfigure}
      \caption{Same as Fig.~\ref{fig:SiO_rot_diags_all_stars} for $^{30}$SiO.}
      \label{fig:30SiO_rot_diags_all_stars}
\end{figure}

\begin{figure}[!h]
   \centering
   \begin{subfigure}[b]{0.33\textwidth}
      \centering
      \includegraphics[width=\textwidth]{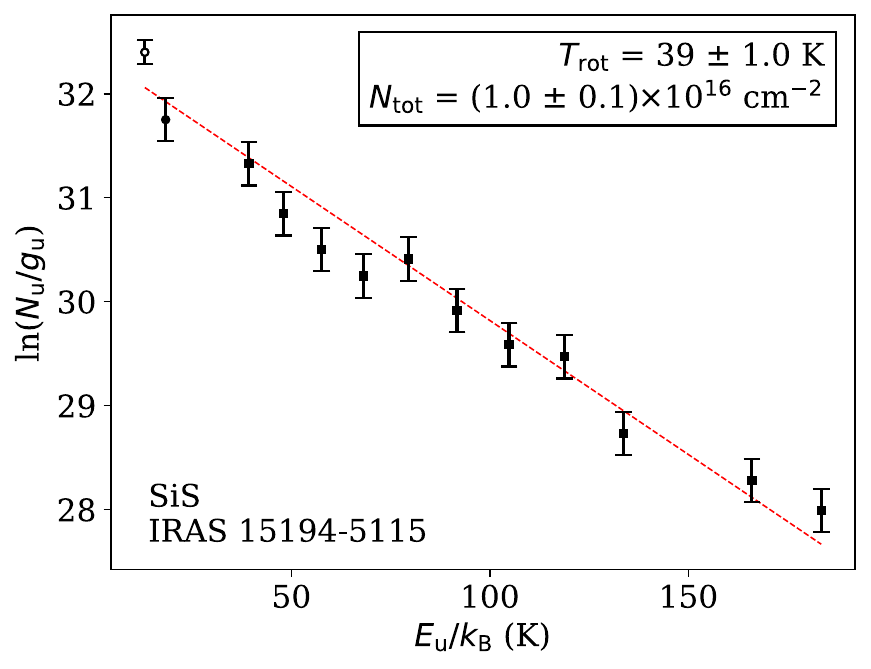}
      \label{subfig:SiS_rot_diag_15194}
   \end{subfigure}
   \begin{subfigure}[b]{0.33\textwidth}
      \centering
      \includegraphics[width=\textwidth]{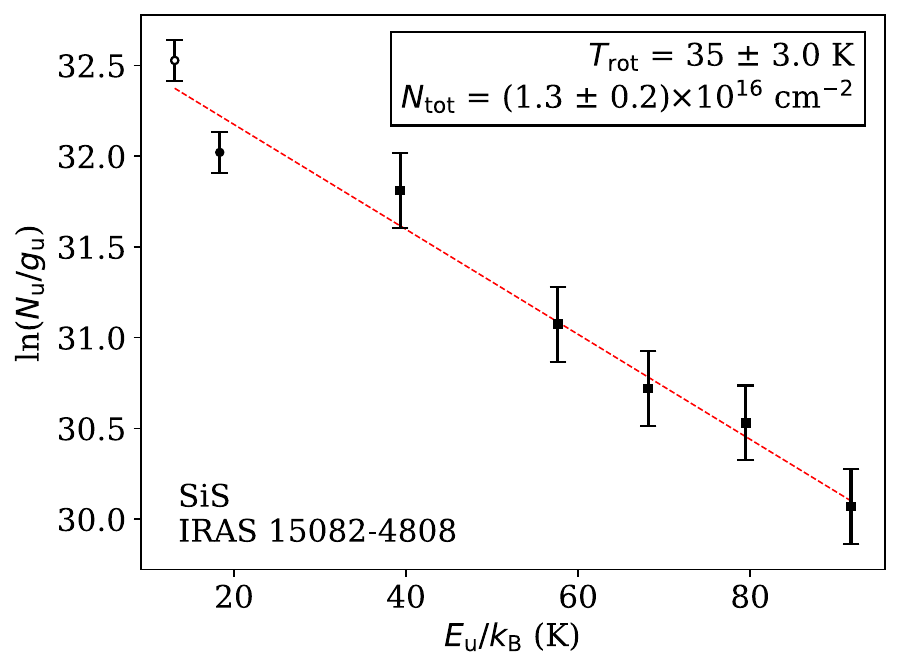}
      \label{subfig:SiS_rot_diag_15082}
   \end{subfigure}
   \begin{subfigure}[b]{0.33\textwidth}
      \centering
      \includegraphics[width=\textwidth]{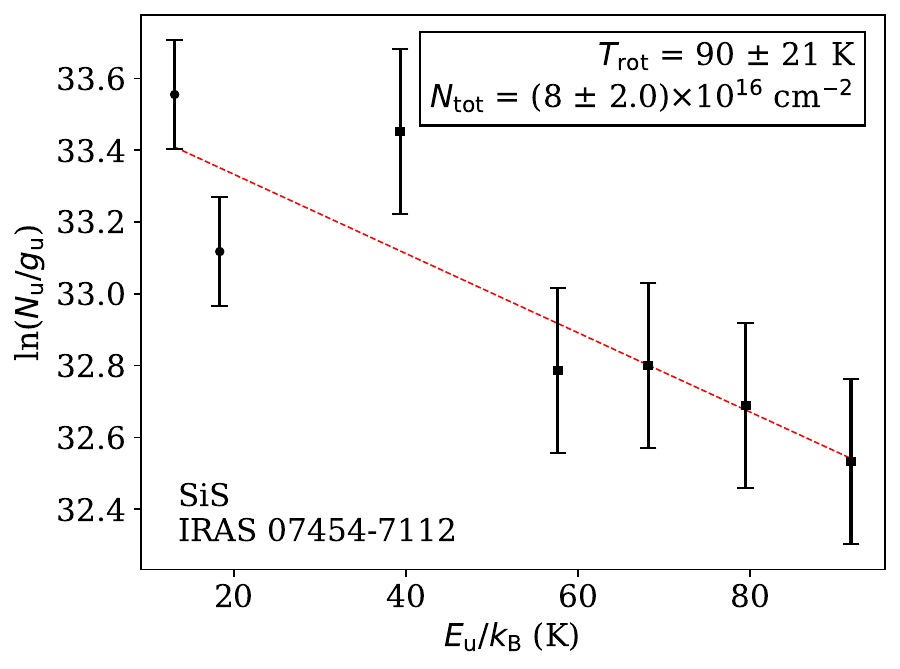}
      \label{subfig:SiS_rot_diag_07454}
   \end{subfigure}
      \caption{Same as Fig.~\ref{fig:SiO_rot_diags_all_stars} for SiS.}
      \label{fig:SiS_rot_diags_all_stars}
\end{figure}

\begin{figure}[!h]
   \centering
   \hspace*{0.325\textwidth}
   \begin{subfigure}[b]{0.33\textwidth}
      \centering
      \includegraphics[width=\textwidth]{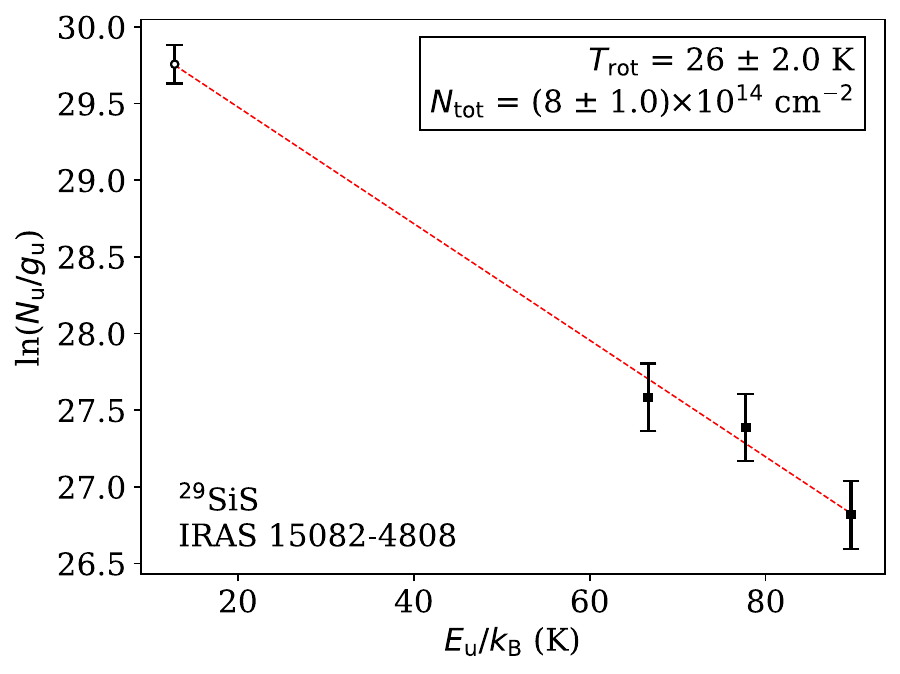}
      \label{subfig:29SiS_rot_diag_15082}
   \end{subfigure}
   \begin{subfigure}[b]{0.33\textwidth}
      \centering
      \includegraphics[width=\textwidth]{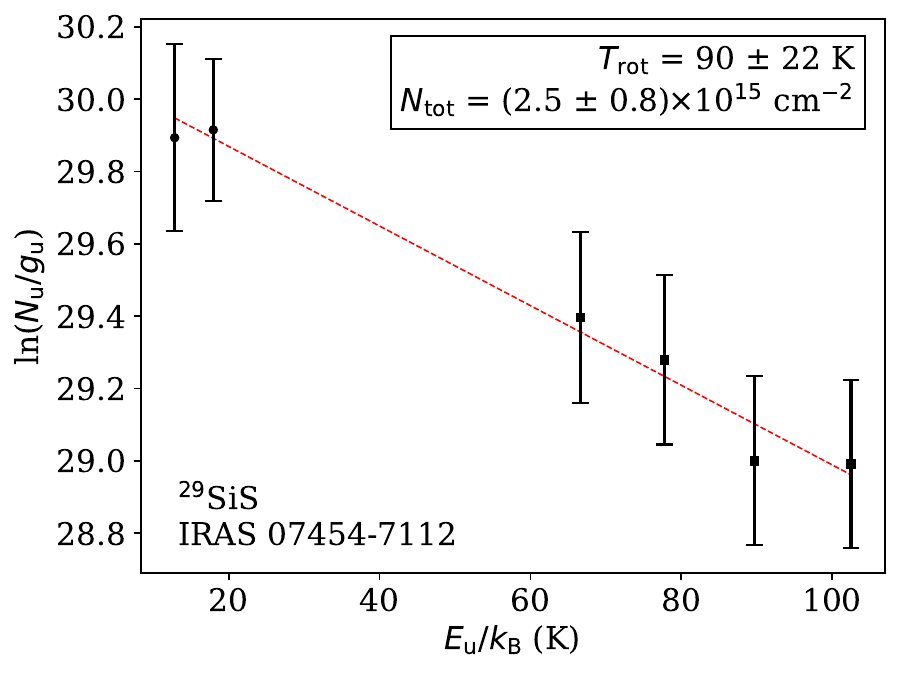}
      \label{subfig:29SiS_rot_diag_07454}
   \end{subfigure}
      \caption{Same as Fig.~\ref{fig:SiO_rot_diags_all_stars} for $^{29}$SiS.}
      \label{fig:29SiS_rot_diags_all_stars}
\end{figure}

\begin{figure}[!h]
   \centering
   \hspace*{-0.025\textwidth}
   \begin{subfigure}[b]{0.33\textwidth}
      \centering
      \includegraphics[width=\textwidth]{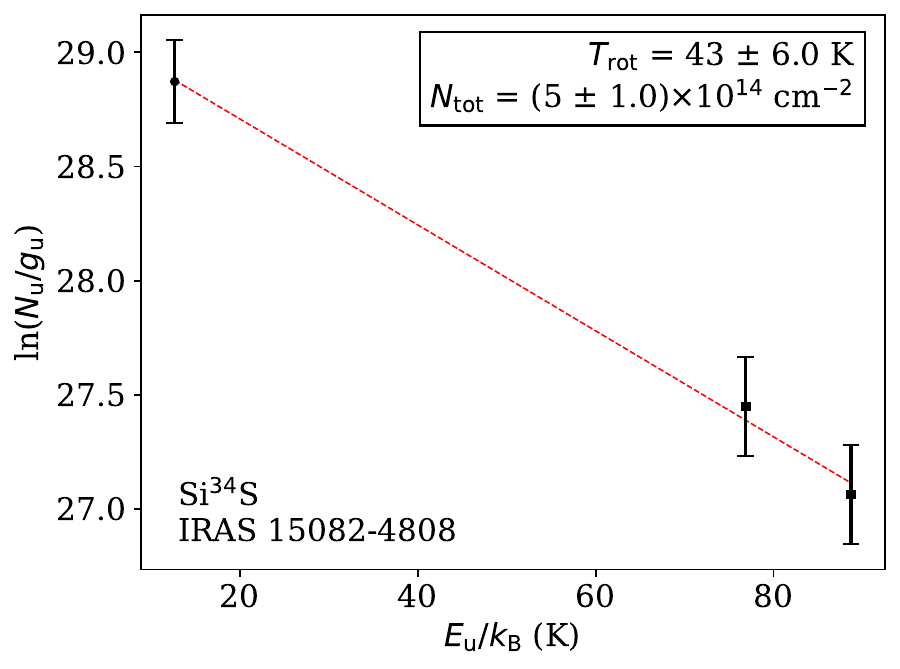}
      \label{subfig:Si34S_rot_diag_15082}
   \end{subfigure}
   \caption{Same as Fig.~\ref{fig:SiO_rot_diags_all_stars} for Si$^{34}$S.}
   \label{fig:Si34S_rot_diags_all_stars}
\end{figure}

\begin{figure}[!h]
   \centering
   \begin{subfigure}[b]{0.33\textwidth}
      \centering
      \includegraphics[width=\textwidth]{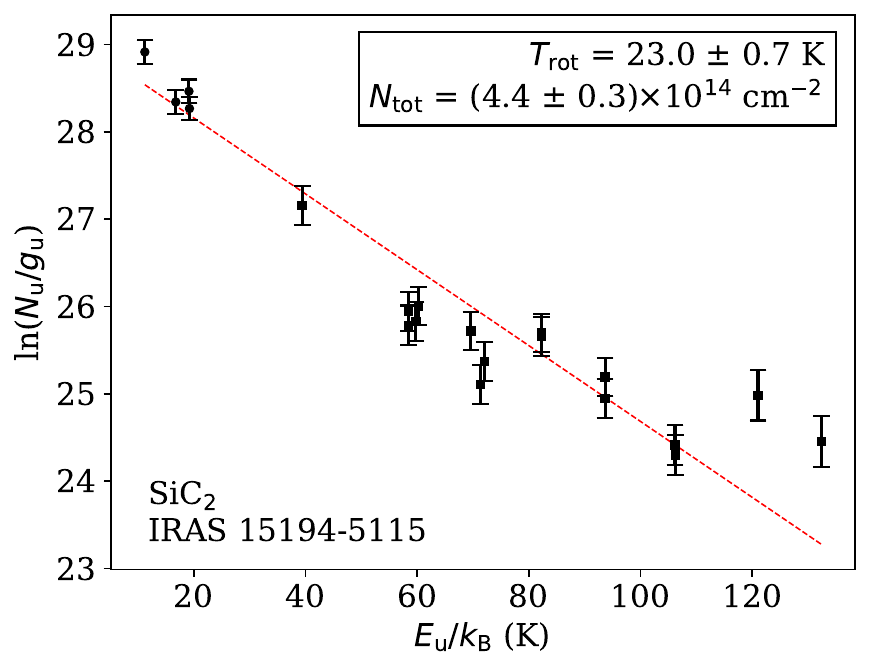}
      \label{subfig:SiC$_2$_rot_diag_15194}
   \end{subfigure}
   \begin{subfigure}[b]{0.33\textwidth}
      \centering
      \includegraphics[width=\textwidth]{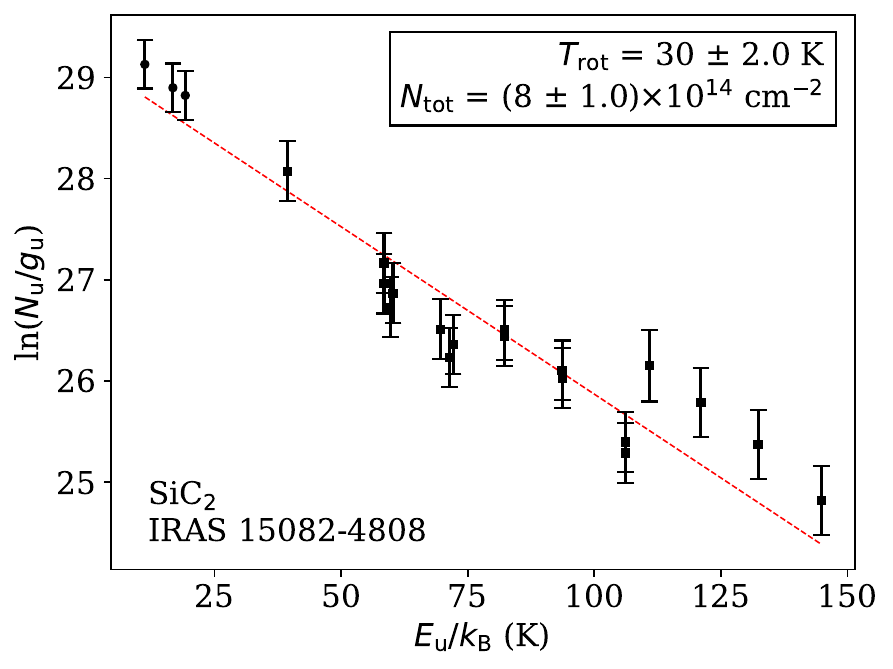}
      \label{subfig:SiC$_2$_rot_diag_15082}
   \end{subfigure}
   \begin{subfigure}[b]{0.33\textwidth}
      \centering
      \includegraphics[width=\textwidth]{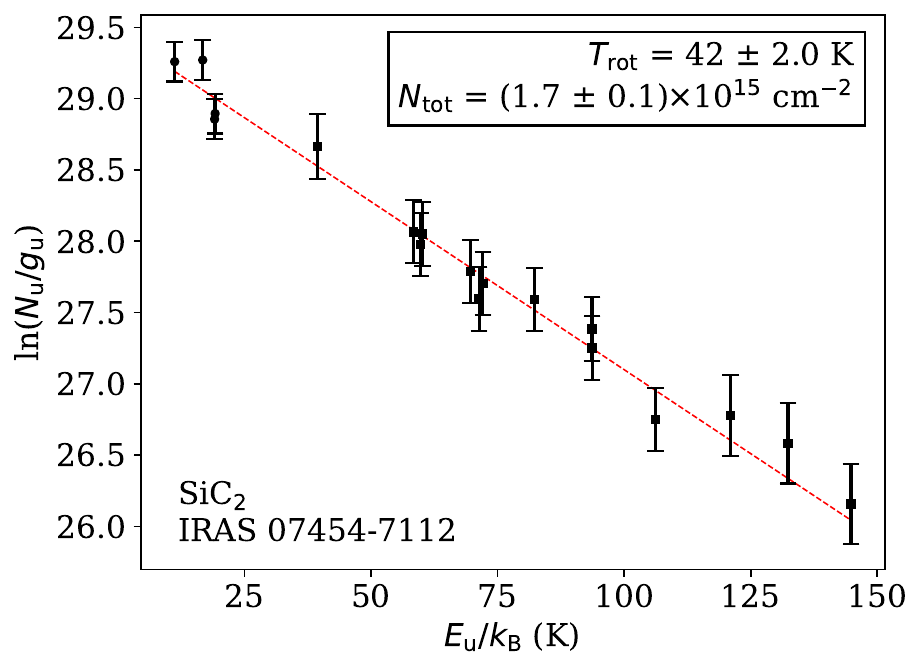}
      \label{subfig:SiC$_2$_rot_diag_07454}
   \end{subfigure}
      \caption{Same as Fig.~\ref{fig:SiO_rot_diags_all_stars} for SiC$_2$.}
      \label{fig:SiC$_2$_rot_diags_all_stars}
\end{figure}

\begin{figure}[!h]
   \centering
   \hspace*{-0.33\textwidth}
   \begin{subfigure}[b]{0.33\textwidth}
      \centering
      \includegraphics[width=\textwidth]{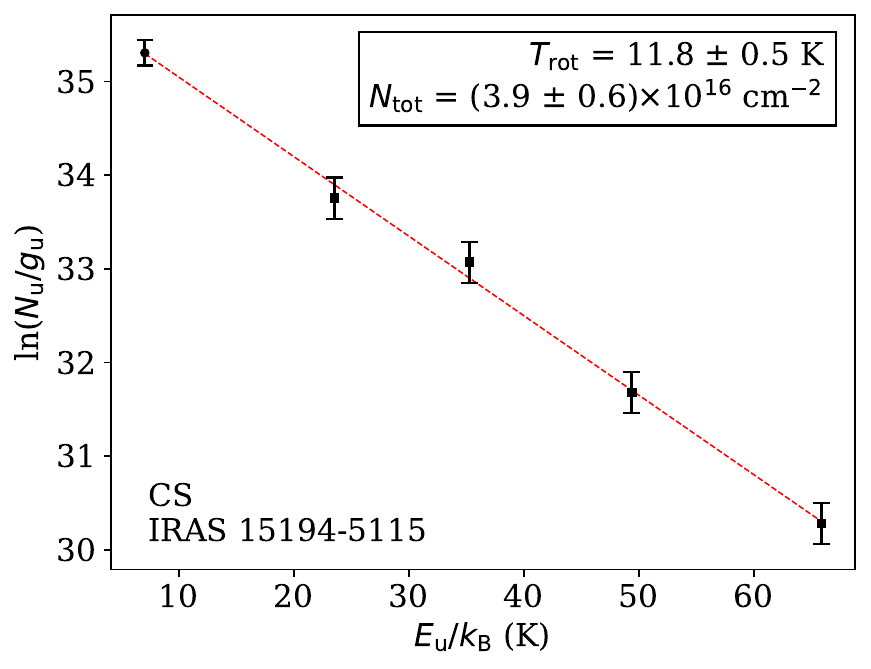}
      \label{subfig:CS_rot_diag_15194}
   \end{subfigure}
   \begin{subfigure}[b]{0.33\textwidth}
      \centering
      \includegraphics[width=\textwidth]{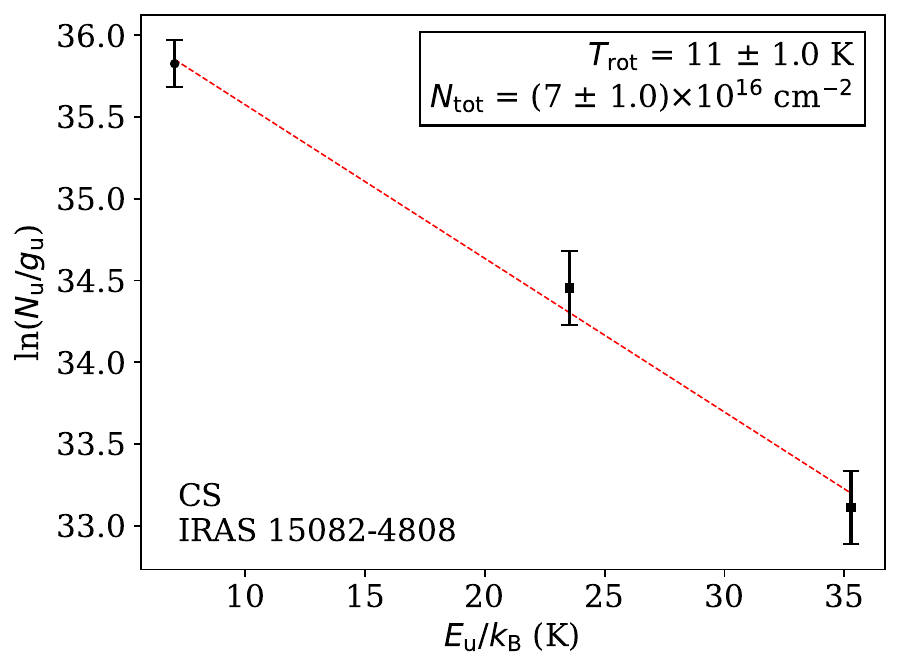}
      \label{subfig:CS_rot_diag_15082}
   \end{subfigure}
      \caption{Same as Fig.~\ref{fig:SiO_rot_diags_all_stars} for CS.}
      \label{fig:CS_rot_diags_all_stars}
\end{figure}

\begin{figure}[!h]
   \centering
   \hspace*{-0.69\textwidth}
   \begin{subfigure}[b]{0.33\textwidth}
      \centering
      \includegraphics[width=\textwidth]{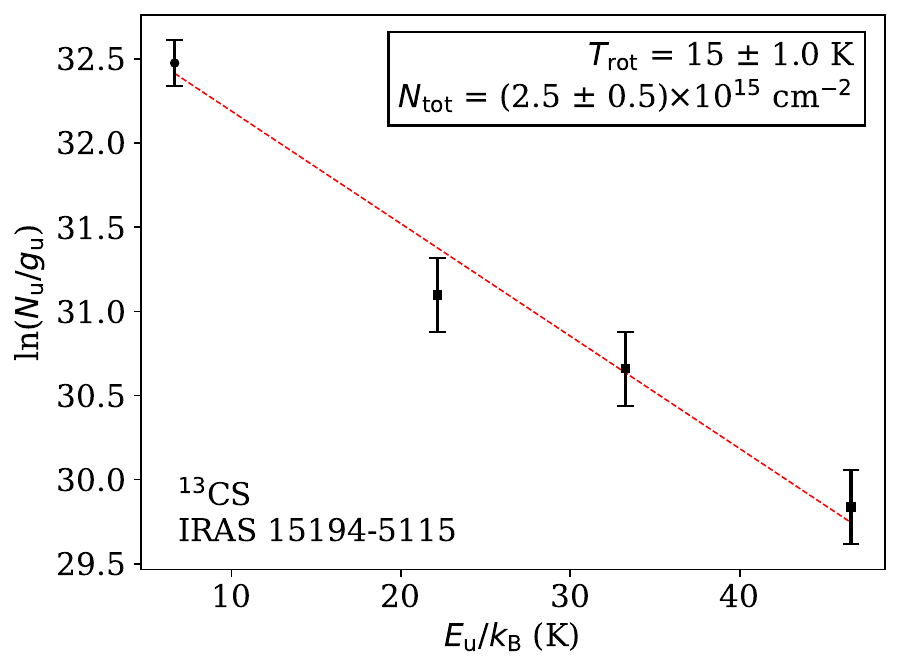}
      \label{subfig:13CS_rot_diag_15194}
   \end{subfigure}
   \caption{Same as Fig.~\ref{fig:SiO_rot_diags_all_stars} for $^{13}$CS.}
   \label{fig:13CS_rot_diags_all_stars}
\end{figure}

\begin{figure}[!h]
   \centering
   \begin{subfigure}[b]{0.33\textwidth}
      \centering
      \includegraphics[width=\textwidth]{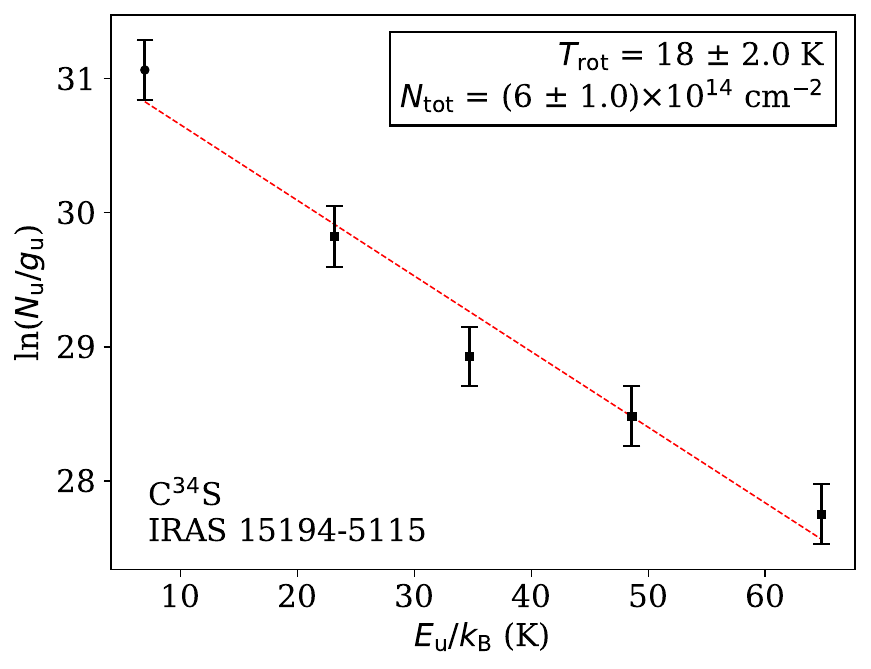}
      \label{subfig:C34S_rot_diag_15194}
   \end{subfigure}
   \begin{subfigure}[b]{0.33\textwidth}
      \centering
      \includegraphics[width=\textwidth]{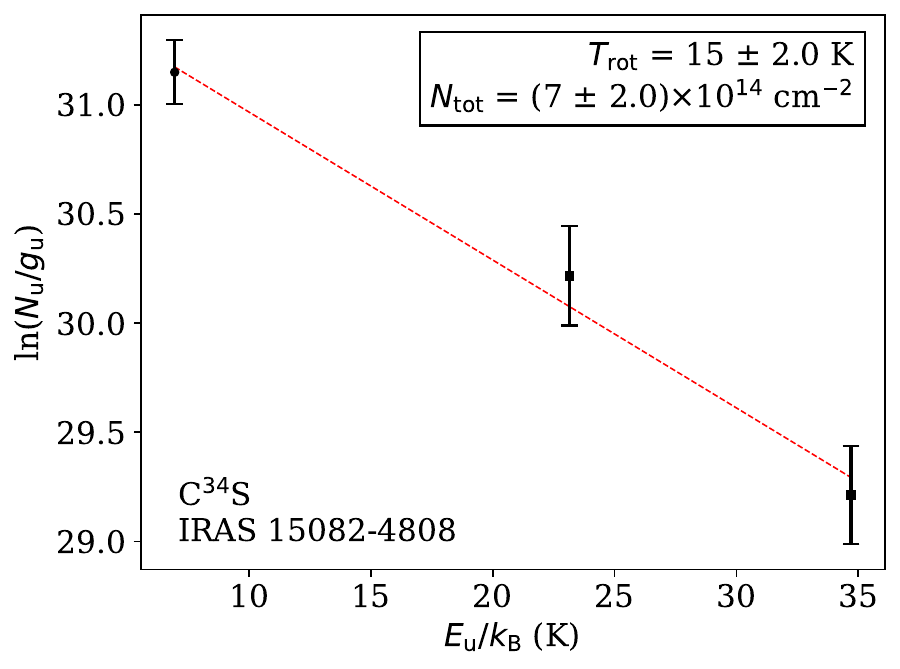}
      \label{subfig:C34S_rot_diag_15082}
   \end{subfigure}
   \begin{subfigure}[b]{0.33\textwidth}
      \centering
      \includegraphics[width=\textwidth]{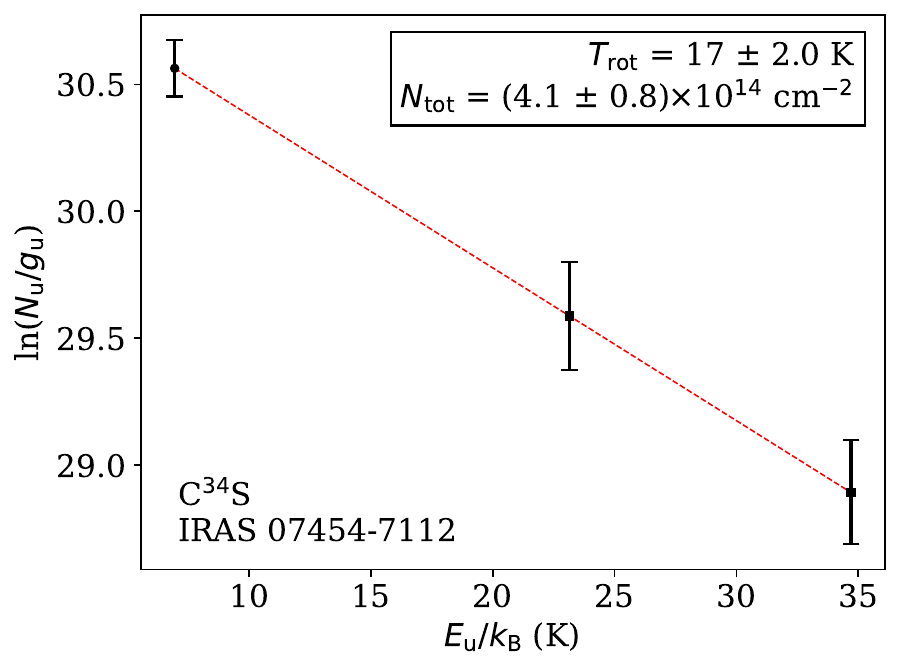}
      \label{subfig:C34S_rot_diag_07454}
   \end{subfigure}
      \caption{Same as Fig.~\ref{fig:SiO_rot_diags_all_stars} for C$^{34}$S.}
      \label{fig:C34S_rot_diags_all_stars}
\end{figure}

\begin{figure}[!h]
   \centering
   \begin{subfigure}[b]{0.33\textwidth}
      \centering
      \includegraphics[width=\textwidth]{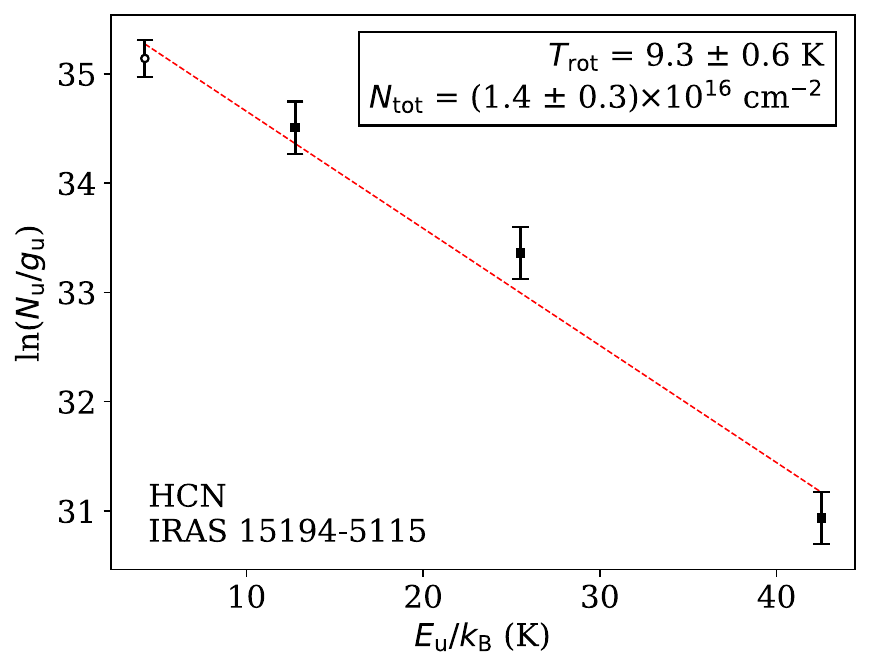}
      \label{subfig:HCN_rot_diag_15194}
   \end{subfigure}
   \begin{subfigure}[b]{0.33\textwidth}
      \centering
      \includegraphics[width=\textwidth]{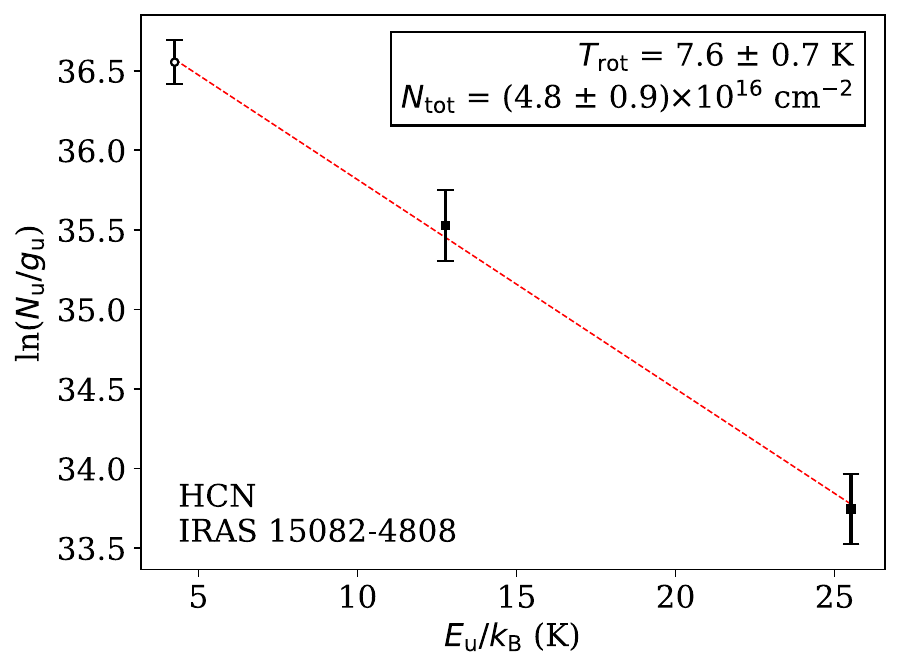}
      \label{subfig:HCN_rot_diag_15082}
   \end{subfigure}
   \begin{subfigure}[b]{0.33\textwidth}
      \centering
      \includegraphics[width=\textwidth]{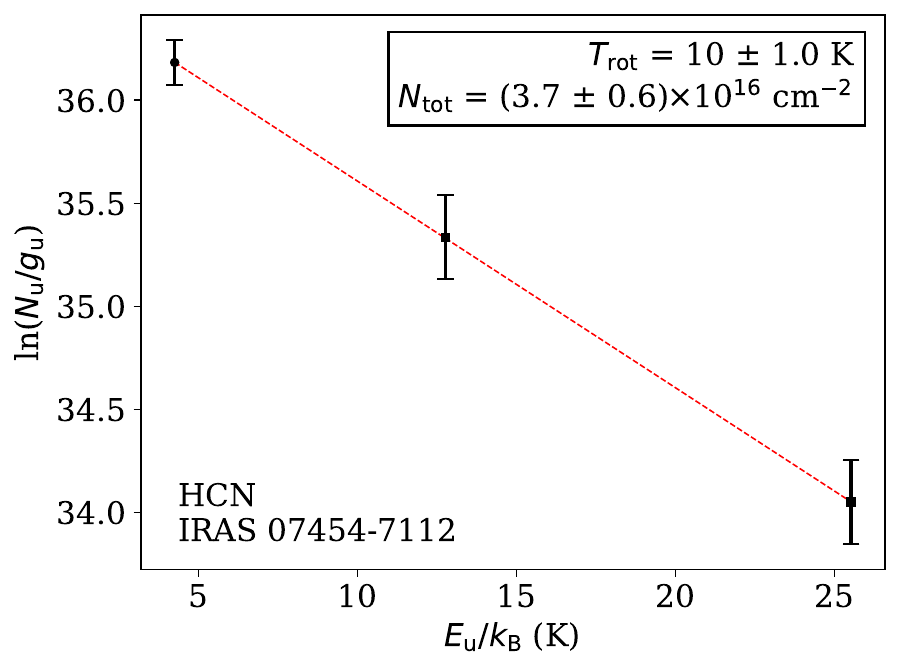}
      \label{subfig:HCN_rot_diag_07454}
   \end{subfigure}
      \caption{Same as Fig.~\ref{fig:SiO_rot_diags_all_stars} for HCN.}
      \label{fig:HCN_rot_diags_all_stars}
\end{figure}

\begin{figure}[!h]
   \centering
   \begin{subfigure}[b]{0.33\textwidth}
      \centering
      \includegraphics[width=\textwidth]{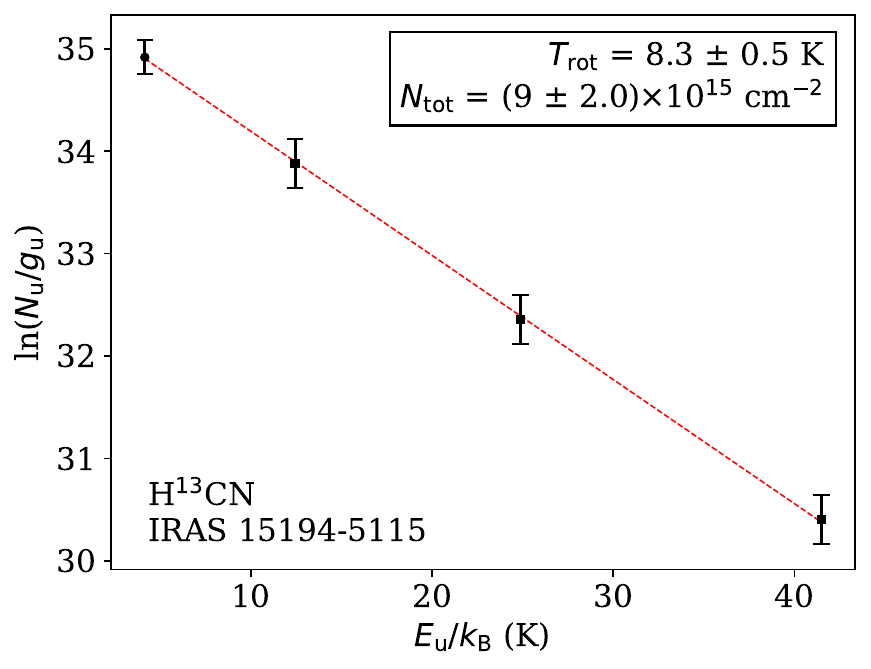}
      \label{subfig:H13CN_rot_diag_15194}
   \end{subfigure}
   \begin{subfigure}[b]{0.33\textwidth}
      \centering
      \includegraphics[width=\textwidth]{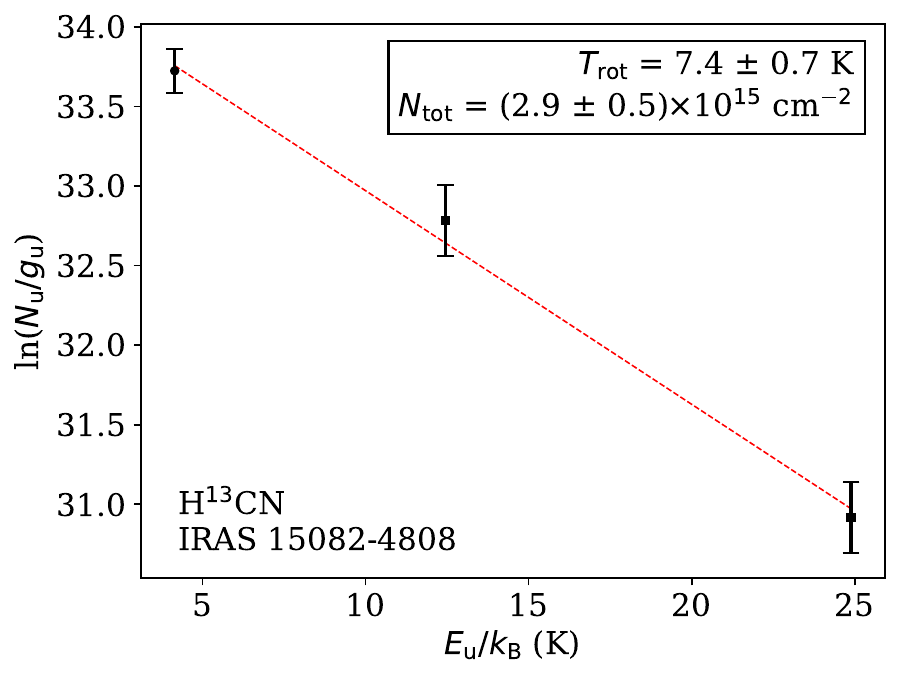}
      \label{subfig:H13CN_rot_diag_15082}
   \end{subfigure}
   \begin{subfigure}[b]{0.33\textwidth}
      \centering
      \includegraphics[width=\textwidth]{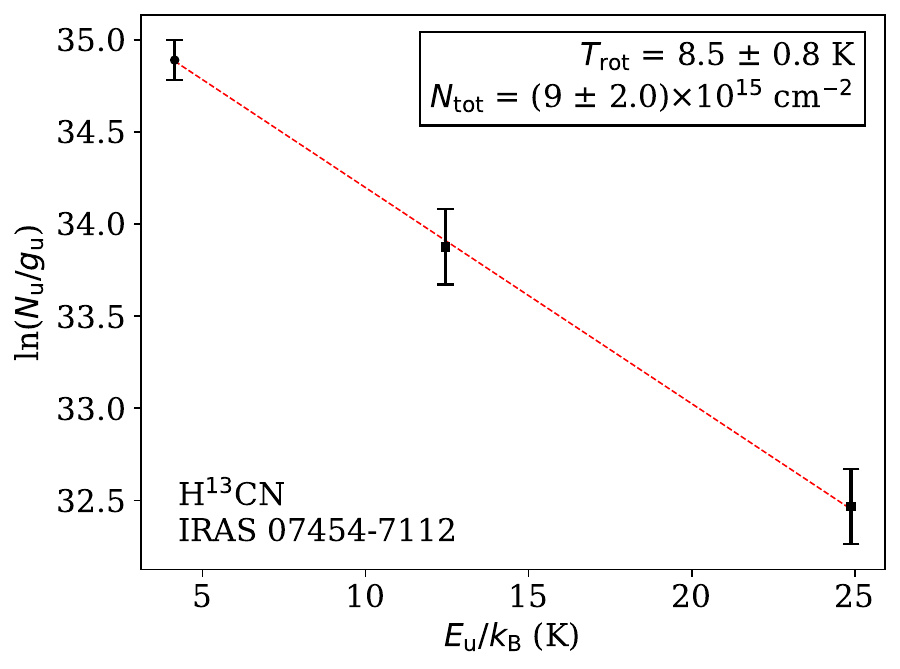}
      \label{subfig:H13CN_rot_diag_07454}
   \end{subfigure}
      \caption{Same as Fig.~\ref{fig:SiO_rot_diags_all_stars} for H$^{13}$CN.}
      \label{fig:H13CN_rot_diags_all_stars}
\end{figure}

\begin{figure}[!h]
   \centering
   \hspace*{-0.69\textwidth}
   \begin{subfigure}[b]{0.33\textwidth}
      \centering
      \includegraphics[width=\textwidth]{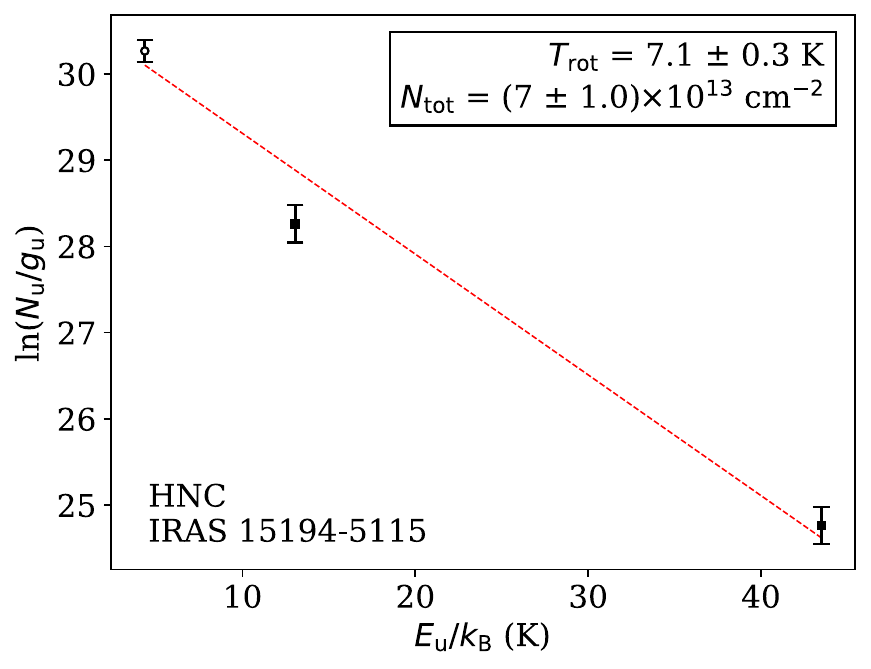}
      \label{subfig:HNC_rot_diag_15194}
   \end{subfigure}
   \caption{Same as Fig.~\ref{fig:SiO_rot_diags_all_stars} for HNC.}
   \label{fig:HNC_rot_diags_all_stars}
\end{figure}

\begin{figure}[!h]
   \centering
   \begin{subfigure}[b]{0.33\textwidth}
      \centering
      \includegraphics[width=\textwidth]{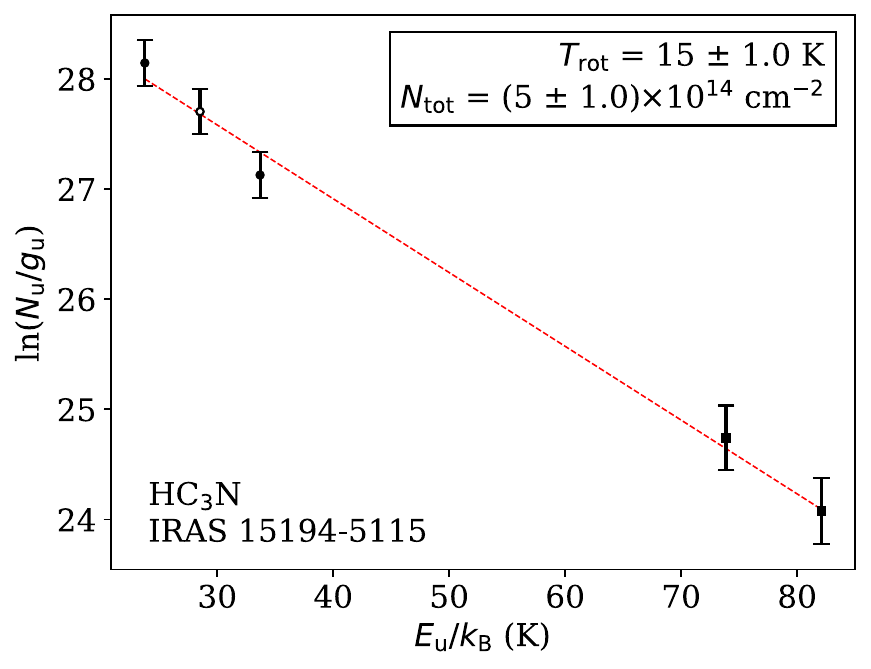}
      \label{subfig:HC$_3$N_rot_diag_15194}
   \end{subfigure}
   \begin{subfigure}[b]{0.33\textwidth}
      \centering
      \includegraphics[width=\textwidth]{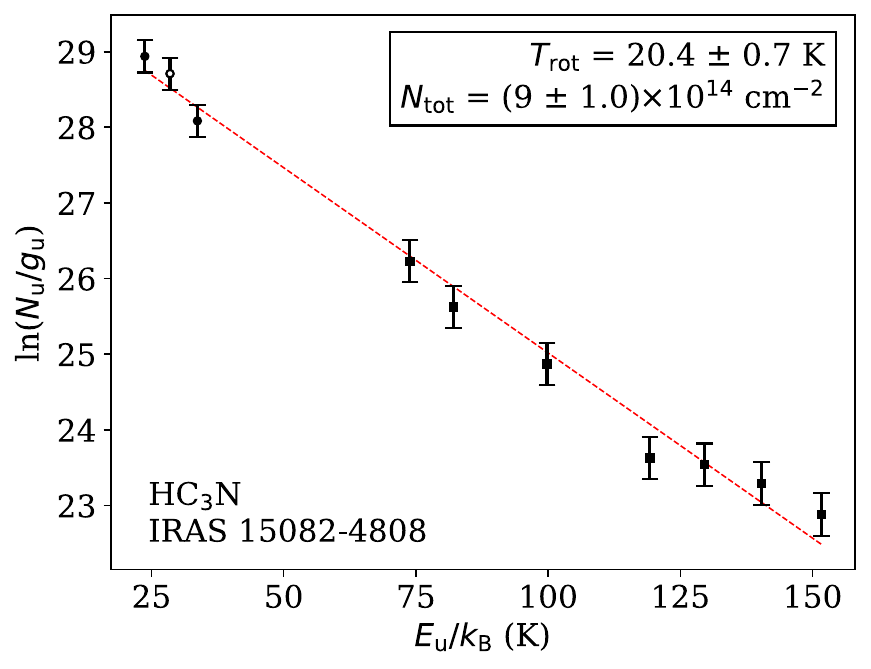}
      \label{subfig:HC$_3$N_rot_diag_15082}
   \end{subfigure}
   \begin{subfigure}[b]{0.33\textwidth}
      \centering
      \includegraphics[width=\textwidth]{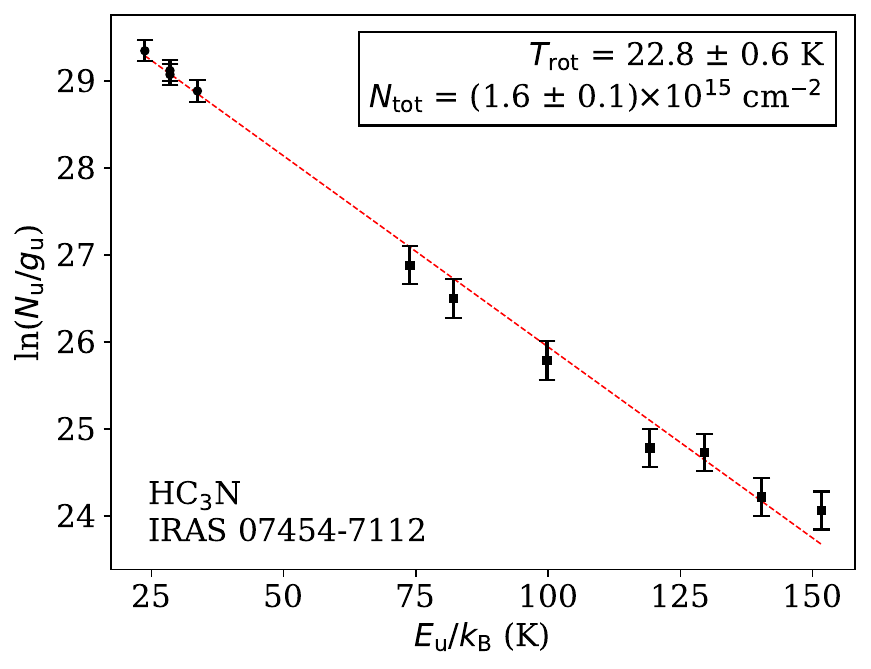}
      \label{subfig:HC$_3$N_rot_diag_07454}
   \end{subfigure}
      \caption{Same as Fig.~\ref{fig:SiO_rot_diags_all_stars} for HC$_3$N.}
      \label{fig:HC$_3$N_rot_diags_all_stars}
\end{figure}

\begin{figure}[!h]
   \centering
   \hspace*{0.33\textwidth}
   \begin{subfigure}[b]{0.33\textwidth}
      \centering
      \includegraphics[width=\textwidth]{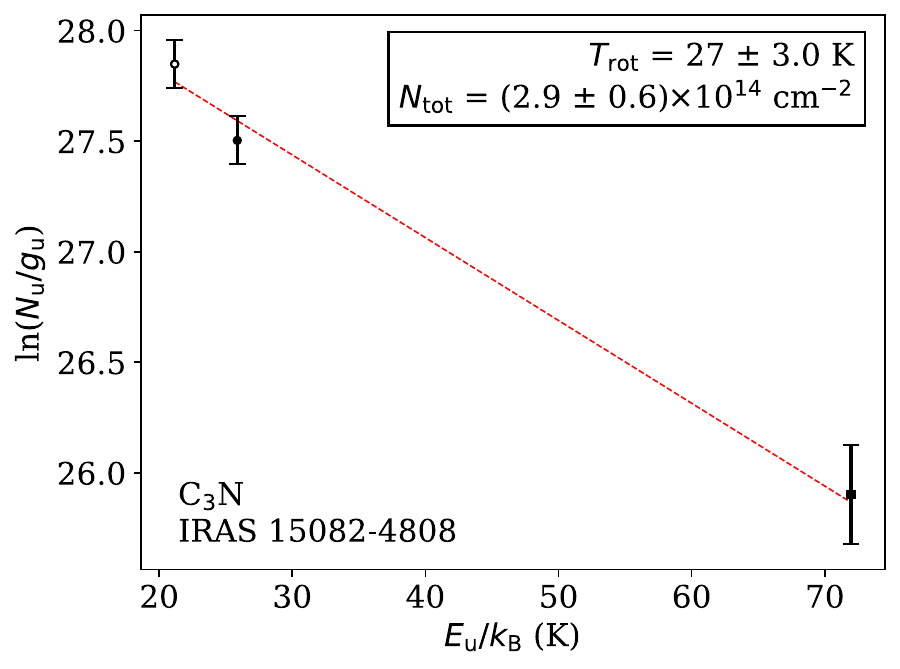}
      \label{subfig:C$_3$N_rot_diag_15082}
   \end{subfigure}
   \begin{subfigure}[b]{0.33\textwidth}
      \centering
      \includegraphics[width=\textwidth]{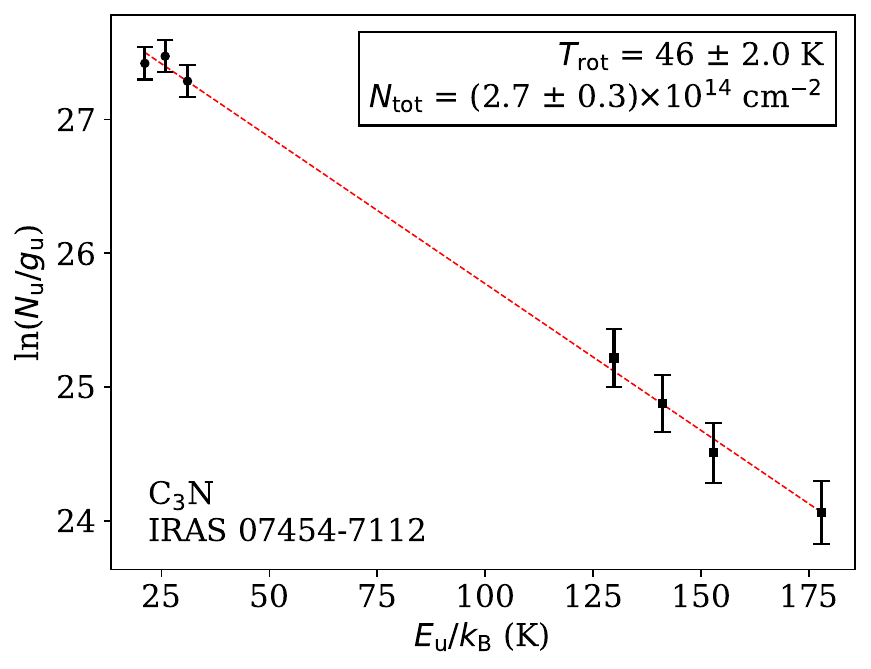}
      \label{subfig:C$_3$N_rot_diag_07454}
   \end{subfigure}
      \caption{Same as Fig.~\ref{fig:SiO_rot_diags_all_stars} for C$_3$N.}
      \label{fig:C$_3$N_rot_diags_all_stars}
\end{figure}

\begin{figure}[!h]
   \centering
   \hspace*{0.33\textwidth}
   \begin{subfigure}[b]{0.33\textwidth}
      \centering
      \includegraphics[width=\textwidth]{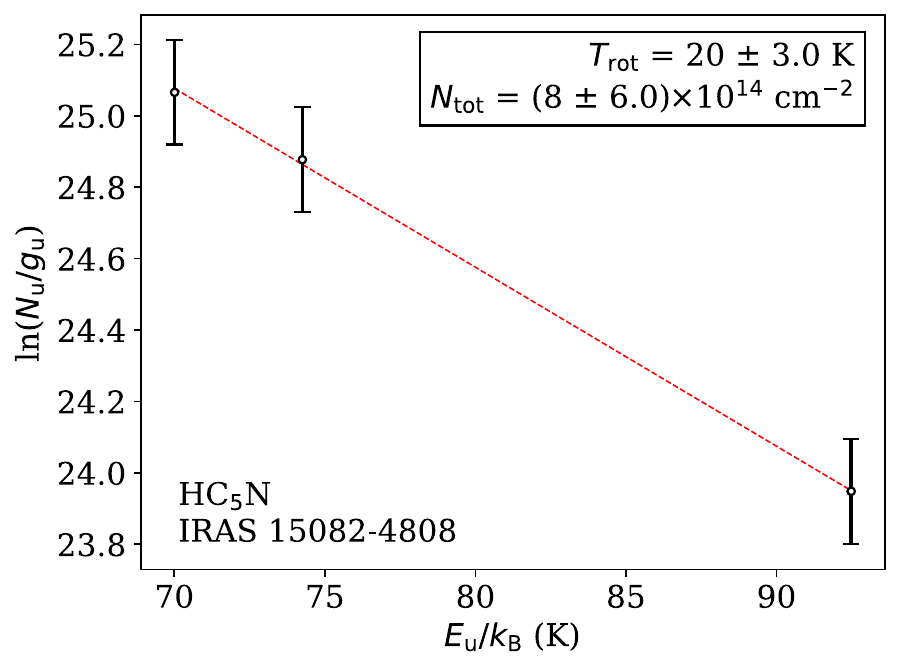}
      \label{subfig:HC$_5$N_rot_diag_15082}
   \end{subfigure}
   \begin{subfigure}[b]{0.33\textwidth}
      \centering
      \includegraphics[width=\textwidth]{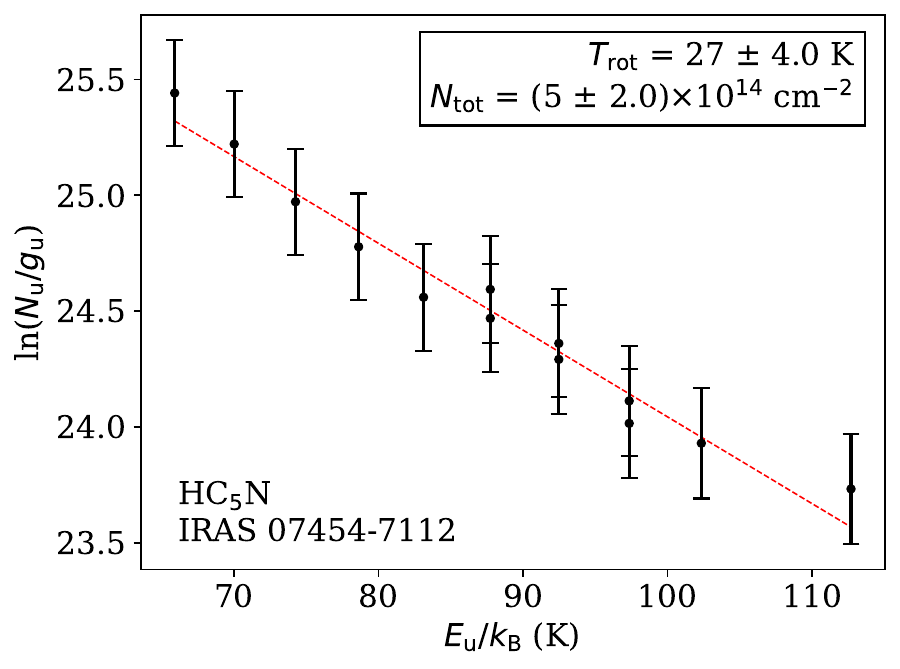}
      \label{subfig:HC$_5$N_rot_diag_07454}
   \end{subfigure}
      \caption{Same as Fig.~\ref{fig:SiO_rot_diags_all_stars} for HC$_5$N.}
      \label{fig:HC$_5$N_rot_diags_all_stars}
\end{figure}

\begin{figure}[!h]
   \centering
   \hspace*{-0.33\textwidth}
   \begin{subfigure}[b]{0.33\textwidth}
      \centering
      \includegraphics[width=\textwidth]{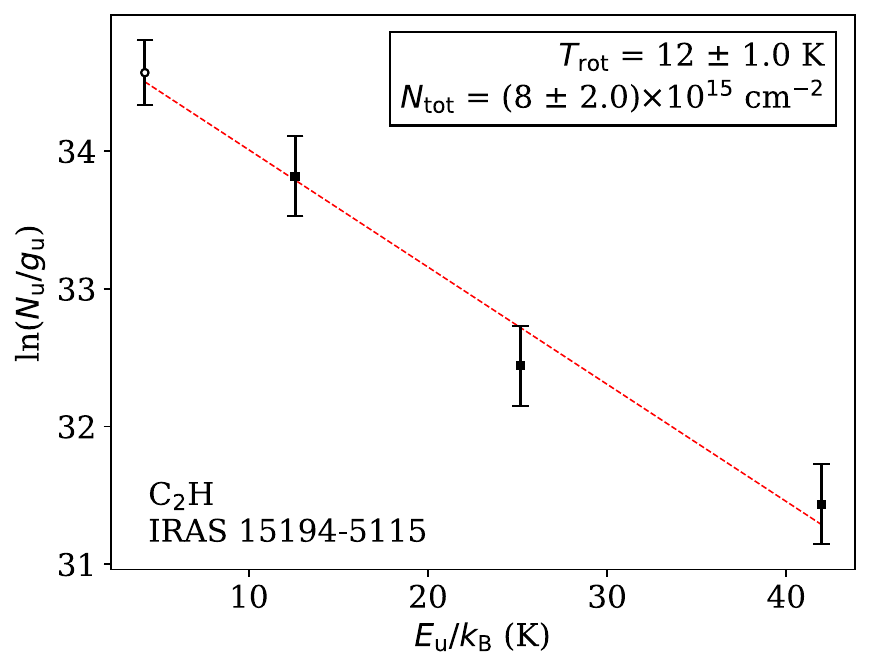}
      \label{subfig:C$_2$H_rot_diag_15194}
   \end{subfigure}
   \begin{subfigure}[b]{0.33\textwidth}
      \centering
      \includegraphics[width=\textwidth]{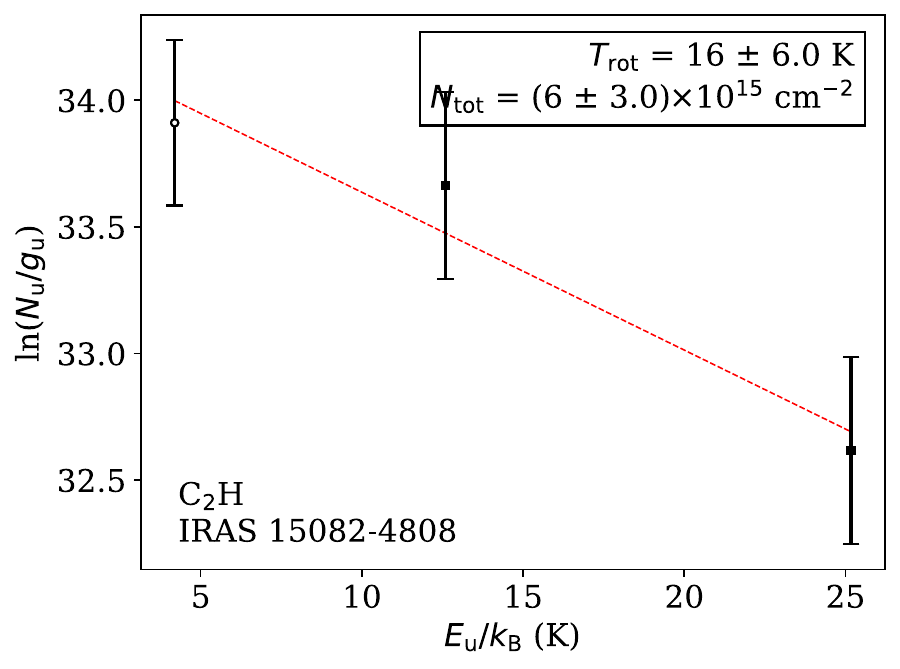}
      \label{subfig:C$_2$H_rot_diag_15082}
   \end{subfigure}
      \caption{Same as Fig.~\ref{fig:SiO_rot_diags_all_stars} for C$_2$H.}
      \label{fig:C$_2$H_rot_diags_all_stars}
\end{figure}

\begin{figure}[!h]
   \centering
   \begin{subfigure}[b]{0.33\textwidth}
      \centering
      \includegraphics[width=\textwidth]{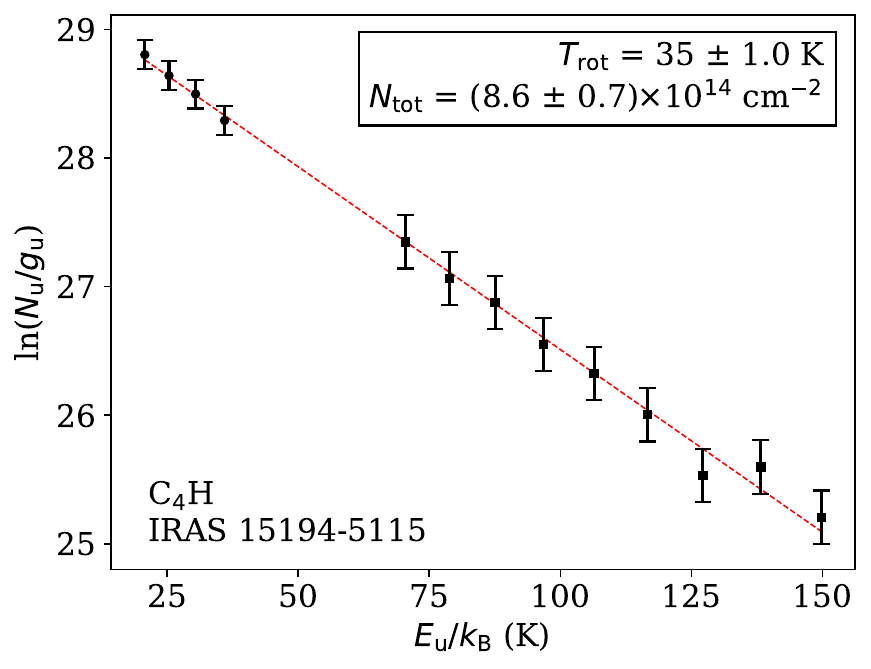}
      \label{subfig:C$_4$H_rot_diag_15194}
   \end{subfigure}
   \begin{subfigure}[b]{0.33\textwidth}
      \centering
      \includegraphics[width=\textwidth]{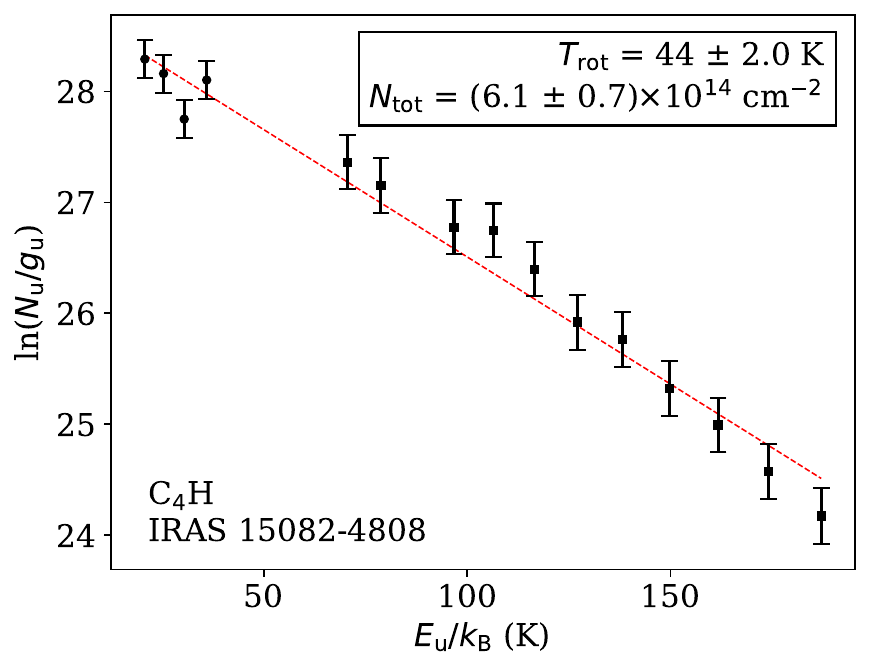}
      \label{subfig:C$_4$H_rot_diag_15082}
   \end{subfigure}
   \begin{subfigure}[b]{0.33\textwidth}
      \centering
      \includegraphics[width=\textwidth]{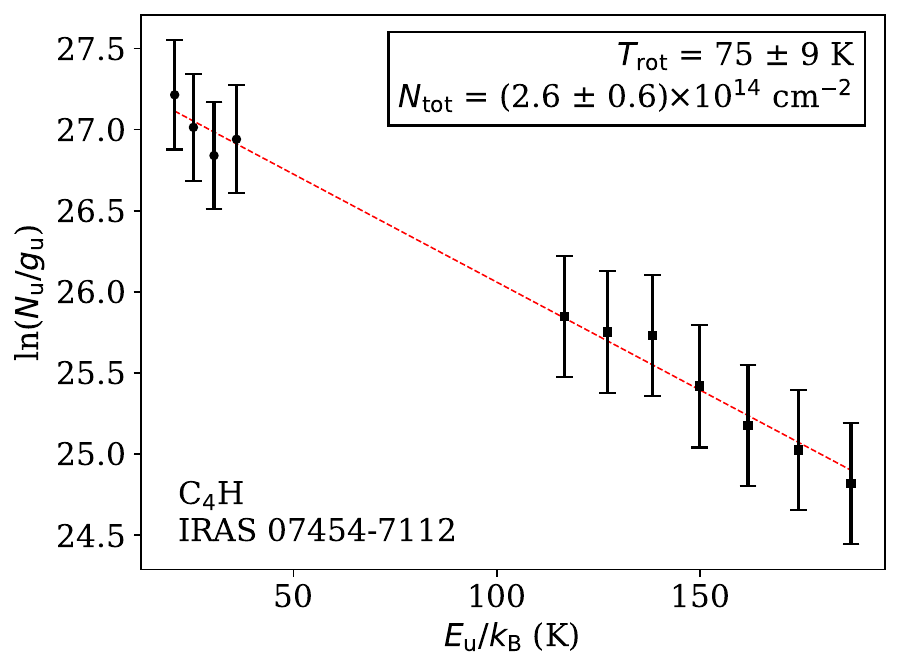}
      \label{subfig:C$_4$H_rot_diag_07454}
   \end{subfigure}
      \caption{Same as Fig.~\ref{fig:SiO_rot_diags_all_stars} for C$_4$H.}
      \label{fig:C$_4$H_rot_diags_all_stars}
\end{figure}

{Using this method, we first obtained estimates of the rotational temperature and total column density for each species without an optical depth correction (C$_{\tau_\nu}$ = 1). From these values, we then calculated a preliminary estimate of the optical depth for each individual line of the species}, using the standard equation
\begin{equation}
   \tau{_\nu} = \frac{c^3g_uA_{ul}}{8\pi\nu^3}\frac{N_{tot}}{\Delta \varv Z(T)}e^{-E_u/k_\mathrm{B}T}(e^{h\nu/k_\mathrm{B}T}-1)
   \label{eqn:optical_depths}
\end{equation}
where $\Delta \varv$ is the full velocity width of the line. {The range of optical depths thus calculated for each species for the three stars are given in Table~\ref{tab:max_optical_depths}.

Employing the correction factors (see Eq.~\ref{eqn:optical_depth_correction_factor}) obtained from the calculated optical depths, we then produced optical-depth-corrected population diagrams, from which optical-depth-corrected rotation temperatures and total column densities were estimated. These values are presented in Table~\ref{tab:N_and_T}. The optical-depth-corrected population diagrams are shown in Figs.~\ref{fig:SiO_rot_diags_all_stars} - \ref{fig:C$_4$H_rot_diags_all_stars}}.

\begin{table*}
   \caption{{Range of optical depths per molecule}}
   \label{tab:max_optical_depths}
   \centering
      \begin{tabular}{lrrr}
    \hline\hline & \\[-1ex]
    Molecule & \multicolumn{3}{c}{IRAS} \\
    \cline{2-4}& \\[-1ex]
     & 15194$-$5115 & 15082$-$4808 & 07454$-$7112 \\
    \hline& \\[-1ex]
    HCN&3.33 - 9.95&12.33 - 21.68&11.55 - 22.44 \\
    H$^{13}$CN&2.34 - 8.26&2.94 - 5.18&6.28 - 11.64 \\
    \hline& \\[-1ex]
    CS&1.16 - 5.85&4.98 - 7.92&3.49 - 4.09 \\
    $^{13}$CS&0.56 - 1.08&--&-- \\
    C$^{34}$S&0.01 - 0.27&0.28 - 0.43&0.20 - 0.35 \\
    \hline& \\[-1ex]
    SiO&0.80 - 4.29&1.98 - 4.23&3.10 - 5.65 \\
    $^{29}$SiO&0.07 - 0.47&0.17 - 0.30&0.25 - 0.49 \\
    $^{30}$SiO&0.04 - 0.06&--&-- \\
    \hline& \\[-1ex]
    C$_2$H&0.51 - 1.21&0.33 - 0.88&0.05 - 0.32 \\
    C$_4$H&0.01 - 0.05&0.01 - 0.03&$\sim$0.01 \\
    \hline& \\[-1ex]
    SiS&0.09 - 0.58&0.49 - 0.87&1.03 - 2.42 \\
    $^{29}$SiS&--&0.03 - 0.07&0.04 - 0.12 \\
    $^{30}$SiS&--&--&0.03 - 0.09 \\
    Si$^{34}$S&--&0.02 - 0.03&0.02 - 0.08 \\
    \hline& \\[-1ex]
    HNC&0.05 - 0.39&--&-- \\
    \hline& \\[-1ex]
    HC$_3$N&0.02 - 0.18&0.01 - 0.28&0.02 - 0.54 \\
    C$_3$N&0.05 - 0.1&0.03 - 0.05&0.01 - 0.04 \\
    HC$_5$N&$\sim$0.01&0.01 - 0.04&0.01 - 0.05 \\
    \hline& \\[-1ex]
    SiC$_2$&0.01 - 0.04&0.01 - 0.03&0.01 - 0.04 \\
    \hline& \\[-1ex]
      \end{tabular}
      \tablefoot{{For each molecule per star, the table lists the optical depth values of the lines with the highest and lowest calculated optical depth}.}
\end{table*}

We note that along with the lines from the ALMA data, we have also included lines detected in our APEX surveys (Table~\ref{tab:apex_line_detections} and Figs.~\ref{fig:APEX_lines_start}$-$\ref{fig:APEX_B7_lines_15194}) in the population diagrams, so that a wider range of energy levels can be sampled.

\subsection{{Abundance calculation from column densities}}
\label{subsubsec:Col_dens_abund_calcs}
{Using the estimates of the radial extents of the emitting regions of different species obtained from the Gaussian fits to the AARPs (see Sect.~\ref{subsubsec:AARPs}), we can estimate the molecular fractional abundances (Table~\ref{tab:f_As}) from the column densities of the respective species given by the population diagrams, using the equation}
\begin{equation}
f_\mathrm{X} = \frac{2 \pi m_\mathrm{H} \varv_\mathrm{exp} N_\mathrm{tot,X}}{\dot{M}} \left(\frac{R_\mathrm{e}^2}{R_\mathrm{e} - R_\mathrm{i}}\right)
\label{eqn:frac_abund_from_N_tot}
\end{equation}
where $\dot{M}$ is the hydrogen mass-loss rate, $\varv_\mathrm{exp}$ the gas expansion velocity, $m_\mathrm{H}$ the mass of the hydrogen atom, and $R_\mathrm{i}$ and $R_\mathrm{e}$ the inner and outer radii of the emitting region, respectively \citep{Olofsson_2003}.

\subsection{{Abundance calculation from single lines}}
\label{subsubsec:Single_Line_Analysis}
Population diagrams cannot be used in the case of molecules for which only one or two lines are detected, or for which the detected transitions are very close together in upper-level energies. In such cases, we obtain a rough estimate of the fractional abundance, using the highest signal-to-noise line available, by assuming it to be constant in the radial range $R_\mathrm{i}$ to $R_\mathrm{e}$ and zero outside it. Assuming an $r^{-2}$ density distribution and a constant, assumed excitation temperature throughout the emitting region, we use the following formula from \citet[][in rewritten form]{Habing_and_Olofsson_2004}
\begin{equation}
   \label{eqn:frac_abund_analytical}
   f_\mathrm{X} = 1.3\times10^{-19} \frac{\varv_\mathrm{exp} D Z(T_\mathrm{ex})}{g_\mathrm{u} A_\mathrm{ul} \dot{M} (R_\mathrm{e} - R_\mathrm{i})} e^{E_\mathrm{l}/k_\mathrm{B} T_\mathrm{ex}} \int S_\mathrm{ul} d\varv
\end{equation}
where $\varv_\mathrm{exp}$ is given in km/s, $\dot{M}$ in $M_\mathrm{\odot}$\,yr$^{-1}$, $D$ in kpc, $R_\mathrm{i}$ and $R_\mathrm{e}$ in cm, and $\int S_\mathrm{ul} d\varv$ in Jy\,km/s. The assumed $T_\mathrm{ex}$ values are 40\,K for centrally-peaked species and 20\,K for shell species (see Sect.~\ref{subsubsec:Results_of_T_rot_estimates} for a justification of these values). {We note that we have not performed optical depth corrections in the case of single-line abundance estimates, as the temperature we use for these calculations is an assumed value, and since the high uncertainties involved anyway deem the abundances thus calculated to be order of magnitude estimates.}

{The figures {below} show the population diagrams (see Sect.~\ref{subsubsec:Rot_Diag_Analysis}) of various molecular species (listed in Table~\ref{tab:N_and_T}) towards the three stars}.

\end{multicols}

\vfill

\vfill

\vfill

\vfill

\vfill

\vfill

\vfill

\vfill

\vfill

\vfill

\vfill

\vfill

\vfill

\vfill

\end{appendix}

\end{document}